\documentclass[final,3p,times,authoryear]{elsarticle}			

\usepackage{mnemonic} 		
\usepackage{subcaption}		
\usepackage{latexsym}          	
\usepackage{amssymb}	
\usepackage{amsmath}	

\usepackage[backref=page]{hyperref}	

\newcommand{\Sun}{\odot}
\newcommand{\Jupiter}{{\rm J}}
\newcommand{\Earth}{\oplus}
\newcommand{\ms}{\,m\,s$^{-1}$}
\newcommand{\kms}{\,km\,s$^{-1}$}
\newcommand{\kmskpc}{\,km\,s$^{-1}$\,kpc$^{-1}$}
\newcommand{\kmsmpc}{\,km\,s$^{-1}$\,Mpc$^{-1}$}

\newcommand{\masyr}{\,mas\,yr$^{-1}$}
\newcommand{\muas}{\,$\mu$as}
\newcommand{\muasyr}{\,$\mu$as\,yr$^{-1}$}
\newcommand{\ddeg}{\hbox{.\hskip-2pt $^\circ$}}   	
\newcommand{\micron}{\,$\mu$m}
\newcommand{\teff}{$T_{\rm eff}$}

\journal{Physics Reports}
\begin{document}
\begin{frontmatter}

\title{Space Astrometry with Gaia: Advances in Understanding our Galaxy}
\author{Michael Perryman}
\ead{mac.perryman@gmail.com}
\ead[url]{https://www.michaelperryman.co.uk}
	\affiliation{School of Physics; University College Dublin}

\begin{abstract}
Gaia is a satellite mission of the European Space Agency which is creating a catalogue of extremely accurate positions, distances and space motions of two billion stars in our Galaxy, along with more than one hundred thousand solar system asteroids, and several million distant quasars, all on the same extragalactic reference system. Complementary information on each object's multi-epoch photometry and spectra provides a vast and unprecedented data base of (model-dependent) fundamental physical quantities, such as each star's mass, age, and chemical composition. I outline the field's historical context, and explain the key principles involved in these space measurements. This is followed by a broad review of the many areas of solar system science, stellar structure and evolution, and topics in Galactic structure, evolution, and dynamics, that are being derived from these data.
\end{abstract}


\end{frontmatter}

\setcounter{tocdepth}{3}	
{\footnotesize
	\null\vspace{-12pt}	
\tableofcontents
}

\section{Introduction}

Gaia is a major scientific satellite developed and operated by the European Space Agency (ESA). It exploits the environmental advantages of space to measure the positions of a large number of stars and other celestial objects with unprecedented accuracy. With studies starting in the early 1990s (and overlapping with the final phases of the pioneering Hipparcos space mission), the project was accepted as part of ESA's scientific programme in 2000. The satellite was launched in December 2013, and operated continuously from its Lagrange Sun--Earth (L2) orbit location until January 2025. Through repeated scanning of the celestial sphere over more than 10~years, its state-of-the-art optical telescope and detector system allows construction of an extremely accurate three-dimensional map of the positions and space motions of some two~billion stars, and more than 100\,000 solar system objects, all referred to an extragalactic reference frame defined by several million quasars and compact galaxies. 

Despite the enormous distances in astronomy, around 1\,pc ($\sim$3~light-years) to the nearest star, and some 8\,kpc (30\,000 light-years) to the centre of the Galaxy, the satellite-based measurements are able to provide accurate distances and space motions of each object observed. Distances are derived from the conceptually straightforward process of triangulation as the Earth (and the Gaia satellite with it) moves in its annual orbit around the Sun. This results in a very small annual oscillation in the (angular) position of each star with respect to the background stars, with an amplitude inversely proportional to its distance. Space motions are derived from their (tiny) secular displacements over time. 

Accurate distances to this huge sample of stars throughout our Milky Way Galaxy permits the calibration of their intrinsic luminosities (and other fundamental properties), which in turn, through numerical models, allows profound insights into their internal structure, and their evolution over time. The accurate space motions allow a vast range of kinematic and dynamical studies of individual stars, as well as the collective stellar structures that populate our Galaxy (including star clusters, spiral arms, gravitational resonant motions, and ancient halo streams). These observations in turn provide a rather detailed chronology of how our Galaxy came into existence some 12--13~Gyr ago.

Complementing these positional measurements are simultaneous satellite-based multi-wavelength optical photometry of all sources, and spectroscopic measurements of some 100~million of the brightest. The multi-colour photometry allows for the detailed characterisation of each star's temperature, radius, mass, and chemical composition, while accurate photometric measurements at multiple epochs (of order 150 throughout the 10-year mission) allow characterisation of each star's variability, a property also offering profound insights into its physical nature. Spectroscopic measurements provide the star's `radial velocity' (along the line-of-sight to the star), as well as further information on its physical properties, and the degree of interstellar extinction along the sight-line to it.

Extensive processing of the satellite data is required to create the catalogues of positions and space motions, and their associated physical properties (temperature, surface gravity, etc.), in a form which can then be used for specific scientific analyses. The global data processing has been entrusted, by ESA, to a European consortium of individuals and institutes, the Gaia Data Processing and Analysis Consortium (DPAC). This data processing task can be considered as a vast iterative solution which takes the many petabytes of satellite data acquired over 10~years as input, and generates the desired physical quantities as output. To date, four main data releases have been made available, each completely superseding the previous: 
Data Release~1 (DR1) in 2016 (based on the first 14~months of satellite data); 
Data Release~2 (DR2) in 2018 (based on the first 22~months);
Early Data Release~3 (EDR3) in 2020 (based on the first 34~months);
and Data Release~3 in 2022 (covering the same data interval as DR3 but with much additional scientific content). 
Two further data releases are foreseen: DR4 in late 2026 which will be based on the first 5.5~years of mission data, and the final DR5 in around 2030 comprising all 10.5~years of mission data.

This article is primarily devoted to an overview of the wide range of scientific results from the mission to date, which are detailed in some 7000 refereed papers published since the first data release in 2016. 
Only one review of the entirety of Gaia's scientific highlights has been undertaken previously, by \citet{2021ARA&A..59...59B},	
which preceded the availability of Data Release~3. 
The aim of this present broad review is to provide an update of Gaia's main scientific discoveries in advance of Data Release~4, to show how the many facets of the Gaia mission interconnect, and to stimulate future exploitation.
The Astrophysics Data System, 
\href{https://ui.adsabs.harvard.edu/}{ADS},
yields almost 12\,000 papers mentioning `Gaia' in their abstract between 2016--2025, just over half of them refereed. It is clearly not possible or appropriate to attempt to include them all. I provide sufficient references to allow further investigations of each field, but remain conscious that there will be important papers or even fields of study that I have not included. 

For readers outside of astronomy, or unfamiliar with the basic concepts, Section~\ref{sec:background} provides a short history of this field of science, and the reasons for making these measurements from space.
The key measurement and operational principles of the Gaia satellite are summarised in Section~\ref{sec:gaia-satellite}. The data processing is outlined in Section~\ref{sec:data-processing}, viz.\ how distances, space motions, and information on multiple systems are derived, and how the photometric and spectroscopic data are used to estimate the star's physical properties (mass, radius, surface gravity, age, chemical abundances, interstellar extinction, etc.).
 
The main scientific review is divided into five sections:
Section~\ref{sec:physical-effects} collects a handful of `physical effects' which are relevant in a few special cases, but which do not sit naturally within the subsequent sections;
Section~\ref{sec:solar-system} details the various solar system studies;
Section~\ref{sec:stellar-structure} covers a wide range of results on stellar structure and evolution (including brown dwarfs and exoplanets);
Section~\ref{sec:galaxy-structure} details results related to Galaxy structure and evolution;
and 
Section~\ref{sec:local-group-cosmology} reviews the impact of Gaia on studies of the Local Group of galaxies, and cosmology.
It concludes with a short description of the future data releases, and plans for the future of space astrometry (Section~\ref{sec:future}).
%
\ref{sec:app-a} provides a list of some of the main acronyms used, and some of the specific notations and units used in this area of astronomy.
\ref{sec:app-b} provides some details of coordinate transformations, position propagation, and some Gaia-related software tools.

\paragraph{Nomenclature}
The vast majority of sources observed by Gaia are stars, and most of these are stars within our own Galaxy. Nonetheless, several million Gaia sources are stars in nearby galaxies in our Local Group, mainly the Large and Small Magellanic Clouds. In the latest data release, DR3, amongst almost 2~billion sources with astrometry, ten million sources are distant compact galaxies, and the more extreme (almost point-like) quasars. Some 100\,000 or more of Gaia's targets are asteroids in our own solar system. Where appropriate, I refer to `sources' observed by Gaia to maintain generality, but I will often refer to `stars', for concision, when any of the other types of source may be implied. The meaning should be clear in context. Incidentally, these numbers will change with the future data releases, DR4 and DR5, which will contain closer to 3~billion sources in total.

\section{Background}
\label{sec:background}

\subsection{The measurement of angles}

The Gaia mission is devoted to the field of astronomy referred to as `astrometry', which deals with the positions of celestial objects: of planets and other solar system bodies, of the billions of stars in our own Galaxy, and of the countless galaxies beyond our own. 
Since recording and refining the positions of the stars and planets was one of the few investigations of the heavens open to the ancients, astronomy and astrometry were largely synonymous until a little more than a century ago, when other types of astronomical investigation, such as spectroscopy, became possible. 

The history of this branch of science extends over more than two millennia. It is a large and multiply-connected field, having its origins in the earliest records of astronomical observations more than two thousand years ago, and extending to the high accuracy observations being made from space today. Over the centuries, improved star positions have led to remarkable and revolutionary advances in understanding our place in the Universe which I will briefly outline. Recording star positions has progressed through naked eye observations, later assisted by optical telescopes, through to the large-scale recording of stellar images on photographic plates from the late nineteenth century, and to the high-efficiency detectors and space-based measurements of the last thirty years. 

Astrometric measurements essentially involve determining the position of a star (or other object) as it appears projected on the celestial sphere. The star's distance being, at least to a first approximation, unknown, positions at any time can be simply and uniquely specified by the two angular coordinates of spherical geometry, precisely corresponding to latitude and longitude on Earth. 
The origin of the coordinate system is a delicate issue in practice, but the principle is straightforward: just as for geographical latitude, one coordinate can be tied to the extension of the Earth's equatorial plane at one moment in time. And, as for geographical longitude and the choice of the prime meridian as its origin, the other is referenced to some arbitrary but well-defined direction in space.

The basis of astrometric measurements, then, is the accurate measurement of tiny angles that divide up the sky. Dividing a circle, whether on paper or on an imaginary sweep of the celestial sky, is a task well-posed in principle.  Practical techniques for doing so aside, it is only necessary to agree on the unit of subdivision. Although scientific users today work and calculate angles in radians, the commonly accepted choice of 360~degrees in a circle, and its sexagesimal subdivisions, was made for us long ago. 
Ascribed to the Sumerians of ancient Babylonia, more than 2000~BCE, it was perhaps guided by the number of days in a year. One degree was subdivided into sixty minutes of arc, and each minute of arc was divided still further into sixty seconds of arc. The choice of sixty rests on the number itself being highly composite: it has many divisors, which facilitated calculations with fractions performed by hand. Astronomical measurements still use fractions of a second of arc (or `arcsec') as units for angular motions projected on the plane of the sky, or for the star's parallax, although more frequently specified in thousandths of an arcsec (milli-arcsec or mas) or, today with Gaia, millionths of an arcsec (micro-arcsec or \muas). 

To help visualise these angles, the Sun and the Moon both cover the same {\it angle\/} on the sky, about half a degree. The much smaller one second of arc corresponds to a linear distance of one meter viewed from a distance of about two hundred kilometers. This very small angle turns out to be a particularly convenient angular measure in astronomy, and it has been used to construct the very basic measure of astronomical distances, the parsec (a portmanteau of `parallax' and `arcsec', and abbreviated as pc). One parsec is defined as the distance at which 1~astronomical unit (au) subtends an angle of 1~arcsec (the `astronomical unit' was originally defined as the mean Sun--Earth distance, although today it is defined as a specific number of meters). One parsec corresponds to approximately 3.26~light-years, a distance unit in common use in popular astronomy, but which I will not generally refer to.

In very round numbers, one second of arc is also the angle to which astronomers can measure, with relative ease today, the position of a star at any one moment from telescopes sited below the Earth's atmosphere. It is the shimmering atmosphere, and specifically the turbulent troposphere, which drives astronomers to build their telescopes at high mountain sites, and indeed which has most recently pushed these measurements to space.

Historically, the human eye imposed its own limit to measuring angles of about one minute of arc. Mainly determined by the small diameter of the pupil through which light enters the eye, this limit is many times worse than that imposed by the atmosphere. 
Until the invention of the telescope, observations by eye could only place far less stringent limits on the accuracy of star positions. The introduction of the telescope, credited to Dutch opticians in the opening years of the 17th century, but more famously improved upon by Galileo in 1609, brought with it two distinct improvements. First, it could detect much fainter objects, revealing vastly more than were visible by eye. The larger diameter of the telescope aperture also gave improved positional accuracies. 

So much for one second of arc. It is a tiny angle, corresponding to the size of a Euro coin viewed from a distance of 5\,km. It is straightforward enough to measure today, but it proved a substantial challenge for the astronomical instrument makers of earlier centuries. 
Space astrometry took a giant leap with Hipparcos in 1989, reaching accuracies of one thousandth of one second of arc. This corresponds, roughly, to the angular size of an astronaut on the Moon viewed from Earth, a golf ball in New York viewed from Europe, the diameter of human hair seen from 10\,km, or the (angular) growth of human hair, in one second, when viewed from a distance of 1\,m. 

The Gaia satellite, launched in 2013 and operated until January 2025, is advancing this by a further factor one hundred, reaching accuracies of order 10~microseconds of arc (10\muas), corresponding to one Bohr radius viewed from a distance of~1\,m.  Such accuracies, naturally, pose extreme engineering challenges in terms of optical quality, detector performance, thermal control, and gravitational and thermal instrumental flexure.

\begin{figure}[t]
\centering
\vspace{-10pt}
\includegraphics[width=0.80\linewidth]{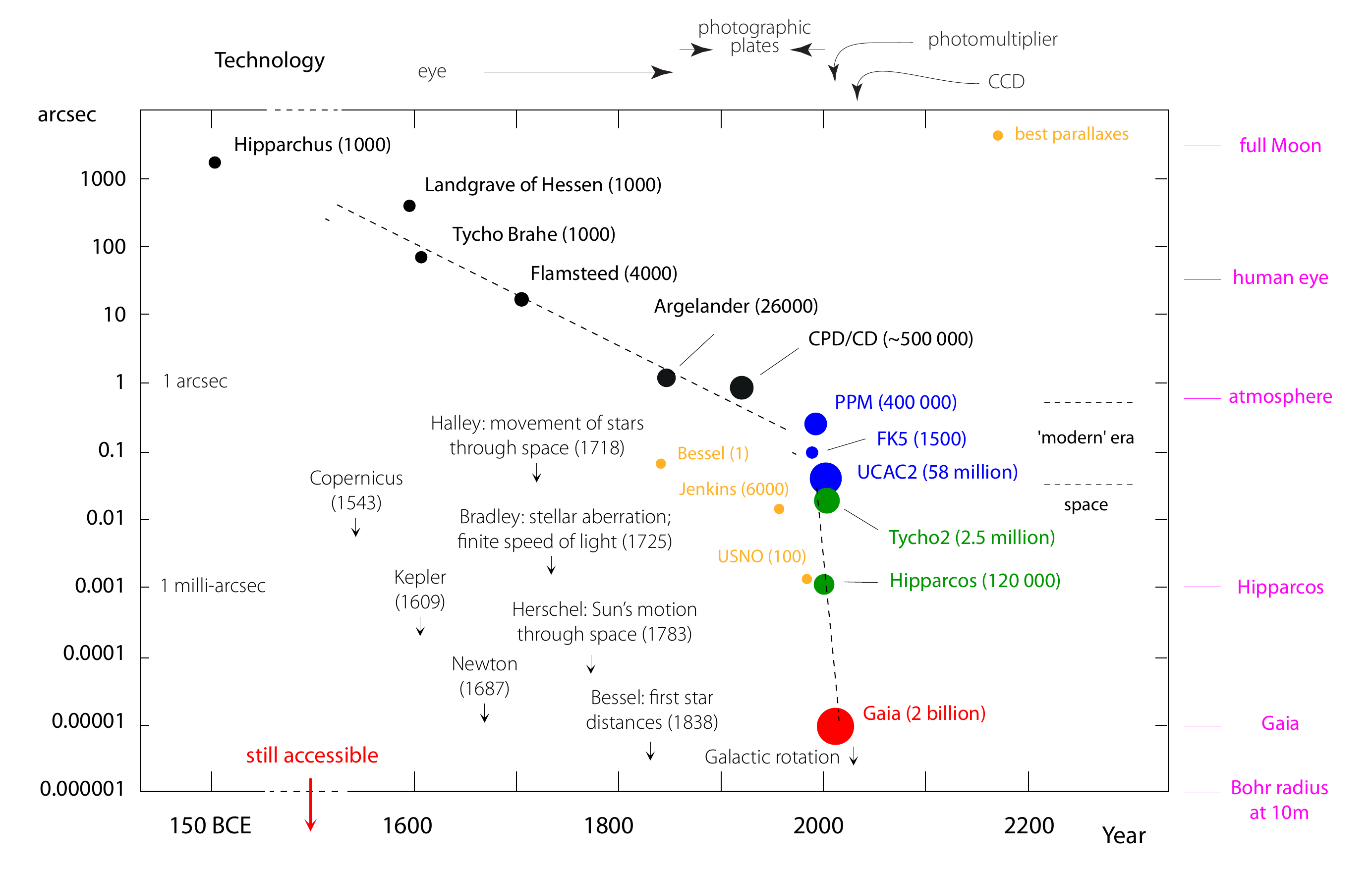}
\caption{The accuracy of star positions (left ordinate, log scale) through history, with some examples of these accuracies at the right. Only a few representative star catalogues are shown, with a circle size proportional to the logarithm of the number of stars. Colours indicate some key historical catalogues (black), a selection of more recent ground-based catalogues (blue), and catalogues of star parallaxes (orange). Accuracies from ESA's first space mission Hipparcos (operational 1989--93) are shown in green, and those from Gaia (2013--25) are shown in red. The relevant technologies are shown on the top axis. Some important scientific figures and their insights making use of the advancing accuracies are also shown.}
\label{fig:accuracy}
\end{figure}

\subsection{History}

Figure~\ref{fig:accuracy} illustrates the historical development of astrometry. The ordinate shows the accuracy on a logarithmic scale, and the abscissae shows time. I will provide a brief and highly selective synopsis of these developments to illustrate the context, and refer the interested reader to a more extended chronology by 
\citet{2012EPJH...37..745P}. 

Amongst the earliest catalogues of star positions were those made by the Greek mathematician and astronomer Hipparchus, of some 1000 stars in around 135~BCE, and the similar sized but more accurate catalogue of Tycho Brahe, in around 1600~CE.
It was the attempts to understand the complex motion of the planets (the word derived from the ancient Greek for `wanderer') which led to the eventual acceptance of Copernicus' heliocentric hypothesis from around 1540 (viz.\ that the Earth orbits the Sun), and to the comprehension and acceptance of Newtonianism. Given the $3\times10^8$\,km measurement baseline of the Earth's orbit around the Sun, a central prediction of the heliocentric hypothesis was that nearby stars should {\it appear\/} to `oscillate' with respect to the background stars during the year. Attempts to detect this `parallax' motion began in around 1600, but took the best part of 250~years until the first parallaxes were measured 
\citep{2013prmc.book.....H}. 	
Major instrumental advances were necessary to reach the accuracy levels required
\citep{1990dcdc.book.....C}.	
Part of the problem was that, since the scale of space -- and the distances to even the nearest stars -- was unknown, the accuracy that would be required to measure stellar parallaxes was itself unknown.

The steady improvement of measurement instruments (astrolabes, transit circles, quadrants, and others), and the advancing capability to accurately `divide the circle', led to improved star catalogues, which were themselves also crucial for the practical task of maritime navigation
\citep{1995ltsl.book.....S}.		
Let me pick out just three important milestones in this long journey.
A remarkable and crucial breakthrough came in 1718, when Edmond Halley, who had been comparing contemporary observations with those that the Greek Hipparchus and others had made, announced that the bright stars Aldebaran, Arcturus, and Sirius were displaced from their expected positions by large fractions of a degree. He deduced that each star had its own distinct velocity across the line-of-sight.\footnote{In the absence of a known distance, a star's space velocity is characterised by the two components of its {\it angular\/} displacement projected on the celestial sphere, referred to in astronomy as the star's `proper motion' (and typically measured in seconds of arc per year, or arcsec yr$^{-1}$).} It was the first convincing experimental suggestion that stars were moving through space.

Soon afterwards, and with instrumental advances reaching an accuracy of a few seconds of arc, measurements of the bright star Gamma Draconis by the Reverend James Bradley, England's third Astronomer Royal, yielded an unexpected surprise: the detection of small systematic positional shifts, of a form very different from that expected from the effects of parallax, and which he eventually correctly attributed as resulting from the addition of the velocity of light to the Earth's velocity as it moves in orbit around the Sun. Announced in 1729 the effect, known as stellar aberration, provided the first direct proof that the Earth was moving through space. His results therefore supported the Copernican theory, that the Sun, rather than the Earth, was the centre of the solar system. It is often cited as one of the most significant discoveries in the history of astronomy.

During the eighteenth century, after Halley's first detection of stellar motions, the movements of many more stars were announced. In 1783 William Herschel found that he could partly explain these collective motions by assuming that, in addition to the Earth's motion around the Sun, the Sun itself was moving through space. 
After many unsuccessful attempts, the first stellar parallaxes were announced in the 1830s. Friedrich Bessel, Wilhelm Struve, and Thomas Henderson all figure in this important development. Bessel received the Royal Astronomical Society's gold medal in 1841, for an achievement which the President John Herschel described as {\it `the greatest and most glorious triumph which practical astronomy has ever witnessed'}. These first stellar distances revealed, at a stroke, the utter vastness of the Universe.

I will pass quickly over the substantial and varied achievements in positional astronomy over the subsequent 150~years. These embraced major advances in telescope size and design, including meridian circles and Schmidt telescopes.  Enormous surveys were undertaken, including the three-part (Bonn, C\'ordoba, and Cape) 19th century Durchmusterungen (only eventually transcribed to computer form in a 15~year effort around the 1980s), the enormous and several decades-long international Carte de Ciel programme, and the vast Palomar--National Geographic and ESO sky surveys. Many catalogues of positions, proper motions, and distances of different sky regions resulted, but all were essentially limited to accuracies of around 1~arcsec or a little better, an accuracy barrier dictated by the Earth's atmosphere. These star positions and motions allowed discussions of, amongst many other topics, the size of the Galaxy, and the distance scale within it, the identification of star clusters and moving groups, the discovery of large numbers of variable stars, and the calibration of stellar luminosities for stars within 10--20~pc of the Sun.

But towards the end of the 20th century, even the most accurate star catalogues, hard won as a result of decades of observations, engineering ingenuity, and intellectual investment, were, in the context of the scale of our Galaxy, extremely modest. 
Before the move to space, the best all-sky {\it positional\/} reference system was provided by the FK5 catalogue, compiled at Heidelberg under the leadership of Walter Fricke
\citep{1988VeARI..32....1F}. 
Extending over 25~years of careful observation and critical analysis, the main catalogue contained just 1535 stars with positions accurate to some \mbox{0.2--0.3~arcsec}. It was published in 1988, just after his death. His work earned him widespread recognition for his contributions to celestial cartography including, in 1981, the Distinguished Service Cross, First Class, of the Federal Republic of Germany. 

A parallel effort was devoted to the determination of star distances, which required only narrow-field measurements, but of the highest accuracy, and which also progressed very slowly in view of the atmospheric barrier and other technical challenges involved. Jacobus Kapteyn in Groningen published a list of just 58 parallaxes in 1901. In 1922 he wrote {\it `I know of no more depressing thing in the whole domain of astronomy, than to pass from the consideration of the accidental errors of our star places to that of their systematic errors'}
\citep{1922BAN.....1...69K}.	
The General Catalogue of Trigonometric Parallaxes, undertaken at Yale University, aimed for a critical compilation of the best available stellar distances. Frank Schlesinger's original version in 1924 listed just short of two thousand stars, establishing a nonetheless frail stellar distance network out to a few tens of light-years
\citep{1924-schlesinger}. 
His life's work brought him the gold medal of the Royal Astronomical Society in 1927, and the Bruce medal (another of the highest honours in astronomy) in 1929. 
Louise Freeland Jenkins brought out a new edition in 1952, with distances for just under six thousand stars, and a supplement in 1963 raised the total to nearly six and a half thousand
\citep{1963-jenkins}. 
A further update, in 1995, of just over eight thousand stars, was compiled by Yale astronomer William van Altena 
\citep{1995gcts.book.....V}. 
It was the catalogue that the world's astronomers consulted near the end of the second millennium if they wanted to know the distance to one of the catalogue's stars. It was also to be the final collection of ground-based parallaxes before those from ESA's pioneering Hipparcos space mission. 

At the time of the push to space, around 1980, progress had run into almost insurmountable problems imposed by the Earth's atmosphere.  The total number of known star distances was certainly respectable, and had been immensely challenging to obtain. But even for the population of stars within our solar neighbourhood it was a small sampling, let alone viewed in the context of the hundred billion or more stars in our Galaxy as a whole. Crucial and niggling were the plethora of discrepancies and errors arising from the shimmering atmosphere. Accuracies were supposedly around one hundredth of a second of arc, but in reality were often much poorer. This made it difficult for astronomers to rely on published values, and problematic to draw wide-reaching scientific conclusions. 

Scientific advances associated with improved astrometric measurements faltered and, in the final decades of the 20th century, astrometry consequently receded in global scientific importance in the face of many other rapidly-advancing branches of observational and theoretical astronomy, such as spectroscopy and cosmology. It took a back seat as observations opened up at other electromagnetic frequencies, such as in the radio, infrared, and X-ray. 

A breakthrough in measuring star distances was brought about by two successive European Space Agency satellites placed above the Earth's atmosphere: the pioneering Hipparcos mission (1989--1993), and the subsequent Gaia mission (2013--2025). Their ultra-high astrometric accuracies have led to major advances in this fundamental science.

\subsection{The move to space}

The main reason for making astrometric measurements from space is to avoid the phase fluctuations caused by Earth's turbulent atmosphere, which impose a limit on star positional accuracies of around 0.1--1~arcsec.  But ground-based telescopes face other major challenges in reaching the positional accuracies ($10^{-5}-10^{-6}$~arcsec) that are targeted today. 
A space platform offers various other important advantages in achieving these demanding accuracies. The Earth itself is not a stable platform, but instead undergoes long-term orbital precession and nutation, and short-term spin variations (the `Chandler wobble') attributed to changes in the mass distribution of its outer core, atmosphere, oceans and crust. The `weightless' environment of space furthermore eliminates effects introduced by the flexing of ground-based telescopes under their own weight as the supporting structures are steered to observe different parts of the sky. 
And the more stable thermal environment of space helps to minimise the tiny thermo-mechanical distortions as ground-based observatories go through their inevitable day and night cycles of warming and cooling. 

Another major complication is that any telescope on Earth can observe only part of the sky at any one time: a telescope in the northern hemisphere only ever sees the northern skies. Even so, it still requires a year to elapse for the entire region to be observable at night. From the ground, a grid of star positions spanning the entire sky can only be constructed from a vast network of thousands of geometrical triangulations from separate telescopes observing accessible portions of the sky at different times. 
Like medieval surveys of the Earth made with early instruments, the result of centuries of celestial cartography pre-Hipparcos was a map of the sky, but one which was greatly warped and distorted. Positions were quite unreliable much below a second of arc, and plagued by unknown errors.  Within this distorted reference system, the determination of stellar parallaxes and proper motions at the angular accuracies demanded was simply impossible.

However, a classical telescope operating above the atmosphere still cannot create a global reference system free of medium- and large-scale distortions. As an analogy, imagine creating a map of the Earth with a 1-m ruler: local measurements might be tied together well, but the distance from Paris to Warsaw would be hopelessly inaccurate.

The solution, first proposed by French astronomer Pierre Lacroute in the 1960s, and studied during the 1970s by the French National Space Agency, CNES, is a scanning telescope with two widely separated fields of view on the sky, superimposed on the same focal plane. This offers two important advantages. First, it unlocks the possibility of constructing a global, and rigid, stellar reference system: stellar space motions referred to this global reference system avoid the local or zonal errors which otherwise plague their kinematic or dynamic interpretation.
 
\begin{figure}[t]
\centering
\includegraphics[width=0.30\linewidth]{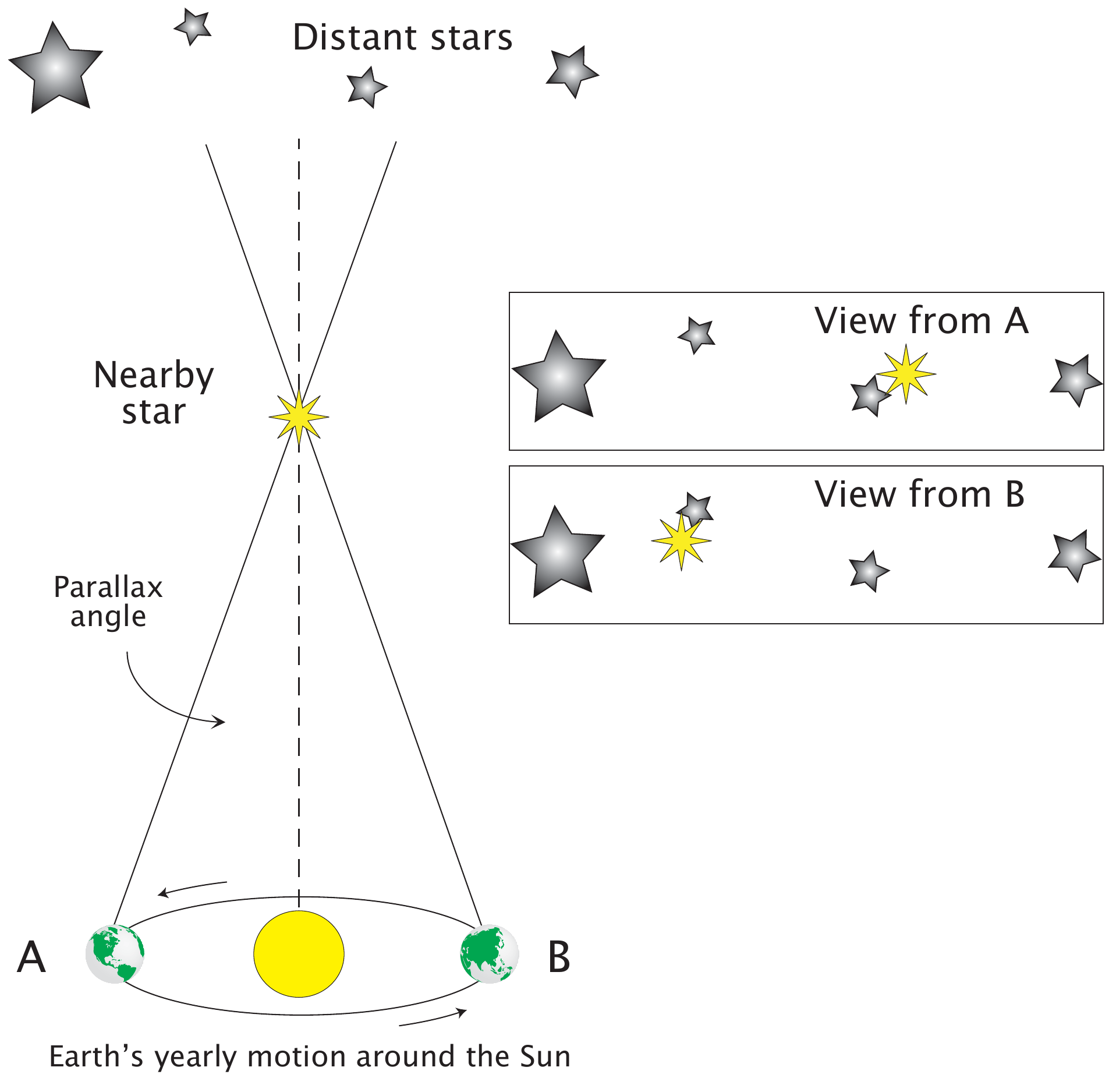}
\hspace{55pt}
\includegraphics[width=0.57\linewidth]{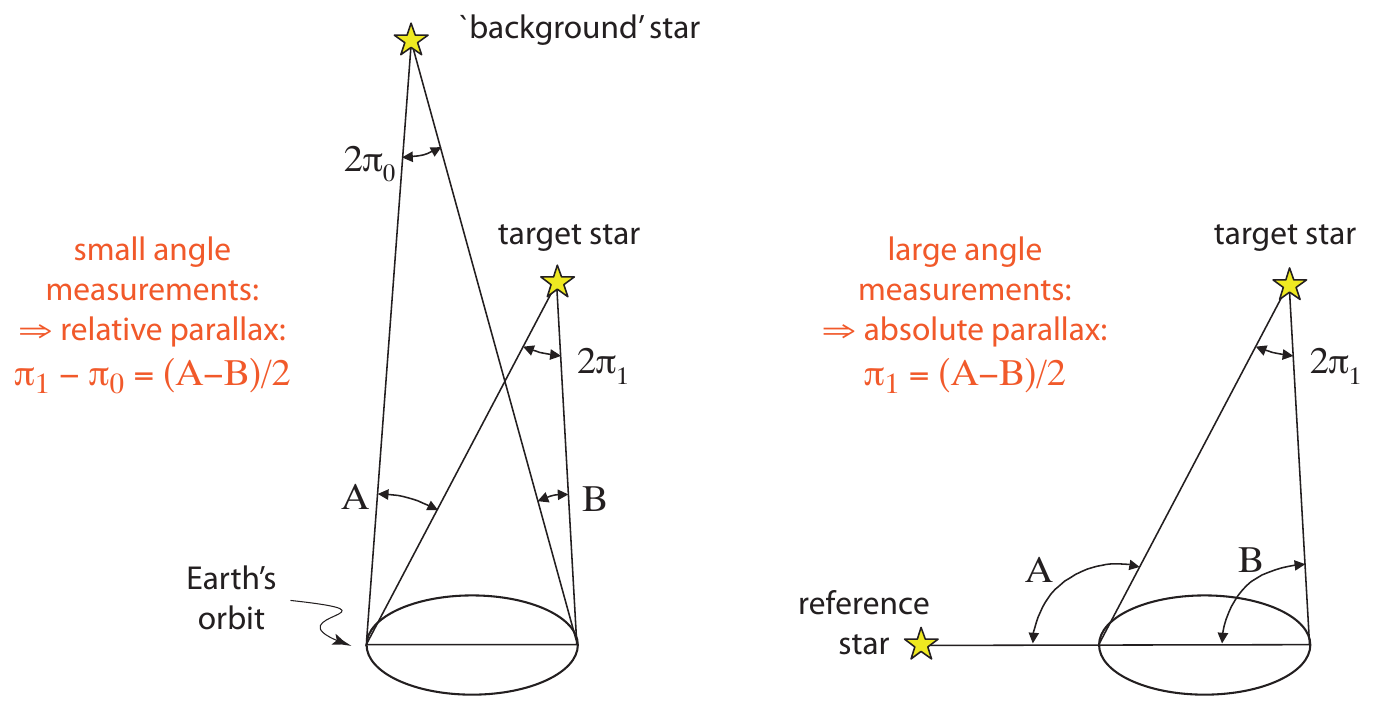}
\caption{Left~(a):~the principle of parallax measurement. Earth's annual orbit around the Sun provides a baseline of $\sim\!\!3\times10^8$\,km, from which a `nearby' star will appear to move with respect to more distant stars. Right~(b):~the principle of `absolute' parallaxes. Narrow-angle measurements can only provide parallax measurements relative to nearby stars on the sky, whose distances are themselves unknown. Large-angle measurements (from the superposition of the two fields) provide the framework for the determination of absolute parallaxes (courtesy: Lennart Lindegren).}
\label{fig:absolute-parallax}
\end{figure}

Second, these accurate wide-angle measurements are also critical in overcoming one of the other major challenges in astrometry, tied to the measurement of a star's trigonometric parallax. A star's distance can only be determined from the relative displacement of its apparent position with respect to one or more background stars, as the Earth (and the satellite) moves in its annual orbit around the Sun (Figure~\ref{fig:absolute-parallax}a). The problem is that we cannot rely on a star's parallax measured with respect to other stars nearby on the sky, but whose distances are themselves unknown (Figure~\ref{fig:absolute-parallax}b). 
Superimposing two widely separated fields allows the problem to be solved. Imagine if we could superimpose an image of the target star with a reference star $90^\circ$ away along the Earth's orbit. There is now no component of parallax for the reference star in this configuration, so that its parallax (or distance) is irrelevant. The situation is reversed 3~months later. Detailed studies showed that the superimposed fields should in fact {\it not\/} be orthogonal, but as long as they are separated by a reasonably large angle, the network of observations made from a continuous scanning of the celestial sphere permits {\it absolute\/} parallaxes to be determined. I go into slightly more detail on this in Section~\ref{sec:basic-angle}.

The same sort of wide-angle measurements cannot be performed from the ground. This is mainly a result of atmospheric turbulence, which becomes worse away from the zenith, and makes wide-angle measurements even more error prone.  It is compounded by gravitational instrument flexure, and other complexities.

\subsection{Hipparcos}
\label{sec:hipparcos}

In the late 1970s, Lacroute's ideas were picked up by Erik H{\o}g in Denmark, and Jean Kovalevsky and Catherine Turon in France. They stimulated wider European interest in such a mission, and (with others) brought it to the attention of the European Space Agency. 
Phase~A studies by ESA in the late 1970s led to the acceptance of Hipparcos, the first space-based astrometric mission, in 1980 (the name, incidentally, was a contrived acronym whose spelling is intentionally different from, but bearing witness to, the important contributions of the ancient Greek astronomer). Its scientific goal was to extend distance measurements to $\sim$100\,pc, to determine luminosities in the upper parts of the Hertzsprung--Russell diagram, and to map distances and motions in the solar neighbourhood. It was centred around a 30\,cm diameter primary mirror (made of zerodur), with two viewing directions separated by $58^\circ$ on the sky, and a photomultiplier tube detector. In a 4-year mission, operated from its non-nominal geostationary transfer orbit, it measured the distances and space motions of the main high-quality catalogue of 120\,000 pre-selected stars, with a limiting magnitude of $V\sim12$\,mag, and with an accuracy of around 1~milli-arcsec. Its lower accuracy component, the Tycho Catalogue, included some 2~million stars. Erik H{\o}g, Jean Kovalevsky and Catherine Turon, with Lennart Lindegren of Sweden, went on to lead the four scientific teams, comprising some 200 European scientists, who worked with ESA (with the author as ESA's project scientist) in the mission's development and execution between 1981--1997.

Hipparcos established the experimental principles picked up on a much larger scale by Gaia, and its crucial and innovative foundations should be underlined.
Physicist Freeman Dyson, in his 1988 book `Infinite in All Directions', wrote that {\it `Hipparcos is the first time since Sputnik in 1957 that a major new development in space science has come from outside the United States'}. 
In his Millennium Essay, 
Malcolm \citet{2001PASP..113....1L} 
wrote: {\it `The bedrock of astronomy remains the compilation of what is out there\ldots\ It is invidious to single out surveys which I find particularly impressive, but I make an exception for the Hipparcos astrometric satellite'}.

Adoption of Hipparcos was widely perceived as a bold but contentious choice by ESA's advisory committees. Some questioned its expected scientific impact, and some doubted its feasibility. Advances below its 1--2~milli-arcsec goals were never entertained at the time --   the rich complexities of dynamical phenomena across the Galaxy's vastness, today being opened up by Gaia, were unknown, beyond reach, and never discussed. The protracted political debate that surrounded the mission's acceptance by ESA is documented in the authoritative 2-volume history of the European Space Agency, 1958--1987 
\citep{2000ESASP1235.....K},
with a summary in my own popular account of the Hipparcos project
\citep{2010mhgs.book.....P}.

The landscape has changed fundamentally in the decades since. In the curious way that science advances across many fronts in parallel, Gaia is, today, central to problems being tackled in stellar evolution, in Galactic structure, in exoplanet science, in solar system studies, and in cosmology.

\subsection{Post-Hipparcos}

The success of Hipparcos, and specifically the scientific interest of astrometry at the milli-arcsec level, motivated the study of further space experiments aiming for an improvement in astrometric accuracy, limiting magnitude, and numbers of stars. Concepts included 
FAME, AMEX, OBSS and SIM in the US;
OSIRIS and LIDA in Russia; 
DIVA in Germany;
and JASMINE in Japan.
Most, with the notable exception of SIM, followed the same basic principles of Hipparcos, based on a continuous sky scanning with two widely-separated fields of view. All, apart from JASMINE, were eventually dropped, for reasons of either cost, technical challenges, or when faced with the advance of Gaia.

Roemer was proposed as an ESA medium-sized mission, by Erik H{\o}g and Lennart Lindegren
\citep{1993IAUS..156...37H, 1994gsso.conf..246H}, 
extending the limiting magnitude and accuracy of Hipparcos to some 100~million stars, complete to $V\sim15.5$\,mag, and with accuracies in the range 0.1\,mas for $V<11$\,mag to 1.5\,mas at $V\sim17$.  For the first time, the mission baselined CCD detectors for the focal plane, although CCDs had been briefly considered during the early Phase~B studies of Hipparcos in around 1982, but promptly discarded due, amongst other issues, to the poor charge transfer efficiency of the early devices. But ESA's scientific advisory teams considered Roemer too modest in terms of its scientific advance compared with Hipparcos. 

Subsequently, the more ambitious Gaia project was proposed to ESA by Lennart Lindegren and Michael \mbox{Perryman}, originally in the form of a small optical interferometer fitting inside the Ariane~5 launch vehicle (\citet{1996A&AS..116..579L}), but subsequently in monolithic form (\citet{2001A&A...369..339P}).   
A 3-year feasibility study was led by the author (as ESA's project scientist) and Oscar Pace (as ESA's study manager), supported by a Science Advisory Group and various industrial teams. Findings appeared as a 380-page report, ESA--SCI(2000)4, in July 2000, which included 100~pages detailing the mission's scientific goals, itself compiled from the contributions of some 100 European scientists. 

As part of these studies, 15 preparatory technologies were identified as being needed to ensure that the satellite could be developed on schedule, and within budget (space industry today rates readiness for flight in terms of the component's `technology readiness level', TRL). Required technology developments were identified, and subsequently undertaken, for
the large silicon-carbide mirrors and ultra-stable payload structure;
the 10-m deployable solar array sunshield;
the micro-Newton reaction system for the fine attitude control;
and inch-worm actuators for telescope refocusing.
Five specific developments were related to advanced CCD performances not available at the time, including 
\mbox{3-side} buttable, small-pixel, high-performance chips; 
the large-area highly-integrated focal plane assembly;
and high-speed, low-noise detection chains.
Other associated developments were required for
the efficient on-board compression algorithms for the science data;
optimisation of the payload data-handling electronics;
and a (non-moving) phased-array antenna suitable for the high data-rate transmission from Gaia's L2 orbit.

Let me try to put the evolution from Hipparcos to Gaia in some perspective. Compared to the 100~billion or more stars in our own Galaxy alone, Hipparcos sampled only a small region, centred on the solar neighbourhood. The Hipparcos catalogue of 120\,000 stars was published in 1997. In printed form, the main catalogue occupied 5~volumes, some 50\,cm of shelf space. 
Gaia's two billion sources, if printed simply in summary form, in the same format as Hipparcos, would require 10\,km of shelf space. In numerical terms it is of the same order as the human population on Earth. And when comprehending its scientific objectives it is perhaps useful to attempt a parallel analogy: just as every person has their own identify, so each star is unique, in terms of mass, size, temperature, age, chemical composition, rotation, magnetic field, and so on. Like a population census, Gaia is akin to having the entire world population available for anthropological study, when previously only a very restricted sample had been available.

The recent and dramatic advances in the accuracy with which these positions (and hence distances) can be measured, and their variation with time (and hence their motions through space), leads to immensely profound insights into the structure, formation and evolution of stars, into the very origin and evolution of our Galaxy, and indeed of the Universe as a whole.

\begin{table*}[t]
\centering
\leavevmode
\footnotesize
\begin{tabular}{lrrrrrccrccc}
\hline
\noalign{\vskip 5pt}
\hspace{23pt}  Tracer & M$_{\rm V}$ & \hfil $\ell$ \hfil & \hfil $b$ \hfil & \hfil
$d$ \hfil & A$_{\rm V}$ & \hfil  V$_{1}$ \hfil & \hfil V$_{2}$ \hfil & $\epsilon_{\rm T}$ \hfil & $\sigma_{\mu_1}$ & $\sigma^{\prime}_{\mu_1}$ & $\sigma^{\prime}_{\varpi_1}$ \\
\hspace{5pt} (d\,=\,dwarf; g\,=\,giant)&  mag  &  deg  &  deg  &  kpc  &  mag  &  mag  & mag  &  km/s  &  $\mu$as/yr  & -- & --\\[4pt]
\hline 
\noalign{\vskip 5pt}
{\bf Bulge:}\\
\qquad $\bullet$ gM 					& --1 	& 0 	& $<20$ 	& 8 	& 2--10 	& 15 		& 20 		& 100 	& 10		&0.01 	&0.10\\
\qquad $\bullet$ horizontal branch		& +0.5 	& 0 	& $<20$ 	& 8 	& 2--10 	& 17 		& 20 		& 100 	& 20		&0.01 	&0.20\\
\qquad $\bullet$ main-sequence turnoff 	& +4.5 	& 1 	& $-4$ 	& 8 	& 0--2 	& 19 		& 21 		& 100 	& 60		&0.02 	&0.6\phantom{0}\\[2pt]
{\bf Spiral arms:}\\
\qquad $\bullet$ Cepheids 			& --4 & all & $<10$ & 10 & 3--7 & 14 & 18 & 7 & \phantom{0}5&0.03 &0.06\\
\qquad $\bullet$ B--M supergiants 		& --5 &  all & $<10$& 10 & 3--7& 13 &17 &7& \phantom{0}4&0.03 &0.05\\
\qquad {$\bullet$ Perseus arm (B)}		& --2 & 140 & $<10$ & 2 & 2--6 & 12 & 16 & 10 & \phantom{0}3&0.01&0.01\\[2pt]
{\bf Thin disk:}\\
\qquad $\bullet$ gK 					& --1 & 0 & $<15$ & 8 & 1--5 & 14 & 18 & 40 & \phantom{0}6&0.01 &0.06\\
\qquad $\bullet$ gK 					& --1 & 180 & $<15$ & 10 & 1--5 & 15 & 19 & 10 &  \phantom{0}8&0.04 &0.10\\[2pt]
{\bf Disk warp:} (gM)					& --1 & all & $<20$ & 10 &1--5 &15 &19 & 10 & \phantom{0}8& 0.04 &0.10\\[2pt]
{\bf Disk asymmetry:} (gM)			& --1 & all & $<20$ & 20 &1--5 &16 &20 & 10 &15& 0.14 &0.4\phantom{0}\\[2pt]
{\bf Thick disk:}\\
\qquad $\bullet$ Miras, gK 			& --1 & 0 & $<30$ & 8 & 2 & 15 & 19 & 50 &10&0.01 &0.10\\
\qquad $\bullet$ horizontal branch		& +0.5 & 0 & $<30$ & 8 & 2 & 15 & 19 & 50 & 20&0.02 &0.20\\
\qquad $\bullet$ Miras, gK 			& --1 & 180 & $<30$ & 20 & 2 & 15 & 21 & 30 &25&0.08 &0.65\\
\qquad $\bullet$ horizontal branch		& +0.5 & 180 & $<30$ & 20 & 2 & 15 & 19 & 30 & 60&0.20 &1.5\phantom{0}\\[2pt]
{\bf Halo:}\\
\qquad $\bullet$ gG 					& --1  & all & $<20$ & 8 & 2--3 & 13 & 21 & 100 & 10&0.01 &0.10\\
\qquad $\bullet$ horizontal branch		& +0.5 & all & $>20$ & 30 & 0 & 13 & 21 & 100 &35&0.05& 1.4\phantom{0}\\[2pt]
{\bf Gravity} (K$_Z$ relation):\\
\qquad $\bullet$ dK					&  +7--8 & all & all & 2 & 0 & 12 & 20 & 20 & 60&0.01 &0.16\\
\qquad $\bullet$ dF8--dG2				&  +5--6 & all & all & 2 & 0 & 12 & 20 & 20 & 20&0.01 &0.05\\[2pt]
{\bf Globular clusters:} (gK)			& +1 & all & all & 50 & 0 & 12 & 21 & 100 & 10&0.01 &0.10\\[2pt]
\qquad $\bullet$ internal kinematics (gK)	& +1 & all & all & 8 & 0 & 13 & 17 & 15 & 10&0.02 &0.10\\[2pt]
{\bf Satellite orbits:} (gM)				& --1 & all & all & 100 & 0 & 13 & 20 & 100 & 60& 0.3\phantom{0}& 8\phantom{.00}\\
\noalign{\vskip 5pt}
\hline 
\end{tabular}
\caption{Main Galaxy components (bold) and some relevant `tracer' populations, together with the astrometric accuracies and limiting magnitudes required in their use as structural or kinematic structures. For each, a typical absolute magnitude ($M_{\rm V}$), Galactic location ($\ell$, $b$), distance ($d$), and extinction $A_{\rm V}$, can be used to estimate the magnitude range to be probed ($V_1-V_2$). Typical transverse velocities and required accuracies ($\epsilon_{\rm T}$, $\sigma_{\mu_1}$) correspond to requirements on accuracies of their proper motions ($\sigma^\prime_{\mu_1}$) and parallaxes ($\sigma^\prime_{\varpi_1}$). The table is based on \citet{1995ESASP.379...95G}.
}\label{tab:tracers}
\end{table*}

\subsection{Scientific goals}

The primary objective of the Gaia mission is to observe the physical characteristics, kinematics and distribution of stars over a large fraction of the volume of our Galaxy, with the goal of achieving a major advance in understanding its dynamics and structure, and consequently its formation and history. 
It is making this goal possible by providing, for the first time, a catalogue which samples a large and well-defined fraction of the stellar distribution, determining the three-dimensional positions and space velocities of every star observed, from which significant astrophysical conclusions can be drawn for the entire Galaxy. 

It is not only by good fortune that, today, Gaia is uncovering so much about the Galaxy's structure and kinematics. The design of the scientific instrument, and its operation in orbit, targeted very specific measurement goals, formulated in terms of limiting magnitude, numbers of stars, and their required astrometric accuracy. 
Complementing its astrometric measurements, Gaia was also designed, from the start, to include accurate multi-colour multi-epoch photometric measurements that would allow each of its billion or more stars to be characterised, in terms of position in the Hertzsprung--Russell diagram, metallicity, and reddening. 
Put simply, measuring the accurate distances and motions of a billion stars in the Galaxy would be a remarkable achievement in itself. But having a meaningful understanding of each star's basic physical properties was also considered paramount.

Understanding our Galaxy requires the accurate measurement of distances and space motions for large and unbiased samples of stars of different mass, age, metallicity, and evolutionary stage. The huge number of stars, high accuracy, and faint limiting magnitude of Gaia is quantifying our understanding of the structure and motions within the bulge, the spiral arms, the disk and the outer halo, and is revolutionising dynamical studies of our Galaxy. 
With today's enormous computing power, the huge stellar content is one of Gaia's great strengths: a vast net that can 
catch fleeting stages of stellar evolution, 
define statistically useful samples of rare star types, 
find tiny features in the Hertzsprung--Russell diagram that point to exotic physics, 
and 
sift through millions of stars to seize on a few that are the fossil record of a galaxy captured billions of years ago.

Table~\ref{tab:tracers}, based on the original compilation by \citet{1995ESASP.379...95G}, gives a summary of the main Galaxy components and sub-populations, together with the astrometric accuracies and limiting magnitudes that were used to drive the mission's technical specifications.
For each sub-population, and its representative kinematic tracers of known absolute magnitude and estimated distance and reddening, the table gives the corresponding range of $V$~magnitudes over which the populations must be sampled, along with the required astrometric accuracy.
From such considerations, it was evident that the Gaia survey should reach at least down to $V=14-15$~mag, and preferably as faint as $V=20-21$~mag, to probe representatives of all of these key populations.


\section{The Gaia satellite}
\label{sec:gaia-satellite}

Gaia was accepted within the ESA scientific programme in 2000 with a target launch date of 2012, and entered Phase~B (the detailed design phase) in 2006. It was launched, from Kourou (French Guyana) on 19~December 2013, and placed in its operational quasi-periodic Lagrange Sun--Earth (L2) orbit, $1.5\times10^6$\,km from Earth. 
After a 6-month commissioning phase, Gaia started routine science operations in July 2014.
It had a {\it specified\/} operational lifetime of 5~years (on which basis component and sub-system failure probabilities were designed), but with a {\it target\/} lifetime of an additional 5~years. 
With its outstanding scientific results demonstrated, ESA's advisory committees approved the extension of mission operations. It ran low on consumables (gas for its attitude control), and scientific operations were duly terminated, on 15~January 2025, having achieved a total data collection period of 10.5~years.

Gaia is a very complex satellite, with multiple state-of-the-art subsystems, and an instrument payload resulting from nearly two decades of wide-ranging studies, optimisations, and prototype developments. It is beyond the scope of this review to go into details of the optical design, the detection system, the on-board data handling, data storage and data transmission to Earth, the choice of orbit, the scanning law and its attitude control. But I will outline some of the key points that are central to the mission's measurement principles, and otherwise refer to a detailed technical description of the satellite that accompanied the first data release in 2016
\citep{2016A&A...595A...1G}.

\paragraph{Telescope and focal plane}
\label{sec:gaia-design}

\begin{figure}[t]
\centering
\includegraphics[width=0.80\linewidth]{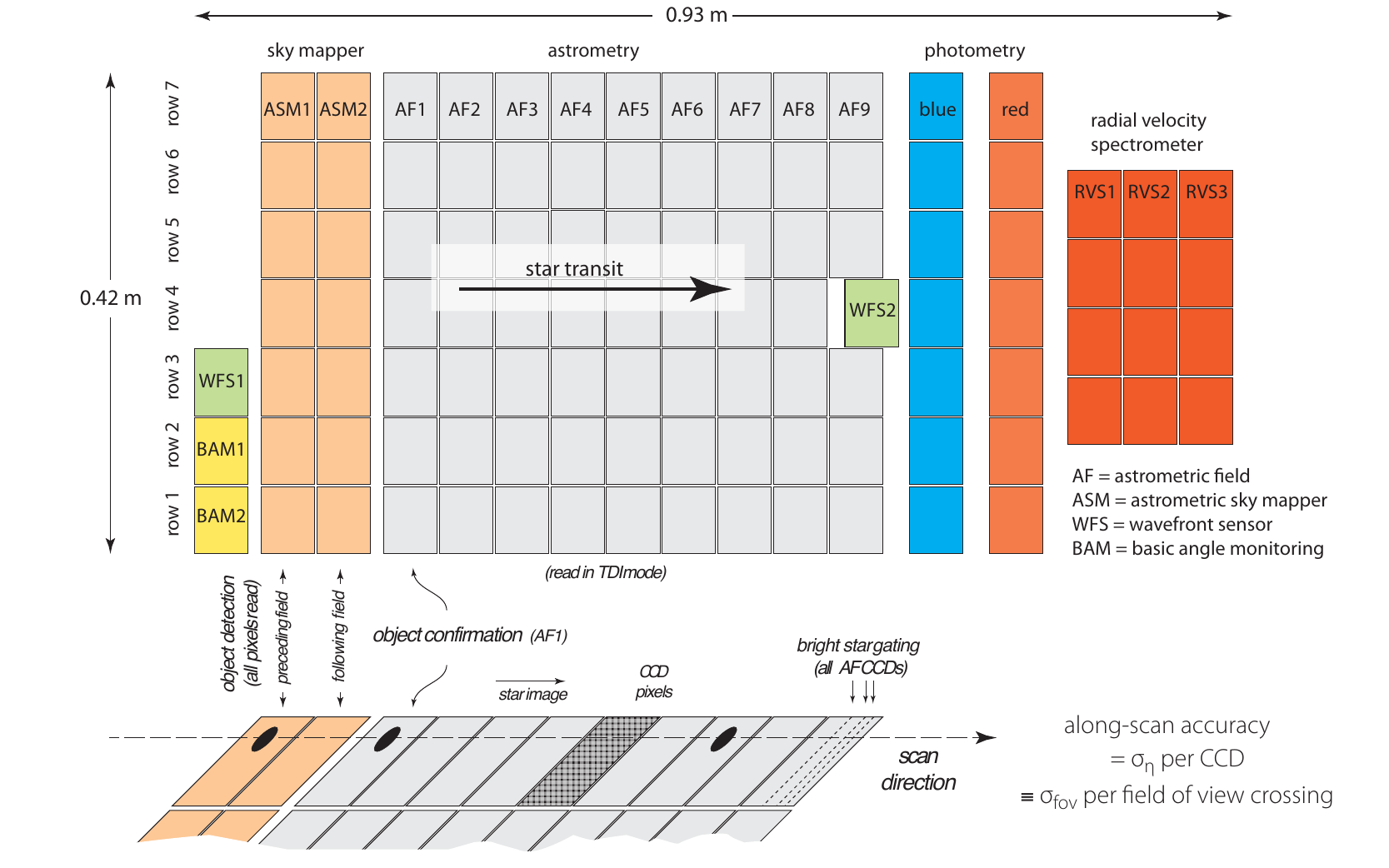}
\caption{Schematic of the Gaia focal plane. The upper figure shows the focal plane viewed from above. Star images pass from left to right as the satellite scans the sky. The lower figure is an inclined schematic of one of the rows. See text for details.}
\label{fig:gaia-fp-concept}
\end{figure}

The Gaia satellite comprises two identical imaging telescopes, mounted on a large high-thermal stability silicon carbide toroidal structure. The two viewing directions are separated by the `basic angle' of 106\ddeg5 (further detailed below), and brought to a superimposed focus in a single focal plane. The focal plane comprises an array of 106 large-format CCD detectors in seven rows, each with its own autonomous control unit (Figure~\ref{fig:gaia-fp-concept}). 
The spacecraft rotates around an axis perpendicular to the two viewing directions, once every 6~hours.  
It takes just over one minute for star images to pass across the entire 1\ddeg5 focal plane in the along-scan direction, where CCDs columns are sequentially dedicated to object detection, astrometry, photometry, and medium-resolution spectroscopy. 

In normal CCD imaging, individual pixels accumulate photoelectrons during an exposure, after which the entire CCD is `read out': pixel columns are successively advanced electronically towards a readout register. 
In contrast, and central to the operation of the Gaia CCDs, is the method of `time-delayed integration', or TDI. Here, star images move slowly across the detector, and all CCD columns are stepped continuously, in phase, towards the readout register. In practice, the scan rate is adjusted to coincide with the clock rate of the CCD readout, with each line of pixels being shifted in a little less than 1~ms. Exposures build up as the satellite scans the sky, the integrated image being read out when it reaches the final readout register. The two fields of view follow a predefined and carefully optimised scanning `law', and the signal of each celestial source is integrated as the image crosses the focal plane. 

Star positions for each focal plane crossing are defined by the image centroid (which can be defined much more accurately than the point spread function). With maximum precision being achieved in the along-scan direction, the primary mirrors (themselves also of silicon carbide) are rectangular, of dimension $1.45\times0.45$\,m$^2$, aligned along scan, which results in an asymmetric point spread function with aspect ratio 3:1. 
The large-format CCDs were specifically developed for Gaia by e2v (now Teledyne e2v). Each is $4.5\times5.9$\,cm$^2$, and comprises 4500 pixels in the along-scan direction, and 1966 pixels in the across-scan direction.
Each pixel is rectangular in the same 3:1 ratio, the $10\times30\,\mu$m$^2$ pixels corresponding to $59\times177$~mas$^2$ on the sky, and a $\sim$2~pixel Nyquist sampling of the point-spread function's full-width at half maximum. 
The image width in the scan direction is about 0.1~arcsec, determining Gaia's ultimate resolution as an imaging system. 

Each CCD includes an anti-blooming drain to avoid charge overflow for bright stars, and a supplementary buried channel to maximise the transfer efficiency of small charge packets. Each CCD can be gated individually to restrict the integration time for bright stars (shown schematically for AF9 in Figure~\ref{fig:gaia-fp-concept}). Special provision is made to minimise the effects of radiation damage (due mainly to prompt solar particles peaking at the time of solar maximum), charge-trapping, and the resulting effects of charge-transfer inefficiency (CTI).

A key feature of the design was to move away from a pre-defined `input catalogue' to determine which stars would be observed (as used for Hipparcos), and to implement a highly sophisticated on-board detection system. This registers all sources which, at each epoch of observation, are brighter than a well-defined magnitude limit (around $V=20-21$~mag). This allows for the systematic observations of, for example, solar system objects, variable stars, and other time-varying transients, all defined by Gaia's own bandpass. 
As the satellite scans the sky, star images first cross the astrometric sky mappers, ASM1 and ASM2, which perform the onboard object detection.  Optical baffles ensure that each sees only one telescope field. In the absence of prior star knowledge, all pixels of the sky mappers are read out, at the expense of a higher read noise, for the preceding and following fields of view, respectively. Thresholding identifies all sources crossing the sky mappers, which are confirmed (and cosmic rays discriminated) through a slower readout of the regions around each candidate star in AF1. 
Fast onboard analysis then produces a list of point-like objects for observation in the following CCDs (in which both fields are superimposed), rejects isolated cosmic ray events, and determines the scan rate about both axes. This information is provided to the attitude control subsystem, which allows the star positions to be predicted for the remaining focal plane crossing.
Confirmed objects are then `observed' by all subsequent CCDs, again in TDI mode at the satellite scan frequency. 

The astrometric field (AF1--AF9) employs a broad photometric band (designated $G$, 350--1050~nm), maximising the astrometric signal-to-noise, {\it and\/} the resulting $G$-band photometric accuracy. 
Each CCD crossing of the astrometric field yields an along-scan accuracy ($\sigma_\eta$), with all 9~yielding an along-scan accuracy per field-of-view crossing ($\sigma_{\rm fov}$). Sources then pass across the blue- and red-optimised photometers, where prisms provide low-resolution spectra used to derive star colours. 
A subset then pass across the $3\times4$ CCD array encoding signals from the medium-resolution Radial Velocity Spectrometer (RVS, Section~\ref{sec:rvs-acquisition}).

Various other tasks rely on measurements in the focal plane. CCD `gating' restricts the integration time for bright stars, and allows the measurement of objects that would otherwise cause detector saturation.  Charge-injection mitigates some of the problems of radiation-induced charge trapping. Two CCDs are dedicated to wave-front sensors which allow telescope focusing. Two further CCDs are dedicated to a specific laser metrology `basic angle monitor' (BAM) system, which monitors any tiny spin-dependent variations in the angle between the two viewing directions.
In practice, the focal plane is of considerable complexity, with very demanding requirements in terms of metrology, thermal control, readout noise, readout processing, and the downstream on-board data handling comprising on-board data storage, and telemetry from the L2 orbit to the ground stations on Earth, all achieved without moving parts.

\begin{figure}[t]
\centering
\includegraphics[width=0.33\linewidth]{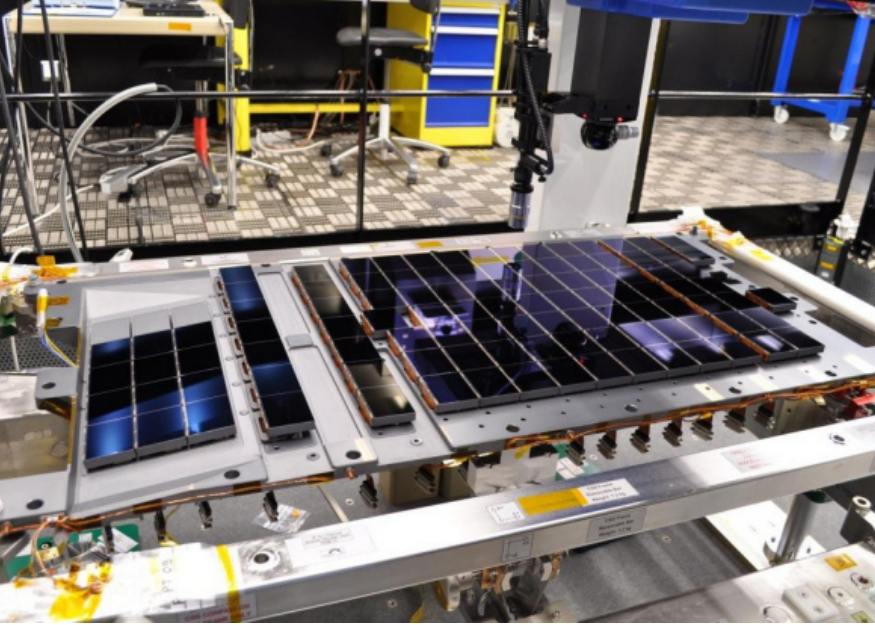}
\includegraphics[width=0.33\linewidth]{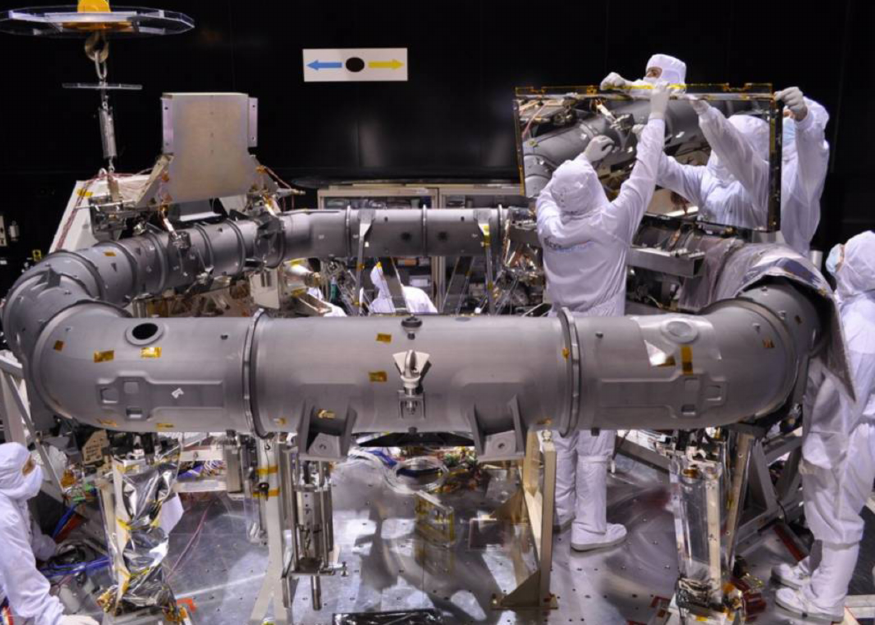}
\includegraphics[width=0.33\linewidth]{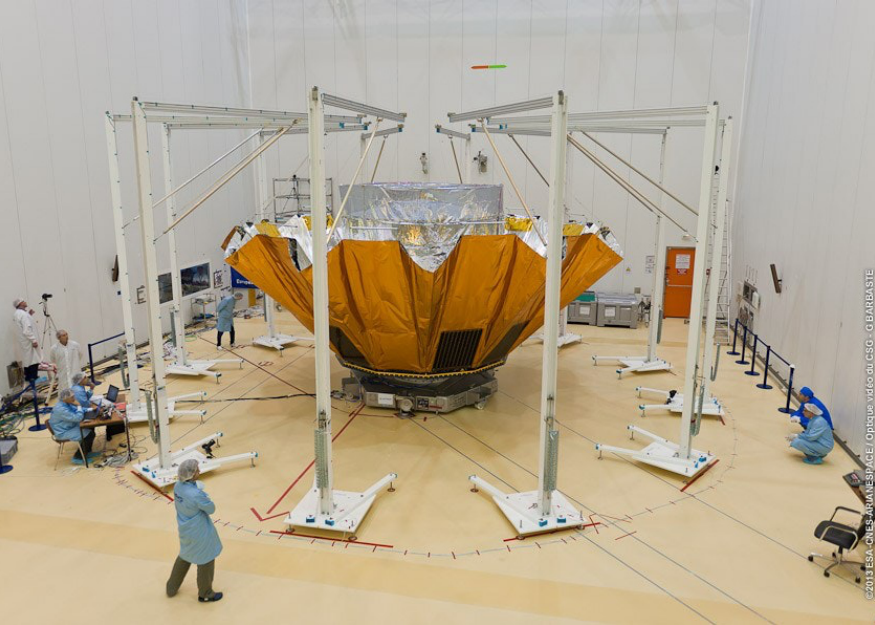}
\caption{Integration of the Gaia satellite: 
(a)~the focal plane assembly (106 large format CCDs occupying an area of nearly $0.5\times1$\,m$^2$);
(b)~the silicon carbide toroidal support structure with one of the primary mirrors being installed;
and
(c)~the integrated satellite with its solar array/sunshield assembly undergoing deployment tests. Courtesy: ESA/Airbus Defense \& Space.}
\label{fig:gaia-satellite}
\end{figure}

Three photographs of the Gaia satellite during integration are shown in Figure~\ref{fig:gaia-satellite}, illustrating 
the focal plane assembly, 
the silicon carbide toroidal support structure with one of the primary mirrors during assembly,
and the integrated satellite with its deployable solar array/sunshield assembly undergoing deployment tests.

\paragraph{Basic angle}
\label{sec:basic-angle}

In the language of space astrometry, and specifically for ESA's space astrometry missions Hipparcos and Gaia, the `basic angle' is the angle between the instrument's two viewing directions on the sky. For Hipparcos, the basic angle was $58^\circ$. For Gaia it is $106\ddeg5$. The two viewing directions, and the angle between them, is a fundamental aspect of these astrometric missions.  What determines the choice, and why is it different for the two missions?

\begin{figure}[t]
\vspace{-5pt}
\centering
\hspace{5pt}
\includegraphics[width=0.56\linewidth]{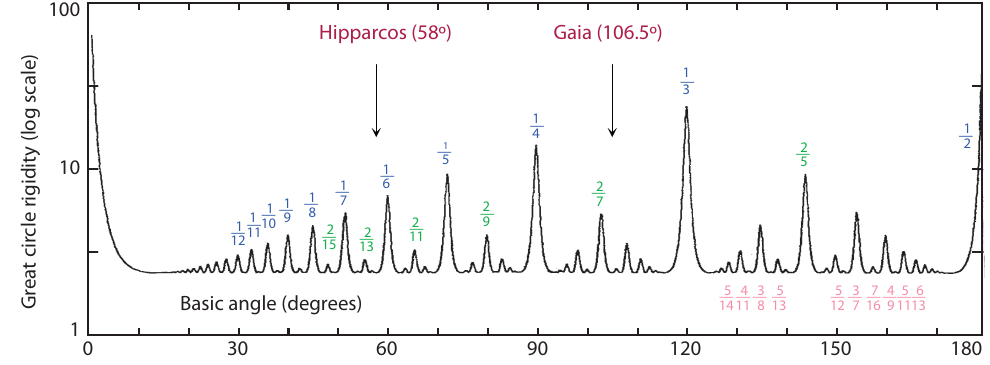}
\hspace{10pt}
\raisebox{5pt}{\includegraphics[width=0.38\linewidth]{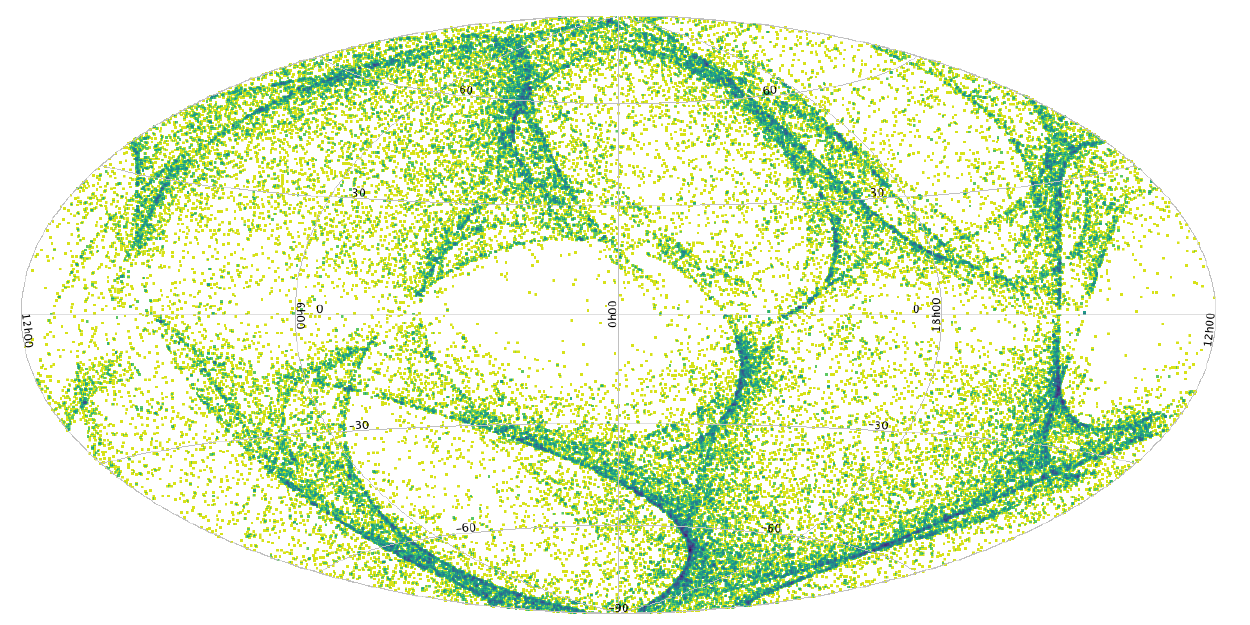}}
\vspace{-5pt}
\caption{Left~(a): rigidity of simulated `great circle solutions' versus basic angle between the two viewing directions; higher peaks correspond to poorer rigidity. Between the peaks are small `valleys' providing good rigidity (courtesy: Lennart Lindegren).
Right~(b): sky distribution of faint EDR3/DR3 sources, $G>22$~mag (Galactic coordinates): higher scan densities reach a fainter magnitude limit (\citet{2021A&A...649A...3R}, Figure~16).
}
\label{fig:distortion-basic-angle}
\label{fig:scanning-law}
\end{figure}

Consider just one of the great circle scans. If the viewing directions were separated by $90^\circ$, a little reflection will reveal that a given star will be connected to a star $90^\circ$ away on the sky as a result of the field superposition. That star, in turn would be connected to another star $90^\circ$ away as the satellite spins, and so on. The end result is that any given target star will be `connected' to only three other fields, separated by $90^\circ$, $180^\circ$, and $270^\circ$. The resulting network of observations will have a poor degree of connectivity, or rigidity. Fields separated by $60^\circ$ perform a little better, but they are only connected to a total of five other regions along any great circle scan.
It follows that small integer fractions of $360^\circ$ (1/2, 1/3, 1/4, 1/5, etc.) are a poor choice if the goal is to create a network of star observations around the great circle with good rigidity. By extension, so too are small integer {\it multiples\/} of these angles, such as 2/3, 2/5, 3/7, etc.\ of $360^\circ$.
More formally, angles to be avoided are $360^\circ\times m/n$, for small integer values of $m$ and $n$.
Detailed studies in the early 1980s (\citet{1981A&A...101..228H}; see also \citet{1998A&A...340..309M}), resulted in the kind of diagram shown in Figure~\ref{fig:distortion-basic-angle}a. Here, the ordinate is a measure of the rigidity of the system as a function of the angle between the two viewing directions. Large values, and in particular the higher peaks, correspond to poorer great-circle rigidity. But between the peaks are small `valley' regions providing good rigidity.
Any of these minima would have worked well for Hipparcos and Gaia, and the choice mainly rested on the physical accommodation of the optical hardware. Underpinned by detailed studies, a value of $58^\circ$ was chosen for Hipparcos, and $106\ddeg5$ for Gaia. Some more intricate subtleties aside, many other choices in the range $40-140^\circ$ would have worked equally well.
In the earliest design studies for Gaia, a system with three viewing directions was also considered, the pairs separated by a carefully optimised $54^\circ$ and $68^\circ$. This would have contributed to a better rigidity of the astrometric solution, but was rejected for reasons of mass and complexity.

\paragraph{Sky scanning}

Gaia continuously scans the sky following a specified scanning `law'. This was designed following the principles pioneered by Hipparcos, and optimised to take account of the somewhat conflicting goals of
maintaining the thermal stability of the payload, 
providing a fairly uniform scanning of the celestial sphere,
allowing a robust `separation' of the astrometric parameters of each star through scans at different celestial orientations,
and minimising the impact of scattered light from the Sun.
More specifically, the satellite's spin axis precesses around the Sun direction at a constant Sun-aspect angle (for thermal stability). At the same time, the satellite revolves around this spin axis -- once every six hours for Gaia. The result is a repeating coverage of the celestial sphere as the Earth (and Gaia, at its L2 orbit with it) moves in its annual orbit around the Sun. Over the 10-year lifetime, the sky is covered reasonably uniformly, each star being observed at some 150 distinct epochs (although with a higher coverage at intermediate ecliptic latitudes), and at many different phase angles. These one-dimensional scans can eventually be used to create a two-dimensional reconstructed image (shown schematically in Figure~\ref{fig:iterative-solution}a). More details are given in \citet{2016A&A...595A...1G}. 

Observations extending over more than a year start to get a better measure of their proper motions, and observations over more than 18--24~months start to get a much better separation between the effects of the star's proper motion through space, and its parallax. As more measurements are accumulated over time, the accuracy on the five astrometric parameters ($\alpha$, $\delta$, $\mu_\alpha$, $\mu_\delta$, $\varpi$) improves.  For more complex binary or multiple stars, and especially in the case of orbital motion, more than 5 or 6 quantities are required to describe the star's motion, and more observations are required to permit their meaningful estimation.
Successive catalogue releases comprise more observations, and as the temporal baseline increases, the accuracy of the astrometric parameters improves as a result of these two effects. More observations also give better constraints on binary systems, on stellar variability, and on the orbits of solar system objects. While the detailed patterns of the sky scanning become less prominent with time, `imprints' can still be seen in some catalogue properties (e.g.\ Figure~\ref{fig:scanning-law}b). Note that, since the scanning motion is symmetric with respect to the {\it ecliptic\/} plane, this symmetry axis is offset in Galactic- or equatorial-based projections.

\paragraph{CCD windows}
\label{sec:windows}

Even at 20~mag, and at the angular resolution of Gaia, most of the sky is rather `empty'. Accordingly, for all of the astrometric, photometric, and radial velocity spectrometer fields, only small windows around the predicted positions of each confirmed object are read out. This avoids the need to read out all CCD pixels, thus allowing a lower readout frequency, and associated lower readout noise, for those sky regions which include a target object. This also avoids having to send all CCD data to the ground, which would have been impractical if not impossible.

The chosen window size, and the binning of pixels in the across-scan direction, uses one of three distinct schemes, dynamically assigned depending on the detected flux. All windows are 12~pixels across scan (2.1~arcsec).  Along scan, they are either 18~pixels (1.1~arcsec) for $G=13-16$, or 12~pixels (0.7~arcsec) for $G>16$~mag. For $G>13$~mag, windows are binned across scan to reduce the CCD readout noise. As a result, bright sources ($G<13$) have two-dimensional images from which both the along-scan and across-scan coordinates can be determined. Fainter stars have only one-dimensional images (of either 12 or 18~samples), from which only the along-scan coordinate can be determined.
Rules determine which windows are used in very dense sky regions, and in conflicts due to close binary stars. Images typically have a different {\it transverse\/} speed for each field, due to the satellite's precessional scan motion. With an integration time of 4.42~seconds per CCD, the transverse motion can reach 4.5~pixels over a single CCD, and can therefore result in significant across-scan smearing. As most samples are binned in this direction, the net effect on the observations is small.
Further details of the onboard detection and windowing are given by 
\citet{2016A&A...595A...3F}	
and
\citet{2018A&A...616A..17A}.	

\section{Data processing}
\label{sec:data-processing}

\subsection{Outline}

The entire scientific data processing (from satellite data reception, to the source catalogues) is undertaken by a single European group, the Gaia Data Processing \& Analysis Consortium, DPAC, distributed geographically, and described pre-launch by \citet{2008IAUS..248..224M}. DPAC consists of nine `coordination units', consisting of more than 400 scientists across 15 countries. These units assume responsibility for all activities involved:
CU1 (system architecture);
CU2 (simulations);
CU3 (core processing);
CU4 (object processing: multiple stars, solar systems objects, etc);
CU5 (photometric processing);
CU6 (spectroscopic processing, including radial velocities);
CU7 (variability processing, including science alerts);
CU8 (astrophysical parameters, derived from all Gaia data);
CU9 (catalogue validation).	
The bulk of the numerical computations takes place at several large data processing centres across Europe (Barcelona, Cambridge, CNES Toulouse, Edinburgh, ESAC Madrid, Geneva, and Turin).

Compressed satellite data, at around 10~Mbps, was sent from ESA's network of 35-m ground stations (in Spain, Australia and South America), via the Mission Operations Centre, to the central processing hub at ESA's ESAC (European Space Astronomy Centre) station outside Madrid.  The ESAC centre received an average of 35~GB of science and `housekeeping' data per day. 
%
Unpacked, decompressed, and with some early processing, this resulted in a daily increment in data volume of 300--500\,GB to be stored in the main data base. This, in turn, resulted in an average size of a `working data base' at ESAC of some 30\,TB. The first processing stages include `source matching', which associate each astrometric, photometric, and spectroscopic observation in the telemetry stream with its unique celestial source. 
Further technical details of the associated hardware and software engineering of this first stage have been given in various reports from around the time of launch
\citep[e.g.][]{
2014ASPC..485..487O,
2014SPIE.9149E..2ES,
2015ASPC..495...47H,
2019ASPC..523..339G}.	

The Gaia satellite data essentially encodes the position of each source along the great-circle scans, at more than 100 distinct epochs during the 10-year mission. This is supplemented by continuous information on the satellite's 3-axis attitude, with respect to the very stars that Gaia is observing, as well as photometric and spectroscopic data coming from the photometric and RVS parts of the focal plane (Figure~\ref{fig:gaia-fp-concept}). 
In a somewhat simplified picture, the overall data processing is organised as a `hub and spoke' system, with the Astrometric Global Iterative Solution, AGIS, undertaken by ESA at ESAC, at the hub. Astrometric results from AGIS, along with all the other data, is sent out to the other data processing centres for the activities under their responsibility. Their results are returned to ESAC, and all processes are iterated, now making use of the intermediate outputs of the other processing tasks as appropriate, including photometric calibration and variability analysis. 

All of these processing units face their own very substantial challenges, including complex calibrations, and huge data volumes. Here I will only outline the principles of AGIS, which is central and critical to the astrometric goals, and which provides a perspective of its complexity, the data interconnectivity, and the rationale underlying the staged catalogue data releases.

\begin{figure}[t]
\vspace{0pt}
\centering
\hspace{0pt}
\raisebox{11pt}{\includegraphics[width=0.262\linewidth]{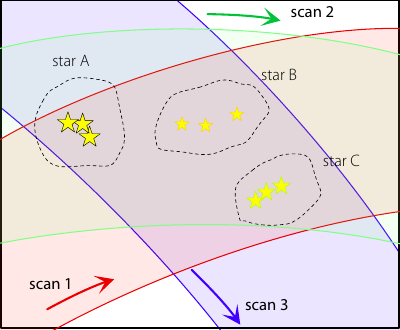}}
\hspace{10pt}
\includegraphics[width=0.34\linewidth]{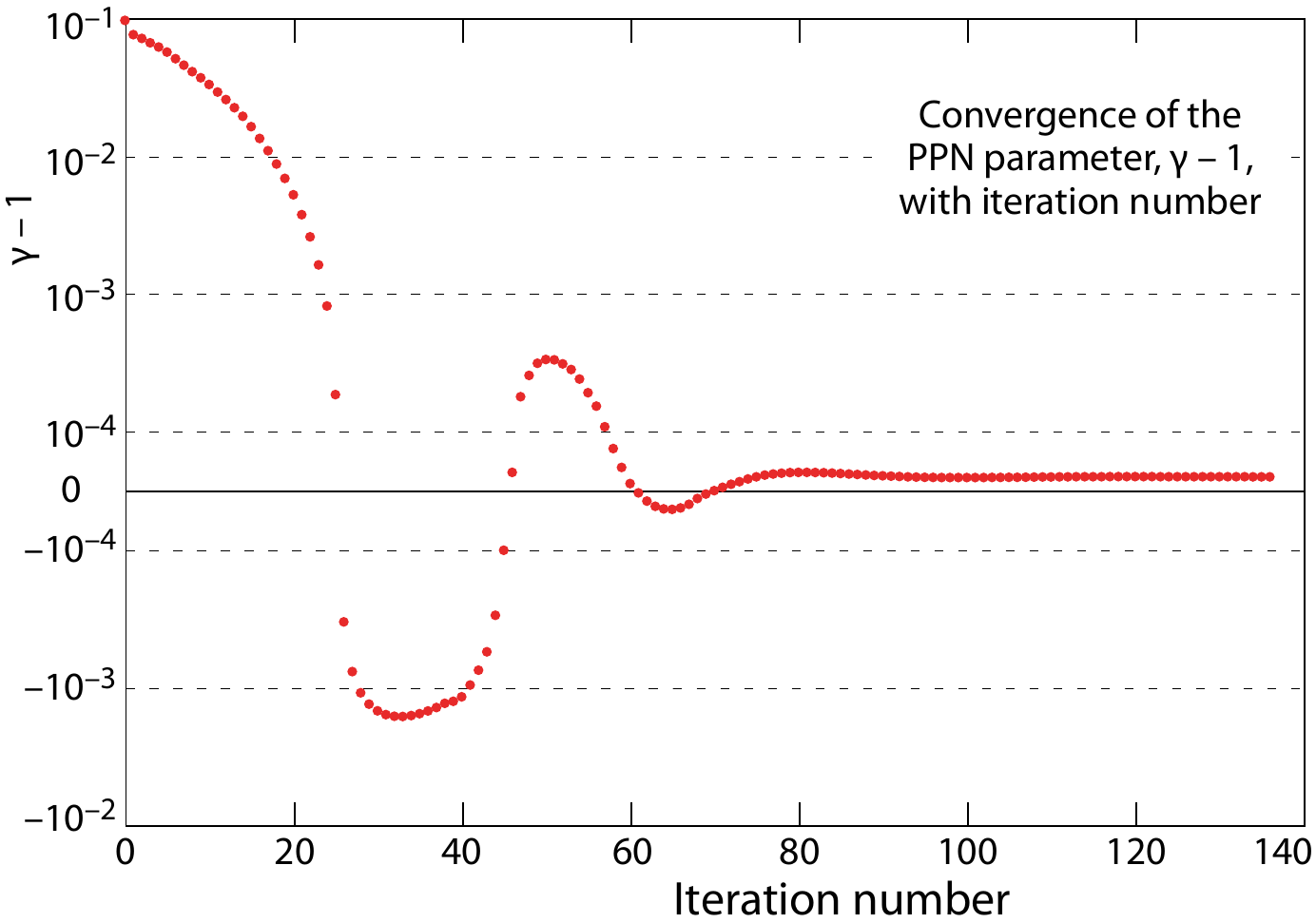}
\hspace{10pt}
\includegraphics[width=0.31\linewidth]{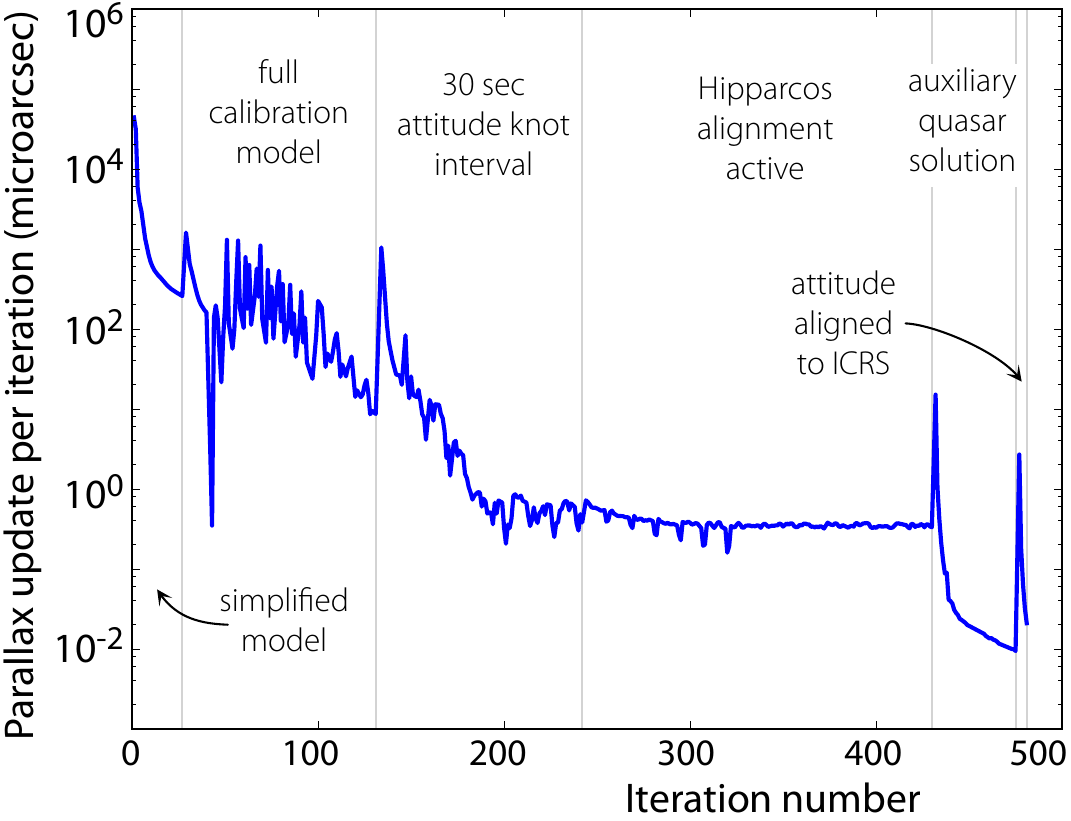}
\vspace{0pt}
\caption{The astrometric global iterative solution, AGIS. 
Left~(a): schematic showing three scans across a small region of sky; depending on the scanning motion across that part of the sky at those particular times, the interval between successive scans may be several hours, or several days or weeks. Between scans, all the stars have moved minutely, through a combination of their true motions through space, and their (apparent) parallax motions due to Earth's annual orbit around the Sun. Over months, and especially over several years, and with many more scans, enough information is collected to allow an estimate of each star's minuscule motion across the sky, along with any other motion that might affect it, such as orbital binary rotation, the effects of unseen planets, or gravitational light bending due to the Sun.
Middle~(b): convergence of the PPN parameter $\gamma$ as a function of iteration number in pre-launch simulations \citep{2012A&A...538A..78L}.
Right~(c): the evolution of parallax updates versus iteration number in the preparation of Gaia~DR1.  The effect of different regimes of the underlying calibration model are evident \citep{2016A&A...595A...4L}.}
\label{fig:iterative-solution}
\end{figure}

\subsection{Astrometric Global Iterative Solution, AGIS}
\label{sec:data-processing-agis}

AGIS is central to Gaia's objectives, and absolutely critical for its ultimate success. At the project's outset, it presented two distinct and substantial challenges: formulating the mathematical description of the astrometric solution, and its technical implementation. The latter posed a formidable task, both in terms of the amount of data to be treated (the total unknowns is of order $10^9$, and the solution treats $\sim\!10^{11}$ observations extracted from $\sim$100~petabyes of raw satellite data), and in the way in which the iterative solution has to be executed, with four key `blocks' (of source, attitude, calibration, and global parameters) being evaluated in a cyclic sequence until convergence. 

The basic challenge is the simultaneous estimation of this very large number of unknowns, representing four distinct types of information: 
(a)~the astrometric parameters for a subset of the observed stars, providing the astrometric reference frame; 
(b)~the satellite attitude, representing the celestial pointing of the instrument axes in that reference frame as a function of time; 
(c)~the geometric instrument calibration, representing the mapping from the CCD detectors to angular directions relative to the instrument axes; 
and 
(d)~a few `global' parameters describing, for example, a possible deviation of space--time from the prescriptions of general relativity.
Although the total number of stars observed is more than two billion, only a subset are used in the astrometric core solution (with the remainder `inserted' into the resulting reference frame). This subset, of some 100~million well-behaved ‘primary sources’, consists of (effectively) single stars and extragalactic sources (quasars) that are sufficiently point-like and stable over time. 
And in practice there are many other complexities, including the colour-dependencies (`chromaticity'), charge transfer inefficiency of the CCDs, and attitude irregularities due to thruster noise and micro-meteoroid impacts.

In its simplest form, four `blocks' are evaluated in a cyclic sequence until convergence. 
The blocks map to the four different types of unknowns: 
the source (star) update, S, in which the astrometric parameters of the primary sources are improved; 
the attitude update, A, in which the attitude parameters are improved; 
the calibration update, C, in which the calibration parameters are improved; 
the global update, G, in which the global parameters are improved. 
The blocks must be iterated because each needs data from the three other processes. For example, when computing the astrometric parameters, the attitude, calibration and global parameters are taken from the previous iteration. These updated astrometric parameters are used for the next A block run.
While this block-iterative solution is intuitive and appealing in its simplicity, its implementation faced many challenges in practice:
it is not obvious, mathematically, that it must converge. And if it does, it is not obvious how many iterations are required, whether the order of the blocks in each iteration matters, or even whether the converged results do, in fact, constitute a solution to the {\it global\/} minimisation problem.
Adding to the complexity is the fact that the core iterative solution also interfaces with all the other (enormous) processing tasks, amongst them the photometric analysis (including variability and `alerts'), the treatment of double and multiple systems, the radial velocity measurements, and the object classification algorithms. 

The first major demonstration solution, pre-launch, was run at ESAC on simulated data
\citep{2012A&A...538A..78L}.
Figure~\ref{fig:iterative-solution}b shows how one of the `global' terms, specifically the estimate of the parameter $\gamma$ (in the PPN formulation of general relativity), eventually converges, from some initial `assumed' value of 1.1, to a stable value close to 1.0, from around iteration 80 onwards. These 135 iterations required $3\times10^{17}$~floating point operations on a 0.65~teraflop/sec IBM 112~CPU cluster over 6~days. 
Figure~\ref{fig:iterative-solution}c illustrates the convergence of parallax updates in Data Release~1 from around 10~milli-arcsec during the initial iterations, to below 1~micro-arcsec by iteration~200. Different regimes of the calibration model were adopted as the iterations proceeded
\citep{2016A&A...595A...4L}.

For Gaia EDR3, the astrometric solution used 32~nodes, each comprising 64\,GB of RAM and 24~Intel-Xeon cores, providing 30~teraflop/s
\citep{2021A&A...649A...2L}. 
The solution for the 14.3~million primary stars used 40~TB of disk storage, with a further 20~TB for the secondary stars. A total of 181 iterations were executed, each outer iteration of the primary sources (performing the source, attitude, calibration and global blocks) taking 90~min. 
The primary star solution processed about $6.5\times10^9$ CCD observations, and determined 71.5~million source parameters, together with 10.7~million attitude, 1.1~million calibration, and 2.0~million global parameters. The secondary star solution processed $78\times10^9$ field-of-view transits, and generated 585 million 5-parameter, and 883 million 6-parameter, solutions.

\subsection{Astrometric parameters}
\label{sec:data-processing-astrometry}

In the most simple case, five astrometric parameters describe the time-dependent position and rectilinear space motion of a given star within some Cartesian coordinate system: two angular coordinates defining their position on the celestial sphere at a given reference epoch (in astronomical nomenclature, right ascension, $\alpha$, and declination, $\delta$, which correspond to latitude and longitude on the Earth's surface), two `proper motion' components which describe the star's space motion again projected on the celestial sphere ($\mu_\alpha$, $\mu_\delta$, expressed in seconds of arc per year), and the star's parallax, $\varpi$ (its angular displacement resulting from the Earth's orbit around the Sun, inversely proportional to its distance). The sixth Cartesian component of the star's position and space velocity is given by its radial velocity, $v_{\rm R}$ (viz., its space motion along the line-of-sight), which is not determinable by classical astrometric techniques, which determines only the star's position projected on the celestial sphere.  Radial velocities can only be obtained spectroscopically, and are being furnished in large numbers by Gaia's own Radial Velocity Spectrometer, RVS, as well as by ground-based radial velocity surveys (APOGEE, GALAH, LAMOST, RAVE, and others).
%
The EDR3 main table, {\tt gaia\_source}, gives astrometry for 1\,811\,709\,771 sources (Table~\ref{tab:data-release-table1}):
585\,416\,709 with \mbox{5-parameter} solutions,
882\,328\,109 with 6-parameter solutions, and
343\,964\,953 with 2-parameter solutions
\citep[][\S5.1]{2021A&A...649A...2L}.

The angular resolution along-scan (the minimum separation between sources with different identifiers) is 0.18~arcsec. However sources separated by $\lesssim0.6$~arcsec generally only have 2-parameter solutions in EDR3/DR3 \citep[][\S5.2]{2021A&A...649A...2L}. The resolving power, or `contrast sensitivity', depends on the separation and relative magnitude between components, and is $\sim$0.5~arcsec for equal magnitude stars ($\Delta G=0$), up to $\sim$1.2~arcsec for $\Delta G=5$~mag 
\citep[][\S4.1.3]{2021A&A...649A...6G}.

\subsubsection{Higher-order astrometric effects}
\label{sec:astrometry-higher-order}
The Hipparcos mission, targeting accuracies at the milli-arcsec level, required a number of refinements in the modelling of astrometric observations that had not previously been relevant, notably the effects of gravitational light deflection by the Sun (which, even for Hipparcos, was a significant effect over the entire celestial sphere), and relativistic stellar aberration. In turn, Gaia has required further advances in data modelling to account for the even more subtle physical effects that are relevant at the micro-arcsec level. Section~\ref{sec:light-deflection} includes more on the effects of light deflection, notably by the Sun and Jupiter, which is taken into account in the astrometric processing, and which also allows new tests of general relativity. Section~\ref{sec:perspective-acceleration} includes details of perspective acceleration, a purely geometric effect which, because of a star's radial motion, results from the changing distance and angle between
the full velocity vector and the line-of-sight. Perspective acceleration has some observable consequences for a small number of nearby stars, as well as in the proper motions of a number of globular clusters with large radial velocities and parallax. It is also relevant in the propagation of source positions from one epoch to another.

The basic assumption that stars (or the barycentres of multiple systems) move with constant velocity with respect to the solar system barycentre requires more careful consideration at the levels of 1\muas. A more complete model (which includes relativistic aberration and gravitational light deflection, relativistic treatment of the satellite motion, and a relativistic formulation for the parallax and proper motion) was given by 
\citet{2003AJ....125.1580K},	
and is the basis for the astrometric processing of the Gaia data, as described by
\citet{2012A&A...538A..78L}.	
A didactic introduction is given by Sergei Klioner in Chapter~5 of 
\citet{2013asas.book.....V}.	

\subsubsection{Goodness-of-fit statistics and binaries}
\label{sec:ruwe}

Amongst many aspects of the data processing that could be mentioned, the quality of the astrometric solutions merits attention.
%
For EDR3/DR3, and in addition to the astrometric quantities and their standard errors (and certain correlations), the solutions provide various goodness-of-fit statistics
\citep[][\S5.1--5.3]{2021A&A...649A...2L}. 
These include three measures determined at the image parameter determination (IPD) stage, and two which characterise the quality of the astrometric solution: 
the renormalised unit weight error (RUWE), and the astrometric excess noise. Both aim to quantify how the motion of the image centre deviates from the motion described by the specified astrometric solution. The RUWE has proven a particularly valuable indicator of the reliability of the Gaia astrometry (e.g.\ the specified parallax). Its construction starts from the unit weight error (UWE), defined as the square root of the normalised chi-square of the astrometric fit to the (along-scan) observations, $[\,\chi^2/(n-n_\text{p})]^{1/2}$, where $n$ is the number of good CCD observations, and $n_p=5$ or 6 is the number of parameters fitted. Because of imperfect calibrations, the RUWE is calculated by including an empirical scaling of the UWE, by functions depending on the source's $G$ magnitude and colour. The result is that RUWE should be $\simeq$1.0 for well-behaved sources, with RUWE~$\gtrsim1.4$ frequently (and sometimes conservatively) taken as indicating the presence of unmodeled binarities.
There is evidence that this single quantity can be used to identify binaries down to separations of $\sim$20~milli-arcsec 
\citep[e.g.][]{2024MNRAS.527.3076D}.	

Inferences exploiting RUWE can be found across the Gaia literature. Examples are
in its dependence on simulated binary properties
\citep{2020MNRAS.495..321P},	
in the parallaxes of stars in globular clusters
\citep{2021A&A...649A..13M},	
as well as in benchmark eclipsing binaries
\citep{2021ApJ...907L..33S},	
hot subdwarfs
\citep{2023ApJ...942..109L},	
subgiants
\citep{2024MNRAS.530.2953Y},	
B/Be stars
\citep{2024MNRAS.527.3076D},	
and in very metal-poor stars
\citep{2022ApJS..263...29X}.	
In a wide-ranging analysis
\citep{2020MNRAS.496.1922B},	
RUWE is shown 
to be consistent with known spectroscopic binaries, 
to increase for massive young stars, 
to be elevated for nearby blue stragglers and blue horizontal branch stars,
and to hint at additional companions to the hosts of extrasolar hot Jupiters.
The high-velocity tail of wide-separation binaries found by
\citet{2019MNRAS.488.4740P}		
may also be largely explicable by such unresolved binaries
\citep{2020MNRAS.496.1922B,		
2020MNRAS.491L..72C}.			
Indeed,
\citet{2020MNRAS.495..321P}
start their abstract with the caution: {\it `Most binaries are undetected'}!

\subsubsection{Distance estimation}
\label{sec:distance-estimation}

By definition, the inverse of a source parallax in arcsec gives its distance in parsec. For nearby stars with parallax errors below $\sim$20\%, this simple prescription serves well.
Parallax {\it errors\/} complicate matters considerably, and particularly so for more distant sources. Firstly, even a strictly normal (Gaussian) error distribution implies the existence, even in a perfectly constructed catalogue, of negative parallaxes, which are evidently non-physical.  Moreover, since a normal distribution of parallaxes results in a non-Gaussian distribution of the (reciprocal) distance errors, it follows that distance estimates derived from the inverse of the parallax are biased. I refer to \citet{2015PASP..127..994B} for a didactic introduction to this problem, and the role of possible priors. Later papers provide further details, along with various prescriptions for improved estimates
\citep{2016ApJ...832..137A,
2016ApJ...833..119A,
2018AJ....156...58B,
2021AJ....161..147B,
2023AJ....166..269B}.
These contributions include catalogues constructed from Gaia EDR3 of 1.47~billion geometric and 1.35~billion photo-geometric distances which adopt a prior constructed from a 3d-model of the Galaxy, which includes interstellar extinction and Gaia's variable magnitude limit
\citep{2021AJ....161..147B}.

\subsubsection{Parallax zero point and parallax bias}

Any {\it systematic\/} parallax errors add further complications. Even small systematics affect estimates of the mean properties of distant populations. Examples include determining the distance scale using Cepheids 
\citep{2021ApJ...908L...6R}
or the tip of the red giant branch 
\citep{2023ApJ...950...83L},
or understanding the dynamics of the LMC, or of halo streams. 

Gaia's two widely-spaced fields of view, and whole-sky revolving scanning (Section~\ref{sec:basic-angle}), in principle allows the measurement of {\it absolute\/} parallax distances. However, even small variations of the `basic angle' between the two fields, and in particular periodic variations caused by heating of the rotating satellite by solar radiation, can lead to variations degenerate with the astrometry, so leading to a {\it global\/} shift of the parallaxes. Identified as a concern for Hipparcos in a technical note by Lindegren (1977), its effect on the Hipparcos parallaxes was estimated to be negligible
\citep{1995A&A...304...52A},		
implying good short-term stability of its basic angle.
For the much higher accuracies targeted by Gaia, the near-degeneracy between a possible basic-angle variation induced by solar heating, and a global parallax offset, is particularly problematic. This led to the inclusion of a dedicated laser-metrology system on board Gaia (the basic angle monitor, or BAM) to measure short-term variations
\citep{2017A&A...603A..45B}.		

Measurements during the first year of operations suggested an amplitude of the relevant Sun-aspect term ($\cos\Omega$, where $\Omega$ is the spin-phase relative to the barycentre) of amplitude 848\muas, corresponding to a parallax zero-point offset of some 700\muas\
\citep{2016A&A...595A...4L}.		
For Gaia DR1, observations were corrected for the basic angle variations based on a simple harmonic fit to the laser-metrology measurements
\citep{2016A&A...595A...2G}.		
The second data release, DR2, allowed for a more detailed analysis. From 500\,000 quasars, which also define Gaia's celestial reference frame, 
a global parallax offset of $-29$\muas\ was estimated 
\cite[][\S5.2]{2018A&A...616A...2L}.	
But plots versus magnitude and colour reveal systematic trends of $\sim$20\muas\ over the relevant data ranges. The plot against ecliptic latitude furthermore shows a roughly quadratic variation, with parallaxes $\sim$10\muas\ smaller towards the ecliptic poles.\footnote{The term `parallax zero-point' refers to the single value arising from unaccounted basic angle variations. The term `parallax bias' embraces corrections dependent on source parameters (notably magnitude, colour, and sky position) originating from current calibration imperfections.}

For EDR3, one million quasars in the range $G=13.4-21$ gave a weighted mean parallax of --21\muas\
\citep{2021A&A...649A...4L}.	
The parallax bias again depends, non-trivially, on magnitude, colour, and ecliptic latitude, and with different dependences for the 5- and 6-parameter solutions (their \S5). 
\citet{2021A&A...649A...4L}	
provided provisional bias functions $Z_5$ and $Z_6$ (see their Figs~21--22), to be subtracted from the catalogue value, as Python implementations at the 
\href{https://www.cosmos.esa.int/web/gaia/edr3-code}{Gaia web pages}
(which in turn points to the gitlab-hosted 
\href{https://gitlab.com/icc-ub/public/gaiadr3_zeropoint}{\tt gaiadr3\_zeropoint}).
More detailed dependencies, including as a function of angular scales, are given in the description of the EDR3 astrometric solution 
\citep{2021A&A...649A...2L},		
while verification of the biases has been examined through comparisons with open clusters 
\citep{2021A&A...649A...5F}.	
An independent treatment of the parallax bias in EDR3, also using quasars and physical binaries, and also revealing both spatial and magnitude dependences, but less so on colour, was given by 
\citet{2021A&A...654A..20G}.	

Data Release~3 (DR3) contains the same source list, positions, proper motions, parallaxes, and broad-band photometry as EDR3, such that the systematic errors present in the astrometry published in Gaia EDR3 carry over to Gaia DR3
\citep[][ \S3.3]{2023A&A...674A..41G}. 	
Accordingly, the parallax bias functions $Z_5$ and $Z_6$ (and the Python implementations) noted above, apply equally to both Gaia EDR3 and DR3.
These bias functions have since been used in many detailed studies
\citep[e.g.][]{2023AJ....166..132E,	
2023A&A...674A..28F,			
2023ApJ...950...83L,				
2023MNRAS.523.2369S,			
2023ApJ...942...12W,			
2024PASA...41...43E,			
2024MNRAS.52711559N,			
2024A&A...690A.327V,			
2024Natur.634..809S}.			

I have noted some 40 papers in the Gaia literature which make reference to their own estimates of the parallax bias: around 20~each for DR2 and EDR3/DR3. They include source samples ranging from Cepheids, globular clusters, eclipsing binaries, seismology, and VLBI, largely with reasonably comparable findings. 

\subsection{Reference frame}
\label{sec:reference-frame}

\paragraph{Background}

Coordinate {\it systems\/} (establishing the principles) and coordinate {\it frames\/} (their practical realisation) are intricate in practice, but the principles are straightforward, and analogous to the geographical framework of longitude and latitude used to define locations on the Earth's surface. In this equatorial coordinate system, astronomers agree on an origin for `right ascension' (the equivalent of longitude) and for `declination' (the equivalent of latitude). 
The origin of right ascension was chosen long ago, by Hipparchus around 130~BCE. This `First Point of Aries', or `vernal equinox', is one of the two points on the celestial sphere at which the celestial equator (the imaginary circle in the same plane as Earth's equator) crosses the ecliptic (Earth's orbital plane around the Sun). In the same way, declination is defined with respect to the Earth's equator, north and south from 0~to $\pm90^\circ$.

The problem gets more complicated because the Earth's spin axis is not inertially fixed in space, but precesses slowly westward about the poles of the ecliptic, with a period of 26\,000 years. This causes the equatorial coordinates of celestial objects to change continuously, by about $1^\circ$ in right ascension over 70~years. The problem is further compounded by the shorter term effects of nutation and polar motion. 
This led to the choice of reference systems which were revised, every few decades, by adjusting the epoch at which the Earth's coordinate system was specified.  Thus, over the past 150~years, astronomers have used reference systems which were successively specified by the 
Besselian epochs B1875, B1900, and B1950, and subsequently the Julian epoch J2000.
As positional measurements improved, the complex motion of the Earth introduced effects which were increasingly difficult to explain, and to account for.  These wobbling terms include not only the Sun and Moon's gravitational torques of precession and nutation, but the many complex effects responsible for polar motion: some internal to the Earth, others forced by climatic and seasonal changes due to oceans, tectonic plate motions, and many others. This led, in turn, to parallel efforts to construct a `dynamical reference system', linked to the observed motion of solar system bodies, whose orbits around the Sun should be largely decoupled from the motion of the Earth.

By the 1990s, radio VLBI measurements became possible, for a few dozen radio stars and quasars, at higher accuracies than were possible using optical measurements from the Earth. In consequence, the celestial reference system adopted by the International Astronomical Union, the ICRS (International Celestial Reference System) moved to one defined at radio frequencies and, in particular, one tied to distant quasars which better represented the ideas of an inertial reference system. By making these measurements from space, Hipparcos and Gaia achieve vastly improved accuracies from above the Earth's perturbing atmosphere. At the same time, measurements from a space platform means that they were freed from the complicating effects of the Earth's spin-axis motion. 

As the definition of the Hipparcos observing programme took shape in the early 1980s, plans were put in place to include stars that could be used, once the catalogue was finalised, to link the rigid reference frame defined by its 120\,000 stars to an extragalactic inertial reference framework, and in the process estimate and correct for any tiny offset in the parallax zero-point. The goal was to determine the global orientation and rotation (or spin) of the coordinate frame defined by the Hipparcos positions with respect to extragalactic sources. The big difficulty was that, because of its limiting magnitude of about 12~mag, only one quasar, 3C~273, could be included in the observing programme, and even that was so faint that it contributed very little to the final link. The effort required to establish this link was substantial. The contributions of several groups over a number of years, and using a variety of less direct techniques, were essential. 
These included interferometric observations of radio stars by radio interferometry (VLBI, MERLIN and VLA); 
observations of quasars relative to Hipparcos stars using CCDs, photographic plates, and Hubble Space Telescope; 
photographic programmes to determine stellar proper motions with respect to extragalactic objects; 
and a comparison of Earth orientation parameters obtained by VLBI and others. 
Combined and suitably weighted, the coordinate axes of the published catalogue were finally believed to be aligned with the extragalactic radio frame to within $\pm0.6$~mas at the mid-catalogue epoch J1991.25. And it was estimated to be `non-rotating' with respect to distant extragalactic objects to within $\pm0.25$~mas/yr
\citep{1997A&A...323..620K}.	

\begin{figure}[t]
\centering
\includegraphics[width=0.60\linewidth]{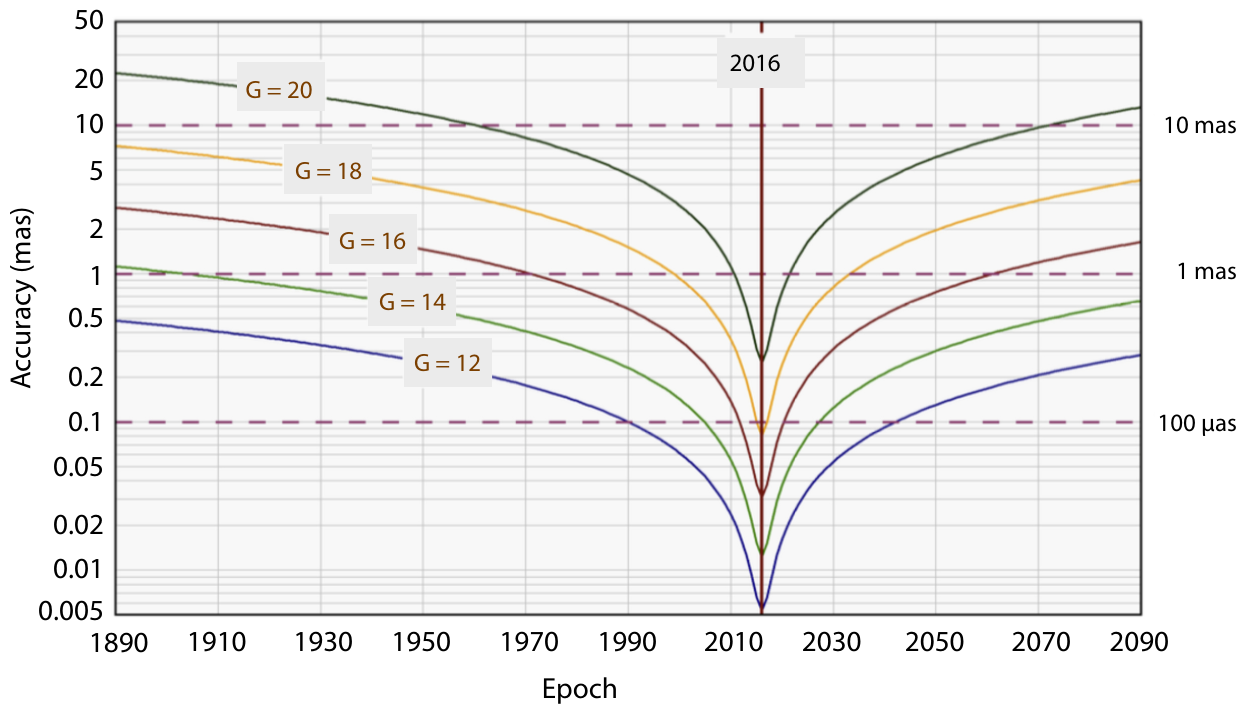}
\caption{Positional accuracy of the Gaia catalogue over time, due to the uncertainties in the proper motions (\citet{2016arXiv160907325H}, Figure~4b).
}
\label{fig:accuracy-vs-time}
\end{figure}

\paragraph{ICRF and Gaia--CRF}

The problem of specifying the reference frame is as central to Gaia as it was for Hipparcos, with the required accuracy correspondingly more demanding. But there is one crucial difference: Gaia's limiting magnitude, at 20--21~mag, allows very large numbers of quasars, all across the sky, to be observed directly.\footnote{
The ICRS (International Celestial Reference {\it System}) is the theoretical framework defining the celestial coordinate system adopted by the International Astronomical Union, while the ICRF (International Celestial Reference {\it Frame}) is its practical realisation, i.e.\ a catalogue of objects materialising the ICRS. 
The original ICRF, based on the VLBI-derived positions of 608 extragalactic radio sources
\citep{1998AJ....116..516M}, 	
was adopted by the IAU in 1998. Later, two extended versions were released: ICRF2 in 2009
\citep{2009ITN....35....1M},
and ICRF3 in 2020
\citep{2020A&A...644A.159C}. 
The latter is adopted by the IAU as the current implementation of the ICRF at radio wavelengths.
The improving reference frames independently being established by Gaia (Gaia--CRF1 accompanying DR1, Gaia--CRF2 accompanying DR2, Gaia--CRF3 accompanying DR3) represent the materialisation of the ICRS in the optical. Comparisons between the radio and optical realisations aid identification of their respective shortcomings.}  The problem was considered in depth during the mission's feasibility study before selection in 2000. Studies indicated that some 500\,000 quasars would be observable, with a mean sky density of about 25~per square degree. Issues of sky uniformity, colour dependency, and possible small structural changes in position were all considered.

The second data release, Gaia~DR2, contains the positions of 556\,869 quasars, extending to $G=21$~mag, and defining a kinematically non-rotating reference frame in the optical, Gaia--CRF2
\citep{2018A&A...616A..14G}.		
A subset have accurate VLBI positions allowing the reference frame axes to be aligned with the radio reference frame. The optical positions for a subset of 2820 sources in common with the ICRF show very good overall agreement with the radio positions.
Gaia--CRF2 was duly superseded by Gaia--CRF3
\citep{2022A&A...667A.148G},		
the updated celestial reference frame for positions and proper motions adopted for EDR3 and DR3 (which contain identical astrometric results). For this purpose, compact extragalactic sources in Gaia DR3 were identified by positional cross-matching with 17 external catalogues of quasars and active galactic nuclei (AGN).
Gaia--CRF3 comprises 1.6~million quasar-like sources, of which 1.2~million have 5-parameter astrometric solutions in Gaia DR3, and 0.4~million have 6-parameter solutions. They populate the magnitude range $G=13-21$~mag, with a peak density at 20.6~mag, at which the typical positional uncertainty is about 1~mas. The proper motions show systematic errors on the level of 12\muasyr\ on angular scales greater than $15^\circ$. 

For the 3142 optical counterparts of ICRF3 sources in the radio (S/X) frequency bands, the median offset from the radio positions is about 0.5~mas, but it exceeds 4~mas in either coordinate for 127 sources.
%
Subsequent examination of the quality of the present Gaia reference frame, Gaia--CRF3, include
comparisons with the radio (VLBI) reference frame, ICRF3
\citep{2023AJ....166....8M,		
2024ApJS..274...28C,		
2024A&A...689A.134L,		
2024AJ....167..229M},		
comparisons with HST proper motions in M31
\citep{2024A&A...692A..30R},	
the effects of colour-dependent variability
\citep{2024ApJS..274...27M}, 	
and positions from pulsar timing
\citep{2023A&A...670A.173L}.	

Each data release has its own specified catalogue reference epoch (listed Table~\ref{tab:data-release-table1}), corresponding approximately to the mean epoch of the observations relevant for each data release. Catalogue positions for each source are those precisely applicable at that specified reference epoch. Note also that the uncertainties in the proper motions of all sources means that positional errors degrade (linearly) with time, away from the catalogue reference epoch, and the accuracy of the global reference frame with it (Figure~\ref{fig:accuracy-vs-time}).

Construction of the Gaia celestial reference frame is founded on the premise that the system of distant quasars does not rotate as a whole. A currently unexplained finding is the different vector spherical harmonic terms for the 1.57~million Gaia-based quasar proper motions when divided into redshift bins. Significant differences are seen between the $z=1-2$ and $z=2-3$ samples, and include a rigid spin, a dipole `glide' along the Galactic rotation axis, and an additional quadrupole term
\citep{2022ApJ...927L...4M,	
2025arXiv250701798M}.	

\subsection{Photometry}
\label{sec:photometry}

\begin{figure}[t]
\centering
\raisebox{10pt}{\includegraphics[width=0.46\linewidth]{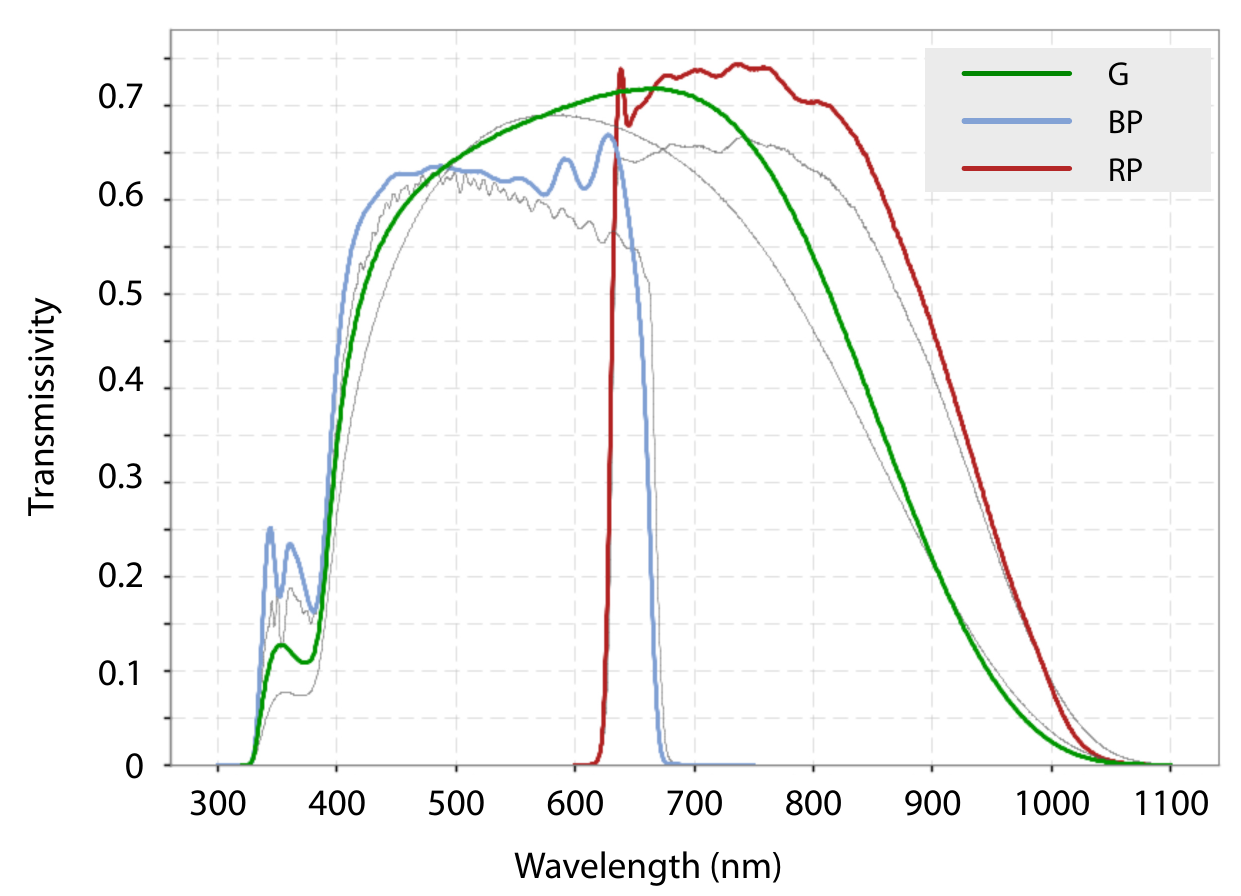}}
\hspace{30pt}
\includegraphics[width=0.44\linewidth]{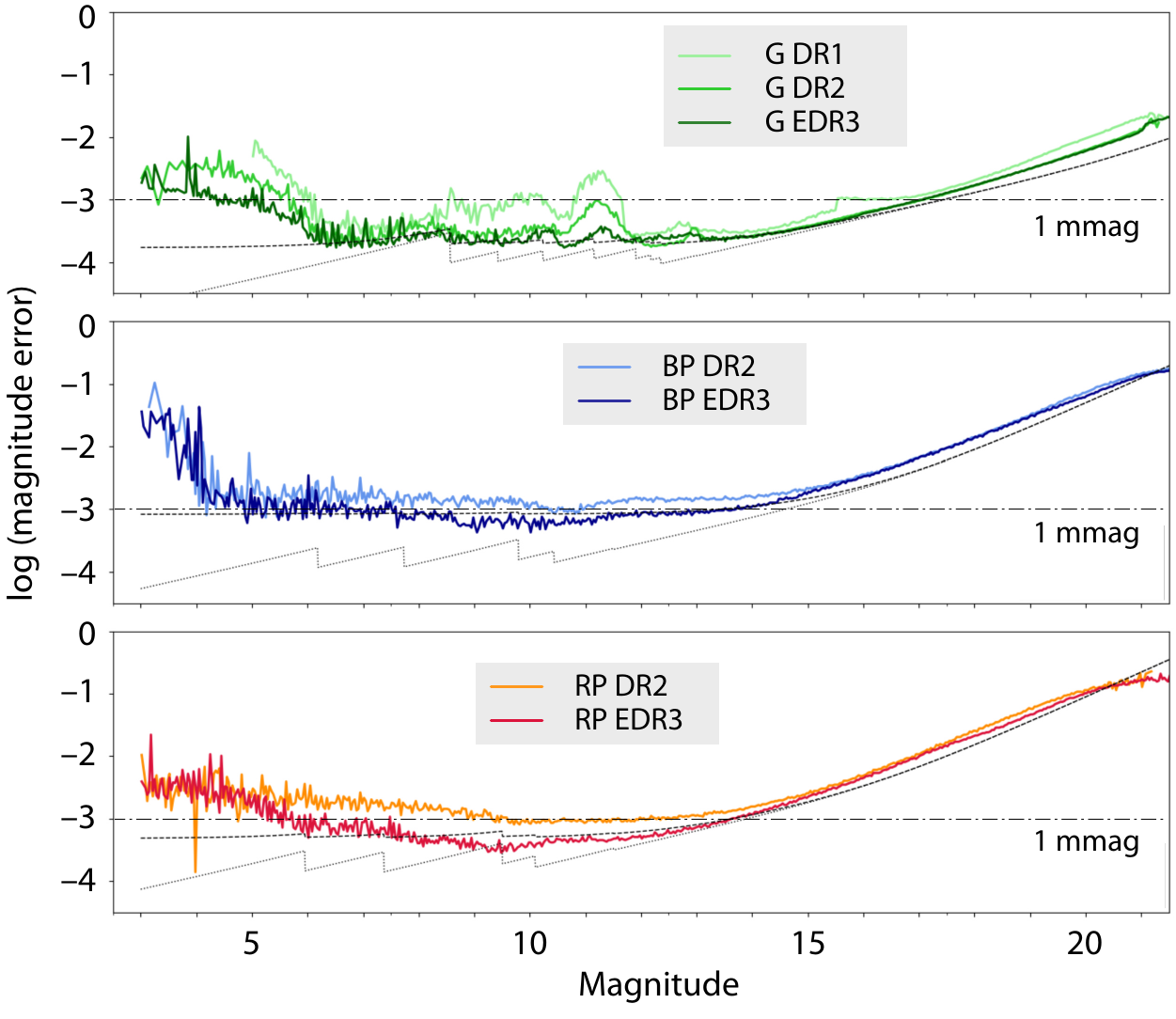}
\vspace{-5pt}
\caption{Left~(a): $G$, $G_{\rm BP}$, and $G_{\rm RP}$ passbands for the Gaia EDR3 photometric system (grey curves are the nominal pre-launch systems).
Right~(b): uncertainties on the weighted means as a function of magnitude. Only sources with 20 or more transits (200 CCD observations in $G$) are included. The Gaia DR1 and DR2 uncertainties are shown for comparison. From \citet{2021A&A...649A...3R}, Figures~24 and 14.
}\label{fig:edr3photometry}
\end{figure}

Gaia observes in three bands, $G$, BP and RP. The primary astrometric field, $G$, maximises photon throughput, and spans the broadest wavelength range 330--1050\,nm, defined by the coated mirror reflectivities and the CCDs (Figure~\ref{fig:edr3photometry}a).
As described in Section~\ref{sec:gaia-design}, the BP and RP data (together often referred to as XP) are acquired as low-resolution spectra ($\lambda/\Delta\lambda\simeq25-100$) over 330--680\,nm and 640--1050\,nm respectively\footnote{
As a historical note, and relevant in the context of its `synthetic photometry' (Section~\ref{sec:synthetic-photometry}), the acquisition of multi-colour photometry in parallel with the astrometric observations was a design goal from the earliest days of Gaia, and was originally foreseen (and extensively optimised) as a set of (eventually 19) broad- and narrow-band filters 
\citep{2006MNRAS.367..290J}. 
Later in the satellite design stage, the dedicated photometric telescope with its associated filters was exchanged for two fused silica prisms which disperse the light entering the main telescope's two fields of view. Their resolution, which matched the earlier filter design, results from the intrinsic dispersion of fused silica, and varies from 3--27\,nm per pixel over 330--680\,nm for BP, and from 7--15\,nm per pixel over 640--1050\,nm for RP.}.
As part of Data Release~3, the spectra have been subject to an extensive validation 
\citep{2023A&A...674A...2D},	 	
and internal/external calibration 
\citep{2021A&A...652A..86C, 	
2023A&A...674A...3M}.		

The photometric processing for Data Release~3, follows the same principles as for previous data releases, and
\citet{2021A&A...649A...3R}	
describe the input data, algorithms, processing, and validation. 
As a result, Gaia DR3 provides mean $G$ magnitudes for 1.8~billion sources, and mean $G_{\rm BP}$ and $G_{\rm RP}$ photometry for 1.5~billion, typically at sub-millimag precision (Table~\ref{tab:data-release-table1}). 
The median uncertainty in the $G$-band photometry is estimated as 0.2~mmag at $G=10-14$, 0.8~mmag at $G\simeq17$, and 2.6~mmag at $G\sim19$, with no trends larger than 1~mmag mag$^{-1}$ (Figure~\ref{fig:edr3photometry}b).
In addition, mean BP and RP {\it spectra\/} are provided for 220~million sources, mostly with $G<17.5$~mag. The final data release, DR5, foreseen around 2030, will extend this spectral library to all 2--3~billion sources.

\subsubsection{Synthetic photometry}
\label{sec:synthetic-photometry}

In principle, the calibrated BP/RP (or XP) spectra can be used to convert the Gaia photometry to any other (optical) photometric system, subject to two (obvious) constraints: that the target passband must be enclosed within the spectral range covered by Gaia's XP spectra (330--1050\,nm), and that the passband's characteristic width must exceed the `line-spread function' of the XP spectra at the relevant wavelength. 
Such synthetic photometry, derived from observed spectrophotometry, is based on the computation of a normalised mean flux obtained by integrating the product of a transmission curve, $S(\lambda)$, and a spectral energy distribution over a given wavelength (or frequency) interval, depending on the adopted photometric system 
\citep[e.g.][]{2005ARA&A..43..293B}. 
It can be used, amongst others, to provide colours within standard systems, for insights into the spectra or colours of unusual stars or stellar systems, and for the validation or re-calibration of existing photometric surveys.
Examples might include, for example, extending several years of Sloan Digital Sky Survey photometry in its five photometric bands ($u, g, r, i, z$) using the Gaia data, or transforming the Gaia photometry to the systems used by Pan-STARRS or the Vera Rubin Observatory. 

The application to Gaia, using the XP spectra, is fully described by
\citet{2023A&A...674A..33G}.
In the Gaia archive, the XP spectra are stored as the coefficients (and their covariances) of a set of orthogonal basis functions, from which the spectral energy distributions can be reconstructed 
\citep{2023A&A...674A...2D,	
2023A&A...674A...3M}.		
A key step in the derivation of the XP synthetic photometry is `standardisation'
\citep[][\S2.2]{2023A&A...674A..33G}. 
For Gaia, this compensates any systematics present in the externally calibrated XP spectra by `tweaking' the Gaia transmission curve to minimise any residual magnitude or colour differences.

They detail a number of important photometric systems which they have reconstructed from the Gaia XP-derived synthetic photometry. Their general conclusions are that existing high-quality photometry can be reproduced within a few per cent over a wide range of magnitudes and colour, for wide and medium bands, and with around milli-mag accuracy. 
For example, in their  external calibration of the XP spectra
\citep[][\S8.4.2]{2023A&A...674A...3M},
they showed that the Hipparcos $H_{\rm p}$, $B_{\rm T}$, and $V_{\rm T}$ photometry, recognised as `a benchmark of excellent precision' \citep{2005ARA&A..43..293B}, 
are all reproduced by the XP synthetic photometry to better than 2.5~milli-mag. 
The work 
\citep{2023A&A...674A..33G}
continues by detailing performances for various other wide-band systems, including
the Sloan Digital Sky Survey,
the Johnson--Kron--Cousins system,
and the PanSTARRS and HST systems.
Amongst narrow-band photometric systems, they give results for 
the Str\"omgren system (widely used for the determination of effective temperature and surface gravity),
the Javalambre J--PAS and J--PLUS surveys,
and the IPHAS H$\alpha$ emission-line survey.

One of their examples is replicating (and demonstrating the robust performance of) the original proposed Gaia photometric {\it filter\/} system: the set of broad- (C1B) and medium-band (C1M) filters, carefully designed to maximise the scientific return in terms of \teff, metallicity, gravity, reddening, and $\alpha$-element abundances
\citep{2006MNRAS.367..290J}. 

To make this synthetic photometry more readily accessible, 
\citet{2023A&A...674A..33G}
provide two derived catalogues. The first is the Gaia Synthetic Photometry Catalogue (GSPC), which includes the majority of the 220~million stars with XP spectra released in Gaia DR3, in 13 passbands, including UBVRI in the JKC system, {\it ugriz\/} in the SDSS system, and two in the HST--ACS/WFC system. The other is the Gaia Synthetic Photometry Catalogue for White Dwarfs (GSPC--WD), comprising 100\,000 white dwarfs with DA/non-DA classification.	

\subsubsection{Calibration of other photometric systems}

Gaia's all-sky, multi-epoch measurements yield photometry which rivals the best available, and can be used to calibrate and correct other photometric surveys. Amongst these are
the 6-colour zero-points of ANU's SkyMapper Southern Survey, SMSS
\citep{2021ApJ...907...68H},
and 200\,000 secondary standards in the Johnson--Kron--Cousins system
\citep{2022A&A...664A.109P}. 	
Work already making use of the GSPC includes correcting patterns in the DESI (Dark Energy Spectroscopic Instrument) Legacy Imaging Surveys
\citep{2023RNAAS...7..105Z},	
corrections for OGLE variables in their measurement of TRGB magnitudes
\citep{2024ApJ...963L..43A},		
recalibration of the 12~passbands of the Javalambre Photometric Local Universe Survey (J--PLUS)
\citep{2024A&A...683A..29L},	
and 
recalibration of narrow-, medium- and broad-band photometry from the Southern Photometric Local Universe Survey (S--PLUS)
\citep{2024ApJS..271...41X}.		
Independent efforts are also ongoing to correct residual systematics in the Gaia DR3 XP spectra using various external spectral libraries
\citep{2024ApJS..271...13H}.	

\subsection{Radial velocities}
\label{sec:rvs}

\subsubsection{Wavelength range}
\label{sec:rvs-wavelength}

Extensive studies were made when deciding to make radial velocity observations onboard Gaia, rather than relying on ground-based observations
\citep{1995ESASP.379..165B,
1995IAUS..166..211P}.
Advantages included guaranteed observations of all sufficiently bright stars, in a coherent system across the celestial sphere, at multiple epochs coinciding with the astrometric and photometric observations (of particular merit for variable and multiple stars), with a superior accuracy due to the absence of atmospheric scintillation, and an improved control of systematics.
Disadvantages included cost, mass, instrument complexity, and the additional risks for the primary astrometric measurements, including thermal and mechanical stability. The scientific case was clear, although the politics of including experiments onboard that could, in principle, be done from the ground, was non-trivial.

The main motivation for acquiring radial velocities for as many stars as possible was to provide the third component of the star's space velocity (given that astrometry can only measure its projected motion across the sky, viz.\ its `proper motion'). Without it, kinematic and dynamical insights of individual objects -- and indeed entire populations -- would be greatly restricted.
In practice, faint and bright regimes can be loosely distinguished. Faint targets will mostly be distant stars, which will be of interest as tracers of Galactic dynamics. Here, the uncertainty in the tangential space motion is dominated by the parallax error, and a radial velocity accuracy of order 5\kms\ was considered acceptable for statistical purposes.  
Brighter stars, $<15$~mag, are of more individual interest. Here the radial velocity would also be valuable for the determination of perspective acceleration (Section~\ref{sec:perspective-acceleration}), while multiple observations would provide an important indicator of multiplicity, in the process averting errors in the star's space motion.

The choice of spectral region, and spectral resolution, was studied in detail as part of the Concept and Technology Study (ESA--SCI(2000)4), and in particular in a series of papers by Ulisse Munari from the University of Padua
\citep{1999BaltA...8...73M,
2001BaltA..10..613M,
2000A&AS..141..141M}.
Most Gaia stars are intrinsically red, and made even redder by interstellar absorption, and a red spectral region was therefore preferred. To maximise the radial velocity signal even for metal-poor stars, strong, saturated lines are desirable. Broad lines also allow the radial velocity to be derived from moderate-resolution spectra: only a sampling of the lines is needed to derive the radial velocity, while oversampling leads to only marginal improvements in accuracy. The chosen spectral resolution was also a tradeoff between information content, signal-to-noise, crowding (overlapping spectra) in the focal plane, and data volume.

Beyond the H$\alpha$ line at 656.3\,nm, three strong spectral features are present in late-type stars: the potassium (K\,{\footnotesize I}) doublet near 768\,nm, the sodium (Na\,{\footnotesize I}) doublet at 819.4\,nm, and the calcium (Ca\,{\footnotesize II}) triplet near 860\,nm. 
The K\,{\footnotesize I} doublet has only a moderate equivalent width ($W = 0.01$--$0.04$\,nm), and a complicated luminosity dependence. It is also a strong interstellar absorption feature (which complicates its use in the case of strong reddening), and it lies at the core of a strong TiO band, which makes it difficult to use for late spectral types.
The Na\,{\footnotesize I} doublet is stronger than the K\,{\footnotesize I} doublet, but again has a complex luminosity dependence, for example being weak in giants. Its two lines are also rather close in wavelength (0.16\,nm), so that high spectral resolution is required. They are similarly affected by TiO bands. Also, both the K\,{\footnotesize I} and the Na\,{\footnotesize I} region have no lines in earlier (B and A) stars, so that no radial velocity information can be derived.

The Ca\,{\footnotesize II} lines are strong throughout most of the Hertzsprung--Russell diagram, with $W \ge 0.3$\,nm for all dwarfs from F8 to M8, and $W \simeq 0.6$\,nm in supergiants. The lines appear around B8, and dominate in M~stars, with a strong luminosity dependence. Being non-resonant, there are no problems with contamination by interstellar components. They are strong enough that useful radial velocities can be derived by cross-correlation even in low signal-to-noise spectra. The Ca\,{\footnotesize II} lines were duly considered an optimum choice
\citep{1999BaltA...8...73M}.
A further advantage of the Ca\,{\footnotesize II} infrared triplet region is that, in early-type stars (B, A and early F), where the Ca\,{\footnotesize II} lines disappear, the hydrogen Paschen series appears instead. These lines are also strong, also allow for radial velocity determination, and are also important for classification 
\citep{1996MNRAS.279...25F}. 
Finally, the 860\,nm spectral region is free from telluric absorption lines, so that follow-up (or complementary) observations from ground can be undertaken in a homogeneous way.

This spectral region also contains numerous other features of astrophysical interest.
First, while the region is not affected by strong molecular bands, several metallic lines are present (N\,{\footnotesize I} 872.89\,nm, Si\,{\footnotesize I} 874.26\,nm, Mg\,{\footnotesize I} 873.60\,nm, Ti\,{\footnotesize I} 873.47\,nm and He\,{\footnotesize I} 873.34\,nm), which allow for a detailed abundance analysis and quantitative spectral classification for the brighter stars, based on the ratios of equivalent widths.
Second, stars with peculiar spectra are also easily detected in this spectral range. For example, the infrared Ca\,{\footnotesize II} spectrum of the proto-typical mass-loosing star P~Cyg shows a characteristic line-profile prominent in the Paschen and He\,{\footnotesize I} lines, easily detected even in low signal-to-noise spectra. Classical~Be stars are also easily found, as well as classical T~Tau stars (which have strong Ca\,{\footnotesize II} emission). Active late-type stars generally display emission cores in the Ca\,{\footnotesize II} triplet.
Third, while there is no resonant {\it interstellar\/} line in the Ca\,{\footnotesize II} triplet region, a medium-intensity diffuse interstellar band (DIB) is present at 862\,nm. The correlation of its equivalent width with absorption is rather tight, suggesting that it could be used as one of the indicators to build a detailed reddening map, especially for high values of interstellar extinction
\citep{1999BaltA...8...73M}.

\begin{figure}[t]
\centering
\includegraphics[width=0.25\linewidth]{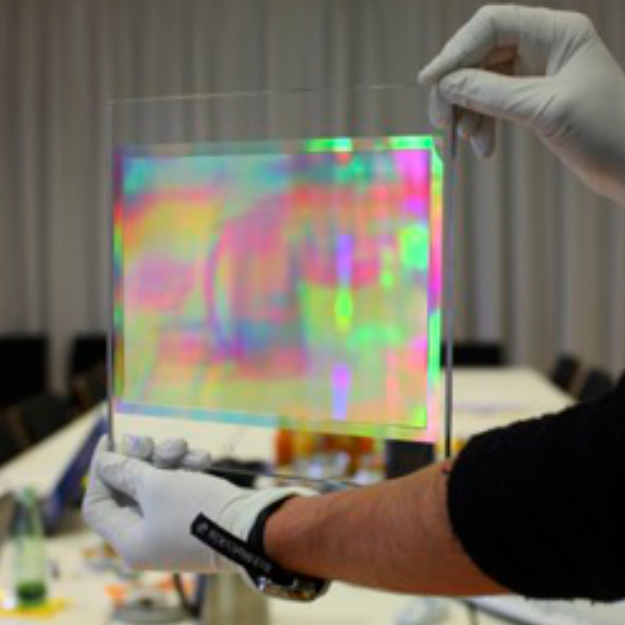}
\hspace{40pt}
\raisebox{-4pt}{\includegraphics[width=0.41\linewidth]{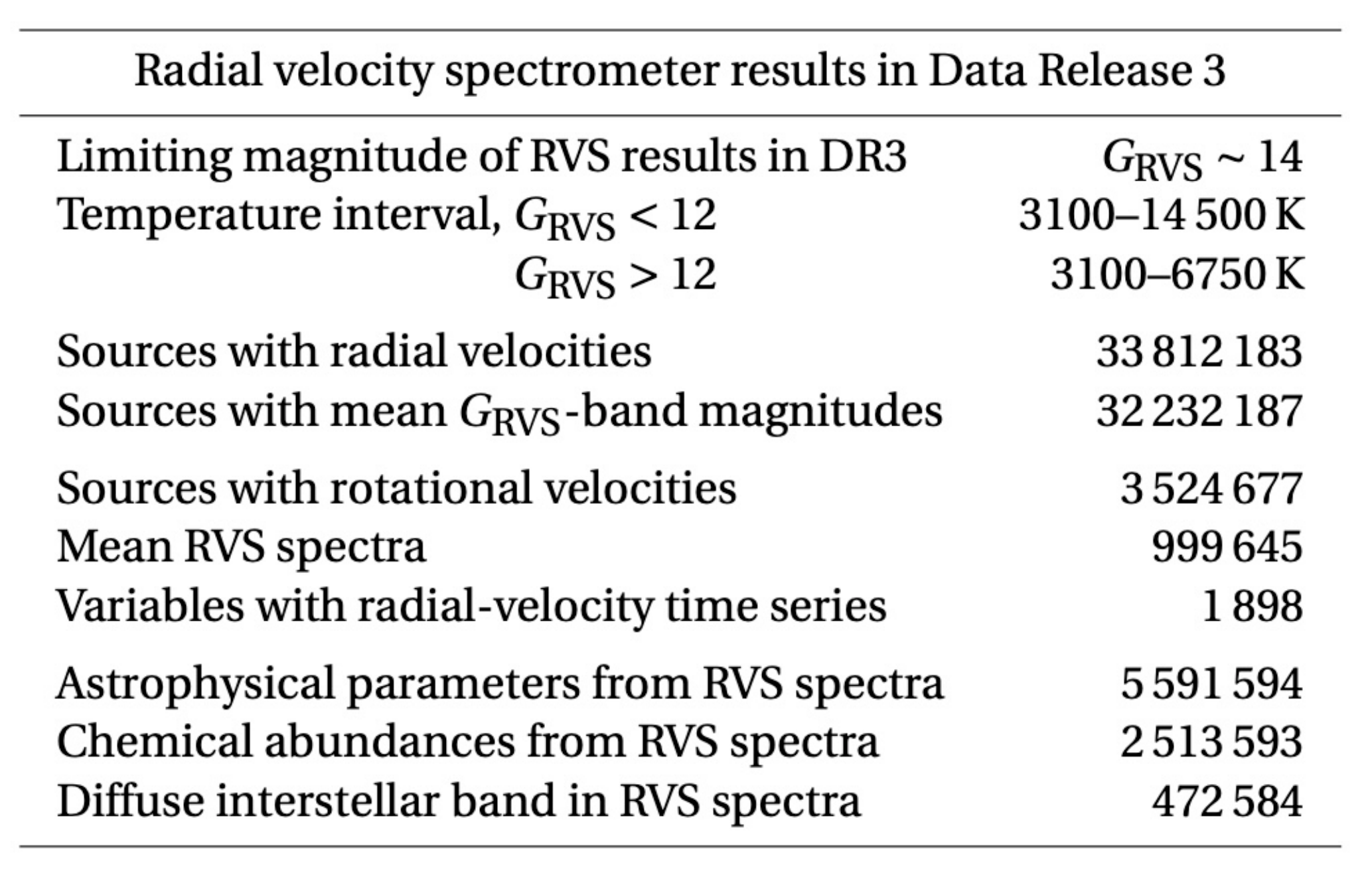}}
\vspace{-5pt}
\caption{Left~(a): RVS grating demonstrator model (Airbus Defence and Space).
Right~(b): summary of the RVS results in Gaia Data Release~3.
}\label{fig:rvs-summary}
\end{figure}

\subsubsection{Acquisition}
\label{sec:rvs-acquisition}

The spectroscopic instrument is highly integrated with the astrometric instrument 
\citep{2016A&A...595A...1G,
2018A&A...616A...5C}.
It uses the same telescopes, a dedicated section of the same focal plane, and the same sky-mapper and astrometric field (AF1) combination for object detection and confirmation. The selection of the fainter objects for observation with RVS is, however, based on an on-board flux estimate derived from the RP photometer, which is collected just before the object enters the spectroscopic instrument field (Figure~\ref{fig:gaia-fp-concept}). 

The radial velocity spectrometer is an integral-field spectrograph, in which the spectral dispersion of all objects in the combined field of view is achieved through an optical module (with unit magnification) in the common path of the two telescopes, actually between the final telescope mirror (M6) and the focal plane. 
This module contains a blazed-transmission grating plate, four fused-silica prismatic lenses (two with flat surfaces and two with spherical surfaces), and a multi-layer interference bandpass-filter plate to restrict the wavelength range to 845--872\,nm (Section~\ref{sec:rvs-wavelength}).
The dispersed light from the spectrometer illuminates a dedicated area of the focal plane containing 12 CCDs, arranged in three strips of four CCD rows (Figure~\ref{fig:gaia-fp-concept}). As a result, objects observed by the spectrometer have 43\% fewer focal plane transits than in the astrometric and photometric fields, typically some 80 over 10~years.
The grating plate, with $R=\lambda/\Delta\lambda \approx 11\,700$ (giving a dispersion of 0.0245\,nm per pixel) disperses images in the along-scan direction, spread over $\sim$1100~pixels (Figure~\ref{fig:rvs-summary}a). The along-scan window size is 1296~pixels, to allow for background subtraction, and window-placement `errors'.
In common with the devices used throughout the focal plane, the CCDs are back-illuminated, with an image area of 4500 lines along-scan and 1966 columns across-scan. Each pixel is 10~$\mu$m $\times$ 30~$\mu$m, corresponding to $58.9\times176.8$~milli-arcsec on the sky. 
As with the astrometric and photometric fields, the CCDs are operated in time-delayed integration (TDI) mode: photoelectrons are integrated over the CCD as the images cross the focal plane, synchronised with the spacecraft spin. 
For the radial velocity spectrometer (as well as the RP photometers) the CCDs are red-enhanced, including an anti-reflection coating centred on 750\,nm. Each focal plane passage corresponds to an integration time of 12.6\,s (4.2\,s for each of the three consecutive RVS CCDs).

For the majority of objects, the spectra are binned on-chip in the across-scan direction over 10~pixels, to form one-dimensional along-scan spectra. Single-pixel-resolution windows (of size $1296 \times 10$~pixels$^{2}$) are only retained for stars brighter than $G_{\rm RVS} = 7$~mag. 
The object-handling capability of the radial velocity spectrometer is limited to about $35\,000$~objects per square degree. In areas exceeding this stellar density, only the brightest objects are allocated a detector window. 
As for the photometers, the data quality is progressively compromised in dense areas by contamination and blending from nearby sources.

\paragraph{Ground processing}
Processing of the RVS data, by Coordination Unit~6, is carried out within an extensive and dedicated processing pipeline, described for Gaia~DR2 by
\citet{2018A&A...616A...6S},
and for Gaia DR3 by
\citet{2023A&A...674A...5K}.
The pipeline takes care of the basic spectroscopic calibrations, including the wavelength scale, geometric calibration of the focal plane, and the treatment of straylight. More Gaia-specific treatment includes calibration of the line-spread function (both along- and across-scan), numerous CCD-related effects (charge-transfer-inefficiency, bias, saturation, and dark current), the flagging of cosmic rays, and the handling of crowded regions and overlapping samples. 
The noise-dominated faint object transit spectra are subsequently `stacked' to derive mission-average radial velocities and rotational broadening effects through cross-correlation techniques. For the brightest subset, epoch spectra and epoch radial velocities are preserved. 
Iterations between calibrations and source parameters are performed entirely within the RVS pipeline.
I say more on the derivation of the astrophysical parameters in Section~\ref{sec:classification-stellar-properties}.

For the vast majority of (faint) stars, the individual spectra are too noisy to derive transit-level radial velocities. As a result, a single, end-of-mission composite spectrum will be constructed by co-adding the spectra collected during all of the RVS CCD crossings obtained throughout the mission lifetime. 
A single, mission-averaged radial velocity is then extracted from this composite spectrum by cross-correlation with a synthetic template spectrum. The cross-correlation method finds the best match of the observed spectrum to a set of predefined synthetic spectra (e.g.\ with different atmospheric parameters) and subsequently assigns the astrophysical parameters of the best-fit template to the observed target. 
For the few million brightest targets, single field-of-view transit spectra will be retained to derive associated epoch radial velocities. For this subset, the radial velocities of the components of (double-lined) spectroscopic binaries is also being estimated. 

Let me emphasise the size of the computational task. Over the 10.5~year mission lifetime, Gaia acquired 50~billion RVS spectra, the equivalent of 160 spectra per second, or 14~million per day. For Data Release~2, 200~million spectra were processed, resulting in radial velocities for 7~million stars. For DR3, 2.8~billion spectra were processed, resulting in 34~million radial velocities and 3.5~million rotational velocities (Table~\ref{tab:data-release-table1}). For Gaia~DR4, underway at the time of this review (mid-2025), 25~billion spectra are being processed.
This bulk processing is undertaken by CNES, Toulouse, where the Hadoop framework for distributed storage and processing was adopted in 2009 to manage the data. The required hardware was built up progressively: in 2014 it comprised 500 processing cores, 2\,TB of RAM, and 360\,TB of storage. Today, it comprises 7000 cores, 45\,TB of RAM, and 9\,PB of storage. Following a sequence of test runs, processing of the 2.8~billion spectra of DR3 extended over 111~days.

\subsubsection{Results in DR3}

Although just over 10~years of data have been acquired, only the first 34~months (July 2014--May 2017) have been processed and released to date, as Gaia DR3 on 13~June 2022. Along with 1.8~billion sources with astrometry, DR3 provides radial velocities for 33~million sources for $G_{\rm RVS}\lesssim14$~mag (Figure~\ref{fig:rvs-summary}b).  When DR4 is released in late 2026, based on 66~months of data, RVS results for the 100~million sources down to $G_{\rm RVS}\sim16$~mag should be available. 

These numbers should be placed in context. When the Hipparcos catalogue of 120\,000 stars was released in 1997, radial velocities were known for just 20\,000. Surveys by Coravel, amongst others (often focused on exoplanets) later measured several thousand more. 
Subsequently, and on a much larger scale, and in the northern hemisphere between 2005--2010, the Sloan Digital Sky Survey's SEGUE extension obtained spectra for 240\,000 stars, with typical radial velocity accuracies of 10\kms\ \citep{2009AJ....137.4377Y}.  SEGUE--2 (2008--2009) observed a further 120\,000 \citep{2022yCat..22590060R}.
Complementing SEGUE in the south, RAVE (RAdial Velocity Experiment) 
was a multi-fiber spectroscopic survey using the 1.2-metre AAO--UK Schmidt Telescope. 
Conducted between 2004--2013, and partly motivated by the prospects of Gaia, RAVE acquired some 574\,000 spectra for around 483\,000 stars
\citep{2017AJ....153...75K}.
The LAMOST--II medium-resolution ($R\sim7500$) spectral survey measured 1\,597\,675 spectra for 281\,515 stars \citep{2019ApJS..244...27W}, achieving radial velocity accuracies of around 1\kms.
The (AAO--HERMES) GALAH+ survey includes 584\,015 dwarfs and giants in the magnitude range $G=11-14$ \citep{2021MNRAS.508.4202Z}. 
And the merged and homogeneous `Survey of Surveys' contains almost 11~million stars with radial velocity precision in the range 0.05--1.50\kms, of which half are exclusively from Gaia \citep{2022A&A...659A..95T}. 

\begin{figure}[t]
\centering
\raisebox{5pt}{\includegraphics[width=0.36\linewidth]{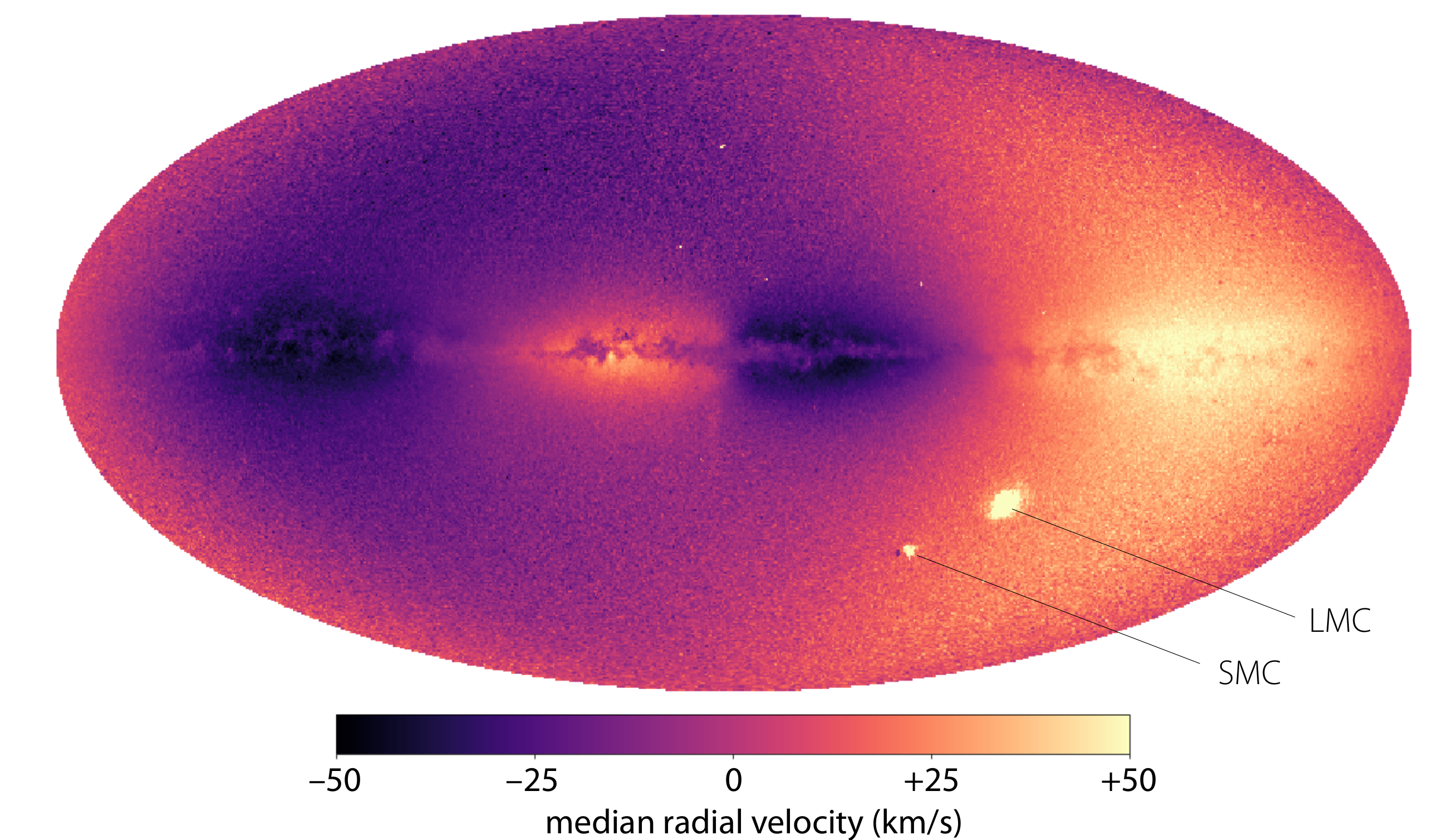}}
\hspace{5pt}
\includegraphics[width=0.29\linewidth]{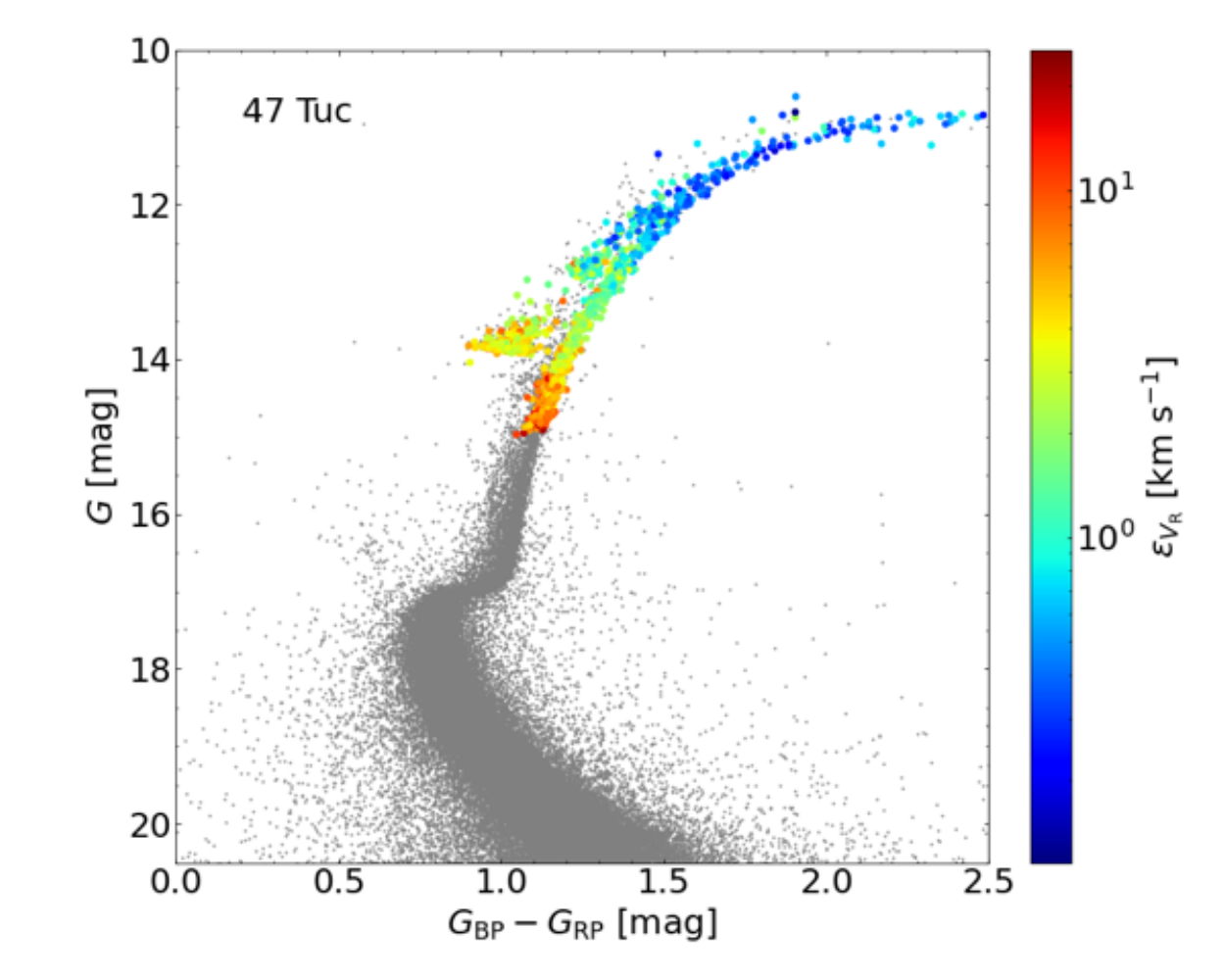}
\hspace{5pt}
\includegraphics[width=0.29\linewidth]{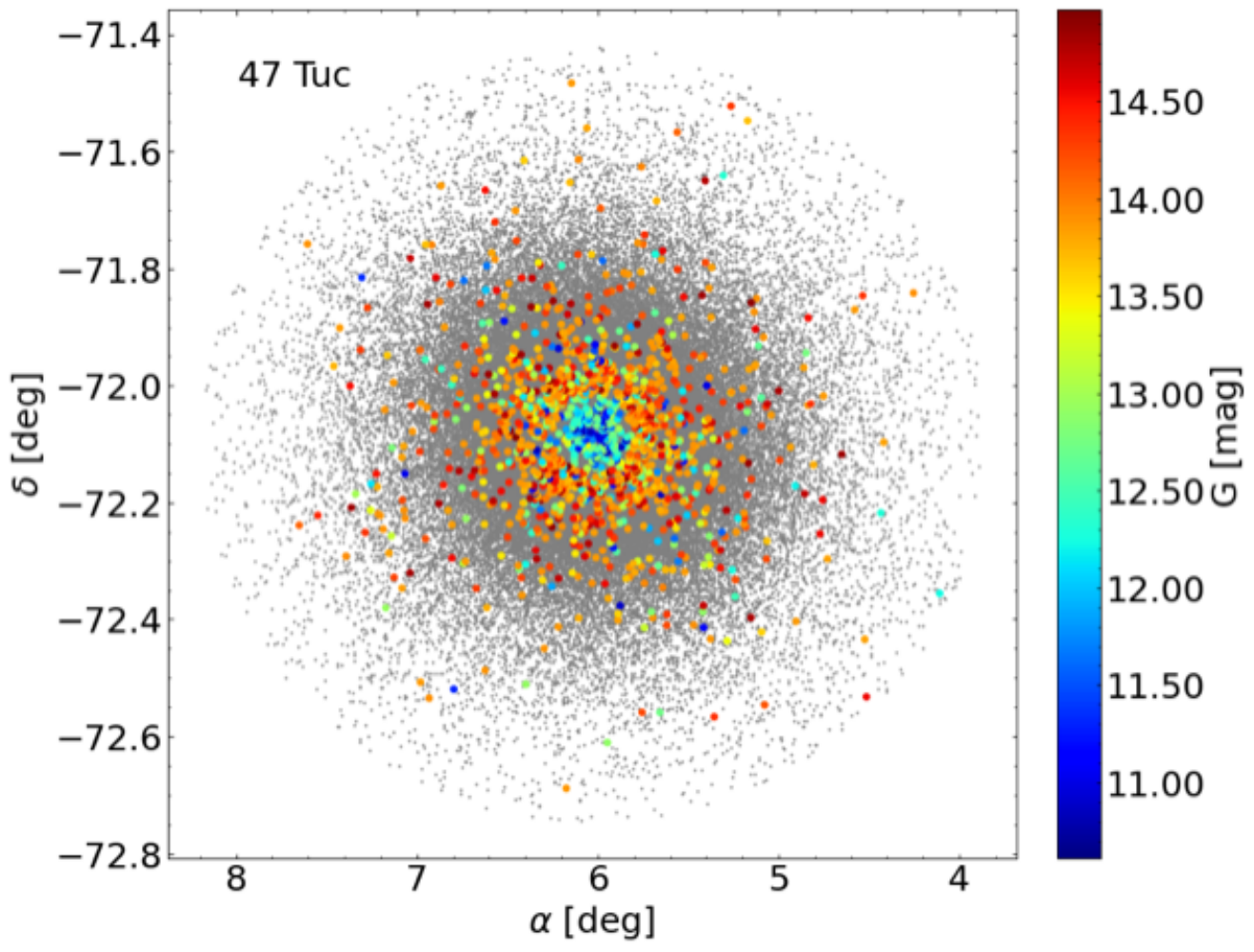}
\caption{Examples of the Gaia DR3 radial velocity results. 
Left~(a): median values as a function of Galactic longitude and latitude.
Middle~(b): radial velocities in the globular cluster 47~Tuc in the colour--magnitude diagram.  Stars with DR3 photometry are shown in grey. Those with a DR3 radial velocity, extending to $G\sim15$~mag, are colour-coded by their formal uncertainty.
Right~(c): sky distribution (in equatorial coordinates) members of 47~Tuc with DR3 radial velocities are colour-coded by their $G$~magnitude. 
From \citet{2023A&A...674A...5K}, Figures~5b and 16a,\,c.}
\label{fig:rvs-dr3-science}
\end{figure}

The following three illustrations of the DR3 RVS content are also taken from \citet{2023A&A...674A...5K}.
Figure~\ref{fig:rvs-dr3-science}a shows the median radial velocities as a function of Galactic longitude and latitude. Clearly seen is the rotation of the Galactic disk, projected along the lines-of-sight, seen as the bright areas (with positive median velocities) and dark areas (with negative median velocities).  Several objects whose radial velocities differ from those of their close environment (and which are sufficiently populated to affect the median value), are visible by their local contrast. In addition to the Large and Small Magellanic Clouds appearing as bright spots in the lower right region of the sky distribution, the Sagittarius dwarf galaxy is visible as a faint quasi-vertical stripe below the Galactic centre. Several globular clusters and other compact objects appear as small dots in the image, including the globular clusters 47~Tuc and Omega Cen.

Amongst particular objects discussed by \citet{2023A&A...674A...5K}, including the open clusters NGC~2516 and Mamajek~4, and the LMC, two figures illustrate the radial velocities in the globular cluster 47~Tuc. In the colour--magnitude diagram (Figure~\ref{fig:rvs-dr3-science}b), stars with a DR3 radial velocity are colour-coded by their formal uncertainties. In the sky distribution in equatorial coordinates (Figure~\ref{fig:rvs-dr3-science}c), stars are colour-coded by their $G$ magnitude. The sample limit is visible as a relatively sharp cut around $G\sim15$~mag. Radial velocities are measured throughout the cluster, including the densest regions of the core.  They cover the upper part of the red giant branch, the asymptotic giant branch, and the horizontal branch. The formal uncertainties are a few \kms\ at the level of the horizontal branch, and a few hundred \ms\ at the tip of the red giant branch.

Looking to the 100~million radial velocities expected in DR4, `deep learning' methods are being developed to best identify spurious velocities, minimise the contamination, and maximise the completeness.

\subsubsection{Redshifts and radial velocities from the BP/RP spectra}

While the RVS is photon-limited at $G\gtrsim16$\,mag, low-resolution BP/RP spectra are acquired for all sources crossing the main focal plane (currently, DR3 provides astrometric solutions for 1.5~billion sources, while low-resolution spectra have been published for a subset of 220~million; the BP/RP spectra for Gaia's full 2~billion or more sources will be made available with DR4 in late 2026).
As previously mentioned, the design, including wavelength range and resolution, was optimised to provide information needed to determine the astrophysical parameters of each source, such as \teff, $\log g$, [M/H], and extinction, across a wide range of spectral type and luminosity class. Processing within the DPAC Consortium (CU8), includes the `astrophysical parameters inference system' pipeline, Apsis 
\citep{2013A&A...559A..74B,
2023A&A...674A..29R},
which runs 13~modules, using different combinations of data and models, to produce astrophysical parameters for stars, galaxies and quasars.
Amongst the modules identifying and classifying the extragalactic sources, the `quasar classifier' module (QSOC) estimates the quasar redshifts using a $\chi^2$ approach, in which the BP and RP spectra are compared to a composite quasar spectrum 
(from SDSS DR12Q; \citep{2017A&A...597A..79P}), 
over trial redshifts in the range $z=0-6$ 
\citep[][\S2.2]{2023A&A...674A..41G}.
A similar approach estimates the redshifts of galaxies using the `unresolved galaxy classifier' module (UGC).
As a result, 6.4~million quasars and 1.4~million galaxies have redshifts given in DR3 
\citep[][Table~\ref{tab:data-release-table2}]{2023A&A...674A..31D}.

\begin{figure}[t]
\centering
\raisebox{7pt}{\includegraphics[width=0.31\linewidth]{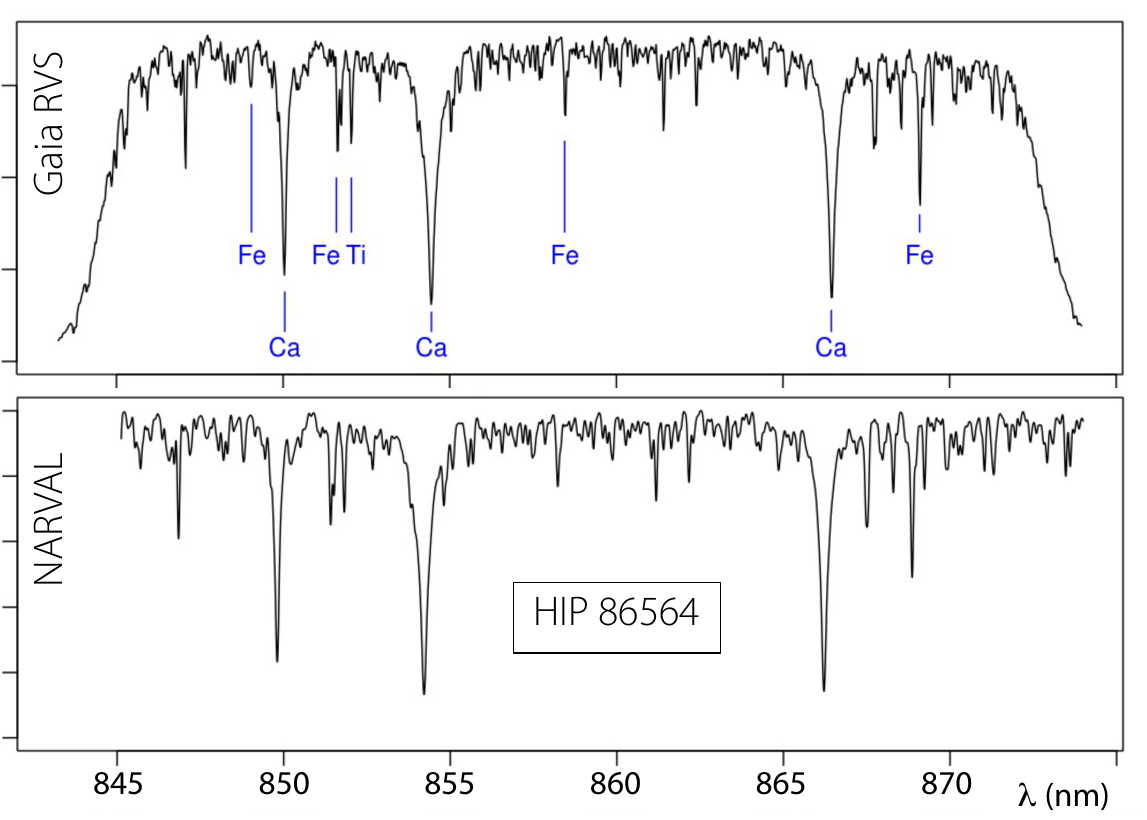}}
\hspace{15pt}
\includegraphics[width=0.31\linewidth]{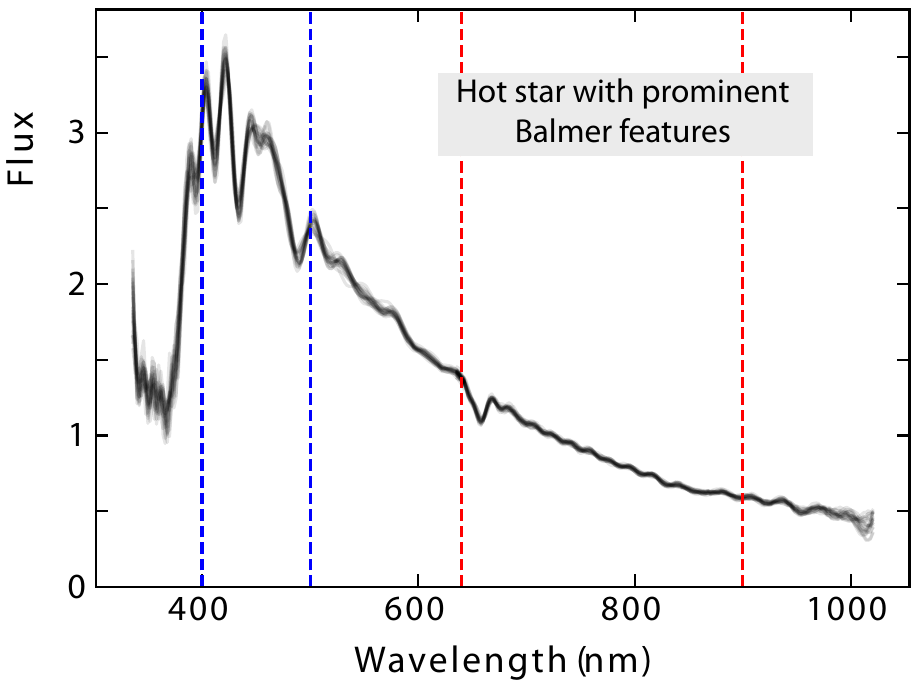}
\includegraphics[width=0.31\linewidth]{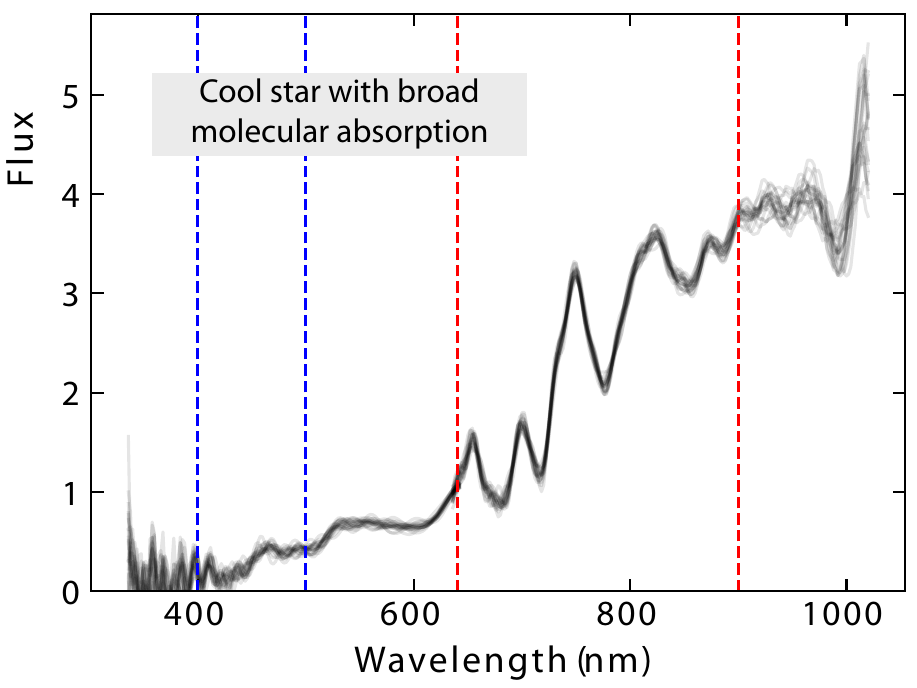}
\vspace{-6pt}
\caption{Left pair, top: the $V=6.7$\,mag K5 star HIP~86564 from a single 4.4\,sec Gaia exposure; bottom: with the NARVAL spectrograph at the Observatoire du Pic du Midi, at the same spectral resolution (\citet{2018A&A...616A...5C}, Figure~16).
Right pair: two example Gaia BP/RP spectra, with the blue and red dashed lines indicating the BP and RP spectral ranges used in the fitting procedure. Left: a hot star displaying prominent Balmer features; right: a red source with broad molecular absorption bands (\citet{2024A&A...684A..29V}, Figure~1).}
\label{fig:radial-velocity-spectra}
\end{figure}

A similar approach has been adopted for determining stellar radial velocities by 
\citet{2024A&A...684A..29V}. 
The principle is straightforward (although of course the details are not): the radial velocity (and its estimated error) is obtained by fitting the BP/RP spectra to a wide range of models based on a grid of synthetic spectra (Figure~\ref{fig:radial-velocity-spectra}). Remaining systematics (mainly functions of magnitude and colours) are corrected using external radial velocity catalogues.
Their most reliable subset comprises 6.4~million sources with uncertainties $<$300\kms, around one quarter of which have no radial velocity yet available from RVS. They also constructed an extended catalogue with all 125~million sources for which they could obtain a valid radial velocity estimate. Typical uncertainties are, of course, far higher than those obtained through precision spectroscopy. Their median uncertainty is $\sim$770\kms, but extends below 100\kms\ (their Fig.~4). The highest accuracy is generally achieved for either very blue ($G_{\rm BP}-G_{\rm RP}\lesssim0.7$) or very red ($G_{\rm BP}-G_{\rm RP}\gtrsim2$) stars.
The scientific value of such imprecise values remains to be seen, but might include searches for black hole binaries or hypervelocity stars
\citep{2024MNRAS.533.2747V}.	

In summary, Gaia DR3 provides radial velocities for 33~million sources to $G_{\rm RVS}\sim14$~mag, with the 66-month based DR4 expected to deliver around 100 million to $G_{\rm RVS}\sim16$~mag in late 2026. Coarse measurements from BP/RP are now also available for 125~million, with the potential of more than a billion with Gaia DR4.

A very different approach to estimating the `missing' radial velocities is to infer them from the five~other {\it astrometric\/} parameters
\citep{2021ApJ...915L..14D}. 
Their 6~million source neural network training set was later used to infer radial velocities for a further 92~million, which they applied to stars in the Enceladus merger using EDR3 
\citep{2023MNRAS.521.1633D}. 
A Bayesian neural network approach has also been used to predict the values for 16~million stars in EDR3
\citep{2022MNRAS.516.3398N}.
These have been tested 
\citep{2024MNRAS.52711559N}, 	
and DR3 astrometry in turn used to predict the missing values for 185 million stars to $G<17.5$. They estimate that the predictions are reliable, to 25--30\kms, for stars within 7~kpc.

\subsubsection{Radial velocity time-series of long-period variables}

Multi-epoch radial velocity observations are acquired by Gaia's radial velocity spectrometer (RVS) for all sufficiently bright sources, $G_{\rm RVS}\lesssim16$, and quasi-simultaneously with the epoch photometry. However, only {\it mean\/} values were generally made available in Gaia DR3. An exception is for the Cepheids \citep{2023A&A...674A..17R} and RR~Lyrae variables \citep{2023A&A...674A..18C}, for which the radial velocity time series were included as part of Gaia DR3. 

Anticipating publication of the full multi-epoch radial-velocity data with DR4 in late 2026, the `Focused Product Release' described by 
\citet{2023A&A...680A..36G}	
provides the radial-velocity time series (and derived variability parameters) for 9614 long-period variables (LPVs) with high-quality observations in the range $G=6-14$~mag, selected from the 1.7~million LPV candidates from DR3 \citep{2023A&A...674A..15L}. 
These have an average of 24 measurements per source (ranging between 12--90), non-uniformly over the 34-months of the DR3 observations. Importantly, the radial-velocity periods are, they show, consistent with those from the multi-epoch $G$, BP, and RP photometry. 
One objective of this particular release was to provide some first insights into how the full radial velocity time-series (as opposed to a single mean value) complements Gaia's time-series {\it photometry\/} in characterising this broad class of variable star.

\begin{figure}[t]
\centering
\raisebox{9pt}{\includegraphics[width=0.33\linewidth]{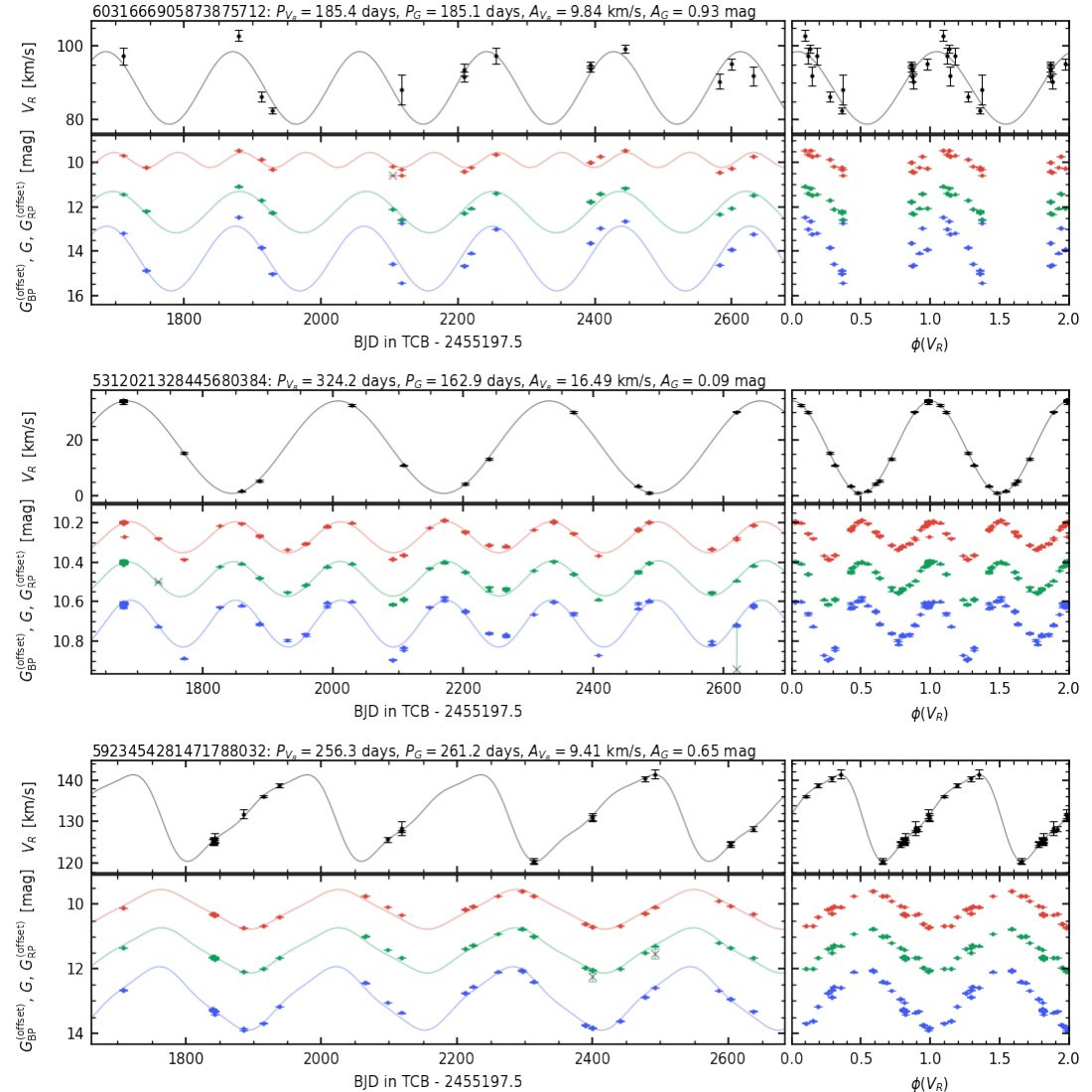}}
\hspace{40pt}
\includegraphics[width=0.324\linewidth]{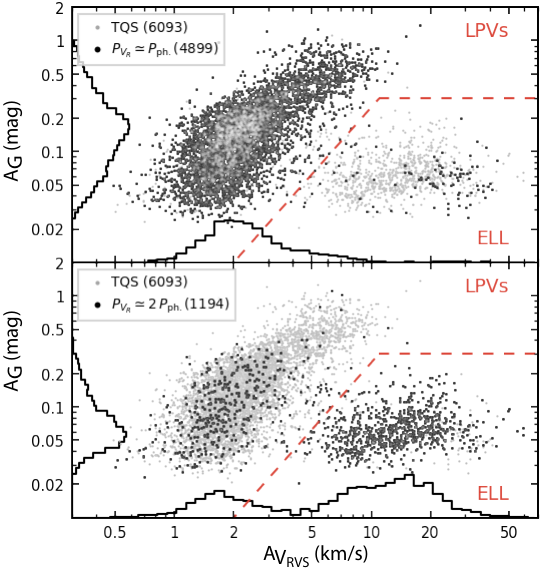}
\vspace{-5pt}
\caption{
Radial velocity time-series of long-period variables.
Left~(a): three examples showing, top to bottom: a source with `mixed consistency' between the radial velocity and photometric time series; a (binary) ellipsoidal variable; and a classical pulsating long-period variable. In each, the radial velocity time-series (in black) is acquired simultaneously with the photometric time series in BP (blue), G (green), and RP (red), each suitably offset, and shown phase-folded in the right panels.
Right~(b): the diagram of photometric amplitude, $A_{\rm G}$, versus radial velocity amplitude, $A_{\rm RVS}$, delineates one region populated by smaller $A_{\rm G}$ and larger $A_{\rm RVS}$ which characterises the ellipsoidal variables (ELL), and a distinct region defined by larger $A_{\rm G}$ and smaller $A_{\rm RVS}$ which characterises the pulsating long-period variables (LPV). From \citet{2023A&A...680A..36G}, Figures~9--13.
}\label{fig:rv-lpv}
\vspace{-5pt}
\end{figure}

LPVs are intrinsic variables with periods range from days to 1000~days or more, embracing pulsating red giants and supergiants, including Mira variables, as well as semi-regular variables generally pulsating in an overtone.  Also included by virtue of their similar light curves are some binary systems, including the ellipsoidal variables (ELL) in which the variability originates from the body's gravitationally distorted shape.
The simultaneity of Gaia's radial velocity and photometric time series assists with various challenges associated with their identification, classification, and physical interpretation (Figure~\ref{fig:rv-lpv}a).
As one example, since the periods of long-period variables extend to several hundred days, obtaining velocity curves at high resolution with a good phase coverage has been an observational challenge, with the result that the total sample of well-observed Mira variables still numbers only a few tens of objects. Over wide ranges of period and metallicity, these suggest a consistent pattern in the velocity variations, with s-shaped velocity curves in the near-infrared, and velocity amplitudes of 20--30\kms.

Another example is the application to ellipsoidal variables, viz.\ close binaries typically comprising a red giant and a main sequence star (Section~\ref{sec:ellipsoidal-variables}). Despite the absence of eclipses, the gravitationally induced elongation of the red giant leads to rotational brightness variations. In turn, this generally results in a different period for the system's photometric variability compared with the orbital period from the radial velocities. But the distinction between the ellipsoidal and long-period variables can be made more robustly than by simply comparing the ratio between the radial velocity and photometric periods. As shown in Figure~\ref{fig:rv-lpv}b, the diagram of photometric amplitude versus radial velocity amplitude clearly delineates the ellipsoidal variables (ELL) from the pulsating long-period variables (LPV).

\subsection{Classification and stellar properties}
\label{sec:classification-stellar-properties}	

\paragraph{The DPAC Apsis modules}

The 34-month Data Release~3 comprises 1.812~billion sources with astrometric solutions to 21~mag, 1.806~billion with mean $G$ magnitudes, 1.5~billion with mean $G_{\rm BP}$ and $G_{\rm RP}$ photometry, along with 219~million mean BP/RP spectra. 
In addition to the astrometric, photometric, and radial velocity data included in the successive data releases, Data Release~3 in 2022 also includes a wealth of `extracted' astrophysical data for each source. 
This work, of source classification and `parameterisation', is undertaken within Coordination Unit~8 of the Gaia DPAC. This involves taking all relevant calibrated data, and extracting key astrophysical quantities based on automated classification techniques.
While these are largely model-dependent quantities, their huge number and homogeneous acquisition and treatment stands in contrast to the observations that have been acquired from the ground over the past century, which have provided a generally heterogeneous collection of stellar properties and chemical abundances for about two million stars in total, with fragmentary sky coverage.

\begin{figure}[t]
\centering
\includegraphics[width=0.60\linewidth]{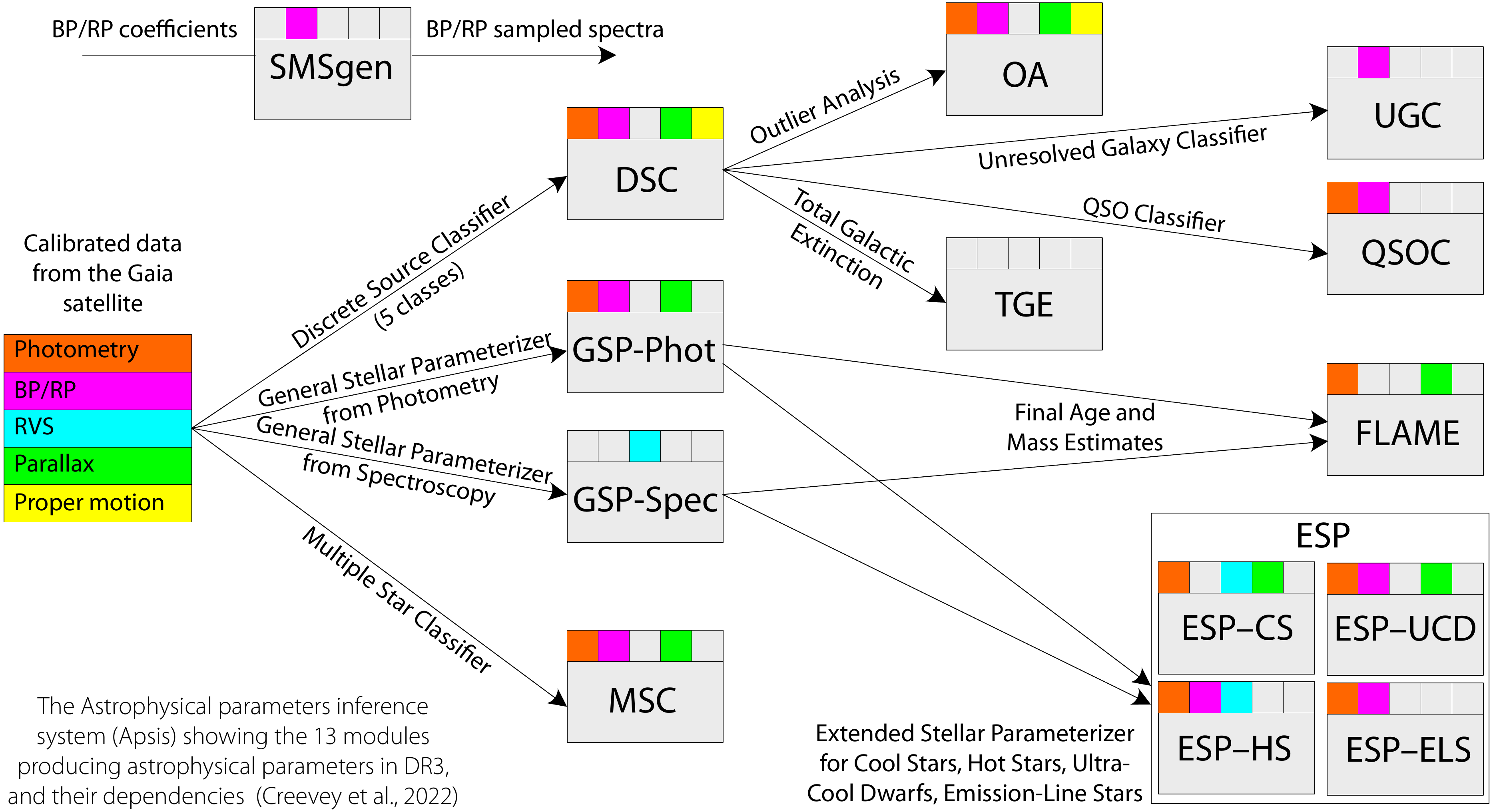}
\caption{Object classification and parameter determination carried out for Data Release~3 by the Gaia DPAC Coordination Unit~8. The 13 Apsis modules are colour-coded to show which of the input data sets are used to generate the various outputs. The input BP/RP spectra for Apsis are in the form of sampled spectra, produced by SMSgen \citep[from][Figure~1]{2023A&A...674A..26C}.
}\label{fig:apsis}
\end{figure}

The `astrophysical parameters' included in Gaia DR3 provide the intrinsic properties of each stellar object (such as effective temperature, age, and chemical composition), as well as other inferred properties such as redshifts of distant sources and object classification.
CU8 does this by taking the mean (calibrated) BP and RP spectra, and the mean (calibrated) RVS spectra, along with astrometry (distances and proper motions) and photometry ($G$, $G_{\rm BP}$, $G_{\rm RP}$), and subjects all these to the `Astrophysical parameters inference system' (Apsis). 
Within the Apsis pipeline, described pre-launch by 
\citet{2013A&A...559A..74B}, 
13~modules use different combinations of input data and/or models to produce astrophysical parameters for stars and sub-stellar objects, as well as galaxies and quasars (Figure~\ref{fig:apsis}). 

The details of these various modules, their outputs, and their validation, are described in a series of papers accompanying Data Release~3. 
Amongst these,
\citet{2023A&A...674A..26C} gives an overview of the Apsis methods and content, along with the detailed data products, and the associated 10~tables of the Gaia archive; 
\citet{2023A&A...674A..28F} focuses on the stellar content (including parameter determination); 
and 
\citet{2023A&A...674A..31D} focuses on the non-stellar content and details of the source classification. 
Three other Gaia papers are module specific:
\citet{2023A&A...674A..27A} describes the derivation of specific parameters from the photometry (GSP--Phot); 
\citet{2023A&A...674A..29R} describes the derivation of specific parameters from the RVS spectra alone (GSP--Spec);
and
\citet{2023A&A...674A..30L} details the derivation of stellar chromospheric activity and mass accretion from the Ca\,{\scriptsize II} infrared triplet in the RVS spectra.

Of the two `General Stellar Parameterizer' modules, 
GSP--Spec uses projection and optimisation methods to match the mean RVS spectra with a large grid of {\it theoretical\/} spectra computed using MARCS models, with a range of atmospheric parameters ($T_{\rm eff}$, $\log g$, metallicity [M/H], [$\alpha$/Fe]) and chemical abundances ([X/Fe]), spanning the full parameter space of Galactic stellar populations
\citep{2023A&A...674A..29R}.
The resulting catalogue of 5.6~million stars includes individual chemical abundances (of N, Mg, Si, S, Ca, Ti, Cr, Fe\,{\scriptsize I}, Fe\,{\scriptsize II}, Ni, Zr, Ce, Nd); the equivalent width of a CN-line;  and the parameters of a diffuse interstellar band.
%
GSP--Phot estimates $T_{\rm eff}$, $\log g$, [M/H], absolute magnitude $M_{\rm G}$, radius $R$, distance, extinctions (A$_0$, A$_{\rm G}$, A$_{\rm BP}$, A$_{\rm RP}$), as well as the reddening E(G$_{\rm BP}$--G$_{\rm RP}$) by forward-modelling the BP/RP spectra, $G$ magnitude, and parallax using a Markov Chain Monte Carlo (MCMC) method.

Of the various `classifiers'
\citep{2013A&A...559A..74B,
2023A&A...674A..31D},
the Discrete Source Classifier (DSC) 
classifies sources probabilistically into five empirical classes (star, white dwarf, physical binary, quasar, galaxy), although it is primarily intended to identify extragalactic sources.
The Unresolved Galaxy Classifier (UGC) 
estimates the redshift of unresolved galaxies. It does this by applying a supervised machine-learning model to their sampled BP/RP spectra. 
The QSO Classifier (QSOC) 
aims to determine the redshift of sources that are classified as quasars by the DSC module. It is based on the cross-correlation between a rest-frame quasar template and an observed BP/RP spectrum, evaluated at a range of trial redshifts. 
The module predicts redshifts in the range $0.0826<z<6.1295$, along with the corresponding uncertainty.
The Multiple Star Classifier (MSC) 
infers stellar parameters by assuming that the BP/RP spectrum is a composite spectrum of an unresolved coeval binary system, and that the two components have a flux ratio in the BP/RP spectrum between 1--5. 
The Total Galactic Extinction module (TGE) 
uses a subset of giants with extinction estimates provided by GSP--Phot as extinction tracers, to construct all-sky maps of the total Galactic foreground extinction. 

In the other processes,
FLAME (Final Age and Mass Estimates) takes the output spectroscopic parameters from GSP--Phot and GSP--Spec, along with astrometry and photometry, to derive the evolutionary parameters: radius, luminosity, mass, and age.
%
ESP--CS computes a chromospheric activity index from the analysis of the RVS Ca\,{\scriptsize II} infrared triplet.
ESP--HS processes the BP and RP spectra, along with the RVS spectra when available, to provide a spectral type for all stars, and stellar parameters for $T_{\rm eff}>7500$\,K. 
ESP--UCD is dedicated to the analysis of ultra-cool dwarfs, and it produces $T_{\rm eff}$ for these stars.  
Finally, ESP--ELS analyses emission-line stars, and provides class probabilities and labels, along with a measurement of the H$\alpha$ equivalent width. 

\paragraph{Numbers for DR3}
The numbers derived by Apsis for DR3 are transformational:
source classification and probabilities for 1.6~billion objects;
interstellar medium characterisation and distances for up to 470~million sources, including a 2-d total Galactic extinction map; 
and redshifts for 6~million quasar candidates and 1.4 million galaxy candidates. 
The astrophysical parameters include various spectroscopic and evolutionary indicators, comprising
$T_{\rm eff}$, $\log g$, and [M/H] (470~million using BP/RP; 6~million using RVS);
radius (470~million), mass (140~million), age (120~million), chemical abundances (5~million), diffuse interstellar band analysis (0.5~million), activity indices (2~million), H$\alpha$ equivalent widths (200~million), and further classification of spectral types (220~million) and emission-line stars (50\,000). 

These catalogues, based entirely on Gaia data, provide the most extensive homogeneous database of astrophysical parameters to date. For example, there are more than 20 times as many sources in the 220~million source XP spectral database as in the largest ground-based spectral survey, LAMOST (although at only 1/20th of the spectral resolution). Such huge numbers of uniform metallicities and other stellar properties, which provided accurate chemical abundances across all stellar populations throughout the Galaxy, are now being used as crucial inputs for studies of star formation, detailed nucleosynthesis modelling, and Galactic chemical and dynamical evolution.

\paragraph{Community-generated catalogues}
The astrophysical parameters generated by DPAC CU8 were never intended be the final word in classification or parameter estimation, and improvements in calibration, and in training sets and algorithms were considered inevitable. 
These will come in part through improved calibration, more so through additional observations as planned for DR4, and also through better modelling of some key spectral lines given the low resolution of the XP~spectra. Furthermore, the use of synthetic model spectra (to match the observed XP spectra) is most likely sub-optimal, leaving parameter estimates sensitive to inaccuracies and omissions in the underlying models.
And at least four other community-generated catalogues of [M/H], \teff, and $\log g$, derived from the XP spectral database, have been made available for wider use since the release of Gaia DR3.

\citet{2023MNRAS.524.1855Z}
used forward modelling to estimate \teff, $\log g$, and [Fe/H] for all 220~million DR3 stars with XP spectra. Their training set used atmospheric parameters from LAMOST, augmented by 2MASS and WISE photometry to reduce degeneracies and yield more precise estimates of \teff\ and reddening. Their catalogue includes \teff, $\log g$, and [Fe/H], along with revised parallaxes and extinctions. It ignores binary stars, and does not cover all parts of the Hertzsprung--Russell diagram, notably white dwarfs.

Building on previous work by 
\citet{2022ApJ...941...45R},
\citet{2023ApJS..267....8A}	
employed a specific machine-learning algorithm, 
\href{https://en.wikipedia.org/wiki/XGBoost}{XGBoost}.
It was trained on 500\,000 stars with stellar parameters from APOGEE, including those with CatWISE 3.4\micron\ and 4.6\micron\ infrared photometry to reduce the degeneracy between \teff\ and reddening. The training set was augmented by some 300 very metal-poor stars from LAMOST 
\citep{2022ApJ...931..147L}, 
and they included the Gaia parallaxes to assist constraints on $\log g$ and [M/H]. 
Although therefore tied to the parameter scale of the APOGEE survey, the resulting catalogue of 175~million stars has a mean precision of 0.1~dex in [M/H] and, obtained as by-products, 50\,K in \teff, and 0.08~dex in $\log g$. They also provide a catalogue of 17~million bright \mbox{($G<16$)} red giants using more conservative cuts to ensure a higher data quality.
\citet{2024ApJ...972..112C} describe an application of this catalogue in identifying the evolution of angular momentum in the Galaxy with metallicity.

\citet{2025ApJ...980...90H}	
used tree-based machine-learning to estimate [M/H] and [$\alpha$/Fe] for 48~million giants and dwarfs 
in low-extinction regions from the DR3 XP spectra. Again, the training set used APOGEE DR17 and the metal-poor stars of 
\citet{2022ApJ...931..147L}. 
It resulted in a mean precision of 0.09~dex for [M/H] and 0.04~dex for [$\alpha$/Fe], with the most reliable values being for giants and metal-rich stars which dominate the training set.
Finally,
\citet{2024MNRAS.52710937Y}		
focused on the much rarer very metal-poor stars, [Fe/H]\,$<-2$, also using XGBoost. For $G_{\rm BP}<16$, they developed classifiers optimised for turn-off stars and for giant stars, finding 11\,000 metal-poor turn-off stars, and 111\,000 or 44\,000 bright metal-poor giants depending on the target purity. For $G_{\rm BP}>16$, they identified 38\,000 additional turn-off candidates, and 41\,000 additional metal-poor giant candidates. Their combined sample of 200\,000 very metal-poor star candidates increased the number of known metal-poor candidates by an order of magnitude.
Investigations by
\citet{2022MNRAS.516.3254W}, 	
meanwhile suggest that metal-poor stars can be identified for $G<16$ using the XP spectra, but that true detections will be overwhelmed by false positives at fainter magnitudes.

\paragraph{Machine learning}

`Machine-learning' is used extensively throughout the Gaia data processing, as well as in the analysis of the large data sets generated, and the recent reviews by
\citet{2019arXiv190407248B},	
and by
\citet{2020WDMKD..10.1349F},	
provide a broad introduction.

\paragraph{Data driven versus physics driven}
As well as providing an unprecedented resource for chemical abundance investigations, the XP data also provide a substantial database for research into classification algorithms, and the trade-offs between `physics-driven' (relying on synthetic stellar spectra) and `data-driven' (based on machine-learning) classification.
Some of this path has already been trodden, for example by APOGEE
\citep[e.g.][]{
2015ApJ...808...16N,	
2019MNRAS.483.3255L,	
2019ApJ...879...69T}.	
As a cautionary example, at least some algorithms which estimate [$\alpha$/Fe] from the XP spectra do so by exploiting known correlations between [$\alpha$/Fe] and other elements, rather than the direct effect of [$\alpha$/Fe] on the spectrum
\citep[e.g.][]{
2021A&A...656A..93G,	
2025ApJ...980...90H}.	

\citet{2025ApJ...979....5L}
argue that physics-driven models inevitably suffer from a `synthetic gap', viz.\ a combination of theoretical and instrumental effects which together produce unresolvable differences between synthetic and observed spectra. Furthermore, data-driven models which depend on `labels' (a generic term here covering \teff, $\log g$, [M/H], and [$\alpha$/Fe]) themselves suffer from `label systematics' which decrease any model's performance.
They demonstrate this by applying unsupervised learning to the XP spectra which (they argue) learns stellar properties directly from the data. They also show that the spectra {\it do} contain meaningful [$\alpha$/Fe] information, by identifying \mbox{$\alpha$-}bimodality in the absence of stellar label correlations. 
They conclude: {\it `Label-dependent models are incapable of exploiting the entire astrophysical information in the XP~data, because they are limited by the availability of stellar labels to train on. Novel data-driven techniques must be developed to tackle this big data problem'.}
In another study, using the same techniques used by Large Language Models for AI, and applied to the XP spectra,
\citet{2024MNRAS.527.1494L}	
argue that {\it `building and training a single foundation model without fine-tuning using data and parameters from multiple surveys to predict unmeasured observations and parameters is well within reach'.}

\begin{table*}[t]
\vspace{-5pt}
\centering
\footnotesize
\begin{tabular*}{\textwidth}[t]{@{\extracolsep\fill} lrrrr} 
\noalign{\vspace{5pt}}
\hline
\noalign{\vspace{2pt}}
 &										Gaia DR1&					Gaia DR2&					Gaia EDR3&				Gaia DR3\\		
 &										&							&							&						\\				
\noalign{\vspace{2pt}}
\hline
\noalign{\vspace{2pt}}
 Observations:\\
 \qquad	-- time period&						Jul~2014--Sep~2015&			Jul~2014--May~2016&			Jul~2014--May~2017&		as EDR3 \\
 \qquad	-- observations duration&				14 months&					22~months&					34~months&				as EDR3 \\
 \qquad	-- reference epoch&					J2015.0&						J2015.5&						J2016.0&					as EDR3 \\
 \qquad	-- catalogue release date&			14~September 2016&			25~April 2018&					3~December 2020&			13 June 2022 \\
  \qquad	-- url: www.cosmos.esa.int/web/&		\href{https://www.cosmos.esa.int/web/gaia/dr1}{\tt gaia/dr1}&				
  	\href{https://www.cosmos.esa.int/web/gaia/dr2}{\tt gaia/dr2}&	\href{https://www.cosmos.esa.int/web/gaia/earlydr3}{\tt gaia/earlydr3}&	
				\href{https://www.cosmos.esa.int/web/gaia/dr3}{\tt gaia/dr3}\\[5pt]
Astrometry:\\
\qquad 	-- total number (3--21~mag)&			1,142,679,769&				1,692,919,135		&			1,811,709,771&				as EDR3\\			
\qquad 	-- 5-parameter solutions&				2,057,050&					1,331,909,727		&			585,416,709 &				as EDR3\\
\qquad 	-- 6-parameter solutions&				2,057,050&					--				&			882,328,109 &				as EDR3\\
\qquad 	-- 2-parameter solutions&				1,140,622,719	&				361,009,408		&			343,964,953 &				as EDR3\\[5pt]
Photometry:\\
\qquad	-- mean $G$ magnitude&				1,142,679,769&				1,692,919,135		&			1,806,254,432&			as EDR3\\			
\qquad	-- mean $G_{\rm BP}$ photometry&		--&							1,381,964,755		&			1,542,033,472&			as EDR3\\			
\qquad	-- mean $G_{\rm RP}$ photometry&		--&							1,383,551,713		&			1,554,997,939&			as EDR3\\	[5pt]		
Radial velocities &							--&							7,224,631			&	   		7,209,831&				33\,812\,183\\[5pt]	
Other:\\
\qquad	-- variable sources&					3,194&						550,737			&			as DR2 &					see Table~\ref{tab:data-release-table2} \\
\qquad	-- known asteroids with epoch data&		--&							14,099			&			\textquotedbl\phantom{xx} &	\textquotedbl\phantom{0000} \\
\qquad	-- reference frame sources&			2,191&						556,869			&			\textquotedbl\phantom{xx} &	\textquotedbl\phantom{0000} \\
\qquad	-- effective temperatures&				-- &							161,497,595		&			\textquotedbl\phantom{xx} &	\textquotedbl\phantom{0000} \\
\qquad	-- extinction and reddening&			-- &							87,733,672		&			\textquotedbl\phantom{xx} &	\textquotedbl\phantom{0000} \\
\qquad	-- radius and luminosity&				-- &							76,956,778		&			\textquotedbl\phantom{xx} &	\textquotedbl\phantom{0000} \\
\noalign{\vspace{2pt}}
\hline
\end{tabular*}
\caption{Key attributes of the Gaia data releases to date. The latest, Gaia DR3, supersedes all previous releases.} 
\label{tab:data-release-table1}
\end{table*}

\subsection{Data releases}
\label{sec:data-releases}

After launch in December 2013, Gaia started its scientific data acquisition in July 2014. For the subsequent 10.5~years, until the end of scientific operations on 15~January 2025, the satellite scanned the celestial sphere, following its carefully designed `scanning law', accumulating additional observations, at different position angles, of all stars down its limiting magnitude of around 20.7~mag. 
As the satellite observations build up over months and years, the accuracies of the five astrometric parameters for each source improve progressively, yielding a `5-parameter solution' (i.e.\ $\alpha$, $\delta$, $\mu_\alpha$, $\mu_\delta$, $\varpi$) which improves with time. In the cases where radial velocities can be well determined, the more complete `6-parameter solution' can be constructed, fully specifying the dynamical phase-space quantities (viz.\ the three positional coordinates, and three velocity components) of each source. 

A very significant fraction of stars occur in gravitationally bound multiple systems, either as binaries, or less frequently in triple, quadruple, or even higher order multiples. In these cases, the 5- or 6-parameter solution is insufficient to describe the space motion of the individual components, and more unknowns, and more observations, are required to characterise them. For example, the uniform space motion of the centre of mass of a binary is described by the 5- or 6-parameter solution, while the orbital motion of the photocentre around the barycentre requires an additional seven Keplerian orbit elements. Multiple systems span a wide range of orbital separations, periods, and mass ratios. As more observations are acquired, over longer time periods, the quality of these solutions improves.

It is appropriate to emphasise again that the global data processing task is iterative: improved astrometric solutions for the 2~billion celestial sources provide improvements in the instrument calibration and the satellite attitude, all of which are fed back into the global astrometric solution. At the same time, the astrometric, photometric, and spectroscopic data is used to characterise the physical properties of each source, information which is also fed back into the iterative solution.
Amongst many reasons for the iterations is the requirement for `chromaticity' calibration, which uses the calibrated multi-colour photometry at each epoch of observation in the subsequent astrometric adjustment. Although the Gaia telescopes are all-reflecting, the image point-spread is not perfectly symmetric, and chromatic corrections based on each star's spectral distribution are required to reach the ultimate accuracies that are targeted.  

In essence, then, each data release is characterised by the data interval that the release covers, underpinned by the massive iterative adjustment of the vast system of equations that specify the source parameters, the instrument calibration model, and the instantaneous attitude of the satellite throughout the observational period. The iterative adjustments of the astrometric solution (AGIS, Section~\ref{sec:data-processing-agis}) are undertaken at ESA's European Space Astronomy Centre (ESAC), near Madrid, and take several months of computations for convergence of each data release.
Another point to emphasise is that while the earlier scientific publications may have made use of DR1, DR2, or EDR3, the content of each new data release completely supersedes all earlier releases.
With this framework, the contents of the various data releases to date can be better appreciated. Table~\ref{tab:data-release-table1} gives the catalogue release date, the relevant data interval, the total numbers of sources, and the numbers with a 5-, 6-, or only a 2-parameter solution. It includes the number of sources with the given photometric measurements, radial velocities, and a number of other key properties inferred from the astrometric, photometric and spectroscopic data.
In the following, I will summarise only a few of the key features of these data releases. Further details are described under the specified url (Table~\ref{tab:data-release-table1}).   

\vspace{5pt}\noindent
{\bf Gaia Data Release~1} (DR1) was based on observations collected between 25~July 2014 and 16~September 2015, and was released on 14~September 2016.
It contained source identifier, positions ($\alpha$, $\delta$) and $G$ magnitudes for sources with acceptable standard errors,
the 5-parameter astrometric solution (position, proper motion, and parallax) for stars in common between the Hipparcos-based Tycho~2 Catalogue (epoch around 1990) and Gaia (epoch around 2015), based on the Tycho--Gaia Astrometric Solution \citep[TGAS][]{2015A&A...574A.115M}.
DR1 was incomplete because of limitations in sky coverage, and data processing constraints. Many bright stars, $G<7$, were missing, as were many fainter stars, sources close to bright objects, high proper motion stars, and stars in areas with very high surface densities (above some 400\,000 deg$^{-2}$).
A series of papers in the journal Astronomy \& Astrophysics in 2016--17 describes the various aspects of the DR1 data release
\citep{
2017A&A...599A..50A,
2016A&A...595A...7C,
2016A&A...595A.133C,
2017A&A...600A..51E,
2017arXiv170203295E,
2016A&A...595A...3F,
2016A&A...595A...2G,
2017A&A...607A.105M,
2016A&A...595A...5M,
2017A&A...605A..52M,
2017A&A...599A..32V},
and others with a more scientific focus.

\vspace{5pt}\noindent
{\bf Gaia Data Release~2} (DR2) was based on observations between 25~July 2014 and 23~May 2016, and was released on 25 April 2018.
It contained the 5-parameter astrometric solution ($\alpha$, $\delta$, $\mu_\alpha$, $\mu_\delta$, $\varpi$) for 1.3 billion sources in the range $G=3-21$. 
The parallaxes and proper motions were, for the first time, based only on Gaia data, and no longer depend on Tycho-2/TGAS.
DR2 also contained mean radial velocities for 7.2~million stars (in the range $G=4-13$), 
$G$ magnitudes for 1.69~billion sources, 
$G_{\rm BP}$ and $G_{\rm RP}$ magnitudes for 1.38~billion sources,  
epoch astrometry for 14\,099 known solar system objects,
effective temperatures for 161 million sources brighter than 17~mag, and 
object classification for more than 550\,000 variable sources (Cepheids, RR~Lyrae, Mira, etc.).
A series of papers in the journal Astronomy \& Astrophysics in 2018--19 describes the various aspects of the Gaia DR2 data release
\citep{
2018A&A...616A..17A,
2018A&A...616A...4E,
2018A&A...616A...1G,
2018A&A...618A..30H,
2019A&A...622A.205K,
2018A&A...616A...2L,
2019A&A...621A.144M,
2018A&A...616A..14G,
2018A&A...616A...3R,
2018A&A...616A...6S,
2018A&A...616A..13G},
and others with a more scientific focus.

\begin{figure}[t]
\centering
\includegraphics[width=0.40\linewidth]{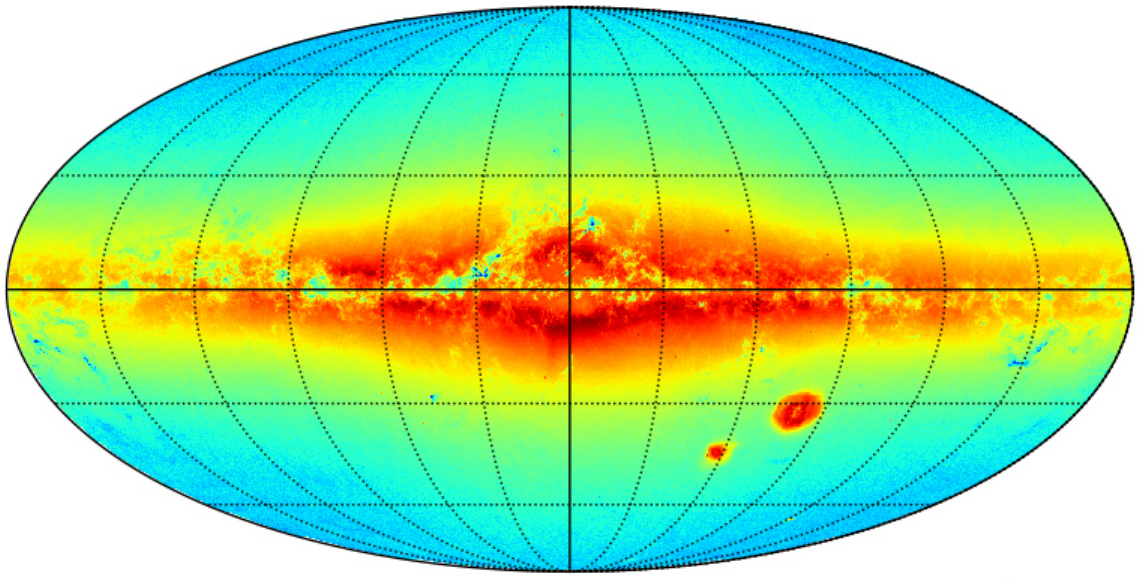}
\hspace{10pt}
\includegraphics[width=0.40\linewidth]{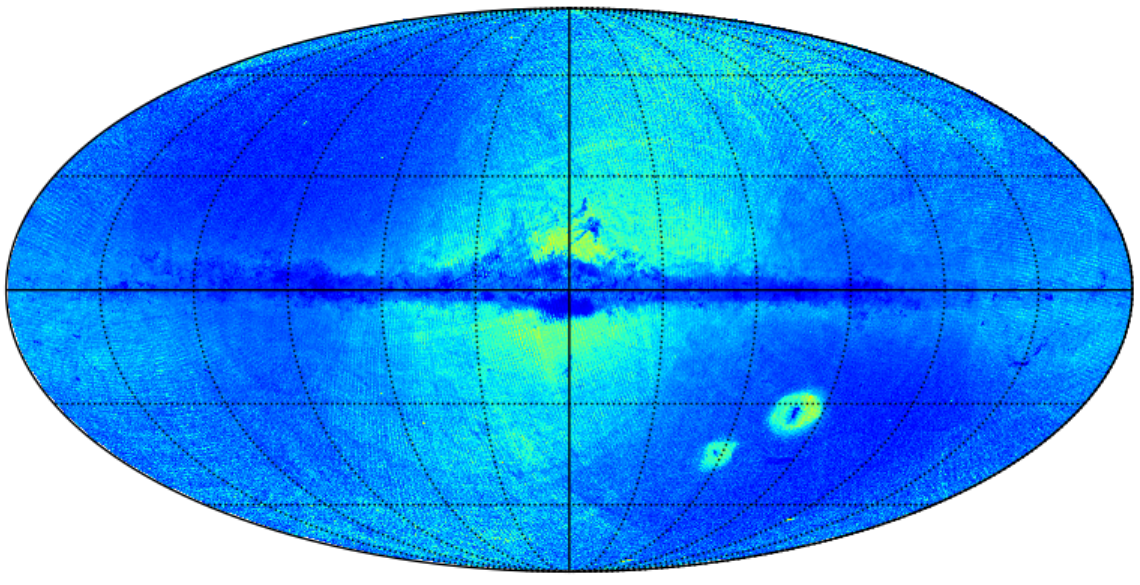}
\caption{Examples of the all-sky characterisation of sources with a 5-parameter solution in Gaia EDR3/DR3 (in Galactic coordinates). Left: density counts, which range from 100 (light blue) to 10\,000 (dark red) stars per square degree.
Right: median parallax error, which ranges from $\lesssim50$\muas\ (dark blue) to $\sim$300\muas\ (yellow). From {\tt gea.esac.esa.int/archive/documentation/GEDR3.}}
\label{fig:edr3-sky-maps}
\end{figure}

\begin{figure}[t]
\centering
\raisebox{7pt}{\includegraphics[width=0.32\linewidth]{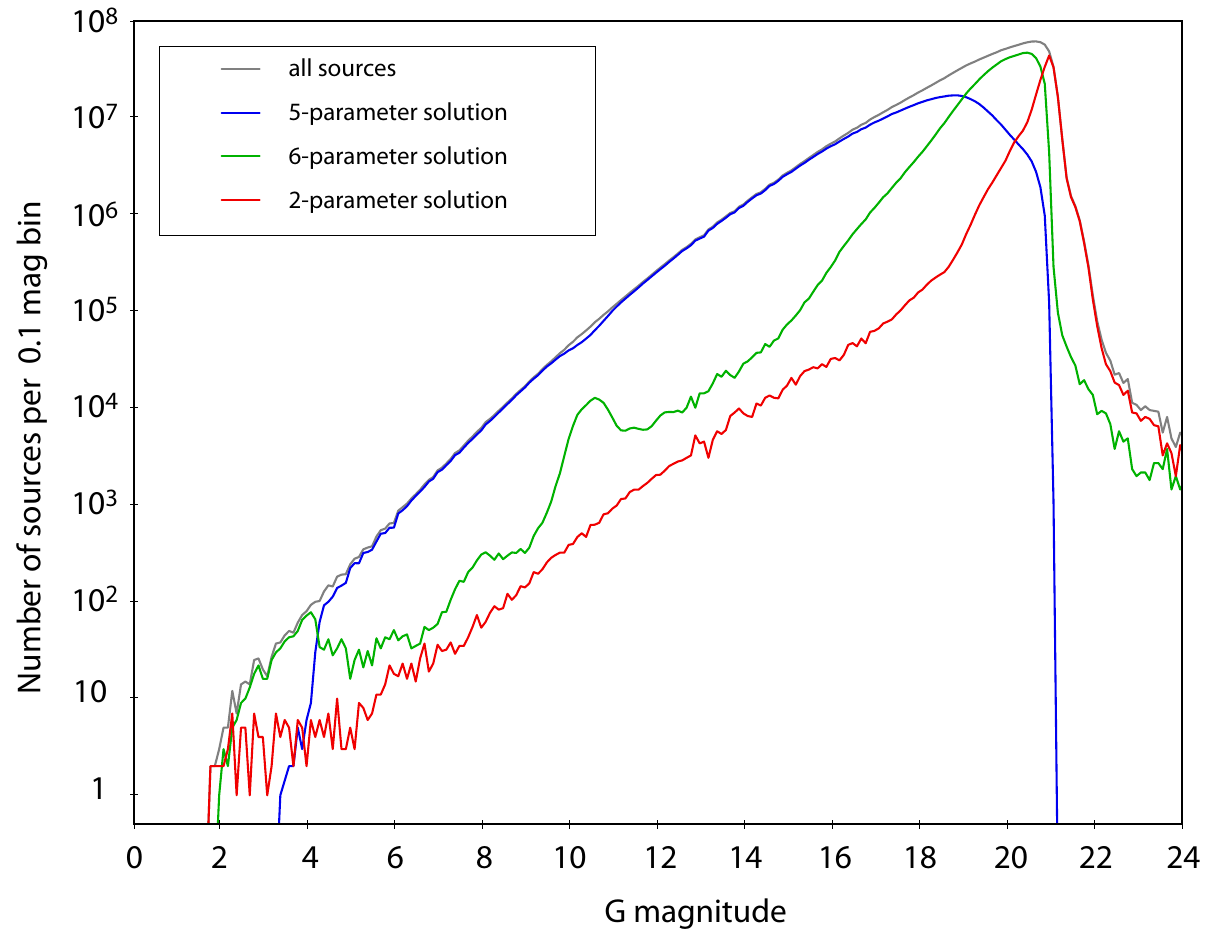}}
\hspace{10pt}
\includegraphics[width=0.64\linewidth]{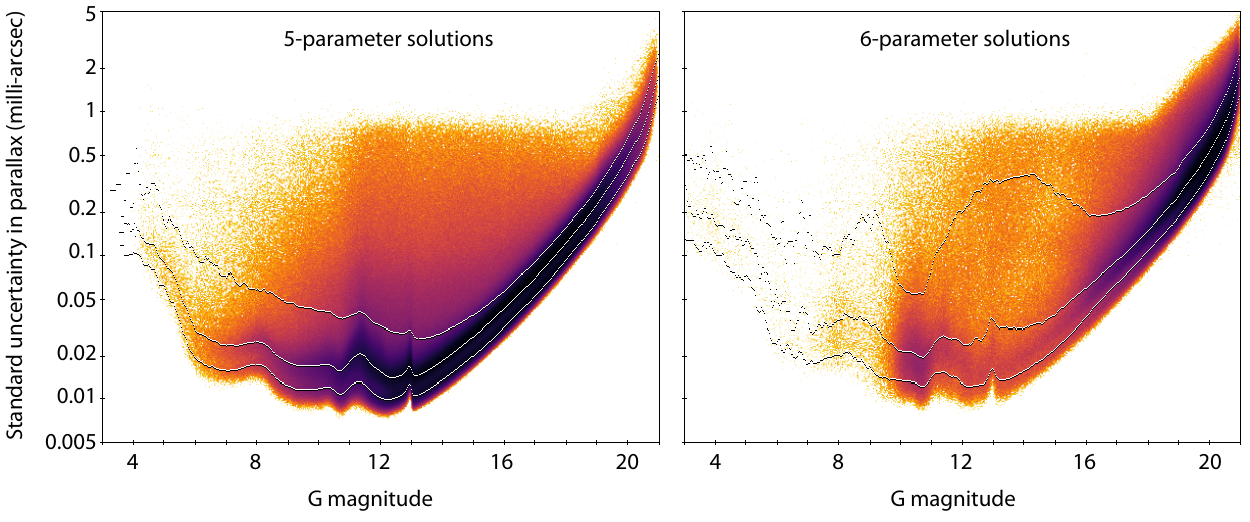}
\caption{Gaia EDR3 astrometry (also applicable to Gaia DR3). Left (a): magnitude distribution, showing all sources (grey), and 5-parameter (blue), 6-parameter (green), and 2-parameter solutions (red).
Right pair (b--c): Uncertainty in parallax versus magnitude, shown separately for the 5-parameter solutions and 6-parameter solutions. Plots include all sources with $G<11.5$, and a decreasing fraction of fainter sources to give a roughly constant number per magnitude interval. The colour scale, from yellow to black, indicates an increasing density of data points. Curves show the 10th, 50th, and 90th percentiles (from \citet{2021A&A...649A...2L}, Figures~5 and~7, where many other details are given).}
\label{fig:edr3-accuracies}
\end{figure}

\vspace{5pt}\noindent
{\bf Early Data Release~3} (EDR3) was based on observations between 25 July 2014 and 28 May 2017, i.e.\ a period of 34 months, and was released on 3~December 2020 (Figure~\ref{fig:edr3-sky-maps}).
It contained 
a 5-parameter astrometric solution for 585 million sources ($G\approx3-21$), 
a 6-parameter solution for a further~882 million sources, 
$G$ magnitudes for 1.806 billion sources, and
$G_{\rm BP}$ and $G_{\rm RP}$ magnitudes for around 1.5~billion sources.
As indications only, 
position uncertainties are 0.01--0.02\,mas ($G<15$), 0.05\,mas ($G=17$), 0.4\,mas ($G=20$), and 1.0\,mas ($G=21$).
Parallax and annual proper motion uncertainties are somewhat comparable (Figure~\ref{fig:edr3-accuracies}).
$G$-band standard errors are around 0.3\,mmag ($G<13$), 1\,mmag ($G=17$), and 6\,mmag ($G=20$).
The precision of the radial velocities is 200--300\ms\ at the bright end. At the faint end, accuracies are around 1.2\kms\ for \teff\,=\,4750\,K and around 3.5\kms\ for \teff\,=\,6500\,K.
A series of papers in the journal Astronomy \& Astrophysics in 2021--22 describes the various aspects of the Gaia EDR3 data release
\citep{
2021A&A...649A...5F,
2021A&A...649A...1G,
2022A&A...667A.148G,
2021A&A...649A...4L,
2021A&A...649A...2L,
2021A&A...649A...3R,
2021A&A...649A..11R,
2021A&A...653A.160S,
2021A&A...649A..10T},
and others with a more scientific focus.

\vspace{5pt}\noindent
{\bf Data Release~3} (DR3), released on 13 June 2022, was based on the same stretch of time and the same set of observations as EDR3, with no new astrometry nor new photometric calibrations. But it is marked many new data products, including non-single stars and solar system objects, inferred from these observations, with associated numbers that are unprecedented in astronomy. 
Precisely as for EDR3, DR3 contains 
the 5-parameter astrometric solution for 585 million sources, with a limiting magnitude of $G\approx21$ and a bright limit of $G\approx3$; 
the 6-parameter astrometric solution for a further~882 million sources,
$G$ magnitudes for around 1.806 billion sources,
$G_{\rm BP}$ and $G_{\rm RP}$ magnitudes for around 1.5~ billion sources,
and cross-matches many other major astronomical catalogues. 
More than 
\href{https://www.cosmos.esa.int/web/gaia/dr3-papers}{30 papers}
in the journal {\it Astronomy \& Astrophysics\/} accompanied and detailed these Gaia~DR3 results, in addition to those focused primarily on the catalogue content 
\citep{
2023A&A...674A...2D,
2023A&A...674A..31D,
2023A&A...674A..13E,
2023A&A...674A...9H,
2023A&A...674A...5K,
2023A&A...674A...3M,
2023A&A...674A..17R,
2023A&A...674A...1G}.

Table~\ref{tab:data-release-table2} summarises the new information contained in DR3, compared with EDR3. 
Notably, radial velocities are provided for more than 33~million sources, and
object classification (as non-single objects, quasars, and solar system objects) are provided for 1.59~billion sources, i.e.\ the majority of the objects observed.
The 10~million identified variables, along with their associated `epoch photometry', have been classified (by `supervised machine learning') into 24 classes, amongst which are 363 microlensing events and 214 planetary transits.
Amongst the solar-system results for 158\,000 sources (including 31 planetary satellites) are orbital solutions and individual epoch observations for 154\,000 objects, along with mean BP/RP reflectance spectra for more than 60\,000 objects.
There are 6.6~million quasar candidates, with redshift estimates for most. Amongst these, the host galaxies are detected for 60\,000, and 15\,000 of these even have estimated surface brightness profiles.
And there are all-sky Galactic extinction maps at four different spatial resolutions (HEALPix levels 6, 7, 8, and 9).

The other major attribute of DR3 is the wealth of astrophysical data that has been distilled from the BP/RP photometry and RVS spectroscopy (Section~\ref{sec:classification-stellar-properties}), notably:
astrophysical parameters ($T_{\rm eff}$, log\,g, [M/H], and reddening) for 470 million objects,
H$\alpha$~emission for 235 million stars,
spectral types for 217~million stars,
evolutionary parameters (mass and age) for 128~million,
activity indices for 1.3~million cool stars, 
emission-line classification for 57\,000 stars,
and spectroscopic parameters for 2.3~million hot stars and 94\,000 ultra-cool stars.

\vspace{5pt}\noindent
{\bf Focused Product Release} (FPR): 	
on 10~October 2023, 
and with DR4 not scheduled until late 2026, the Gaia Data Processing and Analysis Consortium published five papers forming a `Focused Product Release'. This covered:
(a)~specific observations of the globular cluster $\omega$~Cen, resulting in more than half a million new stars observed in its central region
\citep{2023A&A...680A..35G}; 	
(b)~radial velocity time series for 9614 long-period variables with high-quality observations in the range $G=6-14$~mag
\citep{2023A&A...680A..36G};	
(c)~improved solar system astrometry for 157\,000 asteroids using the 66-month time interval covered by DR4
\citep[][\S\ref{sec:solar-system-asteroids}]{2023A&A...680A..37G}; 	
(d)~an improved pipeline treatment for the study of `diffuse interstellar bands' in the RVS data (\S\ref{sec:dib})
\citep{2023A&A...680A..38G}; 	
and 
(e)~a catalogue of secondary sources around quasars
\citep{2024A&A...685A.130G}.	

\subsection{Catalogue access}

Table~\ref{tab:data-release-table1} provides links to the ESA www pages with further details of each of the data releases. For example, following the link to 
\href{https://www.cosmos.esa.int/web/gaia/dr3}{\tt gaia/dr3}, 
also points to the
\href{https://gea.esac.esa.int/archive}{\tt Gaia archive}.
This is the main point of access to the Gaia data (also from the partner data centres: CDS, ASDC, ARI, AIP and Flatiron). Data can be extracted, and downloaded, from the archive using ADQL queries.
The full Gaia DR3 data set ($\sim$10~TB) can also be downloaded from the archive as a set of compressed
ECSV (Enhanced Character Separated Values) files. 

As a simple example, under the `search' tab, individual sources can be interrogated using any of the major catalogue identifiers (as resolved by the CDS SIMBAD/Sesame service), for example Alpha~Ori, Gaia19bvo, GJ~436, HD~12661, Kepler~90, WASP--12, etc. Results include links to the SIMBAD astronomical database, which provides basic data, cross-identifications, bibliography and measurements for non-solar system objects. 
\href{https://simbad.u-strasbg.fr/simbad}{\tt SIMBAD}
may also be queried directly for Gaia sources, sky charts, etc.
Papers making use of the Gaia data for large star samples will frequently include the specific ADQL query used for their sample selection 
(see the Appendixes of, for example,
\citet{2024ApJ...963L..43A,
2024A&A...691A..81D,
2024A&A...691A.242I,
2024A&A...692A.235R}).
ESA's data archive more generally is described by 
\citet{2019ASPC..523..409M}.	

\begin{table}[t]
\vspace{-5pt}
\centering
\footnotesize
\begin{tabular}{lr}
\noalign{\vspace{5pt}}
\hline
\noalign{\vspace{2pt}}
\multicolumn{2}{c}{New results in Gaia DR3}\\
\noalign{\vspace{2pt}}
\hline
\noalign{\vspace{2pt}}
Sources with radial velocities							&	33\,812\,183 \\
Sources with mean G$_{\rm RVS}$-band magnitudes		&	32\,232\,187 \\   
Sources with rotational velocities						&	3\,524\,677 \\[3pt]
Mean BP/RP spectra									&	219\,197\,643 \\
Mean RVS spectra									&	999\,645 \\[3pt]
Variable-source analysis								&	10\,509\,536 \\
Variability types (from machine learning)					&	24 \\
Classified variables									&	9\,976\,881	\\
	\hspace{20pt} Cepheids							&	15\,021 \\
	\hspace{20pt} compact companions					&	6\,306 \\
	\hspace{20pt} eclipsing binaries					&	2\,184\,477 \\
	\hspace{20pt} long-period variables					&	1\,720\,588 \\
	\hspace{20pt} microlensing events					&	363 \\
	\hspace{20pt} planetary transits						&	214 \\
	\hspace{20pt} RR Lyrae stars						&	271\,779 \\
	\hspace{20pt} short-timescale variables				&	471\,679 \\
	\hspace{20pt} solar-like rotational variables			&	474\,026 \\
	\hspace{20pt} upper-main-sequence oscillators			&	54\,476 \\
	\hspace{20pt} active galactic nuclei					&	872\,228 \\[3pt]
Variable with radial-velocity time series					&	1\,898 \\
Sources with object classifications						&	1\,590\,760\,469 \\
Stars with emission-line classifications					&	57\,511 \\[3pt]
Astrophysical parameters (BP/RP spectra)				&	470\,759\,263 \\
Astrophysical parameters (unresolved binary) \hspace{-20pt}	&	348\,711\,151 \\[3pt]
Spectral types										&	217\,982\,837 \\
\noalign{\vspace{8pt}}
\hline
\end{tabular}
\qquad
\begin{tabular}{lr}
\noalign{\vspace{2pt}}		
\hline
\noalign{\vspace{2pt}}
\multicolumn{2}{c}{New results in Gaia DR3 (cont.)}\\
\noalign{\vspace{2pt}}
\hline
\noalign{\vspace{5pt}}
Evolutionary parameters (mass and age)					&	128\,611\,111 \\
Hot stars with spectroscopic parameters					&	2\,382\,015 \\
Ultra-cool stars										&	94\,158 \\
Cool stars with activity index							&	1\,349\,499 \\
H$\alpha$ emission measurements						&	235\,384\,119 \\[3pt]
Astrophysical parameters from RVS spectra				&	5\,591\,594 \\
Chemical abundances from RVS spectra					&	2\,513\,593 \\		
Diffuse interstellar band in RVS spectrum					&	472\,584 \\[3pt]
Non-single (astrometric, eclipsing, etc.)					&	813\,687 \\
	\hspace{20pt} orbital astrometric solutions				&	169\,227 \\
	\hspace{20pt} orbital spectroscopic solutions 			&	186\,905 \\			
	\hspace{20pt} eclipsing binaries					&	87\,073 \\[3pt]
Quasar candidates									&	6\,649\,162 \\
	\hspace{20pt} redshifts							&	6\,375\,063 \\
	\hspace{20pt} host galaxy detected					&	64\,498 \\
	\hspace{20pt} host surface brightness profiles			&	15\,867 \\[3pt]
Galaxy candidates									&	4\,842\,342 \\
	\hspace{20pt} redshifts							&	1\,367\,153 \\
	\hspace{20pt} surface brightness profiles				&	914\,837 \\[3pt]
Solar system objects									&	158\,152 \\
	\hspace{20pt} epoch astrometry (CCD transits)			&	23\,336\,467 \\
	\hspace{20pt} orbits								&	154\,787 \\
	\hspace{20pt} BP/RP reflectance spectra				&	60\,518 \\
	\hspace{20pt} planetary satellites					&	31 \\[2pt]
All-sky Galactic extinction (HEALPix levels)				&	6, 7, 8, and 9 \\
\noalign{\vspace{2pt}}
\hline
\end{tabular}
\caption{Overview of the catalogue contents of Gaia Data Release~3. Amongst these are 33~million radial velocities, 220~million sources with BP/RP spectra, nearly 10~million classified variables, 1.5~billion sources classified, 470~million with astrophysical parameters estimated from the BP/RP spectra (of which 128~million have mass and age inferred from evolutionary models), and 6~million quasars, most with redshift estimates.}
\label{tab:data-release-table2}
\end{table}

\subsection{Population synthesis models and pre-launch simulations}
\label{sec:besancon-model}

Population synthesis models aim to describe a galaxy's observed properties as arising from their constituent stellar population(s), evolving according to well-defined physical processes. An early application was in the interpretation of galaxy colours and spectra
\citep{1972A&A....20..383T,
1983ApJ...273..105B}.
Such models of our Galaxy developed in the early 1980s with the need for star count predictions for operations and observations with the Hubble Space Telescope, while finding numerous applications in Galactic structure, halo modelling, and quasar evolution 
\citep{1980ApJ...238L..17B,
1986ARA&A..24..577B,
1999ascl.soft04001B}. 
Population synthesis models similarly played a key part in preparing for Gaia, and are playing a similar role in preparations for the Vera Rubin Observatory, embracing stars as faint as $r=27.5$~mag 
\citep{2022ApJS..262...22D}. 
Amongst those widely referenced are TRILEGAL 
\citep{2005A&A...436..895G,
2016AN....337..871G},
and the Besançon Galaxy model 
\citep{2003A&A...409..523R,
2012ASSP...26..171R}.
The latter was at the heart of the Gaia Simulation Working Group (which developed into Gaia DPAC Coordination Unit~2) from the early concept definition phase in the 1990s.

The Besançon model is an axisymmetric representation of the Galaxy comprising its four main stellar populations (thin disk, thick disk, bulge, and halo), each with its own star-formation history and stellar evolutionary models. It had already benefitted from the bright star counts from the Tycho~2 catalogue
\citep{2014A&A...564A.102C}.
Based on it, the Gaia Simulator Universe Model was then used to create catalogues of stellar sources (both single and multiple stars), along with nebulae, star clusters, diffuse light, planets, satellites, asteroids, comets, (un-)resolved galaxies, quasars, and supernovae
\citep{2014A&A...566A.119L}. 	
It included distances, magnitudes, spectral characteristics, and kinematics (as well as observation errors), and was used as the basis for the
Gaia Object Generator which generated number counts
\citep[GOG,][]{2014A&A...566A.119L},
the Gaia Instrument and Basic Image Simulator which generated pixel-level images
\citep[GIBIS,][]{2005ESASP.576..417B},
and the Gaia System Simulator which simulated the telemetry stream
\citep[GASS,][]{2010ASSP...14..515M}.
The Gaia Simulator was central in predicting the numbers of stars in the two superimposed viewing directions and the (ecliptic-based) scanning law. This was, in turn, crucial for optimising the CCD readout strategy, the on-board data handling requirements, the on-board memory sizing, and the required downlink data rate from its L2 orbit.
It was also used for tests of the astrometric global iterative solution, providing star distributions for large-scale simulations, choice of primary stars and the effects of binaries, tests of object matching, and so on.

With the successive Gaia data releases have come progressive improvements and insights which have been fed back into these population synthesis models. Initial comparisons of the Besançon model were made with Gaia DR1 by
\citet{2017A&A...599A..50A}.	
More comprehensive results from DR2 were used to construct a revised model (using the PARSEC evolutionary tracks), which also included white dwarfs, the Magellanic Clouds, and a number of open clusters with internal rotation, along with reddening according to revised 3d extinction maps
\citep{2020PASP..132g4501R}.	
Comparisons of the EDR3-based Gaia Catalogue of Nearby Stars (within 100~pc) were used to derive constraints on 
the contribution of high-velocity halo stars, 
the Sun's height above the Galaxy mid-plane, 
and the solar motion 
\citep{2021A&A...649A...6G}.		
In another wide-ranging study based on Gaia~EDR3, the Besançon model was used to construct the gravitational potential of the Galaxy, and the stellar distribution functions of the thin/thick disk components 
\citep{2022A&A...667A..98R}.	
Their model produces densities and kinematics consistent with EDR3, over 6--12~kpc in Galactocentric radius, and $\pm2$~kpc in $z$. It also results in 
a solar motion consistent with other recent studies; 
a thin disk comprising seven age components each with its own velocity ellipsoid; 
an asymmetric drift increasing with age and $z$, and decreasing with $R$;
a tilt of the velocity ellipsoid varying with age, $R$, and $z$;
a thin disk population density exhibiting a significant flare in the outer regions, arising from the variation of the vertical force with $R$;
significant differences between the young thick-disk and old thin-disk properties;
and
different variations with $R$ for the mean azimuthal velocity at the solar position, supporting the idea of a different formation for the thin and thick disks.
Despite good fits assuming an axisymmetric potential, discrepancies are seen towards the anti-centre (where disk substructures have been highlighted), towards the North Galactic Pole (where a vertical wave is seen), and in the model of the stellar halo.
Similar comparisons have been made between Gaia DR2 and the TRILEGAL Galaxy model by
\citet{2021MNRAS.506.5681D}.

Amongst other Gaia related applications, TRILEGAL was used to construct a star count model as a function of magnitude, populating them with planets (according to known dependencies on spectral type, planet mass, and orbits), and so predicting the number of exoplanets detectable with Gaia astrometry over a projected 5- or 10-yr mission
\citep{2014ApJ...797...14P}.
The Besançon Galaxy model was similarly used to predict Gaia's brown dwarf detection rates around FGK stars in astrometry, radial velocity, and transits
\citep{2022A&A...661A.151H}. 	
The Besançon model has also been used in developing unbiased distance estimates from the observed parallaxes (Section~\ref{sec:distance-estimation}), where distance estimates can be improved by adding prior information about the Galaxy
\citep[e.g.][]{2015PASP..127..994B},
and as a constraint on the Galaxy's star-formation history and stellar initial mass function
\citep{2025A&A...697A.128D}.	

\begin{figure}[p]
\centering
\includegraphics[width=0.40\linewidth]{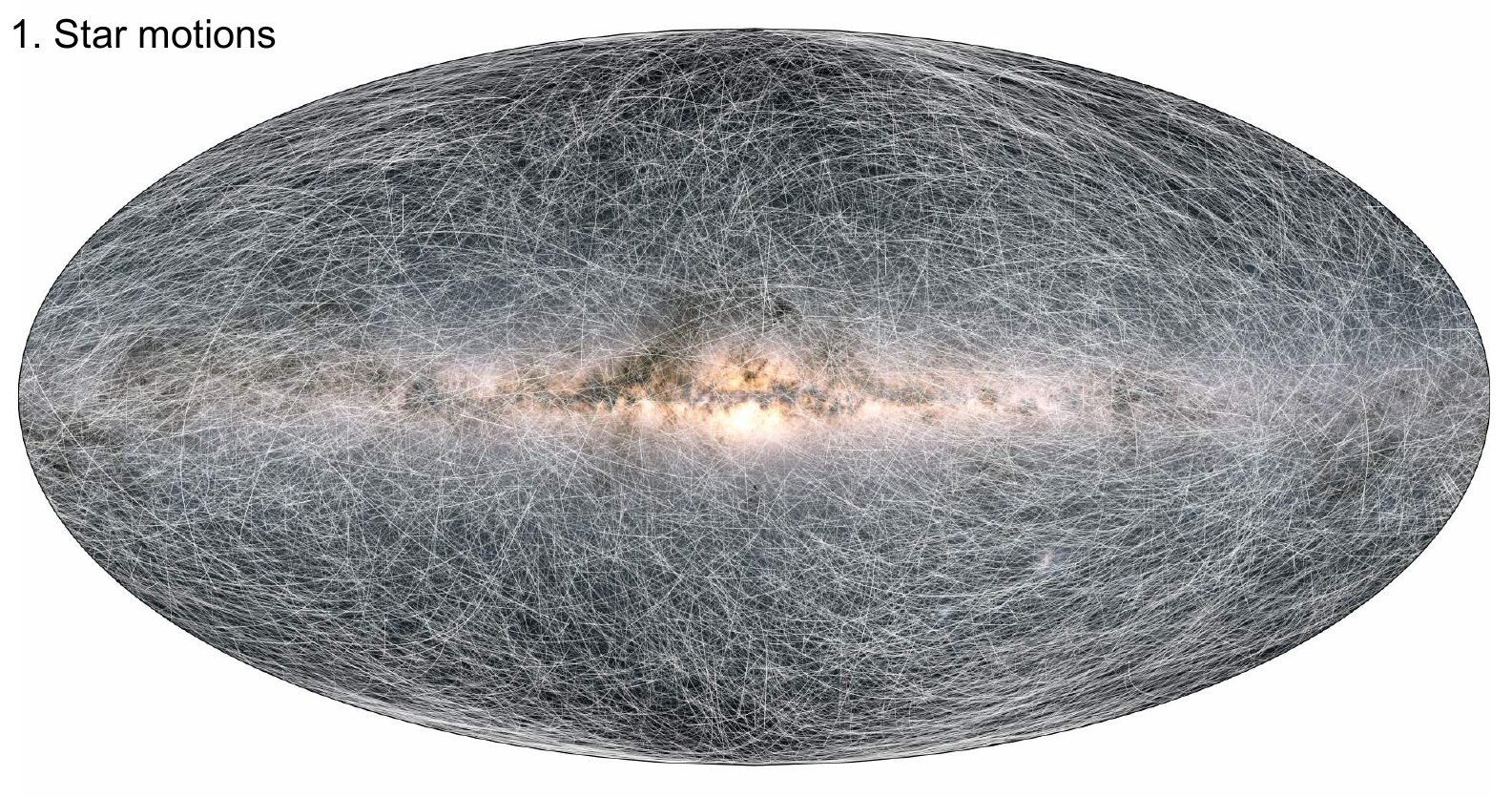} \hspace{20pt}
\includegraphics[width=0.40\linewidth]{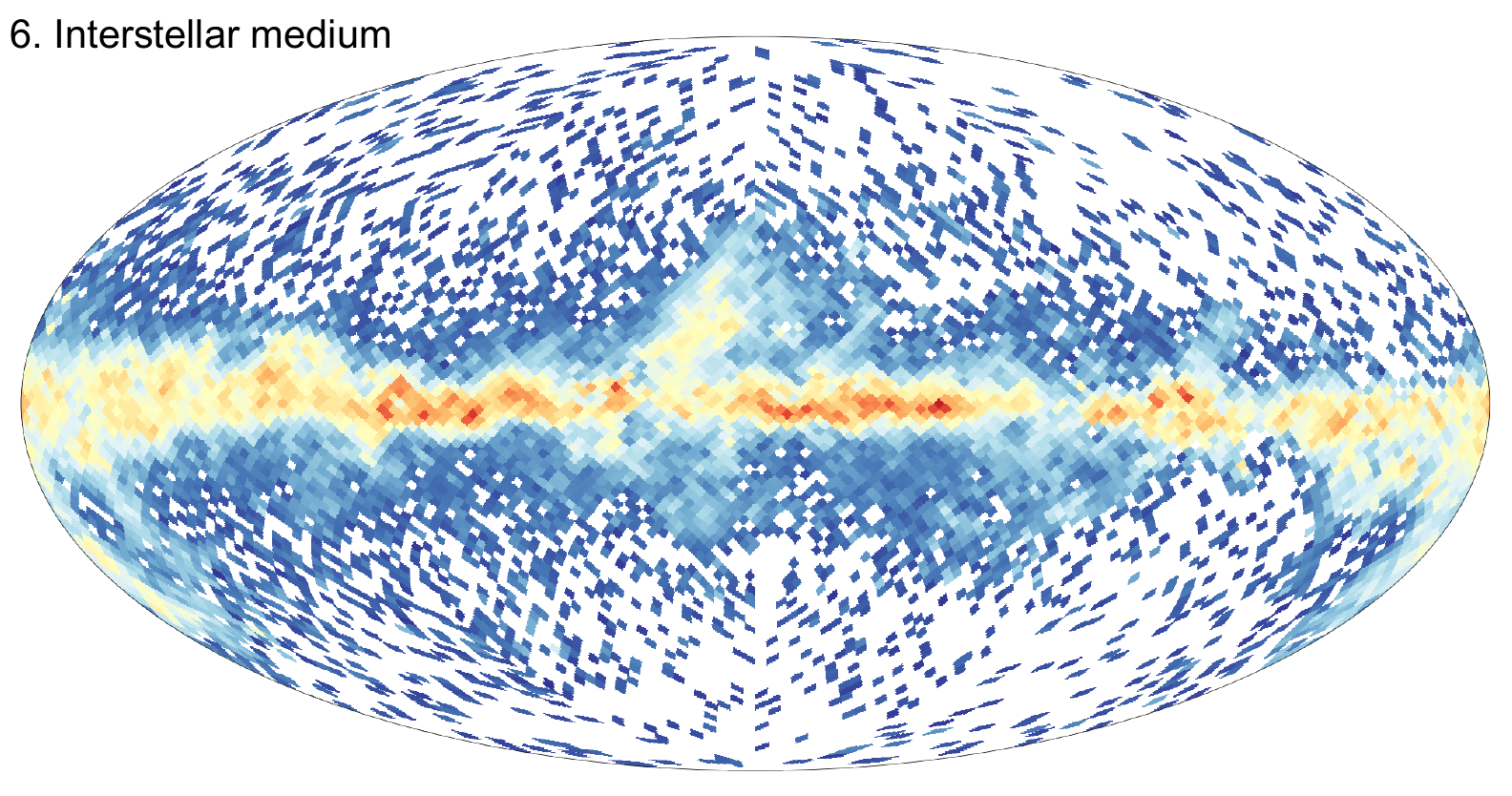}
\includegraphics[width=0.40\linewidth]{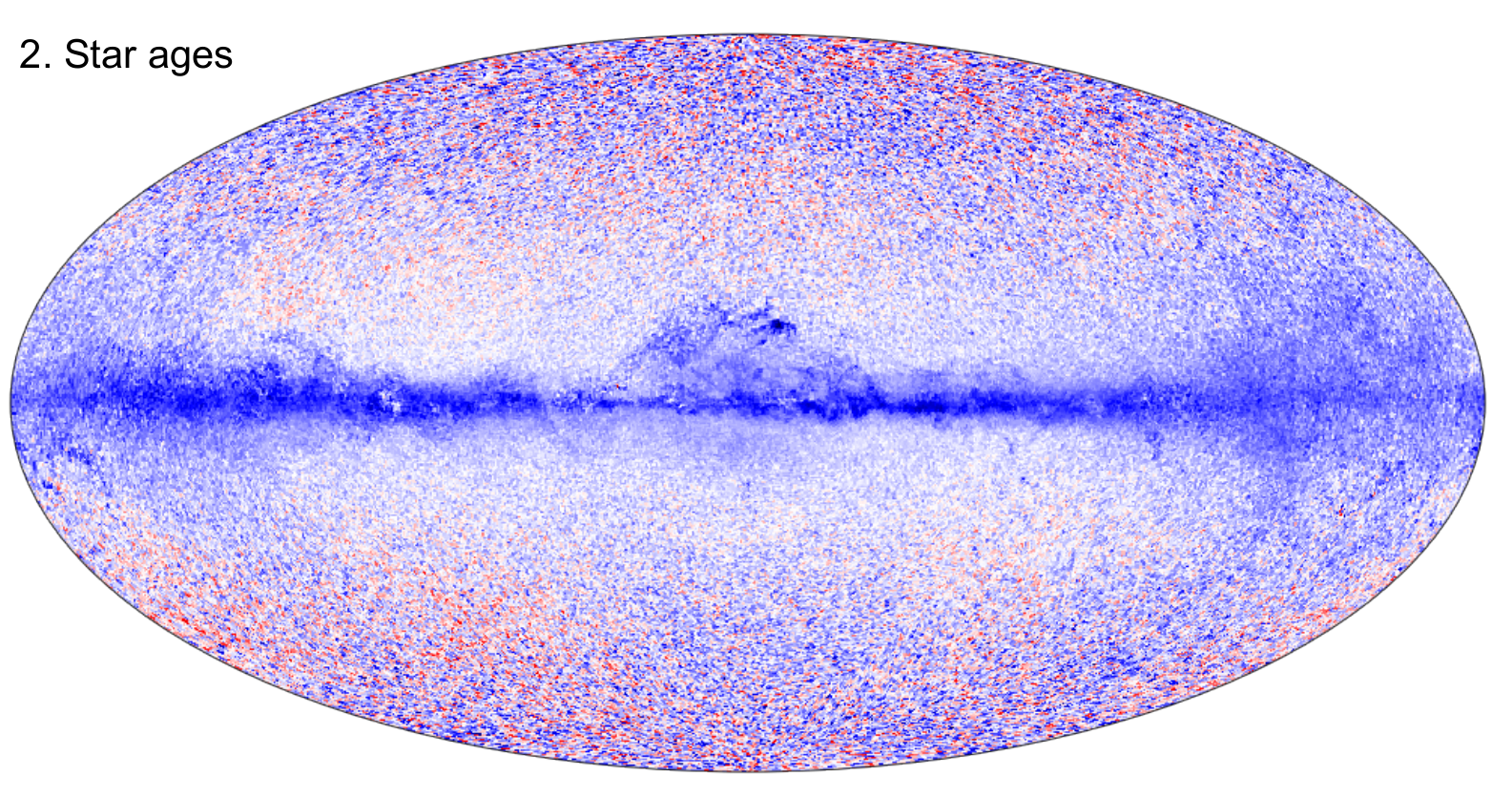} \hspace{20pt}
\includegraphics[width=0.40\linewidth]{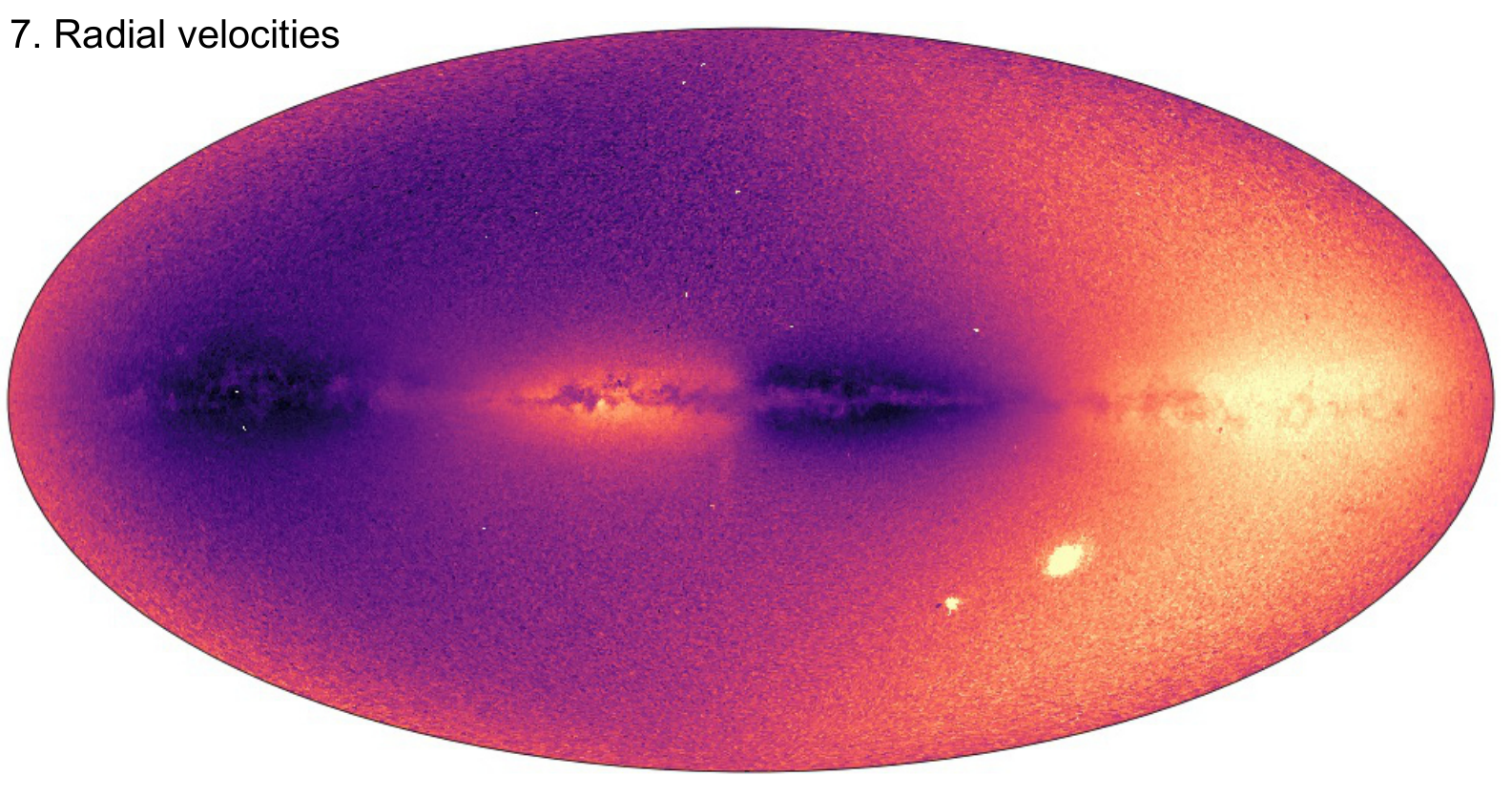}
\includegraphics[width=0.40\linewidth]{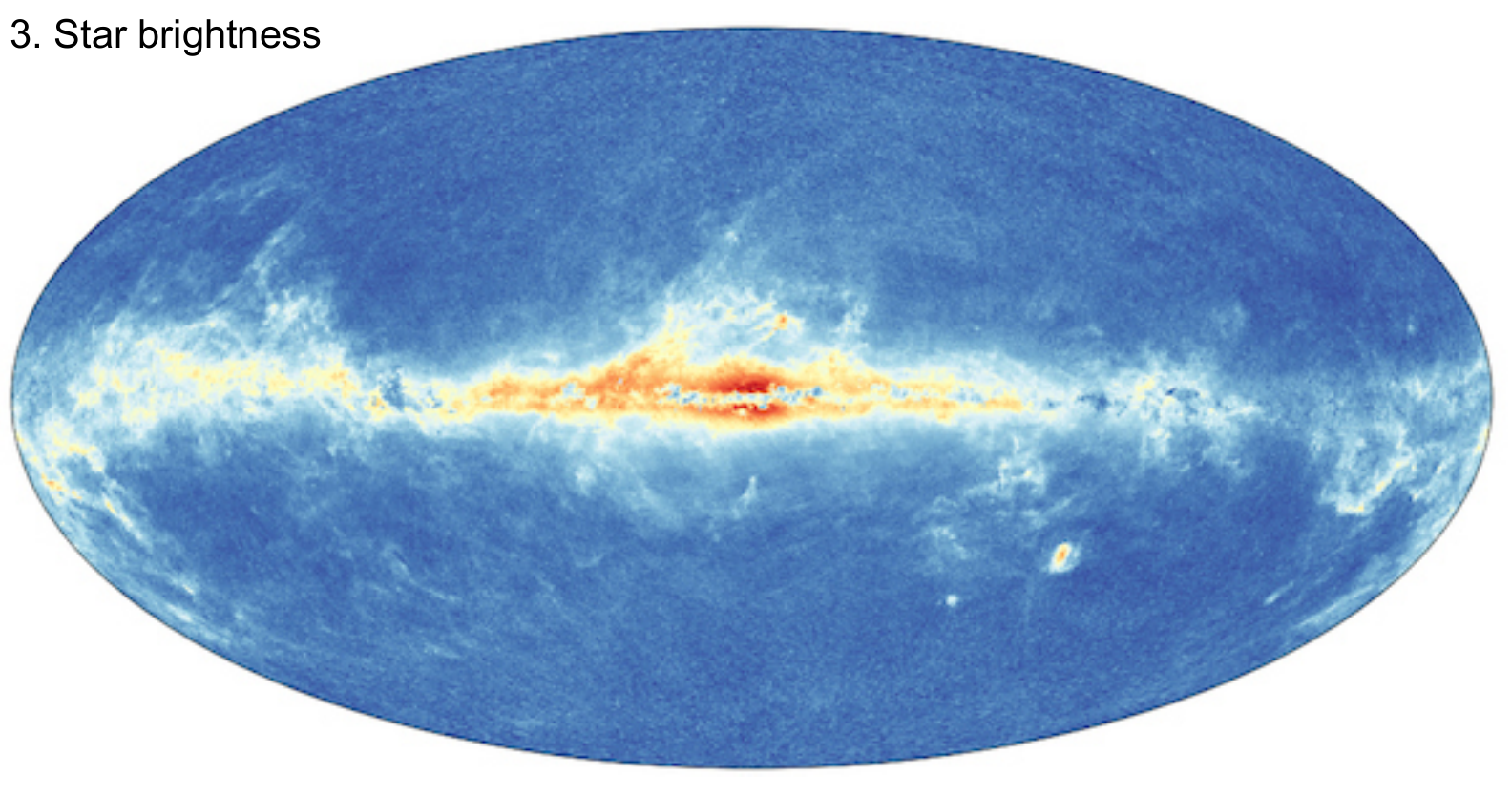} \hspace{20pt}
\includegraphics[width=0.40\linewidth]{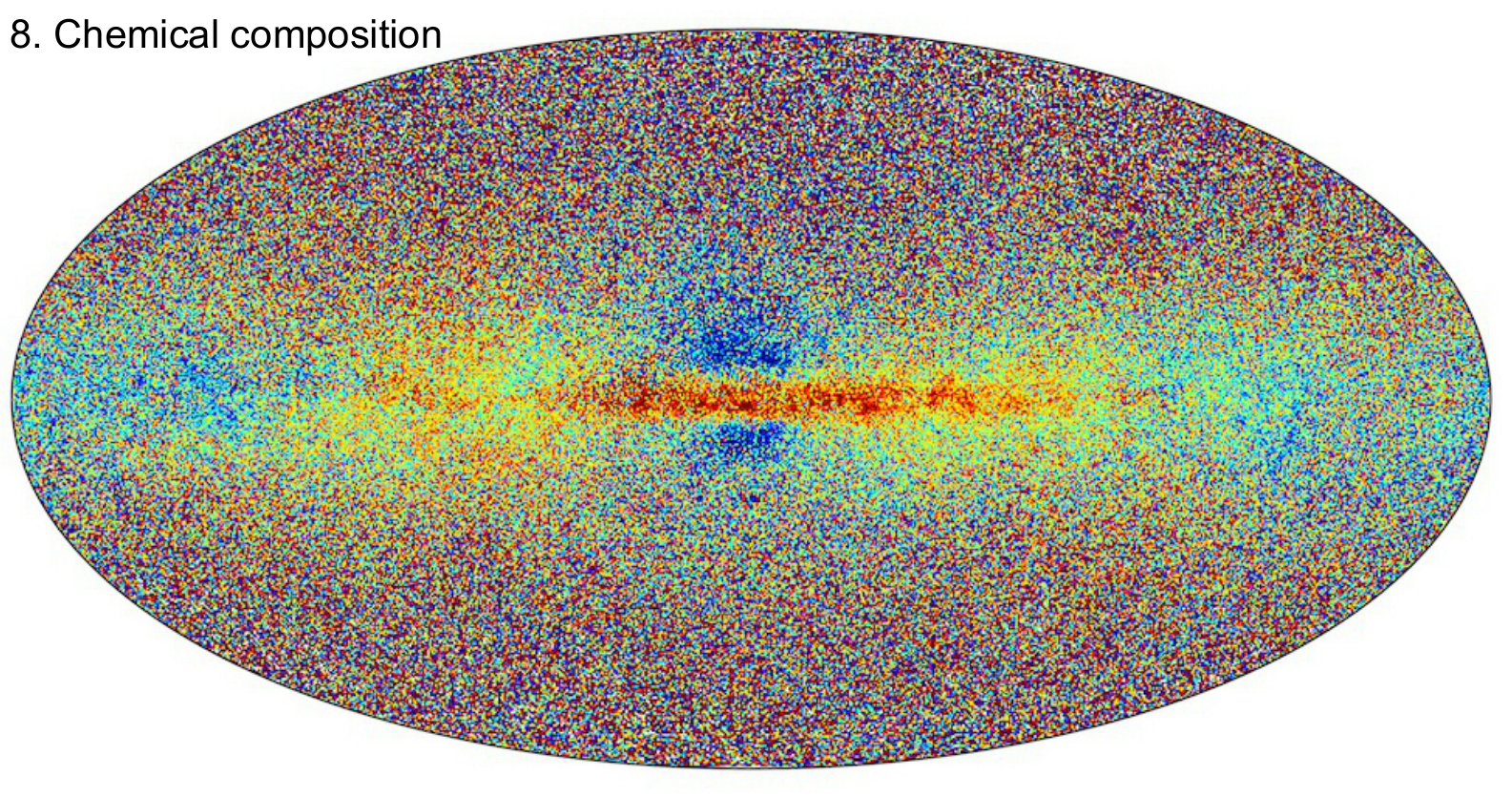}
\includegraphics[width=0.40\linewidth]{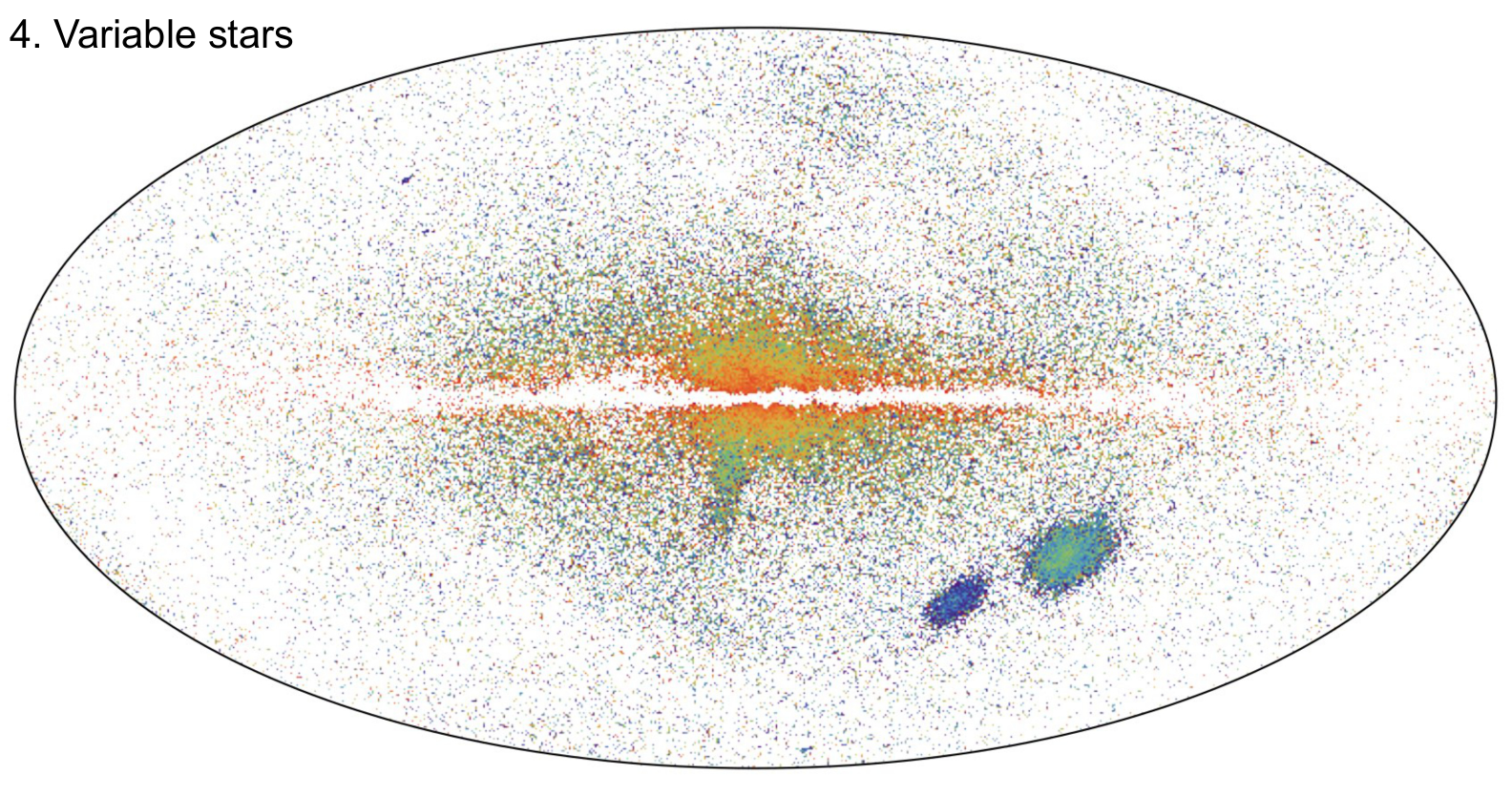} \hspace{20pt}
\includegraphics[width=0.40\linewidth]{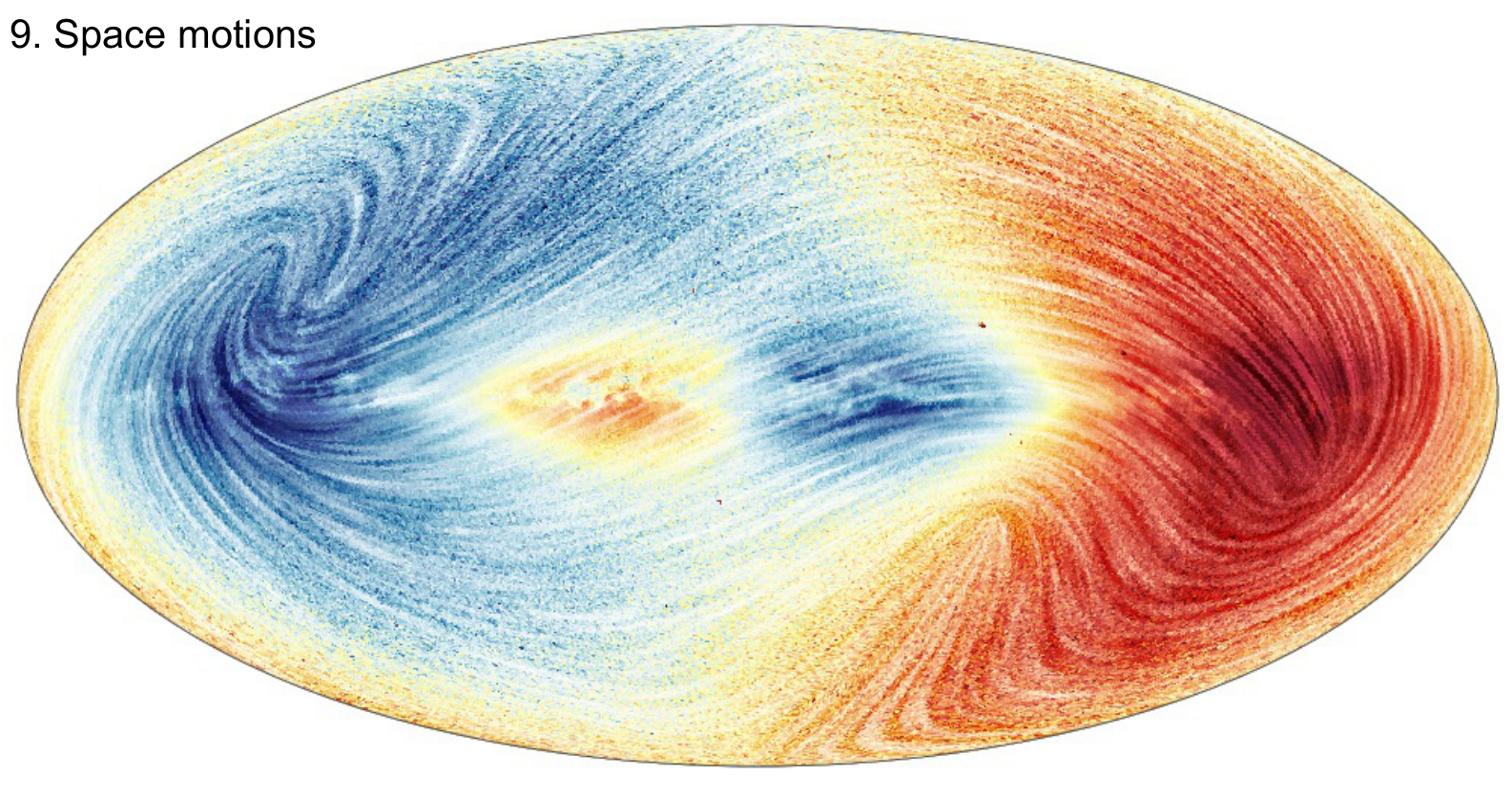}
\includegraphics[width=0.40\linewidth]{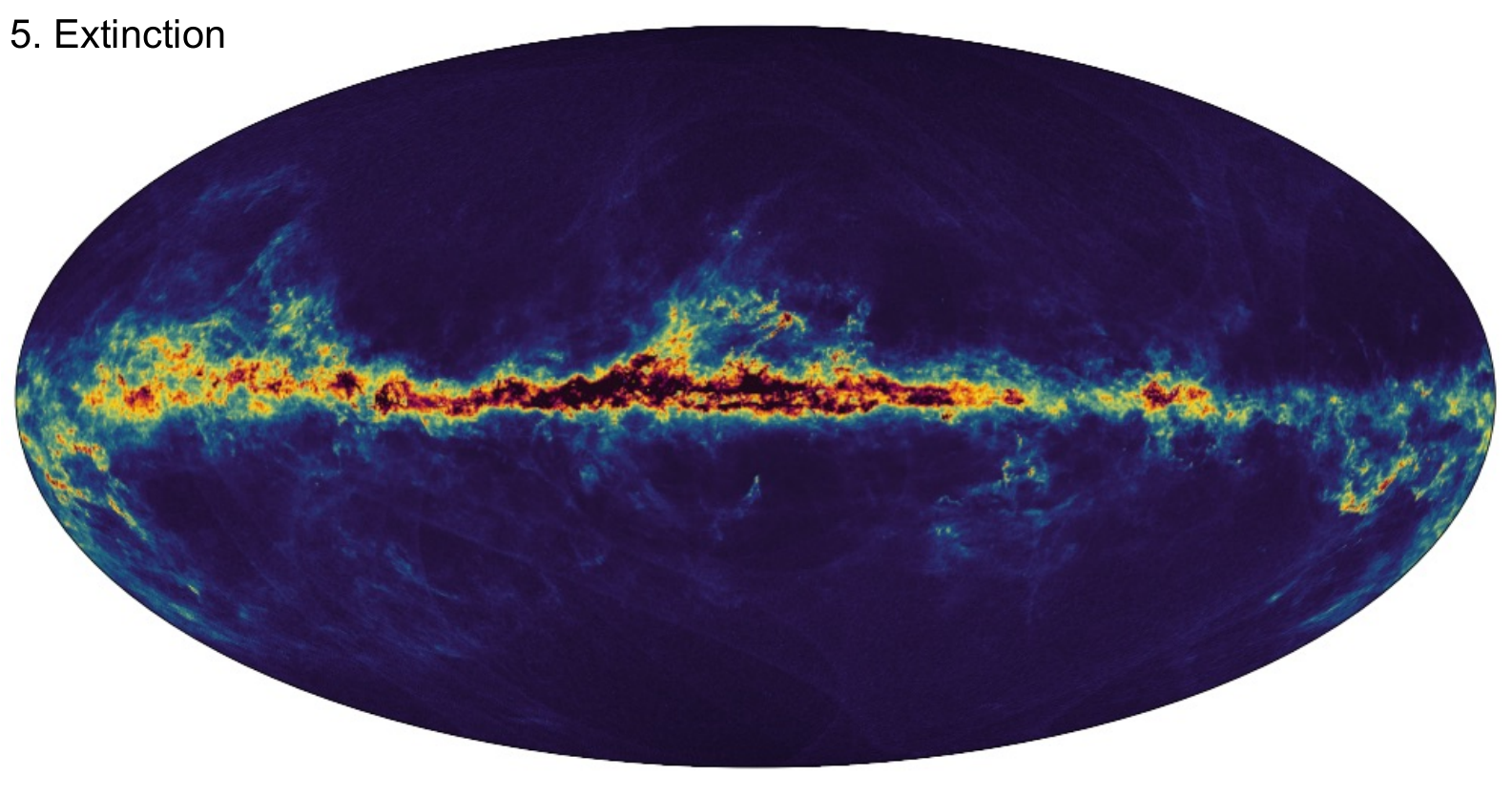} \hspace{20pt}
\includegraphics[width=0.40\linewidth]{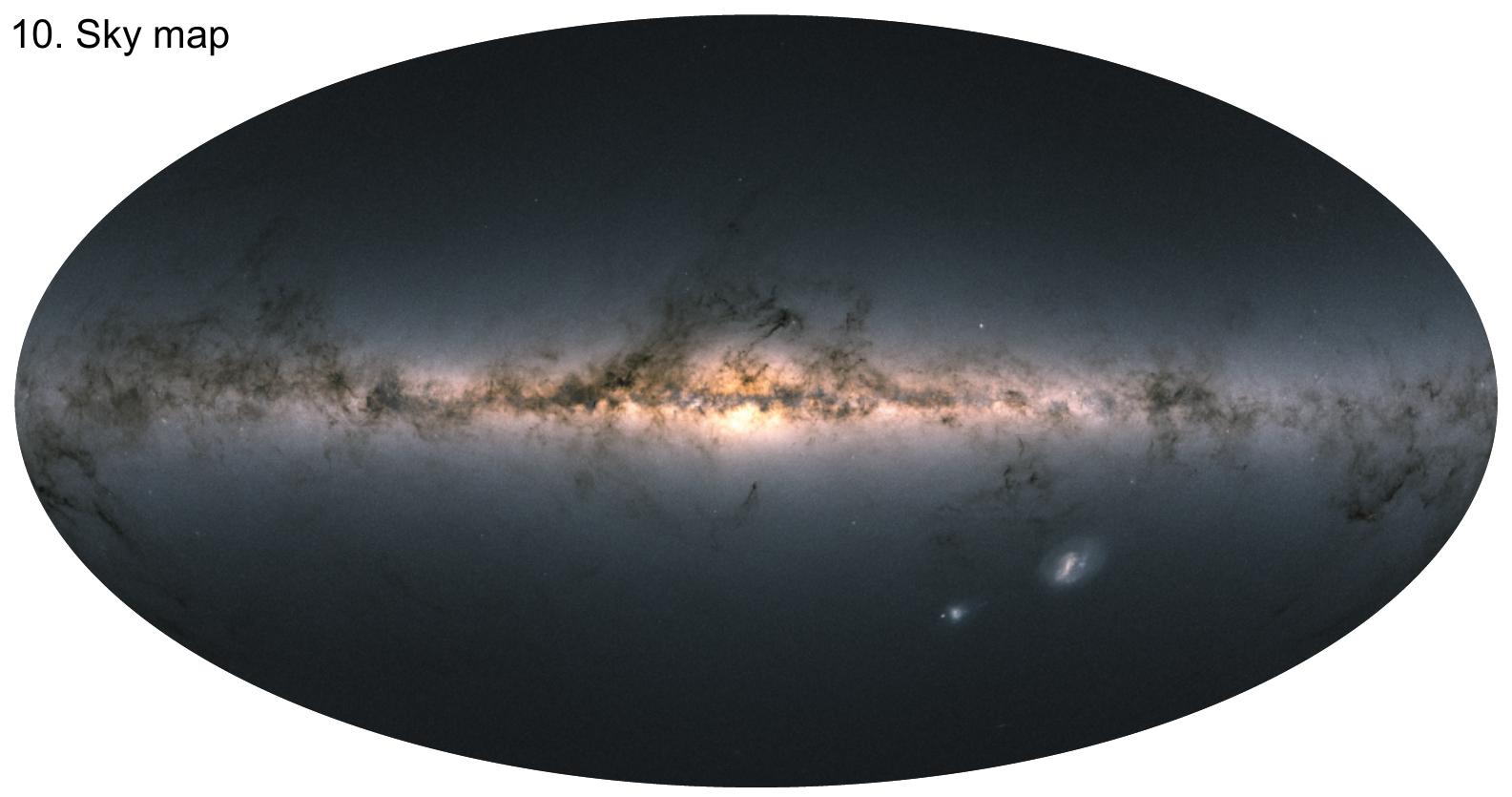}
\caption{Sky projections, in Galactic coordinates, showing various properties of the Gaia DR3 data release (see text for details).
}\label{fig:gaia-maps}
\end{figure}

\subsection{Maps of the Milky Way}

A set of 10 
\href{https://www.cosmos.esa.int/web/gaia/iow_20230613}{maps of the Milky Way},
prepared by members of the Gaia Data Processing and Analysis Consortium (DPAC), were used as the ESA--Gaia `Image of the Week' on 13~June 2023, the first anniversary of the 34-month Data Release~3 (DR3). 
They give a global view of the catalogue contents, projected in Galactic coordinates, according to various criteria reflecting the Galaxy's structure, dynamics, chemical composition and history.
They are included here (Figure~\ref{fig:gaia-maps}) with a concise description below. Further details are given under the `more' link for each description.

{\footnotesize
\noindent
$\bullet$ Star motions (panel~1): trails show the displacement of stars on the sky 400\,000 years into the future
\href{https://www.cosmos.esa.int/web/gaia/edr3-startrails}{[more]}
(A.~Brown, S.~Jordan, T.~Roegiers, X.~Luri, E.~Masana, T.~Prusti, and A.~Moitinho)

\vspace{1pt}\noindent $\bullet$ 
Star ages (panel~2): The average age of stars in the Galaxy, derived from the CU8 FLAME module, based on a random selection of 10~million stars from DR3: blue are the youngest, red the oldest
\href{https://www.cosmos.esa.int/web/gaia/dr3-how-big-or-warm-or-old-are-the-stars}{[more]}
(O.~Creevey, M.~Fouesneau, and the Gaia group at MPIA)

\vspace{1pt}\noindent $\bullet$ 
Star brightness (panel~3): The median $G-G_{\rm RVS}$ colour, highlighting the effect of extinction by interstellar dust. $G_{\rm RVS}$ is derived from the integrated radial velocity spectrometer spectra
\href{https://www.cosmos.esa.int/web/gaia/dr3-how-bright-are-the-stars}{[more]}
(N.~Leclerc, P.~Sartoretti, and the CU6 team)

\vspace{1pt}\noindent $\bullet$ 
Variable stars (panel~4): The 134\,000 RR~Lyrae stars for which an estimate of metallicity was obtained, colour-coded according to metallicity, [Fe/H]
\href{https://www.cosmos.esa.int/web/gaia/dr3-how-do-they-blink}{[more]}
(G.~Clementini, A.~Garofalo, T.~Muraveva, V.~Ripepi, R.~Molinaro, S.~Leccia, and the CU5/CU6/CU7 teams)

\vspace{1pt}\noindent $\bullet$ 
Extinction (panel~5): GSPphot provides extinction estimates for individual stars. This map is based on 470~million sources from DR3, with $A_0$ ranging from 0--4 on the colour scale
\href{https://www.cosmos.esa.int/web/gaia/dr3-what-is-in-between-the-stars}{[more]}
(T.E.~Dharmawardena, and the Gaia group at MPIA)

\vspace{1pt}\noindent $\bullet$ 
Interstellar medium (panel~6): The distribution of diffuse interstellar bands (DIB), colour-coded by the equivalent width of the DIB at 1.8~square degree resolution. Strong DIBs are concentrated towards the Galactic plane
\href{https://www.cosmos.esa.int/web/gaia/dr3-what-is-in-between-the-stars}{[more]}
(H.~Zhao, M.~Schultheis, and the CU8 team)

\vspace{1pt}\noindent $\bullet$ 
Radial velocities (panel~7): The Gaia DR3 radial velocities
\href{https://www.cosmos.esa.int/web/gaia/dr3-do-they-approach-us-or-move-away}{[more]}
(D.~Katz, N.~Leclerc, P.~Sartoretti, and the CU6 team)

\vspace{1pt}\noindent $\bullet$ 
Chemical composition (panel~8): Colours indicate stellar metallicity, [M/H], i.e.\ the mean abundance of all chemical elements except H~and~He. Redder stars are richer in metals. The data is based on all stars in the GSP--Spec database
\href{https://www.cosmos.esa.int/web/gaia/dr3-what-are-they-made-of}{[more]}
(A.~Recio-Blanco, and the CU8 team)

\vspace{1pt}\noindent $\bullet$ 
Space motions (panel~9): Colours show the line-of-sight velocity of stars (blue approaching, red receding). Lines show the proper motions
\href{https://www.cosmos.esa.int/web/gaia/dr3-do-they-approach-us-or-move-away}{[more]}
(O.N.~Snaith, P.~Di Matteo, P.~Sartoretti, N.~Leclerc, D.~Katz, and the CU6 team)

\vspace{1pt}\noindent $\bullet$ 
Sky map of brightness and colour (panel~10): Brightness and colour of the 1.8~billion stars included in Gaia EDR3
\href{https://sci.esa.int/web/gaia/-/the-colour-of-the-sky-from-gaia-s-early-data-release-3}{[more]}
(A.~Moitinho)
}

\subsection{Use in telescope pointing and calibration}

Before entering the more detailed scientific results coming from Gaia, I should mention that the very accurate and very dense positional reference system provided by Gaia is now widely used across ground- and space-based telescopes, both for pointing and calibration. Here, I will give just a few examples.

At ESOC, ESA's satellite operations centre at Darmstadt, Germany, the Gaia catalogue is today used across all satellite operations and instrument pointing. It is incorporated into the latest very high-accuracy star trackers, as well as in the fine guidance sensors which ensure stable long-duration satellite pointing (including use in Euclid, launched in 2023, and in the planned Plato and Ariel missions).
Both the Hubble Space Telescope and James Webb Space Telescope rely on the Space Telescope Science Institute's Guide Star Catalog (GSC) for telescope pointing and instrument calibration. The most recent version, GSC3, uses Gaia DR3 as its base catalogue, which is used for identifying reference stars, and for their image calibration pipeline.
More widely, GSC itself has been adopted for observation planning, preparing finding charts, and for the operation of ground-based telescopes. Amongst these, recalibration of the astrometry for Pan--STARRS1 has been carried out based on Gaia~DR2 for the 1.7~billion objects in PS1/DR2 
\citep{2021AJ....161....6L},
and later based on Gaia~EDR3
\citep{2022AJ....164...73W}.
The infrared Nancy Grace Roman Space Telescope (formerly WFIRST), due for launch in May 2027, makes use of the Gaia entries in GSC both for its guide stars, and for the pipeline image calibration.
For JWST, Gaia has also played a key role in the calibration of the optical distortion for all four instruments
\citep{2021jwst.rept.7716A,
2022SPIE12180E..0YL,
2023ApJ...950..101L,
2024A&A...682A..53P},
as well as in the detailed calibration of NIRSpec's Micro-Shutter Assembly. NIRSpec observations are based on prior NIRCam exposures, which are placed on the Gaia reference frame, thus ensuring that science targets indeed fall within the 200~millli-arcsec wide MSA slitlets.

A prominent example of the Gaia catalogue use in spacecraft navigation was the New Horizons fly-by of Pluto and its moon Charon, in July 2015. Navigation was based on the JPL solar system ephemerides DE430, itself based on the Hipparcos reference frame. But the subsequent fly-by of the trans-Neptunian object Arrokoth on 1~Jan 2019 made use of the Gaia DR2 star positions to adjust the spacecraft's pointing. When the closest flyby images came back, Arrokoth was `framed perfectly'. {\it `None of that would have happened if we hadn’t had the Gaia catalogue'}, Marc Buie, discoverer of Arrokoth, is quoted as saying. {\it `It’s a fundamental rewriting of how we do positional astronomy'.}

In examples of ground-based telescope operations, 
Stephen Potter, Head of Astronomy at SAAO (South African Astronomical Observatory), summarised Gaia's contributions:  {\it `Many of our new observatory operational and scientific developments have directly benefited from Gaia data releases in ways that were simply not possible before. Its high-precision astrometric and spectroscopic data has been transformative for both our operations and our science.'}
For SDSS, the Sloan Digital Sky Survey, Mike Blanton commented that Gaia {\it `has been transformative'}. It now provides a reliable and dense set of guide stars for wide-field spectroscopy across the entire sky, especially for the southern hemisphere. {\it `Before Gaia, many many hours were spent cobbling together various sources of guide stars, ensuring they were on the same system, estimating magnitudes, etc. This instantly became a completely solved problem when Gaia data was available'}. 
Beyond the operational aspects, Gaia is used in star selection for the 3~million star SDSS--V Milky Way Mapper, using cuts on parallax, colour, and absolute magnitude 
\citep{2023ApJS..267...44A}.

At the 4.2-m William Herschel Telescope on La Palma, Gaia is one of the main reasons for building the new fibre-fed multi-object spectrograph, WEAVE, which will take spectra of up 1000 objects simultaneously.  The instrument was built to provide spectroscopic follow-up for several large survey facilities, and in particular Gaia. Use is also made of the Gaia magnitudes of the guide stars to monitor the transparency of the sky while observing.  This allows observations to be extended (or re-scheduled) as required, again promoting more efficient use of the night. More widely, use of Gaia's accurate coordinates, for both targets and guide stars, has made acquisition onto slits and fibres significantly more accurate than before, thereby improving instrument throughput, and telescope efficiency.  
 
The Vera Rubin Observatory (first light in July 2025) presents various challenges for its astrometric calibration. With a 7~mag fainter limit than Gaia, it will observe an order of magnitude more stars. It will use Gaia's 3-colour photometry to transfer Gaia's photometric system to all its objects 
\citep{2019ApJ...873..111I}. 	
And {\it `Astrometric calibration will be based [on Gaia], which will provide numerous high-accuracy astrometric standards in every [LSST] field'}.

\subsection{Animations}

Astrometry, even space astrometry, doesn't easily lend itself to `public outreach'. In terms of visual impact, Gaia (and Hipparcos before it) cannot compete with the images returned by the Hubble and James Webb Space Telescopes, or the drama of planetary fly-by and lander missions. Nonetheless, many explanatory animations and simulated `fly-through' videos (of the solar system, and of the solar neighbourhood and beyond), have been created over the past few years. Some simulate the 3d-distribution of stars in space and, extrapolated over thousands of years, their space motions. Some have been prepared by ESA, and many others by members of the Gaia Data Processing and Analysis Consortium (DPAC), and the members of the various Coordination Units.\footnote{
Planetaria, of course, provide spectacular platforms for visualising the night sky. But a particularly effective (although rarely used) technique is stereoscopic projection using two projectors equipped with orthogonal polarising filters.  Projected onto a large silvered screen (to avoid polarisation loss), with the audience equipped with polarised glasses, the effect is almost like looking up at the night sky, while resolving the depths of space, and preserving star colours. It is visually powerful, but non-trivial to set up and execute. I used this technique to deliver the (Hipparcos-based) Darwin Lecture to the UK Royal Astronomical Society in 1997, and the Invited Discourse at the IAU General Assembly in 2000.} 
A convenient starting point are those in the 
\href{https://www.zah.uni-heidelberg.de/gaia/outreach/gaiasky}{`Gaia Sky'}
category, a suite of videos created by astronomer Stefan Jordan and computer scientist Toni Sagristà (ZAH--ARI, Heidelberg). Stefan Jordan also maintains a compilation of various Gaia YouTube videos and `playlists' (i.e.\ video collections, with some content overlapping), amongst which are:	

{\footnotesize$\bullet$}
\href{https://www.youtube.com/@esagaiamission2542}{Gaia ESA YouTube Channel}: 126 videos

{\footnotesize$\bullet$} 
\href{https://youtube.com/playlist?list=PLTUZKJKqW_n-9dlDBQP7CluQw3IVd4dEC}{Gaia YouTube videos}: 562 videos

{\footnotesize$\bullet$} 
\href{https://youtube.com/playlist?list=PLTUZKJKqW_n-j6R9jAvo5svm1x7w2wjxF}{Gaia DR3 collection}: 115 videos

{\footnotesize$\bullet$}
\href{https://youtube.com/playlist?list=PLTUZKJKqW_n9uI_RlPZAdFcUfMU8_zl9o}{Gaia Sky collection, DR1--DR3}: 132 videos

\noindent
\href{https://en.wikipedia.org/wiki/Gaia_Sky}{\tt GaiaSky}
itself is an open-source multi-platform sky visualisation program, created in the framework of Gaia, and with a powerful range of functionality including 3d representation
\citep{2019-gaiasky}.

\section{Physical effects}
\label{sec:physical-effects}

Measurements made at the micro-arcsec level reveal some specific `physical effects' that are relevant to some of the discussion of solar system science, white dwarfs, and others, and I will consider them together in this section. These include various effects resulting from general relativity (notably light deflection and gravitational redshift), perspective acceleration, stellar and Galactic aberration, and the possible variation of the gravitational constant. I~cover the loosely related phenomenon of gravitational microlensing in Section~\ref{sec:microlensing}.

\begin{figure}[t]
\centering
\includegraphics[width=0.36\linewidth]{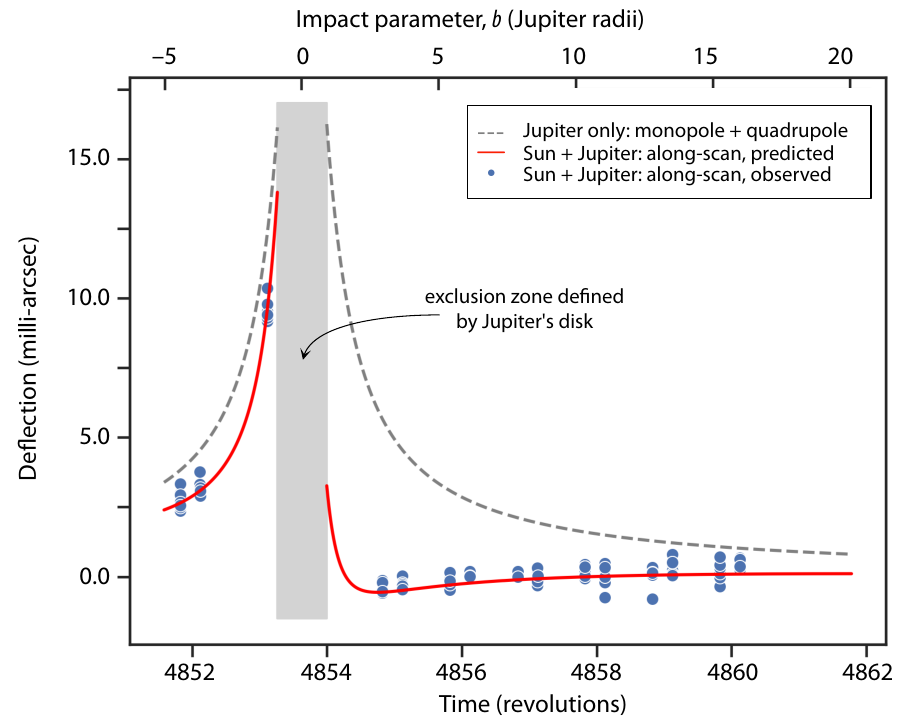}
\hspace{15pt}
\includegraphics[width=0.27\linewidth]{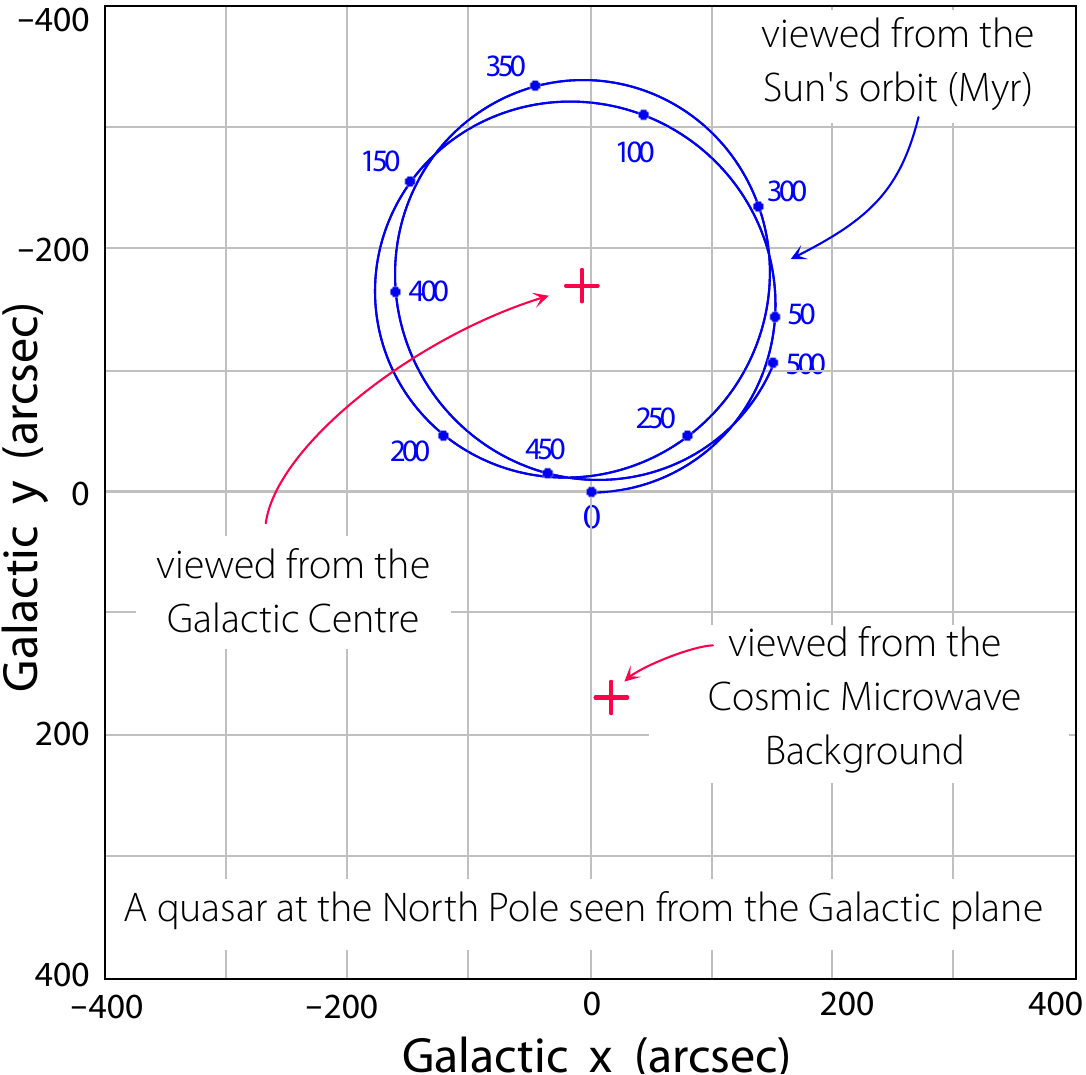}
\hspace{15pt}
\raisebox{10pt}{\includegraphics[width=0.26\linewidth]{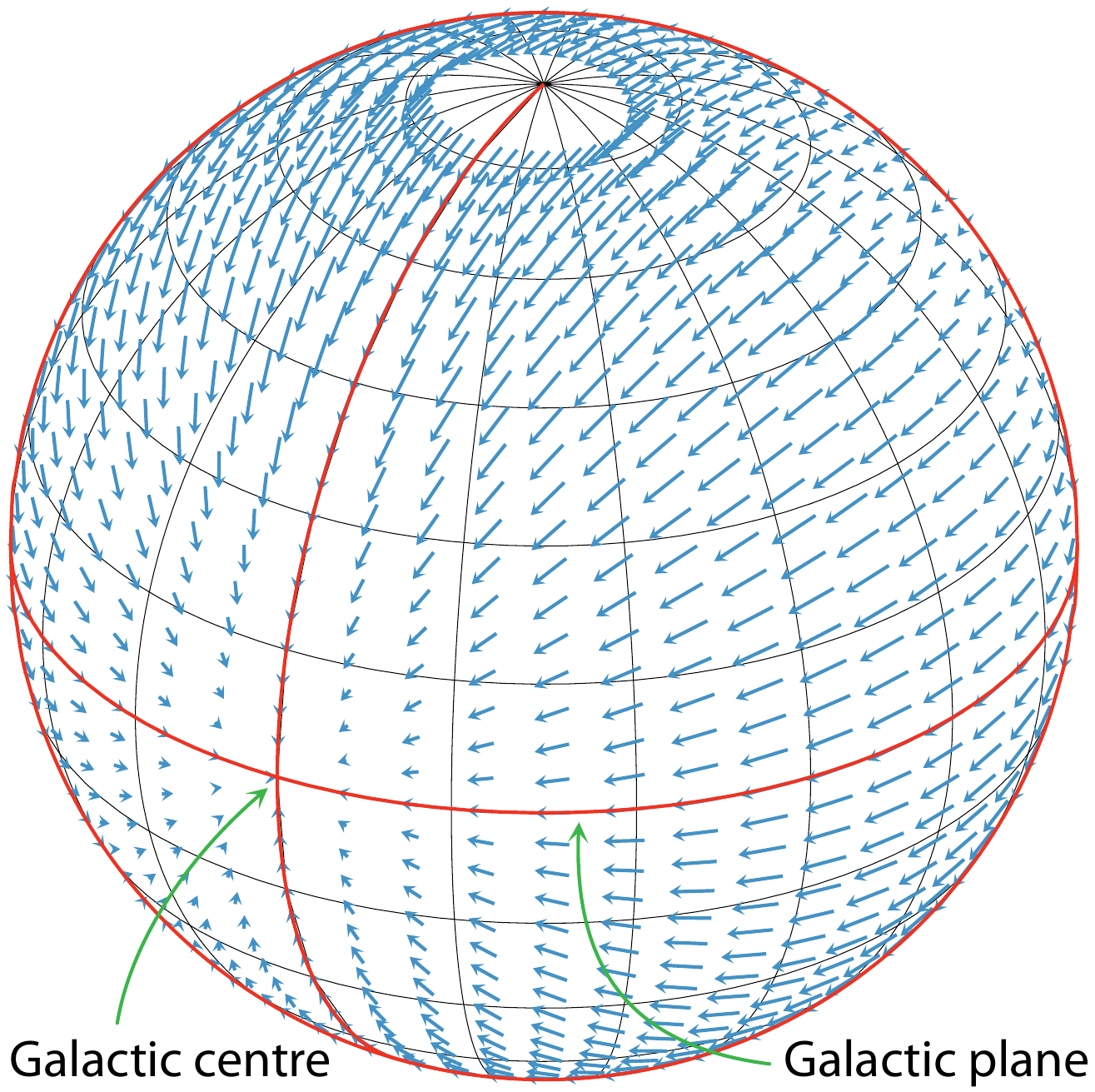}}
\vspace{0pt}
\caption{
Left~(a): Light deflection due to Jupiter, measured by Gaia for the February 2017 grazing incidence observations \citep[][Figure~4]{2022A&A...664A.143A}.
Middle~(b): Galactic aberration over 500~Myr for an observer looking towards Galactic north, showing the apparent path of a hypothetical quasar at the NGP due to the changing velocity of the Sun in its epicyclic orbit around the Galactic centre. Crosses show the quasar position seen by an observer
at rest with respect to the Galactic Centre, and one at rest with respect to the CMB.
Right~(c): the proper motion field of quasar-like objects, in Galactic coordinates, induced by the centripetal Galactic acceleration \citep[][Figures~1 and~2]{2021A&A...649A...9G}.
}\label{fig:deflection-aberration}
\end{figure}

\subsection{Light deflection by the Sun and Jupiter}
\label{sec:light-deflection}

In general relativity, the presence of matter (energy density) distorts spacetime, and the path of electromagnetic radiation is deflected as a result. 
%
Formally, the equations of light propagation are derived from the general relativistic Maxwell equations. For the solar system, they can also be derived in the limit of geometrical optics since wavelength-dependent relativistic effects are much smaller than 1~micro-arcsec. 
For a spherically symmetric gravitational field, the general expression for the deflection angle reduces to the `post-Newtonian' formula
\begin{equation}
\delta\chi = \frac{2GM}{b c^2}\, \frac{(1+\gamma)}{2} \, \cot\left(\frac{\chi}{2}\right),
\end{equation}
where 
$G$ is the gravitational constant,
$M$ is the mass of the deflecting body,
$b$ is the impact parameter,
$c$ is the velocity of light,
and $\chi$ the angular separation between the deflecting body and the source
\citep{2018tegp.book.....W}. 	

Inclusion of the term $\gamma$, which appears in the Parameterised Post-Newtonian (PPN) approximation, is convenient for comparing `metric' theories of gravity and, in particular, assessing how well general relativity matches specific experimental tests
\citep{2018tegp.book.....W}. 
In general relativity $\gamma=1$,   while the classical derivation of light bending yields only the first part of the leading coefficient in this expression, viz.\ the factor~1/2. 
Conventionally, experimental results are expressed in terms of ${1\over 2}(1+\gamma)$. 
Light deflection has been observed, with various degrees of precision, on distance scales of $10^9-10^{21}$~m, and on mass scales from $10^{-3}-10^{13} M_\odot$, the lower range from planetary deflection at 200~arcsec from Jupiter 
\citep{1991AJ....102.1879T},
and the upper range from the lensing of quasars
\citep{1992NuPhS..28..321D}.

For grazing incidence at the solar limb, the expression reduces to $\delta\chi\simeq {1\over 2}(1+\gamma)\,1.7505$\,arcsec. 
In other words, with $\gamma=1$, stars observed close to the solar limb are deflected by an angle of 1.7~arcsec.
But while the scanning of the celestial sphere by both Hipparcos and Gaia precluded observations within several tens of degrees from the Sun direction, the light-bending term remains significant, for both, over the entire celestial sphere. For example, it amounts to about 4\,milli-arcsec even at $90^\circ$ from the Sun, i.e.\ for star light arriving orthogonal to the ecliptic.

The notorious first observational confirmation of general relativity based on the 1919 solar eclipse in Brazil 
\citep{{1920RSPTA.220..291D}, 
{1979Obs....99..195H}},		
was of limited accuracy, while the most recent {\it ground-based\/} astrometric solar eclipse campaign, from Chinguetti (Mauritania) in 1973, yielded ${1\over 2}(1+\gamma)=0.95\pm0.10$
\citep{1976AJ.....81..455J}.
Analysis of the Hipparcos all-sky great-circle abscissae residuals gave ${1\over 2}(1+\gamma)=0.9985\pm0.0015$
\citep{1997ESASP.402...49F}.
At longer wavelengths, the development of radio interferometry and radio VLBI greatly improved on these measurements: a 1995 VLBI measurement of
3C~273 and 3C~279 yielded ${1\over 2}(1+\gamma)=0.9996\pm0.0017$ 
\citep{1995PhRvL..75.1439L},
and a later analysis of 2~million observations of 541 radio sources from 87~VLBI sites gave ${1\over 2}(1+\gamma)=0.99992\pm0.00023$ 
\citep{2004PhRvL..92l1101S}.
And
\citet{2003ApJ...598..704F}
reported VLBA--Effelsberg observations of `light' bending of the quasar J0842+1835 at 3.7\,arcmin from Jupiter.

The geometrical model adopted for Hipparcos (at the accuracy level of 1~milli-arcsec) has required major refinements for the 1~micro-arcsec accuracies demanded by Gaia 
\citep{2003AJ....125.1580K}.
A further complication when considering light bending, whether by the Sun or planets, is that they are not spherical but, due to rotation, better approximated as oblate spheroids. 
Expressing their gravitational potential as the expansion $U=-GM(C_0 r^{-1} + C_1 r^{-2} + C_2 r^{-3}\ldots)$, then defines the monopole ($C_0$), dipole ($C_1$), and quadrupole ($C_2$) moments. 
%
For the Sun and Jupiter, the monopole terms at grazing incidence are 1.7~arcsec and 16.27~milli-arcsec respectively. But while the quadrupole term at grazing incidence is only $\sim$1\muas\ for the Sun, it is a significant 240\muas\ for Jupiter \citep{2003AJ....125.1580K}.
%
A further complication for Gaia is that the geometry of light deflection is similar to that of the parallax displacement, although differing in direction and in their dependence on $\chi$: varying as cot\,$(\chi/2)$ for light bending, and as $\varpi\sin\chi$ for parallax. This allows the two distinct effects to be separated in the observation equations, although they remain highly correlated.

To summarise the situation pre-Gaia, light deflection measurements to date, by whatever means, have confirmed the monopole light deflection predicted by general relativity. The quadrupole light deflection term has not yet been measured. As of early 2025, Gaia results for light-bending by the Sun are not yet available.

While observations close to the Sun were excluded by Gaia ($\chi>45^\circ$), scans very close to grazing incidence were possible for Jupiter, albeit complicated by the effects of scattered light. Stars with very small impact parameters are required, because while light-bending due to the monopole moment falls off as $r^{-2}$, that of the dipole falls off as $r^{-3}$.
Optimisation of the precession and spin phases of the Gaia `scanning law' took account of the predicted configuration of three stars, with $G<15.75$, and within 6~Jupiter radii of its limb, which would take place between 22--26 February 2017. 
These high-cadence observations were duly obtained, with the closest of the three observed at just 7~arcsec from Jupiter’s limb, and the closest bright star, with $G=12.78$~mag, was observed on 25~transits spread over a few days.
The measured deflection, projected in the along-scan direction, was around 10~milli-arcsec. It represented the combined effect of Jupiter and the Sun, is mainly dominated by Jupiter's monopole, and was detected with a signal-to-noise of 50 at closest approach
\citep{2022A&A...664A.143A}. 
It is the closest measurement made to Jupiter's limb in the optical, the highest signal-to-noise at any wavelength, and again provides excellent agreement between the Gaia observations and the predictions of general relativity (Figure~\ref{fig:deflection-aberration}a).

It sets the stage for future efforts to disentangle an along-scan quadrupole component of some 100~micro-arcsec at closest approach, from the much larger monopole deflection of around 10~milli-arcsec. 
%
Two similar Jupiter-grazing events, which occurred in September 2018 and April 2019, have not yet been analysed.

\subsection{Perspective acceleration}
\label{sec:perspective-acceleration}

Classical astrometry ignores a star's radial velocity, i.e.\ its space motion {\it along\/} the line-of-sight. This is because a star's radial motion generally has no effect on its position on the sky, and cannot therefore be determined from angular measurements alone.
Knowing a star's radial velocity is nonetheless important for fully defining its complete space motion. The full space motions of stars are, in turn, essential in understanding kinematics and dynamics, both for individual stars, as well as for groups or populations. 
Radial velocities are determined by measuring the Doppler shift of the stellar spectral lines. Extremely high accuracies can be reached in these velocity measurements: typical large-scale stellar surveys may reach accuracies of around 1\kms, while today's dedicated high-precision radial velocity spectrometers used for exoplanet studies routinely measure stellar velocities, along the line-of-sight, with accuracies of about a few centimetres per second.
But radial velocities are no longer irrelevant in very high accuracy astrometry, where it can affect the determination both of the parallax of a star, and of its proper motion
\citep{1900AN....154...65S,
1999A&A...348.1040D,
2003A&A...403.1077K,
2002A&A...381..446M}.

\begin{figure}[t]
\centering
\includegraphics[width=0.4\linewidth]{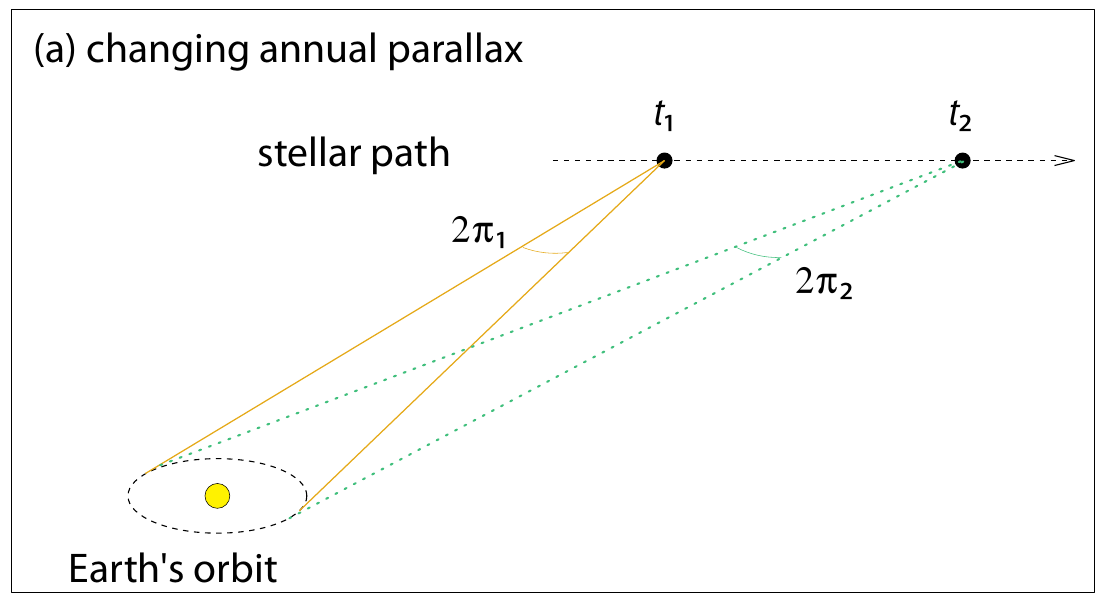}
\hspace{20pt}
\includegraphics[width=0.4\linewidth]{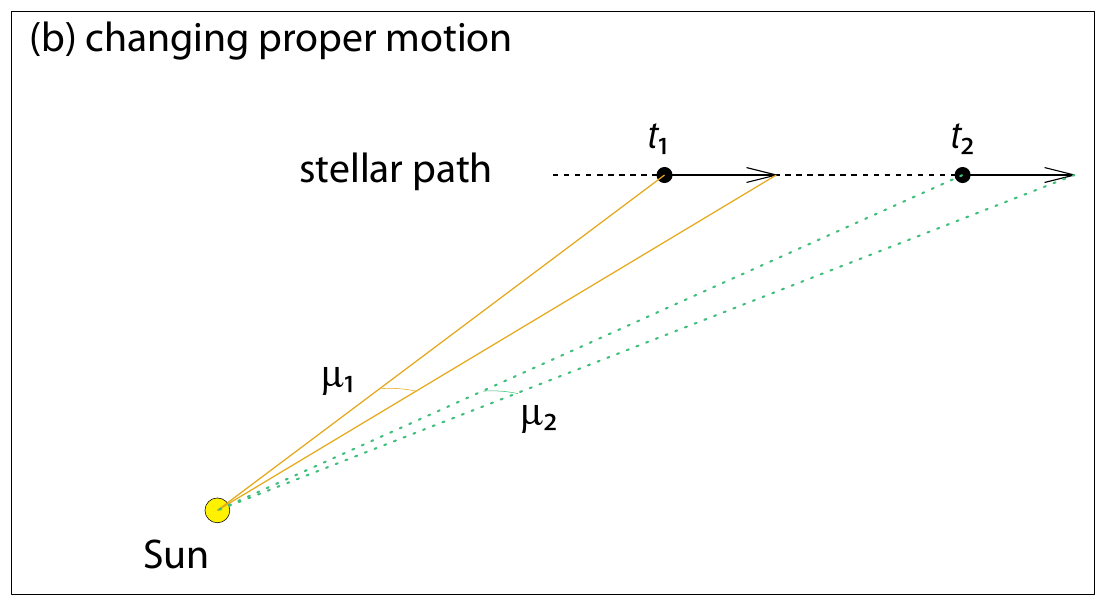}
\caption{Left~(a): a star moving through space, with some {\it radial\/} motion, has a parallax which changes with time, proportional to the product of the radial velocity and the square of the parallax.
Right~(b): for a fixed space velocity, the proper motion varies inversely with distance, and the tangential velocity changes due to the varying angle between the line-of-sight and the space velocity.  The two effects result in a changing proper motion with time, interpreted as an apparent (or `perspective') acceleration of the star's motion on the sky (\citet{1999A&A...348.1040D}, Figure~1).}
\label{fig:dravins}
\end{figure}

The systematic change in trigonometric parallax due to the radial displacement of a star is most easily appreciated from Figure~\ref{fig:dravins}a: a star moving through space, with some radial motion, has a parallax which changes with time, by an amount proportional to the product of the radial velocity and the square of the parallax.  
Similarly, for a fixed space velocity, the angular velocity (or proper motion) varies inversely with the distance to the object. However, the tangential velocity changes due to the varying angle between the line-of-sight and the direction of its space velocity.  The two effects result in a changing proper motion with time, which is interpreted as an apparent (or `perspective') acceleration of the star's motion on the sky (Figure~\ref{fig:dravins}b). The effect is proportional to the product of the star's parallax, proper motion, and radial velocity.  It is always a small effect, but largest for nearby stars with a high proper motion.

At the accuracy of Hipparcos, around 1~milli-arcec, perspective acceleration was of marginal importance, and then only for the nearest stars with the largest radial velocities and proper motions. Nonetheless, it was accounted for in the 21~cases for which the accumulated positional effect over two years exceeds 0.1\,milli-arcec.
For the first Gaia data release, DR1 in 2016, the astrometric accuracies were limited. The effect was simply ignored, by assuming zero radial velocity for all objects
\citep{2016A&A...595A...4L}.
For DR2 in 2018, it was taken into account for just 53 nearby Hipparcos catalogue objects, by using the values of radial velocities taken from the existing literature
\citep{2018A&A...616A...2L}.
Of these, the effect exceeded 0.5~mas for just three stars: Barnard's star (1.975~mas), Kapteyn's star (1.694~mas), and Van Maanen~2 (0.573~mas).
For EDR3 in 2020, the effect was taken into account, where possible, using radial velocities from Gaia’s own radial-velocity spectrometer or, for a small number of nearby stars (mainly white dwarfs) complemented with radial velocities from the literature
\citep{2021A&A...649A...2L}.
Further details are given by 
\citet[][their Table~1]{2014A&A...570A..62B}.

An extension of these principles applies to open (and globular) clusters. Since the cluster stars share the same (average) velocity vector, apart from a (small) velocity dispersion, the cluster's apparent size changes as it moves in the radial direction.  The approach was applied to the Hipparcos data for the Hyades cluster 
\citep{2000A&A...356.1119L}
and to the Ursa Major, Coma Berenices, Pleiades, and Praesepe clusters
\citep{2002A&A...381..446M}.
With Gaia DR2, the effect shows up in the proper motions of a number of globular clusters with large radial velocities and parallax
\citep{2018A&A...616A..12G}.
In their study (Section~\ref{sec:globular-clusters}; their Figure~11), 
NGC~6656 shows significant rotation as well as perspective expansion;
NGC~3201 shows no rotation but significant perspective contraction; 
and NGC~6397 shows halo expansion associated with core collapse.

\subsection{Stellar and Galactic aberration}
\label{sec:aberration}

\paragraph{Stellar aberration}
Stellar aberration describes the apparent displacement of celestial objects from their `true' positions, as a consequence of the velocity of the observer. For an observer on Earth, moving through space with velocity~$v$, then (in the non-relativistic limit) a star's position will {\it appear\/} to be displaced in the direction of the observer's motion by an angle $v/c$. Note that if the observer's motion through space was constant, then the position of all stars would be shifted by the same amount, and it would not be possible to differentiate between their `true' and apparent positions.
But the Earth's motion around the Sun, $v\sim30$\kms, results in a change in a star's apparent position by up to 20~arcsec, depending on its location on the sky. This `annual aberration' term clearly dwarfs the accuracies targeted by Gaia. To account for this effect, both for stars as well as solar system objects, one mission-level requirement was to have continuous and adequate knowledge of the satellite's position and space velocity.

The corresponding satellite ephemeris requirements, at the L2 orbit location 1.5~million km from Earth, are demanding 
\citep{1998mignard-sag}. 
Knowledge of the satellite position is related to the source's observed parallactic displacement, which can be decomposed into the satellite--Earth and Earth--barycentre displacements. This leads to requirements on the satellite position to within $10^{-6}$\,au (150\,km) for the nearest stars, but is further dictated by the continuous requirement of better than 150\,m for typical minor planets in the main asteroid belt. 
Requirements on the satellite's barycentric velocity derive from the need to correct for stellar aberration, within the framework of general relativity, in order to refer the astrometric observations to an observer at rest with respect to the barycentre. The size of the correction is of order $v/c$: in order to compute the correction to within $\delta\alpha=1$~microarcsec ($5\times10^{12}$~rad) the satellite barycentric velocity vector is needed to within $c\delta\alpha$, or $\sim$1.5\,mm\,s$^{-1}$ (for both stars and solar system objects). 

These accuracies are obtained in practice through ESA's satellite ranging, complemented by dedicated optical observations 
\citep{2018vels.confE..31A}, 
and are fully accounted for in the data analysis. 

\paragraph{Galactic aberration}
Similarly, if the observer is accelerating, the star displacements change with time, and a systematic pattern of apparent proper motions, in the direction of acceleration, would become apparent. Two such effects can be identified. One is due to the fact that the Sun is moving in a (roughly) circular orbit around the Galaxy, with an orbital period of 250\,Myr. In addition, the Galaxy itself is in motion within the Local Group of galaxies, and with respect to the cosmological reference frame defined by the Cosmic Microwave Background (Figure~\ref{fig:deflection-aberration}b).
The possibility of discerning the first of these, viz.\ the effects of Galactic rotation, or `secular aberration', has been pursued since the early 1980s in the framework of high-accuracy radio VLBI observations. But with the systematic displacement of the best-placed quasars being only around 100\muas, the effect is not much above the error of individual VLBI positions. The analysis by
\citet{2020A&A...644A.159C}, 
based on 39 radio sources from almost 40~years of VLBI observations, gave an acceleration of $5.83\pm0.23$\muasyr\ in the direction $\alpha=270\ddeg2$, $\delta=-20\ddeg2$.

If stars in our Galaxy provided the only reference frame available, such acceleration terms would probably be impossible to separate from other effects. But with its inertial reference frame materialised by more than a million distant quasars, the Gaia data are accurate enough to discern the {\it acceleration\/} of the solar system due to its 250\,Myr orbit around the Galaxy \citep{1995ESASP.379...99B,
2001A&A...369..339P}.
A first detailed assessment has been made with Gaia EDR3.
\citet{2021A&A...649A...9G}	
used 1\,614\,173 `quasar-like' sources in EDR3 and, from an examination of the proper motion vector field (Figure~\ref{fig:deflection-aberration}c), found a Galactocentric acceleration of $5.05\pm0.35$\muasyr\ in the proper motions (or, in other units, $7.33\pm0.51$\kms\,Myr$^{-1}$, or $2.32\pm0.16\times10^{-10}$\,m\,s$^{-2}$).
This compares favourably with a predicted acceleration of $6.98$\kms\,Myr$^{-1}$ towards the Galactic centre on the basis of the distance and motion of Sagittarius~A*, its central black hole.
From the Cosmic Microwave Background dipole, as measured by the Planck mission, the effect of the Galaxy's motion with respect to the Local Group of galaxies predicts positional displacements of $\sim$200~arcsec. But in the absence of its variation over time, the effect is presently unmeasurable.

\citet{2020arXiv201202169B}	
has argued that this absolute acceleration measurement, combined with relative accelerations from pulsar orbital decay, allows the determination of all of the parameters describing the dynamics of the local Galactic environment, including the circular velocity at the Sun, $V_0=244\pm8$\kms, and its derivative  $V^\prime_0=2\pm9$\kmskpc, the local angular frequency, the Oort constants, and the Sun's motion with respect to the Local Standard of Rest.

\subsection{Gravitational redshift}
\label{sec:gravitational-redshift}

Gravitational redshift is one of numerous physical effects that can modify observed spectroscopic wavelengths and line profiles, in principle affecting the derived radial velocities. For high-accuracy work, for example in exoplanet studies with target accuracies of a few cm\,s$^{-1}$, these small effects (such as atmospheric pulsation, surface convection, stellar rotation, and stellar winds) must be taken into account 
\citep[e.g.][]{1999A&A...348.1040D}. 
And there are three classes of object for which the gravitational effects of the star's mass on the spectral line shifts are being investigated using the Gaia data: white dwarfs, open clusters, and wide binaries. 

According to general relativity, time is dilated in a gravitational potential well, $\phi$, relative to a distant observer.  In the weak field regime, this leads to a spectroscopic velocity offset in the radial $r$ direction
\citep{2018tegp.book.....W}, 	
$\Delta v_r/c=-\Delta \phi/c^2$, which in the Newtonian limit results in a wavelength shift,
$z=\Delta\lambda/\lambda \approx gR/c^2$.
For photons emitted from the Sun, $z=2\times10^{-6}$ (being a Doppler shift of 636\ms).
For the Earth, $z=7\times10^{-10}$, or an equivalent Doppler shift of 0.2\ms. 
In most stars, as for the Sun, the gravitational redshift amounts to a few hundred \ms, comparable to the convective blueshift in their atmospheres. But for white dwarfs, with their much smaller radius and consequently their much higher surface gravity, it reaches tens of \kms. Accordingly the difference between the spectroscopic Doppler shift and the astrometric radial velocity can be used as a proxy for their mass.
Pre-Gaia, gravitational redshifts have been measured directly for white dwarfs
\citep[e.g.][]{1967AJ.....72Q.301G,
2005MNRAS.362.1134B,
2012ApJ...757..116F},
and also for neutron stars 
\citep{2002Natur.420...51C}.
For white dwarfs in binary systems, specifically in the case of Sirius~B, measurements have demonstrated the correspondence between the mass inferred from its gravitational redshift, and from its orbital dynamics
\citep{2018MNRAS.481.2361J}.

Exploiting this effect,
\citet{2019A&A...627L...8P}	
analysed six Hyades white dwarfs using high-resolution UVES--VLT spectra to determine their Doppler shift, and using Gaia~DR2 astrometry to determine their radial velocities independently of spectroscopy
The mass estimates from the resulting gravitational redshifts were systematically smaller, by $0.02-0.05M_\Sun$, than those derived from other methods, which the authors considered to be a significant finding for white dwarf models.
Subsequently, from a sample of 3000 white dwarfs from SDSS and Gaia DR2, the underlying gravitational redshift was used to derive an empirical mass--radius relation across a wide range of stellar masses
\citep{2020ApJ...899..146C}.	
The Gaia data on open clusters allow such studies to be extended, at least statistically, to dwarf and giant stars, exploiting their common (radial) space motions, characterised by accurate astrometry. Objectives include further tests of the equivalence principle, as well as of stellar models
\citep{2019MNRAS.483.5026L,		
2019ApJ...871..119D,			
2022ApJ...929...29G}.			

A similar approach can be applied to wide binaries. The underlying idea is to measure the combined effects of gravitational redshift and convective blue-shift of the spectral lines, by eliminating the contributions from their individual peculiar space motions. 
Co-moving pairs in 3d velocity space, typically wide binaries with $d>25$\,au, were first compiled from Gaia DR1
\citep{2017AJ....153..257O}.
From Gaia DR2,
3741 pairs were identified by
\citet{2019AJ....157...78J},
more than 50\,000 within 200\,pc of the Sun by
\citet{2018MNRAS.480.4884E},
and nearly 100\,000 by
\citet{2020ApJS..247...66H}.
Early attempts to use subsets of these sample to study the contributions of gravitational redshift have been reported
\citep{2022MNRAS.514.1071M,  
2022RNAAS...6..137E}.	

On a Galactic scale, studies using EDR3 suggest that the Galaxy has a nearly flat rotation curve (Section~\ref{sec:rotation-curve}) with a local circular speed 230\kms, declining outwards by $-1.7$\kmskpc\
\citep{2019ApJ...871..120E,
2019ApJ...870L..10M}.
This leads to a local contribution to the gravitational acceleration of $-2.3\times10^{-10}$\,m\,s$^{-2}$, and an associated velocity gradient $-2.4\times10^{-2}$~\kmskpc. In turn, this leads to an {\it apparent\/} radial migration: an apparent radial recession away from the Sun for stars closer to the Galactic centre, and a spurious radial approach beyond 
\citep{2022RNAAS...6...72L}.

\subsection{Variation of the gravitational constant}
\label{sec:g-dot}

A fundamental physical constant is one considered to be universal in nature, and constant in time.
The number of fundamental constants depends on the theoretical framework.  Today, our best description of Nature is governed by two distinct theories: general relativity (for gravity) and the `Standard Model' (for the electromagnetic, weak and strong nuclear interactions). Together these result in 22 `unknown constants', which can be presented as 19 `unknown dimensionless parameters'
\citep[e.g.][]{2011LRR....14....2U}.	
Amongst them are the gravitational constant ($G$), the speed of light ($c$), and the Planck constant ($h$), with all 16~others from the Standard Model (Yukawa couplings, Higgs field, and others). Since these are not predicted by theory, but can only be determined by measurement, due effort is being invested in quantifying their possible rates of change, with the goal of confirming (or invalidating) the associated theory. 

The possibility of a variation of the gravitational constant over cosmological time was introduced as a consequence of an alternative to general relativity by 
\citet{1935rgws.book.....M},	
developing earlier ideas by 
\citet{1917AnP...359..117W,
1919AnP...364..101W}
and 
\citet{1931PCPS...27...15E},
and has been debated and developed since
\citep[e.g.][and others]{
1937Natur.139..323D, 
1937NW.....25..513J,
1961PhRv..124..925B, 
2024Univ...10..404B}. 
Today, observational constraints on $\dot{G}$ have come from three main classes of experiment:
(a)~solar system observations, including the Moon's orbit through lunar laser ranging, 
radar timing to Venus and Mercury, and ranging of the Mariner~10 and Viking spacecraft;
(b)~monitoring of the spin-down of binary pulsars (PSR B1913+16 and PSR B1855+09),
as well as a few isolated pulsars (notably PSR J0437--4715);
(c)~constraints from stellar evolution, specifically based on helioseismology, the ages of globular clusters, white dwarf cooling, and Type~II supernovae.
It is this final group where Gaia can be expected to contribute.

The reason why star observations can be used to place limits on any variation of $G$ is straightforward in principle, and was first outlined by Edward
\citet{1948PhRv...73..801T}
and George
\citet{1967PhRvL..19.1000G}.
The rate of photon energy escape from a star is limited by its surface opacity. In the simplistic case of opacity dominated by Thomson scattering, the stellar luminosity is given by $L\propto G\,^4M\,^3$, and the main sequence lifetime is consequently highly sensitive to the star's mass, but also to the value of~$G$.
The largest effects of a time-varying gravitational constant would therefore be expected for the oldest objects, which is why white dwarfs are important in this context
\citep{2022FrASS...9....6I}. 
And there are two aspects of relevance to Gaia: the white dwarf luminosity function, and pulsation variations in individual objects. I will outline the principles, although neither provide the tightest of observational bounds to date.

For the former, the faintest white dwarfs formed shortly after the birth of the Galactic disk (some 10\,Gyr ago), and their late evolution is dominated by cooling and gravitational contraction under hydrostatic equilibrium. Accordingly, any change in $G$ would modify their energy balance, which in turn would modify their luminosity. In consequence, the age of the disk, which can be estimated from the rapid drop at the faint end of the white dwarf luminosity function, depends on the past value of $G$
\citep{1976ApJ...206..213V}.
Compared to the pre-Gaia upper bound, of $\dot G/G\le -1.8\times 10^{-12}$~yr$^{-1}$
\citep{1995MNRAS.277..801G,
2011JCAP...05..021G},
improved knowledge of the white dwarf luminosity function from Gaia, combined with an age of the disk in the solar neighbourhood, may translate into a more stringent upper bound in the future.

The second area where white dwarfs can contribute concerns their pulsation properties. Again, a secular variation of the gravitational constant would modify their structure and evolutionary time scales, resulting in a change in their pulsation periods
\citep{2022FrASS...9....6I,
2013JCAP...06..032C}.
Estimates of any such period change must, in practice, account for each object's {\it proper motion}, which is where Gaia has made a small contribution for the particularly stable white dwarf G117--B15A
\citep{2021ApJ...906....7K}.
And with theoretical models for white dwarfs predicting that the period change is directly related to the rate of cooling, it follows that any {\it additional\/} source or sink of energy also modifies the rate of cooling and, hence, the rate of change of period. Similar measurement programmes for white dwarfs are therefore also being used to place upper limits on the properties of (amongst others)
axions 
\citep{1992ApJ...392L..23I,
2012MNRAS.424.2792C,
2020A&A...644A.166S,
2022JCAP...02..035D},
and neutrinos
\citep{1994MNRAS.266..289B,
2020PhRvD.102h3007C}.

\section{Solar system}
\label{sec:solar-system}

The solar system is a gravitationally bound system comprising the Sun, eight planets, their moons, a number of minor planets, and a vast number of smaller bodies, including the rocky asteroids and the icy comets. The minor bodies include remnants of its primordial formation, and can be variously classified according to their heliocentric distance, orbit, size, morphology, and composition. 
Here, I will outline Gaia's census of the asteroid population, and describe some of the investigations that have made using the Gaia data to interpret their orbits, their composition, and the associated phenomenon of occultations, viz.\ when a solar system body passes in front of a background star. I also consider the search for solar twins and solar siblings, the search for other stars that have passed (or will pass) close to the Sun in the past or in the future, and what Gaia is informing us about the recent discoveries of interstellar `visitors' or `interlopers'.
The effects of gravitational light deflection by the Sun and Jupiter are detailed in Section~\ref{sec:light-deflection}. Topics related to the Sun's location and space motion within the Galaxy, including the distance to the Galactic centre and the Sun's height above the disk's mid-plane, are covered in Section~\ref{sec:galaxy-structure}.

\subsection{Asteroids}
\label{sec:solar-system-asteroids}

The IAU's \href{https://www.minorplanetcenter.net/iau/mpc.html}{Minor Planet Center} presently lists orbits for more than 200\,000 asteroids. 
These small bodies preserve relatively unaltered materials that date back to the formation of the solar system from the proto-solar nebula 4.567\,Gyr ago 
\cite[e.g.][]{
{2002Sci...297.1678A},
{2015aste.book..471J}}.
Their orbits provide insights into
the early accretion of primordial material from the protoplanetary disk,
the transport of water and other organic material to Earth,
planetary migration in the early solar system,
the existence of complex dynamical resonances,
the characterisation of asteroid `families' (with similar orbits) resulting from asteroid--asteroid collisions,
and mass determination from orbit perturbations.

Gaia's observations of all objects brighter than 20--21~mag is providing a deep, uniform survey of minor planets and other solar system bodies. 
To date, Gaia has provided state-of-the-art orbits of 157\,000 asteroids, reflectance spectra from the BP/RP spectrophotometry, and rotation periods from Gaia's multi-epoch photometry. 
This combination is being used to study
their dynamical and taxonomic properties;
masses from mutual orbit perturbations;
asteroid `families' resulting from collisional fragmentation (which can remain clustered in orbit space, with similar spectra);
space `weathering';
the origin of near-Earth asteroids via solar-radiation driven Yarkovsky migration;
the excess of fast and slow rotators as a result of the YORP effect;
and 
sizes and morphologies from the many occultation events that can now be accurately predicted.
By the end of mission, Gaia is predicted to have observed some 350\,000 asteroids at multiple epochs, many of them new. Hipparcos, with its magnitude limit of 12--13~mag, observed just~48.

Three Gaia catalogues of asteroid data have been released so far.
Gaia~DR2 in April 2018, with observations from July 2014--May 2016, comprised just 14\,000 objects, all with good orbits, and selected to represent most main dynamical classes, including main-belt asteroids (MBA), near-Earth objects (NEO), Jupiter Trojans, and a number of trans-Neptunian objects (TNO).
%
Gaia~DR3, in June 2022 and covering 34~months of observations between July 2014--May 2017, provided astrometry for 157\,000 asteroids, orbits for 154\,787, and BP/RP reflectance spectra for 60\,518
\citep{2023A&A...674A..12T}. 
A special `Focused Product Release' (Section~\ref{sec:data-releases}) provided an intermediate product between DR3 and DR4 (expected in late 2026). It treated the same 157\,000 asteroids as in DR3, but with the epoch astrometry and orbit reconstruction based on the longer data interval of DR4, viz.\ 66~months compared to the 34~months of DR3
\citep{2023A&A...680A..37G}.	

\begin{figure}[t]
\centering
\includegraphics[width=0.47\linewidth]{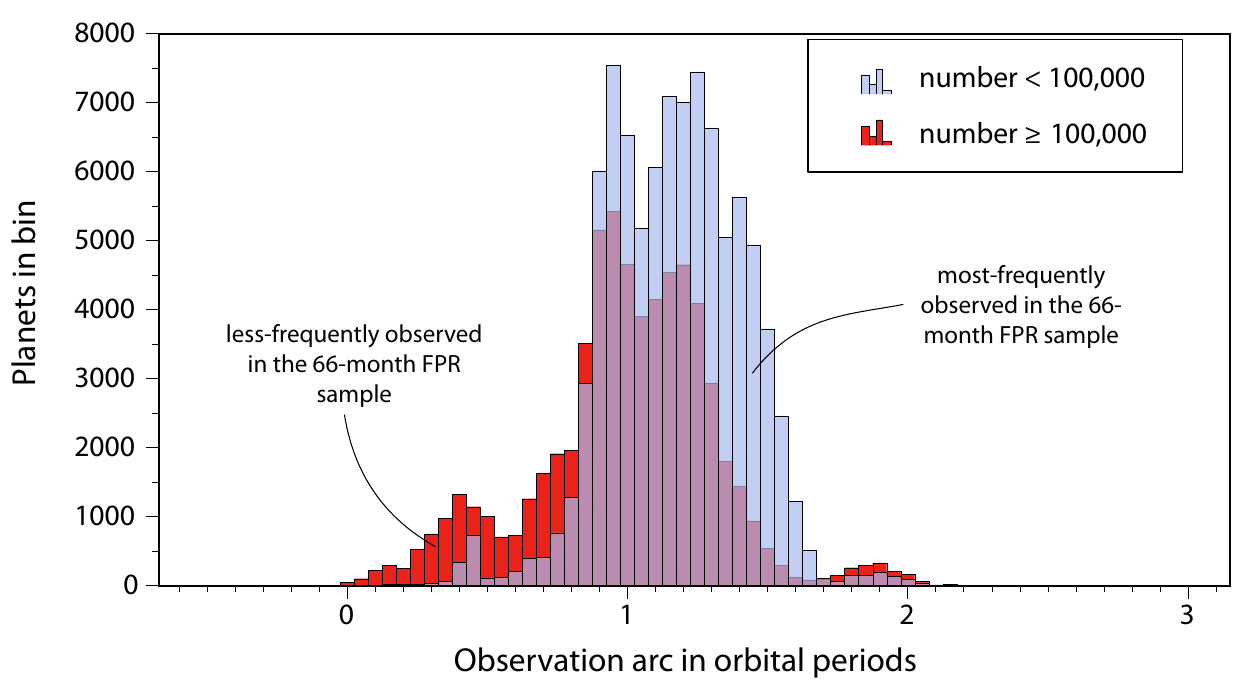}
\hspace{15pt}
\includegraphics[width=0.48\linewidth]{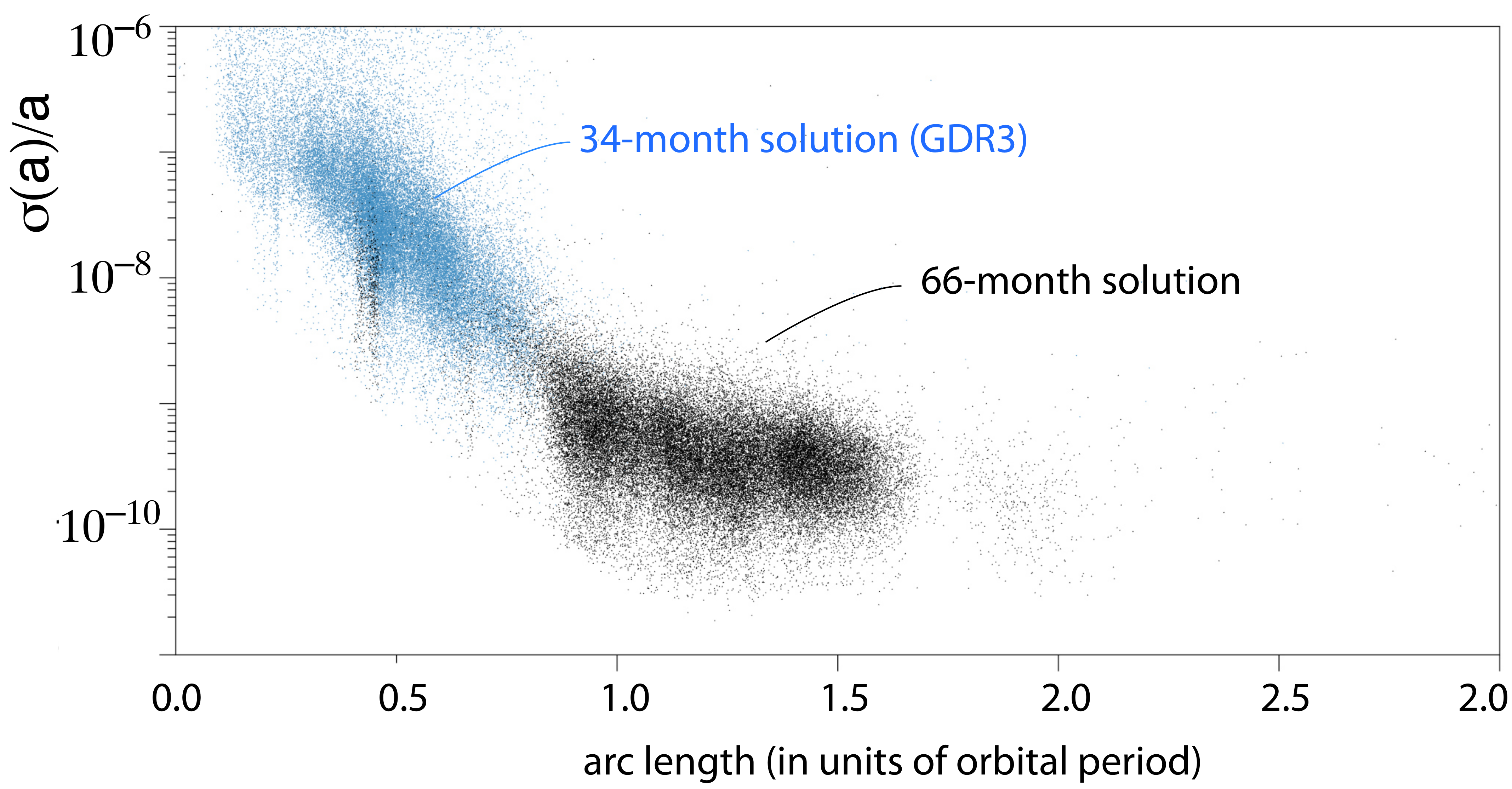}
\caption{Left: solar systems objects in the 66-month `Focused Product Release', showing the arc length measured in orbital periods, for the brightest and most-frequently observed sample (numbers $<100\,000$, in blue), and the fainter and less-frequently observed asteroids (numbers $\ge100\,000$, in red.) The typical coverage is about 1.2~orbital periods for the former, and about 0.5 orbital periods for the latter.
Right: Gaia asteroid orbits: uncertainty of the semi-major axis, $\sigma_{\!\!a}/\!a$, as a function of arc length, for the 34-month Gaia DR3 solution (blue), and for the greatly improved 66-month `Focused Product Release' solution (black). From \citet{2023A&A...680A..37G}, Figures~2 and~15.}\label{fig:fpr-asteroids}
\end{figure}

\subsubsection{Orbits}

Solar system objects are observed at multiple epochs as a result of Gaia's sky-scanning motion. In common with stars and other point-like images, sources are detected as they enter Gaia's two fields of view as long as they are brighter than the detection limit at each specific observation epoch. Positional coordinates in the instrument's focal plane are converted into sky coordinates by using the transformations provided by the astrometric global iterative solution (AGIS, Section~\ref{sec:data-processing-agis}), then converted to the barycentric reference system (with its origin at the solar system barycentre), including the rigorous effects of general relativistic light bending by the Sun. Resulting astrometry yields a single-transit accuracies below 1~milli-arcsec for objects with $G<17.5$, and still below 2--5~milli-arcsec for the faintest observable ($G\sim20.5$). Although this accuracy is essentially in one-dimension (in the direction of the scan motion), the `imprint' of the scanning law largely vanishes when several transits, in different scan directions, are combined. Orbit determination then uses a least-squares adjustment, starting with the previously known orbit elements, and employing the positions of the major solar system bodies from the JPL DE431 planetary ephemerides, to take account of gravitational terms due to the Sun, the major planets, the Moon, and Pluto, and a further sixteen massive main-belt asteroids
\citep{2023A&A...674A..12T}.
New discoveries are published as part of a dedicated
\href{https://gaiafunsso.imcce.fr/index.php}{\tt  follow-up network}
of some 80 observers around the world, leading to further improvements in orbits.

The 100\,000 objects best observed in the 66-month `Focused Product Release' sample have a median of 330 elementary observations over the 66~month period, corresponding to an average of $30\pm10$ distinct `visibility periods' or epochs
\citep{2023A&A...680A..37G}.	
Importantly, the 66-month data covers more than a full orbit for the majority, and this results in a transformational improvement in the accuracy of the orbit reconstruction (Figure~\ref{fig:fpr-asteroids}).
At least for near-Earth asteroids, Gaia is already providing positional accuracies close to that from radar ranging.

The next major improvements will come from the combination of the larger number of observations, the extended arc length, and improved calibrations of the (final) 10-year DR5 solution, whose release is expected around 2030. 
Gaia should eventually lead to an improvement (with respect to the pre-Gaia orbits) of nearly a factor 10 in the orbital elements of typical main-belt asteroids over the first 5~years of the mission, to around a factor 20 for the full 10-year mission
\citep{2018A&A...616A..13G}. 
With this vast network of astrometric measurements of unprecedented accuracy, the Gaia orbit accuracies will, in turn, far exceed those obtained even after decades of orbit-tracking from the ground.
The final accuracies of the Trojan orbits, for example, are expected to be as good as those of the main-belt asteroids today.

Online `visualisations' include the \href{https://www.youtube.com/watch?v=XYir3bQMfgQ}{orbits of 154\,741 asteroids} based on~DR3, colour-coded according to (reflectance) colour, which serves as a proxy for chemical composition. Objects are clearly resolved into the main-belt asteroids, Mars crossers, Jupiter Trojans, near-Earth objects, and the brightest 24 trans-Neptunian objects.
This \href{https://youtu.be/-l42TIevzWg?si=vOZdwvlPPrvHgta7}{orbit of asteroid 18520 Wolfratshausen} illustrates the accuracy of the 66-month Gaia Focused Product Release.

\subsubsection{Masses}

Asteroid masses are known for a dozen or so main belt asteroids, determined from their orbit evolution as a result of gravitational perturbations during close approaches with other minor bodies
\cite[e.g.][]{
{1998A&A...340L...1B},
{1998A&A...334..729V}, 
{2007A&A...472.1017M},
{2020A&A...638A..11P}}.
While no such results are yet available from the Gaia astrometry, it is expected that masses for more than a hundred will be obtained by this method. Masses combined with sizes yield their bulk density, and in turn important constraint on the body's internal structure and composition.

\subsubsection{Reflectance spectra}				
\label{sec:reflectance-spectra}

Complementing the unprecedented astrometric accuracy (and resulting orbits), Gaia's multi-epoch multi-colour photometry provides information about their physical properties, including their shape and rotation 
\citep{2012P&SS...73...52C}, 
and their composition and taxonomic classification 
\citep[e.g.][]{
{2012P&SS...73...86D},
{2022A&A...667A..10K}}.	
%
One of the reasons why these observations are of great interest is that the asteroid population does not remain unchanged over time, but instead various physical processes play a role in their evolution 
\citep{2023A&A...674A..35G}. 
These include 
collisional evolution (which affects the number and size distribution of the main belt asteroids as well as their surface structure); 
space weathering (due to surface irradiation from cosmic rays, the solar wind, and micro-meteorites, which all modify their reflectance spectra); 
and the evolution of the regolith structure and properties due to the continuous thermal cycling of their surfaces.
Such considerations have driven a dozen asteroid flyby missions, as well as numerous ground-based spectrophotometric surveys. Amongst the earliest of these, in the 1980s, were 
an 8-colour survey \citep{1985Icar...61..355Z}, 
a 52-colour survey \citep{1988LPI....19...57B}, 
a 7-colour infrared survey \citep{1993LPI....24..299C},
and numerous others since.
It is estimated that more than 1.5 million spectrophotometric observations of asteroids exist, covering more than 7600 asteroid spectra
\citep{2023A&A...674A..35G}.

\begin{figure}[t]
\centering
\includegraphics[width=0.34\linewidth]{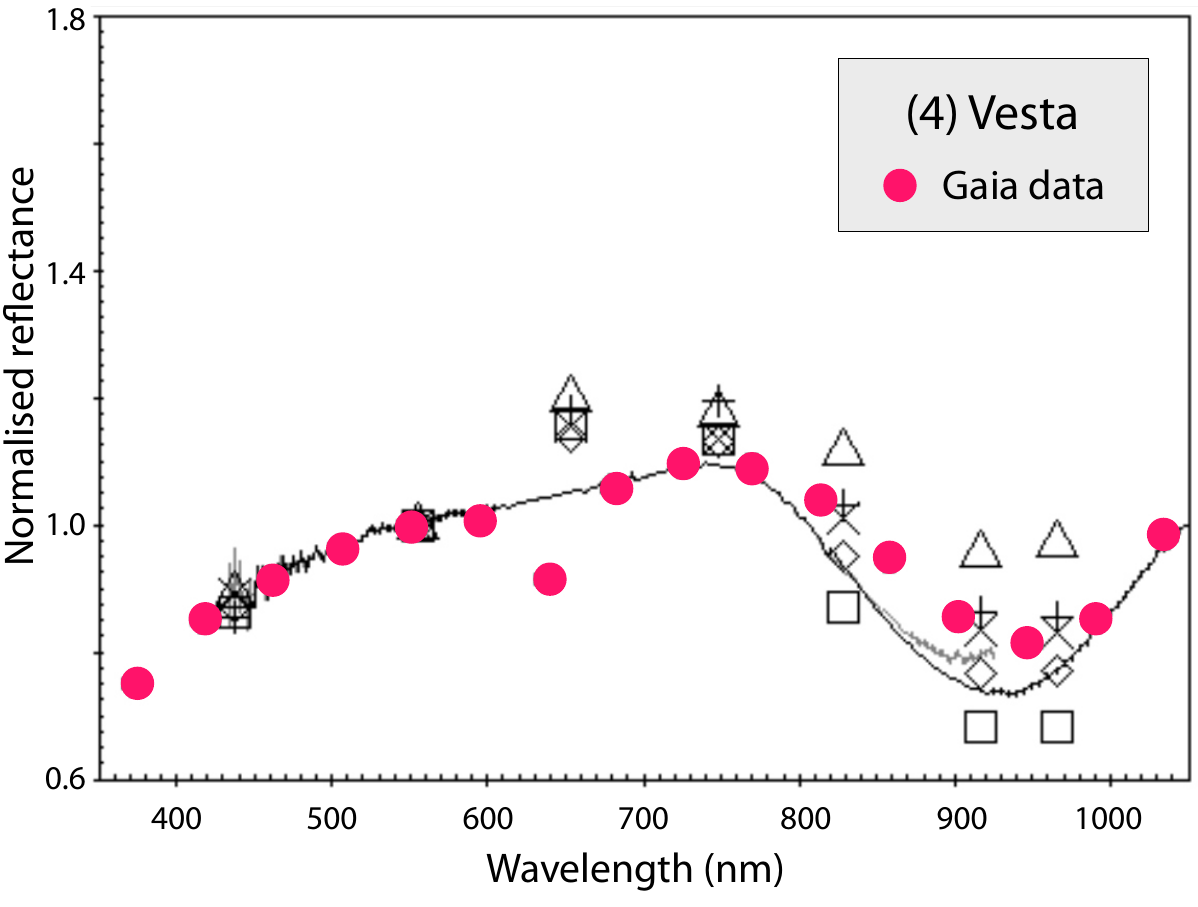}
\hspace{20pt}
\includegraphics[width=0.37\linewidth]{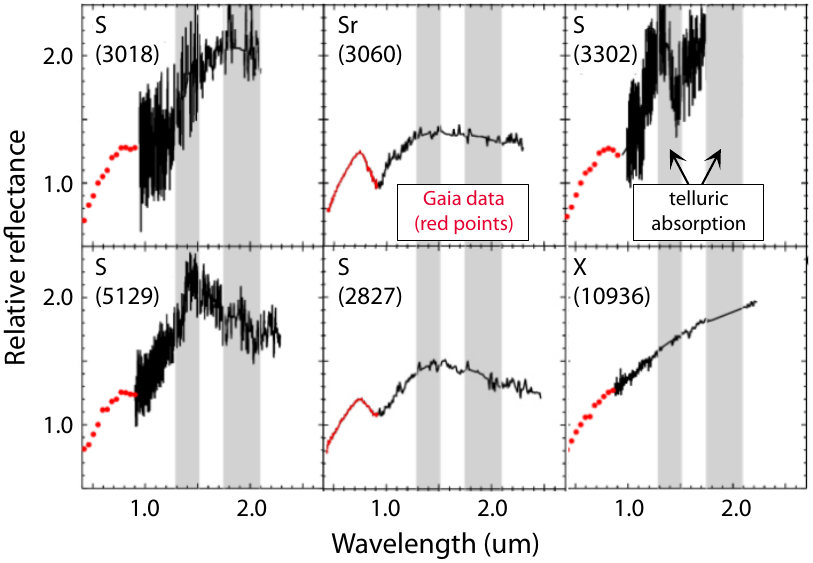}
\caption{Left: reflectance spectrum of asteroid (4)~Vesta from Gaia (red), with other data from ground-based observations and NASA's Dawn mission \citep{2023A&A...674A..35G}. Right: spectra of the S-type family with Gaia's optical spectra in red \citep{2024A&A...682A..64B}.} 
\label{fig:reflectance-spectra}
\end{figure}

Here, Gaia's strengths are its accurate multi-epoch multi-colour photometry, its extensive sky coverage (solar elongations $45-135^\circ$), the large numbers of objects (expected to total around 350\,000), and its broad wavelength range (400--1000\,nm).
%
The multi-colour photometry, designated BP and RP (Section~\ref{sec:photometry}), is achieved by two fused-silica prisms which disperse the spectra over $\sim$45~pixels in the along-scan direction, which are calibrated as part of the overall mission's photometric processing pipeline
\citep{2023A&A...674A...2D}.	
The reflectance spectrum at each epoch is derived by dividing the object spectrum by a solar-type composite, and a mean reflectance spectrum then calculated in 16~spectral bands
\citep{2023A&A...674A..35G}.	
Comparison with ground-based observations under similar illumination geometry show very good agreement across taxonomic class, their examples including (1)~Ceres, (4)~Vesta (Figure~\ref{fig:reflectance-spectra}a), (21)~Lutetia and (433)~Eros.

Numerous studies are making use of the Gaia reflectance spectra, and I will give just a few examples based on Data Release~3. Gaia Data Release~4 will contain more objects, and with twice the temporal coverage.

\subsubsection{Asteroid families}

Asteroid `families' result from collisions, some of which are ancient, others being much more recent 
\citep[e.g.][]{{
2002Natur.417..720N},
{2015aste.book..297N}}. 
Collisional fragments can be ejected at moderate velocity ($\sim$1\kms) and, at least in the main asteroid belt, can remain clustered in orbital space and with similar physical properties.
Here again, improved astrometry from Gaia allows these collisional fragments, with similar orbits around the Sun, to be identified.

\citet{2024A&A...682A..64B} 
studied 263 fragments of an S-type (siliceous or `stony') asteroid recently discovered in the inner main belt, with an age of 4.4\,Gyr, and comprising both planetesimals and collisional fragments 
\citep{2023A&A...676A...5F}.  
Figure~\ref{fig:reflectance-spectra}b shows six of their break-up candidates. The combination of existing and newly-measured spectra extending to the near infrared, combined with Gaia reflectance spectra below 1\,$\mu$m (shown in red), allowed identification of 71 (non-S-type) interlopers, and a final list of 190 S-type members.

The low albedo inner belt C-type (carbonaceous) asteroids, of which there are 11~known families,
have been suggested as the source of the near-Earth asteroids Ryugu (visited by Hayabusa~2 in 2018) and Bennu (visited by OSIRIS-REx the same year). 
Using the Gaia reflectance spectra to characterise these families and probe the origins of Bennu and Ryugu, the Polana and Eulalia groups can be distinguished in the range 370--500\,nm, while the spectra of the Eulalia and Polana families are most similar to those of Bennu and Ryugu respectively
\citep{2023A&A...680A..10D}. 

\subsubsection{Space weathering}

Space weathering is the term used to describe the modification of the reflectance spectra of solar system bodies as a result of surface irradiation from cosmic rays, the solar wind, and micro-meteorites.
Amongst 21\,909 S-type asteroids identified in the Gaia (FPR) reflectance sample, 9225 have been assigned to known families 
\citep{2023A&A...674A..35G}.
Some have ages estimated by backward orbit integration to identify their convergent origin. They showed that the spectral slope increases, and the depth of the 1\,$\mu$m absorption band decreases, with increasing age (their Figure~20), which they attribute to the progressive effects of space `weather'. 

The andesitic meteorite Erg Chech 002 (andesite being a type of volcanic rock, named after the Andes where it was first identified) was found in the Sahara desert in 2020. With a crystallisation age of 4.565\,Gyr, 2.25\,Myr after the solar system origin, it preserves a record of volcanic events and crust formation in primordial planetesimals, although no asteroid analogue had been identified
\citep{2021PNAS..11826129B}.
A search for analogues amongst the Gaia DR3 reflectance spectra found 51 similar main-belt asteroids, and 91 others with the spectrum of a space-weathered analogue
\citep{2023A&A...671A..40G}.	
Near-infrared spectra would provide a more definitive association.

The V-type (basaltic) asteroids are also important in understanding planetesimal formation and evolution in the early solar system. Some 2000 possible V-type asteroids have been identified from their Gaia reflectance spectra, increasing the known number by more than a factor~three
\citep{2023MNRAS.519.2917O}.	

The rare inner-belt T-type (unknown composition) asteroid (596) Scheila underwent a collision in December 2010 
\citep{2011ApJ...740L..11I}.
The Gaia reflectance spectrum (amongst others) has been used to place limits on the effects of surface weathering over a time span of just 10~years
\citep{2022ApJ...939L...9H}.	
Also noteworthy is the discovery of the first olivine-dominated A-type asteroid family
\citep{2024arXiv240210679G}.	

%

\subsubsection{Dynamical effects of solar radiation}			

The orbits of all solar system bodies are mainly controlled by gravitational forces, and in particular by the Sun. Today's state-of-the-art positions and velocities of the Sun, Earth, Moon, and planets
(\href{https://ui.adsabs.harvard.edu/abs/2013arXiv1301.1510F}{INPOP10e}
prepared by the Observatoire de Paris, and \href{https://en.wikipedia.org/wiki/Jet_Propulsion_Laboratory_Development_Ephemeris}{DE430}
by Caltech's JPL) are based on a numerically integrated dynamical model which, for DE430, also includes the perturbative effects of 343 asteroids, viz.\ 90\% of the mass of the main belt.

Solar radiation modifies an object's orbital motion in various somewhat distinct ways, depending on the object's orbit, size, and rotation:
(a)~solar radiation {\it pressure}, imparted by photon momentum, is a tiny force, but one with a large cumulative effect over long periods of time, affecting the motion of small solar system bodies including spacecraft such as Gaia;
(b)~the Poynting--Robertson effect is only relevant for particles smaller than about 1~mm; their motion around the Sun, combined vectorially with the velocity of the incident radiation, results in the particle losing angular momentum and, for small particle sizes, spiralling inwards towards the Sun;
(c)~the Yarkovsky effect is a force acting on a {\it rotating\/} body. It results from the re-radiated thermal emission, which carries momentum, lagging behind the incident radiation in time, thus (perhaps non-intuitively) contributing a component of force tangent to the orbital motion
\citep{
{2005JBAA..115..207B},
{2006AREPS..34..157B}};
(d)~the `YORP' effect is a second-order effect, mainly affecting the body's spin.

\paragraph{Yarkovsky effect}
\label{sec:yarkovsky}

The magnitude of the Yarkovsky effect is dependent on the orbit size and eccentricity, and on its mass, size, shape, spin and composition. But qualitatively, the Yarkovsky force is in the direction of orbital motion for a prograde rotator, causing the semi-major axis to increase steadily. Conversely, a retrograde rotator spirals inwards.
%
Models supplement the standard dynamical model with solar radiation pressure as an extra radial force, $A_1$, and a transverse `Yarkovsky' force, $A_2$, both inversely proportional to the heliocentric radius. Coefficients are determined by a model fit to the observations 
\citep{
1999A&A...344..362V,	
2013Icar..224....1F,		
2015aste.book..509V,	
2018A&A...617A..61D,	
2024A&A...682A..29F}.	
The momentum transfer is minute but relentless, and can result in significant orbit evolution over tens of millions of years. 

The effect complicates orbit determination for Near-Earth Objects, occasionally frustrating their identification 
\cite[e.g.][]{2008AJ....135.2336V}, 
affects the prediction of near-Earth approaches and impact probabilities 
\cite[e.g.][]{
{2002Sci...296..132G}, 
{2009Icar..203..460M},
{2015Icar..252..277V}},	
may explain the origin of high-eccentricity meteorites 
\citep{
1976Icar...29...91P,
1995P&SS...43..787A},
and perhaps the Near-Earth Objects and and Potentially Hazardous Object populations more generally, by delivering main belt asteroids up to 20\,km in size into Earth-crossing orbits 
\citep{2003Icar..163..120M}.
%
It also drives an additional orbital dispersion of asteroid `families' beyond the effects of collisions alone, leading to their characteristic V-shape in diagrams of semi-major axis versus absolute magnitude, which is important for their age estimation
\cite[e.g.][]{
{2001Sci...294.1693B}, 	
{2002aste.book..395B},	
{2014Icar..239...46M},	
{2015Icar..257..275S}, 	
{2017A&A...598A..91V},	
{2020SerAJ.200...25M},	
{2022CeMDA.134...34N}}.	
The Yarkovsky effect provides another possible route to asteroid mass determinations, especially for the Near-Earth Objects
\citep{2003Sci...302.1739C}.

First observational confirmation of the Yarkovsky effect came from radar tracking of the 500-m diameter asteroid (6489) Golevka, from Arecibo, between 1991--2003, when it was found to be 15~km from the position predicted from gravitational interactions with other solar system bodies alone
\citep{2000Icar..148..118V,
2003Sci...302.1739C}.
Studies suggested that the effect might be detectable by Gaia in several tens of NEOs
\citep{
2007AdSpR..40..209T,	
2008P&SS...56.1823D,	
2011MNRAS.413..741M,
2012P&SS...73...47T,	
2015A&A...575A..53D,	
2022Icar..38315040D,	
2024A&A...682A..29F}. 	

Various results have now been reported from Gaia, often in combination with long-term astrometry or radar ranging, leading to revised orbits and identification of impact families 
\citep{
2018A&A...620L...8H,
2023A&A...674A..12T,
2023PSJ.....4..215H,
2023A&A...680A..77D}.
As examples, Gaia DR1 has been used with the Las Cumbres ground network to detect the Yarkovsky effect in 18 out of 36 observed asteroids
\citep{2019Icar..321..564G},
while 
Gaia DR3 (and other) data has been used to derive the orbits of 446 Near-Earth Objects, including 93 Potentially Hazardous Asteroids, 54\,094 inner main belt asteroids, and various Mars crossing asteroids
\citep{2023A&A...680A..77D}. 
The latter determined a significant and improved Yarkovsky term (and associated bulk densities) for 49 Near-Earth Objects, including 10 new detections. 
JPL's \href{https://ssd.jpl.nasa.gov/sb/orbits.html}{Small-Body Database}
includes measurements of the Yarkovsky effect for 234 asteroids
\citep{2023A&A...674A..12T}. 		
All are NEOs, for which small objects and accurate orbits dominate. Some estimates pre-date Gaia \cite[e.g.][]{2020AJ....159...92G}, while others are from Gaia alone or combined with other observations.
Detection difficulty for main-belt asteroids notwithstanding, one notable asteroid without a measured Yarkovsky term today is (35334) Yarkovsky!

\paragraph{YORP effect}
\label{sec:yorp}

The Yarkovsky--O'Keefe--Radzievskii--Paddack effect is a second-order term, influencing the spin rate and spin axis orientation of small {\it irregular\/} asteroids. Like the Yarkovsky force, its effects accumulate over very long time periods, resulting in some rather remarkable changes in the properties of the asteroid population.
The term acknowledges four contributions to the understanding of how solar electromagnetic radiation, through diffuse reflection, absorption, and thermal re-emission, changes a body's angular momentum relative to its centre of mass, and dependent on its detailed properties such as shape and albedo  
\citep{2000Icar..148....2R}. 	
With asteroids being irregular in shape, and having rotation periods of order days, being much shorter than their orbital periods, the YORP effect is the secular change in its rotation state after averaging the solar radiation torques over the body's spin and orbital periods.

The YORP effect is presumed responsible for the many asteroids with very high and very low spin. Those with diameter $\gtrsim$125\,km follow a Maxwell distribution, while 50--125\,km sizes show an excess of fast rotators. The smallest, $\lesssim$50\,km, have an excess of very fast and slow rotators, becoming more pronounced with decreasing size
\citep{
2000Icar..148....2R,		
2007Natur.446..420K,	
2018ARA&A..56..593W}.	
Extreme YORP-driven spin-up may also explain asteroid fragmentation 
\citep{
1969JGR....74.4379P, 	
1975GeoRL...2..365P,	
2014ApJ...784L...8J,		
2020MNRAS.492.2437V},	
and contribute to the formation of binary asteroids
\citep{2008Natur.454..188W},	
and asteroid pairs with very similar orbits
\citep{2008AJ....136..280V}.		
There is indirect evidence for its role in the orbital evolution of asteroids over long periods, most prominently in the clustering of the directions of rotation axes in asteroid families 
\citep{
2002Icar..159..449V,
2006AREPS..34..157B,
2006Icar..182..118V}.

Direct confirmation of the YORP effect came with optical observations of the 100-m diameter asteroid (54509) YORP. Over 4~years, the 12-min rotation period showed a continuously increasing spin rate of $2.0\pm0.2\times10^{-4}$~deg~d$^{-2}$ which could not be explained by gravitational torques
\citep{
2007Sci...316..272L,		
2007Sci...316..274T}.		
Similar results were found for the 1-km sized asteroid (1862) Apollo, where the change is visible in the long-term light curve, amounting to an extra rotation in 40~yr 
\citep{
2007Natur.446..420K,	
2024A&A...682A..93D}.	
Today, the YORP effect has been identified individually for just 12~asteroids	
\citep{
2008A&A...489L..25D,	
2022A&A...657A...5D,	
2024A&A...682A..93D}.	

Although Gaia DR3 has contributed no other direct measurements of the YORP effect so far, the resulting distribution of asteroid spin states are being well-probed by its photometry. DR3 includes accurate photometry for 154\,787 solar system bodies, covering a time interval of 34 months, between July 2014--May 2017. With some 3~million measurements in total, the number of photometric observations per asteroid ranges from a few to several tens.

From the illumination geometry for each observation, parameterised by the sidereal rotation period, the spin axis direction, and a low-resolution convex shape, tens of thousands of trial periods in the range 2--10\,000~h combined with tens of trial pole directions has resulted in spin-states and low-resolution convex shape models for 8600 of the 150\,000 objects
\citep{2023A&A...675A..24D},	
more than doubling the previous numbers 
\citep{2010A&A...513A..46D}.
The findings indicate that small asteroids have poles clustered toward the ecliptic poles, an effect attributed to, and well-explained by, the YORP-induced spin evolution, while the sense of rotation of members of asteroid families are correlated with their semi-major axis. Thus, over the age of the family, orbits of prograde rotators have evolved, due to the Yarkovsky effect, to larger semi-major axes, while those of retrograde rotators have drifted in the opposite direction.

\subsection{Occultations}		

A stellar occultation occurs when a solar system object, such as an asteroid or a planetary moon, passes in front of a star as seen from the Earth, causing a temporary drop in the observed brightness of the star. 
This brightness drop can be used to determine the occulting object's position, along with its size and shape in the direction orthogonal to the occultation. And it can probe other properties of the occulting object, such as binarity, the presence of an atmosphere 
\citep{1990Natur.343..350S}, structures such as rings or moons, or even its topographic features 
\citep{2021EPSC...15..440R}.

The first occultation of a star by an asteroid, Juno, was recorded in 1958. But until the Hipparcos astrometry mission results became available in 1997, the limited accuracy in the knowledge of star positions made it difficult to predict future occultations with any confidence, both in terms of its location on Earth, and the event time. With Hipparcos positions, some 30 such events have been observed every year since. Amongst them, occultations of asteroids, satellites of asteroids, Centaurs and Kuiper Belt Objects, and even Pluto, have all been observed. 

Gaia has revolutionised the field by providing a dense grid of accurate positions of both stars and minor solar system bodies, such that there are, today, greatly improved prospects of identifying, and predicting to within minutes, suitably bright occultation events.
Amateur observers contribute significantly, with one of several large efforts coordinated by the 
\href{https://lesia.obspm.fr/lucky-star}{Lucky Star project}.
Various on-line databases maintain compilations of occultation results
\citep{2019JPhCS1365a2024B,		
2020MNRAS.499.4570H},			
and I will give a few examples of these Gaia-enabled occultation measurements. 

Only a handful of occultations have ever been observed for Jupiter's Galilean moons: of Ganymede (in 1911, 1972, and 2016) and of Io (in 1971), with none of Europa or Callisto. Their rarity has been due to the difficulty in predicting potential occultations due to uncertainties in the positions of background stars. This has changed with Gaia. 
And between 2019--2020, Jupiter was projected against a particularly dense star region, in the vicinity of the Galactic centre, a configuration that will not re-occur until 2031. The high background star density means that the probability of a stellar occultation by the Jovian moons increases dramatically, providing better opportunities to determine their positions, improve their orbits, and measure their shapes independently of satellite probes.  This can help, for example, in the study of tides generated by gravitational forces by Jupiter itself, and in the preparation of space missions targeting the Jovian system, specifically ESA's JUICE mission launched in April 2023, and NASA's Europa Clipper mission focusing on its moon Europa, launched in October 2024. 

The first Gaia-based observations of Europa were made on 31~March 2017, based on Europa's orbit calculated by JPL, and star positions from Gaia~DR1. Together, these predicted that Europa would occult a 9.5~mag star at 06:44 UTC on that day
\citep{2019A&A...626L...4M}.		
Three stations in Chile and Brazil observed the occultation, and measured a brightness drop corresponding to the chord of the satellite that each observed. Taken together, they gave estimates of Europa's major and minor ellipsoidal axes, $1562.0\pm3.6$~km and $1560.4\pm5.7$~km respectively. These values, and the body's resulting oblateness, are in good agreement with those from the Galileo mission images.

\begin{figure}[t]
\centering
\includegraphics[width=0.30\linewidth]{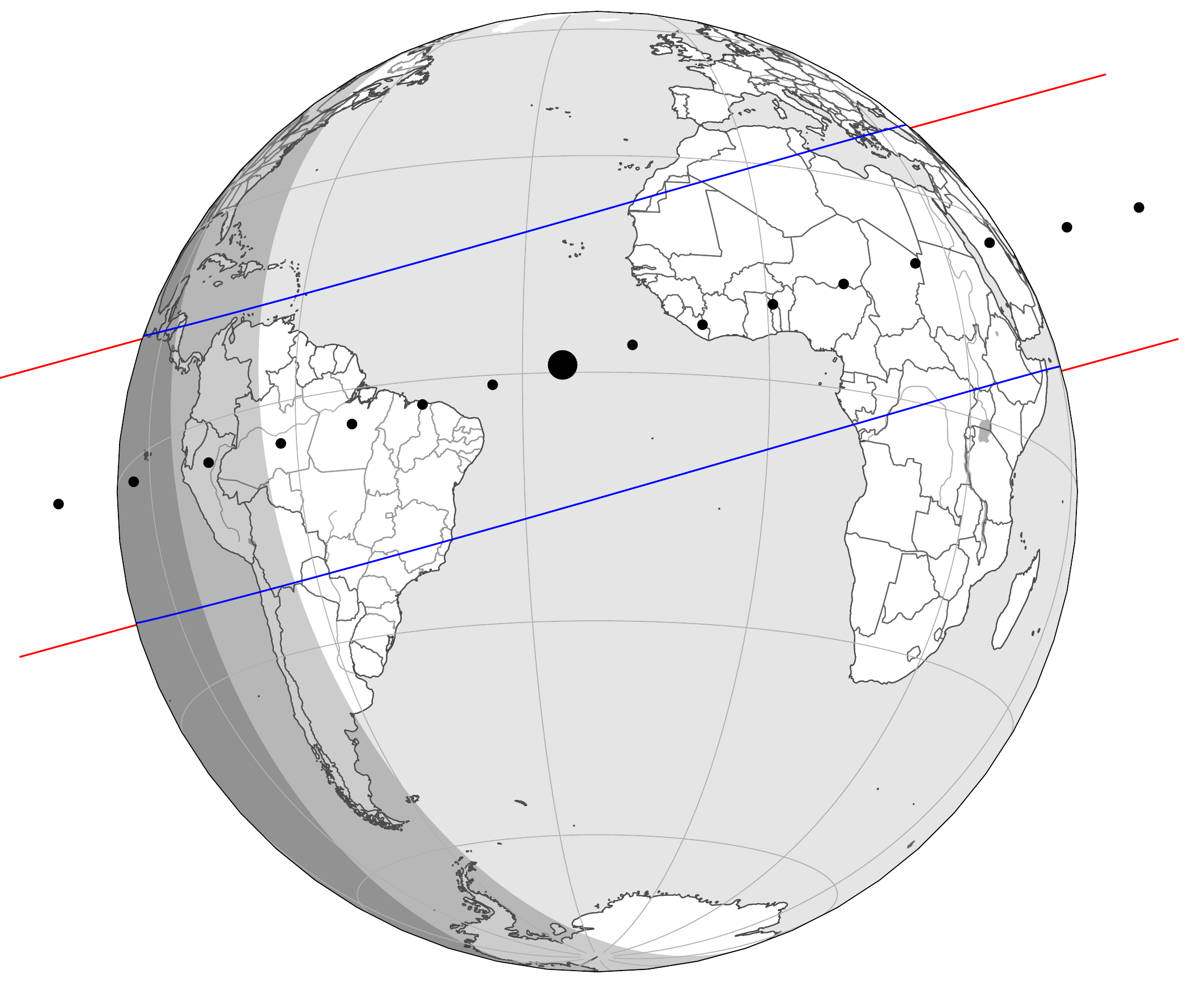}
\hspace{30pt}
\raisebox{-12pt}{\includegraphics[width=0.60\linewidth]{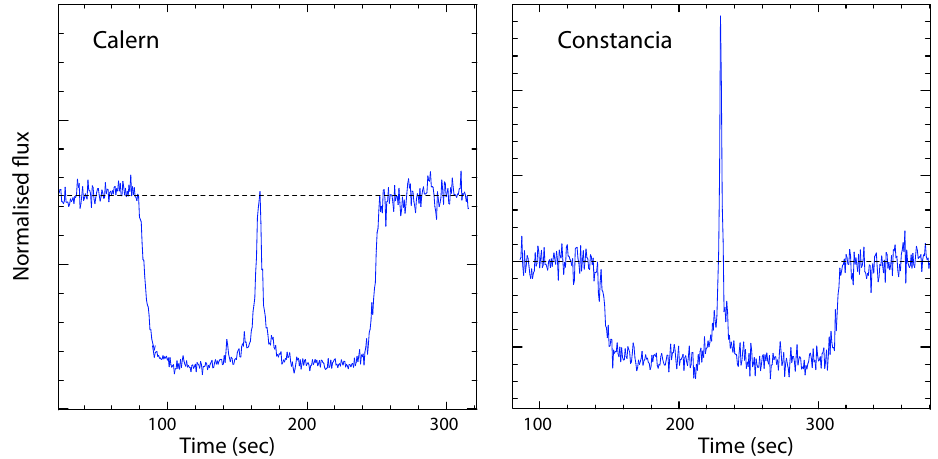}}
\caption{Left: predicted occultation event, based on Gaia astrometry, for the Galilean moon Io, on 2~April 2021 (\citet{2019A&A...626L...4M}).
Right: occultation of Triton and its central `atmospheric flash' observed at Calern and Constancia, 5~October 2017 (\citet{2022A&A...659A.136M}).
}
\label{fig:occultations}
\end{figure}

On 5~October 2017 Triton, the largest of Neptune's satellites, passed in front of a 12.6~mag star. A goal of this particular event was to examine Triton’s atmosphere. During the occultation of a body possessing an atmosphere, stellar rays are refracted and focused by it, creating a `flash' which can be detected by an observer close to the line of `centrality'. This was first observed for Triton in 1990
\citep{1990Natur.343..350S}.	
Using pre-release Gaia DR2 star positions gave Triton's position to 3~milli-arcsec at the time of the event, corresponding to an accuracy of 60~km on the Earth, and to 8~seconds in the event's predicted time (Figure~\ref{fig:occultations}a). 
On 5~October 2017, as predicted, Triton’s shadow swept across Europe and North Africa, and a few minutes later, the US East Coast. More than a hundred observing stations prepared for the event, which had a maximum duration of three minutes, and which was successfully observed at 70 sites, 25~of them detecting the central flash (Figure~\ref{fig:occultations}b).
Its shape and spectral dependence are sensitive to the atmospheric structure, and to the presence of hazes near its surface
\citep{2022A&A...659A.136M}.	
A subsequent occultation by an 11~mag star on 6~October 2022 was visible from India, China and Japan
\citep{2024A&A...682L..24S}.	
Other Gaia-enabled occultation of planets and their moons include studies of
Neptune \citep{2022AGUFM.P32E1864S}
and of Saturn's irregularly orbiting satellite Phoebe \citep{2020MNRAS.492..770G}.

One of the major goals in planetary science is to probe the processes underlying the formation of our solar system, and observational constraints of its outer regions have become an important focus of ongoing research over the past two decades. 
Here, Gaia-based occultations are contributing to the knowledge of 
the dwarf planets 
Pluto \citep{2022DPS....5430703Y}
and
Haumea, along with its satellite Hi'iaka \citep{2022EPSC...16..406F};
the size, shape and various other characteristics of numerous trans-Neptunian objects including
2002~TC$_{302}$ \citep{2020A&A...639A.134O},
2002~MS$_4$ \citep{2021EPSC...15..440R},
2003~VS$_2$ \citep{2022A&A...663A.121V},	
Huya \citep{2022A&A...664A.130S},	
and Quaoar observed from space by CHEOPS \citep{2022A&A...664L..15M};
and similarly for the Centaurs
Chariklo and its rings \citep{2021A&A...652A.141M},
the elongated 2002~GZ$_{32}$ \citep{2021MNRAS.501.6062S},
and (60558) Echeclus \citep{2022EPSC...16..557P}.
The 2017 Gaia-based occultation campaign of 486958 Arrokoth (aka 2014~MU$_{69}$, and formerly Ultima Thule) was important in the planning of the New Horizons flyby 
\citep{2022DPS....5431102B}.

Two other applications of stellar occultations can be mentioned. The first is in the determination of stellar masses through appeal to Kepler's laws, by determining accurate binary separations at a given epoch.  A specific example is for the $V=6.5$\,mag K~giant HIP~36189. This was originally discovered to be a binary system as a result of an occultation by (704) Interamnia in 2003 \citep{2020MNRAS.499.4570H}.  Together with the occultation-based $\rho=13.0\pm0.7$\,mas, with position angle $\theta=231.9\pm4.0^\circ$  \citep{2020MNRAS.499.4570H}, and the orbital period of 2.6-yr determined from Gaia DR3, yields components masses $M_1=3.9\pm2.2M_\odot$ and $M_2=3.5\pm1.6M_\odot$ \citep{2023A&A...674A..34G}.
Finally, it is also possible to probe the occulted {\it star}, as well as the occulting solar system object, and specifically to derive {\it stellar\/} diameters. Here, the occultation methods exploits the diffraction pattern in the shadow cast when a solar system object occults the star, but requires that the photometric uncertainty is smaller than the atmospheric scintillation noise. A recent initiative has been to use the VERITAS Cherenkov telescopes, an array of four \mbox{12-m} optical reflectors located at the Fred Lawrence Whipple Observatory in Arizona
\citep{2019NatAs...3..511B, 	
2020NatAs...4.1164A}. 		
Two such events, using Gaia~DR2 positions and distances, were observed for 
the asteroid (1165) Imprinetta occulting the 10.2~mag star TYC~5517--227--1 on 22~Feb 2018, 
and 
the 88-km diameter asteroid (201) Penelope occulting the 9.9~mag star TYC~278--748--1 on 22~May 2018
\citep{2019NatAs...3..511B}.
In both cases, the diffraction patterns are clearly seen around ingress and egress, yielding accurate angular sizes ($0.125\pm0.022$ and $0.094\pm0.01$ milli-arcsec respectively), and hence linear sizes ($11.0\pm2.0\,R_\odot$ and $2.17\pm0.22\,R_\odot$) from their Gaia-based parallax distances ($820\pm40$\,pc and $215\pm2$\,pc respectively).

\subsection{Stellar fly-bys}

Interstellar space is really very empty, and the mean-free time between star--star collisions in the solar neighbourhood is estimated to be $\sim\!10^{13}$\,yr, some thousand times the age of the Universe. But noticeable effects due to nearby star passages and our Sun can nonetheless occur. In particular, the gravitational effect of the passing star can perturb the somewhat delicate equilibrium of the Oort cloud comets, with the possibility of an increased impact hazard on Earth. 
Evidence that other stars might have come close to the Sun in the past is to see whether there is a correlation between ancient stellar flybys, and increased cratering events, on the Earth, or elsewhere in the solar system.  A compilation of impact cratering events is maintained at the Earth Impact Database, and there have been many attempts to examine whether such impact events are connected to large-scale extinctions on Earth
\citep[e.g.][]{2021MNRAS.501.3350F}.	

Given a description of the Sun's motion through space, and of the distances and motions of other stars in the solar neighbourhood, it is straightforward in principle to track their paths backwards in time, and see whether any close stellar encounters happened in the past, and whether others might occur in the foreseeable future. 
Several such studies, extending over the past and into future, originally made use of the Hipparcos distances and space motions (available since 1997), combined with radial velocities to give their full space motions 
\citep{1999AJ....117.1042G,		
2015ApJ...800L..17M,			
2017MNRAS.464.2290M}.		
For example, GJ~710 will have a closest approach to the Sun of less than 0.4~pc some 1.4~million years in the future
\citep{1999AJ....117.1042G},		
while the closest past encounter was found to be the low-mass binary WISE~J0720--0846 (Scholz's star), at about 50\,000\,au (i.e.\ within the Oort cloud) some 70\,000\,yr ago
\citep{2015ApJ...800L..17M}.		

Numerous similar studies have been made based on the Gaia distances and space motions, refining the flyby parameters of GJ~710
\citep{2016A&A...595L..10B},		
and using much larger star samples
\citep{2017AstL...43..559B,		
2018A&A...609A...8B}.			
A sample of 7.2~million stars from Gaia DR2 included 694 stars with flybys within 5~pc, all occurring within $\pm15$~Myr
\citep{2018A&A...616A..37B}.		
Of these, 26 may pass within 1~pc, and 7 within 0.5~pc. All but one were new discoveries, while GJ~710 remains the largest Oort cloud perturber.  The study estimated that only 15\% of flybys inside 5~pc, and within $\pm5$~Myr, have been identified, mainly due to the absence of radial velocities for the fainter or cooler stars, and implies one encounter within 1~pc every 50\,000 years.
Studies with Gaia DR3 have further refined these orbits and identified other flyby candidates,
including the second closest, HD~7977, and the more controversial white dwarf WD 0810--353
\citep{2022ApJ...935L...9B,	
2022AstL...48..542B,			
2022A&A...668A..14D,		
2022A&A...664A.123D,		
2024A&A...685A.171D,		
2020MNRAS.491.2119W,		
2020A&A...640A.129W}.		


\subsection{Interstellar objects}
\label{sec:interstellar-objects}

The discovery of the first planets beyond our solar system in the 1990s, and the remarkable advances in discovery numbers, their highly detailed observations, and supporting theories and numerical modelling, has transformed our understanding of the formation of planetary systems in general, and of our solar system in particular. From these multidisciplinary and concentrated efforts we now have a much better `big picture' of its formation and subsequent evolution.  
Left over from the final assembly of the planets was a vast range of debris still circling the Sun: rocky asteroids closer to it, the small icy lumps of the Oort cloud comets far out in a vast spherical cloud. 
Gravitational forces between the bigger bodies and the smaller debris particles acted as slingshots, propelling the latter (and sometimes the former) into perturbed and often unusual orbits. Some of these smaller bodies would have been hurled out at very high velocities, often high enough to escape the gravitational embrace of the host system.

Detailed models of the gravitational scattering that takes place during the later stages of planetary formation, in our solar system and in others, implies that interstellar space should have extrasolar planetesimals and comets passing through it, flung out as a by-product of star and planet formation. 
Some of these, from nearby stellar systems, will end up passing through our solar system. These {\it interstellar objects\/} should be recognisable by their extreme hyperbolic orbits, contrasting with the elliptical (bound) orbits of the vast majority of solar system objects.
Serious searches for such objects started around 1990.  The usual way of discovering objects in the solar system is from repeated deep imaging of some part of the sky, repeated days or weeks later to see if any objects might be moving. The changing position of a moving object can be measured repeatedly, and its orbit eventually determined and refined.

The first such interstellar traveller, 1I/Oumuamua, was announced in October 2017. It was discovered with the wide-field survey instrument Pan--STARRS, some 40~days after closest approach to the Sun
\citep{2017Natur.552..378M}. 	
Within three months of its discovery, 30~groups had reported studies of its orbit, and its shape, rotation and surface properties, with much speculation about its origin. This could only be deduced by following its motion backwards in time, taking into account the overall gravitational potential of our Galaxy, and the individual perturbations from stars it has passed along its way. 
In one of the pre-Gaia studies of more than 200\,000 candidate stars, only four candidate progenitors were found
\citep{2018A&A...610L..11D}.
The most promising, HIP~113020 (GJ~876), hosts a four-planet system, with the encounter occurring 790\,000 years ago, with a relative velocity of just 3.9\kms. 
Using the Gaia DR2 data, the closest encounter found by
\citet{2018AJ....156..205B}	
was with the M-dwarf HIP~3757, at a distance of 0.60~pc some 1~Myr ago, but with a relatively high encounter velocity of 24.7\kms.  They also found a more distant encounter, with the G5 dwarf HD~292249, at 1.6~pc and 3.8~Myr ago, but with a lower encounter velocity of 10.7\kms.

Similar orbital tracebacks using Gaia DR2 have been performed for the second interstellar object, 2I/Borisov, discovered by amateur astronomer Gennadiy Borisov in August 2019
\citep{2020A&A...634A..14B}.	
Out of 7.4~million stars, the closest encounter identified occurred 910~kyr ago with the M0V star Ross~573, at a separation of 0.068 pc, nearly an order of magnitude closer than the closest past encounter identified for 1I/Oumuamua. They also concluded that if 2I/Borisov is more than 10~Myr old, their search would be unlikely to find its parent system. 
A third such object, C/2025 N1 (ATLAS) was discovered on 1~July 2025
\citep{2025MPEC....N...12C}.
More about their Gaia-based Galactic orbits, their nature, and the discovery potential with the Vera Rubin Observatory, can be found in recent studies and reviews
\citep[e.g.][]{2023ARA&A..61..197J,	
2024arXiv240802739K,			
2024arXiv240919072P,			
2025AJ....169...78H}.			
%

\subsection{Solar twins and solar siblings}

`Solar twins' are loosely defined as stars being essentially identical to the Sun in all key astrophysical parameters: mass, age, luminosity, chemical composition, temperature, surface gravity, magnetic field, rotation velocity, etc.\
\citep[e.g.][]{1978A&A....63..383H,
1996A&ARv...7..243C,
2017A&A...597A..34M}.
Amongst various reasons for studying them 
\citep{2021MNRAS.504.1873Y},
solar twins may be stars most likely to host planetary systems similar to our own, and may be best-suited to host life forms based on carbon chemistry and water oceans. 
%
%
From the late 1990s, searches used Hipparcos distances, and therefore accurate stellar luminosities, finding that HIP~79672 (18~Sco) and HIP~78399 were amongst the most promising candidates. Today, some 100~solar twins are known. Most recent searches use Gaia to determine accurate absolute magnitudes and photometry, and hence to identify objects whose stellar parameters are most similar to the Sun
\citep{2021MNRAS.504.1873Y}.

In contrast, the hypothetical class of `solar siblings' are stars assumed to have formed at the same time as the Sun, and from the same gas cloud. Unlike solar twins, they do not need to be Sun-like in terms of key properties but, if born in the same gas cloud, they are presumed to have identical ages and initial chemical compositions. 
They are of interest in establishing whether the Sun formed in isolation or, like most stars born in molecular clouds, was one of a cluster, although there are hints that it was one of perhaps a thousand other stars
\citep{2010ARA&A..48...47A}. 
For example, the solar system's relatively sharp outer boundary, at about 30\,au, suggests that the disk of gas and dust from which it formed was truncated by interactions with other cluster members
\citep{2020ApJ...897...60P}. 
Another clue is the large eccentricities and inclinations of the Kuiper belt objects, suggesting that a stellar encounter occurred early in the solar system's history, an event more likely in a denser stellar environment. 
The search for solar siblings aims to gain a better understanding of the conditions under which life developed on Earth
\citep[e.g.][]{2014ApJ...787..154R}.
And this touches on the more controversial idea of `panspermia', in which `spores of life' are ejected and transmitted through space, protected within small rocky bodies on their travels. If the Sun was born in a cluster environment, solar siblings might be especially interesting targets in the search for life.
Pre-Gaia, no reasonably unambiguous candidates had been identified 
\citep[e.g.][]{2014ApJ...787..154R, 2015A&A...575A..51L}.	

In the absence of some more energetic ejection from the birth cluster, solar siblings should be on similar orbits around the Galaxy as our Sun. Accurate stellar ages and precise chemical abundances are essential in making the searches more efficient
\citep{2016MNRAS.457.1062M}.
Given accurate space motions of the Sun and other nearby stars, and given a good model of the Galaxy's mass distribution (and hence gravitational potential), it should then be possible to `reverse' their orbits, and calculate their original common birthplace. 
The most optimistic estimates pre-Gaia suggested that perhaps 10--60 still exist within 100\,pc of the Sun
\citep{2015CeMDA.121..107V}.		
Based on Gaia DR2, HD~186302 was identified as one such candidate, based on it chemical composition, age, and orbital dynamics
\citep{2018A&A...619A.130A}.		
A larger study used Gaia DR2 astrometry for 19\,000 stars with solar values of [Fe/H] from the SDSS--APOGEE DR14 catalogue
\citep{2020MNRAS.494.2268W}.		
From backward integration in time over 5~Gyr, and in various realistic Galactic potentials, they identified some 100 primary candidates, including several of the candidates suggested in earlier studies
\citep{2016MNRAS.457.1062M}.		
Their top candidate, Solar Sibling~1, lies at a distance of $360\pm80$\,pc, and has a dynamical history very close to that of the Sun.  And in contrast with some previous work, they were able to exclude M67 as the Sun's birth cluster.

\subsection{Sun--Earth system}
The search for solar twins connects with the problem of understanding the so-called `Maunder Minimum', a 70-year interval around 1645--1715 when a particularly cold period in Europe appears to have coincided with the almost complete disappearance of sunspots
\citep{1976Sci...192.1189E}.
One of the routes to understanding whether there is a definitive link between the variation in solar irradiance and Earth's climate is to study any changes in activity of other sun-like stars, for which accurate distances are key. In the absence of specific Gaia studies to date, I will simply reference some of the pre-Gaia studies to provide the context
\citep{1990Natur.348..520B,		
1996AJ....111..439H,				
2012ApJ...747L..32L,			
2004AJ....128.1273W}.			

Another, perhaps surprising, application of a space astrometry has been in measuring the Earth's spin-axis motion over the past century or more. The Hipparcos catalogue provided a much-improved reference frame over this extended period, within which historical optical observations made from the ground could be re-interpreted. These extensive catalogues of historical star observations were coordinated by the International Latitude Service established by the International Association of Geodesy in 1900, with preparations for the incorporation of the Hipparcos results coordinated by Commission~19 of the IAU on the `Rotation of the Earth'.  From a number of Hipparcos-based studies 
\citep[e.g.][]{1997A&A...319.1020V,	
2000ASPC..208..239V},			
Figure~7 of \citet{2000ASPC..208..239V} 
shows the polar motion between 1900--1992, including the pronounced beat period, of about 6~yr, between the annual and `Chandler' term.
I am not aware of further re-analysis using the Gaia catalogue, nor whether such analyses would provide new insights.

\section{Stellar structure and evolution}
\label{sec:stellar-structure}

This section on stellar structure and evolution, and the subsequent section on Galactic structure and dynamics (Section~\ref{sec:galaxy-structure}), cover topics that somewhat overlap. Thus, insights being gained into stellar structure and evolution are generally derived from specific stellar samples, for example from well-defined samples in the solar neighbourhood, or from the members of individual star clusters. 
I include in this section a description of the census of nearby stars from Gaia, the census of nearby open clusters and associations, and some of the new insights into the structure of the Hertzsprung--Russell diagram. I also include an overview of results on variable stars, and double and multiple stars.

\subsection{Nearby stars}

\paragraph{Context}

A detailed understanding of the nearby stellar population is central to many areas of astronomy. While `nearby' is an imprecise term, it is often taken to mean the spherical region of space out to (say) 10, 20, or 50\,pc from the Sun. Compared with the scale of our Galaxy, in which the Sun sits some 8\,kpc from the Galactic centre, this region is dominated by stars of our Galaxy's disk.
Surveys of this `nearby' region provide the foundations for defining our Galaxy's stellar luminosity distribution, the local mass density (in both stars and gas), their velocity distribution, the distribution and occurrence of binary and multiple stars, and the occurrence and nature of many other types of objects which comprise our Galaxy, including white dwarfs, brown dwarfs, and exoplanets. 

Before Gaia, it remained an insurmountable challenge to compile anything approaching a complete census of stars in our immediate solar neighbourhood, even out to distances of, say, \mbox{10--20\,pc}. The pioneering ground-based parallax surveys of the early 1900s were successful in identifying nearby bright stars, but problems persisted especially for the lowest luminosity stars, where a complete survey for low-mass stars and brown dwarfs even out to 10--20\,pc remained a challenge.  
Surveys searching for high-proper motion stars in the 1970--80s were successful in detecting potentially nearby stars which were then added to parallax measurement programmes, but they resulted in samples biassed towards high-velocity halo objects. For this reason, early nearby star compilations used spectroscopic {\it and\/} photometric distance estimates to try to identify more nearby candidates. 

One of the first attempts to compile a census of stars in the solar neighbourhood, largely based on trigonometric parallaxes, was led by British Astronomer Royal Richard Woolley, and published as the {\it `Catalogue of Stars within Twenty-Five Parsecs of the Sun'} in 1970. Another, the Catalogue of Nearby Stars (CNS), originally led by Walter Fricke, has been updated and maintained by Heidelberg astronomers for more than 60 years.
%
CNS1, published in 1950, contained 915~single stars and multiple systems within 20\,pc, 
with parallax errors of about 10\,mas.  
CNS2, in 1969, enlarged the distance limit to 22.5\,pc, and contained 1049 stars and multiple systems within 20\,pc.
CNS3, in 1991, extended the census to some 1700 stars nearer than 25\,pc.
CNS4, in 1997, included data from Hipparcos, and accordingly provided the most comprehensive inventory of the solar neighbourhood out to a distance of 25\,pc from the Sun at that time. But Hipparcos could only observe pre-selected stars, contained in its `Input Catalogue'. This extended to about 11--12\,mag, but with completeness only to about 9\,mag.  
%
Although the number of stars within 25\,pc in CNS4 remained largely the same, some important details changed: the Hipparcos measurements identified 119~`new' nearby stars, of which the closest was a high-proper motion star at 5.5\,pc. More significantly, the results implied a considerable shift to larger typical distances, with associated implications for the local stellar mass density.

Other ground-based surveys since then, amongst them RECONS 
\citep[e.g.][]{2018AJ....155..265H,2015AJ....149....5W}
and SUPERBLINK
\citep[e.g.][]{2017AJ....154..118S}, 
have searched for faint nearby stars (notably M~dwarfs and white dwarfs) from infrared measurements or proper motion surveys.
Very broadly, pre-Gaia compilations listed some 5000 stellar systems within 25\,pc, while infrared surveys have identified more than 100\,000 M~dwarfs within 100\,pc.

\begin{figure}[t]
\centering
\raisebox{30pt}{\includegraphics[width=0.42\linewidth]{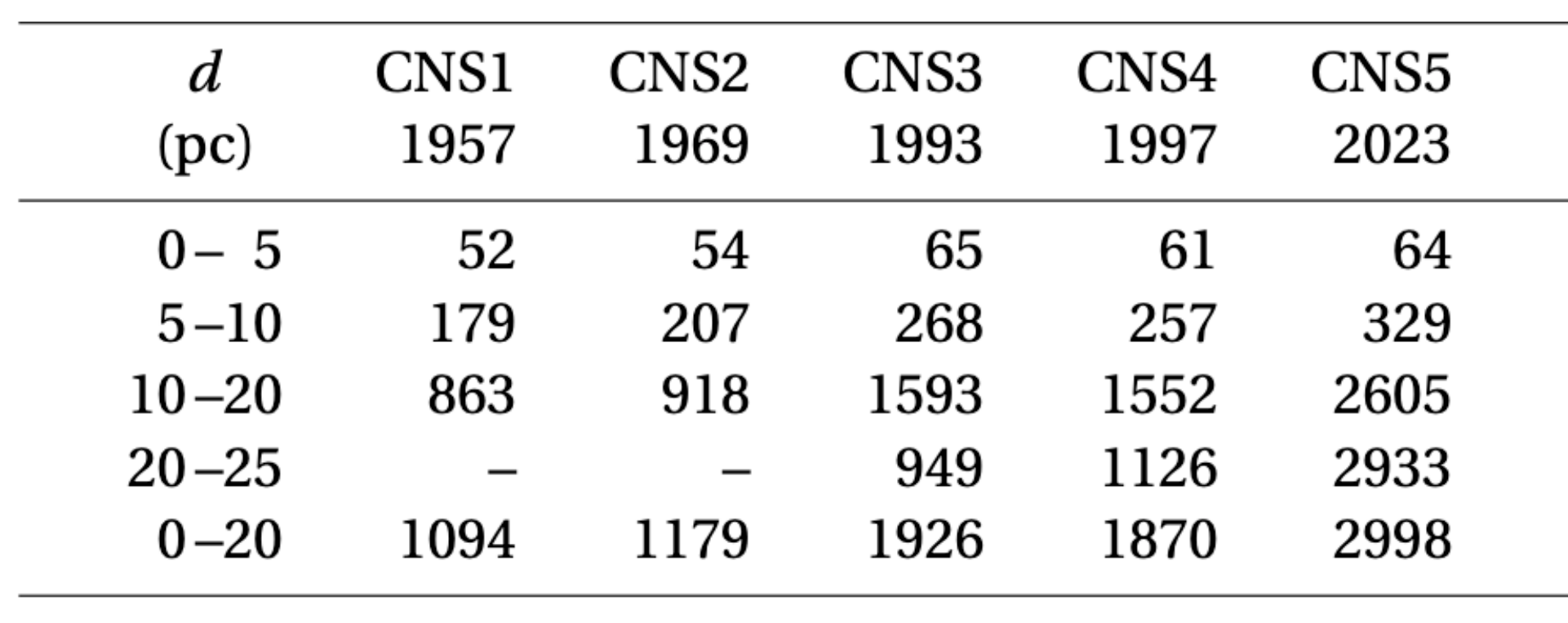}}
\hspace{20pt}
\includegraphics[width=0.46\linewidth]{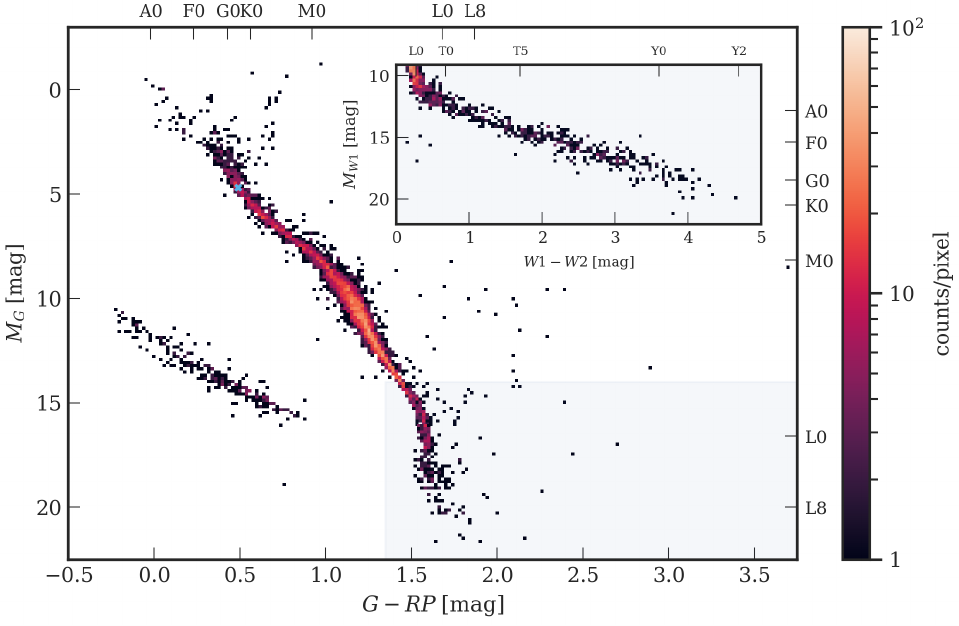}
\caption{Left: successive versions of the Catalogue of Nearby Stars (CNS) as a function of distance; pre-Gaia (CNS1--CNS4) numbers are from \citet{1997ESASP.402..675J}. Right: colour--magnitude  diagram of CNS5 objects in the Gaia EDR3 ($G-G_{\rm RP}$) bands. In the shaded area, the number of main sequence objects not present in Gaia starts to increase in the infrared WISE  bands, shown in the inset \citep[][Figure~5]{2023A&A...670A..19G}.}
\label{fig:cns}
\end{figure}

\paragraph{The Catalogue of Nearby Stars, CNS5}

Continuing these historical developments, the latest version of the Catalogue of Nearby Stars, CNS5, includes the Gaia results, and also aims for completeness to 25\,pc
\citep{2023A&A...670A..19G}. 
CNS5 is based on astrometric and photometric data from Gaia~EDR3 and Hipparcos, and is supplemented with parallaxes from ground- and space-based astrometric surveys carried out in the infrared to reach the lowest luminosities. It contains 5931 objects, of which 5230 are stars (4946 main-sequence, 20 red giants and 264 white dwarfs), and 701 are brown dwarfs (Figure~\ref{fig:cns}).
The much larger Gaia Catalogue of Nearby Stars, GCNS, described below, comprises a much larger sample of 331\,312 objects within 100\,pc. But while expected to be better than 95\% complete for spectral types up to M8 ($M_G=15.7$~mag), GCNS is highly incomplete for later spectral types. A number of bright and red sources are also missing since the GCNS is based only on Gaia EDR3 data.
Despite its much smaller volume, it is the greater completeness of CNS5, for later spectral types well into the brown dwarf regime, that underlies its importance.

An estimate of the completeness of the CNS5 catalogue is important.  For example, using space densities naively computed from the numbers within 25~pc would yield a luminosity function strongly biased due to any incompleteness at the faint end. CNS5 is estimated to be statistically complete down to $G=19.7$~mag (Gaia), and $W1=11.8$~mag (WISE), corresponding to spectral type~L8 \citep{2023A&A...670A..19G}. These authors give the resulting stellar number density as $(7.99\pm0.11)\times10^{-2}$ stars~pc$^{-3}$, with about 72\% of stars in the solar neighbourhood being M~dwarfs.

\paragraph{The Gaia Catalogue of Nearby Star, GCNS}

Gaia Early Data Release~3 (EDR3, December 2020) covered (nearly) the first three years of mission data, and lists nearly two~billion stars with parallax accuracies better than about 1\,mas, even for the faintest. This implies that stars brighter than about 21~mag, and within 100\,pc of the Sun, will be observed and identified as such, and with a distance accuracy of better (and often much better) than 10\%.  

The first Gaia Catalogue of Nearby Star, or GCNS, contains an unprecedented 331\,312 objects within 100\,pc. It includes stars as faint as spectral type M9, i.e.\ with masses down to about $0.1M_\Sun$
\citep{2021A&A...649A...6G}.	
Within our immediate 10~pc horizon, and supplemented by a few well-known stars too bright for Gaia (amongst them Sirius, Fomalhaut, Vega, Procyon, Altair, and Mizar), we now know of 383 stars, all with accurate distances. This includes five companion stars with distances measured for the first time, but not counting a few known unresolved binary systems (notably Procyon, $\eta$~Cas, and $\xi$~UMa). A few very low-luminosity T/Y brown dwarfs are also known, but too faint to be observed by Gaia.
More than 16\,000 resolved binaries are present in the GCNS sample, with some 10\% of F, G, and K spectral types clearly seen to be wide binaries.

In very broad terms (I will say more on some of these topics in subsequent sections), this enormous census is providing the basis for many new insights into our Galaxy's structure and dynamics. It allows characterising the disk thickness for various stellar spectral types, and measuring individual and bulk motions in our region of the Galactic disk (e.g.\ Galactic orbits, integrated over 1~Gyr, show that the most common disk stars follow circular orbits in the Galactic plane, while rarer halo stars visiting the solar neighbourhood have higher eccentricities and inclinations).
It provides our best estimate of the local luminosity function (numbers per cubic parsec) of the various star populations, including main-sequence stars, giants, and white dwarfs. These estimates enter models of star formation and evolution, and of the star formation history of the Galaxy. 
It is providing estimates of the Sun's circular velocity around the Galaxy, its distance from the disk's mid-plane, its vertical velocity with respect to the plane, and (from the vertical motions of nearby stars) the total mass of the Galaxy disk. 
The 100~pc sample contains two well-known open clusters, the Hyades (at a distance of about 47~pc) and Coma Berenices (at about 86~pc). Both the Hyades and Coma Ber clusters stand out in the GCNS as density concentrations in space, as well as in their velocities. The Pleiades is another prominent star cluster, also visible to the naked eye, but slightly more distant (at about 110~pc). 
Several other stellar streams and resonant dynamical structures are clearly visible in the Gaia survey.

\subsection{Stellar rotation}
\label{sec:stellar-rotation}

All celestial bodies rotate: galaxies, stars, and planets amongst them. The rotation of individual stars is inherited from the angular momentum of the gas and dust out of which they formed (albeit subsequently modified by various evolutionary processes), itself ultimately attributed to quantum density fluctuations in the early expanding Universe. 
Since stars are fluid objects, they are neither perfect spheres, nor do they rotate as solid bodies. Instead, their rotation produces an equatorial bulge due to centrifugal forces, and generally involves differential rotation in which the angular velocity is a function of latitude. Differential rotation is believed to have a significant role in the generation of a star's magnetic field, which then interacts with the stellar wind, a process gradually slowing the star's rotation over its lifetime.
As a result, rotation is important in understanding many fundamental stellar properties, including the origin and evolution of angular momentum, the creation and persistence of magnetic fields, magnetic braking and its applications to gyrochronology, and the mixing of chemical elements within their interiors.

There are two principal approaches to determine stellar rotation. The first is from its effect on the stellar spectrum: the fact that different regions of the star are moving at different line-of-sight velocities leads to the broadening of its absorption lines, from which the (projected) rotation velocity can be determined.
The second is from the rotation of active features on the stellar surface, in particular star spots and faculae, leading to significant photometric modulation at the rotation period. 
The compilation of 
\citet{2000AcA....50..509G}
in 2000, derived primarily from line-broadening, contained 11\,000 stars of all spectral types and luminosity classes, albeit restricted to a few nearby open clusters and a few thousand bright stars.
But the study of stellar rotation has been transformed within the past decade. 
The Kepler mission (2009--18) provided a substantial advance exploiting photometric modulation due to surface star spots and faculae. Amongst its 133\,000 main-sequence targets, its high photometric precision and densely-sampled long-duration observations resulted in some 34\,000 rotational periods in the range 0.2--70~days
\citep{2014ApJS..211...24M}.
A further 30\,000 were similarly identified from the extended Kepler K2 mission
\citep{2020A&A...635A..43R}.
Many scientific inferences have followed.

\begin{figure}[t]
\centering
\raisebox{6pt}{\includegraphics[width=0.31\linewidth]{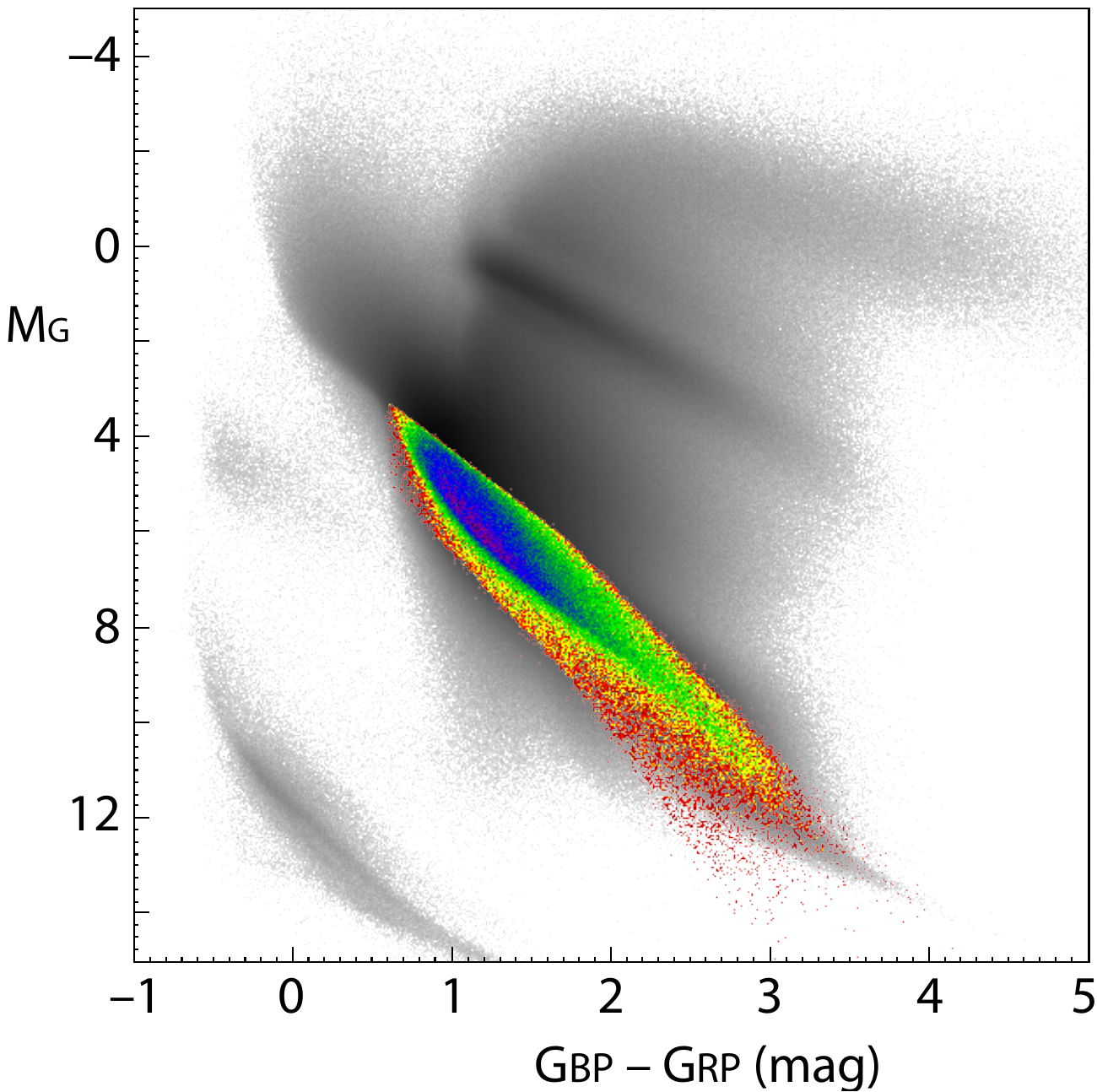}}
\hspace{30pt}
\includegraphics[width=0.45\linewidth]{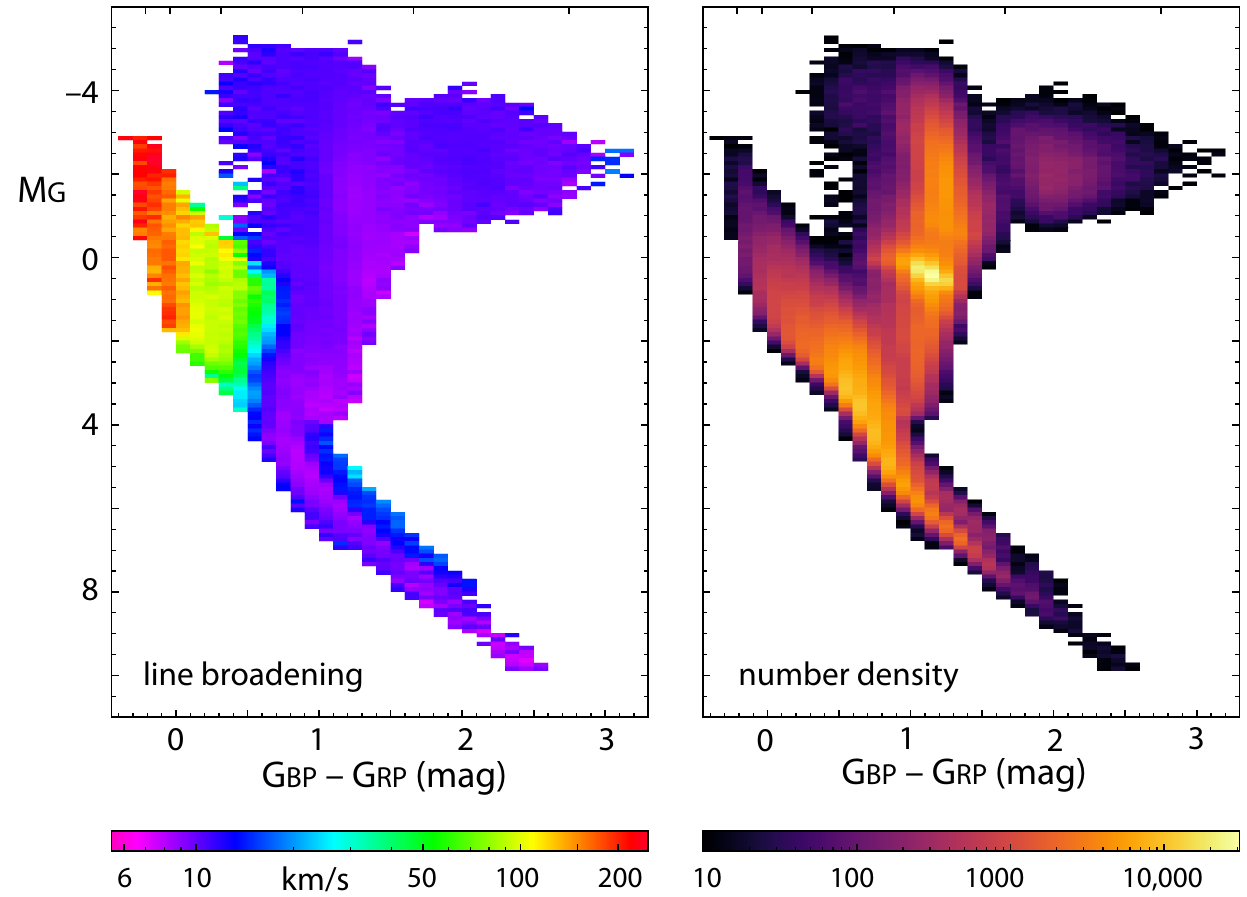}
\caption{Left: the relative density of stars with identified rotational modulation in the Gaia DR2 colour--magnitude diagram, increasing from red to purple \citep[][Figure~1]{2018A&A...616A..16L}.
Right pair: the median amplitude of the line-broadening term (left) and its corresponding number density (right) across the Gaia DR3 colour--magnitude diagram \citep[][Figure~16]{2023A&A...674A...8F}.
}
\label{fig:stellar-rotation}
\end{figure}

Gaia is also determining new rotation periods using these two distinct methods: rotational modulation (from its accurate photometry) and spectral line broadening (from its RVS spectrometry). 
Although Gaia's light curves are not as densely sampled in time as Kepler, the same principles of detection apply. Gaia's advantages include 3-colour time-series space photometry ($G$, $G_{\rm BP}$, $G_{\rm RP}$) covering the whole sky, and to much fainter magnitudes, $G\sim21$. 
Based on the first 22~months of mission data in Gaia~DR2,  
\citet{2018A&A...616A..16L}		 
started with the 500\,000 stars already classified as variable, and identified nearly 150\,000 rotation periods (and modulation amplitudes) in stars where the flux modulation is induced by surface inhomogeneities (star spots and faculae).  Candidate stars (BY~Dra variables) occupy a specific region of the $M_{\rm G}$ versus $G_{\rm BP}-G_{\rm RP}$ colour--magnitude diagram (Figure~\ref{fig:stellar-rotation}, left). They include dwarfs (with amplitudes of milli-mag), sub-giants, and T~Tauri stars (with amplitudes up to several tenths of mag). 
An updated version of the same pipeline was applied to solar-like variables in Gaia DR3
\citep{2023A&A...674A..20D}. 
They started with 474\,000 stars with variability attributed to magnetic activity. Of these, some 430\,000 are newly discovered variables, and some 150\,000 have rotation periods below 1~day, i.e.\ fast rotators poorly sampled by previous surveys. 
Some 150\,000 resulted in a `well-defined' rotation period, in the sense that the period was consistent over restricted data segments, even though the total time-series may lose its overall coherence.

The numbers get even bigger with the results from the RVS spectra, from which Gaia DR3 contains radial velocities for 33~million stars with \teff\ between 3100--14\,500\,K, and $G_{\rm RVS}<12$\,mag. 
In brief, these radial velocities are based on a comparison of the observed RVS spectrum with a library of synthetic spectral models covering a range of temperature, gravity, metallicity and reddening (and `broadened' to match Gaia's along-scan line-spread function, due to the contribution of its optics, the detector, and the satellite spin rate).
A number of detailed physical effects intrinsic to the star can also contribute to broadening of the spectral lines. But in most cases, their spectral impact is sufficiently well described by classical atmospheric modelling, and the spectral line shapes can usually be well replicated by keeping the effective temperature, surface gravity, metallicity, and microturbulence fixed
\citep{2023A&A...674A...8F}.
Ignoring the origin of such second-order effects, the DR3 pipeline also determines an {\it additional\/} line-broadening contribution (`vbroad'), assumed due exclusively to axial rotation, to best fit the observed spectra. 
The resulting catalogue contains 3\,524\,677 stars with \teff\ ranging from 3500--14\,500 K, and $G_{\rm RVS}<12$~mag, and shows reasonable agreement with the values in other catalogues, including GALAH, APOGEE, and LAMOST (Figure~\ref{fig:stellar-rotation}, right pair).
This DR3 line-broadening catalogue accordingly represents, by far, the largest stellar sample for the determination and interpretation of stellar rotation across the Hertzsprung--Russell diagram. 

The many studies now making use of this huge number of stellar rotations include 
studies of the `Kraft break', the mid-F spectral type transition from slowly rotating (cooler) stars to rapidly rotating (hotter) stars, attributed to the disappearance of the outer convective envelope and the decrease in magnetic braking
\citep{2024ApJ...973...28B};		
the connection between high eccentricity and spin--orbit misalignment in binaries
\citep{2024ApJ...975..149M};		
the origin of slow rotators in astrometric binaries 
\citep{2024ApJ...975..153S};		
the relation between rotation and pulsation in Delta~Scuti stars
\citep{2024MNRAS.534.3022M};	
the identification of rapidly rotating red giants
\citep{2024MNRAS.528.3232P};	
and the location of twin binaries in the colour--magnitude diagram
\citep{2025A&A...693A.214H}.		

\subsection{Chromospheric activity}

The Sun's chromosphere lies above its photosphere and below the corona. It is seen as a thin ($\sim\!100$\,km) annulus above the lunar limb during solar eclipse, and its distinctive red colour is dominated by the H$\alpha$ 656\,nm transition. Spicules and prominences, bright features above the limb and dark features on the disk, provide evidence that the chromosphere is out of radiative equilibrium, with a temperature higher than at the top of the photosphere
\citep{1972ARA&A..10...73B}.	
There is a substantial literature on its complex physics
\citep{2019ARA&A..57..189C},		
and its role in coronal heating
\citep{2014LRSP...11....4R},		
and in driving the solar wind 
\citep{2019ARA&A..57..157C}.	
The phenomena collectively known as `solar activity' originate from the cyclic regeneration of the Sun's large-scale magnetic field \citep{2010LRSP....7....3C}, and it is in this context that variations in Ca\,{\scriptsize II} and H$\alpha$ are interpreted.
Similarly, strong and variable magnetic fields in cool stars with convective envelopes 
are driven by analogous stellar dynamos, themselves evidenced by photospheric starspots, chromospheric plages and coronal flares, and by Ca\,{\scriptsize II} and H$\alpha$ emission
\citep{1980ARA&A..18..439L,			
2017ARA&A..55..159L,		
2019ARA&A..57..189C}. 		
Chromospheric activity, characterised by H$\alpha$ and Ca\,{\scriptsize II} variability, thus contributes to understanding stellar magnetic fields, and the resulting variability as a function of stellar type, rotation and age.

Spectral lines used as chromospheric activity diagnostics probe different physical conditions 
\citep[][\S3.2]{2019ARA&A..57..189C}. 		
The H$\alpha$ line is the `classical' diagnostic: the lower level of the ($n=3\rightarrow2$) transition is at an excitation energy of 10.2\,eV, meaning that the opacity and line width is temperature dependent, although with large thermal broadening.
The Ca\,{\scriptsize II} H~and~K lines (396.8/393.4~nm) have the ground state as lower level, and trace the dominant ionisation stage for $T\lesssim13\,000$\,K.
The parameter $R^\prime_{\rm HK}$, relating line emission to bolometric luminosity, facilitates comparison across spectral types
\citep[][\S\,II]{1984ApJ...279..763N}.	
The Ca infrared triplet (850--860\,nm) has metastable lower levels, resulting in temperature-dependent opacities, and with good sensitivity to magnetic field strength.
There is a substantial literature on spectroscopic surveys of chromospheric activity, with numbers boosted by, for example, recent catalogues from LAMOST in these three different lines, including 
a study of 1.1~million solar-like stars using the Ca\,{\scriptsize II} HK lines
\citep{2024A&A...688A..23Z},	
a survey of 560\,000 FGK stars using the Ca~triplet
\citep{2024ApJS..272....6H},	
and 
220\,000 G-type stars showing excess chromospheric activity in H$\alpha$
\citep{2024ApJS..271...60S}.		
%

Gaia's contribution is coming from both the classification of solar-like variability, and the effects of flares and rotating starspots, obtained as part of the wider photometric variability classification undertaken as part of DR3
\citep{2023A&A...674A..14R}, 
and specifically focusing on magnetically active stars, initially in the case of DR2
\citep{2018A&A...616A..16L}, 	
and further refined for DR3 
\citep{2023A&A...674A..20D}.	
The latter analysed a subset of 30~million late-type stars, and derived a catalogue of 474\,000 stars with variability induced by magnetic activity, of which 430\,000 are newly discovered variables. For each, their catalogue gives the rotation period, the $G$~band amplitude of rotational modulation, and the correlation between magnitude and colour variation. They observed three broad clusterings in the amplitude--period diagram: High-Amplitude-Rotators (HAR), Low-Amplitude-Slow-Rotators (LASR), and a new class of Low-Amplitude-Fast-Rotators (LAFR) not seen, for example, in the Kepler data. They concluded that the HAR are dominated by long-lived dark spots, while the LAFR are dominated by bright faculae, characterised by rapidly variable magnetic fields. The LASR show a moderate correlation between brightness and colour variations, implying that dark spots are still the main cause of variability, but that faculae uncorrelated with the spots attenuate the correlation. They interpret the HAR--LAFR gap as evidence for a rapid transition between different magnetic configurations.

The wavelength interval covered by the RVS (845--872\,nm) includes, by design, the Ca\,{\scriptsize II} infrared triplet (Section~\ref{sec:rvs-wavelength}). And it is from these that other insights into chromospheric activity are being derived.
Here, the first stage of the RVS processing employs the `Astrophysical Parameters Inference System', Apsis
(Section~\ref{sec:classification-stellar-properties}).
Of its 13~modules, the Extended Stellar Parametrizer for Cool Stars (ESP--CS) computes a chromospheric activity index by comparing the observed RVS spectrum with a purely radiative equilibrium model. 
\citet{2023A&A...674A..30L}	
then used the excess equivalent width in the triplet cores ($\Delta\lambda=\pm0.15$\,nm) as a measure of chromospheric activity. And, in analogy with the $R^\prime_{\rm HK}$ activity index, they derived a similar $R^\prime_{\rm IRT}$ index. Their resulting catalogue provides an activity index for two million stars. They identified three regimes of chromospheric activity, broadly consistent with previous work on much smaller $R^\prime_{\rm HK}$ datasets. 
The highest activity regime is populated by pre-main-sequence stars (where the excess flux is dominated by mass accretion), and close binary systems (in which magnetic activity may be enhanced by tidal interaction).
Stars with 3500--5000\,K are either very active pre-main-sequence stars, or active main sequence stars with a unimodal activity distribution. 
A dramatic change in activity is found for $T_{\rm eff}<3500$\,K, with a dominance of low-activity stars close to the transition between partially- and fully-convective stars, and a rise in activity into the fully-convective regime.

The Gaia data are, in summary, allowing activity to be characterised as a function of stellar parameters with unprecedented detail, outlining different regimes of chromospheric heating, as well as identifying systems for which emission resulting from mass accretion may dominate.

\subsection{Masses}
\label{sec:stellar-masses}

Masses are one of the most fundamental of stellar properties, crucial in determining their structure and evolution, and everything that follows from them. The few accurate masses that are available underpin the calibration of the model-dependent methods of isochrone or stellar track fitting, which employ models of stellar structure and evolution matched to the observed star properties. Yet ways of determining accurate star masses are limited.
In their 1991 review,
\citet{1991A&ARv...3...91A}, 
listed some 50 accurate masses and, even today, only a couple of hundred are {\it measured\/} to better than 1--2\%
\citep{2021A&ARv..29....4S}.\footnote{
Progress in mass determination in the past four decades is encapsulated in four major reviews.
%
\citet{1980ARA&A..18..115P} gave masses for $\sim$100 eclipsing and visual binary components known to better than 20\% (requiring $\sigma_\varpi/\varpi<7$\%).
\citet{1991A&ARv...3...91A} gave masses, again better than 20\%, for just 45 detached double-lined eclipsing binary systems (90 single stars), covering spectral types O8--M1 on the main sequence, and with two red giants. Compared to \citet{1980ARA&A..18..115P}, data for only 6~systems remained unchanged; improved data were given for 18 systems, and 21 systems were new additions.
\citet{2010A&ARv..18...67T} listed 95 detached binaries containing 190 stars (comprising 94~eclipsing systems, and the astrometric binary $\alpha$~Cen) for which masses {\it and\/} radii were estimated to better than 2--3\%, along with 23 other systems with accurate masses but less accurate radii. 
\citet{2021A&ARv..29....4S} provide the most recent review of mass-determination methods. Their resulting compilation includes masses for 40~detached eclipsing binaries (80 stars) better than 2\% (their Table~2), and for 36~visual binaries (72~stars) better than 3\% (their Table~3).}
Asteroseismology has opened a new, powerful, but nonetheless still model-dependent approach (Section~\ref{sec:asteroseismology}), while other methods can be applied to microlensed systems (Section~\ref{sec:microlensing}), or to stars in specific evolutionary stages.
\citet{2021A&ARv..29....4S}	
have reviewed the various methods that can be used to estimate stellar masses across the Hertzsprung--Russell diagram, shown schematically as a function of mass and current accuracies achieved in Figure~\ref{fig:spectroscopic-binaries}b.
They list just 200 or so stars with relative mass accuracies between 0.3--2\% over the range $0.1-16M_\Sun$, of which 75\% are main-sequence core H-burning, and the remainder cover all other stages.

The most accurate (model-independent) method employs Kepler's third law applied to detached (non-interacting) binary systems. In such a system, each star moves in a closed elliptical orbit, with the centre of mass at one focus. Such a Keplerian orbit in three dimensions is described by 7~parameters: $a,e,P,t_{\rm p},i,\Omega,\omega$, where 
$a$ and $e$ specify the size and shape of the orbit,
$P$~is related to $a$ and the component masses through \mbox{Kepler's} third law, and
$t_{\rm p}$~is the position of the object along its orbit at a particular reference time. 
The three angles ($i,\Omega,\omega$) represent the projection of the true orbit into the observed (apparent) orbit; they depend solely on the orientation of the observer with respect to the orbit. 
For single-lined spectroscopic binaries, radial velocity measures provide the mass function, $f=(M_2\sin i)^3/(M_1+M_2)^2$.  For double-lined spectroscopic binaries, the mass function for both can be established, and hence the mass ratio, but still neither the individual masses, nor the orbit inclination. 
From radial velocities alone, masses can only be determined unambiguously if the system is eclipsing, imposing an explicit value for the inclination, viz.\ $i\simeq0$.

The addition of astrometric measurements changes prospects significantly, because all seven orbit elements become accessible in principle. In practice, different considerations apply according to whether the system is a `visual binary' or an unresolved `astrometric binary', whether the measurements probe all or part of the orbital motion of one or both components, or only that of the photocentre, and whether the orbital solutions also take account of constraints from radial velocity data 
\citep[e.g.][]{2004ASPC..318..123T, 	
2022AJ....163..118A}.			
For example, an orbit solution from the astrometry of (resolved) visual binaries yields the orbit inclination, and hence individual masses. 
But suitable astrometric measurements have not been easily obtained in the past, often calling for speckle or interferometric measurements over many years. 
Over its 3-year mission, Hipparcos provided just 235 orbital solutions for unresolved astrometric binaries
\citep{1997A&A...323L..53L}, 
of which mass ratios could be obtained for about 25 
\citep{1999A&A...341..121S}.

Turning to Gaia, the binary and multiple star processing is particularly complex, dependent on the wide range of systems (orbital period, magnitude difference, variability, etc.), and on the various combinations of data that can be used in the orbit solution (astrometry, photometry, and RVS velocities). 
\citet{2023A&A...674A..34G}
identified 800\,000 binaries in DR3 with orbit or trend parameters, classified as astrometric, spectroscopic, and eclipsing, in various combinations (their Table~1). 
Of these, 165\,500 are astrometric solutions characterising the orbit of the {\it photocentre}, yielding the system's parallax and proper motion, the orbit inclination and its standard error 
(\citet[][Eqs~A6 and A20]{2023A&A...674A...9H}), 
and the `astrometric mass function', which depends on the component fluxes, $F_1$ and $F_2$
(see \citet[][Eq.~2]{2023A&A...674A..34G} and 
\citet[][Eq.~14]{2023A&A...674A...9H}. 
Again, the degeneracies can be broken for eclipsing binaries, or for SB2 spectroscopic binaries.
I am not aware of any synthesis of the best masses available from DR3, but results include several astrometric orbits with $\sin^3 i$ better than 1\%, with the best masses at the level of around~0.3\% 
\citep{2023sf2a.conf..457H}. 

Radial velocities from Gaia's RVS spectrometer are not at the accuracies required for the best mass determinations, and state-of-the-art masses exploiting the Gaia data still largely also rely on ground-based radial velocity measurements. Thus 
\citet{2023A&A...678A..19C}	
combined DR3 astrometry with SB2 data from the Ninth Catalogue of Spectroscopic Binary Orbits (SB9), and APOGEE, to determine masses for 56 systems (43 from SB9, and 13 from APOGEE), and provided an empirical mass--luminosity relation down to $0.12M_\Sun$ (Figur~\ref{fig:spectroscopic-binaries}c).
Similarly, ground-based campaigns are ongoing to acquire radial velocities (along with interferometric or speckle data in some cases) targeting masses at 1\% accuracy when eventually combined with future Gaia astrometry. Amongst these are 70~binaries being observed with OHP--SOPHIE
\citep{2014MNRAS.445.2371H, 	
2016MNRAS.458.3272K, 			
2020MNRAS.496.1355H}, 		
and the follow-up of other suspected SB2s from Gaia 
\citep{2023sf2a.conf..191H}.
Other works are using these fundamental masses, combined with theoretical stellar evolution models, to estimate stellar masses from observed luminosities, based on Gaia $G$-band magnitudes and stellar distances 
\citep[e.g.][]{2020A&A...642A..88L,		
2022Ap&SS.367...37M,	
2023A&A...678A..19C,	
2024PARep...2...41E,	
2024A&A...684A..74M,	
2024A&A...682A..12P}.	

To my knowledge, no results have been published yet from Gaia's resolved orbital binaries, for which masses should also be available.  Data Release~4, in late 2026, will also cover a longer time interval, and will include astrometry at each measurement epoch. At this time, I can offer no useful insight into the expected state of stellar mass determination at the end of the Gaia mission.

\subsection{Initial mass function}

The initial mass function, IMF, is an empirical description of the {\it initial\/} mass distribution of a population of stars resulting from the processes of star formation. It provides direct constraints on
theories of star formation 
\citep[e.g.][]{2014PhR...539...49K},
including the first stars in the Universe 
\citep[e.g.][]{2002Sci...295...93A,
2002ApJ...564...23B,
2009Natur.459...49B},
and on galaxy formation and evolution more generally 
\citep[e.g][]{2009MNRAS.400.1347C,2021MNRAS.508.3226A}.
Early studies by 
\citet{1955ApJ...121..161S}
found that the mass distribution of high mass ($\gtrsim1$ M$_\odot$) stars in the solar neighbourhood is well described by a power-law $N(M) \propto M^{\alpha}$, with slope $\alpha=-2.35$. 
Evidence that the mass function for lower-mass stars in the Galactic disk increases less strongly with decreasing mass has led to the wider use of alternative forms, notably  
the broken power law of
\citet{2001MNRAS.322..231K},
and the log--normal form of
\citet{2003PASP..115..763C}.
Predictions of the form of the IMF are based on theories of star formation. 
In the general picture where star formation results from the fragmentation and competitive accretion in proto-stellar clouds,  detailed models suggest an IMF dependent on chemical composition, and processes such as collisional excitation and dust cooling
\citep[e.g.][]{
1978MNRAS.184...69L,
2001MNRAS.324..573B,
2008ApJ...684..395H,
2025MNRAS.536.1932M}.	
In an alternative formulation, stars self-regulate their masses, balancing accretion and radiative feedback, in which case the IMF is influenced by additional processes dependent on (for example) density, angular momentum, and metallicity
\citep[e.g.][]{
1996ApJ...468..586A,		
2024MNRAS.530.2453C}.		
Observations have resulted in a confusing picture, with debate over the exact form of the IMF, and how it might apply to the first stars which formed in the Universe.
Neither is there agreement as to whether the IMF is universal
\citep[e.g.][]{1989AN....310..127Z,
2002Sci...295...82K,
2004ApJ...604..579P,
2012A&A...545A.147H}, 
or whether it varies across star-forming regions or Galactic environment 
\citep{2007A&A...467..117B,
2008ApJ...675..163H,
2013ApJ...771...29G,
2014MNRAS.444.1957D, 
2017MNRAS.468..319E,
2023MNRAS.526.4004C}.	
In some cases, the IMF has been found to be
`bottom-heavy' in low-mass stars 
\citep[e.g.][]{2010Natur.468..940V,2012Natur.484..485C},
but `top-heavy' in others 
\citep{2022A&A...664A..26P,
2023MNRAS.521.3991B}. 
In their 2010 review, \citet{2010ARA&A..48..339B} 
concluded that {\it `There is no clear evidence that the IMF varies strongly and systematically as a function of initial conditions after the first few generations of stars'}.
A more recent review, including some of the early insights from Gaia, is given by 
\citet{2024arXiv241007311K}.	

Observational characterisation of the IMF of a stellar population starts with a determination of the distribution of {\it observed\/} stellar masses, viz.\ the `present-day mass function'.  But any initial mass distribution is subsequently modified by two main processes, which must be accounted for by models. Stellar evolution results in a present-day mass function depleted in high-mass stars with respect to the IMF, depending on age and star-formation history. In contrast, dynamical evolution resulting from star--star interactions results in a population progressively depleted at its low-mass end. 
Neither is the determination of the {\it present-day\/} mass function a simple task: since stellar masses can generally only be determined by the application of Kepler's third law to a small number (some hundreds) of suitable binary systems (Section~\ref{sec:stellar-masses}), masses are generally estimated from their absolute magnitudes based on theoretical isochrones 
\citep[e.g.][]{2003PASP..115..763C}. 
Absolute magnitudes require stellar distances, which is where Gaia is contributing.
In the past, and today with Gaia, populations widely used for determining the present-day mass function, and hence the IMF, include the Galaxy's disk population as represented by the local solar neighbourhood, and the use of open clusters. Models must account for variations in metallicity and star-formation rate with Galactic latitude,
the increasing contamination of the thick disk with time,
unresolved multiple systems resulting in erroneously bright magnitudes,
and variable extinction. 
A very different approach to constraining the IMF based on transients of stellar origin, and giving comparable estimates of the high-mass slope, is detailed by 
\citet{2024Univ...10..383G}.	

Turning to Gaia, using DR2 magnitudes and parallaxes, 
\citet{2019A&A...624L...1M} 
derived the star-formation history of the solar neighbourhood, and represented the IMF using a three-segment power-law with breaks at fixed masses at $0.5M_{\odot}$ and $1.53M_{\odot}$.
\citet{2019MNRAS.489.2377S} 
used a sample of 120\,000 solar neighbourhood stars with parallaxes, magnitudes and colours also from Gaia DR2, accounting for the population of unresolved binaries, the metallicity distribution, the star formation history, and their variation across the Galactic disk. He found an IMF well represented by a segmented power-law, with a maximum at $M\sim0.15~M_{\odot}$, significant flattening at lower masses, and a slope $\alpha=-1.34\pm0.07$ in the range $0.25-1M_\odot$ (in broad agreement with previous work). Above $1M_{\odot}$ it shows an abrupt decline, with a slope in the range $\alpha=-2.68$ to $\alpha=-2.41$, depending on the inferred star formation history. 
\citet{2021MNRAS.507..398H} 
determined the IMF in the range $0.2-1M_\odot$, and within 250\,pc, according to kinematics and metallicity. The dominant thin-disk population (transverse velocities $v_T<40$\kms) gave an IMF similar to other estimates, described by a broken power law with $\alpha=-2.03$ for $M>0.5M_\odot$, decreasing to $\alpha=-1.34$  for $M\lesssim0.5_\odot$. 
Thick-disk stars ($v_T=60-150$\kms) and those of the high-metallicity `red-sequence' halo, have a similar low-mass slope $\alpha=-1.14$, but steeper at high-mass, \mbox{$\alpha=-2.35$}. The low-metallicity `blue sequence' halo stars have a distinct, bottom-heavy IMF, described by a single power law with $\alpha=-1.82$ over most of the mass range.  They found an IMF of the low-metallicity halo that is similar to the Salpeter-like IMF of early-type galaxies, a stellar population that, like the halo stars, has a high $\alpha$/Fe ratio. Blue-sequence stars, they concluded, are likely the debris from accretion, $\sim$10\,Gyr ago, of the Gaia Enceladus dwarf galaxy (Section~\ref{sec:gse}). 
\citet{2023Natur.613..460L} 
used 93\,000 spectroscopically observed M-dwarfs out to 300~pc (using LAMOST and Gaia DR2), to find a variable IMF that {\it does\/} depend on both metallicity and age. Specifically, the oldest population contains fewer low-mass stars, independent of metallicity. In younger stars, the proportion of low-mass stars increases with metallicity.

Further studies of the IMF in the solar neighbourhood have been made with Gaia EDR3/DR3. 
Using 7~million Gaia DR3 stars with $G<13$,
\citet{2025A&A...697A.128D}	
derived both the local star-formation history, and the IMF, by matches to the Besançon population synthesis model (Section~\ref{sec:besancon-model}). As well as identifying various abrupt changes in the star formation rate (e.g.\ at 1--1.5\,Gyr and 5--7\,Gyr ago), they found (for example) $\alpha=1.45$ in the range $0.5-1.53M_\Sun$, and $\alpha=1.50$ for $M>1.53M_\Sun$. Fits to all-sky Gaia data up to $G=14-17$ are planned.
And, using EDR3,
\citet{2024ApJS..271...55K}	
constructed a sample complete to 20\,pc, and extending into the sub-stellar ($>5M_\Jupiter$) regime. They proposed a 4-component power law, consistent with the long-established values of $\alpha=2.3$ at high masses ($0.55-8M_\Sun$) and $\alpha=1.3$ at intermediate masses ($0.22-0.55M_\Sun$), but dropping to 
$\alpha=0.25$ for $0.05-0.22M_\Sun$ and $\alpha=0.6$ for $0.01-0.05M_\Sun$. This, they conclude, implies that the rate of production diminishes in the low-mass star/high-mass brown dwarf regime, before increasing again in the low-mass brown dwarf regime.

Several Gaia EDR3/DR3 studies have focused on the IMF derived specifically for open clusters.
\citet{2022MNRAS.516.5637E}	
considered 15 nearby clusters, finding a global {\it present-day\/} mass function consistent with a single power-law, with slopes in the range --0.6 to --3 extending from sub-solar to super-solar masses, with a significant correlation between the slope, and the ratio of age to half-mass relaxation time. Their Monte Carlo simulations suggest that all the open clusters analysed could be born with an IMF with slope $\alpha<-2.3$. 
\citet{2024MNRAS.530.4970Y}	
examined the `snake-like' elongation of open clusters (Section~\ref{sec:strings-snakes-pearls}), finding systematic variations in the mass functions along the length of the `snake', and with its `head' conforming to a canonical IMF with $\alpha\sim-2.3$, providing evidence for the delayed formation of massive stars. 
\citet{2024ApJ...966..169P}	
determined the present-day mass function of 93 clusters using Gaia~DR3, finding a single value of $\alpha$ for clusters younger than 200\,Myr, decreasing for older clusters, together consistent with a dynamically evolved Kroupa IMF via the loss of low-mass stars.

\subsection{Distance scale}
\label{sec:distance-scale}

Measuring distances across the vast scale of our Galaxy and beyond has long been a central problem in astronomy. Only for objects within a few tens of parsecs was direct distance determination, through parallax measurements, possible from the ground. Beyond that, estimates had to rely on a distance scale `ladder', constructed from a sequence of indirect and often uncertain measurements relating the closest objects to increasingly distant ones.  
Pre-Hipparcos, even the distance to the Hyades cluster, at around 50~pc, and one of the first `rungs' on the distance `ladder', was still not considered fully secure. That changed with Hipparcos 
\citep{1998A&A...331...81P},
and with it the focus shifted to discussions of the distance to the Galactic centre, and to the Large Magellanic Cloud.
At that time, various `standard candles' were used as luminosity calibrators to estimate the distance to the LMC:
(a)~Population~I objects, including Cepheids, red clump giants, Mira variables, eclipsing binaries, SN~1987a, and main-sequence fitting (and turn-off). Cepheid methods were themselves subdivided into those based on trigonometric parallaxes, main-sequence fitting of open clusters, and the Baade--Wesselink method;
(b)~Population~II objects, including subdwarf main-sequence fitting, horizontal branch parallaxes, RR~Lyrae (pulsationally-unstable horizontal branch stars), globular cluster dynamics, and cooling sequences of white dwarfs in globular clusters. RR~Lyrae were themselves subdivided into the use of statistical parallaxes, Baade--Wesselink method, and double-mode pulsators.
Hipparcos contributed to most of these, except those based on eclipsing binaries, SN~1987a, and globular cluster dynamics. Particular insight came from red clump giants, and globular cluster distances and age estimates from subdwarf main-sequence fitting.

I mention all this in part to recall the insecure status of distance determinations pre-Gaia. But it is also relevant in the context of future steps in improving the accuracy of $H_0$. \citet{2020svos.conf..215C} have suggested that securing the `late Universe' estimates requires: 
(a)~further calibration of the Cepheid and RR~Lyrae distance scales;
(b)~continued investigation of the Gaia systematics and parallax zero-point;
(c)~the development of fine grids of nonlinear convective models for pulsating stars;
(d)~an improved treatment of population effects in Cepheids, RR~Lyrae, long-period variables, and red giant branch stars; and 
(e)~verification of the consistency between the Population~I and Population~II distance scales in the Local Group.

\begin{figure}[t]
\centering
\includegraphics[width=0.30\linewidth]{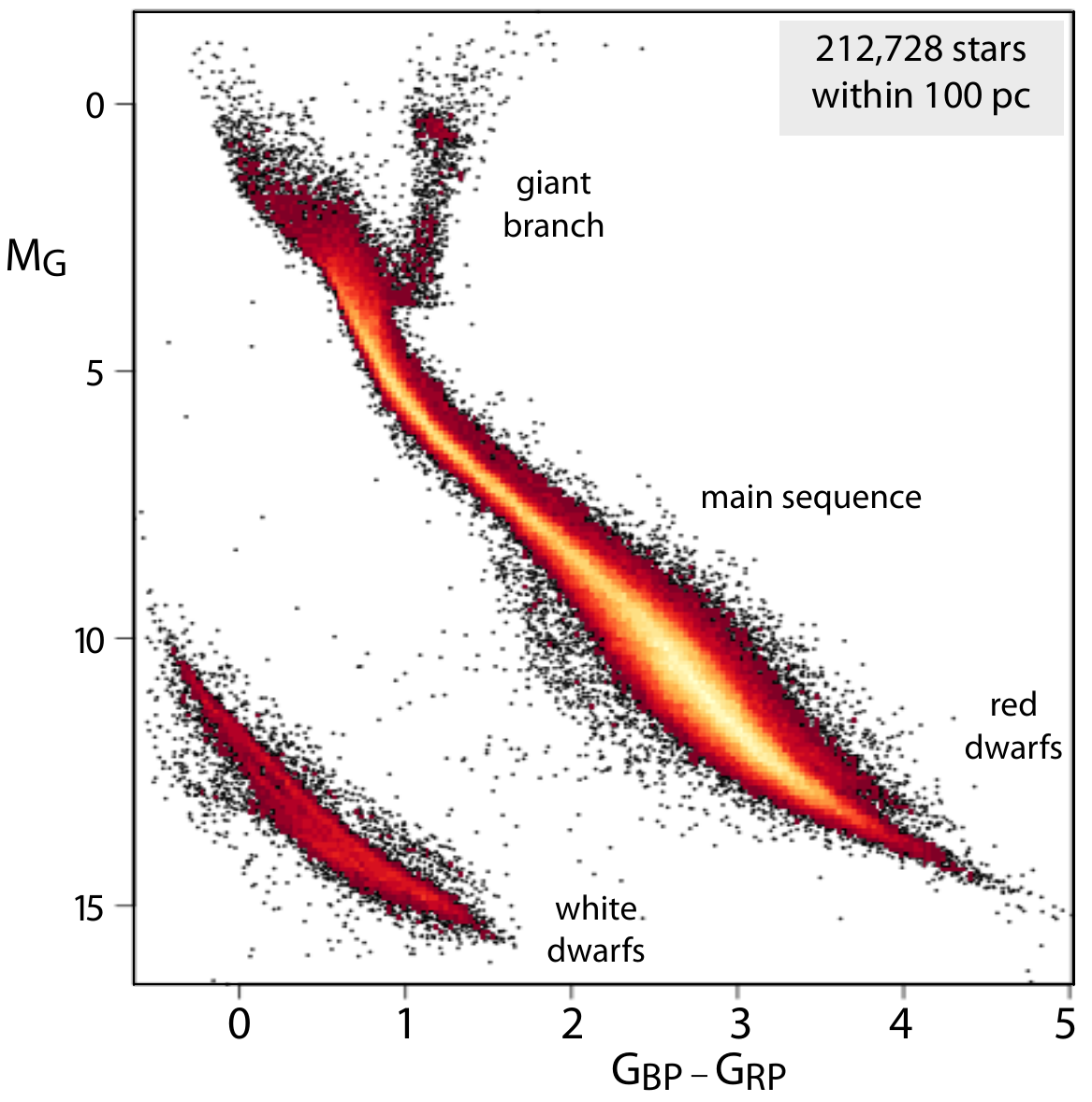}
\hspace{10pt}
\includegraphics[width=0.325\linewidth]{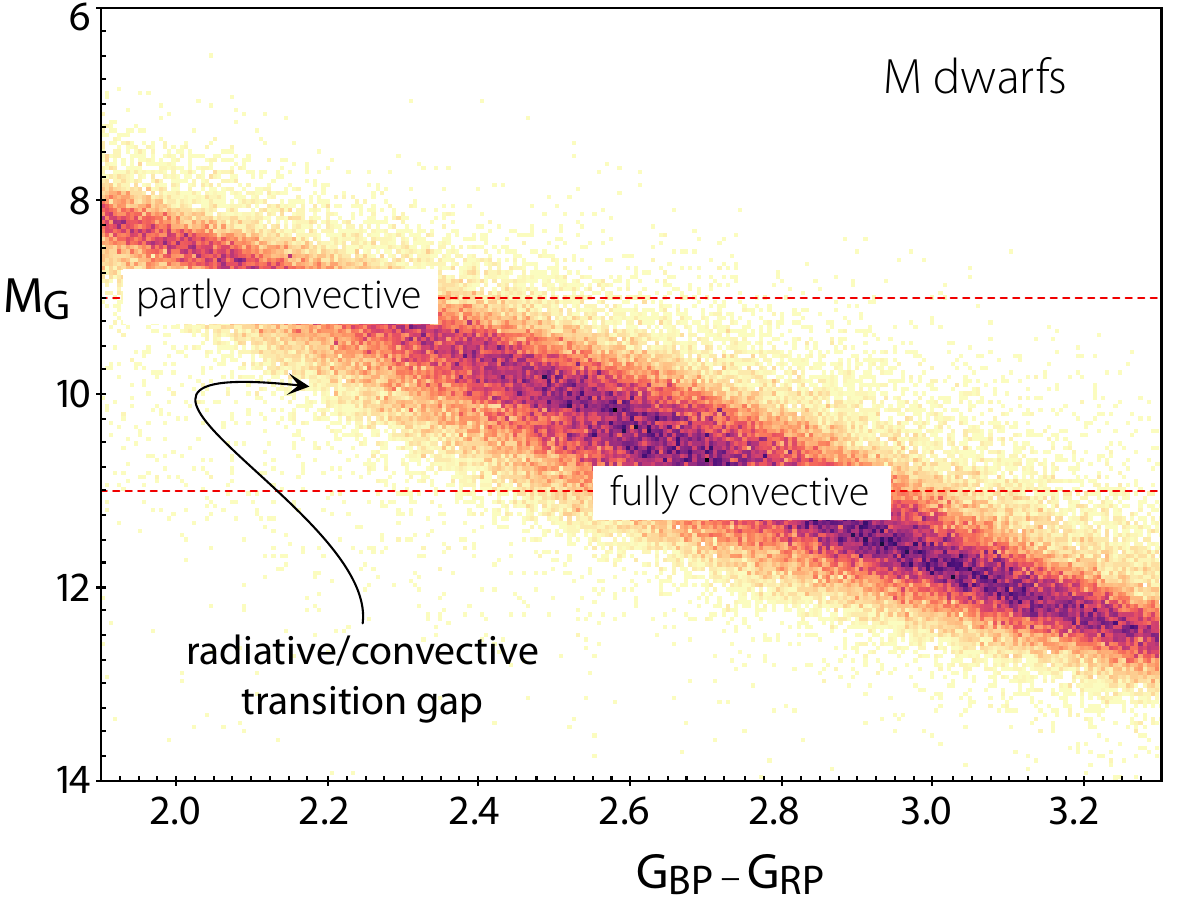}
\hspace{10pt}
\includegraphics[width=0.30\linewidth]{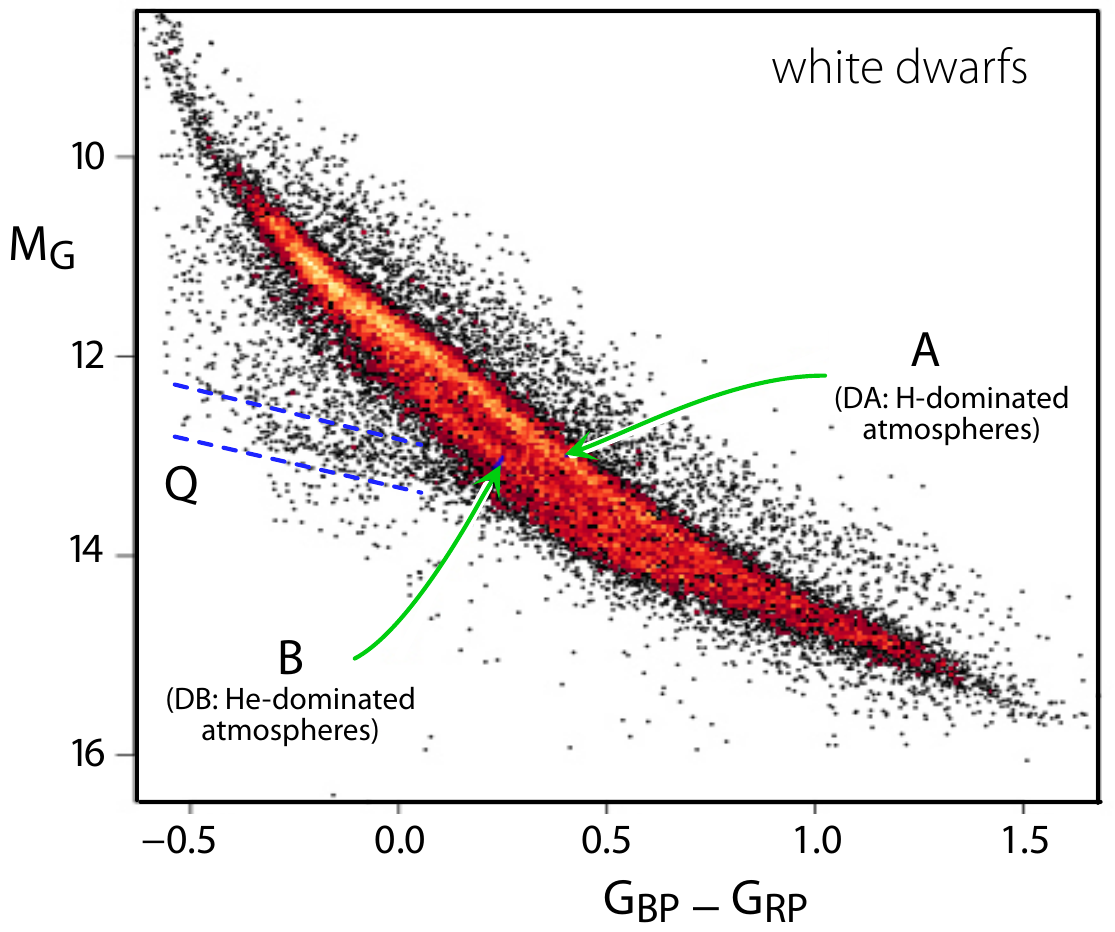}
\caption{Left: solar neighbourhood colour--magnitude diagram for the 212\,728 stars with Gaia DR2 parallaxes $\varpi>10$~mas (\citet{2018A&A...616A..10G}).
Centre: the discontinuity at the location of the radiative--convective boundary of M~dwarfs, due to non-equilibrium burning and mixing of $^3$He in the stellar core (\citet{2018ApJ...861L..11J}).
Right: the white dwarf sequence of the 100~pc white dwarf sample, showing both the prominent bifurcation in the H- and He-dominated atmospheres, and the discontinuity (Q) attributed to core crystallisation (\citet{2018A&A...616A..10G}). 
}
\label{fig:babusiaux-cmd}
\end{figure}

\subsection{Hertzsprung--Russell diagram}
\label{sec:hr-diagram}

The strictly defined census of stars within the solar neighbourhood, including the CNS5 within 25~pc, and the Gaia Catalogue of Nearby Stars within 100~pc, provides for the accurate placement of large numbers of stars in the colour--magnitude ($M_{\rm G}$ versus $G_{\rm BP}-G_{\rm RP}$) analogue of the Hertzsprung--Russell diagram. 
An early demonstration of the fidelity of Gaia's astrometry and photometry was in the DR2-based study of more than 200\,000 stars within 100\,pc 
\citep{2018A&A...616A..10G},	
which presented three striking new features (Figure~\ref{fig:babusiaux-cmd}a).
The first is a clear discontinuity at the location of the radiative--convective boundary of M~dwarfs, today also known as the Jao gap (Figure~\ref{fig:babusiaux-cmd}b), and explained in terms of the non-equilibrium burning and mixing of $^3$He in the stellar core 
\citep{2018MNRAS.480.1711M}.	
The gap arises because convection in the cores of low-mass stars helps mix the intermediate nuclear fusion products, allowing the star to fuse hydrogen more efficiently.  Detailed models 
\citep{2018MNRAS.480.1711M,
2020AJ....160..102J,
2021ApJ...907...53F,
2023ApJ...944..129B,
2023AJ....166...63J}
show that the mixing of $^3$He during the merger of the envelope and core convection zones occurs over a narrow range of masses. This successfully replicates an associated dip in the luminosity function which is responsible for the gap.

The second is the prominent bifurcation in the H- and He-dominated atmospheres of the white dwarf sequence (Figure~\ref{fig:babusiaux-cmd}c). 
Another feature evident in the same sample of white dwarfs, and addressed in several follow-up papers, is a discontinuity in the white dwarf sequence now attributed to ongoing core crystallisation as the white dwarfs cool.
I say more on these two prominent white dwarf features in Section~\ref{sec:white-dwarfs}.

Some other general features of the Hertzsprung--Russell diagram for stars within 100~pc are discussed by  
\citet{2018A&A...616A..10G}.
But deeper insights, and comparisons with stellar evolutionary models, generally follow from the studies of individual open clusters (Section~\ref{sec:open-clusters}), and specific evolutionary states or object classes (white dwarfs, carbon stars, Wolf--Rayet stars, planetary nebulae, supernovae, etc.) which I consider further in Section~\ref{sec:specific-evolutionary-states}.

\subsection{Modelling stellar evolution}
\label{sec:stellar-evolution}

Intimately related to the interpretation of the observational Hertzsprung--Russell diagram is the work ongoing to develop and improve stellar evolutionary models. These aim to reflect the observed distributions of stellar properties which are driven by changes in chemical composition as a result of nuclear reactions. And while the chemical composition, temperature and pressure deep in stellar interiors are out of reach of direct investigation, their observational consequences in terms of stellar luminosity, radii, masses, surface temperature and chemical composition, seismological oscillations, Galactic chemical enrichment, and many others, are nonetheless accessible. 

Development of the numerical codes to calculate evolutionary models began more than 60~years ago with the pioneering work of 
\citet{1958ses..book.....S}
and
\citet{1959ApJ...129..628H}. 
In the decades since, improving models, recent advances in software infrastructure, processing power, and data storage, and substantial observational advances, have led to a progressively deeper understanding of the complex physical processes that occur during the various evolutionary stages. Today, widely cited codes for low- and intermediate-mass stars include ATON, BaSTI, CESAM, DSEP, GARSTEC, GENEC, MESA, Monash, PARSEC, STARS, and YREC. 
Amongst these, MESA (`Modules for Experiments in Stellar Astrophysics') is `open-knowledge' software, with its latest treatments of convection, opacities, nuclear reaction rates, core crystallisation, and starspots described by
\citet{2023ApJS..265...15J}.	
Other recent computational advances include the ongoing development of 3d atmosphere models
\citep[e.g.][]{2019ApJ...882...63J,		
2024Galax..12...87R}.				

Stellar models are constructed by solving certain basic equations: conservation of mass and of energy, hydrostatic equilibrium, and energy transport via radiation, convection, and conduction. These coupled differential equations have explicit boundary conditions at the stellar centre and at the surface. Correctly describing the stellar plasma requires extensive inputs from nuclear and particle physics (e.g.\ opacities and nuclear reaction rates), atomic and molecular physics, thermodynamics, hydrodynamics, and radiative transfer.  
Observational data to constrain these models include bolometric luminosities (from magnitudes, distances, and bolometric corrections), surface chemical composition (from spectroscopy), mass (most rigorously for a small number of favourable binary systems), $T_{\rm eff}$, radius, and oscillation frequencies. Evidently, Gaia is contributing substantially to most of these observational constraints.

Various physical processes are still considered to be poorly or imprecisely known, and not fully accounted for in current models. These include the effects of rotation and rotationally-induced mixing, magnetic fields, tidal effects, gravity waves, and mass loss
\citep{
2019arXiv190210399B,	
2019arXiv190108809D,	
2021FrASS...7...96X},	
certain nuclear reaction rates and cross sections
\citep{2022JPhG...49k0502S},	
and the many complexities of common envelope binary evolution
\citep[e.g.][]{2021MNRAS.507..385H,	
2022MNRAS.512.5462L}.		

I will touch on some of these specific Gaia-based modelling efforts in the relevant parts of this review (e.g.\ for the Hyades cluster, Section~\ref{sec:hyades}). Here, I will only note that Gaia data is being used as tests of models such as
PARSEC \citep{2012MNRAS.427..127B,
2018MNRAS.476..496F},	
COLIBRI for pulsating stars 
\citep{2013MNRAS.434..488M,
2017ApJ...835...77M,	
2019MNRAS.482..929T},	
MESA
\citep{2016ApJ...823..102C,	
2016ApJS..222....8D},		
and
BaSTI
\citep{2024MNRAS.527.2065P}.	

Alongside these developments,
\citet{2025ApJ...979...92W}	
compared the Gaia colour--magnitude diagrams and theoretical model isochrone predictions for three nearby clusters (Hyades, Pleiades, and Praesepe), for both ($G_{\rm BP}-G_{\rm RP}$) and ($G-G_{\rm RP}$), providing empirical colour-corrections for both MESA and PARSEC isochrone models. Applying these correction functions to 31 additional open clusters, and three moving groups, resulted in a significantly improved alignment between the isochrones and observed colour-magnitude diagrams, yielding isochronal ages more consistent with literature values based on the lithium depletion boundary method (Section~\ref{sec:cluster-ages}). 

\paragraph{Convection}

Convection remains a major source of modeling uncertainty which, for certain evolutionary stages, can propagate through to uncertainties in ages and, in turn, understanding the chemical evolution of the Galaxy 
\citep{2000ARA&A..38...35L}.
Most stellar models continue to treat convection according to the so-called `mixing-length theory' 
\citep{2023Galax..11...75J}.	
First applied to stellar interiors in an influential paper on solar convection by 
\citet{1958ZA.....46..108B},
it remains the dominant framework for 1d convective energy transport calculations.
In this picture, a hot `parcel' of fluid within a convection zone, with locally uniform physical characteristics, and in pressure (but not thermal) equilibrium with its surroundings, will rise and expand towards cooler regions. The radial distance over which it travels before losing its defining characteristics is taken as its mean-free path, $\alpha_{\rm ML}$, a dimensionless number measured in terms of the pressure scale-height of the stratified fluid.
`Overshooting', which describes the penetration of convection and mixing beyond the classical Schwarzschild convection cores, further modifies the evolution for $M\gtrsim1.2M_\odot$.

Main-sequence stars are typically modelled over three mass ranges 
\citep{2023Galax..11...75J}: 
$M\lesssim0.5_\odot$ (fully convective);
$M=0.5-1.2M_\odot$ (with radiative cores and convective envelopes);
and 
$M\gtrsim1.2M_\odot$ (with convective cores and radiative envelopes).
The choice of $\alpha_{\rm ML}$ is particularly important for stars with convective outer regions, $M\lesssim1.2M_\odot$, where it can dramatically affect age estimates, since it affects nuclear burning rates, and therefore the main sequence lifetime. Because of the steepness of the red giant branch in the HR diagram, for example, red giants can have very different inferred ages, by several Gyr, depending on the assumed mixing length. 

Testing models in this evolutionary regime has been difficult due to the lack of fundamental masses, and reliable spectroscopic measurements of temperature and detailed abundances for large samples of stars. Gaia’s accurate luminosities for very large numbers of red giants, covering a wide range of metallicity and age, is focusing renewed attention on the need to better characterise $\alpha_{\rm ML}$, and on developing improved models of its dependencies on \teff, $\log g$ and [Fe/H] 
\citep{2021MNRAS.504.3128S,
2023A&A...675A..26W}.	
I say more on progress in characterising $\alpha_{\rm ML}$, and the manifestation of the `convective kissing instability', in the case of open clusters more generally (Section~\ref{sec:clusters-evolution}), and in the case of the Hyades cluster in particular (Section~\ref{sec:hyades}). 
The modelling of convection appears implicitly across many studies, and more explicitly (for example) in  discussions of 
the mass--luminosity relation (Section~\ref{sec:stellar-masses}), 
the distance scale for classical Cepheids (Section~\ref{sec:cepheids-h0}),
and
asteroseismology (Section~\ref{sec:asteroseismology}).

\subsection{Specific evolutionary states}
\label{sec:specific-evolutionary-states}

\subsubsection{White dwarfs}
\label{sec:white-dwarfs}

\paragraph{Context}

White dwarfs are the endpoint of stellar evolution for $\sim$97\% of stars, specifically those below $8-10M_\Sun$, depending on metallicity 
\citep{2013ApJ...765L..43I, 2015ApJ...810...34W}. 
They result from the star fusing H into He (on the main sequence, and later on the red giant branch), then fusing He into C and O (first on the horizontal branch, and later on the asymptotic giant branch), thereafter losing their outer layers to leave only a dense C--O core, representing $\sim$99\% of the total mass.
The remaining 1\% is made of any H/He not fused during earlier evolutionary phases, arranged in a He envelope surrounding the core, and an outer H~envelope, with their thicknesses depending on details of the nuclear burning. The upper layers of the H/He envelope constitute the atmosphere, the only region observable.
Some 20--25\% of white dwarfs consume most of their H during previous evolutionary phases, resulting in H-deficient (and hence He-rich) atmospheres \citep{2005A&A...435..631A}.
Furthermore, if the progenitor is close to the $10M_\Sun$ upper mass limit, it may reach temperatures high enough to fuse C, leading to an O--Ne white dwarf. 

Although very common, white dwarfs have a very low intrinsic luminosity. As a result, reasonably bright objects are relatively nearby, such that even ground-based parallaxes, pre-Hipparcos, were already of reasonably high relative accuracy.  Conversely, more distant objects, beyond say 20--30\,pc, are fainter, and much harder to identify and measure.
Gaia is observing order 500\,000 white dwarfs in the solar neighbourhood, and is making a significant contribution to their understanding, including the mass--radius relation, physics of their cooling mechanisms, their asteroseismic properties, and mass-based inferences on stellar mergers.
The recent review by
\citet{2024NewAR..9901705T}, 
entitled `The Gaia white dwarf revolution' 
notes that more than 400 white dwarf studies had (already by early 2024) made critical use of Gaia data, establishing Gaia {\it `as a fundamental resource for white dwarf identification, fundamental parameter determination and more recently spectral type characterisation'}. 

To understand these Gaia studies, let me summarise a few relevant points:
(a)~masses of field white dwarfs are obtained only indirectly by combining surface fluxes (using distance and radius), spectroscopic \teff\ and $\log g$, evolutionary models, and by appeal to the appropriate mass--radius relationship \citep[e.g.][]{2022PhR...988....1S};
(b)~derived masses follow a narrow distribution peaking at $0.55-0.58M_\Sun$, similar for DA and DB dwarfs 
\citep{2019ApJ...876...67B,
2019MNRAS.482..649O,
2020MNRAS.499.1890M}.
60\% are in the range $0.5-0.7M_\Sun$, but a few extend to $0.2-1.35M_\Sun$ 
\citep{2019MNRAS.488.2892P};
(c)~very low-mass white dwarfs, $\lesssim0.3M_\Sun$ are believed to have formed by mass loss in binary systems \citep{1993PASP..105.1373I}.  The `ultra-massive' objects ($\gtrsim1.1M_\Sun$) are (generally) thought to have O--Ne cores \citep{2007A&A...476..893S};
%
(d)~white dwarfs evolve by radiating their residual thermal energy. The associated cooling can be divided into four luminosity-controlled stages: neutrino cooling (which dominates for the first 20\,Myr), fluid cooling, crystallisation, and Debye cooling.  Over time scales of order 10\,Gyr, it is controlled by the thermal conductivity of the core and envelope, the radiative opacity of its atmosphere, element transport in its interior, and the passage through phase transitions, further complicated by convection, overshooting, and sedimentation 
\citep[e.g.][]{2022FrASS...9....6I}.

Spectroscopic observations of white dwarfs often reveal strong evidence for the `pollution' of their atmospheres due to the infall of debris arising from the processes of exoplanet formation, which I consider further in Section~\ref{sec:exoplanets}. 
White dwarfs are also important for tests of  gravitational redshift (Section~\ref{sec:gravitational-redshift}), for placing constraints on any variation of the gravitational constant (Section~\ref{sec:g-dot}), and in
asteroseismic analyses (Section~\ref{sec:asteroseismology}).

\paragraph{Gaia samples}

The Gaia data have been used to define various white dwarf samples. The Gaia astrometry and photometry can then be used to construct associated colour--magnitude diagrams, to derive mass estimates, and so on.  There are of order 260\,000--500\,000 white dwarf candidates which have been identified by Gaia so far, an order-of-magnitude increase in sample size compared to the knowledge pre-Gaia, albeit with the majority of the more distant not yet having been confirmed spectroscopically 
\citep{2019MNRAS.482.4570G}. 
In statistical studies, there is a trade-off between sample size and completeness, and the samples become progressively more incomplete beyond 30--40\,pc. Quantifying completeness is greatly complicated, for example, by the presence of white dwarfs in binary systems.

Amongst Gaia-based samples variously referenced are
the DR2-based 20\,pc sample, containing 139 systems, nine of which are new detections, with the closest at a distance of 13\,pc \citep{2018MNRAS.480.3942H}.
The EDR3-based 40\,pc sample contains around 1000 candidates
\citep{2019MNRAS.482.4570G,2021MNRAS.508.3877G}, 
not all of which have been confirmed spectroscopically 
\citep{2023MNRAS.518.3055O,2020MNRAS.499.1890M,2020MNRAS.497..130T}. 
Well-defined catalogues to 100\,pc comprise around 13\,000 objects \citep[e.g.][]{2018MNRAS.480.4505J}.
%
A compilation of 531 candidates having large transverse motions relative to the Sun (above 200\kms), are likely to be members of the local Galactic halo population
\citep{2020ApJ...899...83K}.		
To give some idea of the activity in this field, these 40--100\,pc sample papers each have, by early 2025, around 200--400 refereed citations. 

\paragraph{The DA/DB bifurcation}

The bifurcation in the white dwarf sequence in the local colour--magnitude diagram 
(Section~\ref{sec:hr-diagram}, and Figure~\ref{fig:babusiaux-cmd}c) was not predicted by previous models, and has attracted much attention
\citep{2018ApJ...860L..17E,
2018MNRAS.480.4505J,
2019MNRAS.482.4570G}.
It is not simply a separation between H-dominated (DA) or He-dominated (DB) atmospheres, and now appears to be attributable to trace amounts of carbon in the atmospheres added by convective dredge-up
\citep{2019ApJ...876...67B,
2019A&A...623A.177S,
2020MNRAS.492.5003O,
2023MNRAS.523.3363B}.		
New evolutionary models including this effect have recently been presented
\citep{2023A&A...674A.213C}.		

\paragraph{Crystallisation}
\label{sec:white-dwarf-crystallisation}

The faint but significant inclined feature in the local colour--magnitude diagram 
(Section~\ref{sec:hr-diagram}, and labeled `Q' in Figure~\ref{fig:babusiaux-cmd}c)
was seen for the first time in the Gaia DR2 data
\citep{2018A&A...616A..10G},	
and has been subsequently attributed to ongoing core crystallisation as the white dwarfs cool
\citep{2019Natur.565..202T,		
2019ApJ...876...67B}.			
As independently predicted by Abrikosov, Kirzhnitz, and Salpeter in the early 1960s, their cores should slowly `crystallise' as they cool, resulting in a lattice rather than a gas. In this phase transition, the hot plasma fluid (of nuclei and electrons) releases an associated latent heat, providing a new source of energy that delays the object's cooling. There had only been indirect evidence for this theory pre-Gaia, although the details are crucial in estimating cluster ages from these `cooling curves'
\citep[e.g.][]{1987fbs..conf..363W}.
Gaia reveals the crystallisation as a mass-dependent pile-up in the Hertzsprung--Russell diagram as they spend time at this location while they release their latent heat. 
Apart from this statistical feature, there is only indirect evidence for crystallisation in a few specific objects and, again, Gaia is contributing to their understanding
\citep{1997ApJ...487L.191W,
2019A&A...632A.119C}.
Other physical effects that modify white dwarf cooling (including Debye cooling, and convective coupling) are currently included in some stellar models, but are not yet characterised observationally
\citep{2019ApJ...876...67B}.			
The Gaia data on white dwarfs thus provides direct evidence that a first-order phase transition occurs in high-density Coulomb plasmas (a theory that cannot be tested in terrestrial laboratories because of the extreme densities involved).  Meanwhile, it will be some 5\,Gyr until the Sun evolves into a white dwarf, and another 5\,Gyr before it cools enough to form a crystalline sphere.

\paragraph{Mass--radius relation}
\label{sec:white-dwarf-mass-radius}

For a degenerate electron gas at densities $10^6\!-\!10^8$\,gm\,cm$^{-3}$, the equation-of-state leads to a mass--radius relation first derived by 
\citet{1931MNRAS..91..456C},
subsequently refined based on models of degenerate white dwarfs of different chemical composition and envelopes
\citep{1961ApJ...134..683H,
1990PASP..102..954W,
1995LNP...443...41W}.
Although the mass--radius relation remains a largely theoretical construct, it forms an underlying assumption in many related studies.  It enters determination of their masses and luminosity function, and hence a range of applications including the distance to globular clusters, and the age of the Galactic disk and halo by means of their cooling sequences.
Empirical confirmation, and hence observational confirmation of stellar degeneracy, is challenging because of the few white dwarfs with accurately-determined masses and radii, and because their masses are concentrated in a rather narrow interval.  
(I should emphasise that white dwarf masses and radii are estimated only somewhat indirectly, based on stellar surface fluxes, theoretical evolutionary models, gravitational redshifts \citep[e.g.][]{2020ApJ...899..146C}, or white dwarfs in binary systems 
\citep[e.g.][]{1996A&A...311..852S,2017ApJ...848...11B}).
Before Hipparcos, although masses and radii were mostly considered to be consistent with theory, the observational scatter on both was too large to demonstrate a clear correlation
\citep{1996A&A...311..852S}. 
Hipparcos observed just 22 white dwarfs, with the strongest constraints coming from the four objects in binary systems, Sirius, 40~Eri, Procyon, and Stein~2051 
\citep[e.g.][]{1998ApJ...494..759P}. 

With Gaia, larger samples (of order 200) have become accessible for study
\citep{2017MNRAS.465.2849T,
2017ApJ...848...11B,
2019ApJ...876...67B,
2020ApJ...899..146C,	
2023MNRAS.522.3710P}.
Combined with spectroscopic atmospheric parameters, there is a general agreement with the theoretical mass--radius relation (whose detailed form varies with the assumed envelope thickness). 
Of 219 objects considered by 
\citep{2017ApJ...848...11B} using DR1,
73\% were considered consistent with the theoretical relation to within $1\sigma$ (their Figure~13), and so providing strong support for the theory of stellar degeneracy. They identified 15 stars that are better interpreted as unresolved double degenerate binaries. A suggestion that three of their objects fall on the theoretical mass--radius relation for an {\it iron\/} core, were not supported by studies using DR2
\citep[][their Figure~9]{2019ApJ...876...67B}.
A distinct result (from \citep{2019ApJ...876...67B}) is that the observed clump of low-mass, cool Gaia white dwarfs in the mass--\teff\ diagram (at \teff\,$\sim\!5100$\,K and $M\!\sim\!0.56M_\Sun$, their Figure~15) is, they conclude, evidence that the core composition of most is a mixture of carbon and oxygen\ldots {\it `an expected result from stellar evolution theory, but never empirically well established before'.}

One other result is noteworthy. Gaia astrometry, and follow-up with Hubble Space Telescope, gave a mass for the isolated DQ white dwarf LAWD~37, exploiting an {\it astrometric\/} microlensing event in Nov 2019
\citep{2023MNRAS.520..259M}.
Gaia astrometry was used both for the event {\it prediction}, and for the reference frame construction. The derived mass, $0.56\pm0.08M_\Sun$, is in agreement with the theoretical mass--radius relationship and cooling tracks for C--O cores.

\paragraph{Double white dwarf mergers}

White dwarf--white dwarf binaries are a late evolutionary stage of main-sequence binaries.  They are the main source of low-frequency gravitational waves expected in future space missions such as LISA and Tian-Qin (Section~\ref{sec:gravitational-waves}), and they are considered as possible progenitors of some type~Ia supernovae  
\citep{2014ARA&A..52..107M}. 
Models suggests that many should merge within a Hubble time, and that their merger products should therefore exist in the solar neighbourhood, presumably manifest as a single white dwarf of higher mass 
\citep[e.g.][]{
2009A&A...500.1193L,	
2017A&A...602A..16T,	
2018MNRAS.476.2584M, 
2020A&A...636A..31T}.	

One way of verifying the existence of these predicted merger products is to search for their kinematic signatures, making use of the fact that high-mass mergers are in general older than singly-evolved white dwarfs because of their binary evolution phase
\citep{2012MNRAS.426..427W, 
2015ASPC..493..547D}. 
This approach has been limited by the paucity of good white dwarf kinematics, which has changed with Gaia. The EDR3 catalogue of resolved binaries within 1~kpc contains more than a million systems, including 16\,000 white dwarf\,+\,main-sequence binaries, and 1400 double white dwarf binaries
\citep{2021MNRAS.506.2269E}.	
Gaia-based studies of the transverse velocities of white dwarfs as a function of mass suggest a fraction of double-white dwarf merger products of around 20\%
\citep{2019ApJ...886..100C,	
2020ApJ...891..160C,	
2023MNRAS.520..364F},	
and a merger rate 
\citep{2020ApJ...891..160C}	
2--3 times more accurate than previous estimates, supporting the idea that double-white dwarf mergers may contribute to a significant fraction of type~Ia supernovae as a result of the D$^6$ (or `dynamically driven double-degenerate double-detonation') mechanism 
\citep{2010ApJ...709L..64G}.	

In this D$^6$ scenario, the supernova detonation fully disrupts the accreting white dwarf, and possibly releases the donor white dwarf from the binary as a hypervelocity runaway star at its former orbital velocity
\citep{2021ApJ...923L..34B}.	
Using Gaia DR2, 
\citet{2018ApJ...865...15S}	
identified three such candidate D$^6$ remnants, all with Galactocentric velocities $>1000$\kms. These three objects were confirmed with Gaia EDR3 \citep{2021ApJ...923L..34B}, providing new constraints on the mass and radius of the donor white dwarf that becomes the hypervelocity runaway (see also Section~\ref{sec:hypervelocity-stars}).

Gaia is also being used to identify specific massive white dwarfs likely to be such merger products.
\citet{2021MNRAS.503.5397K},	
and later 
\citet{2023MNRAS.518.2341K}, 
searched the Montreal White Dwarf Database 100~pc sample, using Gaia DR2 parallaxes and proper motions to identify 4--5 outliers in transverse velocity (as well as four possible candidate mergers from their magnetic field, and one with rapid rotation). Of 25~`ultra-massive' objects with $M>1.3M_\odot$, they concluded that at least 8~are likely to be the result of mergers.
Another way of statistically identifying merger products was taken by 
\citet{2022MNRAS.511.5984F} 
using Gaia EDR3. 
From white dwarfs within 200~pc, in the range $0.95-1.25M_\odot$, they compared the cooling age distributions to the (Gaia-derived) time-varying star formation rate for main sequence stars, for each of three mass bins.  For white dwarfs in the range $0.95-1.15M_\odot$, they found a cooling age distribution consistent with the star formation rate found from Gaia DR2 by \citet{2019A&A...624L...1M}. For the most massive, $1.15-1.25M_\odot$, the distribution was better matched by a contribution from double white dwarf mergers.

With the growing evidence for the existence of a number of ultra-massive CO white dwarfs arising from double white dwarf mergers, 
\citet{2022ApJ...925...92N} 	
have shown that the merger process naturally produce highly magnetised, uniformly rotating white dwarfs, including a substantial population within a narrow mass range close to the Chandrasekhar mass.
And 
\citet{2022MNRAS.511.5198C} 
have modelled evolutionary (cooling) sequences, accounting for realistic initial chemical profiles, the energy released by latent heat and phase separation during crystallisation, and the energy released by the slow sedimentation of $^{22}$Ne.

\paragraph{White dwarf pulsars}

Amongst white dwarfs in binaries are a new class of `white dwarf pulsars', of which there are just two confirmed examples, AR~Sco and J1912--4410. The former, classified since the 1970s as a \mbox{$\delta$-Scuti} variable, was instead found to be a close binary of 3.56-hr period, with an M~star companion, at a distance of just 116\,pc, and pulsing in brightness with a period of 118\,s, much shorter than the orbital period
\citep{2016Natur.537..374M}. 
The pulsations actually comprise two components of similar frequency, with the higher frequency component (117\,s) associated with the spin period of the white dwarf, and the stronger lower frequency contribution (118\,s) attributed to a re-processed or `beat' frequency. In other words, although the pulsations are driven by the white dwarf’s spin, they mainly originate from the cool star.
Pulsations, also detected in the radio, have a spectrum characteristic of synchrotron radiation.

From the inconsistency between the spin-down power of the compact object and the system's bolometric luminosity,
\citet{1998A&A...338..521I} 	
had already suggested that the close binary AE~Aqr was a white dwarf pulsar. And since the discovery of AR~Sco, three others have been suggested 
\citep{2021arXiv210709913K,	
2021arXiv210903979K}, 		
including the Gaia transient Gaia22ayj 
\citep{2022arXiv220313975K},	
and the radio source GLEAM-X J162759.5--523504.3 
\citep{2022Ap&SS.367..108K}.	

\citet{2023NatAs...7..931P}	
carried out a targeted search for properties similar to AR~Sco, based on non-thermal infrared colours, variability, and location in the Gaia colour--magnitude diagram. They identified J1912--4410 as showing strong optical pulses with a period of 5.3~min, also detected as an \mbox{X-ray} source with eRosita, and identified as a compact binary due to its optical properties inferred from Gaia  
\citep{2023A&A...674L...9S}. 	
The Gaia EDR3 distance of $237\pm5$\,pc allowed 
\citet{2023NatAs...7..931P} 	
to constrain the radius of the \mbox{M~dwarf} from spectral fitting, from which they could estimate $M_1=1.2\pm0.2M_\odot$ and $M_2=0.25\pm0.05M_\odot$, with the companion then inferred to be filling over 90\% of its Roche lobe. The spectral energy distribution, combined with the Gaia distance, again yields a bolometric luminosity $\sim10^{26}$\,J\,s$^{-1}$, well in excess of the total {\it stellar\/} luminosity, $10^{24}$\,J\,s$^{-1}$. 

Various models have been put forward to explain the optical emission, the white dwarf's spin, and the origin of its magnetic field, involving core crystallisation, density stratification, and dynamo creation 
\citep{2021NatAs...5..648S,	
2022MNRAS.514.4111G}. 	
Meanwhile, Gaia distances are allowing accurate determination of their properties, as well as aiding searches for other examples.

\paragraph{Low-mass white dwarfs}

About 150 extremely low-mass white dwarfs, with $M<0.3M_\Sun$, were known before Gaia. Since the Universe is not old enough for these to have formed as the end point of single star evolution, they are taken as evidence for their origin in a common-envelope binary, or following mass-overflow in a multiple system.  Most will merge with their companion over a few billion years, each final merger being a strong source of gravitational waves. Recent theories have predicted a much larger space density of these objects, and the Gaia DR2 data have been used to derive a much-enlarged sample of 5762 extremely low-mass candidates, with $M<0.3M_\Sun$
\citep{2019MNRAS.488.2892P}.		
Other Gaia-based searches are being reported
\citep{2020ApJ...894...53K,
2022ApJ...936....5W}.

\paragraph{The initial-to-final mass relation}
\label{sec:wd-ifmr}

Modelling the relation between the progenitor mass and the final white dwarf mass (the `initial-to-final mass relation', IFMR), must account for the various evolutionary mass-loss processes, and these studies hence provide a probe of the local star-formation and stellar evolutionary history. It can be inferred, by population synthesis methods, from an initial stellar population matched to the observed white dwarf mass distribution. The latest studies are largely based on Gaia data 
\citep{2018ApJ...860L..17E,	
2018MNRAS.480.1300S,	
2021AJ....161..169C,	
2022ApJ...926..132H, 	
2023A&A...678A..20P, 	
2024MNRAS.527.3602C,	
2025ApJ...982...20I}.		

\subsubsection{Carbon stars}

Carbon stars are characterised by atmospheres with more carbon than oxygen. This gives them a deep red appearance, and they were first recognised spectroscopically by Angelo Secchi in the 1860s. They are luminous red giants, on the asymptotic giant branch, and most are long-period variables \citep{2018A&A...618A..58M}.
They arise as a result of carbon being `dredged up' from their core at certain evolutionary phases (`extrinsic' carbon stars arise from mass-transfer in binaries). Specifically, during a series of short-duration thermal pulses, resulting from a complex interplay between cycles of H- and He-shell burning, deep convection `dredges up' core material. It is the phenomenologically-defined `third dredge-up' which brings He, C, and the various s-process elements to the surface, so increasing the C/O ratio. Theory suggests that essentially all stars with initial masses $1.5-4M_\odot$ should go through the carbon star phase, lasting around 300\,000\,yr, before ending as a white dwarf.  
Their mass-loss rates, up to several times $10^{-5}M_\odot$\,yr$^{-1}$, contribute around half the total mass returned to the interstellar medium. Carbon stars are also one of the main sites of heavy-element production ($A\ge90$), through the (slow neutron capture) s-process
\citep{2023EPJA...59...17S}.
Their very high luminosities makes them important for integrated light studies of galaxies, and they are also studied as potential standard distance indicators, capable of calibrating the absolute magnitudes of local Type~Ia supernovae
\citep{2020ApJ...899...67F,		
2020MNRAS.495.2858R,
2021ApJ...923..157L,
2023ApJ...956...15L}. 

The Third Edition of the Catalog of Galactic Carbon Stars listed 6891 \citep{2001BaltA..10....1A}, while a more recent catalogue of Galactic AGB stars lists 11\,209 O-rich and 7172 C-rich carbon stars
\citep{2021ApJS..256...43S}.  
But no individual distances were known before Hipparcos and even then, only a handful with any degree of accuracy
\citep{1998A&A...334..545K,
2001A&A...371..222K}. 
And in their recent review of the present state of knowledge, \citet{2023EPJA...59...17S} noted: {\it `More than 40 years after the pioneering paper by \citet{1981ApJ...246..278I}, the efficiency of the third dredge-up  and the chemical yields from AGB stars are still burdened by heavy uncertainties and disagreements among different authors, mainly due to the lack of a robust theory of convection and mass loss.'}
It is the various uncertainties (on initial mass, mass-loss rates, metallicity, dredge-up efficiency, and convection parameters), poorly constrained by accurate luminosities given their large distances, that propagate through to uncertainties in chemical enrichment models, and thus to the chemo-dynamical history of the Galaxy 
\citep[e.g.][]{2022Univ....8..409G}.

With Gaia Data Release~3, the astrometry, photometry, and mean BP/RP and RVS spectra, have been used to define a sample of 15\,740 carbon stars out of 386\,936 initial candidates
\citep[][\S6]{2023A&A...674A..39G}.	
They demonstrate a strong concentration, not previously evident, towards the Galactic plane, with others in the LMC/SMC, and the Sagittarius stream.
Other Gaia studies are now aiming to characterise their ages, space densities, and luminosity function
\citep{2019MNRAS.482.4726N,	
2020A&A...633A.135A,	
2021MNRAS.506.3669P,	
2022A&A...664A..45A}.	

\subsubsection{Wolf--Rayet stars}
\label{sec:wolf-rayet}

Wolf--Rayet stars represents the final observable He-burning phase in the evolution of massive O~stars of initial mass $\gtrsim25M_\odot$
\citep{1996LIACo..33...39M}. 
At the end of this core-He burning phase, the star explodes as a supernova, and the outer shell of reaction products is expelled into the interstellar medium, while the accompanying core collapse results in either a neutron star or a black hole.
Beneath the dense winds that often hide the stellar surface, Wolf--Rayet stars have $T_{\rm eff}\sim30\,000-150\,000$\,K, $R\sim1-15R_\odot$, and $L\sim10^5-10^6L_\odot$. They have high mass-loss rates of $1-10\times10^{-5}M_\odot$\,yr$^{-1}$, and terminal outflow velocities of 1000--3000\kms\
\citep[e.g.][]{2007ARA&A..45..177C}.
Many Wolf--Rayet (and indeed other O-type) stars are classified as `runaways' (Section~\ref{sec:runaway-stars}), lying well outside their birth places in open clusters and OB~associations, signalled by their space motions of 100\kms\ or more, and by their large distances from the Galactic plane.
Although their characteristic spectra are easily recognisable to large distances (and fall into two main classes: WC/WO and WN), Wolf--Rayet stars are rare, with only around 1200 estimated to exist in our Galaxy;
\citet{2006A&A...458..453V}		
listed just 298, of which 24~were in the open cluster Westerlund~1, and around 60~were in open clusters near the Galactic centre.
Of the 200 or so known in 1990, Hipparcos observed 67, including all with $V\lesssim12$~mag
\citep{1998A&A...331..949M,
1997NewA....2..245V,
1997ApJ...484L.153S}.

Most distances were only poorly constrained pre-Gaia. And major gaps in their understanding arise from uncertainties in the mass-loss rates, angular momentum loss, internal mixing processes, binary fraction, close-binary (including common-envelope) evolution, and stellar mergers
\citep[e.g.][]{2019A&A...625A..57H}.
Today, a number of papers are addressing updates in the physical properties, cluster membership, or runaway status of individual Wolf--Rayet stars using the Gaia distances, e.g.\ 
WR~6 \citep{2020RNAAS...4..213G};
WR~16 \citep{2020MNRAS.495..417C};
WR~36 \citep{2022MNRAS.517.3749R};
and WR~133 \citep{2021ApJ...908L...3R}.	
Others are using the Gaia distances to revisit the bulk physical properties (Galactic distribution, luminosities and mass-loss rates), as well as cluster membership, and consistency with current evolutionary models including their role as progenitors of massive black holes
\citep{2019A&A...621A..92S,	
2019A&A...625A..57H,	
2020MNRAS.493.1512R,	
2020MNRAS.495.1209R}.	
I say more on the Wolf--Rayet stars in the cluster Westerlund~1 in Section~\ref{sec:young-massive-clusters}.

\subsubsection{The oldest stars}
\label{sec:oldest-stars}

Until the Hipparcos results in 1997, there existed an unsettling paradox regarding the age of the Universe. Its expansion age had been estimated at $\sim$11\,Gyr. But some stars had ages, derived from their luminosities and evolutionary models, of $\sim$15\,Gyr. The paradox was (largely) resolved with the new Hipparcos distances of nearby Cepheids. 
\citet{1997MNRAS.286L...1F}
argued that the LMC Cepheids were 10\% further than previously estimated, and thus brighter. 
The overall distance scale was revised upwards by the same amount: globular clusters were more distant than previously thought, their luminosities were therefore larger, and their `turn-off point' in turn implying younger ages. Their analysis revised the age of the Universe up to $\sim$12\,Gyr, and brought the oldest stellar ages down to $\sim$11\,Gyr. 

Today, this age tension has largely eased, and stellar ages determined from high-accuracy astrometric distances and spectroscopy, combined with the latest evolutionary models, generally cap their ages to less than the time that is believed to have elapsed since the Big Bang.
Here, there are two rather direct and distinct measurements that provide the most definitive estimates of the age of the Universe today. One, tied to the `early Universe' estimates of the Hubble constant, is based on measurements of the microwave background radiation by the Planck satellite, and indicates an age of $13.787\pm0.020$~Gyr
\citep{2020A&A...641A...6P}.		
The other is based on observations of the `local' distance scale and expansion rate, which suggest a slightly larger value of the Hubble constant, and hence a slightly younger age for the Universe, both for the younger (Population~I) Cepheids 
\citep{2018ApJ...861..126R},
as well as on the older (Population~II) globular clusters
\citep{2019ApJ...882...34F}.
This age estimate is considered to be consistent with any of the {\it lower limits\/} on its age dictated by the oldest objects within it. 
For example, one such constraint comes from white dwarfs, where the temperature of the coolest, and detailed models of their cooling, provides a lower limit on the age of the Universe.
Another is given by the faintest `turnoff point' of main sequence stars in clusters. The lowest-mass stars that have evolved away from the main sequence set another, independent, minimum value for the age of the Universe.

A few stars still have ages uncomfortably close to this limit. Some, members of the Galaxy's low-metallicity halo population, are passing through the solar neighbourhood, and are sufficiently close, within a few hundred parsecs, that they are particularly well measured. 
%
The most celebrated is HD~140283, informally known as Methuselah, an extremely metal-deficient and high-velocity subgiant. It occupies a location in the HR diagram where absolute magnitude is most sensitive to stellar age, and is one of the 34 FGK-type benchmark stars selected as the `pillars of calibration' for Gaia
\citep{2014A&A...564A.133J,
2020A&A...640A..25K}.
Interest in its age dates back to the work of 
\citet{1956ApJ...124..116B}. 
Later developments are most easily traced by quoting the parallaxes on which the subsequent age estimates are based.
The Fourth Edition of the Yale Parallax Catalogue gave $\varpi=16.4\pm1.6$~mas 
\citep{1995gcts.book.....V}.
Hipparcos gave $17.44\pm0.97$~mas, and the revision by van Leeuwen gave $17.16\pm0.68$~mas
\citep{2010SSRv..151..209V}.
The Hubble Space Telescope Fine Guidance System gave $17.15\pm0.14$~mas \citep{2013ApJ...765L..12B}.
Gaia DR2 (2020) gave $16.114\pm0.072$~mas,
and Gaia DR3 (2022) gave $16.267\pm0.026$~mas.		
Incidentally, the HST parallax, while {\it `five times more precise than that from Hipparcos'} \citep{2013ApJ...765L..12B}, lies $6\sigma$ from the Gaia DR3 value, an illustration of the challenge of absolute parallax determination in the pre-Gaia era.

The ages derived from these improving parallaxes also relied on progressive improvements in spectroscopic abundances and evolutionary models. 
By 2000, models which included observed enhancements in the $\alpha$-elements, and using the Hipparcos catalogue value, provided `strong support' in favour of an age {\it older\/} than 14~Gyr
\citep{2000ApJS..129..315V}.
A decade later, the HST parallax measurement, along with models that included He~diffusion and enhanced O~abundance, and the parallax from \citep{2010SSRv..151..209V}, gave an age of $14.46\pm0.31$~Gyr
\citep{2013ApJ...765L..12B}.
Improved spectroscopic surface abundances (of O, Fe, Mg, Si, and Ca), combined with the HST parallax, still resulted in a conflicting age of $14.27\pm0.38$~Gyr
\citep{2014ApJ...792..110V}. 
Subsequent studies continued to place HD~140283 amongst the very oldest known stars in the Galaxy, and in {\it potential\/} conflict with the accepted age of the Universe:
$13.7\pm0.7$~Gyr \citep{2015A&A...575A..26C};
12.5--14.9~Gyr \citep{2018ApJ...856...10J};
and $12\pm0.5$~Gyr \citep{2021RNAAS...5..117T}. 
But using the latest Gaia DR3 parallax, along with precise interferometric and spectroscopic constraints, the best estimate of the age of HD~140283 is given as 12.3~Gyr
\citep{2024A&A...692L...3G}.	
Without going into further details of the models, the results do depend on the assumed abundances of C, O, and Fe (a solar-scaled mixture results in an age of 14~Gyr), as well as assumptions on the mixing-length parameter invoked to describe convection. 

Today, most would probably consider the cosmic microwave background as providing the most secure estimate of the age of the Universe, with any apparent conflicts pointing to errors in the star's measured properties, or to inadequacies in theoretical models. Future Gaia data releases will be an interesting contribution to this debate.

\subsubsection{The largest stars}

Stars larger than a few hundred $R_\Sun$ are mostly red supergiants, `cool' (\teff$\lesssim4200$\,K), luminous ($L/L_\Sun\ge10^4$), highly evolved descendants of main-sequence stars of initial mass $10-40M_\Sun$, and occupying the upper-right region of the HR diagram
\citep{1974ARA&A..12..215I,	 
2019AJ....158...20M}.		
They are amongst the brightest stars in the infrared, detectable out to Mpc distances, with the most recent pre-Gaia compilation listing some 1400 Galactic candidates
\citep{2014yCat....102023S}.		
The coolest and most luminous have $R\sim1500R_\Sun$
\citep{2005ApJ...628..973L},	
in agreement with the largest radii predicted from MARCS models using improved molecular opacities inspired by the anticipation of the Gaia data 
\citep{2003ASPC..288..331G,	
2003ASPC..298..189P}.		
{\it `It is believed'}, they state, {\it `that stars above this radius would be too unstable and simply do not form'}.
Scaled to the solar system, such photospheres would extend beyond the orbit of Jupiter.

During their short evolutionary phase, lasting 0.2--0.4\,Myr, a star of initial mass $20M_\Sun$ returns some $3-10M_\Sun$ into the interstellar medium 
\citep{1990ApJS...73..769J}.
Their late-stages of nucleosynthesis, stellar winds, and eventual explosions as supernovae therefore represent key processes in a galaxy’s chemical enrichment. But they remain a poorly characterised evolutionary phase, with models failing to match some cool or luminous objects. Challenges include their molecular opacities,
extended atmospheres, 
sonic velocities of the convective layers,
and supersonic velocities in their atmospheres, 
resulting in shocks, photospheric asymmetries, and imprecise radii
\citep{2003AJ....126.2867M,
2009AJ....137.4744L, 	
2009ASPC..412..113W}.

Gaia is providing an improved census, and improved distances, important ingredients for improved models. 
These include a DR2-based sample of 1342 parallaxes, mostly located along the Galaxy's spiral arms, although generally in isolation
\citep{2019AJ....158...20M},	
and a quality sample of 578 used to compare with Geneva stellar evolution tracks, and categorised according to multiplicity, variability, and classification as a runaway as given by the proper motions
\citep{2024MNRAS.529.3630H}.	
While many occur in isolation, some open clusters do host concentrations of red supergiants, amongst them 
$\chi$~Per, NGC~7419, and Westerlund~1, all now with well-determined distances from their mean Gaia DR2 parallaxes 
\citep{2019MNRAS.486L..10D}.	
The largest stars, a number with inferred $R>1500R_\Sun$, have their own literature detailing classification, mass-loss, etc., and with growing insights from Gaia, including Stephenson~2--18, UY~Scuti, VV~Cep, VY~CMa, and WOH~G64.

Their short evolutionary lifetimes means that red supergiants are important as potential core-collapse supernova progenitors, with the Gaia data being used to refine the selection of candidates for observing the progenitor pre-explosion and the early phases of core-collapse supernovae
\citep{2021JPhCS2156a2137M, 	
2024MNRAS.529.3630H}.		
The identification of red supergiants beyond our own Galaxy is also making use of the Gaia parallax and proper motion as membership discriminants, with 
1098 identified in the LMC 
\citep{2024ApJ...965..106Y}, 
and samples of 5498 and 3055 objects identified in M31 and M33 respectively
\citep{2021ApJ...907...18R}. 	

\subsubsection{Blue stragglers}
\label{sec:blue-stragglers}

Found in both open clusters and globular clusters, blue stragglers are stars which are more luminous, and have a higher \teff\ (hence bluer), than the cluster's main sequence turnoff, and their origin remains puzzling. First identified, in the 1950s, in the globular cluster M3 by 
\citet{1953AJ.....58...61S},		
they are inconsistent with standard evolutionary theory, since they should have consumed their nuclear fuel, and evolved to become white dwarfs, long ago. Rather, they are considered to result from collisions, either from binary stars in the process of merging (or recently merged), or a result of mass-transfer in a binary 
\citep{2006MNRAS.373..361M,		
2015ASSL..413.....B}. 			
A compilation in 2007 listed 1887 candidates in 427 open clusters
\citep{2007A&A...463..789A}.		

Gaia is providing the improved location and fidelity of candidates in the colour--magnitude diagram. From Gaia DR2, 
\citet{2021A&A...650A..67R} 		
confirmed just 897 (and 77 yellow stragglers) in 408 open clusters. 
Also from DR2, 
\citet{2021MNRAS.507.1699J} 		
found 868 blue stragglers in 228 clusters. And they established that their numbers {\it increase\/} with cluster age and mass.
Using Gaia DR3, 
\citet{2023A&A...672A..81L} 	
found a further 138.
Their radial distribution provides an important dynamical probe for globular clusters, and Gaia has enabled similar studies for a number of open clusters
\citep{2020MNRAS.496.2402V, 
2021AJ....161...37R}.		
Based on 16 old nearby ($d<3500$\,pc) open clusters,		
\citet{2021ApJ...908..229L} 	
also found that their fractional number increases with age.	
They also found that population synthesis models in which blue stragglers form by mass transfer, dramatically under-produce the numbers in old clusters, as well as overproducing high-mass relative to lower-mass stragglers. 
Ultraviolet observations combined with astrometry and photometry from Gaia DR2/EDR3 now accurately locates blue stragglers in various other clusters
\citep{2022MNRAS.511.2274V,	
2022MNRAS.516.5318P,		
2023ApJ...945...11R,		
2023A&A...676A..47J}, 	
confirming many as post-mass-transfer binaries as inferred from their high ultraviolet flux.
A notable Gaia discovery is a double blue straggler sequence in the ancient (8.5--10\,Gyr) open cluster Berkeley~17
\citep{2023MNRAS.518L...7R}. 	
Parallel sequences have been seen previously in four {\it globular\/} clusters,
the first of which was M30 \citep{2009Natur.462.1028F}, 
the bluer attributed to stellar collisions, and the redder to the results of binary mass transfer. However, the lower densities of open clusters makes formation via collisions unlikely, suggesting a mass-transfer formation channel for both, but with an unexplained offset between them.

\subsubsection{Sub-subgiants}
\label{sec:sub-subgiants}

Subgiants are on the roughly `horizontal' part of the evolutionary track in the transition between the main sequence and giant stages: the start of H-shell burning is accompanied by an increase in radius, and a movement away from the main sequence. They lie just above the main sequence and to the left of the Hertzsprung gap for the more massive stars, and on the slow rise up the lower giant branch for low-mass stars. Their rapid evolution means that they are rare in the solar neighbourhood: Hipparcos observed only~12, including the nearest and brightest, $\beta$~Hyi.

In addition to the more exotic `blue stragglers' and `yellow stragglers' in this region of the HR diagram (Section~\ref{sec:blue-stragglers}), a new class of `sub-subgiants' was found in M67
\citep{1998A&A...339..431B}, 
and a new class of `red stragglers' found in 47~Tuc \citep{2001ApJ...559.1060A}. 
Other examples of both have since been discovered in both open and globular clusters, as well as in the field. Like the blue stragglers, they are unexplained by standard theories of single-star evolution
\citep{2017ApJ...840...66G}.
\citet{2017ApJ...840...66G} attempted to clarify the confusing nomenclature, and recommended the term `sub-subgiant' to refer to stars that are redder than the normal main-sequence and {\it fainter\/} than normal subgiants, and the term `red straggler' to stars similarly redder than the normal red giants but {\it brighter\/} than normal subgiants.
From evolutionary models, three possible explanations for these sub-subgiants have been suggested 
\citep{2017ApJ...840...67L}: 
binary mass transfer, envelope stripping, and magnetic activity.  
New insight has come from Gaia, where the parallaxes and photometry from Gaia EDR3 were used by \citet{2022ApJ...927..222L} to accurately position known RS~CVn (active giant) binaries in the colour--magnitude diagram.  They associated those that fall below a 14\,Gyr, metal-rich isochrone as candidate field sub-subgiants. Out of a sample of 1723 RS~CVn, they found 448 sub-subgiant candidates, a factor seven more than the 65 previously known. They concluded that their ubiquity suggests that they are a normal evolutionary phase for RS~CVn-type systems, rather than rare by-products of dynamical encounters found only in dense star clusters.

A more exotic origin has been suggested by 
\citet{2023ApJ...959..113B}, 
related to the possible existence of very low-mass primordial black holes formed in the first second or so following the Big Bang. These were postulated by \citet{1971MNRAS.152...75H} to obviate the very special initial conditions required to explain the Universe's high degree of isotropy. The central idea is that if dark matter is comprised of primordial black holes at the classical Hawking evaporation limit of $10^{-18}M_\Sun$ ($2\times10^{12}$\,kg), there would be some $10^{30}$ of them in our Galaxy, with an average separation of $\sim$1\,au, raising the possibility of their capture by stars 
\citep{2021PhRvD.104l3031I}.
\citet{2023ApJ...959..113B}
found that evolutionary models for stars containing the lightest black holes are barely influenced. However, in models of the Sun born with a central black hole of mass $10^{-6}M_\Sun$, the Sun would first fade to half its current luminosity over a period of 100~Myr as accretion generates enough energy to quench nuclear reactions. It would then expand into a fully-convective star, where it would radiate for several Gyr with an enriched surface helium abundance, first as a sub-subgiant, and later as a red straggler, before becoming a sub-solar-mass black hole.

\subsubsection{Supernovae and supernovae remnants}
\label{sec:supernovae}

Supernovae are the violent end-points of stellar evolution which lie at the heart of many important questions in modern astronomy. Their detailed taxonomy (they are classified according to their light curves, rather than their inferred physical origin), as well as their possible formation pathways, are both rather involved, and the underlying physics still somewhat uncertain.
In {\it thermal runaway\/} events, low-mass (long-lived) stars accumulate material from a companion star (either a binary companion, or via a white dwarf--white dwarf collision). This can lead to rapid `runaway' fusion, releasing a burst of energy sufficient to disperse the star. Type~Ia supernovae, which follow a characteristic light curve, and which present uniform properties which make them important distance indicators, fall into this class.
In {\it core collapse\/} events, massive (short-lived) stars undergo collapse when nuclear fusion no longer sustains the stellar core against its own gravity. The prompt collapse may cause violent expulsion of its outer layers as a supernova or, if the release of gravitational potential energy is limited, the star may collapse into a black hole or neutron star with little radiated energy. According to the star's metallicity and mass (in the range $10-100M_\odot$), different mechanisms, and different taxonomic classes (amongst them type~I\,b/c and type~II) may result.

Supernovae play a major role in the chemical and dynamical evolution of galaxies. They are the main producers of heavy elements, and hence fundamental for understanding abundances in galaxies and in the intergalactic medium.
As a result of these complex formation pathways, supernovae provide laboratories for studying extreme nuclear processes, being involved in the formation of neutron stars, black holes, millisecond pulsars and gamma-ray bursts. They are sources of gravitational waves, neutrinos, and high-energy cosmic rays. And they, along with their precursors, are implicated in the evolutionary history and accretion rates of interacting binary systems, low mass X-ray binaries, and globular cluster X-ray sources.
The explosion expels much of the remaining stellar material, in the form of a supernova remnant, at velocities up to 10\% of the speed of light. At these highly super\-sonic speeds, a strong shock wave forms ahead of the ejecta, heating the ambient interstellar medium to millions of Kelvin. The shock slows as it sweeps up material, but it can expand over centuries, and tens of parsecs, before it slows to below the local sound speed.
An
\href{https://www.mrao.cam.ac.uk/surveys/snrs}{online compilation}
currently lists 310 supernovae remnants 
\citep{2025JApA...46...14G},
of which only one remnant pulsar, the Crab (at 16.5~mag), is bright enough at optical wavelengths to be observed by Gaia.

In preparing the scientific case for Gaia, early estimates suggested that, with a limit of $\sim$20~mag, the mission could detect supernovae to distances of around 500~Mpc, i.e.\ to redshifts $z\sim0.1$, yielding some 100\,000 discoveries, of all types, over its nominal 5-year lifetime 
\citep{1999BaltA...8...25H}. 
Later pre-launch estimates pointed to numbers closer to 10\,000
\citep{2012Ap&SS.341..163A}.
Actual Gaia discoveries, part of the activities of the Alerts Working Group (Section~\ref{sec:science-alerts}), are based on the photometric light-curves assembled from the satellite's repeated sky scanning. These are used to trigger ground-based observations to follow the light curves over the subsequent weeks, including some events `caught' before maximum.
At the end of the mission, in February 2025, the Alerts database listed over 27\,000 events. While more than 20\,000 remain `unidentified', over 4000 have been classified as supernovae: around 3000 type~I (of which 2800 are type~Ia), and just over 1000 type~II. Some 60~SN events were alerted at magnitudes $<15$~mag.

As well as adding to their known numbers, Gaia parallaxes and proper motions are providing improved distances of known events, their remnant structures and, through kinematic searches for surviving binary companions, the nature of their progenitors.
As a prominent example, the distance to the Crab pulsar and its supernova remnant was widely assumed, pre-Gaia, to be around 1.93~kpc
\citep{1973PASP...85..579T}.
The significantly larger Gaia DR2-based parallax distance, $3.37_{-0.97}^{+4.04}$~kpc
\citep{2019ApJ...871...92F},
would have implications for models of its underlying pulsar physics. 
They also showed that Gaia's multi-epoch photometry already hints at a measurable secular decrease of the pulsar's luminosity, attributed to synchrotron radiation losses.
The distances and kinematics of possible companions of Tycho’s supernova have been derived from DR2 
\citep{2019ApJ...870..135R},
and DR3
\citep{2023A&A...680A..80G}.
From Gaia DR2, and consistent with previous work, no plausible binary companion was found for the Crab, Cas~A, or Vela supernovae
\citep{2019ApJ...871...92F}.	
Nonetheless, for the Vela complex associated with the Vela pulsar, 18~million stars over 450 sq.\ deg.\ were used to reveal the detailed morphology of the cavity shell
\citep{2021A&A...655A..68H}, 
and to estimate the mass of the progenitor star 
\citep{2022MNRAS.511.3428K}.
For the Cygnus Loop, EDR3 parallaxes of stars in or behind the supernova shell have been used to set a precise distance of $725\pm15$~pc, with an uncertainty comparable to its 18~pc radius
\citep{2021MNRAS.507..244F}.
Studies have identified and modelled high-velocity ejecta from specific pre-supernova binary systems, including three hypervelocity white dwarfs, with space velocities 1000--3000\kms\
\citep{2018ApJ...865...15S},
and the runaway X-ray binary HD~153919/4U~1700--37
\citep{2021A&A...655A..31V}.	
Many other searches for the unbound companions of these supernovae progenitors are being carried out with Gaia. Broadly, they aim at estimating both the fraction of progenitors in binaries or triples, and the fraction that survive the supernova explosion
\citep{2019ApJ...871...92F,	
2020MNRAS.498..899N,		
2021MNRAS.507.5832K,		
2021AN....342..553L,		
2023MNRAS.522.2029C,		
2023MNRAS.519.3865K,		
2024MNRAS.535.1315B,		
2024MNRAS.531.4212D,		
2024A&A...691A..63D}.		

\subsubsection{Neutron stars and pulsars}

Neutron stars are the collapsed cores of massive stars ($8-40M_\Sun$), resulting from a supernova explosion.  With radii of only $\sim$10\,km, but with masses $\sim1.4M_\Sun$, their densities, $\sim10^{17}$\,kg\,m$^{-3}$, correspond to those of atomic nuclei. Pulsars are their rapidly spinning, highly-magnetised manifestations. They emit narrow beams of radio emission parallel to their magnetic dipole axis, seen as pulses at the object's spin frequency due to a misalignment of the magnetic and spin axes. 
There are two broad classes: normal pulsars are isolated objects with spin periods $\sim$1\,s. Millisecond pulsars are old ($\sim$Gyr) neutron stars, spun-up during mass and angular momentum transfer from a binary companion; most known still have (non-accreting) binary companions, either white dwarfs or neutron stars. 
%
As of early 2025, the 
\href{www.atnf.csiro.au/research/pulsar}{ATNF pulsar catalogue} 
lists around 3800 (a small fraction of those thought to exist), with more than 80~millisecond pulsars in the Galaxy disk, and some 130 in Galactic globular clusters. The fastest spin-period is the 716\,Hz (1.4\,msec) eclipsing binary pulsar PSR~J1748--2446ad in Terzan~5. The slowest, PSR~J2144--3933, has a period of 8.51\,s.

Distances are important for understanding their physics, providing constraints on the nuclear equation of state, and quantifying energy transport in neutron star winds
\citep[e.g.][]{2018ApJ...864...26J}.
Precision astrometry allows accurate estimates of their spin-down rates (via their luminosities) as well as their (relativistic) binary orbital decay rates.
For radio pulsars, distances can be estimated through the observed pulse dispersion measure, although this relies on models of the Galactic electron density distribution.  Some distances can be inferred from their association with star clusters, from binary spectroscopic parallaxes, from annual variations in pulse arrival times, or from VLBI.
As noted in Section~\ref{sec:supernovae}, other than the Crab, no pulsar is bright enough at optical wavelengths to be visible with Gaia.

Gaia is contributing for pulsars in binary systems, where the brighter companion serves as a distance proxy.
\citet{2018ApJ...864...26J}		
worked with the 2424 (non-globular cluster) objects from the ATNF catalogue, selected the 188 in binary systems, and found 22 counterparts in DR2, of which 12 have significant parallaxes (including white dwarf companions to msec pulsars, and the non-degenerate components of  `black widow' and `redback' systems). Amongst them are specific systems used in tests of alternative theories of gravity (PSR J0348+0432 and PSR J0337+1715), in theories of accretion (PSR J1023+0038) and of particle acceleration (PSR B1259--63). 
Notwithstanding the fact that some pre-Gaia distance estimates are more precise (if not necessarily more accurate) than those from Gaia DR2, they found a generally reasonable agreement between the previous estimates of parallaxes and proper motions with those from Gaia. Their study also identified a number of objects for which pre-Gaia binary and orbit models merit revision and refinement. 
Thus for J0437--4715 (one of the closest known neutron stars), the Gaia distance is 77\% of that derived from orbital derivative models, and provisionally attributed to unmodeled binary reflex motion.
For J2215+5135, the Gaia proper motion ($1.9\pm0.8$\masyr) is much smaller than the pre-Gaia estimate ($189\pm23$\masyr), the latter now considered to be in error.
Further improvements with Gaia DR3 have followed
\citep{2023ApJ...954...89M}.	

While the majority of massive stars have a stellar companion, most pulsars appear to be isolated, implying that most massive binaries break apart due to strong natal `kicks' received in supernova explosions. But the monitoring of newly discovered pulsars has rarely been carried out for long enough to exclude multiplicity. 
A study based on Gaia DR2 targeted an improved estimate of the binary fraction, confirming that the multiplicity fraction is small
\citep{2021MNRAS.501.1116A}. 
Among a number of other applications of these inferred pulsar distances and motions are 
an estimation of optical and X-ray luminosities
\citep{2023MNRAS.525.3963K}, 
a discussion of their space velocities
\citep{2021RAA....21..141Y},	
understanding anomalous accelerations
\citep{2025ApJ...983...62D},	
examining the consistency between the Gaia and pulsar reference frames
\citep{2023A&A...670A.173L},	
and new insights into Galactic structure
\citep{2024PhRvD.110b3026D}.	

\subsubsection{Black holes}
\label{sec:black-holes}

Black holes have a wide range of observable effects across many fields. The most massive, supermassive black holes of $10^6M_\Sun$ or more, lie at the centre of most large galaxies, and are central to the various phenomena underpinning active galaxies and quasars. They are therefore implicit in Gaia's (active) galaxy and quasar survey (Section~\ref{sec:galaxy-quasar-survey}), in the gravitational lensing of quasars (Section~\ref{sec:quasars}), in the dual active galactic nuclei resulting from galaxy mergers (Section~\ref{sec:dual-agn}), and as the source of hypervelocity stars in our own and other Local Group galaxies (Section~\ref{sec:hypervelocity-stars}).
I also mention some speculative ideas related to hypothetical very low-mass primordial black holes in Section~\ref{sec:sub-subgiants}.

There are likely to be many millions of stellar-mass black holes throughout the Galaxy, resulting from standard evolutionary pathways
\citep{2022MNRAS.516.4971S}. 	
Isolated objects represent a particular discovery challenge, but it is their presence in binary systems that offers the best prospects for their discovery and characterisation. In compact systems, black holes are revealed through high-energy radiation as a result of mass-transfer from their non-degenerate companions. In wider-separation binaries, with little accreting matter to signal their presence, they will still result in orbital motion of their companion which may be observable by Gaia. 
I expand on each of these prospects below.

\paragraph{Isolated stellar mass black holes}
\label{sec:isolated-black-holes}

Isolated stellar mass black holes, $5-50M_\Sun$, resulting from the gravitational collapse of massive stars at the end of their lives, are directly discoverable only from photometric microlensing.
The only confirmed object to date, OGLE-2011-BLG-0462 ($7.1M_\Sun$, $d=1.58$\,kpc), was characterised from HST astrometry in 2022
\citep{2022ApJ...933...83S},		
and further confirmed more recently 
\citep{2025ApJ...983..104S}.		
A recent candidate, from the Gaia science alerts system (Section~\ref{sec:science-alerts}), is Gaia18ajz, with a probable lens mass of $4.9M_\Sun$ at $d=1.1$\,kpc 
\citep{2025A&A...694A..94H}.	
Amongst 950 of the OGLE photometric microlensing events observed by Spitzer between 2014--2019, Gaia DR3 proper motions assisted in identifying four `dark remnant' candidates
\citep{2024ApJ...975..216R}.	
Of these, OGLE-2016-BLG-0293 and OGLE-2018-BLG-0483, with lens masses $3-4.7M_\Sun$, are expected to exhibit astrometric microlensing signals detectable by Gaia (Section~\ref{sec:astrometric-lensing}). 
A separate classification of 10\,000 OGLE events, interpreted using Gaia EDR3, yielded 23~black hole candidates
\citep{2025ApJ...981..183K}.
Prospects for detecting primordial black holes with Gaia's forthcoming time-series astrometry have also been assessed
\citep{2023JCAP...05..045V}.		

\paragraph{Interacting binaries}

The discovery space is much larger for stellar mass black holes in binaries. 
In compact (interacting) systems, they are inferred from X-ray emission arising from mass transfer, divided (according to donor mass) into low-mass (LMXB) or high-mass (HMXB) \mbox{X-ray} binaries. 
Today, there are around 20~candidates, including
Cygnus~X1 ($M_{\rm bh}\sim21M_\Sun$ in a 5.6-d binary) 
and
V404~Cyg ($M_{\rm bh}\sim12M_\Sun$ in a \mbox{6.5-d} binary),
both at distances \mbox{2--2.5\,kpc}. 
Cygnus X--1, a high-mass X-ray binary discovered in 1964, provided the first observational evidence for a black hole.  It was characterised (from the 1970s) as comprising a blue supergiant orbiting a $20M_\odot$ black hole at about 0.2\,au. A stellar wind from the supergiant forms a hot accretion disk around the black hole, generating the observed X-rays, with the compact object considered too small to be other than a black hole. A recent study with revised physical modelling
\citep{2021Sci...371.1046M}
included a new VLBI parallax of $0.46\pm0.04$\,mas, in good agreement with their (zero-point corrected) Gaia DR2 value (and indeed with the later Gaia DR3 parallax $\varpi=0.4439\pm0.0149$\,mas, corresponding to $1/\varpi=2250\pm80$\,pc).
Another recent Gaia result suggests that V404~Cyg is part of wide hierarchical triple 
\citep{2024Natur.635..316B}.	

\paragraph{Ellipsoidal variables}

Black holes in binaries may also be manifest, photometrically, through the tidal distortion of the envelopes of their companion star, as extreme examples of the class of ellipsoidal variables (Section~\ref{sec:ellipsoidal-variables}). As I have noted there,
\citet{2023A&A...674A..14R}	
identified 65\,300 candidate ellipsoidal variables in Gaia~DR3, and searches for black hole systems amongst them are ongoing
\citep{2022ApJ...940..126F,	
2023A&A...674A..19G,
2024OJAp....7E..24R,	
2023MNRAS.518.2991S}. 

\begin{figure}[t]
\centering
\raisebox{12pt}{\includegraphics[width=0.19\linewidth]{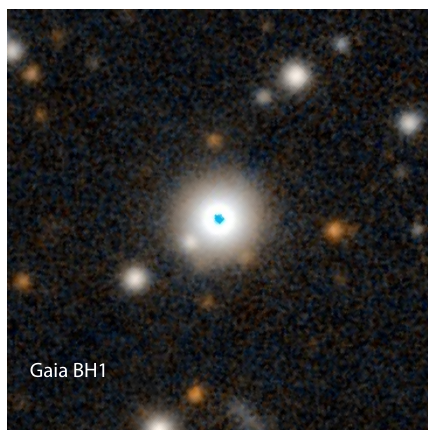}}
\hspace{4pt}
\includegraphics[width=0.32\linewidth]{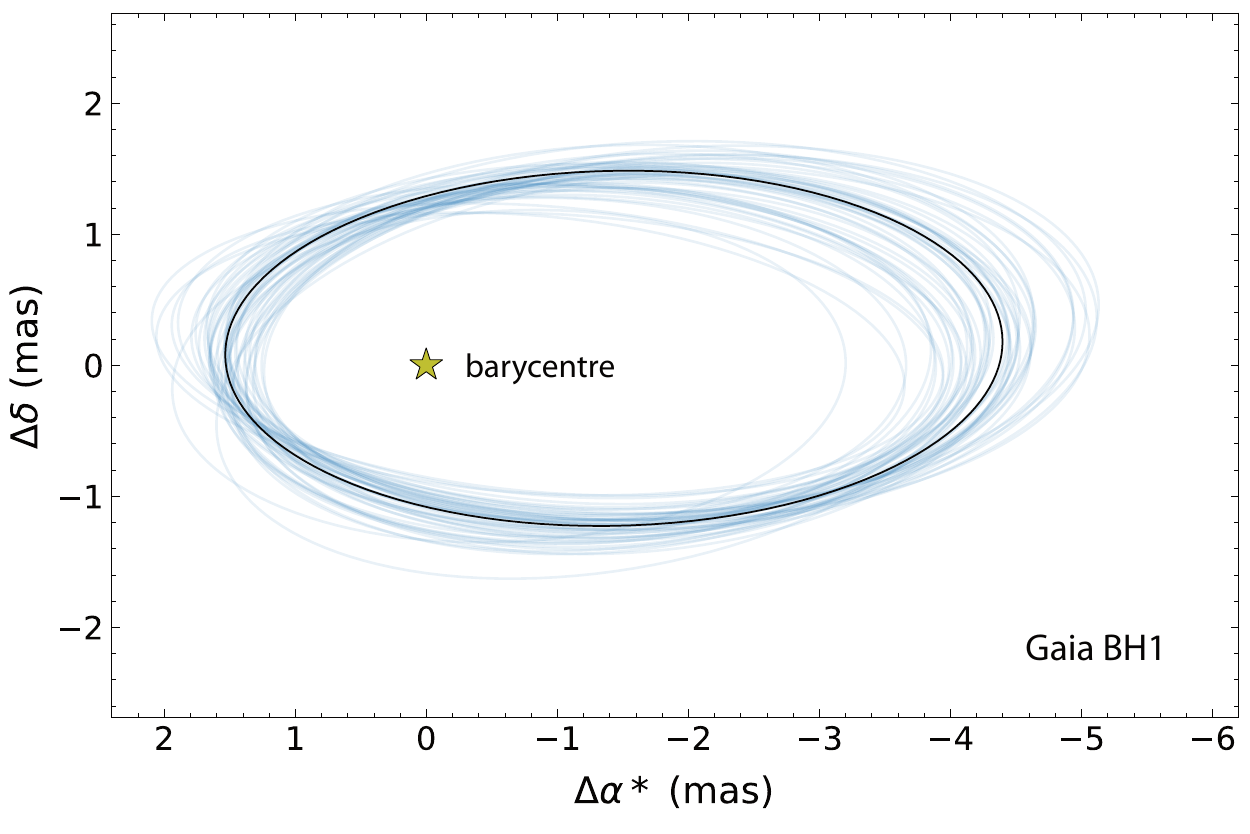}
\hspace{4pt}
\raisebox{20pt}{\includegraphics[width=0.45\linewidth]{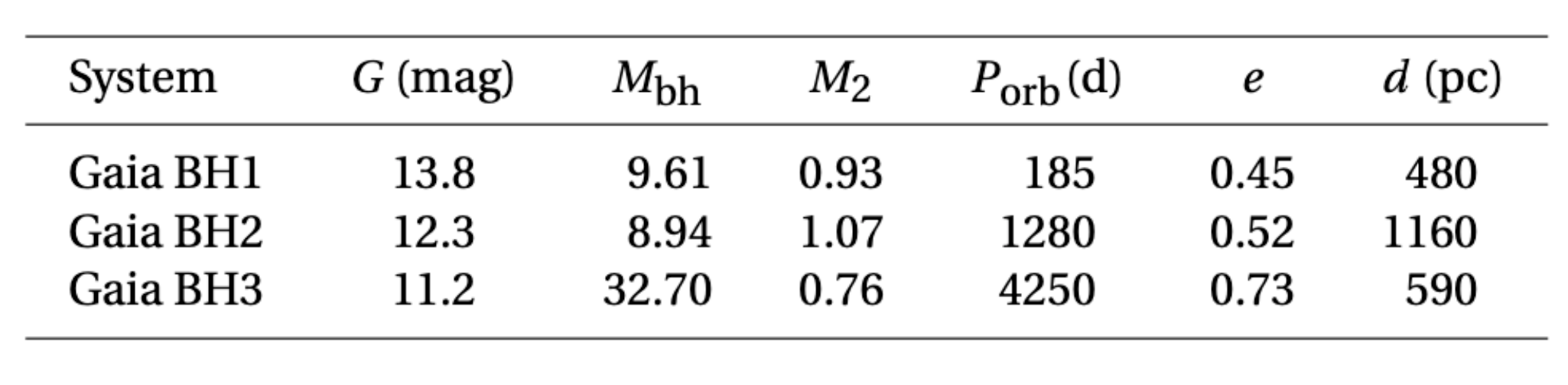}}
\vspace{-15pt}
\caption{Left~(a): $45\times45$~arcsec$^2$ PanSTARRS field of the $G=13.8$~mag binary companion to the Gaia black hole BH1 \citep{2023MNRAS.518.1057E}.
Middle~(b): Gaia DR3 astrometry defines a photocentric ellipse with semi-major axis $a_0=2.98\pm0.22$~mas (solid curve); thin blue curves are orbits drawn randomly from a posterior distribution \citep{2023AJ....166....6C}.
Right~(c): parameters of the black hole binaries BH1, BH2, and BH3.
}\label{fig:gaia-black-holes}
\end{figure}

\paragraph{Gaia's black hole binaries}
Based on the Gaia EDR3 binary catalogues, models soon predicted that many more wide binary systems containing quiescent black holes should be detectable with Gaia astrometry, with prospects of masses for some hundred
\citep{2022ApJ...931..107C,	 
2022A&A...658A.129J}.		
Using the Gaia DR3 catalogue of 100\,000 single-lined spectroscopic binaries 
\citep{2022MNRAS.515.1266E}, 
two papers independently reported the first such binary system, the $G=13.8$~mag Gaia DR3 4373465352415301632 (BH1), comprising a main-sequence sun-like star and a massive non-interacting black hole candidate
\citep{2023AJ....166....6C, 	
2023MNRAS.518.1057E}. 	
Selected for further study on the basis of its high mass ratio, and its location close to the main sequence, follow-up spectroscopy and radial velocity monitoring (from Magellan--E, Gemini--GMOS, VLT--X-Shooter, Keck--HIRES, ESO--FEROS, and Keck--ESI) validated and refined the Gaia orbit solution (Figure~\ref{fig:gaia-black-holes}a,\,b). 
The spectral energy distribution of the visible star is well described by a single star model (a slowly-rotating G~dwarf with \teff\,=\,5900\,K,  $\log g=4.5$, $M_1=0.9M_\odot$). Importantly, no other contribution is required to fit the observed photometry. Joint modeling of the radial velocities and astrometry constrains the companion (black hole) mass to $M_2=9.8\pm0.2M_\odot$. Both papers conclude that the system harbours a massive black hole on an eccentric ($e=0.45\pm0.02$), long-period ($185.4\pm0.1$\,d) orbit, much longer than previously known black hole X-ray binaries. And at a distance of 480~pc, it became the nearest known black hole by a factor of~3. 
Two other similar discoveries followed (Figure~\ref{fig:gaia-black-holes}c): 
Gaia~BH2
\citep{2023MNRAS.521.4323E}, 
and Gaia~BH3
\citep{2024A&A...686L...2G}. 	
All three entered the record books as the nearest known black holes, a distinction previously held by Cygnus~X1 (between 1975--86), and V616~Mon (from 1986 until the discovery of Gaia~BH1 in 2023). Also noteworthy are their large orbital periods, longer than any other black hole binaries. 
Follow-up searches have failed to detect radio or X-ray emission in any of the three
\citep{2023MNRAS.518.1057E,		
2023MNRAS.521.4323E,			
2024ApJ...973...75C,			
2024ATel16591....1G,			
2024ATel16832....1S,			
2024PASP..136b4203R}.			

Gaia~BH3 has attracted particular interest because of its high mass, $33M_\Sun$, more than any other Galactic stellar black hole. This is consistent with the fact that gravitational waves from black-hole mergers have pointed to a population of extragalactic black holes in short-period binaries with masses $30-85M_\Sun$ 
\citep{2020ApJ...900L..13A,
2021ApJ...913L...7A}.
They are, nevertheless, more massive than predicted by most stellar evolution models, in which stars with an initial mass $M>30M_\Sun$ lose most of their mass due to strong winds, resulting in black hole masses below $\sim\!\!20M_\Sun$
\citep{2007ApJ...662..504B,
2008NewAR..52..419V,
2016ApJ...821...38S}. 
That these more massive objects are the remnants of massive metal-poor stars 
\citep[e.g.][]{2010ApJ...714.1217B}
appears to be supported by the low metallicity of Gaia~BH3's companion star, [Fe/H]\,=$-2.56\pm0.11$, and observations that suggest that it is associated with the ED-2 halo stream
\citep{2024A&A...686L...2G,	
2024A&A...687L...3B}.		
The formation of these systems still remains a challenge for standard evolutionary models for isolated binaries, with various other formation channels, including in triple systems or open clusters, noted in the discovery papers
\citep{2023MNRAS.518.1057E,		
2023MNRAS.518.1057E},   		
and developed in a number of subsequent works
\citep[e.g.][]{2023ApJ...958...26H,	
2024PASP..136a4202N,	
2024A&A...690A.144I,	
2024A&A...688L...2M, 	
2022PhRvD.106l3010L,	
2023MNRAS.524.4083P,	
2024ApJ...964...83G,	
2024ApJ...975L...8L,		
2024MNRAS.535L..44G,	
2025OJAp....8E..79T}.	

Other Gaia {\it candidates} include $\beta$~Cyg (Albireo)		
\citep{2018A&A...620L...2B,	
2021MNRAS.502..328D,		
2022A&A...661A..49J},		
and the Gaia~DR3 objects
G5870
\citep{2023ApJ...946...79T},	
and
G3425
\citep{2024NatAs...8.1583W}.	
And future Gaia releases, DR4 and DR5, may identify dozens of similar systems
\citep{2018ApJ...861...21Y,	
2020PASJ...72...45S,		
2022ApJ...928...13S,			
2022A&A...658A.129J,		
2022ApJ...931..107C,		
2023MNRAS.518.1057E,
2023ApJ...953...52S},		
so furthering the understanding of cosmological black hole growth
\citep{2023A&A...673L..10A,	
2023ApJ...944L..31F,		
2023ApJ...943..133F}.		

\paragraph{Intermediate-mass black holes}
Intermediate-mass black holes, $10^2-10^5M_\Sun$, are those too massive to have formed by single star collapse, but lacking the environment to form a supermassive black hole. They may form through massive star mergers in globular clusters 
\citep[e.g.][]{2002MNRAS.330..232C, 
2021ApJ...908L..29G},
or perhaps as primordial objects
\citep[e.g.][]{2018JCAP...05..017B}. 
While none are definitively known, there are numerous candidates: seen as ultra-luminous X-ray sources in globular clusters, as X-ray accretion disks in active galaxies
\citep[e.g.][]{1999ApJ...522..935G,
2007Natur.445..183M,
2018ApJ...863....1C,
2020ApJ...892L..25L},
or as gravitational wave events 
\citep{2021ApJ...908L..29G}.
In globular clusters, Gaia's accurate proper motions are being used to model the velocity dispersion in their outer parts and so infer the central mass profile, as carried out using Gaia EDR3 and HST data for (the core-collapsed) NGC~6397 and (the non core-collapsed) NGC~3201 
\citep{2022MNRAS.514..806V}.	  
There is some Gaia-based evidence for an intermediate-mass black hole in the nearby globular cluster M4 (at 6\,kpc, and of age 12.2\,Gyr), where the velocity dispersion is now particularly well-defined in the outer parts by the Gaia EDR3 data 
\citep{2023MNRAS.522.5740V}.	
The authors conclude that the dark central mass, $800\pm300M_\Sun$, may be either an intermediate-mass black hole, or a compact black hole population. 

\paragraph{Dynamical effects in open clusters}
Models of binary black hole mergers show that, in order to explain recent gravitational wave detections, a significant fraction of stellar-mass black holes must receive only negligible natal kicks. This implies that black holes should be retained even in open clusters with low escape velocities.  
Searches for dynamical evidence of such black holes in the Hyades cluster have been made using Gaia EDR3  \citep{2023MNRAS.524.1965T}.		
They compared N-body models with the radial profile of Hyades members, using masses derived from Gaia DR3
\citep{2022MNRAS.512.3846E}. 	
The observations were best reproduced by models with 2--3 black holes still in the cluster today.
Models that never possessed black holes would have a half-mass radius significantly smaller than observed, while models in which any remaining black holes were ejected within the last 150\,Myr can still reproduce the observations. In these model, the ejected (binary) black holes are at typical distances of 60~pc from the cluster centre, and $\sim$80\,pc from the Sun. 
In half of their models, the black holes were in binary systems with stellar companions. But their periods peak at $\sim\!\!10^3$\,yr, and are therefore unlikely to be found from radial velocity variations. They identified 56 binary candidates on the basis of their large astrometric and spectroscopic errors, none of which were consistent with a massive compact companion.  
Nonetheless, one conclusion of their study is that the nearest stellar-mass black holes to the Sun are in, or near, the Hyades cluster.

\paragraph{Dynamical effects in halo streams}
In the hierarchical merger scenario of galaxy formation, galaxies grow through the mergers of smaller galaxies. Tidal forces slowly disrupt the accreted systems, forming the tidal tails of still-extant progenitors, or debris stellar streams, roughly aligned with the progenitor's orbit. Today, some 100 such ancient stellar streams are known in the Galaxy halo (Section~\ref{sec:halo-streams}). The thickest, up to several kpc in width in the case of Sagittarius, are likely to represent the tidal debris of captured dwarf galaxies. Of the several dozen thinnest streams, $\lesssim100$\,pc, none has a known progenitor, and their small velocity dispersion suggests that they instead originated from (dark matter-free) globular clusters. Models suggest that the presence of stellar mass black holes in the progenitor will affect the stream's morphology and kinematics, such that detailed observations will offer the prospects of distinguishing between the different possible progenitors. 
For the 20\,kpc distant globular cluster Palomar~5, this sort of modelling has been used to suggest that both the cluster's sparseness, and its tidal tails, can be explained by a stellar-mass black hole population today comprising 20\% of its present mass. Possibly formed with a `normal' black hole mass fraction, of a few per cent, stars were lost at a higher rate, and the black hole fraction gradually increased, further enhancing tidal stripping and tail formation
\citep{2021NatAs...5..957G}.
Subsequent N-body simulations have suggested that observations focusing on the cluster's velocity dispersion, and on binaries with periods of $10^4-10^5$ days in its inner and tail regions, will best constrain the black hole population 
\citep{2024MNRAS.527.7495W}. 
Further simulations of streams originating from star clusters with and without black holes suggest that progenitors having a high mass-loss rate are probably rich in black holes
\citep{2025MNRAS.538..454R}. 
Although Gaia has not yet contributed to such stream characterisation, this may be expected with future data releases.

\subsubsection{Planetary nebulae}

Planetary nebulae are the (misleadingly named) dust and gas shells ejected during the late asymptotic giant branch phase for stars of $1-8M_\odot$ (more massive stars terminate in a supernova explosion), the ejecta being ionised by the radiation from the hot, evolving central star. During the red giant phase, the He~core is contracting and heating, and the outer layers are expanding and cooling, the process continuing until core temperatures reach $\sim$$10^8$\,K, initiating core He fusion, and leading to an inert~C and O core with a He- and a H-burning shell.  He~fusion rates are extremely temperature sensitive, leading to unstable pulsational feedback. This successively ejects more of the stellar atmosphere, progressively exposing deeper and higher temperature core regions. At exposed temperatures of \teff\,$\sim30\,000$\,K, the dense ultraviolet photon flux ionises the ejected atmosphere, illuminating the spectacular planetary nebula. As a consequence of these complex processes, planetary nebulae provide important probes of stellar evolution, and of cosmic chemical enrichment
\citep[e.g.][]{2010ApJ...714.1096S}.
But determining accurate distances has always been problematic, and Gaia's contribution was accordingly expected to be significant.

Today, some two thousand Galactic planetary nebulae are known 
\citep[e.g.][]{1992secg.book.....A,
2006MNRAS.373...79P},
amongst some 20\,000 estimated in our Galaxy 
\citep[e.g.][]{2017AstL...43..304B}. 	
Their distances, of a few hundred parsec or more, are typically beyond reach of ground-based parallaxes, although the USNO CCD programme, at $\sim$0.5\,mas accuracy, measured some~20
\citep{2007AJ....133..631H}. 
Accordingly, other distance estimates have been used, including those based on 
surface brightness, interstellar Na~D lines, spectroscopy, 
or assumptions about the nebula's physical radius,	
or on the hypothesis that there is a fixed value of the ionised nebula mass. 
A pre-Hipparcos review gave parallaxes for just 9~objects, with distances between 130--752\,pc, and central stars with $V=12.3-16.6$\,mag
\citep{1997IAUS..180..483P,
1998A&A...337..253A}.
Twenty five years later, a little before the launch of Gaia, it was suggested that out of some 1800 Galactic planetary nebulae then known, only about 40 had distances determined with reasonable accuracy
\citep{2008ApJ...689..194S}.
Meanwhile, many in the Large Magellanic Cloud, typically unresolved from the ground, have been observed with HST. This allows accurate measurement of their apparent radii, in turn providing one of the best pre-Gaia proxies for the planetary nebulae distance scale
\citep{2001ApJ...548..727S,
2008ApJ...689..194S}.

The space motions of the central stars are also of great interest, with 234 proper motions given by
\citet{2008A&A...479..155K}. 
As examples, Sh~2--68 has a tail extending over 45\,arcmin, formed by matter stripped off the main nebula in the course of the central star's trajectory along its Galactic orbit
\citep{2002A&A...381L...9K}.
Distances and proper motions are essential in classifying objects as thin or thick disk
\citep{2004A&A...420..207K},
establishing their vertical distribution and population scale heights
\citep{2017AstL...43..304B}
and, in the case of Sh~2--174, confirming that the central star is located outside the main nebula as a result of its displacement over 10\,000~years
\citep{2004A&A...420..207K}.
Furthermore, the majority of planetary nebulae are not spherically symmetric, and the wide variety of shapes and features are variously attributed to binary central stars (or substellar companions), stellar winds, and magnetic fields. 

Gaia parallaxes provided the first opportunity to establish the distances of a large number of planetary nebulae, and to examine the distance scales that have been previously based on various proxies.
Gaia DR2 resulted in various preliminary findings for several hundred central objects
\citep{2018A&A...616L...2K,
2019A&A...625A.137S,
2019A&A...630A.150G,
2020A&A...638A.103C}.
Broadly, parallaxes compared well with earlier USNO and HST astrometry, while distances inferred by other methods were generally overestimated. 
\citet{2019A&A...630A.150G} inferred that their sample of 1571 central stars suggested completeness to $\sim$2.3\,kpc (with others beyond 4~kpc), along with a scale-height $168^{+27}_{-62}$~pc, and a space density $\rho=64^{+47}_{-18}$~kpc$^{-3}$. They estimated a total of 21\,000 planetary nebulae in our Galaxy, with an implied birth rate $3\times10^{-3}$\,kpc$^{-3}$\,yr$^{-1}$.

Further refinements came with Gaia EDR3
\citep{2021A&A...656A..51G,
2021A&A...656A.110C,
2022RAA....22h5013A}. 
From an initial sample of 2035 nebulae, the central star identification and EDR3 distances were found for 1725, while the 405 most accurate provided Galactic distribution, radius, and kinematic age, and for a subset of 74, their evolutionary state (mass and age) derived from their luminosities and effective temperatures from evolutionary models
\citep{2021A&A...656A..51G}.
Other studies have examined the binary nature of the central stars, and its link to morphology \citep{2019PASA...36...42M,
2020A&A...644A.173G,
2021A&A...648A..95C}.
Based on Gaia DR3, 82 central stars were classified as probable binaries, of which 24 had been previously classified as close binaries, and 58 were new candidates
\citep{2023RAA....23d5006A}.

Four planetary nebulae are known in Galactic globular clusters, and two in the Fornax dSph, their occurrence being of interest both as distance calibrators, and for the wider study of more distant populations. Gaia is confirming their cluster membership, and identifying new candidates
\citep{2020AJ....159..276B,
2019ApJ...884L..15M}.
Several dozen other papers to date discuss individual objects based on the Gaia distances or space motions
\citep[e.g.][]{2018A&A...620A..84B,
2019A&A...626A..10G,
2019MNRAS.482.4951B,
2022ApJ...935L..35F}.

I would conclude this summary by stating that there was much uncertainty, pre-Gaia, in the underlying physics of planetary nebulae due to the absence of accurate distances and space motions. More quantitative testing of models of this important phase of stellar evolution are being enabled by the Gaia data.

\subsubsection{Tip of the red giant branch}
\label{sec:trgb}

In establishing the cosmic distance scale, various `standard candles' have been identified over recent decades, amongst them the Cepheid, RR~Lyrae, and Mira variables, and Type~1a supernovae. While Cepheids are the leading local distance indicators (Section~\ref{sec:cepheids-h0}), the `tip of the red giant branch' (TRGB) offers another promising calibrator. Unlike the Cepheid scale, which only applies to (Pop~I) systems with recent or ongoing star formation, the TRGB method can be used wherever evolved metal-poor red giant branch stars are sufficiently abundant to allow its location to be defined precisely 
\citep{2018SSRv..214..113B,
2021ApJ...908L...5S}.
Specifically, it can reach the galaxy hosts of SN Type~Ia (e.g.\ M101, M106, NGC~1448, IC~1613), and so contribute to determining the Hubble constant
\citep[e.g.][]{2019ApJ...885..141B, 
2021ApJ...915...34H,		
2021ApJ...906..125J,	 	
2023NatAs...7..590H}. 		
The method originated with observations by Baade (1944) and Sandage (1971), the latter suggesting that low-mass stars, evolving up the red giant branch, abruptly halt their increasing luminosity at the moment of core He-ignition, resulting in a sharp discontinuity in the star's evolutionary track. 
\citet{1983ARA&A..21..271I} showed that, for low-mass stars, the bolometric luminosity at this core He `flash' varies by only $\sim$0.1\,mag over ages of 2--15\,Gyr. Recent models continue to confirm the method's potential
\citep{2017A&A...606A..33S,	
2022MNRAS.514.3058S},		
and more quantitative methods have been developed to estimate the tip's location
\citep[e.g.][]{2023ApJ...954L..31S, 
2024ApJ...963L..43A}. 	

Determining $H_0$ involves determining absolute distances to galaxies that host SN~Ia events, but which are also close enough to have their distances measured (whether by TRGB or Cepheids), then using the SN~Ia luminosities to infer distances for a sample of galaxies far enough into the Hubble flow that their peculiar velocities are a small fraction of the cosmological recessional velocities. Complications include the effects of age, metallicity, and extinction
\citep{2021ApJ...919...16F,	
2023JCAP...11..050F, 		
2024IAUS..376...30L,		
2024arXiv240317048L}.		
Establishing the zero-point of the TRGB colour--magnitude diagram requires some more fundamental distance measure. The latest CCHP determination uses a distance modulus of 18.477~mag for the LMC based on 20 detached eclipsing binaries 
\citep{2019Natur.567..200P}, 
to yield an $I$-band TRGB absolute magnitude of $M_I=-4.049$~mag. A similar value is found using the Megamaser-anchored distance to NGC~4258 
\citep{2021ApJ...906..125J}. 
The latest CCHP collaboration value, 
$H_0=69.8\pm0.8\,({\rm stat})\pm1.7\,({\rm sys})$\kmsmpc\
\citep{2019ApJ...882...34F,	 
2021ApJ...919...16F},		 
where estimates include both statistical (`stat') and systematic (`sys') terms,
appears compatible with both Cepheid and Planck values,
while the latest by the CATS collaboration, anchored to NGC~4258, gives $H_0=73.22\pm2.06$\kmsmpc, favouring the Cepheid value
\citep{2023ApJ...954L..31S}. 
\citet{2019ApJ...882...34F} note that their {\it `ultimate goal for the absolute calibration is the geometric parallax measurements for Milky Way red giant branch stars being obtained by Gaia'}.

In the Gaia data, the `tip' manifests itself as a discontinuity in the population's luminosity function, or colour--magnitude diagram. It can be seen as a prominent feature of the $G$ versus $G_{\rm BP}-G_{\rm RP}$ colour--magnitude diagram for the central region of the LMC
\citep[e.g.][Figure~2]{2021A&A...649A...7G}.	
Gaia's primary contribution to the determination of $H_0$ is by determining the TRGB absolute magnitude from the parallaxes of various Milky Way populations.
Early Gaia DR2 results showed that the high Galactic latitude colour--magnitude diagram, drawn from the Galaxy's thick disk and inner halo, is consistent with the CCHP calibrations 
\citep{2019PASA...36....1M}.	
In a wider Gaia EDR3 study of the globular cluster $\omega$~Cen, 
\citet{2021ApJ...908L...5S}	
estimated an $I$-band TRGB magnitude $M_I=-3.97\pm0.06$~mag, fainter by 0.07~mag than that used by \citet{2019ApJ...882...34F}, and yielding $H_0=72.1\pm2.0$\kmsmpc, closer to the Cepheid value.  
Other values have been derived from 
field giants at high Galactic latitude, $M_I=-3.91\pm0.05\,({\rm stat})\pm0.09\,({\rm sys})$~mag
\citep{2022ApJ...939...96L};	
from Gaia synthetic photometry, $M_I=-3.970^{+0.042}_{-0.024}\,({\rm stat})\pm0.062\,({\rm sys})$~mag
\citep{2023ApJ...950...83L};	
from high Galactic latitude halo stars, $M_I=-4.042\pm0.041\,({\rm stat})\pm0.031\,({\rm sys})$~mag
\citep{2023MNRAS.523.2283D},	
and with results from 33 Galactic globular clusters showing a dependence on metallicity for for [Fe/H]\,$>‑1.2$
\citep{2025ApJ...980..218S}.	

Stars near the TRGB are typically regarded as non-variable, but Gaia has illuminated an additional complication. Variability data for LMC stars from OGLE, but using Gaia DR3 to remove foreground and blended stars (and to provide synthetic photometry in the HST--ACS/F814W passband), suggests that all stars near the TRGB are rather small-amplitude red giants (SARG), that follow several period--luminosity sequences
\citep{2024ApJ...963L..43A}.  
They concluded that this variability population diversity affects the TRGB at a level exceeding the stated precision, and finally derived $M_I=-4.025\pm0.014\,({\rm stat})\pm0.033\,({\rm sys})$~mag.

\subsection{Heavy element production}
\label{sec:heavy-element-production}	

\paragraph{Background}
The term `chemically peculiar' generally refers to hot main-sequence stars with distinct patterns of metal and helium abundances, at least in their surface layers. These have generally resulted from post-formation processes, such as diffusion, levitation, magnetic field effects in their outer layers, deep convective `dredge up' at later evolutionary stages, or mass transfer. I have touched on some examples, notably carbon stars, in Section~\ref{sec:specific-evolutionary-states}.
In contrast, the `heavy elements' seen in high-resolution spectra reflect the composition of the interstellar medium from which the stars formed. But an understanding of their origin, especially in the case of the heavier elements, remains highly incomplete
\citep[e.g.][]{2021RvMP...93a5002C,	
2024A&A...686A.295F,	
2025A&A...699A..41S}.	

In brief, elements heavier than iron (found in halo stars, the interstellar medium, dust grains, meteorites, and Earth), are mainly formed by the process of neutron-capture. 
At low neutron densities, this heavy element production occurs via {\it slow\/} neutron-capture (termed the s-process).  The process is `slow' in the sense that there is sufficient time for possible radioactive decay to occur before another neutron is captured.
Extending over thousands of years, it occurs in the He-burning layers of asymptotic giant branch stars, and during the He- and C-burning phases of massive stars, seeded by iron~nuclei from earlier supernovae
\citep{1999ARA&A..37..239B,
2011RvMP...83..157K}. 

High neutron densities enable heavy-element nucleosynthesis by {\it rapid\/} neutron-capture (the r-process).  This involves a succession of neutron captures by heavy seed nuclei. The process is rapid in the sense that the nuclei cannot have time to undergo decay before another neutron is captured. It is inferred to occur over timescales of seconds, and is therefore presumably restricted to explosive environments. Possible sites include core-collapse supernovae, and neutron star--neutron star or neutron star--black hole mergers
\citep{
2014JPhG...41d4008M,	
2019ApJ...875..106C,	
2021RvMP...93a5002C,	
2023ApJ...944..123V,	
2024PhRvL.133x1201W}.	
This NASA
 \href{https://apod.nasa.gov/apod/ap171024.html}{Astronomy Picture of the Day} 
provides a visual synopsis of the likely origin of the elements.
 
Constraints on these processes can be gained from the overall atomic mass distribution in specific objects. High-resolution optical spectroscopy provides access to some 20--30 elements heavier than the iron-group in late-type (FGK) stars. A dozen others (including Ge, As, Se, Cd, Te, Lu, Os, Ir, Pt, Ag, and Pb) are accessible to near-ultraviolet spectroscopy, for example using HST--STIS 
\citep[e.g.][]{1998ApJ...496..235S,	
2005ApJ...627..238C,	
2012ApJ...750...76R}.	
In this section, and selected from many detailed studies, I will look at just two examples of specific heavy elements where insights are benefiting from the Gaia data: cerium, and the actinide elements. 

\paragraph{Cerium and the Galaxy infall history}
Cerium, atomic number 58, is the most abundant element of the lanthanide series, and despite its classification as one of the rare-Earth elements, is not at all rare in the Earth's crust; at 66\,ppm, it lies (just) behind copper, and is far more abundant than lead or tin. Its two most commonly occurring isotopes are $^{140}$Ce, produced in both the s- and r-processes, and $^{142}$Ce, produced only in the r-process
\citep[e.g.][]{2009ApJS..182...51L}.
%
While the formation sites of the r-process elements are still somewhat uncertain (and may include, as noted above, the cataclysmic merging of neutron stars, or of neutron star--black hole binaries), those of the s-process elements are better (if still imperfectly) understood.  Of interest to us here, the `main' s-process elements such as Ce and Nd are produced in low- and intermediate-mass asymptotic giant branch (AGB) stars, making use of neutrons produced by the $^{13}$C\,($\alpha$,\,n)\,$^{16}$O reaction.  
Meanwhile, massive stars ($8-10M_\Sun$) are mostly responsible for the `weak' s-process, producing Sr, Y, and Zr. Here, the neutrons are mainly provided by the $^{22}$Ne\,($\alpha$,\,n)\,$^{25}$Mg reaction in convective He-burning cores and C-burning shells.
Low-metallicity, low-mass AGB stars can produce the elements of the third peak (such as Pb) through the `strong' s-process.

As a result, studying the Ce content of stars across the Galaxy provides a probe of these different production sites. Various studies over the past decade have typically found flat trends of [Ce/Fe] versus [M/H], perhaps with a small decreasing trend for large [Fe/H] values.  Others have suggested a decreasing cerium abundances with age, at least up to about 8~Gyr, both for open clusters and field stars, and an increasing abundance with age for stars older than  8~Gyr.

Gaia~DR3 contains a homogeneous analysis of millions of high-quality stellar spectra, obtained by the radial velocity spectrometer (RVS), reduced and calibrated on the ground within the Gaia Data Processing and Analysis Consortium (DPAC) by Coordination Unit~6, and analysed by the GSP--Spec module of Coordination Unit~8, responsible for the source classification and `parameterisation' (Section~\ref{sec:classification-stellar-properties}).
This leads to abundance estimates for up to 13~individual elements in the atmospheres of several million stars, and the possibility for unprecedented chemical mapping of our Galaxy. 
And among the abundances derived in GSP--Spec are three of the heavy-elements produced by these neutron-capture processes: Ce, Zr and Nd. In common with all Gaia abundances, these are estimated by comparing the mean observed spectra with an extensive grid of synthetic spectra covering a wide range of stellar atmospheric parameters, and with varying abundances. 
The example spectrum shown in Figure~9 of \citet{2023A&A...674A..29R} is centred on the triplet of Ce\,{\scriptsize II} lines around 851.375~nm, and has a derived Ce~abundance compared to iron, [Ce/Fe]\,=\,0.26$^{+0.12}_{-0.08}$~dex.

A first analysis of the Gaia Ce data was made by 
\citet{2023A&A...670A.106C}.
Among the 5.5~million stars parameterised by GSP--Spec, 103\,948 have a derived cerium abundance, more than doubling the numbers previously available. Selection of a high-quality subset according to the flags provided, their resulting sample comprises about 30\,000 Ce abundances distributed in the disk and halo components. The next step is to use the surface gravities determined by the Gaia classification and parameterisation process to establish their nature, and the distances and proper motions to examine their dependencies on location or kinematics within the Galaxy.

From the former, 
\citet{2023A&A...670A.106C}
first established that the sample is mainly composed of stars with $\log(g)<3.5$, that is both red giant branch and asymptotic giant branch stars. 
%
From the astrometry, they found that the closest, being more Ce-poor, more metal-rich, and concentrated within $\pm500$\,pc from the Galactic plane, probably belong to the thin disk population. Stars with higher Ce abundances are at the same time generally metal-poor, and preferentially located at larger distances from the Sun, and at larger distances from the Galactic plane.
%
These hints are confirmed by their kinematic behaviour: the diagram of Ce abundance with respect to metallicity, colour-coded with the eccentricity of their Galactic orbits (their Figure~4a), shows that stars with higher metallicity and low Ce~abundance are on more circular orbits typical of thin disk stars, while those with higher Ce abundances are on more eccentric orbits, and so unlikely to belong to the thin disk population.
%
More detailed examination of their orbital angular momentum also reveals a small number of lower angular momentum halo stars. And, interestingly, eleven of these stars are associated with previously identified halo substructures (Section~\ref{sec:halo-streams}), notably the Helmi~(2), Thamnos~(2), and Enceladus~(7) streams. 
The data was also used to map out the [Ce/Fe]~gradient as a function of Galactic radius, and the vertical gradient as a function of the distance from the Galactic plane.

These distributions of cerium throughout the Galaxy provide support to some of the latest models describing its chemical evolution
\citep[e.g.][]{2024ApJ...965..119S}.	
In their `two-infall' model, designed to reproduce the APOGEE DR16 results of [X/Fe] versus [M/H] abundance ratios (where X stands for various $\alpha$-elements in the solar neighbourhood),
\citet{2021A&A...647A..73S}
concluded that high- and low-$\alpha$ sequence stars were formed by two independent episodes of gas infall. 
More complexity was evident with Gaia DR3. 
\citet{2023A&A...674A..38G}
identified the presence of young metal-poor stars with low [$\alpha$/Fe] values, and
\citet{2023A&A...670A.109S}
then showed that this low-$\alpha$ population itself can be modelled by two sequential infall episodes, resulting in a total of three discrete infall events.
The first, the high-$\alpha$ infall, had an accretion timescale of 0.103\,Gyr. 
The second, termed low-$\alpha$ part~I, had an accretion timescale of 4.110\,Gyr, and with a delay between the first and second of 4.085\,Gyr.
The third, low-$\alpha$ part~II, had an accretion timescale of 0.150\,Gyr, and with a delay between the second and third of 11\,Gyr, started some 2.7\,Gyr ago.

\paragraph{Actinide boost stars}

The actinides are a series of 15 radioactive metallic elements with atomic numbers 89 to 103: actinium (from which they take their name) to nobelium. Naturally occurring (and long-lived) uranium (U) and thorium (Th), and synthetically produced plutonium (Pu), are the most abundant. Like the lanthanides (which include cerium), they form a family with similar if wide-ranging properties, and they play a role in several areas of astrophysics.
Perhaps most importantly, nucleo-cosmochronology, which started with the work of 
\citet{1960AJ.....65R.345F}, 
exploits the very long half-lives of $^{232}$Th, $^{235}$U and $^{238}$U to derive ages within the solar system and beyond
\citep[e.g.][]{1991PhR...208..267C,		
2000PhR...333....1M,		
2001A&A...379.1113G}.	

Determining nuclear-based {\it stellar\/} ages developed from the first detections of Th and U in a number of very metal-poor stars, amongst them the 
`neutron-capture-rich' giant CS~22892--052
\citep[CS: Curti--Schmidt;][]{1995AJ....109.2757M,	
1996ApJ...467..819S,	 	
2003ApJ...591..936S},	
and the r-process rich CS~31082--001
\citep{2001Natur.409..691C,	
2002A&A...387..560H}.	
Incidentally, \citet{2002A&A...387..560H} gave abundances for 44 elements, while \citet{2003ApJ...591..936S} gave abundances or upper limits for~57.
Shorter-lived actinides have also been detected in Galactic cosmic rays, providing constraints on their acceleration sites
\citep[e.g.][]{1998Natur.396...50W,		
2001AdSpR..27..785O,	
2003ApJ...591..228L}.		

As detailed by 
\citet{2020JPhCS1668a2020H},	
the relative level of (lanthanide) Eu to Fe abundance in a star is a proxy for how much r-process enhancement preceded the
star’s formation, compared to the chemical evolution of that gas from supernova events. Stars over-enhanced with Eu relative to Fe compared to the Sun are referred to as `r-process enhanced', and are divided into two categories: 
r-I, with 0.3\,<\,[Eu/Fe]\,$\le$\,1.0 (i.e., a factor 2--10 greater than solar), 
and 
r-II, with [Eu/Fe]\,>\,1.0 (i.e., more than a factor 10 greater than solar). 
As underlined by 
\citet{2020JPhCS1668a2020H},	
these stars are relics of prolific r-process event(s) that occurred before the gas was enriched by supernovae, and are therefore considered tracers of nearly pure r-process events.

Of the variations in elemental abundances of stars enhanced with r-process elements, the variation in the actinide-to-lanthanide ratio is one of the most significant. Stars with an over-abundance of [Th/Eu] relative to the solar system are widely referred to as `actinide-boost' stars, occurring in some 30\% of r-process enhanced stars
\citep[e.g.][]{2020JPhCS1668a2020H}.	
%
By 2014, just six such stars were known:
CS~31082--001 (as above),
CS~30306--132 \citep{2004ApJ...607..474H},	
CS~31078--018 \citep{2008ApJ...681.1524L}, 
HE 1219--0312 \citep{2009A&A...504..511H},
CS~30315--029 \citep{2014A&A...565A..93S},
and 
HE~2252--4225 \citep{2014A&A...569A..43M}.
Other discoveries since include
2MASS J09544277+5246414, identified from LAMOST \citep{2018ApJ...859L..24H},	
and
SPLUS J142445.34-254247.1, identified from the 12-filter S--PLUS Survey
\citep{2023ApJ...959...60P}.	

A key question is then whether such high actinide abundances result from conditions different to those in other r-process enhanced metal-poor stars. For example, does it depend on how neutron-rich the ejecta from an r-process event may be?
While still considering that {\it `the astrophysical production site of the heaviest elements in the Universe remains a mystery'},
\citet{2019ApJ...881....5H}		
concluded that all observed levels of actinide enhancements, including those of the `actinide-boost stars', do not necessarily demand distinct r-process progenitors if such a process (such as a neutron star merger) can generate sufficiently wide variations in neutron flux
\citep{2020JPhCS1668a2020H}.		

Gaia enters the discussions, by providing significant distances and proper motions, with the discovery of three of the latest r-process enhanced stars.
%
For the $G=13.8$\,mag SPLUS J1424--2542, with an estimated age of 10.1\,Gyr,
\citet{2023ApJ...959...60P}   
identified 36 elements, from C to Th. With [Fe/H]\,=\,--3.39, it has the highest known Th/Eu ratio. The Gaia DR3 parallax, $\varpi=0.0796\pm0.0182$\,mas, places it at $1/\varpi=8.13^{+1.41}_{-1.05}$\,kpc, with a (Bayesian) photo-geometric distance 7.82\,kpc. 
From the Gaia proper motions and line-of-sight velocity, they inferred that it belongs to the Galaxy halo, although not to any of the known halo streams.
%
From a comparison of its abundances with Pop~III supernova nucleosynthesis yields, the r- and s-process fractions, and simulations of neutron star mergers, they concluded that its heavy elements resulted from r-process event(s) only, with no contributions from the s-process. 
Remarkably, their models imply that its nascent gas cloud was enriched by at least two progenitor populations: a supernova explosion from a metal-free star of mass $11.3-13.4M_\Sun$, and the aftermath of a binary neutron star merger with masses $1.66M_\Sun$ and $1.27M_\Sun$.

LAMOST J1124+4535 is r-process-enhanced, although not an actinide-boost star.
\citet{2024ApJ...965...79X}	  
derived an age of $\sim$11.3\,Gyr through radioactive-decay dating, on the assumption that the r-process material was produced by a single event immediately preceding the star's formation. The abundances suggest that it was accreted from a disrupted dwarf galaxy similar to the surviving Ursa Minor (UMi) dwarf galaxy. 
From the Gaia DR3 parallax ($\varpi=0.09\pm0.02$\,mas) and proper motion, they infer a distance 7.6\,kpc, and a radial orbit with eccentricity $e=0.53$. The star is associated with an r-II cluster, itself dynamically associated with the LMS--1/Wukong stream
\citep{2023ApJ...946...48H}, 
suggesting that it is the remnant of a disrupted dwarf galaxy with mass similar to that of UMi.

Subaru--HDS spectra of the $V=11.5$~mag, very metal-poor ([Fe/H]\,=\,--2.38), r-process-enhanced ([Eu/Fe]\,=\,0.80), actinide-boost LAMOST J0804+5740, gave abundances for 48 species
\citep{2025ApJ...984L..43L}.	
They evaluated three r-process models: 
a neutron star merger \citep{2015MNRAS.448..541J}, 
a `collapsar' \citep{2024ApJ...966L..37H},
and a magnetorotationally-driven jet supernova \citep{2017ApJ...836L..21N}.
They found that the latter successfully reproduced the actinide-boost signatures.
At a Gaia DR3 distance of $3.05\pm0.18$\,kpc \citep{2022A&A...658A..91A}, and based on the DR3 proper motion and radial velocity, 
\citet{2024ApJ...966..174Z}	
had already shown that LAMOST J0804+5740 is a member of the Gaia Sausage--Enceladus (GSE) stream.  This first such GSE actinide-boost star provides new insights into \mbox{r-process} nucleosynthesis in accreted dwarf galaxies.

From 27~million Gaia DR3 stars, 
\citet{2024MNRAS.527.9767L}	
used their kinematics (and actions) to identify more than 160\,000 {\it ex situ\/} stars, with 0.1\%, 1.6\%, and 63.2\%   members of the thin disk, thick disk, and halo respectively. 
From the same sample,
\citet{2025ApJ...984L..43L}	
found that 2/3 of the known actinide-boost stars are {\it ex situ}, and thus more likely to originate from accreted dwarf galaxies. 
This scenario, in which the r-II stars formed in an extremely neutron-rich environment in which a rare astrophysical event occurred, is supported by the ultra-faint dwarf galaxy, Reticulum~II, where most members are highly enhanced in r-process elements
\citep{2016Natur.531..610J,
2016AJ....151...82R}.	
Other r-enhanced dwarf galaxies accreted by the Milky Way will have deposited many r-II stars in the Galactic halo with similar orbital actions. 

\citet{2023ApJ...946...48H}	
used Gaia DR3 to construct the orbital actions of 161 r-II stars in the solar neighbourhood. 
They found 6 clusters of r-II stars that have similar orbits and chemistry (one was a new discovery). These clusters are therefore inferred to be good candidates for remnants of completely disrupted r-enhanced dwarf galaxies that merged with the ancient Milky Way.

\begin{figure}[t]
\centering
\includegraphics[width=0.31\linewidth]{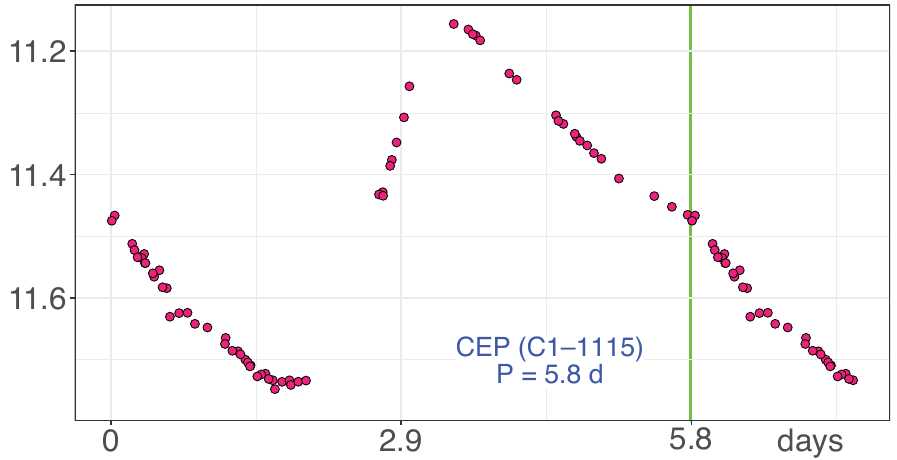}	
\hspace{10pt}
\includegraphics[width=0.31\linewidth]{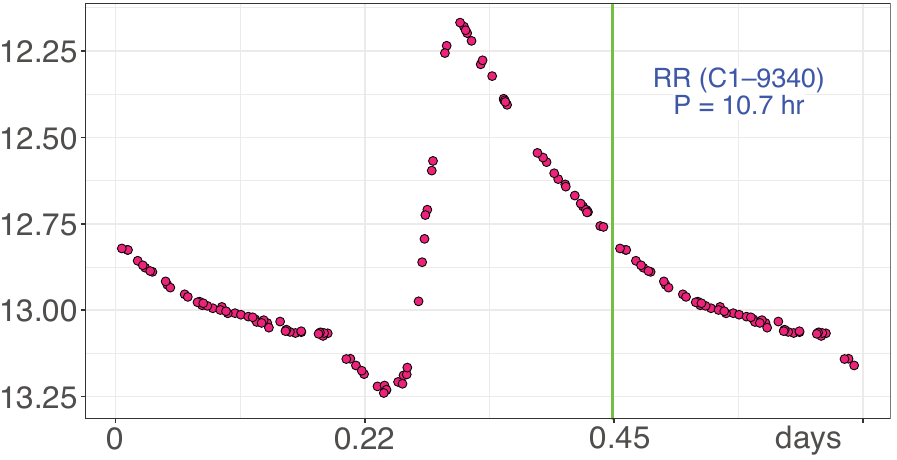}	
\hspace{10pt}
\includegraphics[width=0.31\linewidth]{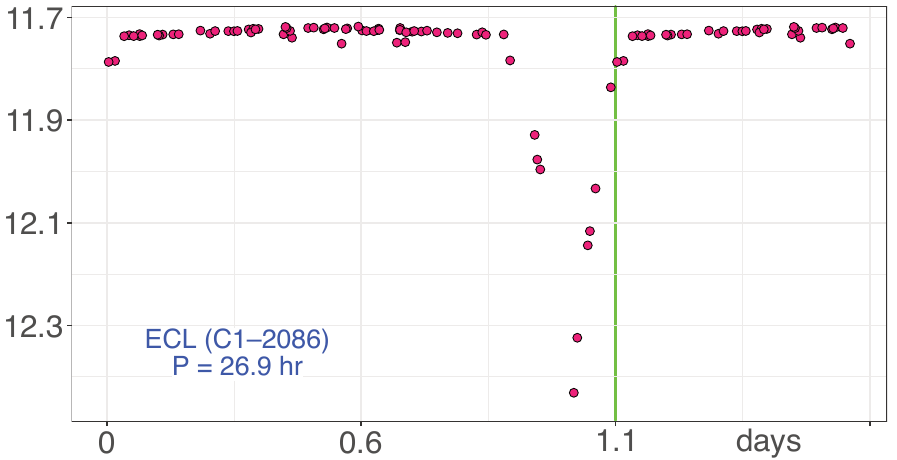} 	
\vspace{-5pt}
\caption{Examples of three (out of more than 10~million) Gaia folded light curves: a Cepheid variable, a RR~Lyrae variable, and an eclipsing binary. The vertical axis is the Gaia $G$ magnitude, the horizontal axis is the observation time (in days) folded at the inferred period (vertical green line). The C1--nnn numbers are the identification numbers assigned within the Gaia--Vari Citizen Science project (Section~\ref{sec:variability-citizen-science}).
}\label{fig:variability-examples}
\end{figure}

\subsection{Variable stars}
\label{sec:variable-stars}

Before Gaia, the number of known variable stars was estimated to be around 150\,000.
As a result of the global variability analysis as part of the Gaia data processing chain
\citep{2018A&A...618A..30H},		
Gaia DR2 released in 2018 (and covering observations between July 2014--May 2016) contained 550\,737 sources classified as variable. These included
228\,904 RR~Lyrae and 11\,438 Cepheids,
151\,761 long-period variables,
147\,535 stars showing rotation modulation, 
8882 $\delta$~Scuti and SX~Phe stars,
and 3018 short-timescale variables (Figure~\ref{fig:variability-examples}).

\begin{figure}[t]
\centering
\includegraphics[width=0.37\linewidth]{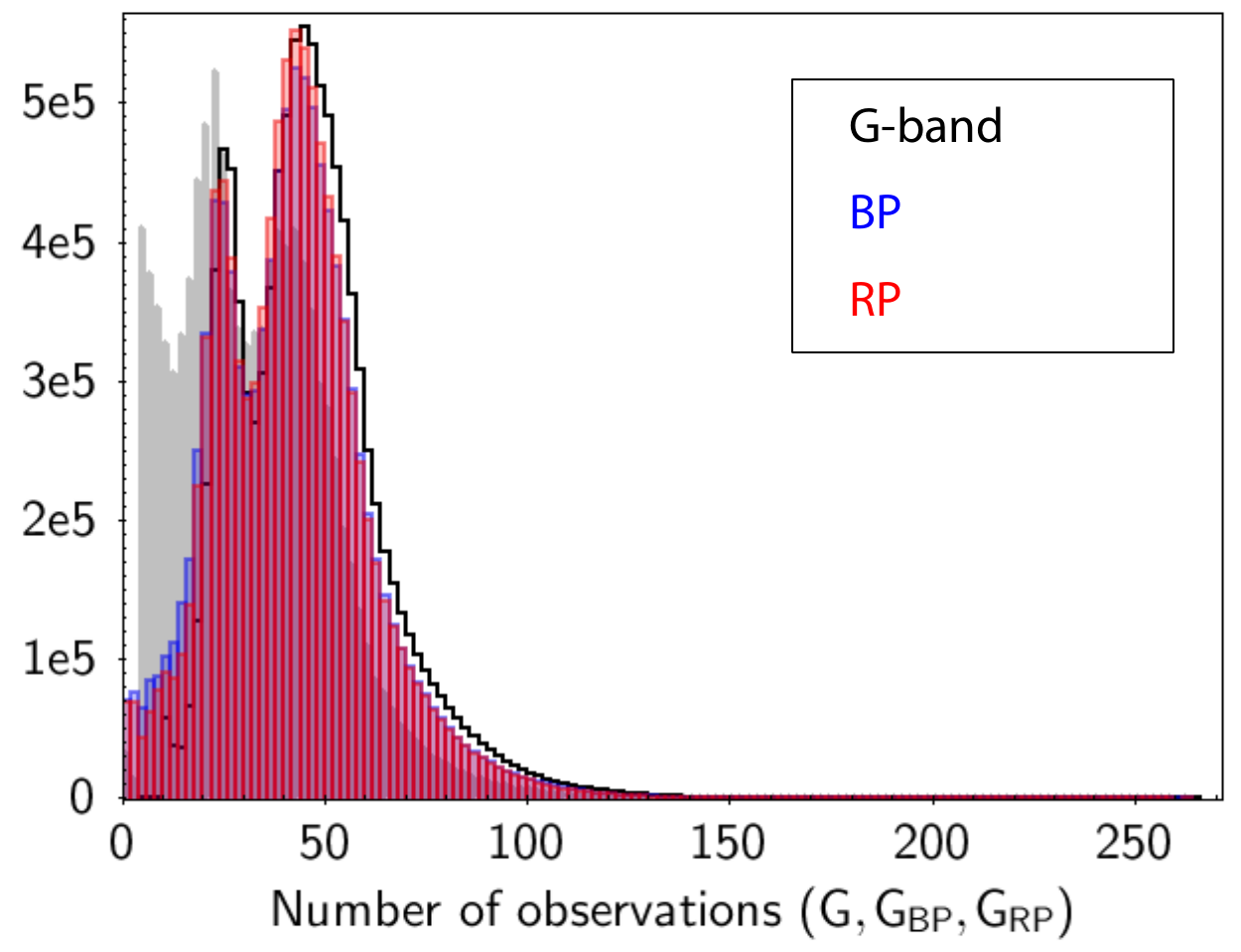}
\hspace{5pt}
\includegraphics[width=0.59\linewidth]{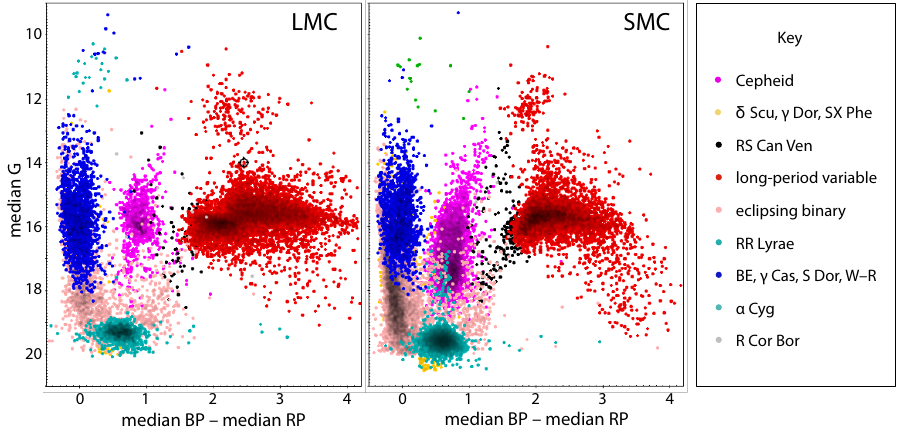}
\vspace{-5pt}
\caption{
Left~(a): number of field of view crossings for the variable stars identified in Gaia DR3, in the $G$ (black), BP (blue) and RP (red) bands. The median numbers are 44, 40 and 41 respectively, although extending up to 265 in the $G$-band. The grey histogram shows numbers for a random sample of DR3 sources, showing that variability analysis favours those with a large number of measurements \citep[from][Figure~1]{2023A&A...674A..13E}.
Right~(b): Gaia DR3 colour--magnitude diagrams for the LMC (left) and SMC (right); \citep[from][Figures 9--10]{2023A&A...674A..13E}.
}\label{fig:variability-lmc-smc}
\label{fig:eyer-variability}
\end{figure}

Gaia DR3, released in 2022 (and covering observations between July 2014--May 2017), included its own variability processing and analysis 
\citep{2023A&A...674A..13E}.
Of the 1.8~billion sources in DR3, 10\,509\,536 were identified as variable (9.5~million stars and one million galaxies/quasars), with associated time series in the photometric bands $G$, $G_{\rm BP}$, and $G_{\rm RP}$ and, in some cases, with the radial velocity time series. The classified variables lie mostly in the range $G=12-21$~mag, with the majority around $G=16-20$~mag, and amplitudes ranging from a few milli-magnitudes to several magnitudes
(Figure~\ref{fig:eyer-variability}a).
After period search and time-series modelling, supervised classification assigned the variables to 24 classes \citep{2023A&A...674A..14R}.
These include
rapidly oscillating Am/Ap,
$\beta$~Cephei,
S~Dor,
Cepheids, 
cataclysmic variables, 
$\delta$~Scuti,
$\gamma$~Dor,
eclipsing binaries, 
ellipsoidal variables,
exoplanets, 
long-period variables, 
microlensing,
RR~Lyrae, 
RS~Can~Ven,
sub-dwarf~B,
supernovae,
solar-like stars, 
slowly pulsating B-stars,
SX~Phe,
symbiotic stars, 
pulsating white dwarfs,
Wolf--Rayet, 
and young stellar objects.
%
Even compared with DR2, the numbers are transformational. There are
297\,778 RR~Lyrae,
16\,141 Cepheids,
1475 $\beta$~Cepheid stars,
2\,352\,775 long-period-variables,
474\,026 showing rotational modulation,
748\,058 $\delta$~Scuti and SX~Phe stars,
1\,934\,844 solar-like,
65\,300 ellipsoidal variables,
7306 cataclysmic variables,
910 white dwarfs,
254 microlensing events,
and 214~exoplanet transits.
\citet{2023A&A...674A..14R} provide further details of the specific modules responsible for the classification of these various variability types.

Variable stars have long been recognised as offering deep insights into stellar structure and evolution. Accurately locating the 10~million variable DR3 sources in the HR diagram reveals the specific occurrences of pulsating, eruptive, and cataclysmic variables, as well as stars that show apparent variability due to stellar rotation or binary eclipses (Figure~\ref{fig:variability-lmc-smc}b).
As detailed in Section~\ref{sec:stellar-rotation}, stellar rotation has been identified for 3~million, and classified according to its origin due to star spots, tidal interactions, or reflection from binary companions (Section~\ref{sec:stellar-rotation}).

The following summarises some of the ongoing Gaia studies for variable stars, but let me stress that, for variable stars in particular, it is far from comprehensive.

\subsubsection{Cepheid variables}
\label{sec:cepheids}

Cepheids are pulsationally unstable stars, located in a narrow region of the upper HR~diagram, with typical periods of 1--30\,days. Classical Cepheids (or $\delta$~Cephei stars) are young high-mass core He-burning supergiants, Population~I objects found in the Galactic plane, notably in spiral arms and open clusters. Type~II Cepheids are low-mass metal-poor Population~II objects found at high Galactic latitudes, in the bulge, and in globular clusters (further sub-divided by period into BL~Her, W~Vir, and RV~Tau-type variables).
The precursors of the classical Cepheids are massive young O and B main-sequence stars. As they evolve rapidly off the main sequence, they pass through a zone, the `instability strip', where their outer atmospheres are unstable to periodic radial oscillations. High-mass stars pass through the instability region at higher luminosities (cooler temperatures) than lower-mass stars, resulting in an instability strip which slants upwards and to the right in the HR~diagram. 

Basic physics considerations lead to the existence of a mass--luminosity relation, and hence also a radius--luminosity relation. However, since neither mass nor radius are easily observable for the majority of stars, the mass--luminosity relation cannot be used to predict luminosities, nor therefore distances. 
The importance of Cepheid variables as distance indicators is that there exists a correlation between period and luminosity, discovered empirically more than a century ago by Henrietta
\citet{1908AnHar..60...87L},			
and subsequently explained theoretically. 
A readable historical review is given by \citet{1969PASP...81..707F}.	
The relationship nevertheless shows a significant scatter about the mean line, even when corrected for reddening, due to the finite (temperature) width of the instability strip. If a colour-term is introduced, the scatter is significantly reduced.
While the Cepheid period--luminosity relation has traditionally provided the most accurate method to derive distances to nearby galaxies, various complications are encountered.

There is an enormous literature on Cepheid variables, and their application to studies of Galactic structure, and the determination of the astronomical distance scale, both within the Galaxy and beyond. 
One main goal has been to establish the slope and zero-point of the period--luminosity relation, such that an observed period yields the object's luminosity and hence its distance.  An important question for their use as distance indicators is whether the relations have the same slopes and zero-points in the LMC/SMC as in the Galaxy. 
The fact that Cepheids can be seen to large distances, and reflect the young Galaxy population, means that they also provide an important tracer of spiral arms, while their proper motions provide an important probe of Galactic rotation. 
Key information is also encoded in the vertical distribution of Cepheids above and below the Galactic plane, and its age dependence. In a simplified picture, Cepheids with a very young age are found preferentially close to the Galactic plane, their assumed birth sites. Evolving in scale height with age as a result of their initial vertical velocity component, they reach their maximum distance and return to the plane after times depending on the local mass density, somewhere in the range of 70--100~Myr.

Hipparcos observed 280~Cepheids, of which 32 are either Type~II (mainly W~Vir) or double-mode. Of the 248 classical Cepheids, 32~are first-overtone pulsators. The mean parallax error of $\sim$1.5\,mas ($d\gtrsim500$~pc) implied limited individual value
\citep{1997MNRAS.286L...1F}.	
The closest is Polaris ($\alpha$~UMi), with $\varpi=7.56\pm0.48$\,mas ($d=132\pm8$\,pc), although it is too bright to appear in current Gaia data releases.
\citet{1997MNRAS.291..683F}	
used 220 Cepheids with Hipparcos astrometry to derive the Oort constants from the first-order expression for Galactic rotation. 

Today, Gaia's high-accuracy parallaxes, combined with the multi-colour multi-epoch precision photometry, makes it extremely powerful for identifying and characterising variability across the entire HR~diagram, and I will consider the following results according to the successive data releases.

\begin{figure}[t]
\centering
\includegraphics[width=0.44\linewidth]{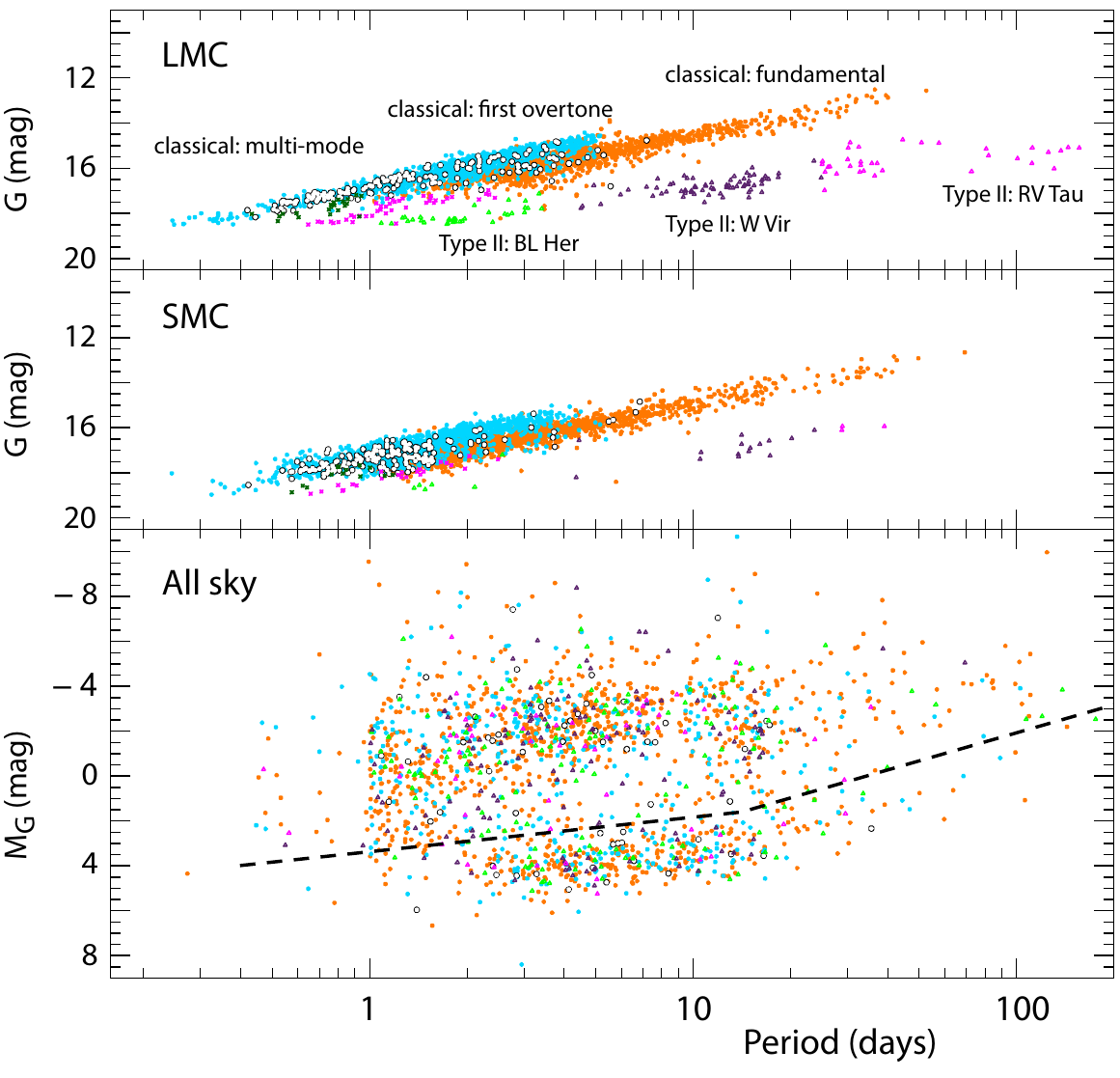}
\hspace{20pt}
\raisebox{3pt}{\includegraphics[width=0.265\linewidth]{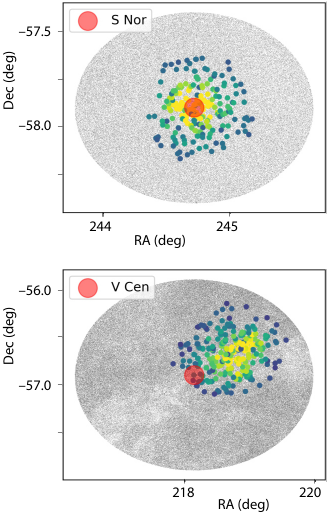}}
\vspace{-5pt}
\caption{Left: the period--luminosity relation for the Cepheids identified in DR2 by 
\citet[][Figure~6]{2019A&A...622A..60C},	
as a function of apparent magnitude for the LMC and SMC, and of absolute magnitude for the others (all uncorrected for reddening). Colours indicate classical Cepheids (sub-divided into fundamental-mode, first-overtone, and multi-mode), and Type~II Cepheids (sub-divided by period into BL~Her, W~Vir, and RV~Tau-type).
Right: sky projections for two of the cluster Cepheids from \citet[][Figure~3]{2023A&A...672A..85C}, showing background stars (grey), colour-coded cluster membership probability (light colours indicating higher probability), and labeled Cepheids (red circles).}
\label{fig:cepheids}
\end{figure}

Already, Gaia DR1 included 599 Cepheids (and 2595 RR~Lyrae stars) in the LMC region, observed at high cadence during the first 28 days as a result of its favourable scanning configuration
\citep{2016A&A...595A.133C}.		
For Gaia DR2 (the first 22~months of the mission), a `Specific Object Study' pipeline was used to validate and characterise Cepheids and RR~Lyrae stars, originally using the period--amplitude and period--luminosity relations in the $G$ band, and subsequently extended to $G_{\rm BP}$ and $G_{\rm RP}$
\citep{2019A&A...622A..60C,		
2019A&A...625A..97R}.			
Gaia~DR2 provides astrometry, mean magnitudes and pulsation characteristics for 9575 Cepheids (Figure~\ref{fig:cepheids}, left), of which 3767 are in the LMC, 3692 are in the SMC, and 2116 are elsewhere (`all-sky'). The majority of those in the Magellanic Clouds were already known from the OGLE survey, although Gaia DR2 includes 118 new objects. The all-sky sample includes Cepheids and RR~Lyrae variables in 87 globular clusters and 14 dwarf galaxies (the Magellanic Clouds, 5 classical and 7 ultra-faint dwarfs), of which 350 are new discoveries. Metallicities derived from the Fourier parameters of the light curves are also given for 3738 fundamental-mode classical Cepheids with periods below 6.3\,d.
A much larger scatter is seen in the all-sky period--luminosity distribution, likely to be a combination of mis-classifications, sources with very high reddening, or the consequences of a simplified treatment of binary/multiple sources. Further and more detailed analyses were subsequently undertaken
\citep{2018A&A...620A.127M,	
2019A&A...625A..14R}.	

In addition to the classical and Type~II Cepheids, the Gaia `Specific Object Study' pipeline 
\citep{2019A&A...622A..60C,		
2019A&A...625A..97R}			
also identifies the less common double-mode Cepheids (which pulsate in two modes, usually the fundamental and first overtone), as well as the shorter-period high-mass `anomalous Cepheids'.
Amongst other studies, the Gaia DR2 Cepheid data was used to characterise the Galaxy's rotation curve 
\citep[e.g.][]{2019ApJ...870L..10M,		
2019MNRAS.482...40K,				
2020ApJ...895L..12A},				
as well the vertical component of the velocity vector
\citep{2019AcA....69..305S},			
and the Galaxy's structure more generally 
\citep{2019Sci...365..478S}.			
Individual masses have been derived from the theoretical mass-dependent `period--Wesenheit' (reddening-free) relations in the various Gaia photometric bands
\citep{2020ApJ...898L...7M}.			
Proper motion anomalies between the Hipparcos and Gaia DR2 positions, caused by close-in orbiting companions, suggest that the binary fraction of classical Cepheids is above 80\%
\citep{2019A&A...623A.116K}.	
%

With Gaia DR3, came a huge quantity of new high-accuracy astrometric and accurate multi-epoch, multi-colour, long-duration photometric information which, despite Gaia's relatively sparse temporal sampling, has allowed for further unprecedented identification of variability across the HR diagram
\citep{2023A&A...674A..36G,	
2023A&A...677A.137M}.		
The former study is mainly focused on the OBAF pulsators on the main sequence, i.e.\ representing the $\beta$~Cep, slowly pulsating~B (SPB), $\delta$~Sct, and $\gamma$~Dor stars, and uncovering a wealth of new information on their nature and extent in the HR diagram.
Dedicated processing of the classical Cepheids, again using the `Specific Object Study' Cepheid/RR~Lyrae pipeline, is detailed by
\citet{2023A&A...674A..17R}.	
It results in a sample of 15\,006 Cepheids (of all types; Table~\ref{tab:data-release-table2}): 4663 in the LMC, 4616 in the SMC, 321 in M31, and 185 in M33. The other 5221 objects fall in the remaining `all sky' subregion, and includes stars in the Galactic field, in open clusters, and in a number of small satellite galaxies. Among this sample, 327 objects were previously known as variable stars but with a different classification, while 474 are new Gaia discoveries.

An important focus of Cepheid studies is to better establish their membership of open clusters and associations (and to identify new open clusters associated with known or newly discovered Cepheids), in order to provide an independent distance. With DR3, different studies have resulted 
in a total of 50 classical Cepheids associated with 45 open clusters
\citep{2022A&A...668A..13H},	
in 43 Cepheid--cluster memberships
\citep{2024AJ....168...34W},		
four classical Cepheids associated with the binary open cluster NGC~7790/Berkeley~58
\citep{2024PASP..136f4201M},		
and the 28.7~day period KQ~Sco in the cluster UBC~1558
\citep{2024PASP..136i4202M}.	
Amongst several studies of the period--luminosity relation
\citep{2024A&A...681A..65T,	
2024A&A...684A.126B,		
2024A&A...684A.170D,		
2024ApJS..275...26M},		
the high-quality sample of
\citet{2023A&A...672A..85C}	
identified 34 Cepheids in 28 open clusters (27 fundamental mode, and 7 overtone), finding three new cluster Cepheids (V0378~Cen, ST~Tau, and GH~Lup), and correcting the host cluster identifications for three others (Figure~\ref{fig:cepheids}, right). They found the fraction of Cepheids occurring in open clusters within $2\,$kpc to be $0.088^{+0.029}_{-0.019}$. Cluster parallaxes could be determined to $\sim$7\muas\ in the range $G=12.5-17$~mag. Their combination of cluster and field Cepheids yields a calibration of the period--luminosity relation for several individual photometric passbands, and provides excellent ($0.3\sigma$) agreement with the `SHoES' team's distance ladder
\citep{2022ApJ...938...36R}. 	
The authors consider the results to be the currently most accurate absolute calibrations of the Cepheid luminosity scale based purely on Galactic Cepheids, as well as the most precise determination of the residual parallax offset.
  
Applied to Galactic structure, other studies have used the Gaia DR3 Cepheids 
in the mapping of the asymmetric disk
\citep{2023A&A...674A..37G},	
as distance markers to trace the oxygen, sulphur, and iron abundance gradients across the disk
\citep{2023A&A...678A.195D},	
estimating the Sun's distance from the Galactic centre
\citep{2023AstL...49..493B},		
and characterising the rotation curve
\citep{2020MNRAS.496.2107C,		
2024MNRAS.529.4681B},	
although the suggestion that galaxy rotation curves can be explained through gravitomagnetic effects without the need for dark matter is certainly contested
\citep[e.g.][]{2023PhRvD.108d4056C}. 
\citet{2024A&A...686A.177A}	
provide 18\,225 radial velocity measures of 258 classical Cepheids between 2010--2022 (and compare them with the Gaia RVS values).
Improved Cepheid samples in the LMC/SMC (and their properties) are variously reported
\citep{2023A&A...672A..65J,	
2023A&A...676A.136G,		
2024MNRAS.527.8671B}.		
Gaia astrometry has been used to rule out suggested ultralow amplitude Cepheid candidates
\citep{2023A&A...676A..28T}.	
The longest pulsation period of all known classical Cepheids, 78.14 days for OGLE-GD-CEP-1884, has been classified as a Cepheid from 10~years of OGLE monitoring, but as a long-period variable in Gaia DR3
\citep{2024ApJ...965L..17S},	
perhaps hinting at the existence of other, so far unidentified, ultra-long period Cepheids in the Gaia data.

\subsubsection{Hubble constant from Cepheids}
\label{sec:cepheids-h0}

The discovery, by Edwin Hubble almost a century ago, that distant galaxies are moving away from us at speeds proportional to their distance, was the first observational evidence for the expansion of the Universe. Hubble's `law' still serves as one of the key pieces of evidence supporting its `Big Bang' origins, today formulated in terms of the $\Lambda$CDM cosmological paradigm. Efforts today include determining the value of $H_0$, the present-day value of the `Hubble parameter' describing the expansion of the Universe with time. 
Over the past half century, many hundreds of research papers have targeted the refinement of $H_0$. What have become known as `late Universe' methods rely on calibrated distance ladder techniques: measuring the redshifts of progressively more distant galaxies, and then determining the distances to them by some other method. For much of the second half of the 20th century, $H_0$ was estimated to lie in the range 50--90\kmsmpc.  Over the past decade or so, data from Cepheid variables and other astrophysical `standards' have converged on a `late Universe' value of around 70\kmsmpc
\citep{2020svos.conf..215C,	
2019CRPhy..20..140M,		
2019NatAs...3..891V}. 		

Since about 2000, `early Universe' techniques have become available. These are, in contrast, based on measurements of the cosmic microwave background (the `echo' from the Big Bang that contains imprints of the Universe's fundamental properties), initially from NASA's WMAP, and more recently from ESA's Planck mission. And it is the discrepancy in the values of $H_0$ characterising the local (`late Universe') cosmic expansion, $H_0=73.2\pm1.3$ 
\citep{2021ApJ...908L...6R}, 
compared to that inferred in the `early Universe' from the Planck satellite measurements, $H_0=67.4\pm0.5$\kmsmpc 
\citep{2020A&A...641A...6P},	
that sits at the heart of the `Hubble tension'.
Both values are very precise, and both are also considered to be very {\it accurate}, such that the disagreement, although small, is considered statistically significant. 
At stake is not so much the precise value of $H_0$ {\it per se}, but whether the different estimates hide some new and exotic physics, either in the understanding of stellar evolution, or in the physics of the early Universe. 

Here, to indicate how Gaia is contributing, I will simply summarise the Cepheid contributions. For some years, the strongest evidence for a high value of $H_0$, $\sim$70\kmsmpc, has rested on an empirical Cepheid-based calibration of the distances to galaxies hosting Type~Ia supernovae. 
\citet{2016ApJ...826...56R}		
based their value of $73.24\pm1.74$ on 19 such Cepheids observed with HST--WFC3, along with a more robust distance to the LMC based on late-type detached eclipsing binaries, HST observations of Cepheids in M31, and new HST-based trigonometric parallaxes for Milky Way Cepheids. 
\citet{2018ApJ...855..136R}		
added new HST-based parallaxes of a further seven long-period Milky Way Cepheids to yield a value of $73.48\pm1.66$\kmsmpc.
\citet{2018ApJ...861..126R}		
used HST photometry of 50 long-period, low-extinction Milky Way Cepheids observed in the same photometric system as extragalactic Cepheids in Type~Ia supernova host galaxies, to derive $73.52\pm1.62$\kmsmpc. For the first time, Gaia parallaxes (from DR2) were included to constrain the distance scale, while simultaneously estimating the global DR2 parallax zero-point offset. 
Finally for this part of the story, 
\citet{2021ApJ...908L...6R}		
used 75 Milky Way Cepheids with HST photometry, now using the greatly improved Gaia EDR3 parallaxes. Applied to the calibration of Type~Ia supernovae, it gave $H_0=73.0\pm1.4$\kmsmpc. Combined with the best complementary sources of Cepheid calibration, they found $H_0=73.2\pm1.3$\kmsmpc, reaching 1.8\% precision, but a $4.2\sigma$ difference with the estimate from the latest Planck microwave background observations. 

As described in Section~\ref{sec:cepheids}, work continues on understanding the zero-point and metallicity coefficients of the Cepheid period--luminosity relation based on the Gaia data
\citep[e.g.][]{2022ApJ...939...89B,	
2022ApJS..262...25D,			
2023MNRAS.520.4154M,			
2021MNRAS.508.4047R}.			
\citet{2020svos.conf..215C}	
argue that the improvement that Gaia should provide in the future should eventually allow $H_0$ be determined to $\sim$1\%, and they spell out various steps necessary to achieve this.
One of the other distance measurement techniques holding great promise is the use of the `tip of the red giant branch', described further in Section~\ref{sec:trgb}.

\subsubsection{RR~Lyrae variables}
\label{sec:rrlyrae}

Globular cluster and other Population~II stars more massive than the Sun have long ago evolved into white dwarfs. In contrast, stars of approximately solar mass are common. After ascending the giant branch, terminating in the helium flash, these stars evolve rapidly onto the `zero-age horizontal branch' with masses around $0.6-0.8\,M_\Sun$, basically comprising a He-burning core and a H-burning shell. Their location along the horizontal branch depends on metallicity, from blue in metal-poor clusters, to red in metal-rich clusters, where they merge into the giant branch to form the region of the (red) clump giants.
RR~Lyrae variables are a subset of the horizontal branch giants, with typical pulsation periods of around 1\,d or less, occurring where the horizontal branch intersects the instability strip
\citep[e.g.][Fig.~3]{2016Ap&SS.361..175A}. 
Like Cepheids, although less luminous, their distinctive light curves allows their detection to large distances, including the Galactic centre in the low-absorption Baade Windows, and in crowded fields.  

There are two major subgroups: the RRab, most relevant to the distance scale, are metal-poor spheroidal component stars, with asymmetric light curves, longer periods ($\gtrsim$0.4\,d), larger amplitudes ($\sim$0.5--1.5~mag), and pulsating in the fundamental mode. 
The less numerous RRc type are old disk component stars, with more symmetric almost sinusoidal light curves, shorter periods ($\lesssim$0.4\,d), smaller variability amplitudes, and pulsating in the first overtone. 
There are also double-mode pulsators, denoted~RRd, which pulsate simultaneously in the fundamental mode and in the first overtone.
Together, RR~Lyrae are important for evolutionary and pulsation models, and as kinematic tracers. An underlying period--luminosity relation has also long given them a role as standard candles for relatively nearby targets, especially within the Milky Way and Local Group. While more common than the Cepheids, there are greater difficulties in accounting for the effects of metallicity, faintness, and blending.
Their typically large distances means that useful trigonometric parallaxes have largely been unavailable, and various other methods have been used for luminosity calibration. These include the use of RR~Lyrae in Galactic globular clusters whose distances have been derived from main-sequence fitting of subdwarfs, the use of statistical parallaxes, and Baade--Wesselink determinations based on interpretation of the colour, light, and radial velocity variations during the pulsation cycle.

Pre-Gaia, several thousand Galactic RR~Lyrae stars were known. Just 179 of the brightest were observed by Hipparcos, which eventually gave useful parallaxes for only a few. Only that for the class's prototype, RR~Lyrae itself, was reasonably accurate, $\varpi=4.38\pm0.59$\,mas, with a weighted mean of $3.87\pm0.19$\,mas later derived from the Hipparcos, HST, and pulsational parallax 
\citep{2002MNRAS.332L..78B}.

The Gaia discoveries are again primarily enabled by the accurate multi-colour multi-epoch photometry, has followed a very similar path to that of the Cepheids (Section~\ref{sec:cepheids}), and I will structure the following according to the results from the successive data releases.  
Gaia DR1 included 2595 RR~Lyrae stars (and 599 Cepheids) in the LMC, which was observed at high cadence during the first 28 days 
\citep{2016A&A...595A.133C}.
For Gaia DR2, covering the first 22~months of the mission, the `Specific Object Study' pipeline was used to validate and characterise Cepheids and RR~Lyrae stars, originally using the period--amplitude and period--luminosity relations only in the $G$ band, and subsequently extended to $G_{\rm BP}$ and $G_{\rm RP}$
\citep{2019A&A...622A..60C,
2019A&A...625A..97R}.	
Gaia~DR2 accordingly provides results, along with mean magnitudes and pulsation characteristics, for 140\,784 RR~Lyrae stars as faint as $G=20.7$~mag: 90\,564 were already known, while 50\,220 are new discoveries.
This huge sample includes objects in the Milky Way disk, bulge, and halo; 46\,443 in the LMC/SMC; 1569 distributed over 87 globular clusters; and 417 distributed over 12 dwarf spheroidal galaxies. The largest numbers are in M3 (159), NGC~3201 (83), Sculptor (176) and Draco (176). 
The accurate multi-epoch photometry resulted in 121\,234 objects whose light curves could be modelled with at least two harmonics, and 67\,681 whose light curves could be modelled with at least three harmonics. 
For the 83\,097 stars with both $G_{\rm BP}$ and $G_{\rm RP}$ photometry, the colour--magnitude diagram shows the different regions occupied by the RRab, RRc, and RRd classes, along with clumps associated with the LMC/SMC, as well as the Draco and Sculptor dwarf spheroidals.
Kepler mission photometry has been used to conclude that the DR2 catalogue has a completeness of 70--78\%
\citep{2018A&A...620A.127M}.
Proper motion anomalies (from a difference between the Hipparcos and DR2 positions, and attributed close-orbiting companions) were found for 13 out of a sample of 198, suggest a binary fraction of at least 7\%
\citep{2019A&A...623A.116K}.

Several groups have used these data to determine improved period--luminosity--metallicity relations, confirming the parallax quality in the process
\citep{2017ApJ...841...84N,
2019MNRAS.490.4254N,
2018MNRAS.481.1195M}, 
and suggesting that they yield a meaningful zero-point of the global parallax system, comparable with quasars, despite their smaller number
\citep{2017JKAS...50....1K}.
The kinematics and spatial distribution of 15\,599 RR~Lyrae stars in the Milky Way bulge are being used to probe its rotation and chemodynamical evolution
\citep{2020MNRAS.498.5629D}.
Various groups have used the bulk distances and proper motions to discern the triaxial structure and kinematics of the Galaxy halo, identifying the inner and outer halos with weak prograde and retrograde rotations respectively
\citep{2018AstL...44..688U},
finding no evidence of tilt or offset of the halo with respect to the Galaxy disk
\citep{2018MNRAS.474.2142I},
but suggesting a scenario in which the bulk of the halo was deposited in a single massive merger event
\citep{2019MNRAS.482.3868I}.
Contributions from the Gaia--Enceladus halo stream can be estimated from their kinematics
\citep{2020MNRAS.492.3408P},
and some 6000--11\,000 RR~Lyrae candidates were found to be associated with the Sagittarius tidal stream
\citep{2020A&A...638A.104R}.
The existence of RR~Lyrae stars well beyond a system's tidal radius provides evidence for tidal disruption and debris stripping, both for 
Galactic globular clusters
\citep{2019MNRAS.483.1737K},
and ultra-faint dwarf satellite galaxies
\citep{2020ApJS..247...35V}.

With DR3, specific processing of the RR~Lyrae again used the (updated) `Specific Object Study' Cepheid/RR~Lyrae pipeline
\citep{2023A&A...674A..18C}.	
This included analysis of the Gaia three-colour ($G$, BP, RP) time-series photometry, along with the epoch radial velocities (which are published for 1096). From 271\,779 initial candidates (Table~\ref{tab:data-release-table2}), a high-quality sample of 270\,905 RR~Lyrae stars was constructed (of which 70\,611 are new discoveries), comprising 174\,947 fundamental-mode (RRab), 93\,952 first-overtone (RRc), and 2006 double-mode (RRd) pulsators. The sky distribution includes variables in 95 globular clusters and 25 Milky Way companions (the Magellanic Clouds, seven dwarf spheroidal galaxies, and 16 ultra-faint dwarf satellites). The catalogue includes an estimate of the interstellar absorption for 142\,660 fundamental-mode RR Lyrae stars (based on the $G$-band amplitude, the $G-G_{\rm RP}$ colour, and pulsation period). Metallicities derived from the Fourier parameters of the light curves are included for 133\,559.

Further studies of these include independent metallicity estimates
\citep{2023MNRAS.525.3486J},		
leading to individual metallicities and distances for 134\,000, as well as mean metallicities and distances to the LMC/SMC, and to 38 Milky Way globular clusters
\citep{2025MNRAS.536.2749M}.	
Detailed studies of the period--luminosity--metallicity and period--Wesenheit--metallicity relations have followed
\citep{2023ApJ...945...83M},		
including use of the parallax-of-pulsation technique
\citep{2024A&A...684A.126B}.		
Application of the sample to studies of Galactic structure
include the distance and dynamics of the bulge region
\citep{2024A&A...684A.176P,		
2024A&A...685A.153P,			
2024AJ....168..139K,				
2025A&A...695A.211P},			
the structure and rotation of the inner and outer halo
\citep{2024A&A...685A.134C,		
2024MNRAS.527.8973C,			
2024ARep...68..664T},			
including suggestions of an accretion origin for some
\citep{2023MNRAS.519.5689M}.	
The DR3 RR~Lyrae sample also provides an excellent probe of the recently-identified (and various unidentified) Galactic phase-space sub-structures
\citep{2025ApJ...979..213S}.		
From 46\,575 of the halo stars (and using a Gaussian prior for the radial velocities), their friends-of-friends algorithm successfully identifies groups moving along similar orbits, amongst which are the Sagittarius stream, the Hercules--Aquila Cloud (HAC), the Virgo Overdensity (VOD), the Gaia Enceladus--Sausage (GES), various other halo streams (Orphan--Chenab, Cetus--Palca, Helmi, Sequoia, and Wukong), as well as the LMC leading arm, and another 18~unknown groups. 

An independent catalogue of 2824 RR~Lyrae stars in 115 Galactic globular clusters (including 1594 fundamental-mode, 824 first-overtone, and 28 double-mode pulsators) is given by
\citet{2024A&A...684A.173C}.		
They found that 77\% of RR~Lyrae in globular clusters are included in the Gaia DR3 Specific Object Study (and 82\% correctly classified), with the majority of the missing sources being located in the crowded cluster centres. They found a number of unexpected results: that 25\% of cluster member stars located within the empirical instability strip are not RR~Lyrae stars, and appear to be non-variable. And they found that 80\% of RRab, 84\% of RRc, and 100\% of the RRd pulsators are located within theoretical instability strip boundaries predicted using MESA models (with metallicity Z\,=\,0.0003, mass $M=0.7M_\Sun$, and helium Y\,=\,0.290). Unexpectedly, they report, a higher Y\,=\,0.357 is required to fully match the location of the RRc pulsators, and a lower Y\,=\,0.220 is needed to match the location of the RRab stars. 
Other studies have used the DR3 sample to focus on specific globular clusters, including
NGC~7006
\citep{2023MNRAS.519.2451A}	,	
Palomar~2
\citep{2023RMxAA..59....3A},		
and NGC~1851
\citep{2024RMxAA..60..381A}.		

\paragraph{The Oosterhoff dichotomy}
The Oosterhoff dichotomy is a long-standing problem in the understanding of Galactic globular clusters. Originating with the work of 
\citet{1919AnHar..78..195B} and \citet{1939Obs....62..104O}, 
it refers to the observational fact that they divide into two distinct groups according to the average period of their RR~Lyrae\,ab stars pulsating in the fundamental mode, $<\!\!P_{\rm ab}\!\!>$. There is a gap between the two mean periods, of about 0.55\,d and 0.65\,d, referred to as the Oosterhoff gap, which divides systems into the classes Oosterhoff~I and~II (Oo\,I and Oo\,II) respectively.
The problem has a long history
\citep[e.g.][]{1973ApJ...185..477V,	
1981ApJ...244L..23S}.
While there is a consensus that it is correlated with the cluster's metallicity, with the more metal-poor (Oo\,II) having larger mean periods
\citep[e.g.][]{2003A&A...410..871C}, 
an explanation has remained elusive
\citep[e.g.][]{2019ApJ...882..169F,	
2021ApJ...919..118F,	
2024A&A...690L..17L}.

Around 20~years ago came prescient hints that the dichotomy may hold a key to understanding the formation history of the halo
\citep[e.g.][]{2004ASPC..310..113C,	
2006ApJ...649L..83S,		
2009Ap&SS.320..261C}.		
A number of Gaia-based studies in this direction have made use of the resulting distances and space motions to isolate the problem
\citep{2021ApJ...919..118F,	
2023MNRAS.525.5915Z,	
2024MNRAS.534.3654P,	
2024A&A...684A.173C}.	
Current understanding is brought together in the Gaia DR3 study by
\citet{2024A&A...690L..17L}.	
They calculated orbits and integrals of motion for the Galactic globular clusters and field RR~Lyrae stars previously attributed to halo streams, with the likely origin of the field RR~Lyrae  matched to their globular cluster $E-L_z$ counterparts. The resulting period--amplitude (or Bailey) diagrams for the inferred {\it in situ\/} stars show a wide and continuous range of metallicities, with no sign of the Oosterhoff dichotomy. The accreted halo RR~Lyrae stars, in contrast, clearly show a much smaller metallicity dispersion, and a clear Oosterhoff gap. 
Although they did not isolate the reason for the Oosterhoff dichotomy, 
\citet{2024A&A...690L..17L} 
conclude that it was `imported' into the Milky Way by these ancient mergers, a finding representing an important advance in our understanding of its origin.

\subsubsection{Mira variables}

Miras are pulsationally unstable AGB stars of initial mass $1.5-4M_\Sun$, with high mass-loss rates, and high luminosities ($\sim\!\!10^3L_\Sun$) due to their large envelopes. They are characterised (and defined) by large variability ($\ge2.5$~mag in $V$, 0.3--1.0 in $K_{\rm S}$), and periods of 100--1000\,d.  They occupy a key stage of stellar evolution, contribute to the heavy-element enrichment of the interstellar medium and, being luminous, are important tracers of Galactic structure. One of the nearest is Mira itself, at $\sim$90\,pc.
Their chemistry is dominated by either C-rich or O-rich species according to the strength of `dredge-up' episodes during the asymptotic giant branch phase, largely reflecting their initial mass and metallicity 
\citep{2018A&ARv..26....1H}.
Both types satisfy their own period--luminosity relations 
\citep{2021ApJS..257...23I}, 
with the O-rich relations typically being tighter in the near-infrared due to effects of significant circumstellar dust in the C-rich variables. This makes the O-rich Mira variables particularly useful distance indicators, both within our Galaxy and beyond. 
They are contributing to the `Hubble tension' debate as an independent Population~I calibrator of the supernova luminosities, currently best served by the classical Cepheids (Section~\ref{sec:cepheids-h0}).
Hipparcos observed some 900 long-period variables, including Miras, and contributed to improvements in the period--luminosity relations, to the physics of the dredge-up episodes, and to an understanding of their pulsation modes 
\citep{1998ApJ...506L..47B,
1999A&AS..140...55B,
1997MNRAS.287..955V}.

The Gaia sample is, of course, vastly larger. Gaia DR2, in 2018, provided an all-sky catalogue of 550\,737 variable stars (with $G$, $G_{\rm BP}$, and $G_{\rm RP}$ photometric time-series), of which 151\,761 are long-period variable candidates having $\Delta G>0.2$~mag,
with one-fifth of these considered to be Mira candidates
\citep{2018A&A...618A..58M}.	
The 34-month DR3 contains 1.72~million long-period variable candidates, including 392\,240 with periods 35--1000\,d, of which more than 40\,000 are identified as Mira variables confirmed by OGLE-IV 
\citep{2023A&A...674A..15L}. 
Segregation into O-rich and C-rich Miras has been demonstrated using the Gaia BP/RP spectra to distinguish the different TiO bandheads 
\citep{2023A&A...674A..15L,
2023MNRAS.523.2369S,
2023MNRAS.521.1462Z}.

In terms of their use as distance indicators, the period--luminosity relation first established for Galactic Miras
\citep{1981MNRAS.196..111R}		
was soon shown to be much tighter for those in the LMC
\citep{1981Natur.291..303G}.		
Many further refinements have come from the microlensing surveys MACHO and OGLE 
\citep[][and references]{2013ApJ...779..167S}, 
and from the Hipparcos parallaxes 
\citep{2000MNRAS.319..759W,		
2008MNRAS.386..313W}.			
The much more extensive Gaia data have been used to revisit the period--luminosity relation (including dependencies on colour and metallicity), both in our own Galaxy
\citep{2023FrASS..1032151S,		 
2023MNRAS.523.2369S}, 		
and in the Magellanic Clouds
\citep{2019ApJ...884...20B}.		
The general findings are that uncertainties in the Galactic period--luminosity relation 
\citep{2023MNRAS.523.2369S}
remain larger than those in the LMC 
\citep{2023FrASS..1032151S,
2023MNRAS.523.2369S}.
Nonetheless, using these relations as anchors for Mira variables in other galaxies, 
\citet{2023MNRAS.523.2369S}	
derived $H_0=73.7\pm4.4$\kmsmpc\ for the Type~Ia host galaxy NGC~1559,
while using HST, 
\citet{2024ApJ...963...83H}		
found $H_0 =72.37\pm2.97$\kmsmpc\ for M101.
As with recent Cepheid determinations, the Mira results suggest a discrepancy in $H_0$ between early and late Universe values.

A further, but exploitable, complication is that the pulsation period of Miras is correlated with their scale height and/or velocity dispersion which, interpreted as a period--age correlation, allows them to be used as age indicators within the Galaxy and beyond 
\citep{1963MNRAS.125..367F,
2020MNRAS.492.3128G}. 
Mira variables in clusters, from Gaia-based membership determinations, confirm this connection 
\citep{2019MNRAS.483.3022G,
2022ApJS..258...43M}. 
\citet{2023MNRAS.521.1462Z}	
used 46\,107 O-rich Mira candidates from Gaia DR3 to derive a period--age relation (with $P$ in~d) 
$\tau\approx(6.9\pm0.3)(1+\tanh[(330-P)/(400\pm90)])$\,Gyr.
Combining these properties, samples of thousands of Gaia Miras are being used as new robust probes of Galactic structure and evolution, including morphology of the bar and bulge
\citep{2020MNRAS.492.3128G,	
2022MNRAS.517.6060S,		
2022MNRAS.512.1857S,
2020A&ARv..28....4N}. 
\citet{2024MNRAS.530.2972S} used Miras from VVV, placed on the Gaia reference frame, to suggest that the bar formed $\sim$8\,Gyr ago, close to the time of the Gaia Sausage--Enceladus infall merger, potentially implying that the bar was tidally-induced.

There are theoretical explanations for the period--luminosity relation which follows from the fact that stars of a given mass, excited by convection, only begin pulsating in the fundamental mode over a narrow range of radii 
\citep{2019MNRAS.482..929T,
2022A&A...658L...1T}. 
But a recurrent topic in the consideration of Mira variables is the finding that the scatter in the Galactic period--luminosity relation is larger than for the LMC and, at the same time, that the Gaia parallax standard errors for Miras are often significantly underestimated, by some 30--80\%. 
This has been attributed in part to (presently uncalibrated) time-dependent chromatic displacements
\citep{2023MNRAS.521.1462Z}, 	
but it is probably compounded by their large physical sizes (e.g.\ Mira, with $R\sim400R_\Sun$, subtends $\sim$40\,mas). This means that many are resolved by Gaia, resulting in significant photocentric motions due to their large convection cells
\citep{2018A&A...617L...1C,	
2021MNRAS.506.2269E,	 	
2022A&A...661L...1C,		
2022A&A...667A..74A,	 	
2022A&A...657A.130M,  		
2022RNAAS...6..142W,		
2023MNRAS.520.3510K}.		

\subsubsection{Delta Scuti variables}

As noted above, the $\delta$~Scuti variables are (largely) radial pulsators, located at the intersection of the instability strip with the main sequence. They are of spectral class A or F, with masses $1.5-2.3M_\Sun$ (depending on metallicity), luminosities $2-50L_\Sun$, moderate to rapid rotators, and are characterised by low-amplitude, short-period pulsations ranging from hours to days
\citep[e.g.][]{2005JApA...26..249G,
2024ApJ...972..137G}.
As detailed by 
\citet[][\S2.1]{2022ARA&A..60...31K},
they actually display {\it `a rich variety of pulsation behaviour, ranging from rare singly periodic stars to stars with hundreds of frequencies spread across the 0--80\,d\/$^{-1}$ range, requiring g~modes, r~modes, p~modes, mixed modes, rotationally split multiplets, nonlinear combination frequencies, and possibly more to understand'}.
Several thousand have been identified by OGLE, Kepler, TESS, ASAS--SN, and others.

Gaia DR2 classified 8882 $\delta$~Scuti stars, including the sub-class of SX~Phe stars
\citep{2018A&A...618A..30H}, 
with at least 12\,000 likely to be present in DR3 
\citep[][Table~3]{2023A&A...674A..14R}.
One objective of the Gaia studies is to characterise the period--luminosity relation as a function of pulsation mode, with a number of detailed dependences and insights emerging 
\citep{
2019MNRAS.486.4348Z,		
2020MNRAS.493.4186J,		
2021PASP..133h4201P,		
2022MNRAS.516.2080B,		
2024ApJ...972..137G,		
2024MNRAS.528.2464R,		
2024RAA....24b5011P}.		
Another goal has been to accurately place them in the HR~diagram, characterise their variability, and explain why many of the stars in the $\delta$~Scuti instability strip do not pulsate. This has led to a redefinition of the empirical instability strip
\citep{2019MNRAS.485.2380M},		
and to the discovery of various detailed dependencies, amongst which the rapid rotators are more likely to pulsate, but with less regular pulsation patterns
\citep{2024MNRAS.534.3022M}.		

\begin{figure}[t]
\centering
\includegraphics[width=0.26\linewidth]{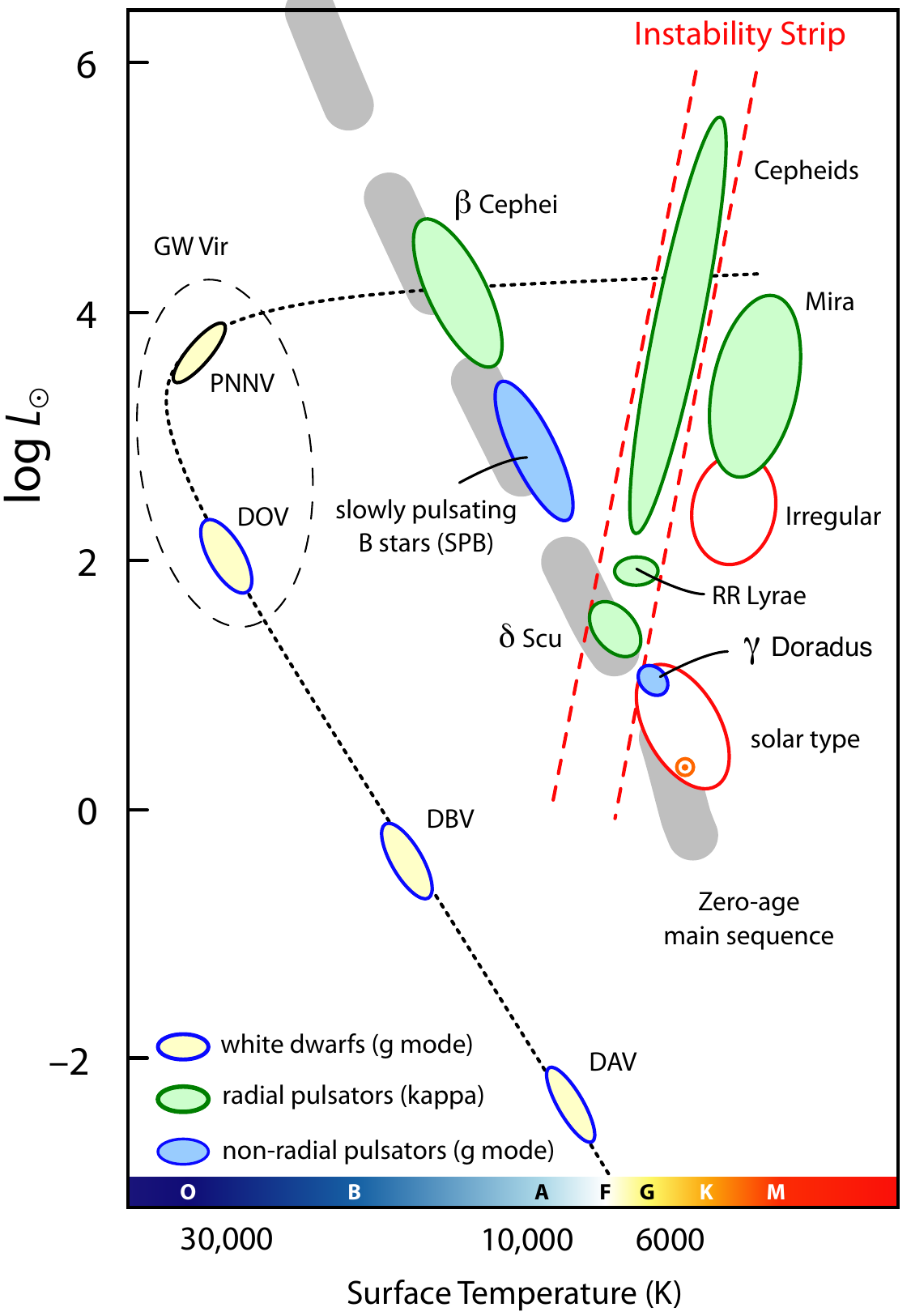}
\hspace{8pt}
\includegraphics[width=0.31\linewidth]{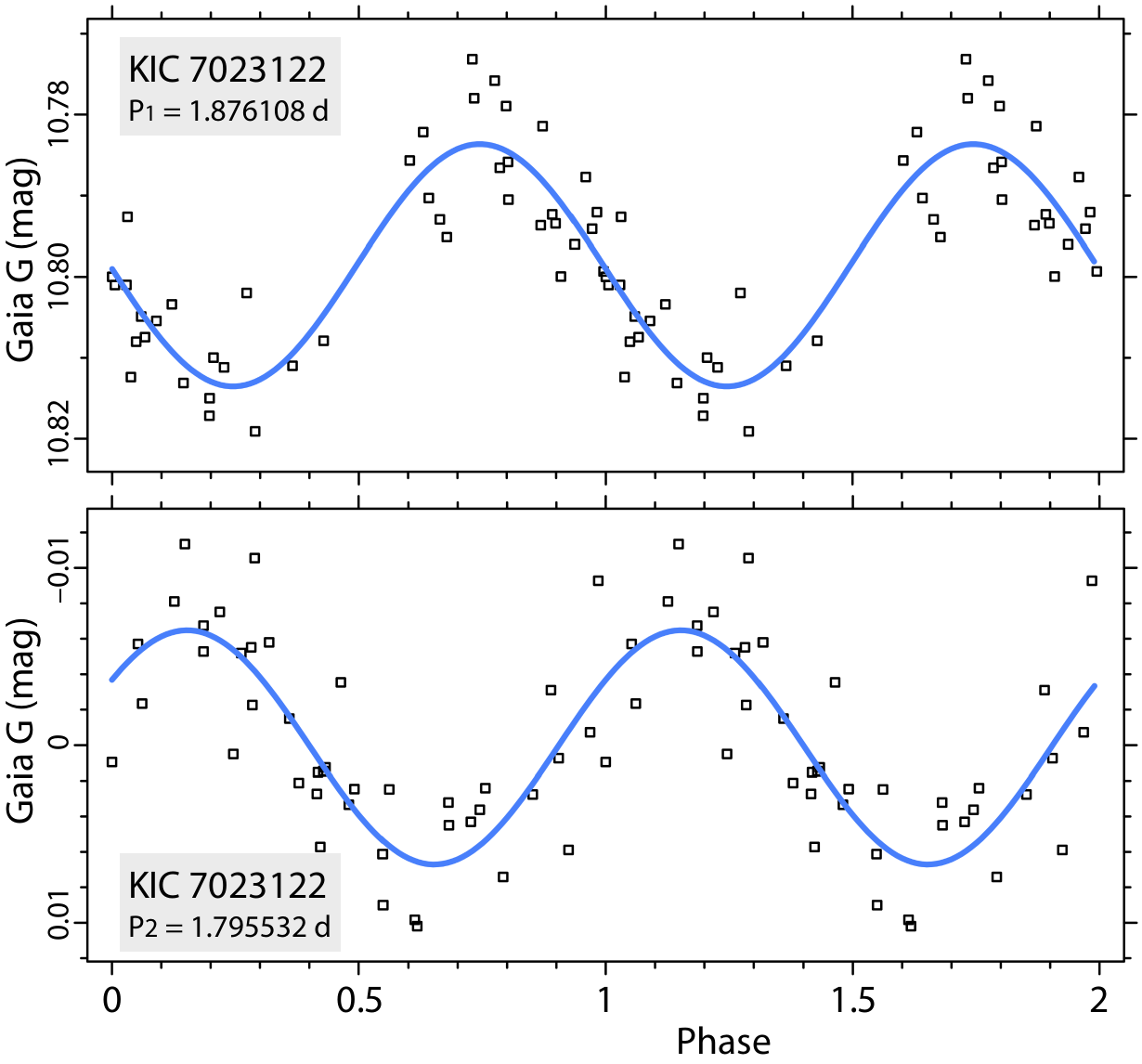}
\hspace{8pt}
\includegraphics[width=0.37\linewidth]{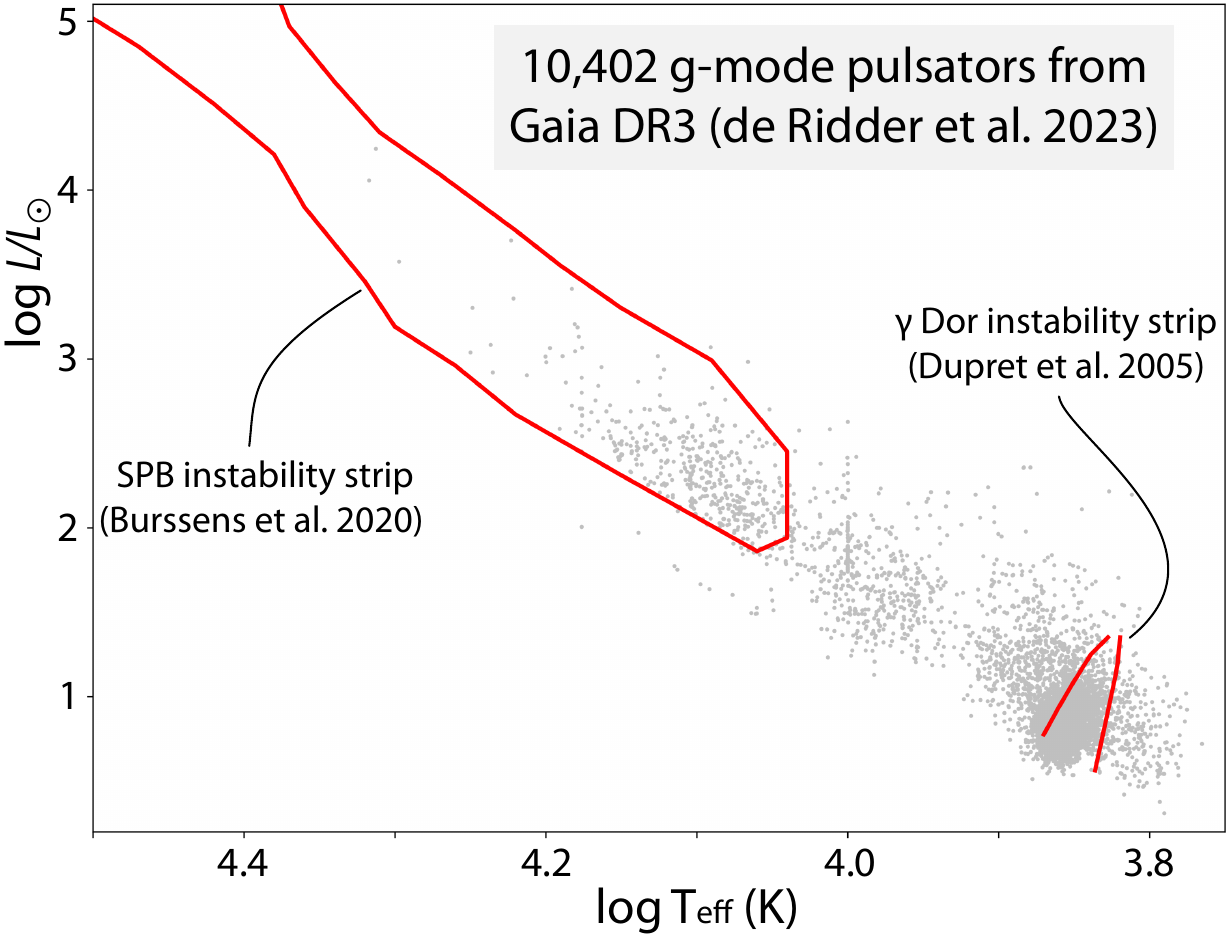}
\caption{
Left:	schematic location of radial (green) and non-radial pulsators (blue) in the HR~diagram (adapted from wikipedia, Instability Strip).
Middle: Gaia light curve of the $\gamma$~Dor pulsator KIC~7023122 and its two g-mode frequencies \citep[from][Figure~B5]{2023A&A...674A..36G}.
Right:  HR~diagram of 10\,402 g-mode pulsators classified as SPB or $\gamma$~Dor stars with a DR3 relative parallax $\sigma_\varpi/\varpi<3$\%. Red lines show the SPB instability strip at high \teff\ \citep[from][]{2020A&A...639A..81B}, and the $\gamma$~Dor instability strip at low \teff\ \citep[from][]{2005A&A...435..927D}, both for solar metallicity \citep[from][Figure~5]{2023A&A...674A..36G}.}
\label{fig:non-radial-pulsators}
\end{figure}

\subsubsection{Non-radial pulsators}
\label{sec:non-radial-pulsators}

\paragraph{Context}
Pulsating variables change in brightness as they cyclically expand, and then contract, on periods ranging from days to weeks or months. Driving these primarily radial pulsations is the `$\kappa$~mechanism': the expansion phase is driven by the blocking of energy outflow by gas with a high opacity, eventually halting as the density decreases, then reversing due to gravity. In this class are the Cepheids, RR~Lyrae, and $\delta$~Scuti (and the related SX~Phoenicis and rapidly oscillating Ap, or RoAp, stars), along with the Mira and $\beta$~Cephei variables (Figure~\ref{fig:non-radial-pulsators}a). The opacity source driving the amplitude variations differs amongst these pulsating sources. 
Falling in the so-called `instability strip' are the $\delta$~Scuti (and SX~Phoe and RoAp) on the main sequence, the RR~Lyrae at the intersection with the horizontal branch, and the highest luminosity Cepheids at the intersection with the supergiants. All these are driven by the changing opacity of He\,{\scriptsize III}.  Mira variables are controlled by ionised hydrogen (H\,{\scriptsize II}), and the more massive $\beta$~Cep (unrelated to Cepheids) driven by changing iron opacity. As a result, these pulsating stars vary over very different timescales, with very different amplitudes (extending up to many magnitudes), and sometimes with multiple harmonic frequency components. 
White dwarfs also pulsate over a range of temperatures and luminosities, grouped into three regions, and are mainly driven by low-degree ($l\le2$) non-radial g-modes. They are excited by opacity variations in the relevant dominant species, viz.\ the DAV (or ZZ~Cet) stars dominated by hydrogen, the DBV dominated by helium, and the DOV and PNNV (together known as GW~Vir variables) dominated by He/C/O \citep[e.g.][]{2022A&A...659A..30C}.

{\it Non-radial\/} oscillations were invoked more than 70~years ago to explain puzzling spectral characteristics of $\beta$~Cephei stars
\citep{1951ApJ...114..373L}. 
The $\gamma$~Dor variables, and the slowly-pulsating B (SPB) stars, while also lying on the main sequence, and also possessing convective cores, are driven by such non-radial gravity-mode (g-mode) pulsations, for which the shape rather than the radius changes.
And while stellar structure theory broadly predicts the occurrence of their pulsation modes
\citep[e.g.][]{2005A&A...435..927D,  2020A&A...639A..81B}, 
a better understanding of their interiors has been limited by their small numbers, the even smaller number for which asteroseismology is available as a probe for investigating their internal physics (Section~\ref{sec:asteroseismology}), and the complexity of the theoretical models required for their more complete understanding.

To give a flavour for the theoretical complexity, the short-period $\delta$~Scuti stars (0.5--6\,hr) are more precisely described by low-order p- and g-modes (pressure/gravity), with the former being mainly driven by the $\kappa$~mechanism in the He\,{\scriptsize II} partial ionisation zone, but supplemented by the so-called $\gamma$~mechanism, related to the details of heat flow during the compression phase
\citep{2005A&A...435..927D}.	
While the longer period $\gamma$~Dor stars (0.4--3\,d) correspond to pulsations in non-radial (and often high-order) g-modes, their driving mechanism is less well understood, and has been partly explained by a convective `flux blocking' mechanism at the base of their convective envelope. To make matters more complex, such theoretical models predict the possible occurrence of simultaneous $\delta$~Scuti p-modes and $\gamma$~Dor g-mode oscillations for the same stellar model \citep{2005A&A...435..927D}.

\paragraph{Gaia results}
Given the importance of asteroseismology for their modelling, the fact that these objects are often multi-periodic, with low amplitudes, generally makes non-radial pulsators difficult to analyse with sparse time series. However, Gaia's accurate multi-epoch photometry, along with the global astrophysical parameters (notably from the Gaia DR3 {\tt GSP--Phot} tables, Section~\ref{sec:classification-stellar-properties}) is allowing them to be detected in very large numbers, and to be accurately located in the Hertzsprung--Russell diagram (Figure~\ref{fig:non-radial-pulsators}b).

Starting with the 450\,605 variable sources in DR3 classified by \citet{2023A&A...674A..14R}, 
\citet{2023A&A...674A..36G}	
identified the dominant oscillation mode in 100\,000 intermediate- and high-mass dwarfs, including 10\,402 non-radial pulsators  of the SPB and $\gamma$~Dor classes. 
Their location in the Hertzsprung--Russell diagram (Figure~\ref{fig:non-radial-pulsators}c), enabled by the accurate Gaia distances, shows that they occupy a much more extended region of the main sequence as compared with the suggested boundaries of the SPB instability strip by \citet{2020A&A...639A..81B}, and of the $\gamma$~Dor instability strip by \citet{2005A&A...435..927D}.
To be clear, the precise borders depend on the physics of internal rotation, gravitational settling, radiative levitation, shear mixing, and others. Nonetheless their study shows that the instability regions cover broader areas of the HR diagram than has been predicted theoretically, in particular between the SPB and $\delta$~Scuti groups, and beyond the $\gamma$~Dor. These pulsators typically have a dominant frequency corresponding to those for fast rotators pulsating in `gravito-inertial' modes 
\citep{2014A&A...569A..18S,
2017MNRAS.467.3864S,
2023A&A...672A.183A}.

The Gaia-based time-series photometry is still subject to instrumental variations not yet fully calibrated. Improvements, and longer time series, will allow more rigorous object classification, better placement in the HR diagram, and more robust identification of frequencies and amplitudes of secondary pulsation modes, all leading to improved ingredients for theoretical modelling.

\subsubsection{Beta Cephei variables}
\label{sec:beta-cephei-variables}
$\beta$~Cephei (BCEP, aka Beta Canis Majoris) variables are the most massive pulsators on the main sequence, named after the class's 3.2~mag prototype. They are core H-burning stars, spectral type \mbox{B0--3 IV--V}, with \teff\,=\,20--30\,000\,K, and mass $7-20M_\Sun$. They are characterised by pulsations of small amplitude (typically 0.01--0.3~mag) and short period (typically 0.1--0.3\,d). The fact that such rapid stellar variations were unprecedented at the time of their discovery led to them being originally ascribed to spectroscopic binary effects
\citep{1902ApJ....15..340F,
1913AN....196..357G}.
From monitoring of other similar objects, long-term amplitude changes were eventually interpreted as a beat effect, and duly explained as the simultaneous presence of radial and non-radial modes with closely separated periods
\citep{1951ApJ...114..373L,	
1955PASP...67..135S}.	
But the physical mechanism responsible for their pulsations remained uncertain for several decades. Only in the 1990s, with newly calculated opacities, were the pulsations explained by the $\kappa$~mechanism, but in this case (and in contrast to the Cepheids) due to the changing opacity of the iron-peak elements.
\citet{2025A&A...698A.253F}	
summarises the progressive chronology in identifying their oscillation modes, and in understanding the mode excitation and mode selection (viz.\ why certain modes are excited to observable amplitudes, and some evolve with time). 

Detailed asteroseismology of $\beta$~Cephei variables started with the 21-year ground-based photometric time series of HD~129929 by 
\citet{2004A&A...415..241A},	
which provided evidence for at least six periods. Similar long-term ground-based photometry extended such detailed modelling to $\nu$~Eri, $\theta$~Oph, 12~Lac, and V2052 Oph, with space-based asteroseismology with MOST, CoRot and Kepler adding a handful of other well-studied examples.
\citet{1993SSRv...62...95S} 
listed 59 secure examples, with others since discovered from surveys such as ASAS--3, KELT, and TESS.
\citet{2024ApJS..271...28S}
concluded that the number of known Galactic $\beta$~Cep stars is about 400. 
Meanwhile, open questions about their properties have persisted. These include whether changes in the dominant pulsation period can be explained as evolutionary effects over their short ($\sim$10\,Myr) main-sequence lifetimes as the H~fuel is depleted, e.g.\ in the case of BW~Vul
\citep{2018JAVSO..46...85C},	
and the discrepancy between the dynamical masses in binaries, and model predictions, e.g.\ for $\sigma$~Sco
\citep{2014MNRAS.442..616T}.	

Of the 1.8~million variable stars in Gaia DR3 (Section~\ref{sec:variable-stars}), 222 were classified as $\beta$~Cep variables, with dominant frequencies in the range $\nu_1=3-8$\,d$^{-1}$
\citep{2025A&A...698A.253F}.	
Given that the Gaia light curves are only sparsely sampled, albeit of high (milli-mag) accuracy, it is not immediately evident that Gaia could contribute to the identification of such asteroseismic pulsation modes.  

But as part of a wider study based on a TESS--Gaia cross-matched (\mbox{g-mode} pulsator) sample comprising  
85\,313 $\delta$~Scuti, 
11\,636 $\gamma$~Dor, 
3426 slowly pulsating B (SPB) stars, 
and these 222 $\beta$~Cep variables,
\citet{2024A&A...688A..93H} 
used the first 2~years of the more densely-sampled TESS photometry to independently classify the Gaia DR3 list of non-radial pulsators, including their dominant and secondary pulsation frequencies. 
They found that the majority of Gaia classifications are indeed consistent with those from TESS, concluding that {\it `the Gaia photometry is exceptionally accurate for detecting the dominant and secondary frequencies, reaching approximately 80\% accuracy in frequency for p- and g-mode pulsators'}. 
%
Indeed, for the higher frequency p-modes, they concluded that it is, as yet, unclear whether the Gaia or TESS data is more accurate for measuring the ‘true’ dominant frequency. 
Their analysis showed that the g-mode pulsators form a continuous group of variable stars along the main sequence across B, A, and F spectral types, implying that the mode excitation mechanisms for all these pulsators {\it `need to be updated with improved physics'}.
A parallel study by
\citet{2024ApJS..271...28S} 
identified 155 $\beta$~Cep candidates using data from TESS and Gaia, of which 83 were confirmed as $\beta$~Cep pulsators. With magnitudes in the range 8--12~mag, amplitudes of 0.1--55.8\,mmag, and pulsation periods 0.06--0.31\,d, the Gaia DR3 parallaxes are mostly between 0.2--0.6~milli-arcsec. 

Asteroseismic modelling (Section~\ref{sec:asteroseismology}) targets the measurement of fundamental stellar properties such as internal rotation, age (in terms of core hydrogen mass fraction), core boundary mixing, and even tidal effects in multiple systems
\citep[e.g.][]{2020FrASS...7...70B}. 
Presently, current theoretical models do not precisely match the observed location of the $\beta$~Cep pulsators in the Hertzsprung--Russell diagram
\citep[e.g.][]{2020A&A...639A..81B},
while numerical computations predict fewer excited modes than those detected from the latest space-based photometry
\citep[e.g.][]{2024A&A...687A.175R}. 

More detailed physics in computing excited modes changes the predicted instability regions. This will further constrain phenomena such as the Coriolis effect due to rapid rotation 
\citep{2017MNRAS.469...13S},	
and radiative levitation due to atomic diffusion, in which the segregation of different elemental isotopes leads to different modes being excited by the opacity mechanism 
\citep{2024A&A...687A.175R}. 

Another of Gaia's strengths for asteroseismology is that, compared with the single 600--1000\,nm TESS passband, Gaia's multi-colour time series (in G, BP, and RP) assists identification of the spherical harmonic wave-numbers ($l$,~$m$) which
characterise the mode's geometry (via the inclination of the pulsation symmetry axis). More specifically, mode-dependent perturbations of the stellar flux, and limb darkening, can be derived from the amplitude ratios at the different wavelengths 
\citep[e.g.][]{
1994A&AS..105..447H,	
2003A&A...398..677D}.	
Further analysis of the Gaia sample, interpreted with more extended TESS-based photometry, is given by
\citet{2025A&A...698A.253F}.	
They identified the mode degrees for 148 stars and, based on grid modelling, derived mass, convective core mass, and ages for 119.

\subsubsection{Citizen Science}
\label{sec:variability-citizen-science}


`Citizen science' is scientific work undertaken by the general public, usually under the direction of professional scientists. It is being used in a wide range of studies, including biology and conservation. Typically, volunteers help to collect or classify data, thereby assisting in various aspects of the scientific process, while helping to inform and educate the wider public.
\href{https://en.wikipedia.org/wiki/Zooniverse}{ Zooniverse}
is one such web portal, with around two million registered volunteers, and 25~space-related projects, amongst them
Galaxy Zoo,
Gravity Spy,
and Planet Hunters.

Within Zooniverse, 
\href{https://www.zooniverse.org/projects/gaia-zooniverse/gaia-vari}{Gaia-Vari},
led by Krzysztof Nienartowicz and Elsa Matias, was launched in June 2022.  Gaia light curves of more than 20\,000 DR3 sources (including eclipsing binaries, Cepheids, RR~Lyrae, and long-period variables), folded at the period from Gaia's automated analysis system, were made available in two main campaigns. 
Around 1900 volunteers took part in the visual inspection of the light curves, providing around 12\,000 comments, for example on whether the original period was (say) wrong by a factor~2, which may occur for eclipsing binaries, or whether the variability type had been incorrectly assigned.  By improving the fidelity of the training set, this work is expected to significantly improve the content of the next Gaia data release, DR4.

\subsection{Science alerts}
\label{sec:science-alerts}


As Gaia scanned the sky, it detected and observed all objects brighter than $G\sim20.7$~mag at that specific field-of-view crossing time, including regular or irregular variables, transients, and (moving) minor bodies within the solar system.
Using Gaia's own increasingly comprehensive data base of variable stars and solar system objects (such that pulsators, regular variables, and eclipsing binaries are largely excluded as alerts), sudden and unexpected brightenings or dimmings could be flagged.
Such changes often point to rare and exotic astrophysical events such as supernovae, microlensing events, and objects experiencing accretion outbursts, all benefitting from prompt and detailed follow-up.

A group at the Institute of Astronomy, Cambridge, leads the satellite photometry processing, and a sub-group handles these `science alerts'
\citep{2021A&A...652A..76H}. 
Following the convention used for supernova discoveries (the prefix SN, followed by the discovery year, suffixed with a one or two-letter designation), successful `transients' are designated GaiaYYaaa, GaiaYYaab\ldots. 
They are published via their own
\href{https://gsaweb.ast.cam.ac.uk/alerts}{Gaia Science Alerts} database, 
as well as the
\href{https://www.wis-tns.org/}{IAU Transient Name Server}.
Most alerts are in the range $G=13-21$, and peak at $G\simeq18.5$~mag.
As one indicator of fidelity, other transients surveys (e.g.\ ASAS--SN, ATLAS, Pan-STARRS, and ZTF) often identify the same event. 

The alerts pipeline has to handle many complications, with more reliable and efficient detection algorithms developed progressively. There are modules optimised for detecting `new' sources, and for bursts in previously known sources. `Bad' transits, including particle events and bright star artefacts, are suppressed by requiring detection in two successive fields of view. 
A final assessment is done by eye. 
Robust alerts are then promptly issued to observers worldwide for follow-up photometric and spectroscopic observations, crucial for object characterisation.

As of 2021, the system had categorised 3484 alerts in 14 classes
\citep{2021A&A...652A..76H}.
At the end of 2024, the database listed 25\,919 alerts, with 7\,141 evens assigned to 23~classes. Amongst these are
4139 supernovae, 40~microlensing events, and 25 tidal disruption events,
along with large numbers of novae, symbiotic stars, young stellar objects (including FU~Ori tars, EX~Lup stars, and dippers), and others.
%
Several dozen have been described individually in the literature, of which I give only a few examples here.

\paragraph{AM\,CVn stars}
Gaia14aae is an AM\,CVn-type binary, one of only 60 or so identified since the discovery of the prototype AM\,CVn (HZ~29) more than 50~years ago
\citep{1967AcA....17..255S}.	
These systems experience what is called `common-envelope evolution'. In this short-lived and incompletely understood evolutionary phase for a wide range of binaries, both components orbit inside a single, shared envelope. Common-envelope evolution occupies a possible end-point for binary white dwarf evolution, and affects the progenitors of Type~Ia supernovae, X-ray binaries and double neutron stars.
The systems are also potentially strong sources of gravitational waves due to their compact orbits 
\citep{
{1967AcA....17..287P},		
{2003CQGra..20S..81N}}.	
Gaia14aae is the only known AM\,CVn-type system experiencing a total eclipse of the central white dwarf. Follow-up studies suggest a degenerate donor, with the system perhaps resulting from a merging double white dwarf, with a much shorter orbital period in the past as a result of mass transfer
\citep{
2015MNRAS.452.1060C,	
2019MNRAS.485.1947G,	
2023MNRAS.519.2567S}.	

\paragraph{Tidal disruption events}
Tidal disruption events (TDE) result from a star passing sufficiently close to a supermassive black hole that tidal forces overcome its self-gravity. Drawn-out material results in a tidal stream that loops around the black hole, some fraction on unbound orbits, the remainder forming an accretion disk, with bursts of electromagnetic radiation (spanning radio to $\gamma$-ray), fading over several months as the material is accreted
\citep[e.g.][]{
1983A&A...121...97C,	
1989ApJ...346L..13E}. 	
Their existence was predicted, 50~years ago, when super\-massive black holes were still only hypothesised, with the first  candidates found as soft X-ray outbursts with ROSAT (1990--99). Today, some 100 such events are known, and they have become important probes of previously dormant supermassive black holes, including their mass and spin
\citep{2020SSRv..216...81A, 
2022ApJ...933...31W}.
The 25 Gaia alerts discoveries, typically identified at $G\sim17-18$~mag, are allowing further individual event and population modelling, while Gaia is also providing the accurate reference frame essential for associating them with previously quiescent galaxies.
Discoveries include Gaia's brightest,
\href{https://gsaweb.ast.cam.ac.uk/alerts/alert/Gaia19bvo}{Gaia19bvo},
alerted at $G=15.22$~mag on 2019 April~13 (historic $G=16.81\pm0.14$~mag).
Follow-up observations, at X-ray and radio wavelengths, suggest a non-relativistic spherical outflow from self-stream intersections, or a mildly collimated outflow from accretion onto the supermassive black hole
\citep{2021MNRAS.500.1673H,
2022MNRAS.511.5328G}.
Gaia's second brightest TDE, 
\href{https://gsaweb.ast.cam.ac.uk/alerts/alert/Gaia19eks}{Gaia19eks/AT2019qiz},
and the closest at the time, was alerted at $G=16.04$~mag on 2019 October~3 (historic $G=18.79\pm0.12$), and in the footprint of a gravitational wave event. Models of the extensive ultraviolet, optical, and X-ray data  indicate a star of mass $\simeq1M_\Sun$ disrupted by a black hole of mass $\simeq10^6M_\Sun$
\citep{
2020MNRAS.499..482N,		
2023MNRAS.525.1568S,		
2023PASP..135c4102K}.		

\paragraph{Microlensing events}
Gaia's 40 alerts classified as manifestations of gravitational microlensing include 
Gaia22dkvLb (Gaia's third exoplanet),	
Gaia19dke (a long-duration event with multiple peaks), 
and Gaia19bld (in which arc-like sub-images have been spatially resolved, and observed to be rotating).
I describe these three systems in Section~\ref{sec:microlensing}.

\subsection{Asteroseismology}
\label{sec:asteroseismology}

\paragraph{Context}
Asteroseismology is the study of stellar oscillations, a field of growing importance which is providing profound insights into internal stellar structure and, through modelling, some of the most basic stellar properties, notably masses, radii, and luminosities
\citep[e.g.][]{2021RvMP...93a5001A}.		
Gaia is making important contributions by
(a)~assessing the fidelity of the underlying asteroseismic models, notably through a comparison between their inferred asteroseismic distances (from their luminosities) and the Gaia parallaxes;
(b)~through the provision of significant variability type classification, and object placement in the HR diagram;
(c)~by discovering significant numbers of new pulsators along with their mode identification;
and
(d)~insights into the various line-broadening mechanisms from the Gaia RVS spectra.

The simplest class of stellar oscillations are {\it radial\/} pulsators, in which the plasma throughout the star oscillates with some specific `fundamental' frequency, or in various overtones, with all points at a given radius oscillating in phase, i.e.\ they remain spherically symmetric, although not in hydrostatic equilibrium. Radial pulsators, which are driven by a varying radiation opacity with temperature (the `$\kappa$-mechanism'), include the RR~Lyrae and classical Cepheids. 
{\it Non-radial\/} oscillations were invoked more than 70~years ago to explain puzzling spectral characteristics of $\beta$~Cephei stars (Section~\ref{sec:beta-cephei-variables}). Subsequent studies showed that non-radial oscillations occur across the HR diagram
\citep{2022ARA&A..60...31K}, 
including $\delta$~Scuti stars, DA white dwarfs, and the Sun itself (Section~\ref{sec:non-radial-pulsators}). This led to the search and discovery of the 5-min oscillations in the Sun, and the development of {\it helioseismology\/} to probe its internal structure 
\citep{1991ARA&A..29..627G, 2002RvMP...74.1073C}. 

In solar-type stars, oscillations are excited by turbulent convection 
\citep[e.g.][\S2.2]{2004SoPh..220..137C}. 
Standing pressure (or acoustic) waves are then trapped between the density decrease toward the surface, and the increasing sound speed toward the centre which refracts the downward propagating wave back to the surface. 
%
Helioseismology observations made from the ground over several decades, with GONG and BiSON, and space observations from SoHO and SDO, have shown that the Sun has a rotation profile with
a rigidly-rotating radiative (i.e.\ non-convective) interior zone;
a thin shear layer, the tachocline, which separates the rigidly-rotating interior and the differentially-rotating convective envelope;
a convective envelope in which the rotation rate varies both with depth and latitude; and
a final shear layer just beneath the surface, in which the rotation rate slows down towards the surface.

Long, uninterrupted high-accuracy space-based photometry, over weeks or months, from CoRoT (launched in 2006), Kepler (2009), and TESS (2018) have transformed the availability of suitable data for asteroseismic analysis. Models characterise each oscillation mode by three integers: the radial order~$n$, the harmonic degree~$l$, and the azimuthal order $-l\le m\le l$
(illustrative animations are given by the 
\href{https://wet.physics.iastate.edu/Video/index.html}{Whole Earth Telescope}).
But since the surface of stars other than the Sun are (typically) unresolved, signal averaging over the stellar surface suppresses information from all but the modes of lowest degree, with $l\le3$. 
Detailed mode properties are determined by the dominant restoring force: either pressure caused by compression and rarefaction (p-modes), or buoyancy caused by density differences affected by gravity (g-modes). 

Amplitudes and phases of the stellar oscillations are largely controlled by the near-surface layers. Frequencies are determined by the bulk sound speed, and the internal stellar density profile. Convective motions within the differentially rotating outer convective zone modify the star's temperature, density and velocity structure, while {\it convective overshooting}, caused by the momentum of cool sinking material into the deeper stable radiative regions, alters the structure of the {\it tachocline}, the transition region between the two. Stellar rotation further influences the fine structure of the frequency spectrum through its variation with stellar radius, and because of the resulting flows and instabilities which are also a function of the star's evolutionary state
\citep{2024A&A...692R...1A}.	
Here, the term `gravito-inertial asteroseismology' refers the study of g-modes in rotating intermediate-mass stars, where the modes are subject to a Coriolis term, as well as the normal buoyancy, as restoring force.
%
Magnetism is another important aspect of stellar structure and evolution which can drastically change the interior rotation, mixing and angular momentum transport
\citep[e.g.][]{2023Ap&SS.368..107B}.	

The detailed interpretation of an observed frequency spectrum proceeds via a comparison with theoretical models. For solar-type oscillations, for example, state-of-the-art asteroseismic models, based on the latest space-based observations, yield uncertainties around 3--5\% on $M_\star$ and 1--3\% on $R_\star$ based on certain `scaling relations'. 
Uncertainties of 5--10\% on luminosities follow from estimates of $R_\star$, and these in turn yield an asteroseismic distance, which can then be compared with the trigonometric distances from Gaia. As a result, the Gaia parallaxes provide a powerful test of asteroseismic (and hence stellar evolution) models. 

\paragraph{Gaia results}
Several studies using the early parallaxes from Gaia DR1 showed that the good agreement between asteroseismic and astrometric distances for solar-like oscillators (both dwarf and subgiants), which had already been demonstrated with the Hipparcos data, also hold true for the more accurate distances from the Tycho--Gaia Astrometric Solution, while the more accurate asteroseismic distances for pulsating red giants from Kepler yielded early insights into systematics in the DR1 parallaxes
\citep[e.g.][]{
2016A&A...595L...3D,
2017A&A...598L...4D,
2017ApJ...844..102H,
2017MNRAS.470L..25Y,		
2017MNRAS.470L..97G, 		
2018MNRAS.476.1931S}. 		
More detailed analyses became possible with Gaia DR2, some including constraints on Gaia's parallax zero-point
\citep[e.g.][]{
2019MNRAS.486.3569H,
2019A&A...628A..35K,
2019ApJ...878..136Z,
2019MNRAS.486.3569H,		
2019MNRAS.489..928S}.		
For example, in a study of 93~dwarfs,
\citet{2018MNRAS.481L.125S}
showed that the asteroseismic radii are 1\% smaller than Gaia radii on average, possibly explained by a negative bias of 30\muas\ in the DR1 parallaxes. They argued that asteroseismic radii are generally accurate to within 1\%, but are perhaps overestimated by 5\% or more at the highest temperatures. 

Slowly-pulsating B~stars (SPB) are a class of rapidly rotating star with masses $3-10M_\Sun$, with convective cores and radiative envelopes.  The extent to which the two mix affects the amount of hydrogen that can be accessed for core H~burning, and therefore their evolutionary paths and lifetimes. While the degree of mixing, and its effect on the boundary layer and the radiative envelope, has so far remained unknown, the Gaia distances again provide strong model constraints. 
In a study of 26~SPB stars,
\citet{2021NatAs...5..715P}
showed that their internal mixing is far from uniform, with some having almost no mixing, while others show levels a million times higher. The mixing shows no clear dependence on the star's mass or age, but does appear to be correlated with the rotation, together pointing to current limitations in the theory of internal mixing in massive stars.

\begin{figure}[t]
\centering
\includegraphics[width=1.0\linewidth]{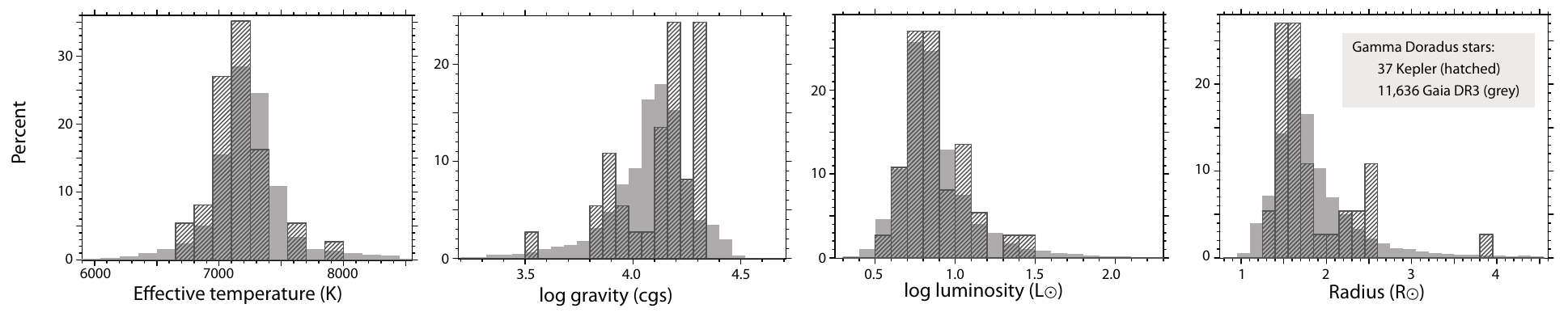}
\caption{Normalised histograms of the Gaia {\tt GSP--Phot} values for 11\,636 Gaia DR3 $\gamma$~Dor stars.  Asteroseismic values derived from Kepler data are shown for 37 $\gamma$~Dor stars (hatched). Panels show, left to right, \teff, $\log g$, $\log(L/L_\Sun)$, and derived $R_\star$ (from \citet{2023A&A...672A.183A}, Figures~1a--4a).}
\label{fig:gravito-inertial-asteroseismology}
\end{figure}

Using Gaia DR3, 
\citet{2023A&A...672A.183A} examined the properties of the 15\,062 g-mode pulsators (a subset of all non-radial pulsators, see Section~\ref{sec:non-radial-pulsators}) classified in Gaia DR3 \citep{2023A&A...674A..36G}.
They started by independently verifying the classification of the g-mode variables, then comparing the observational properties from Gaia with those from Kepler for those stars observed by both, and finally using the resulting metrics to validate the much larger sample of g-mode pulsators, and their astrophysical properties, on the basis of the Gaia data alone.
More specifically, they started with the g-mode pulsators assigned to the classes of SPB stars and $\gamma$~Doradus stars; these are main-sequence dwarfs with $M=1.3-9M_\Sun$, with convective hydrogen-burning cores, and whose internal (differential) rotation properties can be determined from their complex \mbox{g-mode} oscillations. One important finding from the original classification
\citep{2023A&A...674A..36G}	
was that many of the Gaia g-mode pulsators occur {\it outside\/} the `standard' borders of the instability strips for these two classes (Section~\ref{sec:non-radial-pulsators}), a finding ascribed to a combination of inaccuracies in the Gaia effective temperatures, their rapid rotation, incomplete input physics (such as heavy element opacities), or binary-driven instabilities.

For those pulsators whose luminosity ($L$), effective temperature (\teff), and gravity ($\log g$) were determined (and given in the table {\tt GSP--Phot}) as part of the variability classification (Section~\ref{sec:variable-stars}), their properties were compared with those of the 63 {\it bona fide\/} g-mode pulsators (37 $\gamma$~Dor and 26 SPB) observed with Kepler and follow-up spectroscopy 
\citep{2021A&A...650A..58M,
2021NatAs...5..715P}.
For the g-mode pulsators for which an estimate of the spectral line broadening is available from Gaia’s RVS data (Section~\ref{sec:stellar-rotation}), they could test for any connection between the properties of the dominant \mbox{g-mode} (the only mode available from Gaia at this stage of the data processing) and their rotation and/or spectral line broadening. Such studies have been previously hampered by the small sample sizes, or inhomogeneous photometric and spectroscopic data.
%
Following some re-classification of the Gaia DR3 SPB stars to the $\gamma$~Dor class (based on their \teff), resulted in a sample of 15\,062 g-mode pulsators, of which 11\,636 are $\gamma$~Dor, and 3426 are SPB (a major increase over the 37 $\gamma$~Dor and 26 SPB sample from Kepler).
The Gaia and Kepler distributions of effective temperature, luminosity, and radii agree extremely well for the 11\,636 $\gamma$~Dor pulsators, although with gravities somewhat lower than the asteroseismic values (Figure~\ref{fig:gravito-inertial-asteroseismology}). 
The interpretation was more involved for the SPB pulsators, due to the fact that the Gaia sample covers mainly the cooler and less massive class members.

It is also worth emphasising that the accuracy of the {\it dominant\/} frequency determined from the Gaia data is presently limited by the current Gaia DR3 temporal coverage, but nonetheless it is broadly consistent with the much denser Kepler sampling, and with the detection of \mbox{g-mode} frequencies with an amplitude in the Gaia time series $\gtrsim4$\,mmag.

Line-profile variations caused by the g-modes of $\gamma$~Dor and SPB stars occur at the level of several to tens of \kms\ 
\citep[e.g.][]{2006A&A...449..281D}, 
with some of this knowledge gained from spectroscopic follow-up on the $\gamma$~Dor stars discovered with Hipparcos 
\citep{2004A&A...417..189M}. 
In practice, high-resolution multi-epoch spectroscopy is required to distinguish these variations from rotational broadening. But this offers a powerful tool for identifying the spherical wavenumbers ($l$, $m$) of the dominant oscillations. 
\citet{2023A&A...672A.183A} found that the (DR3 RVS) quantity {\tt vsini\_esphs} provides a good estimate of the time-independent spectral line broadening, reflecting the fact that surface rotation is the dominant broadening mechanism in their samples. They also demonstrated that the dominant g-mode frequency is a significant predictor of the Gaia DR3 {\tt vbroad} parameter
\citep{2023A&A...674A...8F}.	

\citet{2024A&A...691A.131M}	
also estimated that the number of g-mode pulsators identified in Gaia~DR3, and thus potential asteroseismic targets, is more than 20 times larger than the Kepler sample. They used \teff\ and $L_\star$ inferred from Gaia to deduce evolutionary and convective core masses, radii, and ages for 14\,000 g-mode pulsators classified from their TESS light curves. As found by 
\citet{2023A&A...674A..36G},	
their new gravity-mode pulsators occupy an extended observational instability region covering masses in the range $1.3-9M_\Sun$. 

Amongst other new asteroseismology insights derived from the Gaia observations are
the effects of microscopic diffusion and isotopic radiative levitation in B-type stars
\citep{2024A&A...687A.175R},	
details of the differential rotation including the near-core rotation frequency for 2497 g-mode pulsators
\citep{2025A&A...695A.214A},	
interior rotation and mixing processes in $\beta$~Cep stars
\citep{2025A&A...698A.253F},	
including the particular case of HD~192575
\citep{2023NatAs...7..913B,
2023Ap&SS.368..107B},	
the fraction of $\delta$~Scuti stars in stellar associations as a function of age
\citep{2024MNRAS.534.3022M},	
the inferred radii of red giants and red clump stars 
\citep{2024A&A...690A.327V,	
2025A&A...693A.159V},	
and insights into the internal structure, crystalline state, and atmospheric depth in white dwarfs
\citep[e.g.][]{2019A&A...632A.119C,
2022RNAAS...6..244B,
2024ApJ...970...27B,
2024A&A...691A.194C}.	

\subsection{Open clusters}
\label{sec:open-clusters}
Open clusters are prominent spatial concentrations of stars, common throughout the Galaxy disk, which represent the sites of relatively recent episodes of star formation. They are of great importance for studies of star formation, stellar evolution, stellar dynamics, and binary star evolution.
Amongst several hundred papers detailing studies of open clusters with Gaia, there is widespread appreciation that Gaia's contribution is proving transformational, with comments such as 
{\it `The Gaia revolution has also allowed for membership studies of open clusters on the Galactic scale including thousands of clusters'} \citep{2023A&A...674A.152F}, and
{\it `It is probably safe to assume that no single instrument will have, in the near future, a transformative impact comparable to Gaia'} \citep{2024NewAR..9901696C}. 
I will organise this review as a brief introduction, the current Gaia census, and their Galactic distribution, ages, kinematics, and chemistry, followed by a look at some specific clusters, notably the Hyades and Pleiades. Let me stress that I am only reviewing the Gaia contributions to their study. And I recall that substantial improvements in quality will come (as with all the topics considered in this review) with the future Gaia data releases, DR4 in late 2026, and DR5 around 2030. 
Recent reviews include further details of Gaia's contribution to the study of open clusters
\citep{2019ARA&A..57..227K,	
2022Univ....8..111C,			
2024NewAR..9901696C}.		

\paragraph{Introduction}

Open clusters (and in contrast to the much larger and older globular clusters) are groups of typically several hundred stars, sites of star formation that were formed from the same giant molecular cloud, and therefore are all of roughly the same age, and initial chemical composition. They are still being formed throughout the disk of our Galaxy, at an estimated rate of one every few thousand years.
Before Gaia, more than a thousand were known within our Galaxy, and many more were thought to exist. A few, specifically the Hyades (at about 45\,pc), the Pleiades (at about 130\,pc), and Alpha Persei (at about 175\,pc), are visible with the naked eye. 
Open clusters are key objects in the study of stellar evolution. Because the cluster members are of similar age and chemical composition, their properties (such as distance, age, metallicity, extinction, and velocity) are more easily determined than they are for isolated stars.

Although loosely bound by gravitational attraction when formed, open clusters slowly disperse as gas (and therefore mass) is stripped by the radiation pressure of their hot young stars. They are further disrupted by close encounters with other cluster stars, and other external structures (such as the bar and spiral arms) as they orbit the Galaxy. They may survive as recognisable clumps for a few hundred million years, with the most massive surviving for a few billion. As stars slowly escape the gravitational field of the cluster, many will still be moving through space on a roughly similar path around the Galaxy, in what is known as a moving cluster or moving group. 

With its survey of all stars down to around 21~mag, its accurate star positions and space motions out to 1000~pc or more, and its unprecedented multi-epoch multi-colour photometry, Gaia is significantly advancing their study. Quality positions allow cluster membership to be refined, space motions convey details of their dynamics and dispersion, and detailed photometry further contributes to classifying membership and chemistry.

\subsubsection{Current census}

Using Data Release~2,
\citet{2018A&A...618A..93C}
examined two extensive pre-Gaia open cluster catalogue compilations 
\citep{2002A&A...389..871D, 2013A&A...558A..53K}.
From more than 3000 candidate clusters, they could confirm only around 1200. 
The conclusion is that many previously suggested clusters are simply `asterisms', including stellar over-densities arising from strong extinction patterns in the direction of the Galactic bulge
\citep{2018MNRAS.480.5242K}.

Many Gaia-based studies have taken up the challenge of identifying {\it new\/} open clusters in DR2 and EDR3/DR3, using the accurate proper motions, and a variety of advanced clustering algorithms, to identify groups of stars that are kinematically coherent in their motion through space, even if they may be spatially sparse
\citep{2023MNRAS.526.4107P}. 
But it is not just the {\it number\/} of clusters that is at stake. A well-defined sample of clusters is needed to understand their distribution as a function of Galactic location, within the plane and the Galaxy's spiral arms, and how their Galactic location evolves with age. For example, excluding false candidates results in a cluster age function in better agreement with theoretical models of their formation and dissolution \citep{2021A&A...645L...2A}. 
But neither is there likely to be a clear answer as to how many new clusters will be found with the Gaia data: at the sparsest levels, even pairs of co-moving stars may point to long-disrupted clusters
\citep[e.g.][]{2019ApJ...884L..42K}.

Amongst the discoveries made with Gaia DR2, a number of papers have each reported more than 200 new clusters:
\citet{2019JKAS...52..145S} reported 207 within~1\,kpc;
\citet{2020A&A...635A..45C} used machine-learning to find 570 with $\vert b\vert<20^\circ$;
\citet{2020A&A...640A...1C} reported 2017 out to 4\,kpc;
and 
\citet{2021MNRAS.504..356D} found 1742, also making use of Gaia radial velocities.
Papers reporting 200 or more new open clusters increased further with EDR3:
\citet{2022A&A...661A.118C} found 628, mostly beyond 1\,kpc;
\citet{2022ApJS..260....8H} found 541 mostly within 3\,kpc;
\citet{2022ApJS..262....7H} reported 836, including 46 with $\vert b\vert>20^\circ$, resulting in a nearly threefold increase in the number at high Galactic latitudes;
\citet{2022A&A...660A...4H} found 703;
\citet{2022A&A...659A..59T} found 467 searching out to 50~pc from the cluster centres;
and
\citet{2023ApJS..264....8H}, using unsupervised machine-learning, discovered 1656 beyond 1.2\,kpc.
With DR3,
\citet{2023ApJS..266...36C} discovered 1179 new clusters within 5\,kpc;
and
\citet{2023A&A...673A.114H} discovered 6272.
The latter used a blind all-sky search, starting with 729 million sources to $G=20$, creating a homogeneous catalogue of clusters including many new objects, and making use of the {\tt HDBSCAN} algorithm \citep{2013-campello}.
And as one example which illustrates the difficulties of cluster identification in the pre-Gaia era as a result of the very high density of background sources, even at moderate distances,
\citet{2021MNRAS.505.1618N} 	
used Gaia DR2 astrometry to identify a new massive young cluster at 2.3\,kpc in the Sagittarius arm with more than 30 stars brighter than $G=13$. They inferred this to be one of the most massive clusters in the solar neighbourhood, and with an initial mass of $\sim10^4M_\Sun$.

The {\it Unified Cluster Catalogue}, online at \href{https://ucc.ar}{https://ucc.ar},
is an ambitious compilation by \citet{2023MNRAS.526.4107P} (with an intuitive interface, and links to SIMBAD), being by far the largest catalogue of open clusters in the Milky Way, with almost 14\,000 clusters listed as of early 2025. Derived from Gaia DR3 using stars as faint as $G=20$~mag, each cluster is processed with their specific probability membership algorithm, incorporating each star's coordinates, parallax, proper motions, and associated uncertainties. Their compilation of these ongoing open cluster discoveries results in a combined list of 24\,983 open clusters, which reduced to 13\,684 {\it unique\/} open clusters after cross-matching, with more than a million probable members identified in total. 
%
Typical clusters have around 50 members, but many of the most populated have several hundred up to several thousand. Most have half-number radii in the range 2--6\,pc, somewhat independent of numerical size
\citep{2022A&A...659A..59T,
2022AJ....164...54Z, 
2023A&A...673A.114H}.
All of these advances allow improved studies of membership, Galactic distribution, age, chemistry, and dynamics.

In their wide-ranging review,
\citet{2019ARA&A..57..227K}
state that star clusters, which cover a huge range of mass, size, and density scales, {\it `remain mysterious'}. {\it `Conceivably'}, they continue, {\it `all stars formed in groups, clusters, or hierarchies, although, for this to be true, most clusters must have dissolved into the Galactic background soon after formation. However, our understanding of when, how, and why stars cluster remains primitive'}.

\subsubsection{Galactic distribution}

Open clusters are found to be concentrated in the Galactic plane, confirming the pre-Gaia picture but with much greater precision and clarity
\citep{2018A&A...619A.155S,	
2021A&A...647A..19T}.		
The youngest are typically located less than 100\,pc from the mid-plane, where the cold gas densities are highest, while the oldest have reached heights of more than 1\,kpc
The inner disk hosts relatively few old clusters, the denser environment presumably leading to higher disruption rates, although the inner population remains poorly constrained as a result of extinction and reddening \citep{2024IAUS..377..107M}. 
Some aspects of their Galactic distribution nevertheless remain uncertain. For example, the dearth of young clusters in the outer disk may be a result of star formation, which is clearly still ongoing at large distances, but which occurs at densities too low to form gravitationally bound clusters 
\citep{2008Natur.455..641P}. 
Old clusters beyond $\sim$12\,kpc may have formed in the inner Milky Way before migrating outwards. And the fact that the orbits of clusters older than 3\,Gyr have larger eccentricities and inclinations than field stars of the same age, suggests that clusters are more likely to survive if their orbits take them beyond the plane for most of the time
\citep{2023A&A...679A.122V}. 
In the outermost regions of the disk, distant clusters tend to be found {\it below\/} the Galactic plane 
\citep{2020A&A...640A...1C}, 
following the disk's known warp
\citep[e.g.][]{2023ApJ...954L...9H}.

The distribution of young clusters has long been considered to broadly follow the expected trace of the spiral arms, although pre-Gaia, distances uncertainties were too large to allow for an accurate characterisation of the spiral structure in the solar neighbourhood. The greatly enlarged Gaia census, along with improved distances and photometry, has brought significant insights into their distribution within 2\,kpc. 
But challenging the picture of a grand-design Milky Way with continuous and well-defined spiral structures, studies find a much more fragmented pattern
\citep{2018A&A...618A..93C,
2019A&A...624A.126C,
2019MNRAS.488.2158M,
2020A&A...640A...1C,
2021A&A...651L..10K,
2021MNRAS.504.2968P,
2023A&A...673A.114H}.
Their distribution and dynamics is also contributing to the long-standing question of whether the spiral perturbations are global and stationary, or local and transient 
\citep[e.g.][]{2020RAA....20..159S}. 
From the space motion of young clusters, the Galaxy's four classical spiral arms appear to have distinct pattern speeds, all of them close to the corotation of the disk
\citep{2021A&A...652A.162C}. 
This supports the idea that they are short-lived structures, rather than long-standing Galaxy-scale density waves.

\subsubsection{Binary clusters}
\label{sec:binary-clusters}

There were already hints in the 1970s that open clusters are sometimes found in pairs
\citep{1976Ap.....12..204R}.		
More recent simulations and observations seem to confirm that such pairs may form together, orbiting each other before gradual mass loss eventually leads to them separating and evolving as two independent clusters
\citep{2016MNRAS.455.3126C,		
2021ARep...65..755C}.			
Several new discoveries and detailed kinematic studies with Gaia DR3 support this picture
\citep{2021ARep...65..755C,		
2025AJ....169...98H,		
2024A&A...686A.225I,	
2025A&A...700A.280L,	
2024PASP..136f4201M,	
2025A&A...693A.317Q,	
2022A&A...666A..75S}.	
Amongst these, of three open clusters near the Aquila Rift cloud, UPK~39 and UPK~41 appear to be a primordial open cluster binary pair, likely to have formed at the same time, with PHOC~39 possibly capturing both in the future
\citep{2022AJ....164..132Y}.	

\begin{figure}[t]
\centering
\includegraphics[width=0.74\linewidth]{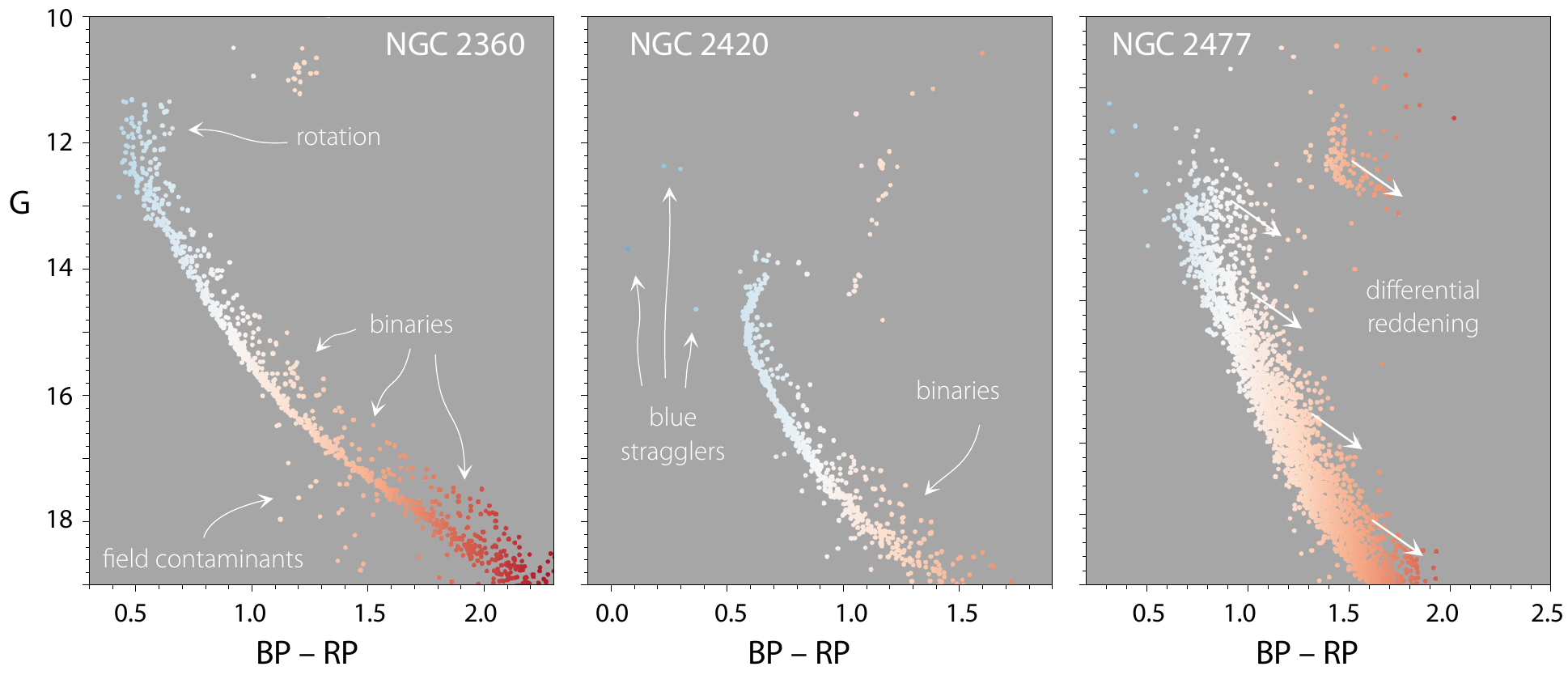}
\vspace{-5pt}
\caption{Gaia colour--magnitude diagrams of three clusters \citep[members from ][]{2023A&A...673A.114H}. Isochrones predict that stars of the same age and composition are aligned on a single line, but models must take account of numerous other effects, including field star contamination, binaries, stellar variability and rotation, differential reddening, and merger products including blue stragglers \citep{2024NewAR..9901696C}.}
\label{fig:open-clusters-cantat}
\end{figure}

\subsubsection{Ages}
\label{sec:cluster-ages}

\begin{figure}[t]
\centering
\includegraphics[width=0.53\linewidth]{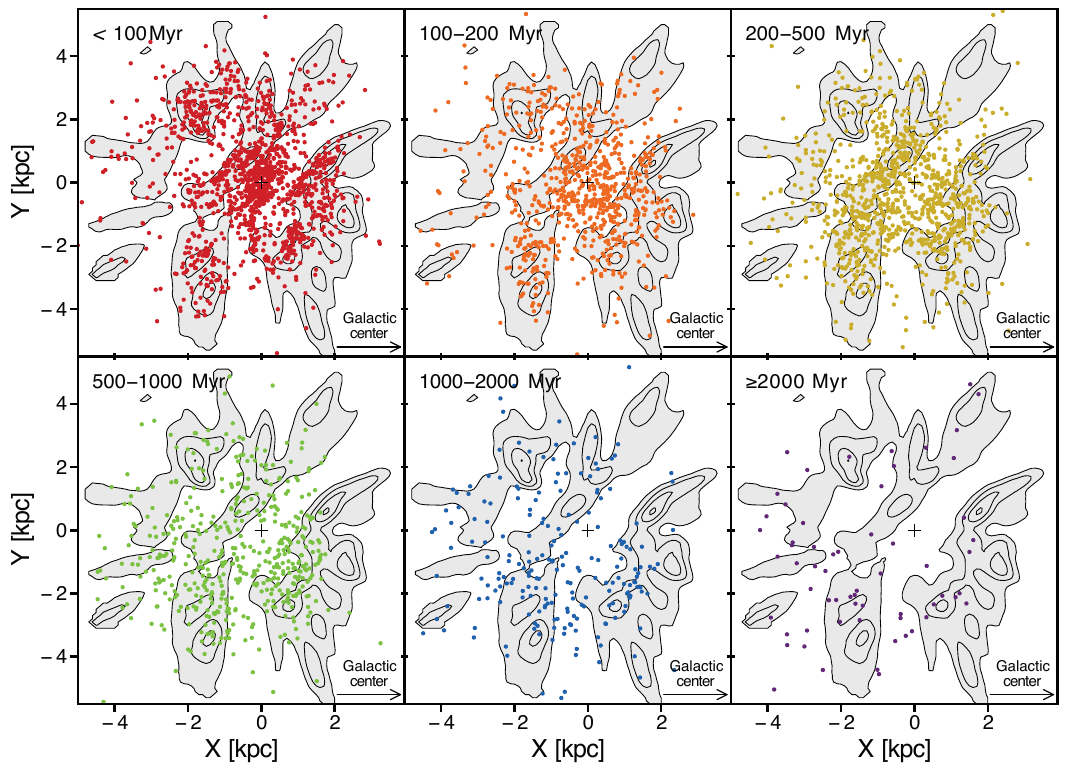}
\hspace{20pt}
\includegraphics[width=0.41\linewidth]{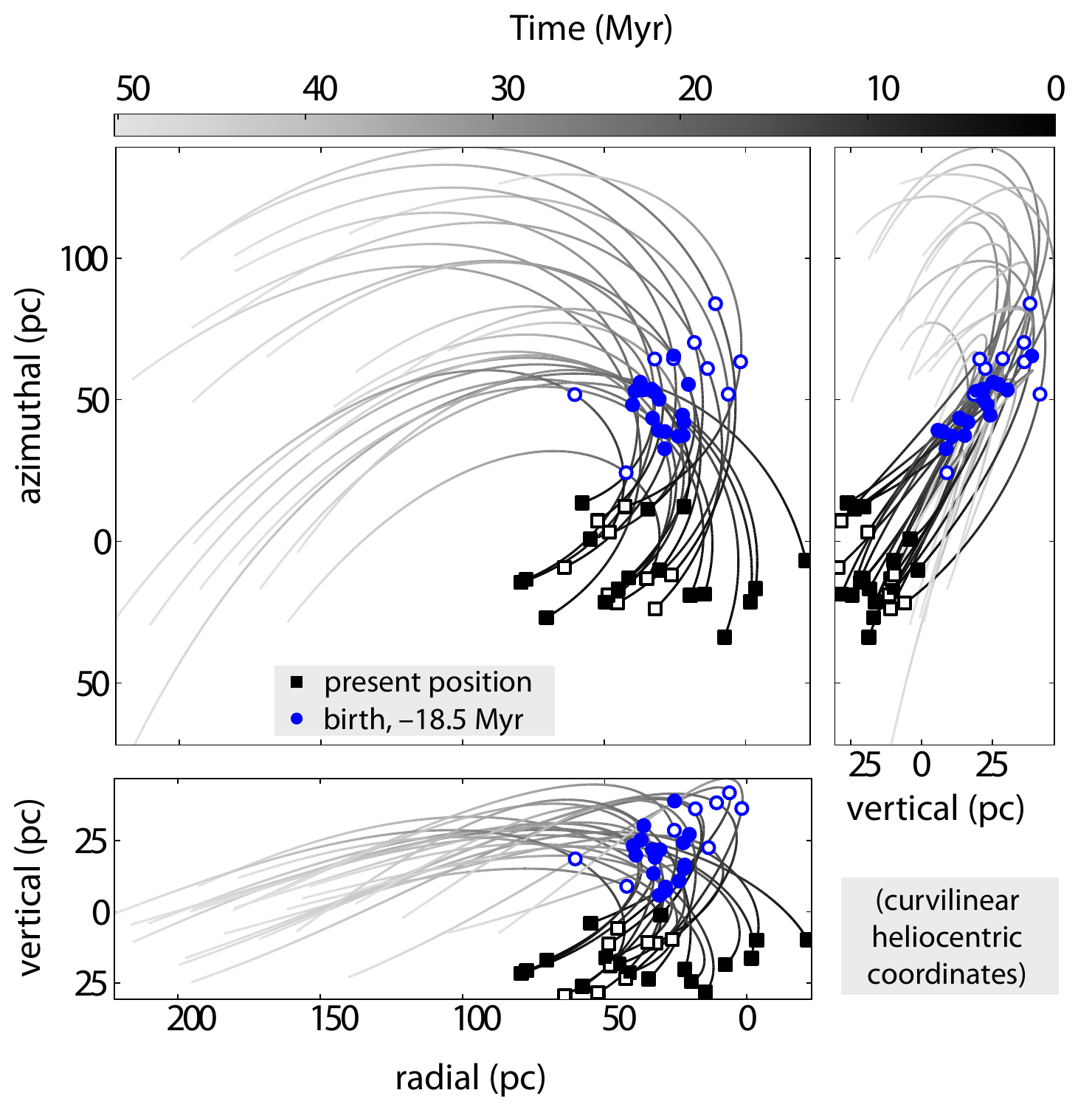}
\caption{Left~(a): open clusters in Galactic $XY$ coordinates, divided in age groups, with an over-density map in grey. Results also reproduce the Galactic metallicity gradient found in spectroscopic surveys (\citet{2024AJ....167...12C}, Figure~14).
Right~(b): dynamical traceback of the $\beta$~Pic association. Orbital projections, in Galactic curvilinear heliocentric coordinates, are shown at the present epoch as black squares, and at the inferred birth epoch, 18.5~Myr ago, as blue circles \cite[from][Figure~3]{2020A&A...642A.179M}.}
\label{fig:clusters-cavallo24}
\end{figure}

Cluster ages are mostly inferred by comparing colour--magnitude diagrams with theoretical isochrones. Accurate ages nonetheless require knowledge of the cluster's metallicity and extinction, as well as requiring sufficient members in key evolutionary phases.  Cluster age determination, based on homogeneous astrometry and photometry, has been transformed by the Gaia data. In some of the first such studies using DR2, Bayesian analysis was used to derive ages, distances, and reddenings of several tens to several hundreds of clusters
\citep{2006ApJ...645.1436V, 	
2019A&A...623A.108B, 		
2020MNRAS.499.1874M,		
2019MNRAS.487.2385M,		
2021MNRAS.504..356D}.		
Studies using EDR3/DR3 aimed to better account for unresolved binaries, field stars, differential reddening, stellar rotation, and the presence of blue stragglers (Section~\ref{sec:blue-stragglers}), by treating each cluster as a mix of single and binary stars, with some field stars (Figure~\ref{fig:open-clusters-cantat}). Applied to 10 clusters using EDR3 gave binary fractions of 30--50\%, best-fit isochrones generally consistent with previous measurements but with more precise ages, and variation of the slope of the mass function as a function of age
\citep{2022ApJ...930...44L}.	
Machine-learning, leading to more robust estimates of ages (and other parameters), has been variously applied:
to 1900 clusters and co-moving groups (comprising 300\,000 sources) within 1~kpc by
\citet{2019AJ....158..122K},	
to about 2000 clusters by 
\citet{2020A&A...640A...1C},	
to some 4000 by
\citet{2023A&A...673A.114H},	
and some 5400 by
\citet{2024AJ....167...12C}.	
The latter authors, for example, argue that their algorithm effectively traces sequences in colour--magnitude diagrams despite photometric errors and outliers, providing greater details of the local Galaxy structure, and also informing about cloud collapse and star formation on larger scales (Figure~\ref{fig:clusters-cavallo24}a). 
Prominent filamentary structures, oriented parallel to the Galactic plane, and some hundreds of parsecs in length, is now recognised
\citep{2019AJ....158..122K}. 
Most of these features lack a central cluster, indicating that the filamentary structure is primordial. Their velocity dispersion increases with age, suggesting a timescale for dynamical heating and disruption of 300~Myr, leaving only individual clusters to be identified at the oldest ages.

There are various other methods used for estimating cluster ages, some relevant only for certain age ranges, but all being advanced by Gaia. 
For ages in the range 20--200\,Myr, the `lithium depletion boundary' provides an independent spectroscopic estimate \citep{2004ApJ...604..272B}. Here, Gaia is probing the consistency between such age estimates, based on improved cluster membership, bolometric luminosities, and effective temperatures
\citep{2019A&A...623A..35L,		
2024NatAs...8..216M}.	
Analyses are taking account of `radius inflation', attributed to enhanced contributions of star spots and magnetic activity in the youngest stars
\citep{2022A&A...664A..70G,		
2022A&A...659A..85F},			
as well as rotational dependencies, with fast rotators generally preserving~Li over longer times 
\citep{2021MNRAS.500.1158J,		
2022MNRAS.513.5727B,			
2023ApJ...952...71S,				
2023A&A...674A.157T, 			
2023MNRAS.523..802J}. 			

A second independent method of age estimation uses gyrochronology. This exploits the decrease of stellar rotation with time, due to magnetic braking, which enables the use of rotation periods as a proxy for age. Here, one goal is to provide ages of main sequence stars, as well as clusters too young or too sparse to host the types of evolved stars more robustly used as age markers
\citep[e.g.][]{2003ApJ...586..464B,
2015Natur.517..589M,
2024ApJ...962...16D}.
The field {\it `\ldots has benefited from a tremendous boost enabled by Gaia'} 
\citep{2024NewAR..9901696C}, 
with Gaia astrometry providing improved membership, and Gaia photometry already providing rotation periods for 3~million stars 
\citep{2023A&A...674A...8F}.
Data from DR2/DR3 is providing these gyrochronology-based age estimates for 
the Hyades and Praesepe \citep{2019ApJ...879..100D}, 
$\alpha$~Per \citep{2023AJ....166...14B}, 		
M~67 \citep{2023A&A...672A.159G},			
along with some 20~other open clusters to date.

A third independent age estimate, especially relevant for young or sparse clusters and associations, uses Gaia's high-accuracy proper motions to yield age estimates based on `dynamical traceback', which is considered further in Section~\ref{sec:traceback-ages}.
Finally, ages can be inferred, somewhat indirectly, from asteroseismology: the detection of solar-like oscillations (in Kepler, as well as in Gaia photometry) provide global seismic parameters such as the `large separation' and the frequency of maximum oscillation power which, combined with the effective temperature, can be used to derive stellar masses through so-called `scaling relations'.  In turn, the mass can be used to provide an indirect estimate of age through evolutionary models
\citep[e.g.][]{2019A&A...623A.108B}.

\subsubsection{Traceback ages}
\label{sec:traceback-ages}

An independent age estimation, especially relevant for young or sparse clusters and associations, and yielding what are referred to as dynamical or traceback ages, makes use of their expanding space motions. Assuming that the stars were formed at a time when the association was most spatially concentrated, tracing back their space motions can establish their birth epoch. In most cases, this yields ages broadly consistent with those from stellar evolution models, but not always precisely so. 
In practice, dynamical traceback models have variously employed linear trajectories, epicyclic approximations, or orbit integration within a specified Galactic potential. One crucial limitation has been knowledge of the stars' space motions, due to uncertainties in their distances, proper motions, and radial velocities 
\citep[e.g.][]{2002ApJ...575L..75O,
2003ApJ...599..342S,
2004ApJ...609..243O}. 
Today, the Gaia astrometry and radial velocities are providing a major advance in the determination of traceback ages for several associations and moving groups. 

One of the nearest (at about 40~pc), richest and today most intensively studied of these nearby associations or moving groups, $\beta$~Pic was discovered through the identification of two companions to the A~star $\beta$~Pic \citep{1999ApJ...520L.123B}. 
Subsequent identification of other co-moving stars has led to several hundred possible members known today, with some hosting disks, exoplanets, and exocomets. 
Of some two dozen ages for the $\beta$~Pic moving group tabulated by \citet[][their Table~6]{2020A&A...642A.179M}, pre-Gaia estimates have centred around 20~Myr 
\citep[e.g.][]{2014MNRAS.445.2169M}, 
but with Li depletion estimates as high as $25\pm3$~Myr 
\citep{2016A&A...596A..29M}, 
and more uncertain dynamical traceback ages ranging from 11.5~Myr
\citep{2002ApJ...575L..75O} 
to $31\pm21$~Myr 
\citep{2007ApJS..169..105M}.
Gaia astrometry has now been used to refine group membership, based on
DR2 
\citep{2020AJ....159..166U},	
and
subsequently DR3 
\citep{2024MNRAS.528.4760L}. 
The latter resulted in 106 single and resolved companions, and 47 unresolved binaries, but still with a wide model-dependent age range:
$23\pm8$~Myr from Dartmouth magnetic models fit to the lithium depletion boundary,
$33\pm10$~Myr fit to the Gaia $M_G$ versus $B_{\rm P}-R_{\rm P}$ colour--magnitude diagram, 
and $11\pm4$~Myr as best fit to the 2MASS--Gaia $M_{K_S}$  versus $B_{\rm P}-R_{\rm P}$ colour--magnitude diagram.

New insights have come from the association's Gaia-based dynamical traceback age.
\citet{2020A&A...642A.179M}	
used DR2 astrometry, supplemented by new ground-based as well as Gaia-determined radial velocities, to determine the accurate space motions for their selected 26~{\it bona fide\/} members. They used a specific axisymmetric (bulge/disk/halo) Galaxy potential to integrate the equations of motion, propagated backward in time to --50~Myr. They defined the dynamical age as the time at which the members of the association were most concentrated in space (Figure~\ref{fig:clusters-cavallo24}). They found a dynamical traceback age of $18.5^{+2.0}_{-2.4}$~Myr, broadly consistent with the most robust estimates from isochrone or lithium depletion models, and a size (defined by the trace covariance matrix) of 7~pc at birth. And they concluded that, with Gaia, the observational uncertainties no longer dominate the uncertainties in the  age, although it does remain sensitive to the definition and selection of association members.
Further studies using Gaia DR2 were made by 
\citet{2019MNRAS.489.3625C}. 
Using Gaia DR3,
\citet{2023ApJ...946....6C}		
concluded that the radial velocities include biases, due to gravitational redshift and convective blueshift, of order 0.6\kms. Their chosen sample of 25~stars (out of their full sample of 76 members) then yields a corrected age of $20.4\pm2.5$ Myr.
  
Other Gaia-based dynamical traceback studies have been made for the Orion star-forming region, a system of complex morphology at a distance of 400~pc, with multiple stellar populations, and star formation having taken place over an extended period of some 10~Myr 
\citep{2018AJ....156...84K,	
2021ApJ...917...21S}.		
Others include 32~Ori
\citep{2022AJ....164..151L},		
TW~Hya 
\citep{2023AJ....165..269L},		
Upper Scorpius
\citep{2021MNRAS.507.1381S}, 	
Ophiuchus
\citep{2022A&A...667A.163M},	
Cepheus Far North
\citep{2020A&A...637A..43K,
2022ApJ...941...49K},	
Fornax--Horologium
\citep{2022ApJ...941..143K},	
and others
\citep{2019ApJ...870...32K, 	
2024A&A...683A.165P}.		
Amongst these, the dynamical age of Tucana--Horologium ($\sim$40\,Myr) is consistent with both isochronal and Li-based ages
\citep{2023MNRAS.520.6245G}, 
while for Upper Scorpius the dynamical age is younger than from its colour--magnitude diagram
\citep{2022A&A...667A.163M}.

Based on a comparison of the dynamical traceback and isochrone ages for six young stellar associations,
\citet{2024NatAs...8..216M}	
showed that there is a systematic difference between the two, with the traceback ages consistently younger than the isochrone ages by an average of $5.5\pm1.1$~Myr. They concluded that the discrepancy arises because the two methods have a different time origin: if the star cluster is gravitationally bound before the dispersion of the parent gas cloud, the zero-point of the expansion time scale is a few Myr after that of the colour--magnitude diagram method.  In other words, the dynamical traceback `clock' starts when a stellar cluster or association begins to expand after expelling most of the gas, whereas the isochronal `clock' starts earlier when most stars form; for example, when most of the envelope material has collapsed
onto the disk, and the central protostar becomes observable in the infrared 
\citep[e.g.][]{2003A&A...398.1081W}. 
As a result, the age difference between the two methods provides important clues about the cluster formation and gas dispersal. In particular, the age difference appears to represent an observational measurement of the duration of the embedded phase, and the timescale of gas dissipation. 
And the oldest age, corresponding to the first star to leave the cluster, generally provides a better match to the isochronal age than the traceback method
\citep{2024A&A...683A.165P}.

\subsubsection{Evaporation and tidal tails}
\label{sec:tidal-tails}

Open clusters formed from the collapse of giant molecular clouds, and they travel on roughly circular orbits through the Galactic disk. They are progressively disrupted as a result of internal dynamics, and gravitational encounters with other molecular clouds and the wider Galactic potential.
It was the dearth of clusters older than 500\,Myr that led 
\citet{1958ApJ...127...17S}	
to suggest that clusters are `dissolved' by the very clouds from which they formed.

As described above, Gaia's distances and proper motions are allowing the identification of many more clusters, along with rigorous members and age estimation. The motions of their individual stars also allows measurement of their Galactic orbits
\citep[e.g.][]{2018A&A...619A.155S}, 	
as well as details of their dissolution, bulk rotation, and internal dynamics
\citep{2024NewAR..9901696C}.

\begin{figure}[t]
\centering
\raisebox{8pt}{\includegraphics[width=0.45\linewidth]{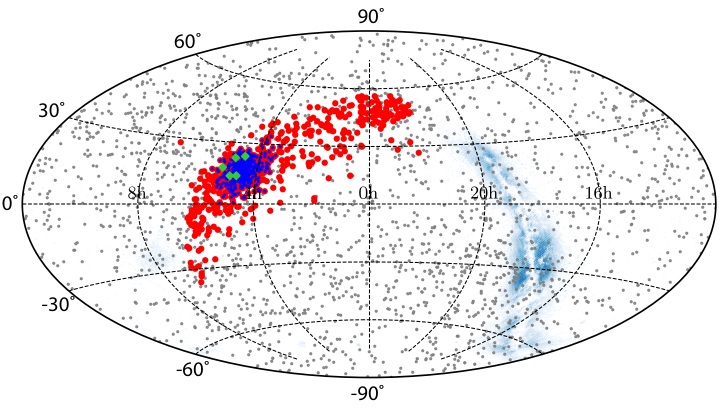}}
\hspace{20pt}
\includegraphics[width=0.38\linewidth]{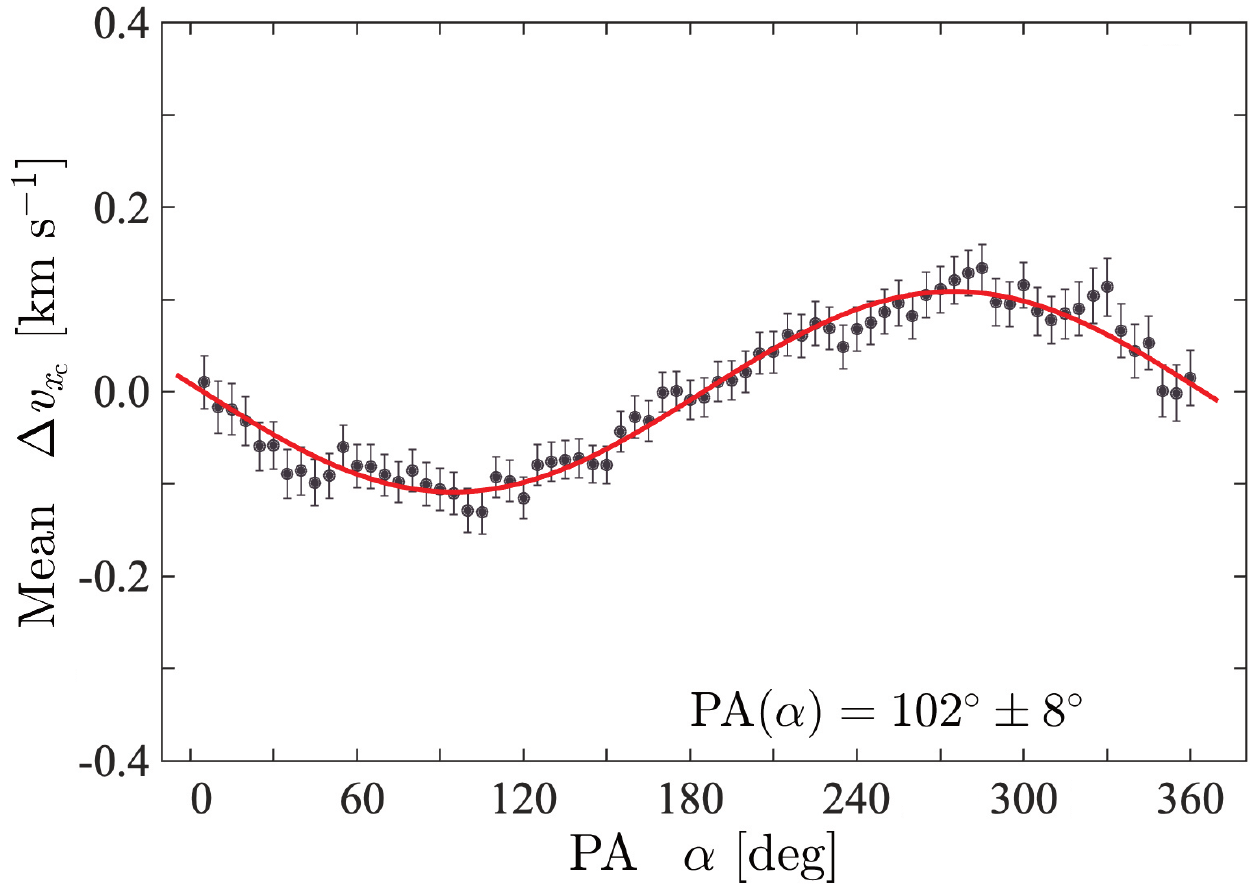}
\vspace{-5pt}
\caption{Left: Hyades cluster members from Gaia EDR3, in equatorial coordinates. Grey dots denote all 3055 initial candidates. Small blue dots denote 510 of the 515 DR2 candidate members that are confirmed by Gaia EDR3, corresponding to the central cluster region (green diamonds denote the five deprecated DR2 members). Filled red circles indicate the 920 sources that survived an ad hoc density filter aimed at suppressing non-members and highlighting the classical cluster and its tidal tails. The cloudy light blue structure in the background denotes the densest part of the Galactic plane in the direction of the Galactic centre (\citet{2021A&A...649A...6G}).
Right: Mean rotation velocity of the Pleiades for stars within one tidal radius of the cluster centre, based on Gaia DR3. Error bars (grey) and the best-fitting sine function (red) are also shown (\citet{2024ApJ...963..153H}).}
\label{fig:hyades-pleiades}
\end{figure}

As a result of two- or three-body encounters within the cluster, stars can reach escape velocity, slowly diffusing, or `evaporating', into the field population. Preferential escape through the cluster's Lagrange points leads to the creation of `tidal tails' in both the leading and trailing directions along their orbit
\citep{2000MNRAS.318..753F,
2008MNRAS.389L..28G,
2010ARA&A..48..431P}.
Tidal tails were first discovered around {\it globular\/} clusters
\citep{1995AJ....109.2553G}, 	
and prominent examples include Palomar~5
\citep{2001ApJ...548L.165O}	
and NGC~5466
\citep{2006ApJ...637L..29B}.	
But the low-density tidal tails of {\it open\/} clusters only became evident with Gaia.
For the Hyades, elongation was already found in Gaia DR1 
\citep{2018MNRAS.477.3197R}, 	
and later revealed in detail with DR2  
\citep{2019A&A...623A..35L,	
2019A&A...621L...2R,		
2019A&A...621L...3M,		
2020MNRAS.498.1920O}.		
Studies with EDR3 extended these tails to $\pm$400\,pc (Figure~\ref{fig:hyades-pleiades}a), and showed that their orientation relative to the cluster's bulk motion constrain its initial rotation
\citep{2021A&A...647A.137J}.	
Spatial concentrations in both tails are consistent with epicyclic overdensities which, in turn, provide information both on the cluster properties, and also on the Galactic potential
\citep{2009MNRAS.392..969J,	
2010MNRAS.401..105K,		
2012MNRAS.420.2700K}.		
Models suggest that the cluster is close to final dissolution, with only some 30~Myr of its existence remaining
\citep{2020MNRAS.498.1920O}.	

There is a growing literature on many other clusters with tidal tails revealed by Gaia's astrometry and photometry, with numerical modelling used to interpret their asymmetries and mass-loss rates. These include
Praesepe
\citep{2019A&A...627A...4R, 	
2023A&A...677A.163A};		
the Pleiades 
\citep{2023A&A...677A.163A};	
the rapidly dissolving Ruprecht~147 
\citep{2019AJ....157..115Y}; 	
the disk-shocked M~67 
\citep{2019A&A...627A.119C}; 	
Coma~Ber 
\citep{2019ApJ...877...12T}; 	
NGC~752 
\citep{2021MNRAS.505.1607B,	
2022MNRAS.514.3579B}; 		
UBC~274 
\citep{2022A&A...664A..31C}; 		
the Gaia discovery COIN--Gaia~13 
\citep{2022RAA....22e5022B};	
and Blanco~1
\citep{2020ApJ...889...99Z,	
2023A&A...677A.163A,		
2024ApJ...977..103S}.		
More than a hundred other nearby open clusters now have either known tidal tails, or an elongated morphology consistent with their tidal disruption
\citep{2021A&A...656A..49H,	
2021ApJ...912....5H,			
2021A&A...645A..84M,		
2022MNRAS.517.3525B, 		
2022ApJ...928...70C,		
2022ApJ...939...94M,		
2022A&A...659A..59T,		
2022MNRAS.517.3613K,		
2023A&A...671A..88P,		
2024ApJ...961..113F,		
2024A&A...691A..28K,		
2024ApJ...970...94K,			
2024MNRAS.532..622N}.		
In the Gaia-discovered `colliding' clusters IC~4665 and Collinder~350, each may be driving the other's tidal disruption
\citep{2022MNRAS.511L...1P}.

\subsubsection{Rotation and expansion}

Some clusters are expected to display bulk rotation, whether inherited from their progenitor cloud, or provided by impulsive interactions with massive structures, or resulting from the long-term action of tidal forces. Each is likely to result in distinct patterns, perhaps with a dependency on cluster age. Indeed, an open question in star formation concerns the angular momentum that newly formed clusters possess after emerging from their parent cloud. For example, 
\citet{2017NatAs...1E..64C}		
found an alignment of stellar spins and binary orbital spins. While this would support a scenario in which clusters are born with some net angular momentum, propagating down to stellar scales, and with the imprint surviving over several Gyr, other studies suggest that turbulent scrambling dominates
\citep{2023ApJ...944...39H}.	

Evidence for rotation based on the proper motions from DR2 was reported for the ancient (8\,Gyr) NGC~6791, for which the estimated inclination angle was broadly consistent with the mean inclination determined for its constituent stars
\citep{2019MNRAS.483.2197K}.	
With DR2/EDR3, bulk rotation was found in Praesepe 
\citep{2020AN....341..638L,	
2022ApJ...938..100H},	 	
and in eight other open clusters, with expansion identified in~14, and contraction evident in two
\citep{2023A&A...673A.128G}.
With the availability of Gaia DR3, astrometry and radial velocities were used to determine the mean 3d rotation velocities of the Pleiades, $\alpha$~Per, and Hyades clusters, within their tidal radii, of $0.24\pm0.04$, $0.43\pm0.08$, and $0.09\pm0.03$\kms\ respectively 
\citep[][see Figure~\ref{fig:hyades-pleiades}b]{2024ApJ...963..153H}.	
A wider study of 1379 open clusters with Gaia DR3 identified spin signatures in 10~clusters and 16~candidates, with expansion found in~18, and contraction in three
\citep{2024A&A...687A..89J}.		
The expansion rate was compatible with theoretical estimates based on the expulsion of residual gas, while the spin-axis orientation is independent of the cluster's orbital angular momentum. These authors concluded that at least 1\% of clusters are born rotating, or have undergone strong interactions since. 

\subsubsection{Mass segregation}

The structure of an initially spherically symmetric system evolves through small changes of velocity during two-body encounters, termed `relaxation'. And this is accompanied by mass segregation: massive stars and binaries sink to the centre, while low-mass stars are elevated to the halo region where they leave the cluster, or `evaporate', more easily 
\citep{1969ApJ...158L.139S,	
1995MNRAS.277.1522K,		
2007MNRAS.374..703K,		
2008gady.book.....B} \S7.5).	
Pre-Gaia evidence for mass segregation was reported in Praesepe and NGC~6231
\citep{1998A&A...333..897R,		
2013MNRAS.434.3236K},			%
as well as the Hyades 
\citep{2011A&A...531A..92R},		
although possibly attributable to the formation of more massive stars towards the cluster centres.
With Gaia, mass segregation is observed in 
Coma~Ber 
\citep{2018ApJ...862..106T},	
in Ruprecht~147 
\citep[][Fig.~10]{2019AJ....157..115Y}, 
in Czernik~3 
\citep{2020MNRAS.498.2309S}, 	
in ASCC~92 
\citep{2023MNRAS.519.6239P},
and in several of the 15~clusters using EDR3
\citep{2022MNRAS.516.5637E}.
But there is no over-riding consensus: in a study of 773 clusters using Gaia EDR3,
there was {\it `\ldots\, no significant evidence that clusters lose and segregate mass with age'}
\citep{2023MNRAS.525.2315A},
while in a study of 60~clusters using Gaia DR3, the core radii do appear to decrease with age
\citep{2023MNRAS.522..956A}.
Other studies underline the difficulty of disentangling the many effects at work
\citep{2023A&A...679A.122V,
2024A&A...687A.101A,	
2024A&A...687A.291H}.

\subsubsection{Binary star fraction}

The evolution of open clusters is also influenced by its binary population, including their own disruption by passing perturbers. Various studies have used Gaia DR3 to probe the effects of binarity.
For 78 clusters, 
\citet{2023A&A...672A..29C}	
found a binary fraction ranging from 15--60\%, hints of a correlation between the total fraction of binary stars and the central density, and with a radial distribution depending on cluster age.
Comparable ranges of binary fractions were found for 202 clusters by
\citet{2023A&A...675A..89D}, 
and for 16~clusters by 
\citet{2023ApJS..268...30L}.	 
\citet{2023AJ....166..110P}	
found a decreasing binary fraction with increasing cluster age (depending strongly on stellar density), only limited evidence for mass segregation, but clear evidence for the early disruptions of binary stars, with the binary fraction depending strongly on radial distance.
Several other studies along these lines have been reported
\citep{2024ApJ...962...41C, 
2024AJ....168..156C,	
2024ApJ...971...71J,	
2025MNRAS.536..471A}.	

\subsubsection{Evolutionary models}
\label{sec:clusters-evolution}

The scientific importance of open clusters is underpinned by the assumption that their stars started life at the same time, formed from the same giant molecular cloud, so are of similar age, and with similar initial chemical composition. These two boundary conditions provide strong constraints on models of stellar structure and evolution. Gaia's rigorous membership lists, and accurate distances and photometry, allow for the construction of clean and accurate colour--magnitude diagrams for main-sequence fitting, for the identification and characterisation of stars in specific evolutionary phases, and to identify deviations from existing state-of-the-art models.
The study of chemical abundances in open clusters, and their use in models of stellar structure and evolution, is a huge topic, and I will give only a couple of `pointers' most directly related to Gaia.

Chemical compositions of individual stars are deduced from high-resolution spectroscopic observations. Although this is not the main thrust of the Gaia mission, Gaia~DR3 itself actual provides (model-dependent) stellar parameters and chemical abundances for 5.6~million stars derived from its Radial Velocity Spectrometer
\citep{2023A&A...674A..29R}.
This has provided {\it `a powerful and precise chemo-dynamical view of the Milky Way with unprecedented spatial coverage and statistical robustness'}, including its immediate application to 503 open clusters
\citep{2023A&A...674A..38G}. 	
And in the context of these wider spectroscopic studies, Gaia-based memberships are assisting recent or planned large-scale spectroscopic surveys focused on Galactic chemo-dynamics, including 
the Gaia--ESO survey \citep{2013Msngr.154...47R}, 
APOGEE \citep{2017AJ....154...94M}, 
GALAH \citep{2015MNRAS.449.2604D},
LAMOST--LEGUE \citep{2012RAA....12..735D}, 
RAVE \citep{2020AJ....160...82S}, 
4MOST \citep{2023Msngr.190...13L}
and
WEAVE \citep{2024MNRAS.530.2688J},
and several other smaller programmes.

I will say more on the particularly favourable case of the Hyades open cluster in Section~\ref{sec:hyades}, and only note here that there remain differences between the exquisite colour--magnitude diagram provided by Gaia
\citep{2023AJ....165..108B},  
and the predictions of state-of-the-art evolutionary numerical models, such as MIST and PARSEC. 
\citet{2025ApJ...979...92W}	
derived {\it empirical\/} colour corrections based on the latest MIST and PARSEC models, and the Gaia $G_{\rm BP}-G_{\rm RP}$ and $G-G_{\rm RP}$ colours of the Hyades, Pleiades, and Praesepe. Applying these empirical corrections to 31 other clusters and three moving groups gave significantly better agreement between the isochrones and observed colour--magnitude diagrams, and isochrone ages consistent with the `lithium depletion boundary' method.

Clearly, some physics is missing. In response, various theoretical efforts are ongoing, targeting (for example)
a better concordance in the different estimates of the cluster's mean metallicity, 
more secure estimates of the He~abundance (where the primordial abundance is imprinted in the lowest mass stars), 
updated input physics including opacities \citep{2016A&A...585A...7K}, 
an improved treatment of the degeneracy between metallicity, [Fe/H], the He-to-metal enrichment, $\Delta$Y/$\Delta$Z, and the convective mixing length parameter, $\alpha_{\rm ML}$ \citep[e.g.][]{2021MNRAS.501..383T}, 
an improved treatment of convection in general and the contribution of the so-called `kissing instability'
\citep{2012ApJ...751...98V},
the problem of superadiabatic convection in the outer layers of partially convective stars \citep{2001MNRAS.320...66C},
and the possibility of `radius inflation' identified in some of the Hyades stars
\citep{2019ApJ...879...39J}. 	
Efforts are also ongoing to better understand the effects of rotation, stellar activity and variability, and to eliminate still unidentified (more extreme) brightness- and mass-ratio binary systems.

The `turnoff' region of the Hertzsprung--Russell diagram, the region where stars begin to leave the main sequence and evolve into giants, presents various puzzles, all being advanced by Gaia. 
One is the `extended main sequence turnoff', a broadening of the turnoff region seen in many clusters
\citep[e.g.][]{2018ApJ...869..139C}. 	
Once tentatively attributed to an age spread, Gaia studies show that the redder stars often correspond to fast rotators, whereas bluer stars rotate more slowly 
\citep{2018MNRAS.480.3739B,	
2018ApJ...869..139C,		
2018ApJ...863L..33M,		
2019NatAs...3...76L,			
2022ApJ...938...42H,		
2023MNRAS.524..108G}. 		
For the Galactic cluster Stock~2, the effect is attributed to differential reddening
\citep{2021A&A...656A.149A}.
The extended turnoffs in the Magellanic Clouds also appear to involve stellar rotation, but combined with significant age spreads 
\citep[e.g.][]{2014ApJ...797...35G}, and effects of stellar variability \citep{2016ApJ...832L..14S}. 
`Isochrone clouds' have been invoked to describe a coeval population of various masses, and fractional core masses, due to different internal mixing profiles
\citep{2019MNRAS.482.1231J}.	
But 
\citet{2019A&A...632A..74J}		
concluded that a {\it `global theoretical interpretation is still lacking'}.

\subsubsection{The Hyades cluster}
\label{sec:hyades}

The Hyades is the nearest moderately rich open cluster, at a distance of $\sim$45\,pc, and an extension on the sky of $20^\circ$ or more.  It has 1000--2000 members, a total mass of $300-400M_\odot$, an age of 600--800~Myr and, as revealed by the Gaia census, two prominent `tidal tails'. 
Throughout much of the 20th century, it occupied an important role in studies of Galactic structure, the chemical evolution of the Galaxy, and in the determination of the Population~I distance scale. It has been used as the basic observational material for several fundamental relationships in astrophysics, including the location of the main sequence in the Hertzsprung--Russell diagram, the mass-luminosity relationship, and the basis for the determination of luminosities of supergiants, OB stars, and peculiar stars in clusters. Without entering these historical details, I will only emphasise here the important role that Gaia is playing in better defining the cluster's main sequence, and the implications for stellar evolution models.

Hipparcos astrometry was used to make the first `space-age' study of the cluster, providing improved determination of its distance, structure, dynamics, and age \citep{1998A&A...331...81P}. That work identified 282 members (13~new), and a well-defined main sequence giving a He-abundance Y\,=\,$0.26\pm0.02$, an age of $625\pm50$\,Myr, and a centre-of-mass distance of $46.34\pm0.27$\,pc.
With the availability of Gaia DR2, 1764 objects were found to share the cluster's mean proper motion
\citep{2019A&A...623A..35L}, 
more than doubling the pre-Gaia census of around 800 members, including a handful of white dwarfs, and a dozen brown dwarfs. From 381 objects within the cluster's 9\,pc tidal radius, they derived a mean distance of $47.03\pm0.20$~pc, and a mean space velocity of $46.38\pm0.12$\kms.
%
The Hipparcos-based cluster age, $625\pm50$\,Myr, derived from its main sequence and model isochrones which include convective overshooting \citep{1998A&A...331...81P}, had been largely confirmed from stellar binaries \citep{2001A&A...374..540L}, its white dwarf cooling age \citep{2019A&A...623A..35L}, and the lithium depletion at the stellar/sub-stellar limit \citep{2018ApJ...856...40M}. However, the effects of rotation \citep{2015ApJ...807...58B}, and enhanced convective overshooting 
\citep{1988A&A...193..148M,
2012A&A...547A..99T} 
leave open the possibility of a greater age. 

\begin{figure}[t]
\centering
\includegraphics[width=0.34\linewidth]{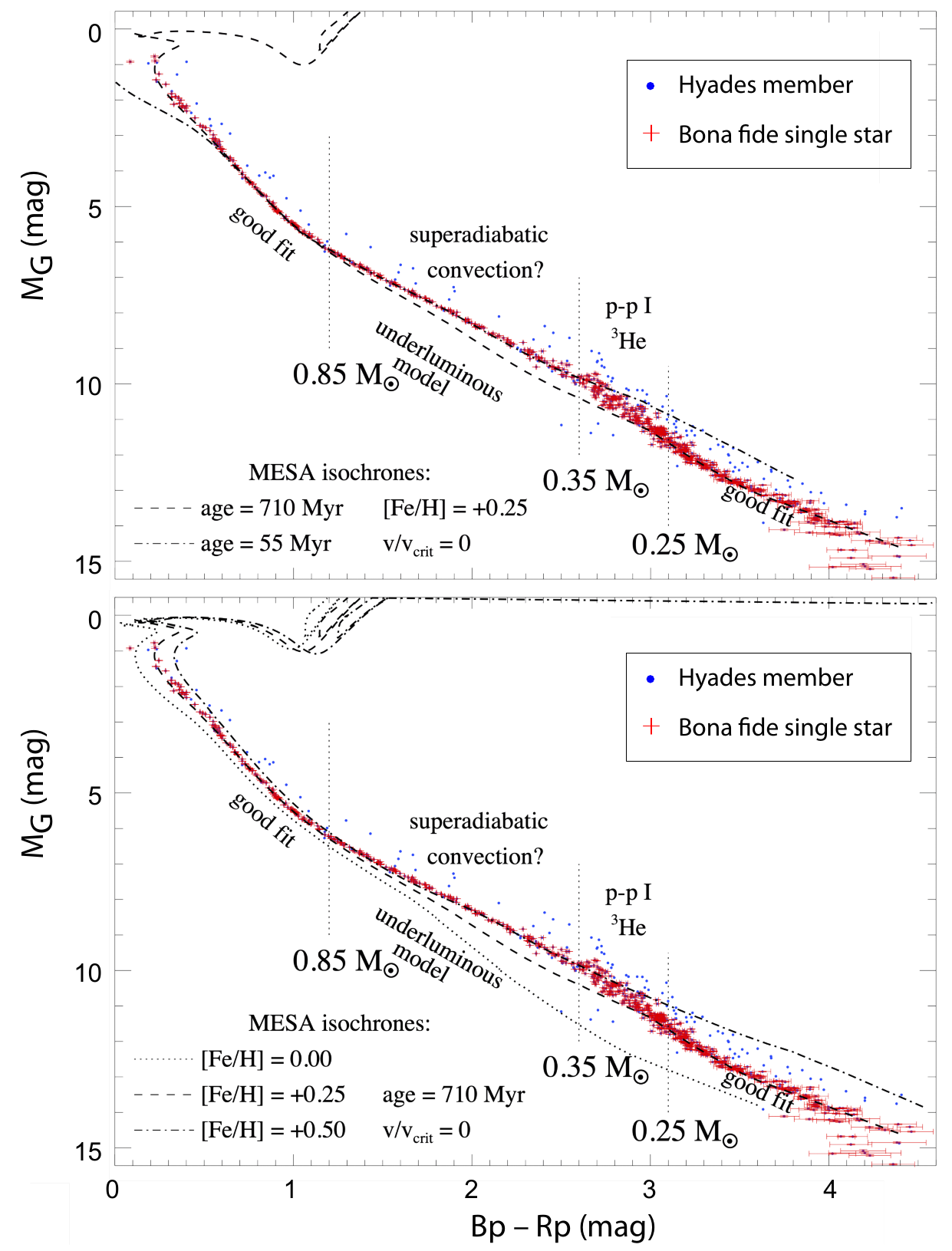}
\hspace{20pt}
\includegraphics[width=0.34\linewidth]{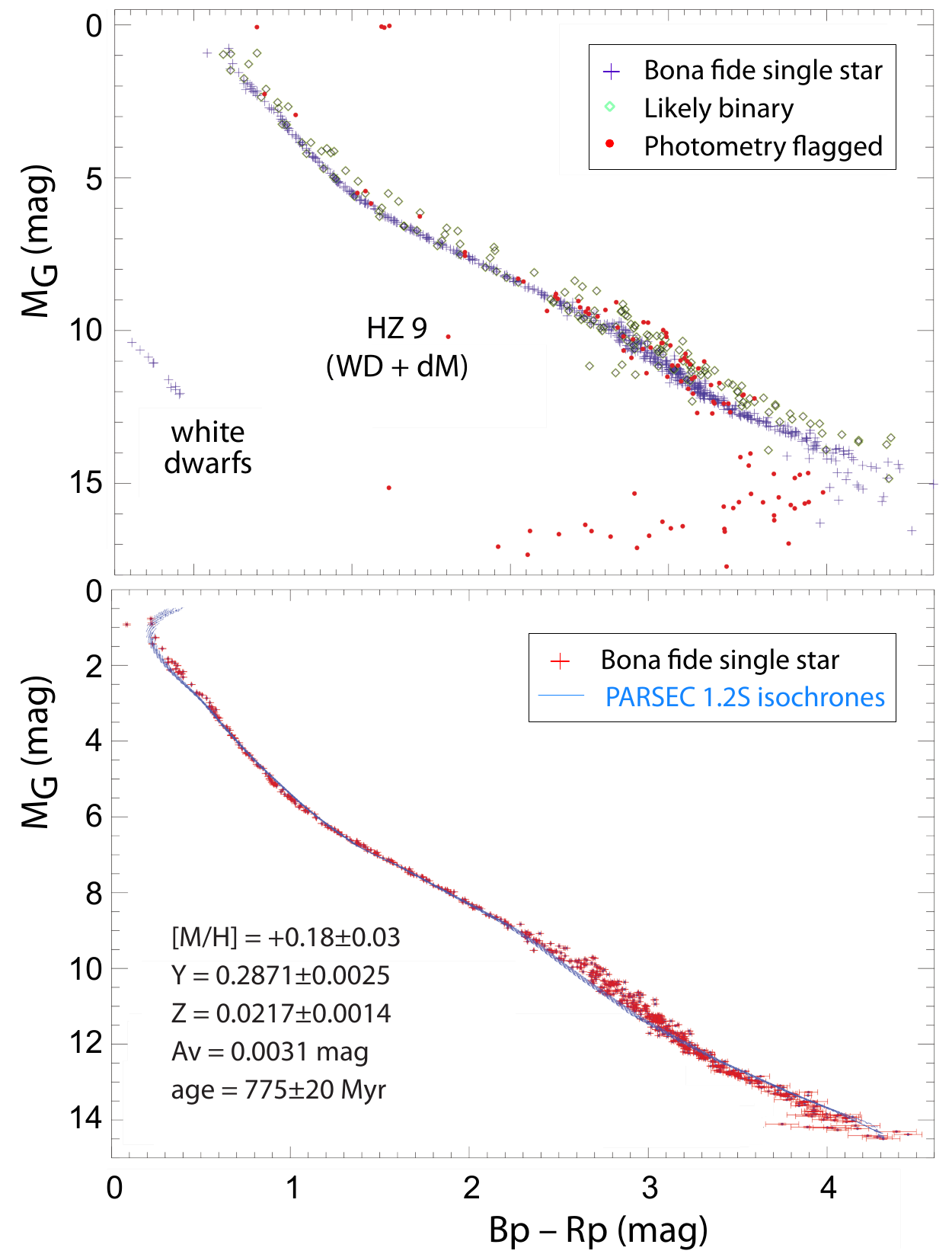}
\caption{Left pair (from \citet{2023MNRAS.518..662B}). Top: Hyades EDR3 colour--absolute magnitude diagram, showing stars from the Gaia Catalogue of Nearby Stars (GCNS, blue dots), and {\it bona fide\/} single stars (red crosses). MESA isochrones of 710~Myr provide a good fit to the upper and lower main sequence, but under-predict luminosities for $0.25-0.85M_\Sun$. Bottom: the effect of metallicity for the 710~Myr isochrones.
Right pair (from \citet{2023AJ....165..108B}). Top: Hyades DR3 color--absolute magnitude diagram, showing the results of their classification into {\it bona fide\/} single main-sequence stars and white dwarfs, likely binary and multiple systems, and stars with problematic photometry in at least one of the Gaia bands.
Bottom: PARSEC isochrones for [M/H]\,=$+0.18\pm0.03$, age $775\pm20$~Myr, and $A_V=3.1$\,mmag
provide a very good fit to the observed main-sequence. In the colour range BP--RP~=~2.4--3.2~mag ($0.22-0.40M_\Sun$) stars are systematically brighter than predicted by the isochrones.
}\label{fig:hyades-cmd}
\end{figure}

The mean metallicity has also been subject to debate, with pre-Gaia estimates lying in the slightly super-solar range [Fe/H]\,=\,0.05--0.14~dex. 
%
Part of the problem is the degeneracy between [Fe/H], the He-to-metal enrichment, $\Delta$Y/$\Delta$Z, and the convective mixing length parameter, $\alpha_{\rm ML}$, with the latter also dependent on the part of the main sequence being modelled. Using the PISA evolution models, and estimates of [Fe/H]\,=\,0.14--0.18, and the Gaia DR2 photometry and parallaxes, mean values of $\alpha_{\rm ML}=2.01\pm0.05$ and $\Delta$Y/$\Delta$Z\,=\,$2.03\pm0.33$ were derived 
\citep{2021MNRAS.501..383T}.
%
A key point to emphasise is that ongoing improvements in the observational definition of the Hyades main sequence has led to improvements in stellar models, such as the effects of superadiabatic convection in the outer layers of partially convective stars \citep{2001MNRAS.320...66C}, and the incorporation of updated input physics, including opacities \citep{2008Ap&SS.316...25D}.
Further improvements came with improved astrometry and photometry from Gaia EDR3, also leading to a more rigorous identification of {\it bona fide\/} single stars
\citep{2023MNRAS.518..662B}. 
%
This provided a particularly tightly defined stellar sequence (Figure~\ref{fig:hyades-cmd}, left pair), revealing specific inadequacies in the theoretical models. The non-rotating MESA models \citep{2016ApJ...823..102C} for [Fe/H]\,=\,+0.25, and an age of 710~Myr, provide a good fit for $M>0.85M_\odot$, and for $M<0.25M_\odot$. But for intermediate masses, the models systematically under-predict the observed luminosity. 
They suggest that for (partially convective) stars with $M>0.35M_\odot$, superadiabatic convection with $\alpha_{\rm ML}$ following the solar model is unlikely to be optimal. For $M<0.35M_\odot$, the increased scatter might be due to the (unmodelled) `convective kissing instability', an effect driven by variations in the $^3$He energy production rate due to instabilities at the convective core--envelope boundary 
\citep{2012ApJ...751...98V,
2023MNRAS.525.6005M}.	
For a Hyades-like stellar population, the application of solar-scaled models to sub-solar mass stars could, they conclude, result in a significant underestimate of the age, or to an overestimate of the metallicity. 

Model fitting to the even `tighter' main sequence based on Gaia DR3 was subsequently reported 
\citep{2023AJ....165..108B}. 
To address the discrepancy between the observations and theoretical isochrones in the range $0.25-0.85M_\odot$, they based their updated models on the PARSEC stellar evolution code.
%
Figure~\ref{fig:hyades-cmd}, right pair, shows their colour--magnitude diagram classified according to {\it bona fide\/} single main-sequence stars, probable binary and multiple systems, and stars with questionable photometry in at least one of the Gaia bands. 
The second shows the $\sim$600 single-star candidates, and the best fitting PARSEC isochrones, with age $775\pm20 $~Myr. While the fit is improved compared with MESA~1.2, over the colour range $G_{\rm BP}-G_{\rm RP}=2.4-3.2$~mag ($\sim0.22-0.40M_\odot$), stars are systematically brighter (or redder) than the isochrone predictions.

\subsubsection{The Pleiades cluster}
\label{sec:pleiades}

The Pleiades, in the constellation Taurus, has been recognised as a group of stars since antiquity, and has long played an important role in establishing the distance scale -- dating back more than a century to the work of Lewis Boss, Jacobus Kapteyn, Willem de Sitter, Henry Plummer, and Ejnar Hertzsprung. The cluster contains more than 1000 members, with a total mass of about $800M_\Sun$. Its light is dominated by young, hot blue stars, some dozen of which can be seen with the naked eye. The cluster has a core radius of 2--3~pc, and a tidal radius of $\sim$13~pc. 
Pre-Hipparcos, various less-direct methods had all converged on a distance of 130--135~pc. 
But, in 1999, the Hipparcos parallaxes gave a mean distance of $118.3\pm3.5$~pc, 7\% `shorter' than the previous consensus
\citep{1999A&A...341L..71V}.
Later work, including
optical interferometry of the binary star Atlas \citep{2004A&A...425L..45Z},
infrared main-sequence fitting \citep{2005A&A...429..887P}, 
HST--FGS observations \citep{2005AJ....129.1616S},
VLBI \citep{2014Sci...345.1029M},
consistently argued that the Hipparcos distance must be erroneous. 
The problem did not disappear: a re-analysis of the Hipparcos data a decade later led to only a small revision, from 118 to 120~pc
\citep{2009A&A...497..209V}.

The first Gaia results, DR1 in 2016, included the Tycho--Gaia Astrometric Solution (TGAS), a subset of about 2~million stars incorporating the Hipparcos and Tycho-2 positions centred at 1991.25.
Parallaxes for 152 Pleiades members gave a mean distance of $133.7\pm0.6$~pc ($7.48\pm0.03$\,mas), in agreement with the non-Hipparcos values 
\citep{2017A&A...601A..19G}.
Otherwise, their results for 19 open clusters, ranging from the Hyades (at 47~pc) to IC2422 (at 440~pc) were in agreement with those from Hipparcos.
Other analyses followed with Gaia DR2.
Uing 1594 cluster stars, an order of magnitude more than DR1, gave a mean parallax to the cluster centre of $7.34\pm0.27$~mas, or $136.2\pm5.0$~pc
\citep{2018RNAAS...2..150A}.
Other studies included 
1248 members yielding a mean distance of $135.15\pm0.43$~pc, and an age of $132\pm26$~Myr from its single white dwarf LB~1497
\citep{2019A&A...628A..66L},
and 1454 members yielding  $136.0\pm0.1$~pc
\citep{2019PASP..131d4101G}.
The papers cited here, let me stress, have much more to say about the Pleiades cluster than simply its distance. Gaia reveals, for the first time, the cluster's depth, allowing each star to be pinpointed in space with respect to its centre. It identifies stars that are slowly escaping from the cluster, gradually merging into the background population over millions of year. And it clarifies its age and stellar content from its highly accurate colour--magnitude diagram. 

Taken together, it is clear that all methods, whether ground- or space-based, now converge on a mean distance of $\sim$136~pc, with the exception of the various Hipparcos estimates, which placed it significantly, but erroneously, some 10--15~pc nearer. 
%
While the distance to the Pleiades provided, by far, the largest controversy in the Hipparcos results, it should be placed in context. At 135~pc distance, the mean individual parallax is $\sim$7.4~mas. Hoping to establish the distance with a 10\% error was always going to be at the limit of the Hipparcos accuracies. Substantially compounding the problem, the stars are also much closer together on the sky than Hipparcos was designed to probe, almost certainly leading to correlated astrometric errors
\citep[e.g.][]{1998ApJ...504..170P,
1999ApJ...523..328N},
and possibly an associated bias in the satellite attitude effectively leaving the central part of the cluster tied with only limited rigidity to the rest of the catalogue
\citep{2002AJ....124.3299M,
2022RMxAA..58..315M}.	

Today, the Gaia parallaxes have been pushed to such great distances that the Hyades and Pleiades, for many decades key steps in unravelling the cosmological distance scale, are no longer relevant in this endeavour -- a century of ingenuity and experimental perseverance has been consigned, at a stroke, to history. 
Open clusters will remain of central importance for evolutionary and dynamical studies of stellar systems. But the Hyades and Pleiades have probably made their final appearance in the lengthy and ingenious history of the distance scale in astronomy.

\subsection{Associations}
\label{sec:associations}

In contrast to the higher stellar density and gravitationally bound open clusters, {\it associations\/} were viewed as looser groups of some 10 or more physically related stars, often of large spatial extent
\citep[e.g.][]{1991ASPC...13....3L}.	
OB~associations are characterised by their overdensity of O and B~stars (the even younger T-associations comprise T~Tauri stars). High masses and luminosities imply that the stars are young and short lived, and so implicitly associated with sites of recent star formation. They are unbound, and survive as recognisable groups only for a short time, of order 25--50\,Myr.

A number of OB~associations within 150~pc of the Sun have long been known
\citep[e.g.][]{1914ApJ....40...43K,
1921MeLuS..26....3R,
1929PUAms...2....1P,	
1946PGro...52....1B},
including Scorpius--Centaurus (with subgroups Lower Centaurus Crux and Upper Scorpius), Taurus--Auriga, Hercules--Lyra, and Tucana--Horologium. 
OB~associations have also been found in the Large Magellanic Cloud and the Andromeda Galaxy.
Their large extent on the sky has traditionally prevented accurate kinematic membership determination for any but the brightest stars. 
Many studies were already made with the Hipparcos data, which provided a major improvement in their kinematic detection, resulting (for example) in improved astrometric members for 12~young stellar associations out to a distance of 650\,pc
\citep{1999AJ....117..354D}.		

In the `monolithic formation scenario', the unbound OB~associations found in the solar neighbourhood and beyond were significantly more compact (and gravitationally bound) at the time of their formation, subsequently expanding (through the expulsion of residual gas as a consequence of stellar feedback) into the configurations seen today
\citep[e.g.][]{2012MNRAS.426.3008K,	
2012MNRAS.419..841K}. 				
Over the past few years, this model has seemed somewhat in contradiction to observations of present-day star formation, which appears to proceed over a wide range of environments, including both large-scale hierarchical structures and isolated young stellar objects
\citep{2020MNRAS.495..663W}. 

The first studies of around 50 OB~associations using the early Gaia data already seemed in contradiction with the idea of monolithic formation. The problem was that none of the associations exhibited any significant evidence for an expanding velocity field. Indeed, their bulk kinematic properties were more consistent with randomised velocity fields, suggesting that they were formed in a highly sub-structured state with multiple small-scale star formation events rather than a single, monolithic burst 
\citep{2017MNRAS.472.3887M,	
2018MNRAS.475.5659W,			
2018MNRAS.476..381W,			
2019A&A...621A.115C,			
2020MNRAS.493.2339M}.		
In a much larger study using Gaia DR2 
\citep{2020MNRAS.495..663W}		
their starting point was the Galactic OB star catalogue, \mbox{GALOBSTARS} \citep{2003AJ....125.2531R}, containing some 16\,000 OB-star candidates.  Further selection according to Gaia colours resulted in a set of 11\,844 OB~stars out to distance of 3--4\,kpc. Their cluster-finding algorithm resulted in the discovery of 109 OB~associations out to about 3\,kpc. With typically 10--50 OB~stars in each of these, they then searched for additional, lower mass members from the wider DR2 catalogue. This resulted in a typical total of about 6000 stars in each association. From the detailed kinematics, they concluded that a simple monolithic cluster, that subsequently underwent gas-expulsion driven expansion, can be ruled out as an origin of the associations, but neither are the motions consistent with purely random velocity fields. Only with a combination of small-scale localised expansion events, along with positional substructure and a randomised but sub-structured velocity field, were they able to reproduce the kinematic properties of these OB associations. 
In other word, the Gaia results suggest a scale-free, hierarchical picture of star formation, in which stars are formed across a continuous density distribution throughout molecular clouds, rather than exclusively within clusters, and in which OB associations are formed {\it in situ\/} as relatively large-scale and gravitationally-unbound structures. 

From a substantial detailed literature, let me give some further details of the Gaia-based studies of the nearest and most well-studied association, Scorpius--Centaurus (Sco--Cen, aka Sco~OB2), and its subgroups including Lower Centaurus Crux and Upper Scorpius. 
At 130\,pc, and extending over 2000~sq.~deg., Hipparcos identified over 400 members  
\citep{1999AJ....117..354D}. 	
Gaia is providing a more rigorous membership, and the detection of lower-mass stars, and has brought the latest census to around 15\,000  
\citep{2018ApJ...868...32G,		
2018AJ....156...76L,		
2018A&A...614A..81R,	
2018MNRAS.476..381W,	
2019A&A...623A.112D,	
2020AJ....160...44L,		
2021MNRAS.507.1381S,	
2022AJ....163...24L,		
2023MNRAS.522.1288B,	
2023A&A...678A..71R, 	
2023A&A...677A..59R,	
2023MNRAS.519.3992Z,	
2024ARep...68..878B}.	
Amongst these, within Lower Centaurus Crux,
\citet{2018ApJ...868...32G}		
reported {\it `\ldots the revelation of a large moving group containing more than 1800 intermediate- and low-mass young stellar objects and brown dwarfs that escaped identification until Gaia DR2'}.
Gaia DR2 allowed the discovery of 11\,000 pre-main-sequence members
\citep{2019A&A...623A.112D},	
along with eight kinematically distinct components consisting of 8185 stars distributed in dense and diffuse groups, each with an independently fit kinematic age
\citep{2023MNRAS.519.3992Z}.	
From EDR3, it is estimated that the complex contains 10\,000 members with masses $\gtrsim0.01M_\odot$.
\citep{2022AJ....163...24L},		
while DR3 has resulted in more than 13\,000 young co-moving objects (19\% of sub-stellar mass), organised into 37 co-moving groupings
\citep{2023A&A...677A..59R}.		
\citet{2021MNRAS.507.1381S}	
concluded that star formation in Sco--Cen lasted more than 10~Myr, proceeding in small groups that, after a few Myr, dissolve in the field of the older population, but retaining memory of their initial structure.
A detailed star-formation chronology is given by 
\citet{2023A&A...678A..71R}.	

Adding to this complex picture, and amongst their identified structures of pre-main sequence stars selected through the use of $\log g$,
\citet{2025ApJ...984...58H}	
drew attention to a population of more than 1000 stars, east of Sco--Cen, which they named Ophion. While this overdensity had been noted in the Gaia DR3 study by 
\citet{2023A&A...677A..59R},		
what makes it unusual is that, despite a coherent age of $\sim$20\,Myr, and at a common distance of $\sim$200\,pc, it has negligible kinematic coherence. Specifically, it has a large velocity dispersion $>20$\kms, without any obvious expansion pattern, therefore appearing to be fully `disrupted', but at the same time still persisting as a spatial overdensity. 
Given the fragmented morphology in this region of Sco--Cen, and the presence of several apparent `bubbles' in the distribution of young stars, they infer that these features were likely driven by one or more supernovae.
Their conclusion is that this Ophion star-forming region lost mass rapidly during the gas-expulsion phase, possibly through a combination of both supernovae and tidal interactions, and resulting in the acceleration of the bulk of the stars to very high velocities. 
That the region remains spatially coherent would simply be because it has been caught in the earliest stages of its rapid and complete disruption. 
And it is noteworthy that cluster search algorithms relying on both spatial and kinematic coherence would not have found it.

On the present understanding of associations, 
\citet{2020NewAR..9001549W} 
concludes:
{\it `It is clear now that OB associations have considerably more substructure than once envisioned, both spatially, kinematically and temporally. These changes have implications for the star formation process, the formation and evolution of planetary systems, and the build-up of stellar populations across galaxies.'}

I will finish this section with a summary of the present understanding of the formation, evolution, and eventual disruption of star clusters by quoting verbatim from the wide-ranging review `Star Clusters Across Cosmic Time' by
\citet{2019ARA&A..57..227K}.	
They state: ``The current literature suggests a picture of this life cycle with several phases:
(a)~clusters form in hierarchically-structured, accreting molecular clouds that convert gas into stars at a low rate per dynamical time until feedback disperses the gas;
(b)~the densest parts of the hierarchy resist gas removal long enough to reach high star formation efficiency, becoming dynamically-relaxed and well-mixed. These remain bound after gas removal;
(c)~in the 100\,Myr after gas removal, clusters disperse moderately fast, through a combination of mass loss and tidal shocks by dense molecular structures in the star-forming environment;
(d)~after $\sim$100\,Myr, clusters lose mass via two-body relaxation and shocks by giant molecular clouds, processes that preferentially affect low-mass clusters and cause a turnover in the cluster mass function to appear on 1--10\,Gyr timescales;
(e)~after dispersal, some clusters remain coherent and detectable in chemical or action space for multiple Galactic orbits.''

\subsection{Young massive clusters}
\label{sec:young-massive-clusters}

Young massive clusters are dense aggregates of young stars, loosely defined as younger than about 100~Myr and more massive than about $10^4M_\Sun$, and which may form the {\it `fundamental building blocks of galaxies'}
\citep{2010ARA&A..48..431P}. 
While only a few exist in our Galaxy and the Local Group, they are abundant in starburst and interacting galaxies. They typically host a wide range of exotic stars and unusual binaries, and the nearest are therefore of particular interest for studying the stellar mass function, stellar evolution, and stellar dynamics. 

The archetypal young compact cluster Westerlund~1 is thought to be the Galaxy's most massive young star cluster. Discovered in 1961 by Bengt
\citet{1961PASP...73...51W},
it lies at a distance of 3--4~kpc, comprises some 200 members, has a total mass (including gas and dust) of $\sim\!\!60\,000M_\Sun$, a radius of $\sim$1~pc, and an age of $\sim$5~Myr
\citep{2005A&A...434..949C}.
It contains many rare, evolved, high-mass stars, including
four red supergiants, 
six yellow hypergiants (with initial masses $20-60M_\Sun$), 
one of the largest known stars (W237, with $R\sim1200R_\Sun$), 
24 Wolf--Rayet stars (Section~\ref{sec:wolf-rayet}), 
numerous OB supergiants,
an unusual supergiant which may be the remnant of a recent stellar merger 
\citep{2005A&A...434..949C},
and the anomalous X-ray pulsar CXO J164710.20--455217, a slow rotating neutron star inferred to have formed from a high-mass ($M\gtrsim40M_\Sun$) progenitor 
\citep{2006ApJ...636L..41M}.
While hosting some of the most massive stars in the Galaxy, its great distance means that its detailed properties have remained poorly known. Studies are hampered by high interstellar absorption: while its brightest stars have $V\sim14$~mag, none are brighter than $B\sim19$~mag. 
None of its stars were observable by Hipparcos (with its limit of $V=12-13$~mag), but various recent studies have converged on a distance of $\sim$4~kpc
\citep[e.g.][]{2006A&A...446..279C,
2007A&A...468..993K,
2008A&A...478..137B}.
This is supported by distances of its known double-lined spectroscopic binaries 
\citep{2021ApJ...912...16B,
2022MNRAS.517.3749R}.

Gaia has provided the first parallax distances. Most work has found a cluster distance close to the 4~kpc of previous studies, variously reported as 
$d=3.87^{+0.95}_{-0.64}$~kpc from DR2
\citep{2019MNRAS.486L..10D},  
and with 
$d=4.12^{+0.66}_{-0.33}$~kpc
\citep{2021ApJ...912...16B},
$d=4.23^{+0.23}_{-0.21}$~kpc
\citep{2022A&A...664A.146N}
and
$d=4.06^{+0.36}_{-0.34}$~kpc
\citep{2022MNRAS.516.1289N}
all estimated from Gaia EDR3.
But significantly smaller distances were also found from DR2,
$d=2.6^{+0.6}_{-0.4}$~kpc
\citep{2020MNRAS.492.2497A}, 	
and from EDR3, 
$d=2.8^{+0.7}_{-0.6}$~kpc
\citep{2021RNAAS...5...14A}.
The closer distance would have significant implications in terms of lower stellar luminosities, lower turnoff mass, lower total mass, and increase in age. And here the story becomes more involved. If Westerlund~1 formed through a single burst of star formation, its age is well constrained to around 5~Myr from the presence of both red supergiants (which only form at $\sim$4~Myr), and Wolf--Rayet stars (which fall off sharply after $\sim$5~Myr), as well as by at least two of the eclipsing binaries. Yet ages of around 10~Myr are estimated for both the cool supergiant W75 
\citep{2021ApJ...912...16B},
and the cluster's red supergiants 
\citep{2022MNRAS.516.1289N}.
One interpretation is that star formation actually proceeded over 5--12~Myr 
\citep{2021ApJ...912...16B}.  
This could be consistent with recent rotating and non-rotating (Geneva) models by
\citet{2022MNRAS.511.2814Y},
who tailored their developments to the Gaia observations of Westerlund~1, and who could broadly reproduce the observed population of red supergiants, yellow hypergiants, and Wolf--Rayet stars. 
The improved accuracies of Gaia DR4 and DR5 will provide further insights.

\subsection{Strings, snakes, and pearls}
\label{sec:strings-snakes-pearls}

The gas in giant molecular clouds, such as Orion, is often found in the form of extended filaments
\citep[e.g.][]{1987ApJ...312L..45B,	
2018A&A...610A..77H,			
2018AJ....156...84K,				
2021ApJ...919...35Z}.			
Perhaps mirroring this sort of star-forming environment, one of Gaia's new open cluster discoveries, Meingast--1
\citep{2019A&A...622L..13M},  
aka the Pisces--Eridanus stream
\citep{2020A&A...638A...9R,	
2020A&A...639A..64R},		
is a massive elongated co-moving population, at a distance of only 100\,pc, which extends $120^\circ$ on the sky, viz.\ several hundred~pc in length. Identified in velocity space, it is hidden amongst hundreds of thousands of unrelated field stars. 
It appears to be a dissolving open cluster with an age of around 120~Myr, similar to the Pleiades
\citep{2019AJ....158...77C,	
2020BAAA...61R..81A,		
2020MNRAS.496.2422H}.		

Even more curious is the Gaia DR2-based discovery of 1901 distinct groups within 1\,kpc, together comprising 288\,370 stars
\citep{2019AJ....158..122K}.	
Many appear to be filamentary structures that they refer to as `strings', coherent both spatially and kinematically, oriented parallel to the Galactic plane, and with some spanning hundreds of pc in length. One of the largest, $\alpha$~Per, extends over $120^\circ$.
Each group can be characterised with a single isochrone, spanning ages of $10^7-10^9$~yr. Their velocity dispersion increases with age, and suggests a timescale for dynamical heating of $\sim$300\,Myr. This timescale is also consistent with the age at which the string population begins to decline, while the population in more compact groups continues to rise. This all suggests that various dynamical processes are disrupting the more weakly bound primordial string populations, leaving only individual clusters to be identified at later times.
Extended this search to a distance of 3\,kpc yields a catalogue (referred to as Theia; \citep{2022AJ....164..137K}) of 8292 separate low-density filamentary or streams, comprised of hundreds to thousands of stars and spanning hundreds of~pc
\citep{2020AJ....160..279K}.	

Other similar young and highly elongated stellar structures in the solar neighbourhood have been discovered in the Gaia data (and sometimes referred to as `snakes'), notably in the area of the Vela~OB2 association
\citep{2018MNRAS.481L..11B,	%
2020ApJ...904..196T,		%
2022MNRAS.513..503W},		%
in the  Orion star-forming region
\citep{2019MNRAS.489.4418J},
and elsewhere
\citep{2020MNRAS.491.2205B,	
2020ApJ...904..196T}.		
Somewhat distinct from these filaments are the stellar `pearls': distinct clusters that follow similar Galactic orbits, and identified as overdensities in action-angle space 
\citep{2022ApJ...928...70C}. 
While these are not, it seems, coeval systems originating from the same molecular gas cloud, their existence points to recent star formation which is strongly clustered along an orbit. 

In summary, Gaia is revealing considerable and largely unexpected complexity in the various physical and kinematical structures in these young star-forming regions, and an unambiguous picture of how they arise and evolve remains unclear
\citep{2022MNRAS.511.2829M, 
2022AJ....163..275A,			
2022ApJ...936..160Z}.	 	
I will finish with a quote from 
\citet{2023ASPC..534...43Z}, 	
to which I refer for a more substantial review:
{\it `This new 3d view of our solar neighbourhood in the age of Gaia shows that star-forming regions once thought to be isolated are often connected on kpc scales, causing us to reconsider models for the arrangement of gas and young stars in galaxies.'}	

\subsection{Molecular clouds}
\label{sec:molecular-clouds}

Closely related to the topic of stellar associations and the birthplaces of young stars (and of interstellar extinction, Section~\ref{sec:extinction}) is that of molecular clouds, the regions where star formation is actively occurring. I will give a brief background, explain why their 3d structure is relevant, and illustrate how Gaia is contributing.

Molecular clouds are regions of the interstellar medium dense enough (10--30\,cm$^{-3}$) to form molecular gas. This results in `dark nebulae' that obscure light from background stars, and in whose densest regions star formation can occur. {\it Giant\/} molecular clouds (GMC) extend to sizes of 10--200\,pc, and span masses $10^4-10^7M_\Sun$.
Maps of the Galaxy in CO, a constituent more easily detectable than the more dominant H$_2$, show that such molecular gas is largely confined to the mid-plane of the Galaxy, with a scale height 50--60\,pc. Since ultraviolet photons from associated star formation transform molecular material back to the atomic state, their close association with spiral arms in turn suggests that molecular clouds form and dissociate on timescales $\lesssim$10~Myr.
Parallaxes for relatively nearby stars make it possible to derive cloud distances and even their 3d structures, important in understanding how star formation is affected by properties such as turbulence and magnetic fields 
\citep[e.g.][]{2001ApJ...562..852H,	
2012ARA&A..50..531K,			
2014ApJ...782..114E,			
2015ARA&A..53..583H}.			

The principles underlying Gaia's contributions can be seen in the GMC archetype, Orion~A, the most active local star-forming region. From it, many key observables, including star-formation rates and history, multiplicity, initial mass function, and protoplanetary disk populations, have been derived. And estimates of cloud mass, physical size, and star formation, all depend critically on accurate distance measurements.
The Orion Nebula Cluster, the richest cluster at the northern end of Orion~A, lies at $\sim$400~pc. Pre-Gaia observations, while suggesting a distance gradient across the cloud 
\citep{2014ApJ...786...29S,	
2017ApJ...834..142K},		
could say little about its shape or 3d structure.
\citet{2019A&A...627A..57J} 
used Gaia DR2 to identify three populations corresponding to three separated episodes of star formation, with ages 4.5, 2.1, and 1.4\,Myr, interpreted as the result of gas-inflow driven star formation, modulated by the repeated ejection of ionizing stars 
\citep{2018A&A...612A..74K}.	
And
\citet{2025MNRAS.541.1753S}	
used these results, the Gaia inferred properties of the Hyades and Pleiades clusters, and N-body simulations, to demonstrate that the Pleiades may have been similar to the Orion Nebula Cluster about 100\,Myr ago, and will be similar to the Hyades around 700\,Myr in the future.

The Gaia DR2 parallaxes for a number of young stellar objects (YSO, Section~\ref{sec:yso}), with ages $\lesssim3$\,Myr, can now be used as tracers of the cloud's 3d shape
\citep{2018A&A...619A.106G,		
2018AJ....156...84K}.	
The former, for example, used a sample of 682 YSO  to trace distances of young objects within the Orion~A cloud, concluding that Orion~A is not the straight filamentary cloud seen in projection, but instead a cometary-like cloud with aspect ratio 30:1, and two distinct components: a denser and enhanced star-forming `head', and a lower density 75~pc long `tail'. The true extent, 90~pc rather than the projected 40~pc, makes it the largest molecular cloud in the local neighbourhood. 

\citet{2019ApJ...879..125Z}	
also used DR2 parallaxes to derive distances and extinctions to tens of thousands of stars towards several dozen molecular clouds within 2~kpc.  They gave average distances (accurate to 5--6\%) to most of the major named clouds in the major CO~survey \citep{2001ApJ...547..792D}, including Perseus, Orion~A, Taurus, Ophiuchus, California, and Cepheus. An 
\href{https://faun.rc.fas.harvard.edu/czucker/Paper_Figures/stellar_post_comp.html}{interactive version} of their Figure~1 shows their distance--extinction relations towards the Perseus cloud as characterised before and after Gaia, with the most probable distance and extinction to each star indicated with a red cross.
\citet{2019A&A...624A...6Y}	
similarly used DR2 parallaxes and $G$-band extinctions to derive distances to 52 high Galactic latitude ($\vert b\vert\ge10^\circ$) molecular clouds, 13 for the first time, and again with errors of typically 5\%.

The two-volume {\it Handbook of Star Forming Regions\/}\ \citep{2008hsf1.book.....R,2008hsf2.book.....R} contains 60~star forming regions within 2~kpc. Although well studied, their distances have remained poorly constrained pre-Gaia. Amongst them, several have pre-Gaia estimates that varied by more than a factor two (e.g.\ Circinus, Coalsack, NGC~2362, IC~5146).
\citet{2020A&A...633A..51Z}	
derived DR2-based star distances and extinctions in sightlines towards these clouds. They used an implementation of the method pioneered a century ago by
\citet{1923AN....219..109W},	
viz.\ by fitting a line-of-sight dust model to identify the distance at which a discontinuity appears in the stellar reddening. Comparison with maser (parallax) distances results in consistent estimates, with 10\% scatter, across their entire distance range of 150--2500~pc.

Several studies have used the Gaia distances and photometry in various combinations to probe the 3d structure of several of these molecular clouds, resulting in  
resolved dust maps of clouds out to 400~pc with a resolution of 1~pc
\citep{2020A&A...639A.138L},
the discovery of skeleton-like `spines' of each cloud
\citep{2021ApJ...919...35Z},	
and various other works employing extinctions, dust-mapping, and Gaia DR2 distances to recover the molecular cloud distances and structures 
\citep{2019A&A...624A...6Y,	
2020A&A...641A..78B,		
2020A&A...641A..79H,		
2021A&A...655A..68H}.		
One approach is to create dendrograms of the hierarchical structures applied to the surrounding dust regions, providing estimates of the volume, mass, and density for each cloud and its associated substructures
\citep{2023MNRAS.519..228D,	
2024ApJ...961..153C}. 	

Gaia is also contributing to an understanding of the internal dynamics of molecular clouds.
\citet{2022MNRAS.513..638Z} 
studied 150 young stellar objects with $d\lesssim3$\,kpc, finding their associated clouds to be elongated, and oriented parallel to the disk mid-plane. As probed by these young objects, the turbulence is isotropic, and the 2d velocity dispersion is related to size by $\sigma_v\propto r^{0.67}$. The turbulent energy dissipation rate decreases with Galactocentric radius which, they suggest, can be explained if the turbulence is driven by cloud collisions. This is a topic closely related to the problem of the origin of wide binaries
\citep{2023ApJ...949L..28X}, 	
further detailed in Section~\ref{sec:wide-binaries-origin}.

\subsection{Young stellar objects}
\label{sec:yso}

Young stellar objects (YSOs) represent the earliest stages of star formation. The term embraces the {\it protostar\/} phase, which starts with the gravitational collapse from the parent molecular cloud (lasting of order 0.5\,Myr while the object is still accumulating mass), and the subsequent {\it pre-main-sequence\/} phase (PMS), which starts with the exhaustion of the infalling gas.
The PMS phase continues until contraction, and the associated temperature increase, initiates H~fusion. Residual gas and dust is blown away, the object becomes visible optically, and the star settles onto the zero-age main-sequence (ZAMS).
As the disk material is depleted, its infrared emission decreases, such that YSOs can be classified in evolutionary stages based on their infrared spectral index 
\citep[classes~I, II and III; due to][]{1987IAUS..115....1L}. 	
Pre-main-sequence stars are classified by mass as T~Tauri stars 
\citep[$M\lesssim2M_\Sun$, with ages $\lesssim10$\,Myr;][]{1945ApJ...102..168J}, 
or Herbig Ae/Be stars 
\citep[$2-8M_\Sun$;][]{1960ApJS....4..337H, 1994A&AS..104..315T}, 
with more massive stars contracting too rapidly to be visible as PMS objects.
YSOs are associated with many other early evolutionary phenomena and phases 
\citep[e.g.][]{1985ARA&A..23..267L}: 
circumstellar and protoplanetary (proplyd) disks, jets and bipolar outflows, masers, Herbig--Haro objects (with associated nebulosity), and dippers, as well as the episodically accreting FU~Ori and EX~Lup (aka EXor) variables.

The first large-scale application of Gaia to the identification of YSOs came with a study using DR2 
\citep{2019MNRAS.487.2522M}.	
They constructed a cross-matched catalogue of 103~million objects from Gaia and WISE (in the infrared), and used machine learning to assign each object to four classes: YSOs, main-sequence, evolved stars, and extragalactic objects. This yielded 1.1~million YSO candidates at 90\% probability. 
With Gaia DR3, the Konkoly Catalogue of 12\,000 optically selected YSOs within 2\,kpc, KYSO
\citep{2023A&A...674A..21M}, 	
based the {\it Handbook of Star Forming Regions\/}
\citep{2008hsf1.book.....R,2008hsf2.book.....R},
was created as a training set for the variability processing, whereby the 12.4~million DR3 variables were assigned to 25 classes, including the class of YSOs
\citep[][their \S4.24]{2023A&A...674A..14R}. 
This yielded 79\,375 YSOs of various classified sub-types, including 
dipper stars (DIP),
eruptives such as FU~Ori-type variables (FUOR),
pulsating PMS stars (PULS\_PMS), 
Herbig~Ae/Be types (HAEBE), 
and T~Tauri star (TTS), among which are classical (CTTS), weak-lined (WTTS), and late~G to early~K type pre-main-sequence (GTTS) stars.
Some 40\,000 of these had not been previously catalogued as YSOs.
Given the relatively low completeness and high contamination expected for optical observations alone, detailed validation by comparison with numerous literature catalogues has been carried out
\citep{2023A&A...674A..21M}.	
They found that most candidates are indeed found along lines-of-sight to known star-forming regions and the Galactic mid-plane, with minimal contamination and high completeness. 

\begin{figure}[t]
\centering
\includegraphics[width=1.00\linewidth]{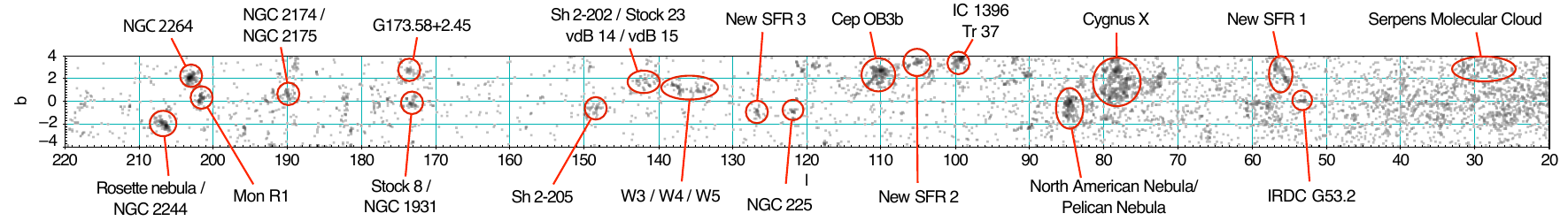}
\vspace{-15pt}
\caption{The 6504 (probable) Young Stellar Objects of Class~II (i.e.\ showing both optical and infrared emission) identified from Gaia EDR3, and shown in Galactic coordinates. Several well-known star forming regions and young clusters are identified \citep[][Figure~21]{2023MNRAS.521..354W}.
}\label{fig:yso}
\end{figure}

An ADS search in early 2025 returned some 150 refereed papers on Gaia--YSO studies, and the following can only give a flavour of the many ongoing investigations.
The largest catalogue of mid-infrared (3--9\,$\mu$m) YSOs in the Galactic plane is from the Spitzer--IRAC wide-area mapping. In their study of 120\,000 of these YSO candidates, 
\citet{2021ApJS..254...33K}		
used DR2 distances to associate these YSOs with the Local Arm, the Sagittarius--Carina Arm, and the Scutum--Centaurus Arm. 
In a classification exercise using a `naive Bayes classifier', and data from WISE, UKIDSS, 2MASS and IGAPS, along with photometric variability from Gaia EDR3,  
\citet{2023MNRAS.521..354W}		
identified 6504 candidate Class~II YSOs, with good sensitivity to different evolutionary stages (Figure~\ref{fig:yso}). 

Between the end of the AGB phase and the onset of the planetary nebula phase, the AGB dust shell becomes optically thin, resulting in infrared colours similar to those of YSOs because of their similar cold, dense circumstellar dust envelopes. With YSOs and AGB stars commonly identified from their infrared two-colour diagrams 
\citep[e.g.][]{2014ApJ...791..131K,
2021ApJS..256...43S}, 
misclassification can arise in the significant overlap regions 
\citep[e.g.][]{2021ApJ...916L..20L}. 
\citet{2024JKAS...57..123S}		
showed that distance and extinction data from Gaia DR3 is highly effective in distinguishing these very different stellar classes.
Gaia-based studies of YSOs in specific star-forming regions include
the $\rho$~Oph region \citep{2021A&A...652A...2G},	
Orion \citep{2023JApA...44...92N},	
and in the Taurus--Auriga molecular complex \citep{2024A&A...681A..72G}.	
Gaia parallaxes and proper motions are also being used to study young stars in the regions of Herbig Haro objects and flows, for example 
in Canis Major \citep{2019A&A...630A..90P},	
around the BBWo~192E nebula \citep{2020MNRAS.498.5109M},	
in the Dobashi 5006 Dark Cloud \citep{2023Ap.....66...52M},	
and in the Taurus B18 cloud \citep{2023ApJ...943..182D}.		

Another source of Gaia YSO discoveries is the Science Alerts pipeline (Section~\ref{sec:science-alerts}). As of mid-2025, this database included 27\,000 objects, of which nearly 300 had been classified as YSOs (with many others likely amongst the 20\,000 classed as `unknown'). These include the FU~Ori variables, PMS stars displaying abrupt changes in magnitude and spectral type, and attributed to large-scale accretion episodes with mean time between events of 10\,000 years or more. Pre-Gaia, only around 10~were known, with the Gaia Alerts discovering a number of others
\citep[e.g.][]{2018ApJ...869..146H,	
2020ApJ...899..130S,	
2019AJ....158..240H}.	
The related class of EXors, named after the prototype EX~Lupi, are indistinguishable from T~Tauri stars apart from repeated luminosity increases, a year or so in duration, attributed to episodic accretion. Again, a number of new Science Alerts discoveries are being characterised by Gaia astrometry
\citep[e.g.][]{2022ApJ...941..165P,		
2022ApJ...926...68G,		
2022ApJ...927..125C,		
2023MNRAS.524.3344N,		
2023MNRAS.524.5548S,		
2024ApJ...967...41G}.		

\subsection{The Local Bubble}
\label{sec:local-bubble}

\paragraph{Context}

Our Sun is thought to lie just inside the boundaries of a warm, low-density, partially-ionised interstellar cloud ($T\!\sim\!7000$\,K, $n\!\sim\!0.1$\,cm$^{-3}$).  The solar heliosphere results from the expanding high-velocity solar wind within this cloud, and extends from the Sun to about 100--150\,au in the direction of its motion through space. The termination shock, marking the transition to the wider interstellar medium, was crossed by the spacecraft Voyager~1 in December 2004. 
This and other nearby interstellar clouds are themselves located in a low-density region of the interstellar medium, referred to as the local cavity.  This is partially filled with hot ($\sim$10$^6$\,K), low-density coronal gas, 
about 100~parsec in size, and referred to as the Local Bubble 
\citep{1987ARA&A..25..303C}.	
Detectable in soft X-rays, the Local Bubble was originally attributed to stellar winds from OB~stars in the nearby Sco--Cen association, or to one or more nearby supernova explosions within the last 10~Myr. 

The size and shape of the tenuous Local Bubble, and its precise origin, have been much debated, and many different observations have contributed to our knowledge of it today. Its dimensions can be determined directly by measuring the extent of the hot emitting bubble gas, or more indirectly by tracing the absence of absorption by neutral interstellar gas. 
But the its detailed morphology rests on our knowledge of the distances towards the respective line-of-sight stellar targets towards which the level of absorption is measured. Before Hipparcos, reasonably accurate distances from ground-based parallax measurements extended out to only about 20~pc.
Hipparcos distances out to 200--300\,pc allowed the Local Bubble absorption characteristics to be defined with much greater certainty
\citep[e.g.][]{1998A&A...333..101W}.
Mapping over 1000 lines-of-sight revealed `interstellar tunnels' of different widths which connect the Local Bubble to surrounding cavities
\citep{1999A&A...346..785S}
supported the model in which expanding supernova-driven bubbles interact and merge to form large-scale interstellar cavities
\citep{1974ApJ...189L.105C}.

Various attempts have been made to trace the origin of the supernovae that might have been responsible for the Local Bubble, with attention focusing on nearby OB~associations as sites of recent star formation, of which the nearest, the Scorpius--Centaurus OB~association, is currently about 130\,pc from the Sun. 
Hipparcos data were used to extrapolate backwards in time both the Sun's position, and those of the various association members, to show that the Sco--Cen association, and especially the Lower Centaurus Crux subgroup, was much closer to the Sun 5\,Myr ago, making it a likely source of the few supernovae needed to produce the Local Bubble
\citep{2001ApJ...560L..83M}.
Other Hipparcos-based studies also concluded that the Local Bubble could have been excavated by some 20 supernova explosions 10--20\,Myr ago, some of them as close as 40\,pc
\citep{2002PhRvL..88h1101B,
2002A&A...390..299B,
2006MNRAS.373..993F}.
Nearby supernova explosions are also of great interest in understanding their consequences for the geological record on Earth, their possible effects on life, and the existential threats to life on Earth in the unlikely event of a nearby supernova explosion occurring today
\citep{2004PhRvL..93q1103K}.

This picture of the Local Bubble has remained largely unchanged over the past two decades. It is clear that the Sun lies within a cavity of low-density, high-temperature plasma surrounded by a shell of cold, neutral gas and dust. But the precise shape and extent of this shell, the cause and timescale of its formation, and its relationship to nearby star formation remained uncertain, because of inadequate knowledge of the properties of the local interstellar medium.

\paragraph{Gaia contributions}

The Gaia EDR3 distances and space motions are providing a revised picture of the 3d positions, shapes, and motions of dense gas and young stars within 200\,pc of the Sun. This has enabled some striking conclusions to be made about the Local Bubble.
\citet{2022Natur.601..334Z}
selected stars from young clusters out to 300--400\,pc with age $\lesssim20$\,Myr, including all young clusters associated with star-forming regions currently near the surface of the Local Bubble (including the Lupus, Ophiuchus, Chamaeleon, Corona Australis, and Taurus Molecular Clouds), along with older stellar populations in the Sco--Cen association. Their sample totalled more than 1200 stars, with an average of nearly 40~stars per cluster.
They then performed a dynamical `traceback' of each cluster's motion through the Galaxy, along with that of the Sun, using a numerical model of the Milky Way's gravitational potential, from 20~Myr ago to the present day. They then modelled the expansion of the Local Bubble, using hydrodynamic simulations to study the dynamical evolution of super-bubbles driven by clustered supernovae in a uniform interstellar medium.  
Based on the momentum injection required by supernovae to sweep up the total shell mass ($1.4\times10^6M_\odot$), given its present expansion velocity (around 6--7~km/s), they estimate that 10--20 supernovae were required to form the Local Bubble. This agrees with earlier results, from their stellar membership and an adopted Initial Mass Function, that the Upper Centaurus Lupus and Lower Centaurus Crux clusters have produced some 14--20 supernovae over their lifetimes.

This new Gaia-era 3d model of the Local Bubble’s is seen more forcefully in their 
\href{https://faun.rc.fas.harvard.edu/czucker/Paper_Figures/Interactive_Figure1.html}{interactive model},
which illustrates the inner surface of neutral gas and dust, and the 3d shapes and positions of local molecular clouds delineated at a resolution of about 1~pc (the Sun shown as a yellow cross at the centre).  
%
%
Their first major finding was that every well-known molecular cloud within about 200~pc of the Sun lies on the {\it surface\/} of the Local Bubble. These `surface' clouds include every star-forming region in the Sco--Cen association (Ophiuchus, Lupus, Pipe, Chamaeleon, and Musca), as well as Corona Australis and the Taurus Molecular Cloud, the latter lying 300~pc away from Sco--Cen on the opposite side of the bubble. 
The one exception, the Perseus Molecular Cloud, at a distance of 300~pc, has likely been displaced by the recently discovered Perseus--Taurus Superbubble, also facilitated by Gaia
\citep{2021ApJ...919L...5B}, 
containing Taurus on its near side and Perseus on its far side.

Going further, they found that the young stars show an outward expansion mainly perpendicular to the bubble's surface. Tracebacks of these young star motions support a scenario where the origin of the Local Bubble was a burst of star formation, followed by death of the most massive stars as supernovae, taking place near the bubble's centre beginning 14\,Myr ago. 
The expansion of the Local Bubble created by the supernovae swept up the ambient interstellar medium into an extended shell that has now fragmented and collapsed into the most prominent nearby molecular clouds. This, in turn, provides robust observational support for the theory of supernova-driven star formation.

\subsection{Extinction}
\label{sec:extinction}

A recurrent theme in the Gaia literature is interstellar extinction. This is important in the determination of stellar parameters (e.g.\ [M/H], \teff, and $\log g$) through spectral modelling (Section~\ref{sec:classification-stellar-properties}), in the interpretation of some key evolutionary states, and in describing the distribution of gas and dust in the solar neighbourhood. 

\paragraph{Context}

Light is absorbed and scattered by gas and dust along the sight line from star to observer, and most stellar spectra are affected by at least some interstellar dust.  The details depend on the properties of dust grains along a line-of-sight, and measurements also convey information about their composition and size distribution. 

Interstellar extinction is rather a smooth function of wavelength, albeit with superimposed absorption features due to particular chemical species. These include the ultraviolet (217\,nm) bump, diffuse interstellar bands (with two seen in the Gaia RVS spectra; see Section~\ref{sec:dib}), and others beyond Gaia's response in the infrared.
Extinction in a given direction can be inferred by comparing an observed spectrum with its closest atmosphere model. In the solar neighbourhood, extinction in the $V$-band averages some  0.7--1.0~mag kpc$^{-1}$, but it is much higher in specific regions, notably in the Galactic plane, and especially towards the Galactic centre.
Since blue light is more strongly attenuated than red, extinction also causes objects to appear redder as well as dimmer. This `reddening' is often simply characterised by an object's {\it colour excess\/} in some specified photometric system, e.g.\ $E_{B-V}=(B-V)_{\rm obs}-(B-V)_0$. 
The parameter $R\equiv A_V/E(B-V)$ characterises the ratio of total to selective extinction. Ranging between 2.2--5.8 for
sight lines where ultraviolet extinction has been measured, a mean relationship $A_V=3.1\,E(B-V)$ is frequently used
\citep[e.g.][]{1989ApJ...345..245C,		
1999PASP..111...63F,				%
2007ApJ...663..320F}. 				%

There is a huge literature on extinction, where historical advances have also benefitted from ultraviolet and infrared observations
\citep{1930PASP...42..214T, 	
1949PGro...53....1V,			
1950AnAp...13..367S, 		
1952ApJ...116..575M,		
1952ApJ...115..103C, 		
1969ApJ...157..611G,		
1980AJ.....85...17M, 			
1980A&AS...42..251N, 		
1992A&A...258..104A}.		
For extragalactic sources (and in interpreting the CMB), a 2d map of {\it integrated\/} dust extinction and reddening is usually sufficient.  Here, the maps of 
\citet{1998ApJ...500..525S}, 	
based on IRAS 100\micron\ and COBE--DIRBE 100\micron\ and 240\micron\ data, are still some of the most widely used.

\paragraph{Gaia contribution}
Around the time of the Gaia launch, 
\citet{2014A&A...561A..91L} noted that {\it `3d maps of the Galactic interstellar matter are a potential tool of wide use, but accurate and detailed maps are still lacking'}. Gaia is changing this.  Accurate parallaxes provide the means of quantifying the dependence of extinction on distance for millions of sight lines in the Galaxy. And in addition, reddening and extinction {\it for each star\/} can be estimated from Gaia's low-resolution BP/RP spectra, and from the RVS spectra. In the following, the referenced papers provide substantially more details, and generally also provide links to their derived extinction maps.

Using DR2 parallaxes and Gaia's integrated (BP/RP) photometry, supplemented by other photometry as noted, the following products have been described:
$G$-band extinctions ($A_G$), $E(G_{\rm BP}-G_{\rm RP})$, and reddening maps for 88~million sources 
\citep{2018A&A...616A...8A,
2019A&A...631A..32L};		
$E(G-K_{\rm S})$, $E(G_{\rm BP}-G_{\rm RP})$, and $E(H-K_{\rm S})$ for 56~million stars in the Galactic plane, with an angular resolution of 6~arcmin, using infrared photometry from 2MASS and WISE. These reveal, for example, the disk warp, and complex structures associated with the Sagittarius, Local, and Perseus arms
\citep{2019MNRAS.483.4277C};
the dust distribution in a $6\times6\times0.8$~kpc$^3$ volume centred on the Sun using 2MASS photometry
\citep{2019A&A...625A.135L};	
reddening maps for 799~million stars with $\delta>-30^\circ$ and out to a few~kpc, using Pan-STARRS1 and 2MASS photometry
\citep{2019ApJ...887...93G}		
extinctions for 137~million stars with $G<18$, using Pan-STARRS1, 2MASS, and AllWISE photometry
\citep{2019A&A...628A..94A}	
reddening maps for 2~million stars using LAMOST photometry
\citep{2022ApJS..260...17S}.	
With EDR3, several improved and updated maps, based on the same general methods, were constructed and made available
\citep{2022A&A...658A..91A,	
2022A&A...661A.147L,		
2022A&A...664A.174V}.		

Substantial improvements came with the availability of Gaia DR3, in part from the improved parallaxes, but also with the availability of the mean BP/RP (aka XP) spectra for 220~million stars (Section~\ref{sec:classification-stellar-properties}).
Accordingly, parallaxes, atmospheric parameters, 2MASS and Gaia~EDR3 photometry were used to compute ages, masses, and reddenings for 5~million stars with RVS spectra
\citep{2023A&A...669A.104K}. 	
Another model using Gaia, 2MASS and ALLWISE was given by \citep{2024MNRAS.535.2149O}.
Revised extinctions were also derived through improved reductions of the XP spectra (Section~\ref{sec:photometry}), resulting in alternative catalogues of 
220~million stars
\citep{2023MNRAS.524.1855Z},	
175~million stars
\citep{2023ApJS..267....8A},		
and 48~million stars
\citep{2025ApJ...980...90H}.	
Using the XP~spectra on a slightly different metallicity scale (based on empirically calibrated theoretical models), reddening and metallicity was estimated for 80~million main-sequence stars, modelling foreground extinction out to approximately 3~kpc
\citep{2024ApJS..272...20A}			
Gaia, Pan-STARRS1, SkyMapper, 2MASS, and WISE photometry for 100~million stars was used to derive 3d extinction maps in $V$ and $G$ within 2~kpc (resolution 3.6--11.6~pc transverse and 50~pc radial; see their Fig.~10), a 3d differential extinction map, and a 2d map of total Galactic extinction for Galactic latitudes $\vert b\vert>13^\circ$, and with a precision in $A_V$ of 0.06~mag
\citep{2023AstL...49..673G}.
An analytical model, treating the 3d dust distribution as a superposition of three overlapping layers, is given by 
\citep{2023arXiv230203871G}.	
The XP spectra also allow determination of extinction at their corresponding spectral resolution, $R\sim20-100$.
These were used to construct a `universal extinction' versus wavelength curve, derived from the Gaia data without reference to any previous model, and with each wavelength interval modelled separately
\citep{2023MNRAS.524.1855Z}.	
The resulting smooth extinction curve (their Fig.~15) agrees reasonably well with the $A_V=3.1\,E(B-V)$ model of
\citep{1989ApJ...345..245C}.	

\paragraph{Grain growth}

X.~\citet{2025Sci...387.1209Z}	
used the low-resolution BP/RP spectra to measure extinctions for 130~million stars, finding clear spatial structure on many length scales. In the Galactic plane, $|z|<400$\,pc, within 2.6\,kpc of the Sun (their Fig.~1), they found spatial correlations between $R(V)$ and several large-scale structures, including the Radcliffe Wave 
\citep[Section~\ref{sec:gould-radcliffe};][]{2020Natur.578..237A}, 
the `Split' 
\citep{2019A&A...625A.135L},
and the Carina--Sagittarius arm.

There is, they found, interesting physics encoded in these diagrams. Rather than increasing continuously with density (as a simple consequence of growing grain size), they found that $R(V)$ is above average in very diffuse regions, but then {\it decreases\/} as the dust density {\it increases}. But in very dense regions, such as the Orion Nebula and dense cores in Taurus, $R(V)$ increases sharply. 
Their explanation involves the two dominant mechanisms of dust grain growth in the interstellar medium: {\it accretion\/} of elements from the gas phase onto the grain surfaces, and {\it coagulation\/} of grains that collide and stick together, the latter occurring preferentially at higher densities. While both processes increase the average grain size, their effects on the grain-size {\it distribution}, and thus $R(V)$, are very different
\citep{2012MNRAS.422.1263H}. 

In their simplified picture, accretion deposits a layer of equal thickness onto each grain, such that the fractional increase in surface area is larger for smaller grains. The increasing contribution of the small grains causes a steepening of the extinction curve, and hence a decreasing $R(V)$.
Coagulation, in contrast, transforms pairs of small grains into larger grains, decreasing the relative abundance of small grains. The extinction curve becomes flatter, thus increasing $R(V)$.
Their results, they argue, are consistent with a picture in which accretion is the dominant mechanism of grain growth in the less dense regions of interstellar clouds, leading to the observed decrease in $R(V)$ with increasing dust density. In the densest regions, coagulation takes over, causing $R(V)$ to increase rapidly.

The work has already had some wider consequences. 
The observed extinction curves now require additional dependencies than $R(V)$ alone
\citep{2025ApJ...988....5G}.	
The findings offer the prospects of characterising magnetic field dependencies
\citep{2025ApJ...979...12H},	
and insights into the growth of polycyclic aromatic hydrocarbons as a consequence of gas-phase accretion
\citep[X.][]{2025ApJ...979L..17Z}.	

\begin{figure}[t]
\centering
\hspace{-8pt}
\raisebox{2pt}{\includegraphics[width=0.30\linewidth]{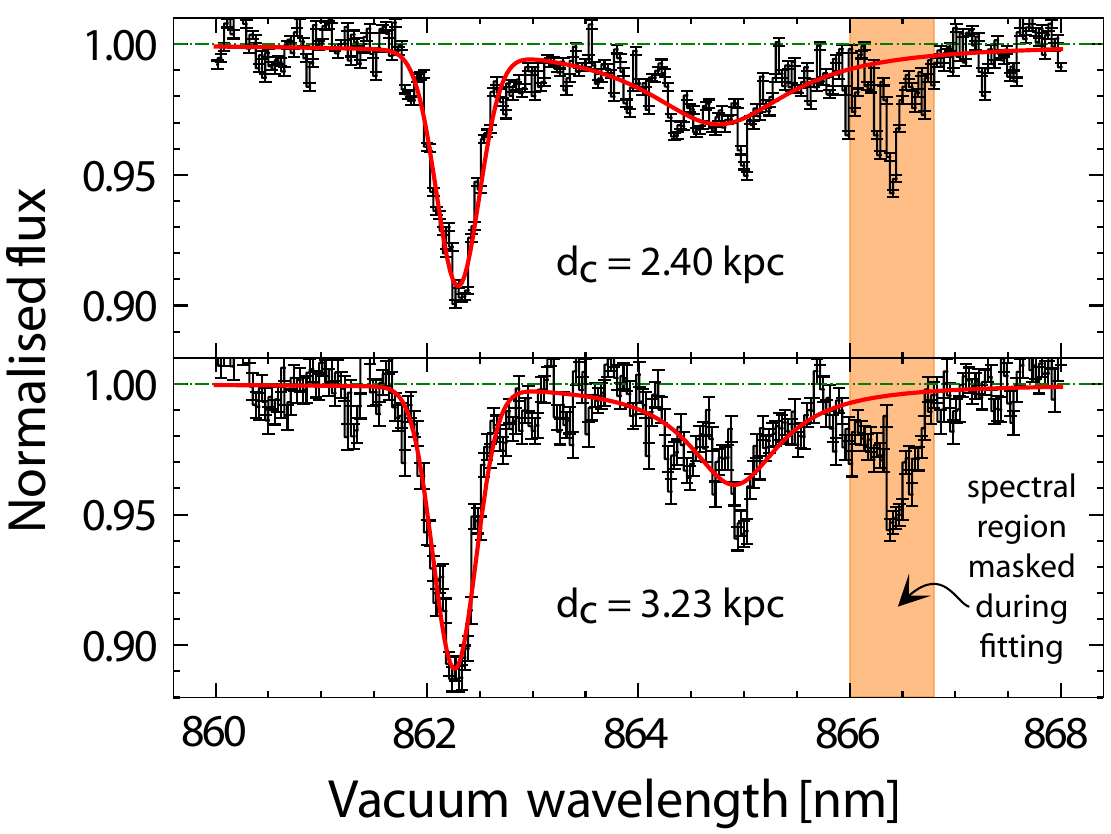}}
\hspace{10pt}
\includegraphics[width=0.66\linewidth]{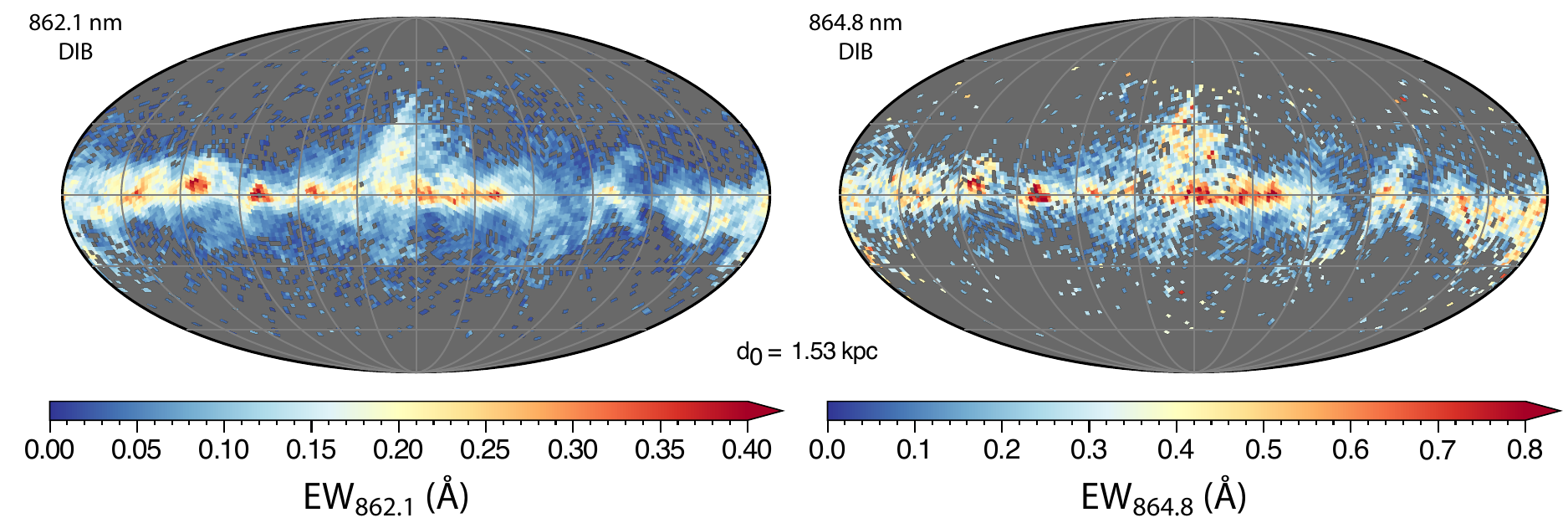}
\vspace{-5pt}
\caption{Left: fits to the 862.1\,nm and 864.8\,nm diffuse interstellar bands (red lines) in stacked Gaia RVS spectra for two distance volume elements in one Galactic direction.
Right pair: Galactic distributions of the two diffuse interstellar bands, at a distance $d=1.53$~kpc. Other distance distributions are given in the referenced paper \citep[from][Figures~7 and~20]{2023A&A...680A..38G}.}
\label{fig:dib}
\end{figure}

\subsection{Diffuse interstellar bands}
\label{sec:dib}

Diffuse interstellar bands (DIBs) are absorption features in stellar spectra which are superimposed on a smoother wavelength-dependent interstellar extinction (Section~\ref{sec:extinction}). They were first reported over a century ago, but their origin remains unclear. Several hundred bands have been discovered, at ultraviolet, visible and infrared wavelengths 
\citep[e.g.][]{2019ApJ...878..151F}.
And while there is a general consensus that they are probably due to large and complex molecules in the Galactic interstellar medium, 
only one carrier has been securely identified: ionised buckminsterfullerene (C60$^+$), with five absorption bands in the near-infrared \citep{2020JMoSp.36711243L}.
For Gaia, the choice of wavelength range for the radial velocity spectrometer was driven by a number of considerations (Section~\ref{sec:rvs-wavelength}). Amongst these, and within the adopted wavelength range (845--872\,nm), was a known medium-intensity DIB at 862\,nm, first identified in 1975. The correlation of its equivalent width with absorption was known to be rather tight, suggesting that it could be used to build a detailed reddening map, especially for high values of interstellar extinction 
\citep{1999BaltA...8...73M}.

Within the Gaia data processing, the 862\,nm DIB is characterised as part of the GSP--Spec module (Section~\ref{sec:classification-stellar-properties}), using a Gaussian profile fit for cool stars and a Gaussian process for hot stars. In addition to the equivalent widths and their uncertainties, the tabulations in DR3 list their characteristic central wavelength, line width, and various quality flags
\citep{2023A&A...674A..29R}. 
The first major study of this feature, from Gaia DR3, contained 476\,117 stars with valid DIB measures (by an order of magnitude, the largest homogeneous full-sky sample obtained to date), with a high-quality sample comprising 141\,103 stars \citep{2023A&A...674A..40G}.		
They showed that the 862\,nm line provides an excellent tracer of the spiral arm structure, with equivalent widths generally well-correlated with estimates of interstellar reddening (Section~\ref{sec:extinction}), with both showing larger values in the Galactic plane. They also drew attention to some notable differences, in particular that the scale height of the 862\,nm carrier is smaller than that of the dust, and that it is concentrated within the inner kpc from the Sun. Another striking difference between the 862\,nm DIB carrier and dust distributions is that the former features are present in the Local Bubble around the Sun, even though this region is known to contain almost no dust. 
Its rest-wavelength in the Galactic anticentre direction, $\lambda_{0}=862.086\pm 0.0019$\,nm, is the most precise determination to date. And based on this rest-wavelength, the Galactic rotation curve within 1--2\,kpc from the Sun shows a remarkably good agreement between the 862\,nm diffuse interstellar band velocities and the CO gas velocities, reinforcing the suggestion that the associated carrier(s) could be related to gaseous macromolecules.
 
Although the line at 862\,nm was the only confirmed DIB in Gaia's RVS spectral range at the time of its consideration, a weaker feaure at 864.8\,nm was also considered as a candidate
\citep{2008A&A...488..969M}.	
From the 8458 RVS spectra in DR3, its existence was confirmed, with both equivalent widths being correlated
\citep{2022A&A...666L..12Z}.	

As one of the `Focused Product Release' studies (Section~\ref{sec:data-releases}), a new pipeline applied to Gaia DR3 (DIB--SPEC) processed all 6.8~million RVS spectra, measuring both DIB signals, at 862.1\,nm and 864.8\,nm, out to 4\,kpc, stacked in `volume pixels' (or `voxels') to improve the signals
\citep{2023A&A...680A..38G}. 	
The resulting sky maps (Figure~\ref{fig:dib}) show a reasonable correlation between the dust reddening found from stellar absorption and the equivalent widths of both DIBs, allowed the detection of the 862.1\,nm feature inside the Local Bubble ($\lesssim200$\,pc), and traced the DIBs in the outer spiral arm, and beyond the Scutum--Centaurus spiral arm.
This significant advance in mapping these two specific DIBs with Gaia, and out to much larger distances, nevertheless leaves their origin unknown.

\subsection{The Gould Belt and the Radcliffe Wave}
\label{sec:gould-radcliffe}

\paragraph{Context}

It was pointed out by John
\citet{1847raom.book.....H}		
in the 1840s that the distribution of bright stars is asymmetric about the Galactic plane.
In the 1870s, Benjamin
\citet{1879RNAO....1.....G}
concluded that a prominent group of bright stars was aligned along a great circle at about $20^\circ$ to the Galactic plane. The (expanding) structure has since been known as Gould's Belt, and many later studies of bright stars, OB~associations and young clusters have confirmed its existence
\citep[e.g.][]{1974AJ.....79..456S,
1997FCPh...18....1P,
2014Ap.....57..583B}.
In a little more detail, the O--B5 stars, supergiants, associations, young clusters, molecular clouds and star-forming regions within 1\,kpc of the Sun populate two flat systems inclined by about~$20^\circ$. Those of the Galactic belt, aligned nearly along the Milky Way, are characterised by a fairly random distribution in space and age. The Gould Belt, more compressed vertically, hosts most of the O--B2 stars and youngest stellar groups near to the Sun, and has an age of around 30~Myr.  It has an ellipsoidal shape with a semi-major axis of about 500\,pc and a semi-minor axis of about 340\,pc, the Sun being located 150--250\,pc off centre. 

The formation of the Gould Belt system has long been debated. Suggested origins include
collisions of high-velocity clouds with the Galactic disk 
\citep{1992A&A...261...94C,
1994A&A...281...35C};
supernova explosions and/or star formation driven by stellar winds originating in the central Cas--Tau association
\citep{1982A&A...112..195O,
1994A&A...281...35C};
relation to the Lindblad Ring, an expanding H\,{\scriptsize II} feature 
\citep{1973A&A....24..309L};
a high-speed, oblique collision between a gas cloud of $10^6M_\odot$ and a dark matter clump of $10^7M_\odot$
\citep{2009MNRAS.398L..36B}; 
or simply as a random spatial distribution of a few prominent OB~associations (Orion, Perseus~OB2, and Scorpius--Centaurus--Lupus) mimicking an `inclined belt' in the sky
\citep{1988ApJ...333..826F,
1994A&A...286...60L,
2015A&A...584A..26B}.	
However, the detailed physical arrangement of these local associations and gas clouds has remained uncertain.

\paragraph{Gaia contribution}
The distances from Gaia offer a somewhat different picture, and support the suggestion that the Gould Belt results from a 2d projection effect rather than a physical ring
\citep{2015A&A...584A..26B}.	
Using Gaia~DR2, the {\it three-dimensional\/} distribution of young stars and star-forming regions within 500~pc of the Sun could be examined
\citep{2018A&A...620A.172Z}.	
They found that younger stars (mainly in Sco--Cen, Orion, Vela, and Taurus) are organised in dense, compact clumps, and are surrounded by older more diffuse regions (such as Cepheus, Cassiopeia, and Lacerta). Most importantly in this context, they found that the 3d density maps show no evidence for the existence of a ring-like Gould Belt structure.

The next step in unravelling the true nature of the distribution of young stars and star-forming regions within a few hundred parsec of the Sun makes use of {\it The Star Formation Handbook}, a compilation of around 60 of the most important low- and high-mass star-forming regions within 2\,kpc, distributed across the northern and southern skies 
\citep{2008hsf1.book.....R, 2008hsf2.book.....R}.
High-resolution multi-wavelength observations of these nearby regions has provided much of our knowledge of how molecular gas is transformed into stars. And although accurate distances are critical for understanding the star- and planet-formation processes, distances pre-Gaia were obtained inhomogeneously, and often inaccurately, on a cloud-by-cloud basis.

Today, using stars in front of and behind these complexes, and Gaia-based extinction maps, Gaia DR2 parallaxes yield the distances of many of the nearby molecular dust clouds with 5\% accuracy. Comparison with distances derived from maser parallax measurements suggests consistent distances with better than 10\% scatter for clouds spanning their entire distance range of 150\,pc to 2.5\,kpc
\citep{2019ApJ...879..125Z,2020A&A...633A..51Z}.
These distances were then used to establish their 3d structure
\citep{2020Natur.578..237A}. These authors found a narrow and coherent 2.7-kpc arrangement of dense gas in the solar neighbourhood, at an angle of $30^\circ$ to the Galactic $y$-axis, running along a significant length of the local (Orion) arm.  The structure comprises the majority of nearby star-forming regions, has an aspect ratio of about 1:20, and contains $3\times10^6M_\odot$ of gas. While containing many of the clouds associated with the Gould Belt (notably Orion, Perseus, Taurus, and Cepheus), it is also inconsistent with them being part of a physical `ring'. Named by its discoverers at the Radcliffe Institute (Cambridge, MA), this `Radcliffe Wave' is undulating, with its 3d shape well described by a damped sinusoidal wave on the plane of the Milky Way, with an average period of 2\,kpc, and a maximum amplitude of 160\,pc.

Having been discovered from the spatial distribution of molecular clouds, the Radcliffe Wave has been confirmed, both in vertical positions and in vertical velocities, by several subsequent studies, and in several different stellar populations, all making use of the Gaia data 
\citep{2021arXiv211104685D,	
2022A&A...660L..12T,	
2022AstL...48..434B, 	
2022Ap.....65..579B,		
2022MNRAS.517L.102L,	
2022A&A...664L..13S, 	
2022ApJ...936...57T}.	
%
Line-of-sight velocities in gas ($^{12}$CO) and 3d velocities of the young stellar clusters from Gaia have confirmed the vertical oscillations, and show that the most massive star-forming regions spatially associated with the Radcliffe Wave (including Orion, Cepheus, North America, and Cygnus~X) move as if they are part of an oscillating wave driven by the gravitational acceleration of the Galactic potential
\citep{2024Natur.628...62K}.	
These observations support the suggestion that it served as the birthplace for the Upper Centaurus Lupus and Lower Centaurus Crux associations, home to the supernovae that generated the Local Bubble (Section~\ref{sec:local-bubble}) about 15~Myr ago
\citep{2022Natur.601..334Z},	
as well as providing insights into the Galaxy's gravitational potential, including the local baryonic and dark matter density.
Using Gaia-based orbits,
\citet{2025A&A...694A.167M}	
estimated that the solar system intersected the Radcliffe Wave between 18.2--11.5\,Myr ago, with the closest approach, 14.8--12.4\,Myr ago, coinciding with the middle Miocene climate transition on Earth.

The origin of the Radcliffe Wave remains a matter of detailed investigation and considerable speculation. Most studies interpret the wave as arising from a gravitational perturbation of the Galactic disk, due to an external impactor such as a dwarf satellite galaxy of the Milky Way
\citep{2022Ap.....65..579B},	
although it has also been attributed to the Kelvin--Helmholtz instability, appearing at the interface between the Galactic disk and the halo rotating at different velocities
\citep{2020Natur.583E..24F}. 	
Both interpretations currently face their own specific challenges
\citep{2024Natur.628...62K}.	

\subsection{Binary systems}
\label{sec:binary-systems}

\paragraph{Context}
Over the past century, binary (and multiple) star systems have been discovered through a variety of methods. They are loosely classified, according to observational criteria, as {\it visual\/} binaries (where both components are `visible'), {\it astrometric\/} (where the companion's presence is revealed by the photocentric motion), {\it spectroscopic\/} (revealed as orbital variations of spectral lines), or {\it eclipsing\/} (where the light from one component is eclipsed by the orbital motion of the other). Other manifestations of binaries include cataclysmic variables, ellipsoidal variables (including heartbeat stars), variability-induced movers, and others.
With its combination of accurate multi-epoch astrometric, photometric, and radial velocity measurements, and high angular resolution (Section~\ref{sec:data-processing-astrometry}), Gaia is discovering and characterising most of these types, simultaneously and uniformly, and in substantial numbers
\citep{2024BSRSL..93..170M}.	
New insights include 
the discovery of large numbers of new eclipsing binaries and ellipsoidal variables,
the high frequency of equal-mass `twin binaries', 
the measurement of stellar masses (which underpin estimates of stellar structure, age, and composition),
the orbital characterisation of very wide binary systems,
the discovery of large numbers of sub-stellar mass companions, including brown dwarfs and exoplanets,
the discovery of systems with a large mass function including those with neutron star or black hole companions,
and many others.
Brown dwarfs (many in binary systems) and exoplanets (as substellar companions) are considered in Section~\ref{sec:brown-dwarfs} and Section~\ref{sec:exoplanets} respectively.

\subsubsection{Processing}

The processing of non-single stars, embracing the astrometric, photometric, and spectroscopic processing, is carried out within Coordination Unit~4 of the Data Processing and Analysis Consortium. It is described in detail in Chapter~7 of the DR3 documentation 
\citep{2022gdr3.reptE...7P}.
For the latest data release, DR3, an overview of the astrometric binary star processing is described by
\citep{2023A&A...674A...9H},	
with other papers describing orbit determination for astrometric binaries in general
\citep{2023A&A...674A..10H},	
and the specific treatment for eclipsing binaries
\citep{2023A&A...674A..16M},	
and for SB1 spectroscopic binaries
\citep{2025A&A...693A.124G}.	

Processed results for non-single star are made available as four tables
\citep[][\S2.1]{2023A&A...674A..34G},	
organised by type of solution (or period range) rather than by kind of binaries:
{\tt nss\_two\_body\_orbit} contains the orbital parameters for all three categories, astrometric, spectroscopic or eclipsing binaries, all being unresolved;
{\tt nss\_acceleration\_astro} contains accelerations or derivatives of it for sources that have an astrometric motion better described using a quadratic or cubic rather than a linear proper motion;
{\tt nss\_non\_linear\_spectro} are trend (long period) solutions of spectroscopic binaries;
{\tt nss\_vim\_fl} are sources displaying photocentre displacement due to the photometric variability of one component of an otherwise fixed binary.
Table~\ref{tab:data-release-table2} includes a summary of non-single star types in DR3. Amongst 813\,687 non-single star solutions are 
169\,227 orbital astrometric solutions 
186\,905 orbital spectroscopic solutions 
and 87\,073 eclipsing binaries.
A breakdown of the non-single star solutions is given in 
\citep[][Table~2]{2023A&A...674A..34G}, 	
which lists, for example, 
246\,947 acceleration (7-parameter) solutions, and 91\,268 `jerk' (9-parameter) solutions.

Perhaps as many as half of all stars occur in binary or multiple systems. They occur in a wide variety of configurations, and display a vast range of behaviour and physical phenomena. The interaction between close components significantly influences their evolution, and gives rise to a great variety of structural and dynamical phenomena. As a result, it is challenging to attempt even a synopsis of the studies that are being made with the Gaia data. Some example applications are given by 
\citet{2023A&A...674A..34G},	
including the occurrence of binaries in the red giant branch and asymptotic giant branch,
the occurrence of compact companions (such as white dwarfs), 
the discovery of new binary ultracool dwarfs,
the nature of the brown dwarf desert around solar-type stars using true rather than minimum masses,
and the occurrence of higher-order multiplicities.
I will pick out just a few of these topics below.


\subsubsection{Resolved binaries}

Resolved binaries appear in as separate catalogue entries in the various data releases, and their occurrences are detailed in both 
the Fifth Catalogue of Nearby Stars out to 25\,pc \citep{2023A&A...670A..19G}, 
and the Gaia Catalogue of Nearby Stars out to 100\,pc \citep{2021A&A...649A...6G}. 
An important sample derived from EDR3 is the catalogue of resolved binaries within 1~kpc 
\citep{2021MNRAS.506.2269E}.
Starting with all sources within 1\,kpc (parallaxes $\varpi>1$~mas), selecting pairs with a projected separation less than 1~pc (i.e.\ an orbital period of $\sim 10^8$~yr, beyond which the Galactic tidal field is likely to result in their disruption), parallaxes consistent within $3\sigma$, and proper motions consistent with a Keplerian orbit resulted in a sample of 1\,256\,400 binary systems with a 90\% probability of being physically bound. 
%
In absolute numbers, this is by far the largest catalogue of high-confidence binaries of any type. For example, as of January 2024, the online version of 
the Washington Double Star Catalog
\citep[WDS,][]{2001AJ....122.3466M}	
listed 157\,012 double star entries (inclusive of the Hipparcos discoveries).

Combined with Gaia's multi-colour photometry allows each system component to be placed in the Hertzsprung--Russell diagram, and thereby classified as (for example) {\it both\/} components being main sequence (877\,416 systems), main sequence plus white dwarf (16\,156), or both are white dwarfs (1390). 
The catalogue also contains about 10\,000 high-confidence binaries in which the primary is a giant (about half of these are in the red clump), along with 130 giant--giant binaries, and some 13\,000 binaries in which one component is a subgiant.

\begin{figure}[t]
\centering
\includegraphics[width=0.70\linewidth]{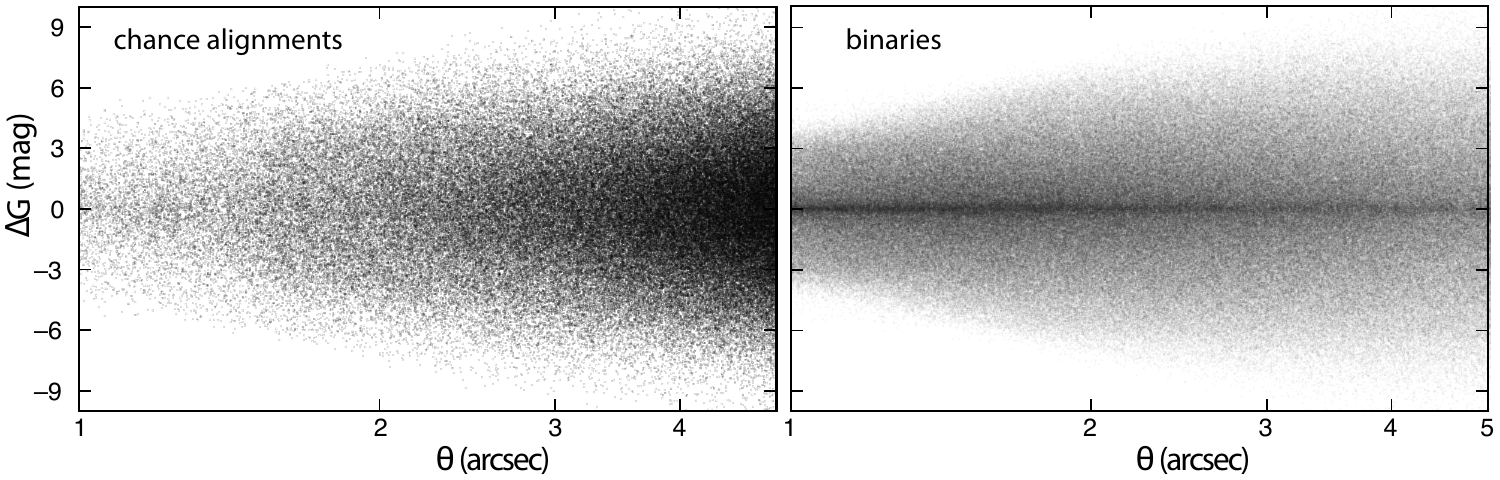}
\caption{From the catalogue of 1.3~million resolved binaries, this shows the magnitude difference between the two stars in chance alignments (left) and bound wide binaries (right) as a function of angular separation. The sign of $\Delta G$ is randomised. The excess of equal separation binaries, $\Delta G\sim0$, independent of separation, stands out prominently (\citet{2021MNRAS.506.2269E}, Figure~9).}
\label{fig:twin-binaries}
\end{figure}

\paragraph{Twin binaries}

An interesting and unexplained property of binary stars is the excess population of equal-brightness (and presumably equal-mass) `twin' binaries, first seen in ground-based studies of spectroscopic binaries \citep{1974AJ.....79..967T,1979AJ.....84..401L, 2000A&A...360..997T}. Such a population has been invoked to explain the `binary sequence' in the colour--magnitude diagrams of some open clusters, such as in Praesepe \citep{1999A&A...352..479M} and the Pleiades \citep{1999A&A...346...67K}. And in a very different context, an equal-mass companion to our Sun in the solar birth cluster at a separation of 1000~au could have increased the likelihood both of forming the observed population of outer Oort Cloud objects, and of capturing Planet Nine
\citep{2020ApJ...899L..24S}.

Extending these studies to {\it resolved\/} binaries has been complicated by various selection effects that plague the construction of complete binary samples. Gaia's uniform astrometric survey provides a new opportunity for further insights.
A sharp excess of equal-mass `twins', statistically significant out to separations of 1000--10\,000\,au was identified in a sample of 42\,000 wide main-sequence binaries from Gaia DR2
\citep{2019MNRAS.489.5822E},
being even more prominent in the much larger visual binary catalogue based on Gaia EDR3 
\citep{2021MNRAS.506.2269E}.
Figure~\ref{fig:twin-binaries} shows the magnitude difference, $\Delta G$, as a function of separation, for both chance alignments, and true binaries (for which both components have consistent parallaxes and proper motions). The excess population with $\Delta G\simeq0$ is clearly apparent. 
%
For separations more than 100--200\,au, viz.\ beyond the typical maximum sizes of observed circumstellar and circumbinary gas disks, they suggest that the most plausible scenario is that twin binaries formed at closer separations, and their orbits were subsequently widened by dynamical interactions in their cluster environments 
\citep[e.g.][]{2000MNRAS.314...33B, 2020MNRAS.494.2289A}. 
Further insight could come from knowledge of the alignment of their spin vectors, and how this varies with separation 
\citep[e.g.][]{2020A&A...642A.212J}. 
A subsequent finding was that essentially all wide twins are on very eccentric orbits, $e=0.95-1$, in turn implying pericentre distances of order 10\,au
\citep{2022ApJ...933L..32H}. 
This is consistent with a scenario in which twins are born in circumbinary disks, and are subsequently widened via dynamical interactions, either in their birth environments, or subsequently by tidal torques from the Galactic disk, or scattering by passing stars and molecular clouds 
\citep{2023MNRAS.524.3102M}.
The twin binary population is also evident in the colour--magnitude diagram of the catalogue of spectral line-broadening from DR3
\citep{2025A&A...693A.214H}.		

\subsubsection{Triple systems}
\label{sec:triple-systems}

One of the objectives in characterising the considerable complexity of binary and higher multiplicity stellar systems is to understand the underlying processes of star formation, for example whether it is the result of core fragmentation or gravitationally unstable disk accretion, and the subsequent structural and dynamical evolution of multiple systems.
Among field objects, the multiplicity and width of the orbital period distribution are known to be steep functions of the primary mass, i.e.\ the likelihood of being in a binary or multiple system increases as the component masses increase 
\citep{2013ARA&A..51..269D}.	
Most higher-order systems are therefore triple, with four or more components being less common 
\citep{1997A&AS..124...75T,2001IAUS..200...84T}.

Despite their importance in constraining theories of star formation, even a relatively complete census of triple systems has proved challenging. 
\citet{2010ApJS..190....1R} defined a sample for solar-type stars within 25~pc, with a total number of only~56 systems. Extension to smaller masses \citep{2019AJ....157..216W}, or larger distances \citep{2014AJ....147...86T}, has been hampered by the lack of a uniform coverage of the parameter space in the full range of periods and mass ratios.
A fairly complete Hipparcos sample of F and G dwarfs within 67~pc contained 4847 primary stars, with 2196 known stellar pairs; some of them belonging to 361 systems ranging from triples to quintuples \citep{2014AJ....147...86T}.
In a subsequent analysis, 
\citet{2014AJ....147...87T}
found a multiplicity rate ($n\ge2$) of 46\%, a log--normal period distribution with median of 100\,yr, and a nearly uniform mass-ratio distribution, independent of period. The fraction with $\ge3$ components was $0.13\pm0.01$, and the fractions with $n=1,2,3,\ldots$ components were 54\,:\,33\,:\,8\,:\,4\,:\,1. 

Turning to Gaia, the EDR3 million-star {\it binary\/} compilation noted above \citep{2021MNRAS.506.2269E} actively {\it removed\/} triple systems from their binary star catalogue, because it becomes far more complex, and requires more extensive astrometric data, to identify triple systems with any degree of statistical certainty, and even more so for quadruple or higher multiplicity systems.
%
A specific effort to identify triple systems {\it resolved\/} by Gaia was made by \citep{2022ApJ...926....1T}, who started with the Gaia Catalogue of Nearby Stars, within 100\,pc, and based on common proper motions and other criteria, identified 392 hierarchical triple systems (i.e., those in `nested' orbits, and with limited mutual interaction), along with 30 quadruples and one quintuple, $\xi$~Sco.
Given its selection uniformity, and its size, this sample is unprecedented in its unbiased census of wide triples, and a number of specific properties in terms of primary mass, orbital alignment, and eccentricities were described. Future Gaia releases will consolidate the 100~pc triple star sample by adding close resolved binaries, and providing astrometric and spectroscopic orbits for many.

This sample of relatively wide triples represents only a small fraction of {\it all\/} such hierarchical systems, whose inner pairs are mostly unresolved by Gaia. While their periods and mass ratios are unknown, their existence can be inferred from the increased astrometric noise, or their variable radial velocities.  The number of triple systems within 100~pc has been estimated as around 10\,000 
\citep{2022ApJ...926....1T}.
A list of 8000 candidate multiples derived from wide binaries in the Gaia Catalogue of Nearby Stars were selected as those where one (or both) components have excessive astrometric noise
\citep{2023AJ....165..180T}.
High-angular resolution observations from the ground should clarify their nature.

One goal here is to resolve the long-debated mechanism underlying the formation of very close binaries, for which many orbits are inferred to have decreased by a factor 10--100 since formation. Specifically, many main-sequence stars have companions at a few stellar radii, far closer than predicted by star-formation models. These compact binaries are often accompanied by a more distant tertiary, but with characteristics of the inner and outer binaries notably distinct from the general binary population \citep[e.g.][]{2007ApJ...669.1298F}. This has led to suggestions that their formation may be associated with the Kozai--Lidov resonance mechanism. This somewhat non-intuitive effect in hierarchical triples was discovered by \citet{1962AJ.....67..591K} for asteroid orbits, and independently by \citet{1962P&SS....9..719L} for artificial and natural satellite orbits. If the inner binary orbit is initially circular, there is a critical inclination between inner and outer binaries such that the orbit of the inner binary does not remain circular as it precesses: both the eccentricity of the inner binary and the mutual inclination execute periodic oscillations. 
\citet{2023MNRAS.518.1750H} used Gaia DR3 to search for resolved tertiary companions (with separations of $10^3-10^4$~au) to close (eclipsing, spectroscopic, and astrometric) binary systems. He found that the wide tertiary fraction increases with decreasing orbital period of the inner binary. With the inferred eccentricities of wide tertiaries consistent with a thermal distribution, and similar to those of wide binaries at similar separations, his results appear to exclude the Kozai--Lidov mechanism as an explanation for most wide tertiaries, at least at $>10^3$\,au.

Amongst the Gaia triple systems \citep[e.g.][]{2020AJ....159..265T} is the first resolved triple white dwarf system, J1953--1019 \citep{2019MNRAS.483..901P}. All three members have pure-hydrogen (DA) atmospheres, masses $0.60-0.63M_\odot$, and cooling ages of 40--290\,Myr, consistent with coeval evolution.

\subsubsection{Quadruple systems}
\label{sec:quadruuple-systems}

Two stable configurations of quadruple systems can exist, as conveniently illustrated by the `mobile' diagrams introduced by \citet{1968QJRAS...9..388E}.
In this system, Hiererachy~2 corresponds to two binaries which orbit their common centre of mass. Hiererachy~3 corresponds to a triple system itself orbited by a fourth component. Unstable and short-lived trapezium systems are also observed in star-forming regions.
Incidentally, amongst known quadruple systems are the bright stars
Capella, actually a pair of giants orbited by a pair of red dwarfs,
and Mizar, a visual binary whose resolved components are themselves both binaries.
But it is the great range of geometries, separations, and mass and magnitude differences, demanding a range of observational methods, that underlies the great difficulties in achieving any degree of survey completeness. For example, the inner binaries may be spectroscopic, eclipsing, or resolved; and the binary pairs may be very compact, or in wide orbits that are essentially unbound.
This range of properties also substantially complicates their measurement with Gaia: while a single star's motion can be characterised with just 5~astrometric parameters (two~position and two~proper motion components, and the parallax; or 6~including the radial velocity), and in principle over just 3--4~years, multiple systems require many more observations for their characterisation, and a more extended temporal baseline, crucially depending on the relevant orbital periods.

Two main mechanisms are considered to be responsible for multiple systems: fragmentation and capture. Formation by fragmentation of bound gas into multiple components includes `turbulent fragmentation' during the initial large-amplitude perturbations, or `disk fragmentation’ during the later stages, usually associated with the growing importance of centrifugal effects.
Formation through the mutual capture of initially unbound stars, includes cases where the excess kinetic energy of an initially unbound pair is transferred to a third star, and others where it is dissipated either via tides in the stellar envelope or in extended star--disk interactions.
The review of current theories by \citet{2019MmSAI..90..386C} emphasises the role that Gaia can be expected to play in distinguishing rival hypotheses, for example by constraining eccentricity distributions.
There are many other complications. For example, in a quadruple with two binaries on a long mutual orbit, each binary acts as a Kozai--Lidov perturber on the other \citep{2013MNRAS.435..943P}. The binaries can then experience eccentricity oscillations, and be driven to high eccentricity. Orbital `flips' and collisions are similarly enhanced in quadruples, with quadruples perhaps being a source of some triple systems \citep{2019MNRAS.482.2262H}.

Gaia is contributing by discovering more (a few tens) of such quadruple systems
\citep[e.g.][]{2022ApJ...926....1T,
2022MNRAS.511.3881F,
2023AJ....165..256Y,		
2023AJ....165..160T},	
and helping to characterise a number of others.
Amongst these, the HD~74438 system is a candidate progenitor of sub-Chandrasekhar type~Ia supernovae through white dwarf mergers
\citep{2022NatAs...6..681M}.	
And a recent Gaia discovery from EDR3, the white dwarf HD~190412~C, at just 32~pc, is bound to the triple system HD~190412, thus representing a new Sirius-like system in the solar neighbourhood
\citep{2023MNRAS.523.4624V}. 
Its location in the  \teff--mass diagram implies that it is undergoing crystallisation (Section~\ref{sec:white-dwarf-crystallisation}), making it the first confirmed crystallising white dwarf whose total age can be externally constrained from the properties of the bound triple.

\subsubsection{Wide separation binaries}		

As evident from the preceding sections, binary stars, as well as triple or even higher multiplicity star systems, are common. They form, in the turbulent gas clouds of dense regions of the interstellar medium, over a very wide range of separations.  Binaries with very close separations can eventually spiral in and merge, while those formed with wider separations, and less weakly bound, can eventually be broken apart, either due to close stellar encounters (for example in star clusters), or as a result of numerous distant passages that incrementally pull on the binary, and it slowly evolves from being bound to being unbound. Indeed, the wider their separation, the more easily they are disrupted, whether by passing stars, molecular clouds, or the Galaxy's spiral arms. Indicative survival times are of order a billion years at 0.1~pc separation (20\,000~au, or 0.3~light-years), and perhaps around 100 million years at 0.5~pc (100\,000~au, or 1.6~light-years).
Although there is no precise definition, binaries with separations of 100--2000\,au are often referred to as `wide' binaries, and those above about 30\,000\,au as `ultra-wide'. At these enormous separations, the two stars will be very widely separated on the sky, by a degree or more, with very long orbital periods, but sharing an almost identical space motion over millennia. Not surprisingly, the wider a binary is, the more difficult it is to identify as a physical pair -- and this has been a major barrier to discovering and studying wide binaries in the past.
As an example of the situation pre-Gaia, a programme aimed at identifying wide binaries in the Galactic halo found less than 100 candidate pairs from the Sloan Digital Sky Survey
\citep{2018MNRAS.480.4302C}.	

A wide separation binary can be recognised if both components share identical, or very similar, distances and space motions, and Gaia is identifying many thousands of very wide and ultra-wide binaries in this way. 
Already, Gaia DR2 allowed the identification of more than 50\,000 wide binaries, with separations 50--50\,000~au, and within 200\,pc
\citep{2018MNRAS.480.4884E}.	
While its sheer size is impressive, more interesting is the make-up of the resulting binary population. More than 50\,000 are pairs of main-sequence stars, with typical ages $\sim$3\,Gyr, while more than 3000 comprise a main-sequence primary and a white dwarf secondary, and 400 systems consist of white dwarf--white dwarf pairs. 
In terms of numbers versus separation, a marked difference between the three populations is evident. The main sequence binaries are reasonably consistent with a single power-law of slope over separations 500--50\,000~au, while the main sequence--white dwarf, and white dwarf--white dwarf binaries, show distinct breaks at 3000 and 1500~au respectively.  These distributions can be explained if the white dwarfs receive a `kick' of about 0.75\kms\ during their formation, presumably due to asymmetric mass-loss. 
Another catalogue of 99\,203 wide binaries was constructed from stars with proper motions above 40~\masyr, selected from Gaia DR2
\citep{2020ApJS..247...66H}.	
Their search, extending to $1^\circ$ separation, includes systems with separations up to 10\,000--100\,000~au.

Stars born together should be chemically homogeneous, and wide binary systems provide an opportunity to test this assumption. For 25 systems comprising main-sequence stars of similar spectral type in Gaia DR2, chemical abundances from high-resolution spectroscopy found that 20~pairs (80\%) were found to be homogeneous in [Fe/H] as well as in all other elemental abundances
\citep{2020MNRAS.492.1164H}.	
Other work has used the Gaia wide binary sample 
to compare their (statistically inferred) eccentricity distribution with predictions based on models of dynamical interactions within their natal cluster environment
\citep{2020MNRAS.496..987T};	
as a probe of the possible existence of Massive Compact Halo Objects (MACHOs) in the Galactic halo
\citep{2020ApJS..246....4T};	
in understanding the frequent occurrence of asymmetrical nebulae and bipolar lobes in planetary nebulae
\citep{2020A&A...644A.173G};	
and the role of binary systems in the formation of `hot Jupiter' type exoplanets
\citep{2020MNRAS.497.2250H}.	

\paragraph{Origin of wide binaries}
\label{sec:wide-binaries-origin}

An understanding of the origin of these very wide ($\gtrsim10^3$\,au) binaries remains elusive. In 2010, 
\citet{2010MNRAS.404.1835K}	
wrote that {\it `their origin has long been a mystery'}. 
Still today, it {\it `remains a mystery'\/}
\citep{2023ApJ...949L..28X},		
and indeed {\it `a major puzzle'} 
\citep{2023MNRAS.524.3102M}.	
%
Simulations find that they cannot form by fragmentation of a single molecular cloud core, nor in typical clusters 
\citep{2019MmSAI..90..386C,		
2020MNRAS.496.5176D}.			
%
Other possible formation processes include
the fragmentation of turbulent molecular cores
\citep{2007prpl.conf..133G},		
cluster dissolution 
\citep{2010MNRAS.404.1835K,		
2011MNRAS.415.1179M,
2023ApJ...955..134R},	
dynamical `unfolding' of compact triple systems
\citep{2012Natur.492..221R},		
formation in adjacent molecular cloud cores 
\citep{2017MNRAS.468.3461T},	
trapping in the tidal tails of stellar streams	
\citep{2021MNRAS.501.3670P},	
%
or from 3-body encounters
\citep{2024ApJ...970..112A}.		
Observations nonetheless suggest that the components are likely to have formed in the same star-forming region. Evidence (including studies based on Gaia DR2) includes
the large fraction of wide pairs of young stars in low-density star-forming regions 
\citep{2017MNRAS.468.3461T},	
the chemical homogeneity of wide binary components
\citep{2020MNRAS.492.1164H,		
2019ApJ...871...42A},			
the metallicity dependence of the wide-binary fraction 
\citep{2021MNRAS.501.4329H},	
and N-body simulations 
\citep{2001ApJ...555..945K}.	

One of the suggested formation mechanisms is that they arise as a natural consequence of star formation in the turbulent interstellar medium. In this scenario, stars form within the high-density filaments and cores of molecular clouds at the collisional interfaces of converging highly supersonic turbulent flows
\citep{2009ApJ...692..364F, 	
2018MNRAS.480.3916M,		
2018PASJ...70S..53I,		
2019ApJ...878..157X,		
2023ApJ...949L..28X,		
2024MNRAS.532.2425H}.		
Gaia observations have provided support for this through two independent avenues. 
The first is by measuring the eccentricity distribution of wide binaries, which has been possible for the first time with Gaia, and which finds an excess of very eccentric `superthermal' orbits
\citep{2020MNRAS.496..987T,		
2022MNRAS.512.3383H}.			
Given that such a distribution does not arise as a result of Galactic tides
\citep{2023MNRAS.524.3102M},	
nor as a result of stellar encounters
\citep{2024MNRAS.532.2425H},	
the latter authors concluded that the eccentricity distributions measured by Gaia favour their origin via turbulent fragmentation. A testable prediction is that the power-law distribution of eccentricities should be a monotonically decreasing function of binary age.

The second piece of evidence favouring a turbulent origin of wide binaries is that, in this model, the turbulent velocities of the gas are expected to be imprinted on those of newly formed stars. And indeed, recent Gaia observations find that the velocity differences and spatial separations of young stars statistically follow the power-law velocity scaling of interstellar turbulence. This has been demonstrated for 
1439 young stars in the Orion Molecular Cloud
\citep{2021ApJ...907L..40H}, 	
for a total of 3000 stars in Ophiuchus, Taurus, Perseus and Orion
\citep{2022ApJ...934....7H},	
and for 15\,149 young stellar objects in 150 associations younger than 3~Myr within 3\,kpc
\citep{2022MNRAS.513..638Z}. 	
The latter authors found that their associated clouds are elongated, and oriented parallel to the disk mid-plane. As probed by young stellar objects, the turbulence is isotropic, and the 2d velocity dispersion is related to size by $\sigma_v\propto r^{0.67}$. The turbulent energy dissipation rate decreases with Galactocentric radius which, they suggest, is explained if the turbulence is driven by cloud collisions.

\paragraph{Ultra-wide binaries and MOND}
\label{sec:ultra-wide-binaries}

Modified Newtonian Dynamics (MOND) is a theory which attempts to account for anomalous long-range gravitational effects (notably the flat rotation curves of galaxies) without invoking the current dark matter paradigm
\citep{1983ApJ...270..365M}.			
The most straightforward versions have been excluded by various observations, although in the absence of a convincing detection of dark matter, new tests are still undertaken. 
For binary separations larger than about 5000\,au, the individual stars have sufficiently small orbital accelerations (below about $10^{-10}$\,m\,s$^{-2}$) to provide a direct probe of MOND-like theories. 
Studies using Gaia DR1 and DR2, using large samples of binaries with separations up to 40\,000\,au
\citep{2019MNRAS.482.5018E,		
2019IJMPD..2850101H,			
2019MNRAS.488.4740P},			
concur that the distribution of the component velocity ratios, beyond $\sim$7000\,au, has a very long non-Newtonian tail
\citep[e.g.][Fig.~11]{2019MNRAS.488.4740P}.		
But it was unclear whether this is evidence for some form of MOND
\citep{2019IJMPD..2850101H},
or explicable as stars born in the same cluster and currently undergoing a chance close `flyby' 
\citep{2019MNRAS.488.4740P},
or as a population of hidden triple systems
\citep{2020MNRAS.496.1922B,		
2020MNRAS.491L..72C}.			
Studies using more accurate samples of wide-separation binaries from DR3 all continue to confirm the non-Newtonian distributions at low accelerations, but the explanations differ. Proponents of a MOND-like explanations have confirmed their earlier conclusions
\citep{2023ApJ...952..128C,	 
2024ApJ...960..114C,		
2023MNRAS.525.1401H}, 	
while others offer a more conventional explanation in terms of undetected close binary companions
\citep{2023OJAp....6E...4P,	
2024MNRAS.527.4573B}.		

\subsubsection{Ellipsoidal variables}
\label{sec:ellipsoidal-variables}

Ellipsoidal variables are close binaries with orbit inclinations that are too small to yield eclipses, but whose components are nonetheless distorted by their mutual gravity. Their quasi-sinusoidal light curves are a result of the changing projected areas and surface brightnesses of the distorted stars. The effect is in addition to certain other photometric signatures of close non-eclipsing systems, specifically the reflection of each star’s light from the other's surface, and the relativistic Doppler beaming of the light due to their orbital velocities \citep[e.g.][]{2011MNRAS.415.3921F}.
The semi-amplitude of the ellipsoidal effect depends on the stellar radii and binary separation, but even for the closest systems, $a\lesssim15R_\star$, it is only of the order of a few per cent \citep[e.g.][]{1993ApJ...419..344M}.
A compilation in 1985 listed some 20~ellipsoidal variables, and a similar number of suspected systems \citep{1985ApJ...295..143M}.
Their number has increased substantially with accurate photometric monitoring by the Kepler mission, and with some 15\,000 most recently found with TESS \citep{2023MNRAS.522...29G}.

Interest in ellipsoidal variables is partly motivated by efforts to further understand stellar evolution. Binary systems, particularly those with short orbital periods, are the progenitors of cataclysmic variables, supernovae, and gravitational wave-generating neutron star and black hole mergers. The complex evolutionary processes that lead from an initial main-sequence pair to the final merging stages remain poorly understood, and are best-probed by large uniform samples of various small-separation systems. 
Ellipsoidal variations have also been observed in a number of star--planet systems, in which tidal distortion of the star by the planet results in a periodic flux modulation due to changes of the star's visible surface area as the stellar tide, created by the planet, rotates in and out of view \citep{2009Sci...325..709B,2010A&A...521L..59M,2012ApJ...747...82C}.

As part of their extensive classification of variable stars identified as part of the Gaia DR3 processing, 
\citet[][\S4.10]{2023A&A...674A..14R}		
identified 65\,300 candidate ellipsoidal variables, although probably somewhat `contaminated', for example, by W~UMa eclipsing binaries.
But an area that is generating particular interest today, and based on Gaia data, is the role of ellipsoidal variables in the search for `unseen' compact secondaries, either black holes, neutron stars, or perhaps white dwarfs, using an approach that has been applied to similar photometric searches with OGLE 
\citep{2021MNRAS.504.5907G}.	
The idea is that, given an observed ellipsoidal amplitude, and the primary star's mass and radius, a minimum mass ratio of the binary can be estimated. A binary with a minimum mass ratio significantly larger than unity might be a candidate for having a dormant compact-object companion. Given that, in most cases, the primary mass and radius are not well known, 
\citet{2021MNRAS.504.2115G}	
used, instead, the idea of a `modified minimum mass ratio', mMMR, assuming that the primary fills its Roche lobe. The modified minimum mass ratio is always smaller than the minimum mass ratio, which is, in turn, smaller than the actual mass ratio. Therefore, binaries with mMMR~$>1$ are candidates for hosting a compact secondary. 
Applied to Gaia DR3,
\citet{2023A&A...674A..19G}		
identified 6306 short-period ellipsoidal candidates with relatively large-amplitude modulation in the $G$-band, mMMR~$>0.5$, and a subset of 262 with mMMR~$>1$, for which the compact-secondary probability is higher. However, given the two main underlying assumptions, that the light curve is attributable to ellipsoidal variations only, and that the primary is on the main sequence, they emphasise that follow-up radial velocity observations are needed to verify their true nature.
Other similar searches have been undertaken with Gaia DR3
\citep{2022ApJ...940..126F,		
2023MNRAS.518.2991S}.			
But to date, follow-up spectroscopy has resulted in only a low yield of massive companions. 
\citet{2023MNRAS.524.4367N}	
selected 14 targets for follow-up spectroscopy from
\citet{2023A&A...674A..19G}		
obtaining spectra at 2--10 epochs for each. From the semi-amplitudes of the radial velocity curves, they derived minimum companion masses of $M_{2, \rm min}\le0.5M_\odot$ in all cases. Seeking alternative explanations for the ellipsoidal-like variability, they could best reproduce the light curves and radial velocities with models for unequal-mass contact binaries with starspots. Some may also be detached main-sequence binaries, or even single stars with variability due to pulsations or starspots.

\paragraph{Heartbeat stars}
\label{sec:heartbeat-stars}
The related `heartbeat stars' are detached binaries with eccentric orbits, $(e\gtrsim0.3)$, and short periods, typically 1--100\,d. They typically display very small amplitude variations, 1--2\,mmag (although reaching 40\% for MACHO 80-7443-1718 \citep{2021MNRAS.506.4083J}), which are driven by tidal distortion, reflection and Doppler beaming. These effects are most prominent near periastron, where they combine to generate the characteristic electrocardiogram-like `heartbeat' signature 
\citep{2012ApJ...753...86T}.  	
Many oscillate throughout the orbit, attributed to tidally excited oscillation modes within one or both stars. A key signature of tidally excited oscillations is their occurrence at integer multiples of the orbit frequency 
\citep{2017MNRAS.472.1538F},	
the largest corresponding to resonances with the star's normal modes (e.g.\ for KOI--54, the dominant oscillations are at the 90th and 91st orbital harmonics
\citep{2011ApJS..197....4W}). 	
They are attracting interest as probes of tidal dissipation, internal structure, orbit circularisation, and short-period binary formation
\citep{1995ApJ...449..294K, 
2016ApJ...829...34S,
2018ApJ...854...44M,	
2020A&A...640A..16T}.	

A study making use of the Gaia data is for the 180 TESS systems identified by
\citet{2025ApJS..276...17S}.	
Their phase-curve fits (using {\tt eBEER}) revealed a degeneracy in $M_\star$, $R_\star$, and \teff\ using only photometry. Gaia magnitudes and distances were used to estimate absolute magnitudes, and improved estimates of \teff\ and $R_\star$. Gaia then strongly constrains the values, and yields masses in agreement with main-sequence predictions based on the Gaia temperatures. They also found agreement between their estimates of eccentricity, $e$, and argument of periastron, $\omega$, compared with those from supplementary radial velocities
\citep{2016ApJ...829...34S}. 	
Finally, by examining the Gaia `renormalised unit weight error' (RUWE, where large values might suggest the presence of a tertiary, Section~\ref{sec:ruwe}), they confirmed that the orbital precession measured for a number of systems (which reaches $9^\circ$/yr in the case of TIC 451708707) is indeed most likely driven by tides and not, for example, by Kozai--Lidov oscillations from a tertiary companion.

\subsubsection{Cataclysmic variables}

Cataclysmic variables are compact interacting binaries containing a white dwarf accreting from a donor star overflowing its Roche-lobe. In most systems, the companions are low-mass, late-type stars. As a class, they allow for the study of (non-)equilibrium accretion disks, themselves relevant for the understanding of \mbox{X-ray} binaries, black holes and active galactic nuclei
\citep{1995cvs..book.....W}.	
If the white dwarf is non-magnetic, the outflowing donor material forms an accretion disk around it. 
For larger magnetic fields, the accretion flow follows the field lines and accretes onto the white dwarf at its magnetic poles, resulting either in `intermediate polars' with truncated disks ($B\sim 10^2-10^3$\,T), or in `polars' in which the accretion disk is fully suppressed ($B\gtrsim10^3$\,T). 
As in many area of astronomy, complete samples to well-defined limits are important for detailed studies. For example, their space density, $\rho$, places strong constraints on models of their formation and evolution
\citep{2020MNRAS.494.3799P}.
Early binary population synthesis studies 
\citep[e.g.][]{1992A&A...261..188D,
1996ApJ...465..338P}
found $\rho=0.2-2.0\times10^{-4}$\,pc$^{-3}$, with more recent models suggesting $\rho\sim1-2\times10^{-5}$\,pc$^{-3}$
\citep[e.g.][]{2015ApJ...809...80G, 2018MNRAS.478.5626B}.
These predicted values are systematically larger than those inferred from observations, a result often taken as suggesting shortcomings in theoretical models. However, accurate space densities have been hampered by the small size of X-ray selected samples, as well as by the lack of accurate distances.  

At the time of Gaia DR2 in 2018, the most extensive X-ray samples were drawn from the ROSAT Bright Survey, yielding 15 or so non-magnetic cataclysmic variables with distances 100--500\,pc, and from the 70-month Swift Burst Alert Telescope (BAT) yielding a similar number of intermediate polars, with slightly larger distances in the range 500--1000\,pc
\citep[][Tables~2--3]{2018A&A...619A..62S}.\footnote{
Larger X-ray samples are becoming available with the Spektrum-X-Gamma mission's eROSITA all-sky survey, operational 2019--2022 \citep{2021A&A...647A...1P}. Compared to ROSAT, this provides a factor~25 sensitivity improvement at low X-ray energies, broader energy coverage (0.3--10\,keV), and better spatial resolution. Spectroscopic identification, classification and detailed follow-up to determine orbital periods, are ongoing.}
Based on the distances of these small sample members from Gaia DR2, most with accuracies better than 5\%, 
\citet{2018A&A...619A..62S} concluded that most of the distances to the intermediate polars (in particular) had been somewhat under-estimated in the past. He gave revised upper limits for their space densities of 
$\rho< 1.1 \times 10^{-6}$\,pc$^{-3}$ for the non-magnetic objects (assuming a Galactic scale height of 260\,pc),
and 
$\rho< 1.3 \times 10^{-7}$\,pc$^{-3}$ for the long-period intermediate polars (assuming a scale height of 120\,pc),
even lower than the previous estimates.

\citet{2020MNRAS.494.3799P} 
presented a volume-limited sample, defined by their Gaia DR2 parallaxes, composed of 42 objects within 150~pc, which they estimated as $\sim$80\% complete. It includes two new Gaia discoveries.  
Amongst their findings was the high fraction of {\it magnetic\/} cataclysmic variables, at 36\%. This is in contrast with the absence of magnetic white dwarfs in the detached population of cataclysmic variable {\it progenitors}, underlining the fact that the evolution of magnetic systems has to be included in the next generation of population models.
Their resulting space density, $\rho=4.8\pm0.7\times10^{-6}$\,pc$^{-3}$, is (again) significantly lower than predicted by binary population synthesis studies.
From a much larger sample of 1587 cataclysmic variables, many from the AAVSO database, along with distances from Gaia DR3,
\citet{2023AJ....165..163C}
derived local space densities of all objects, and magnetic systems, of $6.8\pm1.2\times10^{-6}$ and $2.1\pm0.5\times10^{-6}$\,pc$^{-3}$ respectively, again broadly confirming previous estimates. They suggest that the ongoing discrepancy with population synthesis models arises from cataclysmic variables which remain undetected so far, notably systems with a very low mass-loss rate, as well as those in the period gap.
They derived the exponential scale heights of all CVs, and the magnetic systems, of $375\pm2$\,pc and $281\pm3$\,pc, respectively, considerably larger than those suggested in previous studies. And they found that a simple scaling, of $\sim$500, relates the luminosity function of all cataclysmic variables with that of white dwarfs.
Further insights on the occurrence and evolution of cataclysmic variables continue to build on the photometry and distances from Gaia DR2 and DR3, and the location of the various sub-types in the Hertzsprung--Russell diagram
\citep[e.g.][]{2020MNRAS.492L..40A,	
2022ApJ...938...46A}.		

Considerable physical insights (and various open questions) are being leveraged from these complete Gaia samples, along with their distances and photometry, related to the details of the accretion process and their associated evolution, the different efficiencies of magnetic wind braking and gravitational wave radiation in removing angular momentum from the binary orbit, and the attendant phenomena of the `period gap', the `period minimum', and the `period bounce'
\citep{2020MNRAS.494.3799P,
2020MNRAS.491.5717B}.

\subsubsection{Spectroscopic binaries}

Spectroscopic binaries are unresolved (short-period) systems revealed by the varying Doppler motion of their spectral lines. Single-lined systems (SB1) are characterised by the lines of one component, while double-lined systems (SB2) show the lines of both.
%
Pre-Gaia numbers were given by the Ninth Catalogue of Spectroscopic Binary Orbits (2009) which listed around 4000 orbits 
\citep{2009yCat....102020P}, 
and the more recent APOGEE compilation of $\sim$1000 others
\citep{2020ApJ...895....2P}.	
With Gaia, the first assessment of stellar multiplicity (viz.\ `non-single stars') derived from DR3 \citep{2023A&A...674A..34G}	
includes 186\,905 SB1+SB2 systems based on Gaia's RVS spectra; details of the spectroscopic binary processing are given by 
\citet{2025A&A...693A.124G}. 	
This compilation includes the Keplerian elements, including period, eccentricity, and solution quality, but not the individual radial velocity measurements.  
This was assessed as comprising 181\,327 SB1, and a further 5000 or so SB2
\citep{2022MNRAS.517.3888B}.	
This huge increase in numbers, by a factor~30, represents an unprecedented data set for studies of short-period binaries, including
binary frequency as a function of primary mass
\citep{2010ApJS..190....1R,2016AJ....151...85T,2017ApJS..230...15M},
the mass--ratio distribution
\citep{1992ApJ...394..592M, 2015A&A...575L..13B, 2017MNRAS.472.4497S, 2019MNRAS.487.3356S},
and the eccentricity--period relation
\citep{2009A&A...498..489J,	
2013A&A...556A...4D},		
properties important in understanding binary formation and evolution.
\citet{2022MNRAS.517.3888B} 
used two external radial velocity sources for an independent validation of the Gaia orbits: 17\,563 stars from LAMOST DR6, and 6018 from GALAH DR3.  They constructed a function that estimates the reliability of the Gaia orbits, eliminating some spurious systems, and resulting in a `clean' Gaia sample of 91\,740 SB1 systems. They provisionally attributed a paucity of short-period binaries with low-mass primaries to observational bias.
Further spectroscopic monitoring of a very restricted number of Gaia sources, including those in the Gaia--ESO spectroscopic survey, points to the regions of parameter space currently poorly defined by Gaia, for example as a result of the limited 34~month temporal coverage of Gaia DR3
\citep{2024A&A...684A..74M,	
2024A&A...690A.276V,		
2023AJ....165..220T}.		

\paragraph{Orbit circularisation}
An important process in binary star evolution is orbit circularisation, driven by tidal friction. Both `equilibrium' tides (due to the companion’s gravitational attraction, leading to a bulge rotating at the orbital period) and `dynamical' tides (from tidally-excited, gravity-driven oscillations near the core--boundary surface) extract energy from the orbit, resulting in a secular decrease of the orbital period and eccentricity, and stellar rotation rates tending to spin--orbit synchronisation 
\citep[e.g.][]{1981A&A....99..126H,
2008EAS....29...67Z,
2014ARA&A..52..171O}.
Early studies of spectroscopic binaries showed circularisation `cut-off periods', with shorter period binaries having more circular orbits
\citep{1984IAUS..105..411M,
1988ApJ...326..256M}.
More recent studies have been generally based on the inference that tidal circularisation operates mainly on the main sequence
\citep{2002A&A...386..222W,
2005ApJ...620..970M}.

\begin{figure}[t]
\centering
\raisebox{-2pt}{\includegraphics[width=0.30\linewidth]{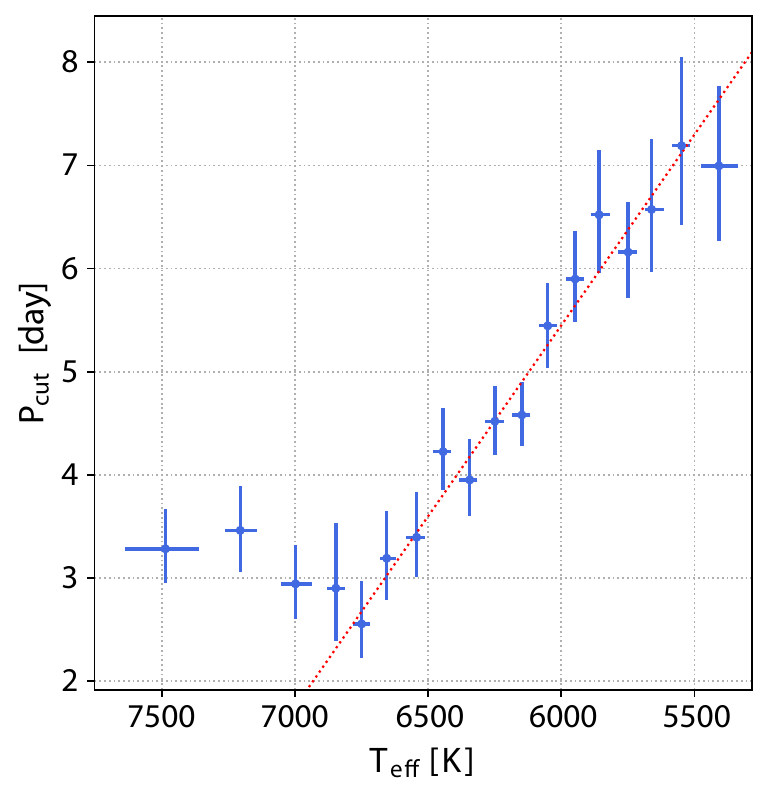}}
\hspace{10pt}
\raisebox{-6pt}{\includegraphics[width=0.385\linewidth]{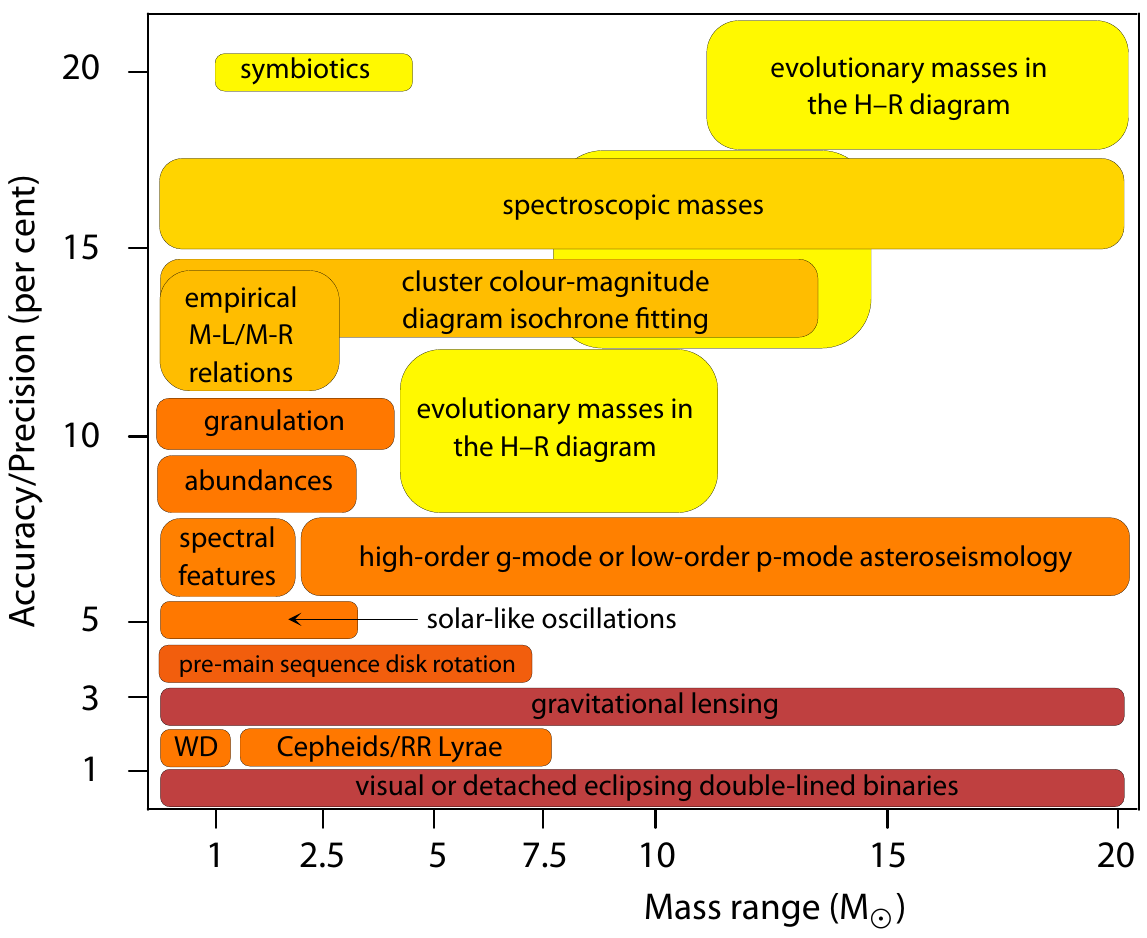}}
\hspace{10pt}
%
\includegraphics[width=0.235\linewidth]{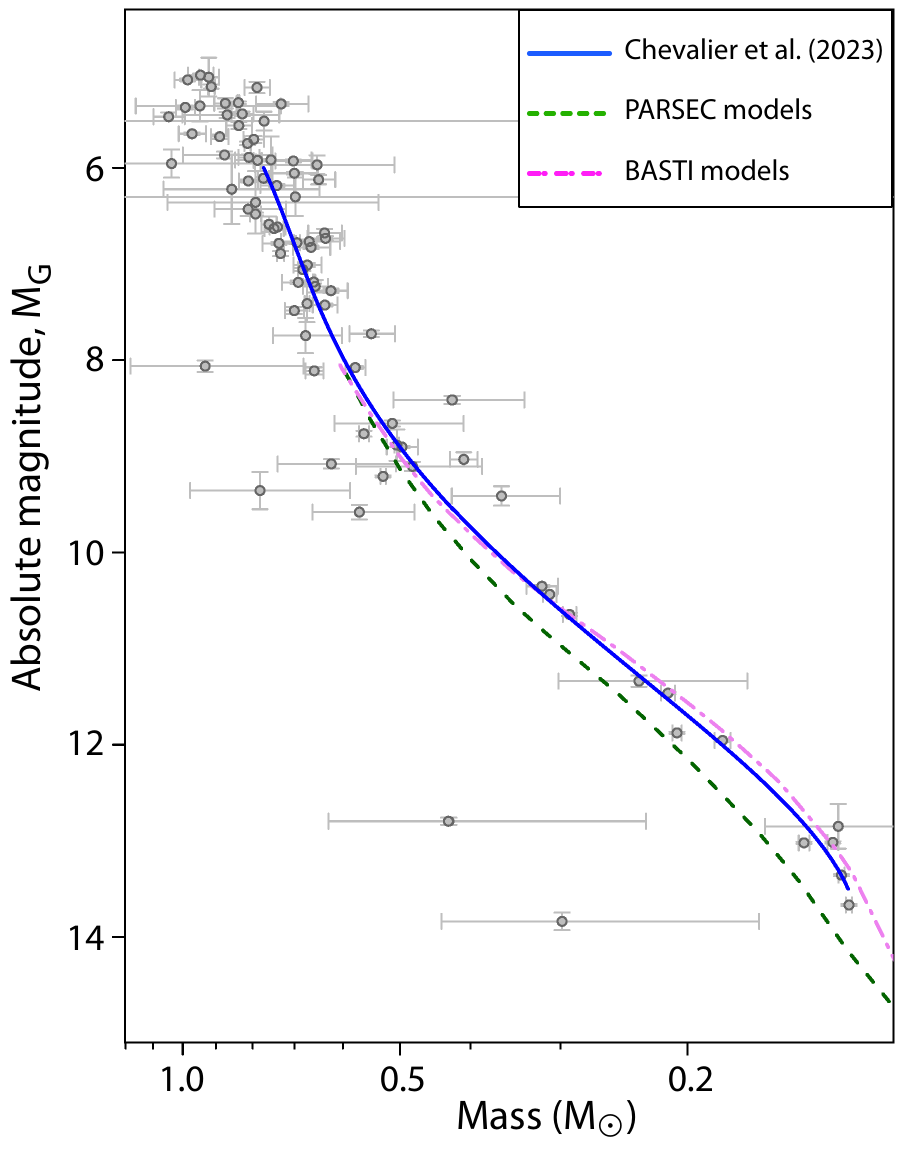}
\vspace{-5pt}
\caption{
Left (a): cut-off in orbital period of spectroscopic binaries versus temperature, with the best-fit linear model in red \citep{2023MNRAS.522.1184B}. 
Middle (b): precision of the various mass determination methods as a function of mass: darker colours are less model-dependent, with the red regions providing the most accurate masses (\citet{2021A&ARv..29....4S}).
Right (c): mass--luminosity relation showing the Gaia values derived from a combination of SB2 binaries and non-single star astrometric solutions (grey), along with solar metallicity main-sequence isochrones from PARSEC (green), from \citet{2015A&A...577A..42B} (red), and BaSTI \citep{2024MNRAS.527.2065P} (pink), the latter deviating only below $0.12M_\Sun$ \citep{2023A&A...678A..19C}.
}
\label{fig:spectroscopic-binaries}
\end{figure}

\citet{2023MNRAS.522.1184B} 
selected 17\,000 main-sequence systems, focusing on A, F and G-type primaries. From the period--eccentricity dependence, they found that the cut-off period does not depend on the stellar age but, instead, varies with stellar temperature (and therefore mass), decreasing linearly from 6.5\,d at \teff\,$\sim5700$\,K to around 2.5\,d at 6800\,K (Figure~\ref{fig:spectroscopic-binaries}a).
Following earlier models \citep{2008EAS....29...67Z}, they assumed that circularisation is determined by turbulent dissipation (by convective envelopes in cool stars, and radiative envelopes in hotter stars). But their findings on the cut-off period for F- and G-stars are inconsistent with the circularisation taking place during the star's {\it main-sequence\/} lifetime. Their favoured explanation is that the eccentricity distribution was determined, instead, during their {\it pre-main-sequence\/} phase, when the stars were much larger, and the circularisation processes correspondingly much faster
\citep{1984IAUS..105..411M,
1989A&A...223..112Z,
2011MNRAS.411.2804K}.
A subsequent study of eight open clusters supports this idea
\citep{2023MNRAS.524.3978M}.
The finding is important for modelling the evolution of short-period binaries, including
close triple systems,
cataclysmic binaries, 
novae, 
X-ray binaries, 
hot Jupiters, 
and the origin of chemically peculiar objects such as C~stars and Ba~stars.

\subsection{Brown dwarfs}
\label{sec:brown-dwarfs}

Brown dwarfs are loosely characterised as objects having masses below the hydrogen-burning limit, i.e.\ below around $0.07M_\Sun$.  Although predicted on theoretical grounds in the early 1960s 
\citep{1962AJ.....67S.579K},
it was more than 30~years before the discovery of the first, 
Teide~1 in the Pleiades cluster
\citep{1995Natur.377..129R}
and, soon after, a faint object orbiting the nearby star GJ~229
\citep{1995Natur.378..463N,
1995Sci...270.1478O}.
With their relatively low effective temperatures, they are faint at visible wavelengths, emitting most of their energy in the infrared. 
Subsequent searches for isolated brown dwarfs have accordingly been made from infrared surveys such as DeNIS, 2MASS and WISE, and (given their proximity) taking advantage of their high proper motions. Searches for faint brown dwarf companions to main sequence stars and white dwarfs include multi-colour imaging and coronagraphy surveys, and radial velocity monitoring for close companions.

As progressively cooler objects have been discovered, new spectral classes have been introduced, according to their effective temperature and spectral features. Accordingly, brown dwarfs comprise late~M, L, T, and the coolest~Y (\teff\ $\sim500-600$\,K), together often referred to as `ultra-cool dwarfs'. Today, several thousand are known, including the nearest, the L/T-type binary Luhman~16 which, at a distance of 2~pc, is the third closest system to the Sun.
One area of active brown dwarf research is their role as the link between planets and low-mass stars, with ambiguity still remaining about whether brown dwarfs tend to form like high-mass planets, say above $13M_\Jupiter$ 
\citep[e.g.][]{2009MNRAS.392..413S},
or through interstellar cloud fragmentation like low-mass stars 
\citep[e.g.][]{2001AJ....122..432R}.
They also present great challenges for atmospheric and evolutionary modelling as a result of their complex chemistry and cloud physics 
\citep[e.g.][]{2001RvMP...73..719B}.

Brown dwarfs are numerous and ubiquitous, with estimates suggesting that our own Galaxy contains some \mbox{25--100 billion}
\citep{2017MNRAS.471.3699M}.
They are faint (and physically complex) objects, and their lifetimes are also remarkable: due to their slow hydrogen fusion, they are expected to remain convective over most of their lives, surviving for many times the current age of the Universe, with the lowest masses persisting for trillions of years
\citep{2005AN....326..913A}.
A curious phenomenon that has been predicted, and which Gaia may be able to test, is that brown dwarfs more massive than the H-burning limit may well exist
\citep{1992ApJ...393..258S}.
To form such an `overmassive' brown dwarf, mass must be added sufficiently slowly to a `traditional' brown dwarf -- but one old (and cool) enough to prevent their cores from igniting when they cross the H-burning limit. Two different possible formation scenarios have been suggested
\citep{2019ApJ...871..227F,
2022ApJ...932...91M},	
and the latter detail how future Gaia data releases might demonstrate their existence.

Gaia contributions fall into two distinct areas: isolated brown dwarfs, and brown dwarfs in binary systems.

\paragraph{Isolated brown dwarfs}
Studies during the preparation of Gaia's scientific case in the 1990s suggested that, since their luminosities fade rapidly with age to very faint absolute magnitudes, objects with $V<20$~mag will be strongly biased towards very young objects, and those of higher mass. The youngest might be visible out to 400~pc, while the oldest would only be detectable if very nearby.  Some 20\,000 detections with Gaia were given as an indication, although with much uncertainty. Despite most being at the faint end of Gaia's survey, their proximity would  ensure a high parallax precision. Excellent positioning in the Hertzsprung--Russell diagram would then allow the determination of ages and masses by sequence fitting, in turn providing an accurate picture of their recent formation history and mass function.

With Gaia DR1,
\citet{2017MNRAS.469..401S}
identified 321 L/T dwarfs, representing 45\% of known L/T dwarfs with $G<20.3$\,mag. They also estimated the Gaia magnitude of previously known objects, predicting that 1010~L and 58~T objects would have $G<21.5$, with 543~L and 10~T having $G<20.3$. 
With Gaia DR2,
\citet{2018A&A...619L...8R}
identified 3050 later than M7, 647~L, and 16 T~dwarfs, being 61\% of known objects with $G<21.5$ and 74\% with $G<20.3$.
\citet{2019MNRAS.485.4423S}
identified 695 M8--T6 objects with accurate parallaxes and proper motions, finding that 100 are in 47 multiple systems, of which 20 were new. Their sample formed the training set for the ultra-cool dwarfs data processing work package (ESP--UCD).
\citet{2020A&A...637A..45S}
found a further 50 new candidates within 20~pc, objects with relatively small proper motions and low tangential velocities, and concentrated towards the Galactic plane. They include three M7--M8, five M8.5--M9.5, four L0--L1, and seven L4.5--L6.5.
\citet{2020MNRAS.494.4891M}
identified a further 10 new ultra-cool dwarfs in seven wide binaries, including an L1+L2 system with a projected separation of 959~au. 
\citet{2021AJ....161...42B}
constructed a volume-limited sample of 369 L0--T8 dwarfs out to 25~pc, defined exclusively by parallaxes, thereby providing the most precise space densities.
With Gaia DR3, came the inclusion of the mean low-resolution $B_{\rm P}/R_{\rm P}$ spectra for 220~million sources
\citep{2023A&A...674A...1G}.	
With this,
\citet{2023A&A...674A..39G}	
started with 94\,000 ultra-cool candidates, from which they constructed a high-quality sample of 31\,000, deriving radii, luminosities, and bolometric fluxes using Gaia parallaxes combined with infrared (mostly 2MASS and WISE) photometry.

\begin{figure}[t]
\centering
\includegraphics[width=0.68\linewidth]{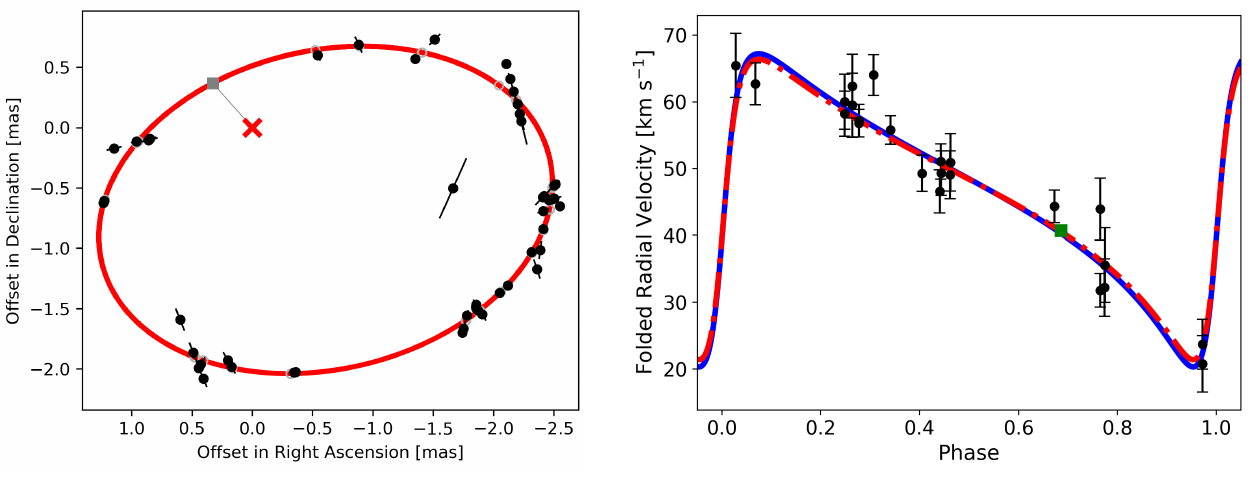}
\vspace{-5pt}
\caption{The orbit of Gaia DR3 5136025521527939072, detected as both an astrometric and single-lined spectroscopic binary ({\tt nss\_solution\_type} = AstroSpectroSB1), probably comprising a dormant neutron star ($G=12$~mag, $M_1=1.2M_\Sun$, $M_2=1.5M_\Sun$, $P=536$\,d), illustrating the DR3 data quality.
Left: along-scan residuals of the mean epoch astrometry (black circles), and the modelled astrometric orbit (red), the orbit focus (red cross) and periastron (grey square).
Right: phase-folded radial velocity data, together with the estimated astrometric (red) and spectroscopic (blue) orbit elements. The single Observatoire de Haute Provence--Sophie radial velocity (green square), was obtained 4.5~yr after the end of the Gaia data segment used. It was not part of the fit, but confirms the quality of the predicted orbit \citep[from][Figure~43]{2023A&A...674A..34G}.}
\label{fig:arenou23-orbit}
\end{figure}

\paragraph{Brown dwarfs in binaries}
Due to the large reflex motion they induce in their host stars, radial velocity surveys were expected to discover brown dwarf companions to solar-type stars with relative ease.  A prominent feature of the early radial velocity exoplanet discoveries in the 1990s, in contrast, was the general absence of close-in ($a<3-4$\,au) substellar objects with $M\sim10-80M_\Jupiter$, a paucity referred to as the {\it brown dwarf desert}. 
Later radial velocity surveys of several thousand stars confirmed this pattern, finding brown dwarf companions out to $P\!\sim\!10$\,yr, but yielding only a small number, of order 100, in the mass and separation range characterising the `desert' 
\citep[e.g.][]{2016A&A...588A.144W}.

Notwithstanding their relative rarity, the detailed study by
\citet{2022A&A...661A.151H}
showed that Gaia's astrometric precision should allow for the detection of (unseen) companions down to Jupiter masses, allowing the efficient detection of large numbers of brown dwarfs. Additionally, Gaia's low-accuracy multi-epoch radial velocity measurements for $G_{\rm RVS}<12$~mag should provide additional detections for the more massive, while a further small sample will have detectable transits in the Gaia photometry. 
From astrometry alone, they predicted 30--40\,000 detections out to several hundred parsecs for the nominal 5~year mission, and 45--55\,000 for a 10-year mission. 

In view of the progressive and iterative nature of the data analysis, in particular regarding the astrometric solutions for binary and multiple systems, it will clearly be some years before Gaia's full discovery potential is realised. But numerous insights are becoming available.
In their wide-ranging paper on stellar multiplicity based on Gaia~DR3,
\citet{2023A&A...674A..34G}	
made a first analysis of the occurrence of 
ultra-cool dwarf binaries (comprising both very low-mass stars and brown dwarfs) 	
and sub-stellar companions to main sequence stars (comprising both brown dwarfs and planets). 
For the former, they examined the orbital solutions for the Gaia `ultra-cool dwarf sample' defined by 
\citet{2019MNRAS.485.4423S}.
Emphasising that around 75\% of that sample are fainter than the $G<19$ cut-off adopted for processing with the non-single star pipeline, they found (targeted) orbital solutions only for previously known orbital systems. 
They obtained 13 such orbits, of which only 5 have $G<19$. Three have previously published astrometric orbit solutions, for which the Gaia results are in reasonable agreement.
\citet{2023A&A...680A..16U} 
used DR3 astrometry to clarify the masses of 31 systems with existing radial velocity-based orbits.
For their sub-stellar companions, 
\citet{2023A&A...674A..34G}
confirmed that the Gaia DR3 astrometric performance reaches the sensitivity to detect substellar companions around a statistically significant number of stars, providing measurements of their three-dimensional orbital architectures and true masses for the first time.
They identified 1843 brown dwarfs (and 72 exoplanets) in their catalogue of companion masses. Only 10 of the brown dwarf companions were already known (from ground-based radial velocity exoplanet surveys). Their Table~11 provides a comparison of periods and eccentricities from Gaia and from the literature, along with the estimates of the minimum mass (from radial velocities) and true mass (from the Gaia astrometry).
Selecting HD~77065\,b as a favourable (but tolerably representative) example, estimates from radial velocity work are 
$P=119.113\pm0.003$\,d, $e=0.694\pm0.0004$, and $M\sin i=41\pm2M_\Jupiter$ 
\citep{2016yCat..35880144W}. 
The corresponding Gaia DR3 values are 
$P=119.1\pm0.2$\,d, $e=0.70\pm0.04$, and with the measured inclination $i=42\pm3^\circ$ yielding a true mass $M=64.2\pm5.1M_\Jupiter$. The semi-major axis of the astrometric orbit is $a_0=1.04\pm0.07$~milli-arcsec. An example of the quality of the astrometric orbits (although not for a brown dwarf) is shown in Figure~\ref{fig:arenou23-orbit}. 

\citet{2023A&A...674A..10H} fitted the sources in the 34~month astrometric time series of Gaia DR3 with a single Keplerian astrometric-orbit model with 12~free parameters, and an additional jitter term. They determined solutions for 1162~sources, loosely classified as 
1093 low-mass stars, $>120M_{\rm J}$,
52 as brown dwarfs, $20-120M_{\rm J}$,
and 17 of planetary mass, $<20M_{\rm J}$,
thus providing {\it `a positive outlook for the expected harvest from the full mission data in future data releases'}.
An alternative classification approach used the value of RUWE and/or the Hipparcos--Gaia proper motion anomaly to identify 9698 planet-hosting candidates $<13.5M_{\rm J}$ with semi-major axis 1--3~au, many of which may eventually prove to be of brown-dwarf or stellar mass
\citep{2024arXiv240916992K,	
2024arXiv240916993K}.		
More on the robustness of the expected orbital solutions is given by
\citet{2024OJAp....7E.100E}.	

\subsection{Exoplanets}
\label{sec:exoplanets}

\paragraph{Context}
The first exoplanets, planets around stars other than our own, were discovered in the mid-1990s, through repeated high-accuracy radial velocity observations of nearby stars. Since then, in a burst of intense activity, nearly 6000 exoplanets have been discovered as of mid-2025. They can be discovered through the periodic radial velocity variations of the host star as the planet(s) move in orbit around it. And in analogous way, high accuracy astrometric observations of a candidate host star can reveal the component of the star's motion around the system's barycentre projected on the celestial sphere. Exoplanets can also be detected through high-accuracy photometric observations if the planet passes across the face of the star, a configuration that shows up as a periodic decrement in the starlight as the planet transits, once every orbit.  Exoplanets have also been discovered through gravitational lensing, direct imaging, eclipse timings, along with other less-direct means \citep{2018exha.book.....P}. 

Three of these methods: astrometric motions, photometric transits, and gravitational microlensing are relevant to Gaia, and the first (of many future) discoveries have been reported. In addition, distances and space motions from Gaia are helping to characterise known planets, and provide further insights into theories of planet formation.

As of mid-2025, more than 1000 of the nearly 6000 known planets have been discovered from a variety of ground-based radial velocity measurement programmes. More than 200 have been discovered by microlensing, and nearly 100 from direct imaging. The bulk, more than 4000, have been discovered from their photometric transit signals: around 500 from ground-based transit measurement programmes, 30 from the pioneering CoRoT space mission, 4000 from NASA's Kepler mission, and more than 600 from the TESS mission, with several thousand other TESS exoplanet candidates awaiting confirmation. The exoplanet systems detected by Kepler, in particular, reveal a wide range of properties, with multiple planets, orbital resonances, and masses ranging from many multiples of Jupiter's mass, down to a few hundred planets with masses comparable to Earth.

\subsubsection{Astrometric detections}

The discovery of exoplanets by astrometry has a surprisingly long history: as early as 1855, Captain W.\,S.\,Jacob, at the East India Company's Madras Observatory, who had been measuring the astrometric orbit of 70~Oph, erroneously reported that anomalies made it `highly probable' that the system contained a planetary body. The method has continued with a checkered history to this day, with claims and refutes, from the 1930s onwards, surrounding possible planets orbiting Proxima Cen, 70~Oph, 61~Cyg, and Barnard's Star. 
%
%
As of May 2025, the 
\href{https://exoplanetarchive.ipac.caltech.edu/}{NASA Exoplanet Archive}
lists just 5~astrometric discoveries:
DENIS--P J0823--49 from ground-based measurements (with VLT--FORS2), a $28M_\Jupiter$ object orbiting an ultracool L~dwarf of mass $0.07M_\Sun$ at a distance of 20~pc
\citep{2013A&A...556A.133S};
GJ~896, a $2.3M_\Jupiter$ planet orbiting a binary star system inferred from radio observations with the VLBA
\citep{2022AJ....164...93C};
HIP~66074\,b, one of the first exoplanet candidates detected by Gaia astrometry
\citep[][referred to there as Gaia--3\,b]{2023A&A...677L..15S};
and the most recent confirmed candidates Gaia--4\,b and Gaia--5\,b
\citep{2025AJ....169..107S}.	

\begin{figure}[t]
\centering
\includegraphics[width=0.32\linewidth]{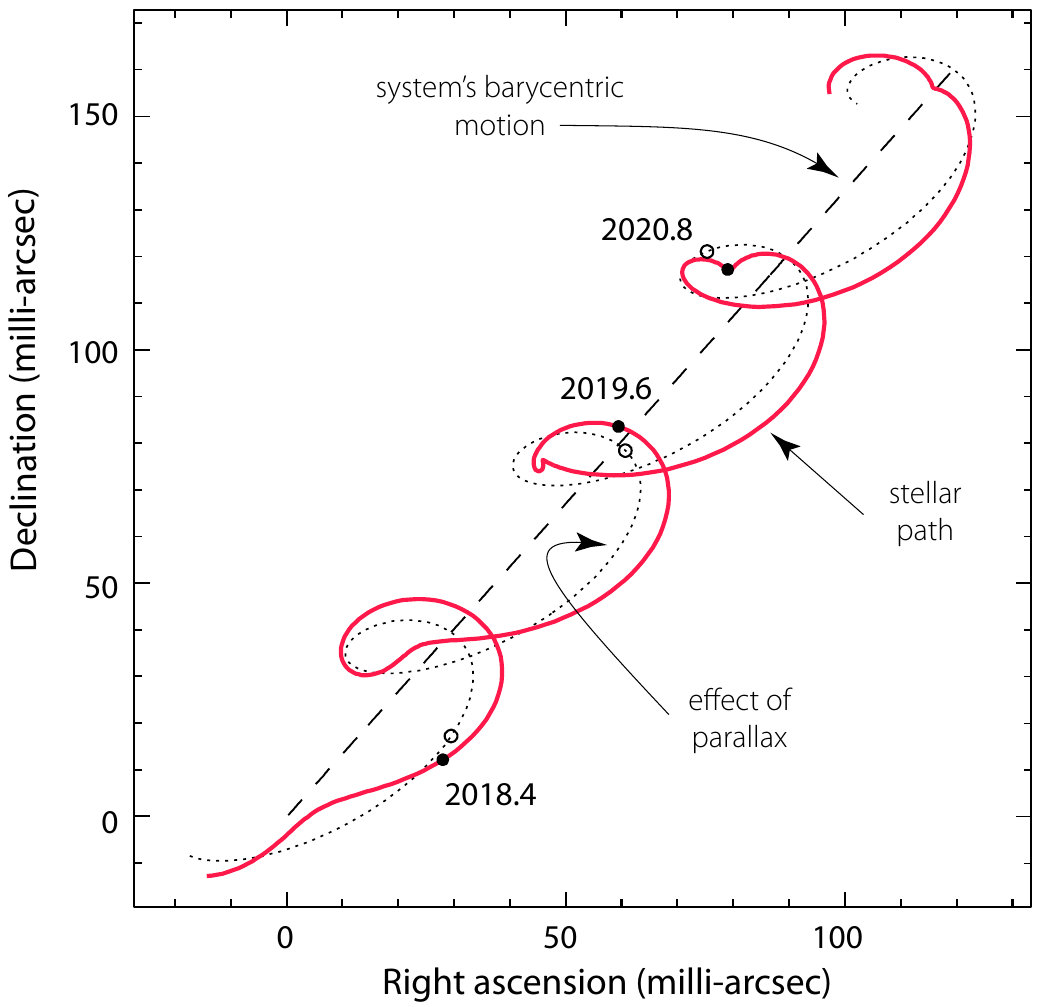}
\hspace{20pt}
\raisebox{3pt}{\includegraphics[width=0.32\linewidth]{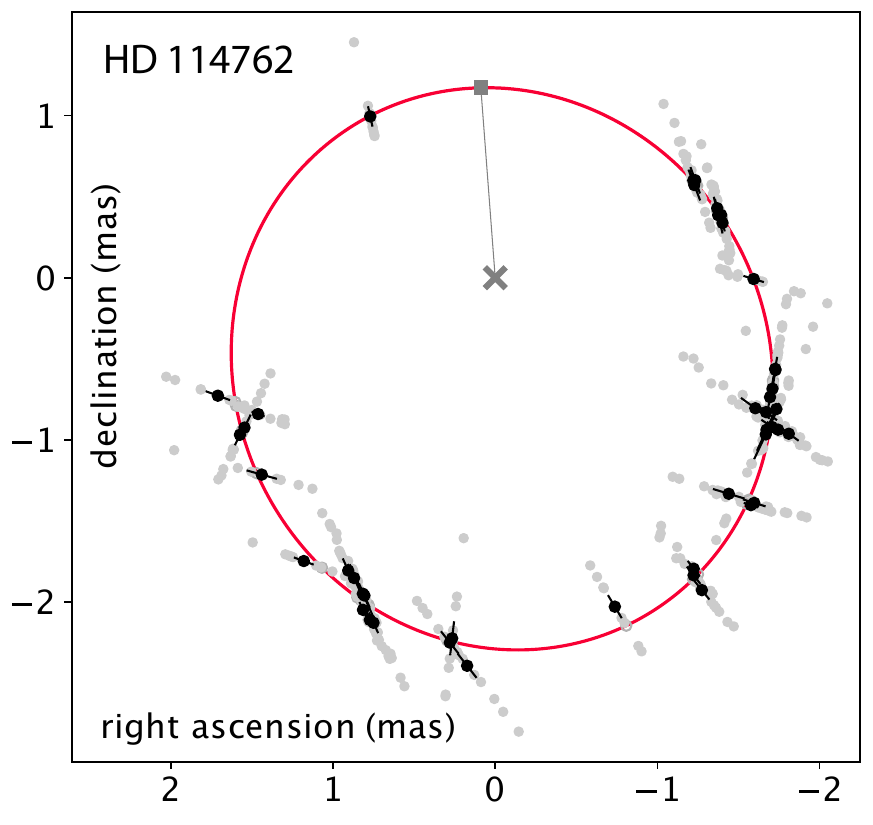}}
\caption{(a)~Schematic showing the simulated motion on the sky of a star--planet system. The system barycentre movies linearly through space. The effect of the star's parallax is shown as a dotted curve. The red curve shows the motion of the photocentre as a result of the star's proper motion, parallax, and orbiting planet. 
(b)~The detailed treatment of orbital binary systems by Coordination Unit~4 includes a number of impressive and compelling examples of the fitted orbits, including the 7.15~mag HD~114762, with $P=83.74\pm0.12$~day, $e=0.32\pm0.04$, and parallax $\varpi=25.36\pm0.04$~mas \citep[][Figure~12]{2023A&A...674A..10H}.
}
\label{fig:exoplanets}
\end{figure}

As part of the Gaia data processing, Coordination Unit~4 processes all non-single stars
\citep{2023A&A...674A..34G}.
Data Release~3 contains 800\,000 solutions with either orbital elements or trend parameters (whether for astrometric, spectroscopic and eclipsing binaries, or some combination).
Of these, 130\,000 are full orbit solutions, while 300\,000 show non-linear astrometric motion, i.e.\ indicative of partial orbit coverage
\citep{2023A&A...674A..10H}.
All will be substantially improved in future data releases as more data is included, and as the calibration models improve. And many of the present non-linear solutions will progress to full orbit solutions as the temporal baseline improves (Figure~\ref{fig:exoplanets}a).

Amongst sub-stellar companions, this processing resulted in 1843 brown dwarf and 72~exoplanet {\it candidates}
\citep{2023A&A...674A..34G}. 
Of these, only 10 brown dwarfs and 9 exoplanets were previously known from radial velocity surveys, all with a reasonable agreement between their inferred orbital periods. Validation of these candidates is not straightforward, and amongst their candidates, 
\citet{2023A&A...674A..34G}
list just two with {\it validated\/} orbits implying the presence of new planetary companions:
HIP~66074 with $P=297\pm2.8$\,d, $e=0.46\pm0.17$, $a_0=0.21\pm0.03$~mas, and $M_{\rm p}=7.3\pm1.1M_\Jupiter$
\citep{2023A&A...677L..15S}; 	
and HIP~28193 with $P=827\pm50$\,d, $e=0.07\pm0.10$, $a_0=0.25\pm0.02$~mas, and $M_{\rm p}=5.3\pm0.6M_\Jupiter$.
Their sub-milli-arcsec semi-major axes are especially noteworthy, with their smallest candidate having $a_1=0.14\pm0.04$~milli-arcsec (Figure~\ref{fig:exoplanets}b).
Amongst their other candidates are NASA's other astrometric listing DENIS--P J0823--49, along with earlier radial velocity discoveries including GJ~876, HD~114762, HD~162020, and HD~164604. Others add to the very rare class of giant planets orbiting white dwarfs, including WD~0141--675 which, at 9.8~pc, is one of the metal-enriched systems suggestive of the capture of planetary debris.

Despite the relatively small number of published candidates today, the prospects for future astrometric discoveries with Gaia are substantial. 
Detailed simulations found that some 8000 giant exoplanets (with masses $1-3M_{\rm J}$, orbiting F, G, and K-type stars) should be detectable by Gaia astrometry
\citep{2008A&A...482..699C}.		
Later estimates extended the host stars to a broader range of spectral types, distances, and magnitudes, using the best estimates of exoplanet frequency distributions, and detailed simulations of the Gaia instrument and its scanning of the sky
\citep{2014ApJ...797...14P}.	
This work predicts that some 20\,000 Jupiter-mass long-period planets should be discoverable out to distances of 500\,pc for the nominal \mbox{5-yr} mission (including some 1000--1500 around M~dwarfs out to 100\,pc), rising to some 70\,000 for the \mbox{10-yr} mission. The planets that Gaia should discover in large numbers -- Jupiter-mass planets, in Jupiter-like orbits -- will not be the sort of habitable planets that are so keenly targeted by exoplanet scientists today. But they will clearly signal planetary systems that are, perhaps, very much like our own, with a Jupiter-like sentinel orbiting far out from the star capturing potentially hazardous objects left over from the system's formation.
Amongst these should be 25--50 transiting planets with periods in the range 2--3~years, perfectly placed for detailed studies of their atmospheres through follow-up transit measurements from the ground. 
The first bulk release of Gaia's astrometric exoplanet discoveries, perhaps of order 1000 candidates, will presumably come with DR4, with the others not until Data Release~5 around 2030. Final numbers will be dependent on the quality of the final instrument calibrations, as well as on the actual form and multiplicity of the relevant exoplanet population.

\subsubsection{Transit detections}

While predominantly an astrometric mission, various photometric transit detections are also expected with Gaia, best considered under three categories: planets already known from other discovery techniques (notably radial velocity) with transits discoverable {\it a posteriori\/} in the Gaia photometric data; transiting planets discovered from the Gaia photometry alone; and Gaia astrometric discoveries subsequently found to transit (from Gaia or other photometry). 
Despite an accuracy of $\sim$1\,mmag per transit at $G\lesssim14-16$ (Figure~\ref{fig:edr3photometry}b), standard methods of searching parameter space for transits is poorly-suited to Gaia-like data sets with long time baselines and limited temporal coverage. 
A directed follow-up strategy for discovering transits in low-cadence photometric surveys has been detailed for Hipparcos and Gaia 
\citep{
2011MNRAS.415.2513D,	
2012ApJ...753L...1D,	
2013MNRAS.428.3641D}.	
\citet{2012ApJ...753L...1D} took into account the scanning law, Galactic models, and detection limits to $\lesssim16$\,mag, and concluded that the low cadence and relatively small number of measurements gives a limit on detectable orbit period of $P\lesssim10$\,d, and a resulting number of expected discoveries from Gaia {\it photometry\/} of between one thousand and several thousand.

The first photometric transit discoveries with Gaia were announced in May 2022
\citep{2022A&A...663A.101P}.	
This made use of a version of the Box-Least-Square method, applied to a set of stars prioritised by machine-learning classification. Radial-velocity follow-up observations were performed using the LBT--PEPSI spectrograph, confirming Gaia's first two exoplanets, Gaia--1\,b and Gaia--2\,b, as hot Jupiters. They are included in the 
\href{https://exoplanetarchive.ipac.caltech.edu/}{NASA Exoplanet Archive}.

A regular transit-type signature cannot be taken as unambiguous evidence for a transiting planet without excluding the many `false-positive' signals that plague both ground and space searches. These include
stellar binaries with grazing eclipses,
background eclipsing binaries,
eclipsing binaries in a hierarchical triple system,
and eclipsing binaries presenting only secondary eclipses.
Because Kepler's pixel size is large, 4~arcsec\,$\times$\,4~arcsec, binaries within this sky area, and even unassociated background stars and eclipsing binaries, can be superposed on the candidate, and must be excluded.
Because Gaia identifies binaries down to 0.7~arcsec separation, it can assist with candidate rejection. Gaia quantifies the goodness-of-fit for each of the 1.46~billion sources with a full astrometric solution in Data Release~3  (Section~\ref{sec:ruwe}), allowing many false candidates to be swiftly excluded.
The problem is still more acute for TESS, where the pixel size is 21~arcsec\,$\times$\,21~arcsec, and the assistance of Gaia is proving correspondingly more powerful.

\subsubsection{Microlensing detections}
\label{sec:exoplanet-microlensing}

One further Gaia exoplanet currently appears in the 
\href{https://exoplanetarchive.ipac.caltech.edu/}{NASA Exoplanet Archive}:
Gaia22dkvLb, included in August 2024. It was a gravitational microlensing event (Section~\ref{sec:microlensing}), 
discovered as part of the 
\href{https://gsaweb.ast.cam.ac.uk/alerts}{\tt Gaia Science Alerts} 
processing (Section~\ref{sec:science-alerts}), part of the Gaia Data Processing and Analysis Consortium (DPAC), which processes the real-time photometric data from the scanning satellite, and flags unexpected magnitude increases, allowing rapid follow-up by ground-based photometry 
\citep{2021A&A...652A..76H}.
For microlensing candidates, the goal is to initiate a rapid and dense sampling of the rising light curve (while the lensing event is still ongoing), with the aim of detecting light-curve perturbations signalling the existence of an orbiting planet.
This specific
\href{http://gsaweb.ast.cam.ac.uk/alerts/alert/Gaia22dkv}{event announcement},
on 2022--08--16 at 22:40:56, drew attention to a bright, Galactic plane source with historic magnitude $G=13.18\pm0.02$, as it brightened by 0.6~mag to an `alert magnitude' of $G=12.53$~mag.
The object designation, Gaia22dkvLb, denotes the Gaia science alert identifier, Gaia22dkv (itself following the convention GaiaYYaaa, with YY denoting the event year, and a 3-letter encoding of the event sequence in that year), with the suffices Lc adhering to the microlensing exoplanet nomenclature in denoting the planet~b associated with the lensing object~L.

Taking account of the many complexities in this kind of analysis, the discovery paper
\citep{2024AJ....168...62W}	
concluded that the lens is a $M_{\rm L}=1.15^{+0.16}_{-0.08}M_\Sun$ star at a distance $D_{\rm L}=1.27^{+0.43}_{-0.25}$\,kpc, orbited by a planet of mass  $M_{\rm p}=0.59^{+0.15}_{-0.05}M_{\rm J}$ with a projected orbital separation $a=1.41^{+0.76}_{-0.36}$\,au, and an orbital period $P=2.96\pm0.20$\,yr.
The host star and planet masses are typical of exoplanet systems, including microlensing discoveries, but there are other properties that make it more unusual: the planet orbits within the snowline, where liquid water might be present, and the system is bright enough for radial velocity follow-up. And while ground-based surveys focus on the direction of the Galactic bulge to provide enhanced lens and source densities, and hence improve alignment probabilities, Gaia22dkvLb (found as a result of Gaia's all-sky scanning) lies in a non-bulge direction, although still in the Galactic plane. Only one other non-bulge microlens planet is known, TCP~J0507+244Lb, found as an alert from ASAS--SN
\citep{2018MNRAS.476.2962N}. Both reached $V<12$~mag at peak.  All-sky rates for such bright events are very low, with only $\sim$0.1~events~yr$^{-1}$ predicted
\citep{2008ApJ...681..806H}.	

\subsubsection{Orbit inclination}

A crucial issue for radial velocity discoveries is that these measurements are sensitive only to the component of the star's motion {\it along\/} the line-of-sight. It follows that the planet mass can only be estimated to within the uncertainty of $\sin i$, where $i$ is the (unknown) inclination of the orbit to the line-of-sight. Accordingly, radial velocity discoveries are accompanied by their quoted `minimum' mass. But the true mass, which can only be provided by two-dimensional measurements on the plane of the sky, may be much larger.
I will give just two examples of Gaia's contributions from a rapidly expanding literature.
The first is HD~92987\,b, with a minimum mass of $17M_\Jupiter$, and a semi-major axis of 9.6~au, and considered as a promising candidate for direct imaging
\citep{2019A&A...625A..71R}.
But the $2087\pm19$\ms\ astrometric signal reveals that the orbiting companion is not close to its minimum mass, but instead a $0.25M_\Sun$ star viewed at a near-polar inclination of $175^\circ$
\citep{2021AJ....162...12V}.
A second example is the historically interesting HD~114762\,b, with a minimum mass of $11M_{\rm J}$ (Jupiter masses). It was originally classified as a `probable brown dwarf'
\citep{1989Natur.339...38L}, 
although it has been considered as one of the first possible exoplanet discoveries. 
Examination of the excess astrometric noise from Gaia vindicates the discoverers' conservative claim, and yields an inclination of about $i=6^\circ$, and hence a companion mass of around $100M_{\rm J}$, placing it firmly in the brown dwarf regime 
\citep{2019A&A...632L...9K}. 

\begin{figure}[t]
\centering
\includegraphics[width=0.5\linewidth]{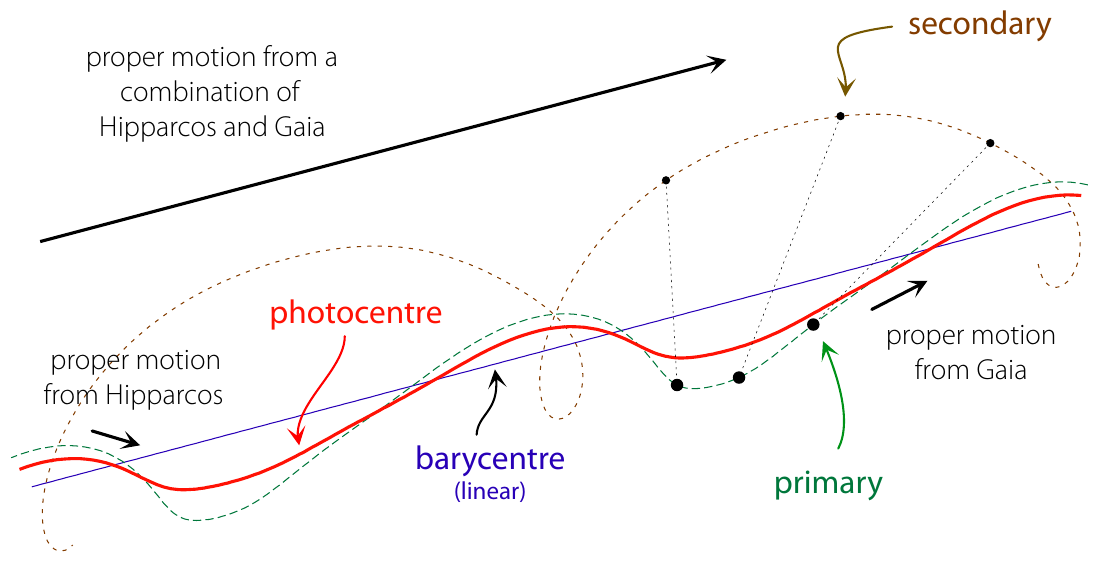}
\caption{Schematic orbital motions of the components of a binary system (star--star, or star--planet) about the barycentre cause the photocentre (red curve) to trace a non-linear path on the sky. The type of solution permitted by a series of astrometric observations (including an `acceleration' solution, or full orbital solution) depends on the length of the observation interval compared with the system's orbital period. The measurements by Hipparcos (1989--1993), and by Gaia (2014--2025), sample different segments of the photocentric motion (from \citet{1997ESASP.402...13L}).}
\label{fig:pm-hipparcos-gaia}
\end{figure}

\subsubsection{Pointers to imaging detections}

Exoplanet imaging represents a considerable technical challenge because of the close (angular) proximity of planets to their host stars, and the very small ratio of the planet-to-stellar brightness.
A number of important objectives nevertheless motivate its pursuit, including 
discovering long-period planets in wide orbits ($a\gtrsim20$\,au) that cannot easily be probed by other techniques;
characterising orbital motion;
studying formation mechanisms and planet--disk interactions in young protoplanetary disks in which planet formation is ongoing;
and 
as a first step towards a far-future goal of resolved spatial imaging of an exoplanet surface. 
As of mid-2025, some 150 planets, in around 80~systems, have been discovered through imaging.
Amongst them, Jason Wang's
\href{https://jasonwang.space/orbits.html}{time-lapsed movies}		
illustrate planet orbital motion in $\beta$~Pictoris, Fomalhaut, 51~Eridanus,
and the 4-planet system HR~8799.

Gaia is contributing to these goals by identifying where and when to search for suitable imaging candidates. Specifically, for planets with orbit periods of 20--30 years or more, space astrometry can exploit the 30-year interval between the Hipparcos and Gaia proper motions, because each samples the star's reflex motion at very different epochs in a long-period planet orbit (Figure~\ref{fig:pm-hipparcos-gaia}). These `accelerating systems' can then be used to determine masses for previously imaged planets, or as `dynamical beacons' to pinpoint where a new planet must lie, and so increasing the prospects of discovery compared to `blind' imaging surveys.
Using this approach, direct imaging the companion to the Sun-like star HD~33632 which, at a projected separation of 20\,au, induces a $10.5\sigma$ Hipparcos--Gaia astrometric acceleration, gives a dynamical mass of $46\pm8M_{\rm J}$
\citep{2020ApJ...904L..25C}.
Others such applications have been reported
for $\pi$~Mens		
\citep{2020A&A...642A..31D};
in establishing a dynamical mass of $61\pm4M_{\rm J}$ for the companion to HD~984
\citep{2022AJ....163...50F};
finding a $28M_\Jupiter$ companion to HIP~21152 in the Hyades cluster
\citep{2022ApJ...934L..19K};
for the discovery of 10~companions out of 25~`accelerating' stars
\citep{2022MNRAS.513.5588B};
and an even larger statistical survey
\citep{2022A&A...657A...7K}.
With more observations, and future Gaia releases, many more imaging discoveries lie in the future.

\subsubsection{White dwarf pollution}

A remarkable connection between the observed spectra of white dwarfs, and the formation and survival of exoplanets, has been assembled over the past two decades. In his history of the advances in understanding the heavy element pollution of white dwarf photospheres,
\citet{2015ASPC..493..291Z}
suggests that the first observational evidence for the existence of exoplanet systems actually came more than a century ago from one such `polluted' star, van Maanen~2 
\citep{1917PASP...29..258V,1919AJ.....32...86V},
although van Maanen did not appreciate its significance, nor even that he had observed a white dwarf.
Today, vMa~2 is considered to be the prototype of the white dwarf DZ~spectral class, with high-resolution spectra showing that its atmosphere is deficient in hydrogen, and yet at the same time rich in the refractory elements, with lines of Fe, Ca, and Mg.

Other white dwarfs with photospheric metals were duly discovered, with as many as 16 terrestrial-like heavy elements in the case of GD~362
\citep{2007ApJ...671..872Z}.
It was soon appreciated that their intense gravity should cause any heavy elements to sink rapidly 
\citep{1945AnAp....8..143S},
with elements heavier than helium pulled into their interiors on time scales far shorter than their cooling ages
\citep{1979ApJ...231..826F}.
A replenishing source for these `polluting' metals is therefore required. Initially, accretion from the interstellar medium was favoured 
\citep[e.g.][]{2010MNRAS.404.2123F},
although further observations 
\citep[e.g.][]{1987Natur.330..138Z},
and theoretical work 
\citep[e.g.][]{2003ApJ...584L..91J}
shifted the accretion paradigm to rocky asteroidal debris. 

In parallel, dust disks around white dwarfs were first discovered in the early 1990s, their presence being inferred from their large mid-infrared spectral excess
\citep{1990ApJ...357..216G}. 	
Major advances in their discovery and characterisation has been made from space, notably with Spitzer, and more recently WISE, and some 100 or more dust-disk systems are known today.
The model currently favoured attributes the heavy element pollution in white dwarf atmospheres to the accretion of rocky planetesimals which have been scattered into the tidal disruption radius of the white dwarf, where they are torn apart into dusty debris that subsequently spreads out to form a circumstellar disk. Some of this material can be perturbed by a planetary body, to be intermittently accreted by the white dwarf
\citep[e.g.][]{2016NewAR..71....9F}.

Gaia is contributing to the discovery of white dwarfs with circumstellar disks also comprising a gaseous component, which is revealed by the presence of the Ca\,{\footnotesize II} 860\,nm triplet within Gaia's RVS spectrometer's spectral band. 
In the first statistical study using the Gaia data, 
\citet{2020MNRAS.493.2127M}
selected 7705 single white dwarfs from the Sloan Digital Sky Survey, and with Gaia magnitudes brighter than 19~mag. They identified five gaseous disk hosts, all of which had been previously discovered. 
Restricting the 260\,000 white dwarfs identified in the Gaia DR2 release
\citep{2019MNRAS.482.4570G}
to those brighter than $G=18.5$~mag, using the Gaia distances to constrain the models of the spectral energy distribution, and 
with infrared data from large area surveys (including 2MASS, UKIRT, VHS, and WISE),
resulted in a further six white dwarfs with gaseous debris disks, bringing the total number known to just~21
\citep{2021MNRAS.504.2707G}.
Every Ca\,{\footnotesize II} triplet emitting system that has both multi-epoch observations and asymmetric emission profiles shows morphological variability, and those that have been studied in detail have shown that the variations are periodic in nature, and well described by the precession of a fixed intensity profile in the disk. 
The period of this precession can span the range of years to decades, and this range is supported by theoretical studies that show the effects of general relativistic precession and pressure forces on the gaseous disk.

\subsubsection{Characterisation of known systems}

Many studies of exoplanets are now using Gaia distances to contribute to an understanding of the properties of their host stars. 
By determining the distances and space motions of their host stars, Gaia is precisely characterising their host star luminosities, their location in the observational Hertzsprung--Russell diagram and, via their inferred radii, the physical radii of the transiting planets themselves. 

Searching the occurrence of `exoplanets' and `Gaia' in the abstracts of ADS papers over 2018--2025 yields nearly 400 refereed papers. As one generic example, the Gaia--Kepler Stellar Properties Catalog lists the stellar properties of 186\,301 Kepler stars, homogeneously derived from isochrones and broadband photometry, Gaia DR2 parallaxes, and spectroscopic metallicities
\citep{2020AJ....159..280B}. 
Their median catalogue uncertainties are 112\,K for \teff, 0.05~dex for $\log g$, 4\% for $R_\star$, 7\% for $M_\star$, 13\% for $\rho_\star$, 10\% for $L_\star$, and 56\% for stellar age. These constraints on stellar properties enable trends in exoplanet properties to be examined as a function of stellar mass and age.

Accurate radii are important in characterising the planet's properties. Together with the planet's mass, the radius provides an estimate of its average density, allowing a broad categorisation of its nature, i.e.\ whether the planet is a gas giant (like Jupiter or Saturn), an ice giant (like Uranus or Neptune), or a rocky planet (like Earth or Mars). Combined with information about their periods and host star properties, this leads to insights into both the physics of planetary atmospheres and interiors, and the physics of planet formation and evolution. Masses of transiting planets can be constrained by radial velocity measurements, while the planet radius depends, through the transit measurements, on the assumed mass and radius of its host star. Aside from direct measures from interferometry, spectroscopic indicators of surface temperature and luminosity can be used to infer the radius using the Stefan--Boltzmann law: $L=4\pi R^2\sigma T_{\rm eff}^4$\,, where $L$ is the bolometric luminosity, $R$ the radius, and $T_{\rm eff}$ the effective temperature. And this requires knowledge of the distance to the star in order to convert fluxes to luminosities. 

As described in Section~\ref{sec:classification-stellar-properties}, the Gaia project's `astrophysical parameters inference system' 
\citep[Apsis,][]{2018A&A...616A...8A,	
2023A&A...674A..27A}	
provides detailed and homogenous (albeit model-dependent) astrophysical information for each star, derived from Gaia's three-colour photometry and its RVS (radial velocity spectrometer) spectra. Bundled with Gaia DR3, \mbox{Apsis} provides (model-dependent) stellar radii, along with \teff, $\log g$, and [M/H], for 470~million stars. From knowledge of the host star radius, the planet radius can be derived from the transit depth, $\Delta F$, which is determined by the planet/star radius ratio, $\Delta F=(R_{\rm p}/{R_\star})^2$. 

Using Gaia DR1 distances to derive the radii of the planet-hosting stars,
\citet{2017AJ....153..136S}
derived the radii of 116 exoplanets with uncertainties $\sim$10--20\%. 
Using DR2 parallaxes,
\citet{2018ApJ...866...99B}	
gave revised radii for 177\,911 Kepler stars. Their $\sim$8\% precision was a factor 4--5 improvement over previous estimates. From these stellar radii, they estimated the radius of 2123 \mbox{Kepler} planets (and 1922 candidates), confirming a gap in the radius distribution of close-in planets, those between Earth--Neptune size
\citep{2020AJ....160...89P}.	
This general picture was confirmed in a 1000 planet sample
\citep{2018AJ....156..264F},	
in which stellar radii improved from 11\% to 3\% precision (with good agreement with the Apsis results, their Figure~3), with planet radii improved to 5\% precision. 

Further improvements in numbers and accuracies came with Gaia DR3.
\citet{2023arXiv230111338B}	
presented a homogeneous catalogue of 7993 planet-hosting stars (3248 from Kepler, 565 from K2, 4180 from TESS), and their total of 9324 transiting planets. 
They used isochrone fitting and Gaia DR3 parallaxes, photometry, and metallicities to compute \teff, $\log g$, masses, radii, stellar densities, luminosities, ages, distances, and $V$-band extinctions, finding residual scatter (compared to interferometry and asteroseismology) of 2.8\%, 5.6\%, 5.0\%, and 31\% between their \teff, radii, masses, and ages and those in the literature. 
They determined radii for 4281 Kepler, 676 K2, and 4367 TESS planets, with the `planet radius gap' being less prominent in the K2 and TESS samples than in the \mbox{Kepler} sample alone. And they identified a clear radius inflation trend in their large sample of hot Jupiters.

A number of studies are making use of the Galactic motions of exoplanet host stars.
\citet{2023AJ....165..262Z} 	
used Gaia DR3 proper motions and radial velocities to show that stars making large vertical excursions from the Galactic plane host fewer super-Earths and sub-Neptunes. This sort of study has led to further insights into planet occurrence as a function of Galactic kinematics and stellar age
\citep{2024arXiv240910485B,	
2025ApJ...979..120H}.		

One of the areas in which Gaia is contributing, and which illustrates the complex foundations upon which formation theories depend, is that of the origin of the perplexing class of `hot Jupiters'. These are massive gas giant planets, somewhat like Jupiter or Saturn, but which orbit surprisingly close to their host star, typically within about 0.1\,au (Mercury orbits at just under 0.4\,au), corresponding to extremely short periods of around 3--9\,d.
For reasons related to the condensation of volatiles versus temperature, {\it in situ\/} formation of these hot Jupiters seems implausible, and the consensus is that they must have arrived at their present locations from their region of formation beyond the snow line (beyond around 3\,au), where solid material is inferred to be abundant in the form of condensed ices. 
In turn, there are two main hypotheses as to how they arrived at their present locations: either via disk-driven inward migration, or through the generation of a highly eccentric orbit with small pericentre, followed by its tidal damping and orbital circularisation. And there are at least three plausible mechanisms which might generate such high-eccentricity orbits: planet--planet encounters leading to orbit `scattering', resonant orbit changes due to a distant stellar companion (the Lidov--Kozai mechanism), or as a result of close stellar encounters, perhaps in a young star cluster.

An early approach to this problem used the Gaia DR2 data to identify old co-moving stellar groups around exoplanet host stars, finding that system architectures have a strong dependence on local stellar clustering in position--velocity `phase space'
\citep{2020Natur.586..528W}.	
These authors found that hot Jupiters preferentially exist in phase-space overdensity regions (for example, generated by resonances and instabilities driven by the Galaxy's central bar or its spiral arms, or by the passage of satellite galaxies), suggesting that their extremely `tight' (short-period) orbits arise from environmental perturbations, rather than from inward migration or planet--planet scattering. Later work has further developed this sort of picture
\citep{2021AJ....162...46D,	
2021arXiv210906182K,
2021ApJ...911L..16L,		
2022MNRAS.509.1010R,		 
2021ApJ...913..104R}.		 

\subsubsection{Habitability and SETI}

The search for habitable planets, and for life on other worlds, is a large and growing discipline, particularly enabled by the space observatories Kepler (2009--2018), the Transiting Exoplanet Survey Satellite (TESS, launched in 2018), and James Webb Space Telescope (launched in 2021). 
With the general consensus among biologists that carbon-based life requires water for its self-sustaining chemical reactions, the search for habitable planets has focused on identifying environments in which liquid water is stable, preferably over billions of years. The presence of liquid water in turn implies that the atmospheric pressure and ambient temperature will be in ranges that promote a rich variety of organic reactions.
Amongst a vast literature on exoplanet habitability, the 
\href{https://phl.upr.edu/hwc}{Habitable Worlds Catalog}, 
which replaced the Habitable Exoplanets Catalog in January 2024, today lists some 70 potentially habitable worlds out of nearly 6000 known exoplanets. Of these, 29 are more likely to be rocky planets capable of surface liquid water, while the other 41 might include water worlds or mini-Neptunes, with a lower likelihood of habitable conditions. 
Somewhat parallel efforts include the Catalog of Habitable Zone Exoplanets
\citep{2023AJ....165...34H},
and the TESS Habitable Zone Catalog
\citep{2021AJ....161..233K}. 
All these studies use Gaia distances and broadband magnitudes to determine stellar effective temperatures, radii, and masses, as part of their candidate selection. 

In the same way that an exoplanet orbit must be aligned with the observer's line-of-sight to experience a transit, only a subset of favourably located stars in our region of the Galaxy could see Earth as an exoplanet, transiting slowly across the face of the Sun. The Earth's `transit zone' is simply a band around the Earth's ecliptic plane, projected onto the celestial sphere. Various studies have used Gaia to identify such stars
\citep{2016AsBio..16..259H,	
2018MNRAS.473..345W,		 
2020MNRAS.499L.111K}; 	
the latter authors used Gaia DR2 to identify more than 500 stars within 100\,pc that are in a position to observe a minimum 10-hr long observation of Earth's transit across the face of our Sun. 
And as part of its extended mission, TESS has also searched for transiting planets in the ecliptic region that could already have found life on our transiting Earth
\citep{2021AJ....161..233K}.

In October 2019, 
\href{https://breakthroughinitiatives.org/news/27}{\tt Breakthrough Listen}
began a collaboration with TESS scientists to look for signs of advanced extraterrestrial life. This includes expanding their original target list (adding more than a thousand `objects of interest' identified by TESS, and aided by Gaia), and refining their analysis strategy, for example, by including knowledge of planet alignments to predict when transmissions are more likely to occur
\citep{
2020MNRAS.498.5720W,
2023PASP..135c4201D,
2023AJ....165..114Z}.

A number of studies more explicitly related to SETI are making use of the Gaia data
\citep{2024AcAau.225....1W}. 	
The famous `Wow!'\ signal of 15 August 1977, detected at Ohio State University’s `Big Ear' radio telescope, was a strong 72-second anomalous signal at 1420~MHz -- the emission frequency of neutral hydrogen, which physicists Philip Morrison and Giuseppe Cocconi speculated might be the preferred medium of extra-terrestrial communications.
Observatory director John Kraus described it in a letter to astronomer Carl Sagan: {\it `The Wow signal is highly suggestive of extraterrestrial intelligent origin, but little more can be said until it returns for further study'}.
Kraus and others searched for stars that could be the source of the signal, writing {\it `We checked star catalogs for any Sun-like stars in the area and found none'}.
Today, the Gaia catalogue is far more extensive than the star catalogue that Kraus had access to in the 1970s. Amateur astronomer Alberto Caballero used Gaia DR2 to identify a possible stellar origin: out of 66 G~and K~stars in the likely sky region, one of them has sufficient information to infer that it closely resembles the Sun, i.e.\ the same temperature, radius, and luminosity. The object, 2MASS 19281982--2640123, lies in the Sagittarius constellation, at a distance of 550\,pc
\citep{2022IJAsB..21..129C}.
It is not yet known whether this, or other stars in this part of the sky, are accompanied by planets. Other explanations have recently been proposed
\citep{2024arXiv240808513M}.	

Various other phenomena, probably with a purely physical explanation, have aroused interest as possible alien techno-signatures. Amongst these are the interstellar object Oumuamua (Section~\ref{sec:interstellar-objects}), and the curious `Boyajian's star' (KIC--8462852), discovered in the Kepler data from its unusual light-curve 
\citep{2016MNRAS.457.3988B}.
The star shows dimming up to 20\%, lasting between \mbox{5--80\,days}, and with an irregular cadence and unusual profile. Considerable speculation accompanied the unusual light curve, with interpretations ranging from occulting clouds of exocomets, to a `swarm of alien megastructures' -- and hence an outstanding SETI target
\citep{2016ApJ...816...17W}.
Searches for other stars showing this sort of behaviour include a study using Gaia EDR3, which revealed a number of similar sources occupying a restricted region of the Hertzsprung--Russell diagram, and relatively close on the sky
\citep{2022AJ....163...10S}.
Boyajian’s star, incidentally, lies at a Gaia~DR3 distance of $444\pm2$\,pc, or $2.2545\pm0.0099$~mas. The author concluded that {\it `the dippers in the clump and other stars in the same region would be appropriate targets for SETI searches'}.

In early discussions of SETI, Soviet astronomer Nikolai Kardashev developed the idea of the `Kardashev scale' as a method of measuring a civilisation's level of technological advancement, based on the amount of energy it is able to use 
\citep{1964SvA.....8..217K}. 
The classification rests on the premise that any civilisation's energy needs grow continuously with time. His Type~II civilisation is one that has developed to the point that it requires, and can harness, most of the energy emitted by its parent star. And one possible manifestation of such a civilisation could be the presence of a (hypothetical) megastructure around the host star, usually referred to as a `Dyson sphere' as later popularised by physicist Freeman Dyson 
\citep{1960Sci...131.1667D,
2020SerAJ.200....1W}.
Practical searches began with the launch of the infrared IRAS satellite in 1983
\citep{1985IAUS..112..315S,
2000AcAau..46..655T,
2004IAUS..213..437J,
2009ApJ...698.2075C}, 
and continued with searches using Kepler, Spitzer, WISE and others
\citep{2005ApJ...627..534A,
2014AcAau.105..512T,
2014ApJ...792...26W,
2021MNRAS.507.3761C}.
In all this work, a continuing difficulty has been to distinguish anomalous Galactic stars from the more ubiquitous extragalactic infrared (AGN) sources. Instead of searching for sources with an infrared excess, Gaia is contributing by searching for under-luminous stars (due to the `missing' luminosity intercepted by the Dyson sphere) as a result of their discordant trigonometric and spectroscopic parallaxes
\citep{2015ApJ...810...23Z,
2016ApJ...816...17W}.
A number of Gaia-based searches are now being undertaken, with a handful of `interesting candidates' identified
\citep{2018ApJ...862...21Z,
2022ApJ...924...78H,
2022MNRAS.512.2988S,
2024MNRAS.531..695S}.

Finally in this section, I will mention some search strategies which benefit from the availability of high-accuracy astrometry, given that SETI is confronted by multiple dimensions of potential search space. Accordingly, searches for (intentional or accidental) signals clearly benefit from insights as to what wavelengths and types of signal to search for, along with where (and when) to look. And when transmitting a signal over large distances, it may well be more efficient to send a brief beamed signal than one which is continuous and omnidirectional, but this again requires that the receiver can figure out where and when to look.
The problem can be simplified if the signal transmission and reception are considered of mutual interest to both parties. In that case, appeal can be made to `game theory' to formulate a cooperation in which, although the players cannot communicate, they can still establish strategies which are superior to purely random searches. The idea is to establish 
`focal points', aka `Schelling points' or `beacons', as preferred search locations in space or time
\citep{2018haex.bookE.186W}.		 
Again, the topic has an extensive literature, with various astronomical beacons having been suggested as temporal markers, involving synchronisation with supernovae, neutron star mergers, or $\gamma$-ray bursts. The latter, for example, provide plausible natural `synchronisers' because of their luminosity, occurrence rate, isotropic distribution, large distance, and short duration.

The availability of the Hipparcos astrometry in 1997 generated renewed interest in this possibility. 
In one approach, a transmitting civilisation propagates an `announcement signal' downstream (i.e., $\sim180^\circ$) away from a $\gamma$-ray burst, immediately upon receiving the burst signal
\citep{1999PASP..111..881C}. 
The searching civilisation (Earth) then monitors (angularly proximate) upstream target stars for a synchronised emitted signal at the time lag set by the additional path length.
In this configuration, a planet-hosting star 20\,pc from Earth, and at an angle of $1^\circ$ from the direction of a burst event seen from Earth, will transmit an event which would be detected on Earth 3.63\,d later, and with an arrival time that can be predicted to $\pm1.8$\,h. The timing uncertainty drops to just 60\,s in the case of $10\,\mu$as Gaia-level parallaxes. For a number of hypothesised transmitting stars, the errors on these time delays remain modest, below 1~day, for offset angles up to $5^\circ$.
A generalisation of this approach exploits the concept of a `SETI ellipsoid'
\citep{1994Ap&SS.214..209L}.	
Several studies using the Gaia data, mostly using the supernova SN~1987A as primary beacon, and some sifting through 10~million stars from the Gaia DR3-based variable star catalogue, then searching for `technosignatures' as modulations of the variability parameters, are now being carried out
\citep{2022AJ....164..117D, 	
2024AJ....167..101C,		
2023AJ....166...79N}.		

\subsection{High-velocity stars}

Very broadly, disk stars in the solar neighbourhood move with respect their neighbours with relative space velocities of around 10\kms. A very small fraction have much higher velocities, and sufficiently distinct from the overall stellar velocity distribution, that they must have been imparted by specific formation processes. 
`Runaway stars' have velocities above about 30\kms. There are two main theories for their origin, and both appear to operate: the binary-supernova scenario, BSS, and the dynamical ejection scenario, DES, in which a star is ejected via dynamical interactions in a young, compact cluster 
\citep{2016A&A...590A.107O,
2020ApJ...903...43D}.
`Hypervelocity stars', moving with space velocities of 500--1000\kms\ or more, cannot be explained by either the BSS supernova or DES ejection mechanism, and point to an even more extreme formation scenario.

\subsubsection{Runaway stars}
\label{sec:runaway-stars}
The discovery of runaway stars followed, more than 60~years ago, from the realisation that many young massive O~stars 
\citep{1961BAN....15..265B, 1986ApJS...61..419G, 1987ApJS...64..545G},
Be~stars, and Wolf--Rayet stars, the final He-burning phase for initial mass $\gtrsim25M_\Sun$
\citep{1980A&A....85..201M, 1998A&A...331..949M}
lie well outside their likely birth places in open clusters and OB~associations, with some velocity vectors `pointing back' to their likely origin. 
The conventional definition of a `runaway' adopts a 30\kms\ 3-d space velocity threshold \citep[e.g.][]{1986ApJS...61..419G}, while unbound stars with a lower velocity are frequently referred to as `walkaways'. 
And they are to be distinguished from `fast halo stars', which often simply reflect the different rotation states of the Galaxy disk and halo
\citep{2021ApJS..252....3L}.

In the binary-supernova scenario, runaway stars result from the evolution of massive binary systems \citep{1973NPhS..242...71V,1973NInfo..27...86T}. 
Simply stated, the more massive star evolves fastest, while wind-driven mass-loss (assisted by Roche-lobe overflow in the closest massive binaries) makes O~stars evolve into lower-mass (and themselves evolving) Wolf--Rayet stars. At the end of the Wolf--Rayet phase, in both cases, the star explodes as a supernova. 
If the first supernova explosion is symmetric, the binary system remains bound, leading to the class of high-mass X-ray binaries. If the first supernova is asymmetric, the binary system may disrupt, depending on the magnitude and direction of the extra kick velocity \citep{1982ASSL...98..417D}.  In either case, with the supernova explosion duration being very short compared to the orbital period, the star receives a recoil velocity, and becomes a runaway, with velocities reaching 200\kms\ for the `tightest' and most massive pre-supernova binaries. 
If the system has not already separated after the first supernova, the system will normally become unbound as a result of the second, producing two high-velocity, single pulsars. In rare cases, the binary can survive this second supernova, producing a binary pulsar 
\citep{1985ASSL..120..207D,1998A&A...331..949M}.

Runaway stars have an important role in the context of $\Lambda$CDM cosmology. Certain inconsistencies in large-scale simulations have been attributed to challenges in modelling the `baryon cycle', i.e., how galaxies accrete and expel their gas
\citep{2017ARA&A..55...59N}.		
Stellar feedback (in which stars influence the surrounding environment through the injection of energy, momentum, and mass), is an important factor, involving contributions from protostellar jets, stellar winds, supernovae, and ionising radiation.
Runaway stars can travel hundreds of parsecs into the low-density interstellar medium, depositing momentum and energy, and affecting supernova rates and gas densities
\citep{2009ApJ...695..292C,	
2020MNRAS.494.3328A,		
2023MNRAS.526.1408S,		
2023MNRAS.521.2196A,		
2024A&A...681A..28A}.	

Pre-Hipparcos studies of runaway stars were based largely on radial velocities, while Hipparcos provided higher accuracy proper motions to allow the orbits of many local O and B stars to be traced back to their likely birthplaces in nearby associations and young clusters 
\citep[e.g.][]{1998A&A...331..949M, 2000ApJ...544L.133H, 2001A&A...365...49H}.
Today, studies make use of Gaia's much improved distances and space motions to establish the origin of thousands of runaway stars, with the goal of distinguishing between the BSS/DES mechanisms, and their relation to known supernovae, supernova remnants, and high-mass \mbox{X-ray} binaries.
Several dozen studies, starting with Gaia DR2 and later making use of Gaia EDR3/DR3, are shedding light on the frequency and nature of ejection events in young clusters, including runaways originating in the dense Orion Trapezium nebula, and in the Large and Small Magellanic Clouds
\citep[e.g.][]{2020MNRAS.495.3104S,2018A&A...616A.149M,2018A&A...619A..78L,2019ApJ...884....6M,2020ApJ...903...43D,2020ApJ...900...14F,2022A&A...663A..39B,2021A&A...645A.108R,2023A&A...679A.109C,2022MNRAS.517.1946K,2018AJ....156..265M,2021AstL...47..224B}.
\citet{2023A&A...679A.109C},
for example, used DR3 astrometry to identify 106 O-type (42 new) and 69~Be-type (47 new) runaways. 
Together, some 200 or more new runaway stars have been discovered with Gaia, and the origin of many identified. Dynamical ejections appear to dominate in some cases.

Runaway stars often create spectacular `bow shocks' as they plough through the interstellar medium, a number having been imaged in the radio, as well as by Spitzer, HST--ACS, JWST, and WISE
\citep{
2015A&A...578A..45P,	
2016ApJS..227...18K,	
2018A&A...616A.149M,	
2022A&A...663A..80M}.	
In a sample of O and Be runaways selected on the basis of Gaia DR3, for example, 
\citet{2025A&A...694A.250C}	
used WISE infrared imaging to demonstrate a clear relation between the bow shock morphology and the object's space motion determined by Gaia (their Fig.~4).

\paragraph{Runaway stars in R136}

Gaia is providing some particularly interesting insights into runaway stars associated with the massive young cluster R136 in the Large Magellanic Cloud, which comprises some 100 O~and Wolf--Rayet stars within 5\,pc (20~arcsec) of the cluster centre
\citep{1998ApJ...493..180M,	
2013A&A...558A.134D}.
The first two runaways associated with R136 were provisionally identified using radial velocities from the VLT--FLAMES Tarantula Survey
\citep{2011A&A...530A.108E, 	
2010ApJ...715L..74E,		
2012ApJ...746...15B},		
both of which were subsequently confirmed from their high proper motions in Gaia~DR2
\citep{2018A&A...619A..78L,	
2019MNRAS.482L.102R}.		
Another 10 candidates, extending out to \mbox{35--400\,arcsec} (10--100\,pc), were identified from two-epoch (2011--14) HST observations
\citep{2018AJ....156...98P},	
also later confirmed by Gaia
\citep{2021NewA...8201455G,	
2024NewA..10602128T}.		
A major advance in characterising the runaway population came with Gaia DR3
\citep{2024Natur.634..809S}.	
Starting with 80\,000 sources with $G<18$\,mag and within $1^\circ$ of R136, these authors identified 55 massive runaway stars, reaching projected distances of up to 460\,pc, and increasing the number of known runaways escaping from the cluster core by an order of magnitude. 
They classified the runaways as dynamical ejections falling into two distinct groups. The first comprises massive stars ejected isotropically (with kinematic ages peaking around 1.8\,Myr) which, they show, are consistent with dynamical interactions during and after the birth of R136. The second group comprises 16 stars with kinematic ages around 0.2\,Myr, ejected preferentially in a northerly direction, which they attribute to the effects of a later cluster interaction. 
Theorists have been swift to added such a `sub-cluster ejection scenario' as a plausible cluster ejection mechanism in addition to BSS and DES
\citep{2024A&A...690A.207P}.	

\subsubsection{Hypervelocity stars}
\label{sec:hypervelocity-stars}

Even more extreme velocities than runaway stars, in a related but more extreme environment, were predicted by \citet{1988Natur.331..687H}. He showed that a close encounter between a tightly bound binary, and a $10^6M_\Sun$ black hole, causes one component to bind to the black hole, with the other ejected at up to 4000\kms (i.e.\ 1\% of the speed of light).
The first such `hypervelocity star', HVS1, was discovered, from SDSS, to have a radial velocity of $\sim$800\kms\ \citep{2005ApJ...622L..33B}. At a distance of $\sim$55~kpc, 30~kpc above the disk and 60~kpc from the Galactic centre, its space motion was consistent with Hills-type ejection from the $4\times10^6M_\Sun$ black hole (Sgr~A*) at the Galaxy centre \citep{2015ARA&A..53...15B}.
 
Following this discovery, a wide-field spectroscopic survey at the MMT has discovered a few dozen other bound and unbound candidates from their extreme radial velocities, with masses in the range $2.5-4M_\Sun$, and at distances varying between 50--100\,kpc from the Galactic centre. 
Other surveys have been made with the Chinese LAMOST telescope
\citep{2017ApJ...847L...9H},	
and a number followed-up with the Hubble Space Telescope 
\citep{2015ApJ...804...49B}.	
As a result of the enormous distances travelled on their journey through the Galaxy halo, in different directions and now at a vast range of distances from their origin, hypervelocity stars probe the Galaxy's gravitational potential, making them possible tracers of the matter distribution in the Milky Way 
\citep{2005ApJ...634..344G,	
2006ApJ...651..392S}.		

Amongst objects likely originating close to the Galaxy centre are HVS~22, with a velocity of some 1500\kms. In contrast, HVS3, with a velocity $>800$\kms, appears to come from the Large Magellanic Cloud
\citep{2018A&A...620A..48I,	
2019MNRAS.483.2007E}.		
This large velocity, consistent with the Hills mechanism, suggests that the LMC itself harbours a massive black hole of at least $4\times10^3-10^4M_\Sun$.
Other discoveries, many from Gaia, appear to have different origins, suggesting that other extreme ejection mechanisms might also be at work. In addition to the tidal disruption of binary stars by a massive black hole in the Galactic centre, these include
single star encounters with a binary black hole 
\citep{2003ApJ...599.1129Y,2006ApJ...651..392S,2007MNRAS.379L..45S}; 
interaction between a globular cluster with a single or binary black hole 
\citep{2016MNRAS.458.2596F};
three-body processes in globular clusters
\citep{2023ApJ...946..104W, 	
2023ApJ...953...19C};	
and the `dynamically driven double-degenerate double-detonation' (D6) mechanism
\citep{2018ApJ...865...15S}.

Since the DR2 availability in April 2018, Gaia is providing proper motions and distances of previously known candidates with unprecedented accuracies, helping to pinpoint their origin and quantifying their production rate, while also searching for other examples. 
Amongst the first discoveries from Gaia DR2, some were believed to originate from the Galactic centre, while others are not 
\citep{2018AJ....156...87L,		
2019ApJS..244....4D,		
2019A&A...627A.104D, 	
2019MNRAS.490..157M,		
2021ApJS..252....3L}.	 	
Discoveries included three hypervelocity white dwarfs, conjectured to be the companions to primary white dwarfs that exploded as Type~Ia supernovae 
\citep{2018ApJ...865...15S}.		
The orbit of at least one can be traced back to an old supernova remnant, strengthening the idea that some hypervelocity stars may originate from extreme binary supernova ejection events 
\citep{2019MNRAS.489..420R}.	
Subsequent searches have used EDR3
\citep{2021MNRAS.503.1374M,	
2021A&A...646L...4I,		
2023AJ....166...12L},		
and most recently DR3
\citep{2022MNRAS.515..767M,	
2023ApJ...944L..39L,	
2023OJAp....6E..28E,	
2023MNRAS.518.6223I, 	
2023RNAAS...7..267P,	
2024MNRAS.533.2747V}.	

Hypervelocity stars, whose space trajectories have been tracked by Gaia and which have been considered likely to originate from {\it beyond\/} the Galaxy, include 
the aforementioned HVS3 from the LMC \citep{2019MNRAS.483.2007E},
as well as others from the LMC 
\citep{2023ApJ...952...64L}; 		
and others from the Sagittarius dwarf spheroidal galaxy
\citep{2019MNRAS.490.5647M,	
2022ApJ...933L..13L}.			 
Models of hypervelocity stars generated in Andromeda suggest that their number, in {\it our\/} Galaxy, might be between 10--4000
\citep{2024MNRAS.529.3816G}.

\subsection{Gravitational microlensing}
\label{sec:microlensing}
%

\paragraph{Context}
Under certain conditions, light rays from a distant object (the source) are bent by the gravitational potential of a foreground object (the lens) to create source images which are distorted (and possibly multiple), and which may be highly focused and hence significantly amplified. Its manifestation depends on the fortuitous alignment of the background source, intervening lens, and observer. 
Different regimes are generally recognised, depending on whether effects are discernible at an individual object level ({\it strong lensing\/}), or only in a statistical sense ({\it weak lensing\/}). Strong lensing can be further divided, somewhat subjectively according to telescope resolution, into {\it macrolensing\/} (resulting either in multiple resolved images, or in `arcs' in which the source is both sheared and magnified) and {\it microlensing\/} (in which discrete multiple images are unresolved). 
If the foreground lens is of complex structure (whether a cluster of galaxies, a binary system, or a star orbited by one or more planets), the background source may show a more complex light curve resulting from the time-varying magnification as the alignment changes. 

Microlensing events have (almost exclusively) been discovered from repeated long-duration {\it photometric\/} monitoring, mainly from dedicated ground-based surveys
(e.g.\ MOA
\citep{2008ExA....22...51S},
OGLE 
\citep{2009ASPC..403..110U}
and KMT
\citep{2016JKAS...49...37K})
and, since 2013, with Gaia. Once a microlensing candidate is discovered, generally from its rising light curve, the goal is to initiate a rapid and dense follow-up sampling of the detailed light curve while the lensing event is still ongoing. The aim is then to characterise the nature and geometry of the event including, for example, identifying light-curve perturbations signalling the existence of an orbiting planet.
The rarity of such alignments has motivated ground-based surveys to focus on the Galactic bulge, where alignment probabilities benefit from much higher stellar surface densities. 

Tens of thousands of (photometric) microlensing events have been discovered in the past 30~years. Amongst these, and out of some 6000 exoplanets discovered in total, some 200 have been found from microlensing.  
All but 3 of these have been by the ground-based surveys, and all but one of the previous microlensing planets (TCP~J0507+244, aka Kojima--1) lie in the direction of the Galactic bulge. 
The literature is substantial, with various reviews covering 
microlensing in general 
\citep[e.g.][]{
1996ARA&A..34..419P,		
2006glsw.conf..453W,		
2015IJMPD..2430020R},		
and on planetary aspects in particular 
\citep[e.g.][]{
2005NewAR..49..424G,
2012ARA&A..50..411G,
2013RPPh...76e6901G,
2016ASSL..428..135G}.		

Gaia's microlensing discoveries are coming from the variability analysis of the multi-colour multi-epoch photometry (Section~\ref{sec:variable-stars}, with the specific Gaia processing for these microlensing events described by
\citet{2023A&A...674A..23W}), 
with others from the `science alerts pipeline' (Section~\ref{sec:science-alerts}). 
%
Gaia~DR3 includes around 400 photometric microlensing events identified in the variability analysis (Section~\ref{sec:data-releases} and Table~\ref{tab:data-release-table2}), while the science alerts pipeline has identified around~40 (Section~\ref{sec:science-alerts}).

For the microlensing events of relevance to Gaia, the source is typically a star at 5--8\,kpc, and the primary lens is a single point mass of order $1M_\Sun$, at a distance 1--2\,kpc. The two (magnified) images of such a background source then have a separation of order 1\,mas, well below typical telescope resolution. 
Gaia is contributing in several important ways, detailed below:
(a)~it provides the possibility of measuring the {\it astrometric\/} displacement of the image centroid, as well its photometric magnification;
(b)~the full-sky survey means that discoveries are not restricted to the Galactic bulge. This has led to the  discovery of only the second non-bulge direction microlensed exoplanet Gaia22dkv, and the remarkable long-duration multi-peaked Gaia19dke, both discovered from the science alerts pipeline, and both described further below;
(c)~it can provide the proper motion of the foreground object in these lensing events, which can be used to characterise the lens star and, under certain circumstances, can lead to the determination of the lens mass, and hence the unique possibility of determining the mass of an isolated star;
and 
(d)~the accurate large-scale proper motions can be used to {\it predict\/} future microlensing events. 

\paragraph{Astrometric microlensing}
\label{sec:astrometric-lensing}

In addition to the {\it photometric\/} manifestation of microlensing, time-varying magnification of the unresolved images also leads to a small time-varying displacement of their photocentre, typically by a fraction of a milliarcsec. Several studies of the effect and its detectability were made before the launch of Gaia, with one objective being to measure the lens-star mass
\citep{
1995A&A...294..287H,	
1995AJ....110.1427M,	
1996ARA&A..34..419P,	
1996ApJ...470L.113M,	
1996AstL...22..573S,		
1998ApJ...502..538B,	
1999MNRAS.304..845H, 	
2016ApJ...823..120N,	
2000ApJ...534..213D, 	
2002ApJ...568..717H, 	
2017IJMPD..2641015N}.	

Given that the latest Gaia data release, DR3, provides only the mean astrometry for each source, and does not yet include the individual multi-epoch astrometric measurements, specific Gaia astrometric microlensing results are not yet available. 
Gaia should also be able to detect astrometric lensing {\it independently\/} of any photometric signature
\citep{2000ApJ...534..213D}. 	
Pre-launch, 
\citet{2002MNRAS.331..649B} 	
estimated that there will be 1300 photometric microlensing events for which Gaia will measure at least one data point on the amplified light curve, with some 25\,000 sources having a significant variation of the centroid shift, and 2500 for which the mass can be recovered with an error of better than 50\%. The high-quality events are dominated by disk lenses within a few tens of~pc, and source stars within a few hundred~pc. 
Predictions for black hole and neutron star events are given by
\citet{2024MNRAS.531.2433S}. 
Astrometric microlensing specifically related to planet detection has also been investigated
\citep{
1999ApJ...522..512S, 	
2002MNRAS.329..163H, 	
2002ApJ...564.1015H, 	
2002ApJ...573..825A,	
2003ApJ...597.1070H},	
Although the planet's astrometric perturbation is short in duration, the amplitude can be large for Jovian planets, the duration for $\delta\theta>10\,\mu$as being of order days. 

One astrometric event {\it has\/} been measured. Based on predicted alignment events by 
\citet{2011A&A...536A..50P},	
the effect was first measured with HST by 
\citet{2017Sci...356.1046S}. 
From positional measurements at 8~epochs between 2013--15, they measured a 2~milli-arcsec shift in the position of a background star as the nearby (52\,pc) white dwarf Stein~2051B passed in front: Gaia astrometry was used both for the event {\it prediction}, and the reference frame construction. Models gave a white dwarf of $M=0.67\pm0.05M_\Sun$, and further confirmation of the physics of degenerate matter.

\paragraph{Event prediction}
The Gaia astrometry can be used to {\it predict\/} future astrometric microlensing events by identifying foreground high-proper motion stars which will pass by a (projected) background star in the coming months or years
\citep{2011A&A...536A..50P,	
2012ApJ...749L...6L}. 	
The fact that a high proper motion star crosses a large area of sky over a relatively short time provides more opportunities for chance alignment, especially in the Galactic plane. High proper motion stars are also preferentially nearby, with a relatively large $\theta_{\rm E}$ and hence large astrometric cross section.

From a list of 148\,000 stars from Gaia DR2 with proper motions larger than 150\masyr,
\citet{2018A&A...615L..11K}	
searched for closest approached of background stars close to their paths, calculating dates, separations, astrometric shifts, and event magnifications. They detected ongoing events by two nearby high-proper motion stars. Luyten~143--23 had a predicted closest separation of 108~mas in July 2018 with a shift of 1.7~mas, and another in March 2021 with a closest separation of 280~mas, and an expected shift of 0.7~mas. Ross~322 had a predicted separation of 125~mas in August 2018, and an expected shift of 0.7~mas.  Although the first of the Luyten events was not observed as part of the Gaia sky scanning, the other two were. Results are (I believe) awaited.

Other such predictions have been made
\citep[e.g.][]{2018A&A...617A.135M,	
2018MNRAS.478L..29M,		
2020MNRAS.498L...6M}.		
All 1.7~billion stars in GDR2 were searched by 
\citet{2018A&A...618A..44B},	
who predicted 76 microlensing events between July 2014 and July 2026. Nine of these are caused by the white dwarf LAWD~37, and another five by the white dwarf Stein 2051~B. As noted above, the latter event was actually measured, using HST, by \citet{2017Sci...356.1046S}.
\citet{2024MNRAS.527.1177S}		
used Gaia DR3 to predict a total of 4500 astrometric microlensing events between J2010.0 and J2070.0 with displacements greater than 0.1~mas.
The method also has applicability to more exotic lens types, including pulsars
\citep{2018ApJ...866..144O},	
and black holes
\citet{2018MNRAS.476.2013R}.

\paragraph{Gaia discoveries}

Amongst Gaia's microlensing events, one (discovered from the science alerts processing) has been suggested to be a possible isolated stellar-mass black hole, Gaia18ajz (Section~\ref{sec:isolated-black-holes}).
I will say more on three others here, also discovered from the science alerts processing:
Gaia22dkv, a bright event whose astrometric motion may be detectable with Gaia~DR4;
Gaia19dke, a long-duration event, with prominent {\it multiple\/} peaks arising from a {\it single\/} unresolved lens;
Gaia19bld, with rotating arc-like sub-images.

\begin{figure}[t]
\centering
\includegraphics[width=0.28\linewidth]{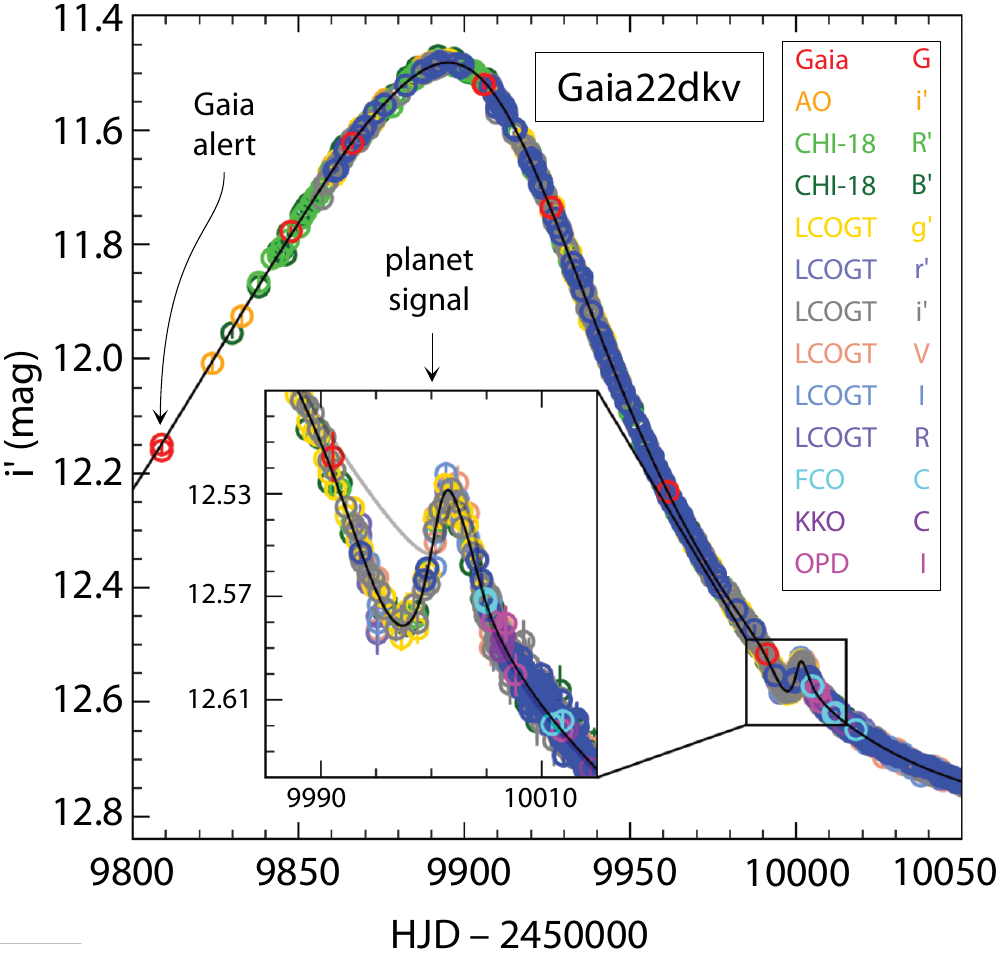}
\hspace{10pt}
\includegraphics[width=0.27\linewidth]{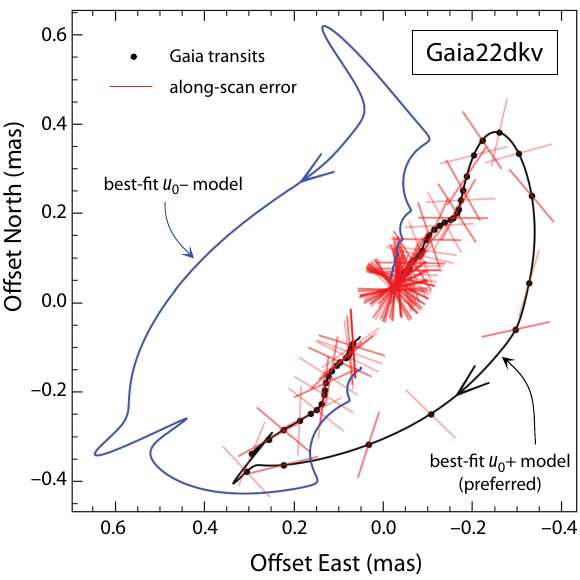}
\hspace{10pt}
\raisebox{2pt}{\includegraphics[width=0.38\linewidth]{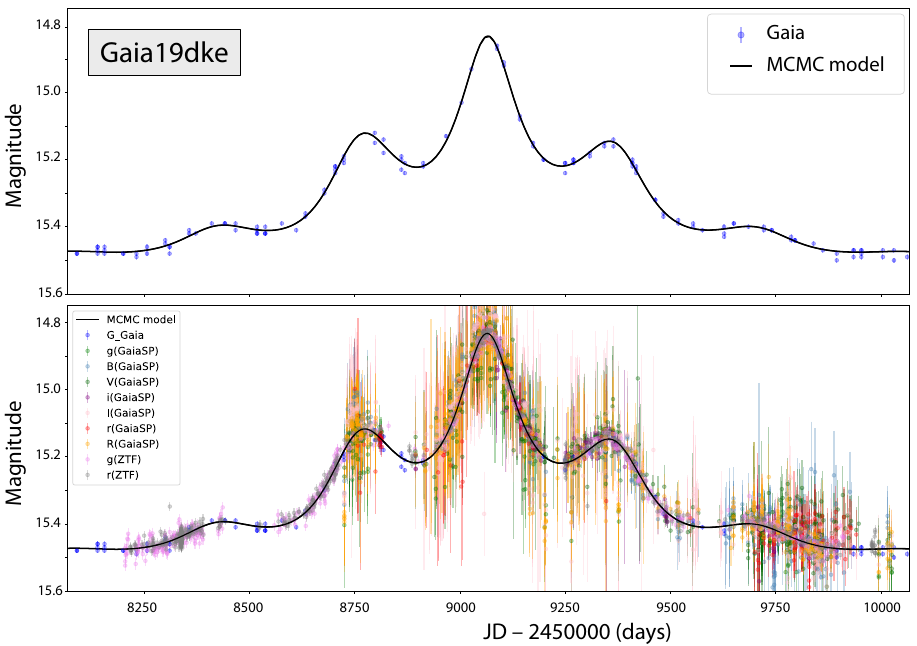}}
\caption{
Two Gaia science alerts microlensing events.
Left: the exoplanet system Gaia22dkv, showing follow-up photometry over 250~days, from several telescopes (all transformed to~$i^\prime$), with the model fit superposed (\citet{2024AJ....168...62W}). 
Middle: the host star's simulated astrometric motion (black curve), which may be measurable when the Gaia epoch astrometry is available (\citet{2024AJ....168...62W}).
Right: the multi-peak event Gaia19dke, actually due only to a single unresolved lens moving with similar transverse velocity to the source (\citet{2024AcA....74...77M}).	
}
\label{fig:gaia22dkv}
\end{figure}

\href{http://gsaweb.ast.cam.ac.uk/alerts/alert/Gaia22dkv}{Gaia22dkv}
is a bright, Galactic plane source with historic magnitude $G=13.18\pm0.02$, which brightened by 0.6~mag to an `alert magnitude' of $G=12.53$~mag. Subsequent observations, and the complex analyses that accompany these microlensing events (Figure~\ref{fig:gaia22dkv}a), found that the lens is a $M_{\rm L}=1.15M_\Sun$ star at a distance $D_{\rm L}=1.27$\,kpc, orbited by a planet of mass $M_{\rm p}=0.59M_{\rm J}$ with a projected orbital separation $a=1.41$\,au, and orbital period $P=2.96$\,yr
\citep{2024AJ....168...62W}. 
The host star and planet masses are typical of exoplanet systems, including microlensing discoveries. But three other properties make it more unusual: the planet lies within the `snowline' where liquid water might be present, it is an unusually bright host star, making it especially suitable for radial velocity follow-up, and it lies in a non-bulge direction. The latter, a discovery enabled by Gaia's all-sky scanning, is important for characterising the exoplanet population as a function of Galactic location.
Furthermore, based on these system parameters and the known epochs of the Gaia measurements
\citep{2024AJ....168...62W},	
simulations suggest that the {\it astrometric\/} motion of the host star should be measurable from the Gaia epoch astrometry which will be made available in future data releases (Figure~\ref{fig:gaia22dkv}b, black curve). This will constrain the microlens parallax, $\varpi_{\rm E}$, break the current ($u_0+/-$) model degeneracy, and so directly yield the lens mass and distance.

Many complexities enter the characterisation and modelling of microlensing events 
\citep[e.g.][]{2024arXiv240706689R}.		
Two key parameters are the angular `Einstein radius' of the event, $\theta_{\rm E}\propto M_{\rm L}^{1/2}$, and the associated linear Einstein radius, $R_{\rm E}=\theta_{\rm E} D_{\rm L}$, where $M_{\rm L}$ and $D_{\rm L}$ are the lens mass and distance. 
The total magnification of an event varies with time due to the relative transverse motion between source, lens, and observer. 
For a relative source--lens transverse velocity, $v_\perp$, a typical event time scale is then given by the Einstein radius crossing time, $t_{\rm E}=R_{\rm E}/v_\perp$. 
For a bulge source (at 8\,kpc), and a $1M_\Sun$ lens half way to the source, this `Einstein time' is around~35\,d.
Longer duration events are therefore either more massive, and/or with a small relative velocity between source and lens. 

The longest recorded event, \mbox{OGLE-1999-BUL-32} (MACHO-99-BLG-22), had $t_{\rm E}=640$\,d
\citep{2002MNRAS.329..349M}.
Long-duration events found in the Gaia alerts include
\href{https://gsaweb.ast.cam.ac.uk/alerts/alert/Gaia18cbf}{Gaia18cbf},
$t_ {\rm E}=491$\,d 
\citep{2022A&A...662A..59K}, 		
and
\href{https://gsaweb.ast.cam.ac.uk/alerts/alert/Gaia19dke}{Gaia19dke},
$t_ {\rm E}=159$\,d \citep{2024AcA....74...77M}.
Figure~\ref{fig:gaia22dkv}c shows the $G$-band photometry for Gaia19dke over five years (other photometry, including ZTF, is shown in the lower plot). 
%
Modelling shows that the event, with its prominent {\it multiple\/} peaks, is actually due only to a single unresolved lens, with mass $M_{\rm L}=0.50M_\Sun$ at distance $D_{\rm L}=3.1$\,kpc, and with the source distance $D_{\rm S}=4.9$\,kpc
\citep{2024AcA....74...77M}.
Given that long events arise from large $M_{\rm L}$ and/or small $v_\perp$, it follows that in the latter case, good source--lens alignment can extend over many months. 
%
The time scale of lensing events tends to increase with Galactic longitude \citep{2019ApJS..244...29M}, 
due to the fact that, away from the bulge, both lens and source are generally located in the disk, and are often therefore moving with similar transverse velocities.
Multiple peaks can then result from small variations in this general alignment as the Earth (and in this case Gaia) move in their annual orbit around the Sun
\citep{2002MNRAS.336..670S}.	
Such sources, being found in Gaia's all-sky scanning, are less common in surveys confined to the Galactic bulge.

My third example is
\href{http://gsaweb.ast.cam.ac.uk/alerts/alert/Gaia19bld}{Gaia19bld}.
This was announced in April 2019, as it brightened, at $G=14.44$\,mag. 
The Gaia light curve is well defined, with nearly 200 observations between 2014--23, and reaching a peak of $G=10.5$\,mag, three months after the alert was issued.
%
Photometry from the Las Cumbres Observatory established it as a microlens event, with spectroscopy showing features consistent with a K3 supergiant
\citep{2019ATel12948....1R}.
Following the first resolved imaging of a microlensing event, using the VLTI--GRAVITY interferometer for the ASAS-alerted non-bulge system TCP~J0507+2447		
\citep{2019ApJ...871...70D},			
the importance of Gaia19bld (with $t_{\rm E}\sim107$\,d) is that, for the first time, the rotating arc-like sub-images have been resolved using the ESO VLT--PIONIER interferometer, using baselines up to 128\,m
\citep{2022NatAs...6..121C}.	
%
From the image separation, they derived the angular Einstein radius $\theta_{\rm E}=0.0818\pm0.0020$~mas, in agreement with the Earth--Spitzer (microlens parallax) value $\theta_{\rm E}=0.0823\pm0.0018$~mas 
\citep{2022A&A...657A..18R}.
Combining $\theta_{\rm E}$ and the microlens parallax $\varpi_{\rm E}$ yields the microlens mass to unprecedented accuracy, $M_{\rm L}=1.147\pm0.029M_\Sun$. The model fits also provide distance estimates to both the source, $D_{\rm S}=8.4\pm1.5$\,kpc, and the lens, $D_{\rm L}=5.5\pm0.6$\,kpc.

\section{Galaxy structure and dynamics}
\label{sec:galaxy-structure}

The main structural components of the Milky Way Galaxy include the flat disk, its most conspicuous component, which contains some hundred billion stars of all types and ages orbiting the Galactic centre. The Sun is located about 8.5~kpc from the centre. The disk displays spiral structure, and also contains interstellar material, mainly atomic and molecular hydrogen, and a significant amount of dust. 
The inner~kpc of the disk also contains the bulge, which is less flattened, appears to contain a bar, and consists mostly of fairly old stars.  At its centre lies a supermassive black hole of about $3\times10^6M_\Sun$.  
The disk and bulge are surrounded by a halo of some billion old and metal-poor stars, as well as around 140 globular clusters, and a small number of satellite dwarf galaxies. The entire system is (currently assumed to be) embedded in a massive halo of dark material of unknown composition and poorly known morphology.
The distributions of stars in the Galaxy are linked through gravitational forces, and through the star formation rate as a function of position and time. The initial distributions are modified, often substantially, by small and large scale dynamical processes: these include instabilities which transport angular momentum (bars and warps), and mergers. 
Knowledge of all of these structures, including their kinematics and dynamics, and their origin and evolution, has always been limited by knowledge of stellar distances and space motions.
\citet{2016ARA&A..54..529B}	
provide a wide-ranging review of the state of knowledge of the Galaxy pre-Gaia.

\paragraph{Structure and dynamics of the solar neighbourhood}

A century ago, the Oort--Lindblad model
\citep{1927MNRAS..87..553L,
1927BAN.....3..275O}
laid the foundations for our view of the Galaxy disk as being in a state of differential rotation around the Galactic centre, leading to its empirical description based on the Oort constants. Within this rotating framework was evidence for considerable sub-structure, including clusters, moving groups, and spiral arms.
As we know today, the Galaxy's rotation curve (Section~\ref{sec:rotation-curve}) has an innermost part $R\lesssim3$\,kpc in almost solid body rotation. The rotation velocity rises outwards, is roughly constant at $R\sim R_0$, and is fairly flat or with a slow decline at larger radii, implying the presence of invisible or dark matter in the outer parts.  

Expressions for the radial and transverse velocities relative to the Sun, $v_R$ and $v_T$, in particular valid for $d\ll R_0$, i.e.\ characterising differential Galactic rotation in the solar vicinity, can be formulated in terms of the Oort `constants' $A$ and $B$ 
(\citet{1981gask.book.....M}, Equation~8--15,16 and 8--19,20).  	
These have units of frequency, and are usually expressed in \kmskpc. 
The local value of the angular rotation rate given by $\Omega_0=A-B$, and its local derivative by $\Omega_0^\prime=-(A+B)$.
Physically, and in analogy with fluid dynamics, $A$ describes the azimuthal shear of the velocity field, while $B$ describes its vorticity.
Extending these ideas to a non-axisymmetric system, specifically in the presence of spiral density waves and the central bar, but still restricted to motions in the plane, the Oort constants will vary with azimuth, and the resulting numerical values further depend on the type and distance of the tracer stars studied. 
A more general expression for the velocity field employs {\it four\/} such constants, quantifying the local divergence ($K$), vorticity ($B$), and azimuthal ($A$) and radial ($C$) shear of the velocity field 
\citep{1942psd..book.....C,		
2003ApJ...599..275O}.		

A more complete deformation tensor, including motions out of the plane, and assuming only that it can be represented by a continuous smooth flow, was first described as a first-order Taylor series expansion by Ogorodnikov (1932) and \citet{1935MNRAS..95..560M}.
\citet{1973MNRAS.161..445C},
for example, showed that the local proper motions were not well described by the Oort--Lindblad model, finding a significant expansion component, which he attributed to spiral density waves and, within 300\,pc, to the burst of recent star formation which gave rise to the Gould Belt.
The Hipparcos distances and proper motions provided further evidence of a more complex picture, and some first attempts were made to represent the global field of tangential velocities in terms of vector spherical harmonics. This  provided more tantalising hints of the complex stellar velocity patterns in the solar neighbourhood
\citep{2000A&A...354..522M,	
2007AJ....134..367M}.		

Gaia has brought a step-change in the observational description of our Galaxy's structure and kinematics. As described in Section~\ref{sec:stellar-structure}, major progress has been made in characterising the solar neighbourhood, including the identification and characterisation of many new open star clusters and associations. 
The nature of various nearby moving groups and dynamical streams has long been debated. As I will detail in this section, Gaia has confirmed that some are `evaporating' open clusters, some are associated with various dynamical resonances, while others are the tidal debris of accreted satellite galaxies (Section~\ref{sec:moving-groups}). 
By providing, for the first time, accurate 6d phase-space distributions (positions and space motions) of vast numbers of stars within a few~kpc from the Sun, Gaia has revealed that the Galactic disk is both deeply structured and strongly perturbed.
New insights are becoming available on the morphology and dynamics of the central bar and the Galaxy's spiral arms.
Newly discovered phase-space structures include the Radcliffe Wave (Section~\ref{sec:gould-radcliffe}), large numbers of stellar streams in the Galaxy halo (Section~\ref{sec:halo-streams}), and the remarkable Gaia phase-space spiral (Section~\ref{sec:phase-space-spiral}). 
Amongst other advances are an increasingly coherent picture of our Galaxy's three-phase formation, involving spin-up, merger, and cooldown (Section~\ref{sec:age-formation}).
All of these rich structural features continue to provide support for the $\Lambda$CDM cosmological paradigm (Section~\ref{sec:cosmology}).

\begin{table}[t]
\centering
\footnotesize
\begin{tabular}[t]{lllcccc}
\noalign{\vspace{-2pt}}
\hline
\noalign{\vspace{2pt}}
Source& Reference& Stars& \multicolumn{3}{c}{Solar motion wrt LSR (\kms)}&  Total \\
&&& $u_\odot$& $v_\odot$& $w_\odot$&  $V_\odot$ \\
\noalign{\vspace{2pt}}
\hline
\noalign{\vspace{2pt}}
{\bf Pre-Hipparcos:}\\
\qquad Compilation	&		 \citet{1981gask.book.....M} 			&various 	& $\phantom{0}9\phantom{.00\pm0.00}$& $12\phantom{.00\pm0.00}$ & $7\phantom{.00\pm0.00}$ & 16.5  \\ 
\qquad APM-based	&		 \citet{1995MNRAS.277..820E} 		&various 	& $\phantom{0}7.3\phantom{0}\pm1.5\phantom{0}$& $13.9\phantom{0}\pm2.3\phantom{0}$& $8.8\phantom{0}\pm2.2\phantom{0}$& 18.0 \\ 
\noalign{\vspace{5pt}}
{\bf Hipparcos:}\\
\qquad Oort--Lindblad &		\citet{1997MNRAS.291..683F}		&Cepheids& $\phantom{0}9.3\phantom{0\pm0.00}$&  $11.2\phantom{0\pm0.00}$&  $7.61\pm0.64$& 16.4 \\
\qquad\qquad\textquotedbl&	\citet{1998AJ....115.1483M}			&Cepheids&  $10.62\pm1.20$& $16.06\pm1.14$& $8.60\pm1.02$& 21.1 \\
\noalign{\vspace{2pt}}
\qquad Ogorodnikov--Milne&	\citet{1998AJ....115.1483M}			&O--B5 stars&  $11.59\pm0.49$& $13.39\pm0.48$& $7.12\pm0.44$& 19.1 \\
\qquad\qquad \textquotedbl&	\protect{\citet{2000A&A...354..522M}}	&A0--A5 dwarfs& $\phantom{0}9.92\pm0.25$& $10.71\pm0.26$& $6.96\pm0.21$& 16.2 \\
\qquad\qquad \textquotedbl& 	\qquad \textquotedbl					&F0--F5 dwarfs& $          11.46\pm0.37$& $11.16\pm0.37$& $7.02\pm0.41$& 17.5 \\
\qquad\qquad \textquotedbl& 	\qquad \textquotedbl					&M0--M5 giants& $\phantom{0}7.37\pm0.61$& $20.29\pm0.63$& $6.85\pm0.66$& 22.6 \\
\qquad\qquad \textquotedbl&  	\citet{2000RMxAA..36...97B}		& all Hipparcos &	$10.30\pm0.06$& $19.13\pm0.05$& $7.09\pm0.04$& 22.8 \\
\noalign{\vspace{2pt}}
\qquad Vectorial harmonics&	\citet{2007AJ....134..367M}			&non-binary& $\phantom{0}9.9\phantom{0}\pm0.2\phantom{0}$& $15.6\phantom{0}\pm0.2\phantom{0}$& $6.9\phantom{0}\pm0.2\phantom{0}$& 19.7 \\ 
\noalign{\vspace{2pt}}
\qquad Spiral-density wave&	\protect{\citet{1999A&A...341...81M}}	&Cepheids&  $\phantom{0}7.8\phantom{0}\pm1.3\phantom{0}$&  $13.6\phantom{0}\pm1.4\phantom{0}$&  -- & -- \\
\qquad\qquad \textquotedbl&	\protect{\citet{2001ApJ...546..234L}}	&Cepheids&  $\phantom{0}8.8\phantom{0}\pm1.0\phantom{0}$&  $11.9\phantom{0}\pm1.1\phantom{0}$&  -- & -- \\
\noalign{\vspace{2pt}}
\qquad Other&	 			\citet{1998MNRAS.298..387D}		&dwarfs&  $10.0\phantom{0}\pm0.36$&  $\phantom{0}5.25\pm0.62$&  $7.17\pm0.38$& 13.4 \\
\qquad\qquad \textquotedbl&	\citet{2001AN....322...15B}			&K0--K5 giants& $\phantom{0}9.0\phantom{0}\pm0.5\phantom{0}$& $21.0\phantom{0}\pm0.5\phantom{0}$& $7.7\phantom{0}\pm0.4\phantom{0}$& 24.1 \\ 
\qquad\qquad \textquotedbl& 	\citet{2005ApJ...629..268H}			&dwarfs & $10.1\phantom{0}\pm0.5\phantom{0}$& $\phantom{0}4.0\phantom{0}\pm0.8\phantom{0}$&  $6.7\phantom{0}\pm0.2\phantom{0}$& 12.8 \\ 
\noalign{\vspace{5pt}}
{\bf Gaia:}\\
\qquad Cepheids (DR1)&		\citet{2017AstL...43..152B}			& Cepheids& $\phantom{0}7.90\pm0.65$& $11.73\pm0.77$& $7.39\pm0.62$& 16.0 \\
\qquad White dwarfs (DR2)&	\citet{2019MNRAS.484.3544R}			& local white dwarfs &  $\phantom{0}9.5\phantom{0}\pm1.2\phantom{0}$&  $\phantom{0}7.5\phantom{0}\pm1.2\phantom{0}$& $\phantom{0}8.2\phantom{0}\pm1.2\phantom{0}$&	14.6 \\

\qquad GCNS (EDR3)&		\protect{\citet{2021A&A...649A...6G}}		&$G<13$\,mag&	$11.3\phantom{0\pm0.00}$&	$\phantom{0}6\phantom{.0\pm0.00}$&	$7\phantom{.00\pm0.00}$&     14.6\\ 
\qquad\qquad \textquotedbl& 	\protect{\citet{2022A&A...667A..98R}}	&Besançon model&	$10.79\pm0.56$&		$11.06\pm0.94$& 	$7.66\pm0.43$&	17.3 \\ 
\qquad\qquad \textquotedbl& 	\citet{2023Univ....9..252G}			&$d<100$\,pc&  	$10.1\phantom{0}\pm0.1\phantom{0}$&  $22.8\pm0.1\phantom{0}$&  $7.8\phantom{0}\pm0.1\phantom{0}$& 26.1 \\	
\noalign{\vspace{2pt}}
\qquad Halo streams (EDR3)&	\citet{2020arXiv201205271M}		&$d=3-30$\,kpc&	$\phantom{0}8.88\pm1.21$&  $241.91\pm1.67$&  $3.08\pm1.08$& -- \\
\noalign{\vspace{2pt}}
\hline
\end{tabular}
\vspace{-5pt}
\caption{Solar motion with respect to the Local Standard of Rest \citep[from][Table~9.2]{2009aaat.book.....P}. $u_\odot$ is toward the Galactic centre, $v_\odot$~in the direction of Galactic rotation, and $w_\odot$ toward the North Galactic Pole. GCNS is the Gaia Catalogue of Nearby Stars.
The $v_\odot$ component of \citet{2020arXiv201205271M} includes the contribution of the Local Standard of Rest, for which recent (Gaia-based) estimates are in the range $240-260$\kms\ \citep[e.g.][]{2018ApJ...867L..20H, 2020arXiv201202169B}. 
}\label{tab:solar-motion}
\end{table}

\subsection{Solar motion}
\label{sec:solar-motion}

Gaia is the latest step in the long journey of trying to understand the Sun's detailed location and motion within the Galaxy. Key quantities are the Sun's distance from the Galactic centre, its height above the Galaxy mid-plane, its motion with respect to the some Local Standard of Rest, and the motion of this Local Standard of Rest around the centre of the Galaxy. 
These quantities are also coupled with the effects of the Galaxy's rotation curve, viz.\ how the rotation of the Galaxy depends on distance from the Galactic centre. Observationally, these quantities are interconnected, but I will consider them separately in the following treatment.\footnote{
The following concepts are generally implicit in discussions of the Sun's location and motion:

\vspace{1pt}\noindent
(a) Distance of the Sun from the Galactic centre, $R_0$. It is a working hypothesis that the Galactic centre (defined by the black hole, Sgr~A*, itself assumed to coincide with the Galaxy's barycentre) defines the origin of an inertial coordinate system. This is clearly only an approximation for dynamical motions related to the barycentre of the local group of galaxies. 

\vspace{1pt}\noindent
(b) Circular velocity, $\Theta(R)$, usually measured in \kms, is the velocity of an object moving in a circle of radius $R$, in the Galactic plane and about the Galactic centre, for which centrifugal force balances the Galaxy's gravity. The simplest definition assumes axisymmetry.

\vspace{1pt}\noindent
(c) The `solar neighbourhood' is a loose concept considered to be a volume centred on the Sun, of arbitrary size much smaller than the Galaxy, but containing a representative subset of its population. It may range, for example, from a sphere of radius 10\,pc for the faint, common white dwarfs or M~dwarfs, out to 1\,kpc or more for the brighter and rarer O~and~B stars.

\vspace{1pt}\noindent
(d) The `Local Standard of Rest' (LSR) is the velocity of a hypothetical group of stars in strictly circular orbits at the solar position. Its practical definition is again complicated by the wide choice of stars and stellar types that can be chosen to represent it.

\vspace{1pt}\noindent
(e) The `solar motion', $\mathbf{v}_\odot=(u_\odot,v_\odot,w_\odot)$, can then be determined with respect to a range of the Galaxy's stellar and interstellar constituents. Most frequently, it is estimated with respect to the Local Standard of Rest, being the difference between the Sun's velocity and that of the reference system which, by definition, moves around the Galaxy with circular velocity $\Theta_0\equiv\Theta(R_0)$.
} 

The Sun's velocity with respect to the Galactic centre is traditionally described as the solar motion with respect to the local solar neighbourhood added to the velocity of the Local Standard of Rest.
Table~\ref{tab:solar-motion} is a compilation of some of the pre-Hipparcos, Hipparcos, and Gaia determinations of the solar motion. 
Of the Gaia-based determinations, three have used the Gaia Catalogue of Nearby Stars within 100\,pc, itself based on Gaia EDR3 
\citep{2021A&A...649A...6G}. The solar motion is derived, as usual, by model fitting to the velocity distributions of representative stars. 
\citet{2021A&A...649A...6G} 
based their determination on bright stars, $G<13$\,mag, while 
\citet{2023Univ....9..252G} employed a more complete sample. 
\citet{2022A&A...667A..98R} adjusted their (Besançon) dynamical model of the Galaxy disk (Section~\ref{sec:besancon-model}) to the Gaia EDR3 data.


A very different approach was taken by
\citet{2020arXiv201205271M}.	
They used EDR3 to measure all three components of the Sun's velocity with respect to the Galactic halo, as represented by 17 stellar streams over distances of 3--30\,kpc. Their method (which is independent of a Galactic potential model) is based on the assumption that, in low-mass streams, stellar proper motions should be directed along the stream structure in a non-rotating rest frame of the Galaxy, such that any observed deviation arises due to the Sun's own reflex motion. 
Being in broad agreement with past measurements by other techniques suggests that the inner Galaxy, and in particular the disk, is not moving with respect to the inertial frame defined by the halo streams.

\subsection{Distance to the Galactic Centre}
\label{sec:distance-galactic-centre}

The Galactic centre is the rotational centre of the Milky Way, associated with the supermassive ($\sim4\times10^6M_\Sun$) black hole, and the related compact radio source Sgr~A*. Its distance, $R_0$, is a fundamental quantity entering models of the Galaxy's rotation, its structure and dynamics, in estimates of its mass and of its major components, and models of its chemical and dynamical evolution. 
At 7.5--8.5\,kpc, it is too distant for significant trigonometric parallax determinations. Interstellar dust limits most studies to infrared and radio wavelengths, while the large and variable extinction complicates the use of secondary distance indicators.
Starting more than a century ago
\citep[e.g.][]{1918ApJ....48..154S},
estimates have been made from the space density of objects believed to be distributed symmetrically around the Galactic centre, viz.\ globular clusters, RR~Lyrae stars, Mira variables, and H$_2$O masers. 
From the mean of 25~determinations between 1974--86 (in the range 6.7--10.5\,kpc),
\citet{1986MNRAS.221.1023K}
gave $R_0=8.54\pm1.1$\,kpc, while
\citet{1993ARA&A..31..345R}
also combined different estimates to derive $R_0=8.0\pm0.5$\,kpc. 
Based on models of Galactic rotation and the Oort constants,
\citet{1998MNRAS.297..943O}
gave $R_0=7.1\pm0.4$~kpc. 
The Hipparcos-based estimates, also indirect, included $R_0=8.5\pm$0.5~kpc based on Cepheid kinematics
\citep{1997MNRAS.291..683F},
$R_0=9.3\pm$0.7~kpc from a revised RR~Lyrae period--luminosity relation for globular clusters 
\citep{1998AJ....115..204R},
and $R_0=8.2\pm$0.2~kpc from the red clump giants in Baade's window detected by OGLE
\citep{1998ApJ...503L.131S}.
In my own review of the Hipparcos results in 2009
\citep{2009aaat.book.....P},
I concluded that {\it `\ldots estimates for $R_0$ still lie in the rather broad range 7.5--8.5\,kpc'}, with {\it `a value of $R_0=8.2$\,kpc being suggested as reference'}.

\begin{figure}[t]
\centering
	\begin{subfigure}{0.38\linewidth}
	\includegraphics[width=\linewidth]{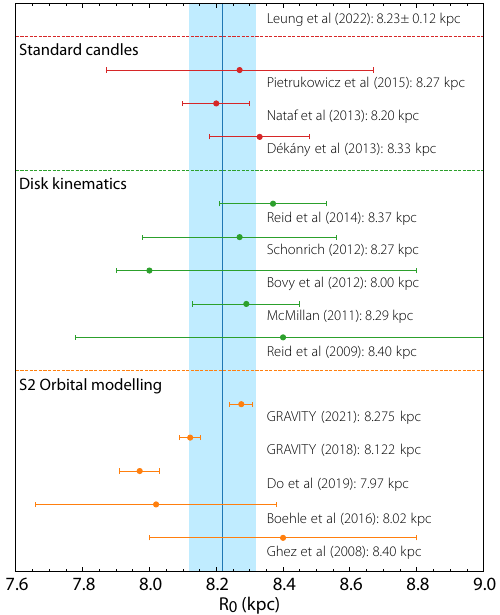}
	\end{subfigure}
	\hspace{10pt}
	\begin{subfigure}{0.58\linewidth}
	\includegraphics[width=\linewidth]{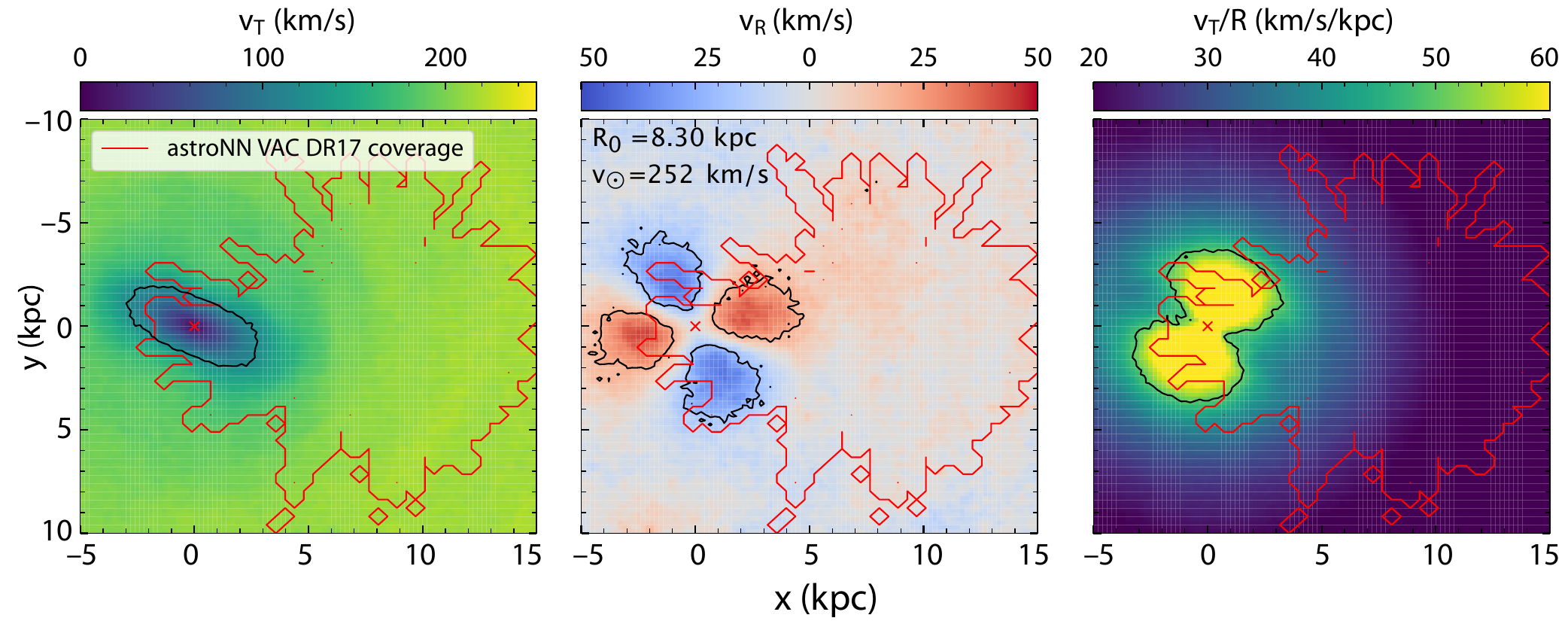}\\[5pt]
	\phantom{0000000}\includegraphics[width=0.66\linewidth]{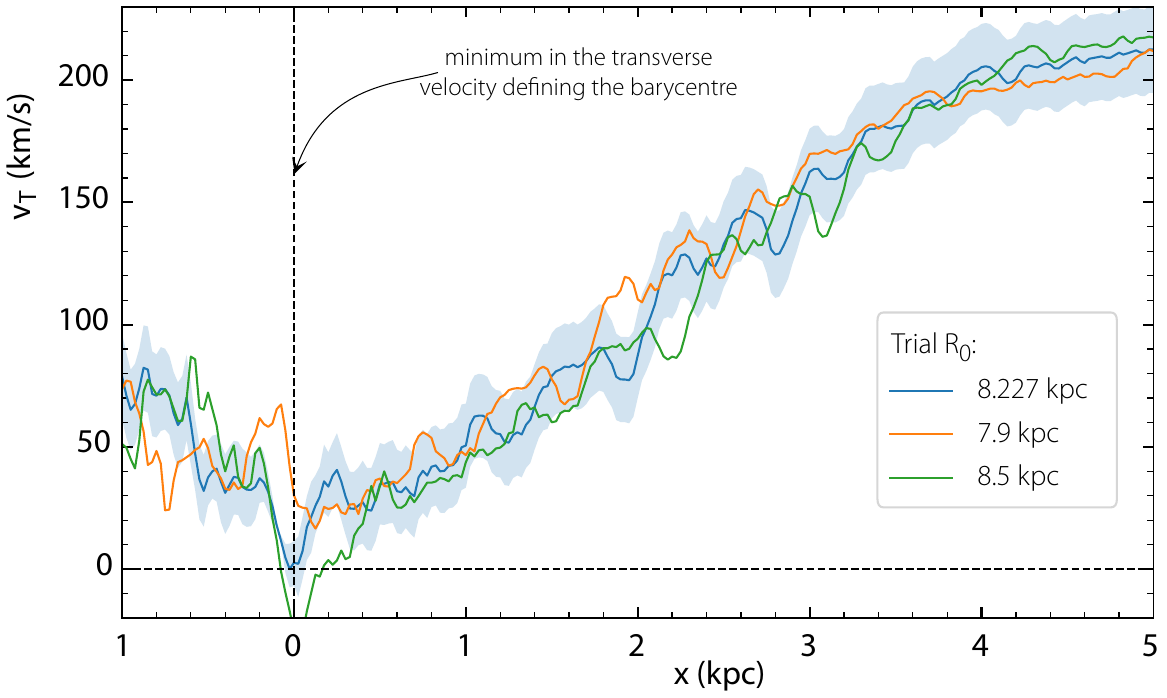}
	\end{subfigure}
\caption{
(a, left): recent estimates of $R_0$, taken from \citet{2023MNRAS.519..948L}, with their value included at top. 
(b, right upper): simulated kinematic maps of the bar and disk from \citet[][Figure~6]{2023MNRAS.519..948L}. The red contours define the region of their APOGEE/Gaia sources. The best match between these simulations (here with $\Omega_{\rm bar}=24.5$\kmskpc) and their data (their Figure~2) yields their value of $R_0$. 
(c, right lower): median tangential velocity $v_{\rm T}$, calculated for three values of $R_0$ including their optimal value, $R_0=8.23$\,kpc \citep[from][Figure~12]{2023MNRAS.519..948L}.}
\label{fig:galactic-centre}
\end{figure}

From around 2000, dynamical studies of the Galactic centre based on high-resolution adaptive optics imaging in the infrared provided more direct estimates.  From improved proper motions and radial velocities for a hundred stars within the central few arcseconds of Sgr~A*,
\citet{2000MNRAS.317..348G}
gave $R_0=7.8-8.2\ (\pm0.9)$\,kpc.
The orbital parallax of the Galactic centre star~S2, around the black hole, based on astrometry and radial velocities, resulted in $R_0=7.9\pm0.4$~kpc from the GRAVITY collaboration
\citep{2003ApJ...597L.121E}.
Two other estimates were given by
\citet{2008ApJ...689.1044G}:
$8.0\pm0.6$~kpc for an unconstrained Keplerian orbit, or $8.4\pm0.4$~kpc if the black hole is considered at rest with respect to the Galactic centre
\citep[cf.][]{2020ApJ...892...39R}.
%
Some of the most recent estimates (Figure~\ref{fig:galactic-centre}a) have been based on standard candles, notably RR~Lyrae variables
\citep{2013ApJ...776L..19D}		
and OGLE clump giants
\citep{2013ApJ...769...88N};		
on disk kinematics
\citep[e.g.][]{2014ApJ...783..130R};	
and on the improved orbit modelling of S2. These values, $8.275\pm0.009\,(\pm0.033)$~kpc from GRAVITY	
\citep{2018A&A...615L..15G,		
2021A&A...647A..59G},			
and $7.971\pm0.059\,(\pm0.032)$~kpc from Keck speckle imaging
\citep{2019Sci...365..664D}		
are the most precise to date, but are inconsistent. A separate question is whether Sgr~A* indeed coincides with the Milky Way barycentre.

A new approach, exploiting the Gaia data, assumes that the motion of stars in the central {\it bar\/} region, itself dynamically linked to the massive disk, directly defines the disk--bar barycentre. The first such study used APOGEE DR16 (the all-sky, high-resolution near-infrared spectroscopic survey) with Gaia DR2
\citep{2019MNRAS.490.4740B}.
An update by \citet{2023MNRAS.519..948L}
is based on APOGEE DR17 (providing the radial velocities) and Gaia EDR3 (providing the proper motions), with the most accurate distances extracted from a neural-network analysis, trained on Gaia parallaxes of less-distant stars. Together, they provide a 6d phase-space data set of the Galactic bar region.
Figure~\ref{fig:galactic-centre}b shows the region covered in the Galactic $x-y$ plane (red line), superimposed on simulations of the disk--bar system by 
\citet{2017MNRAS.464..702K}. 
The bar, rotating here at a pattern speed of 24.5\kmskpc\ at an angle of $25^\circ$, is seen as a density enhancement in the distribution of transverse velocities, $v_{\rm T}$ (left), as a quadrupole pattern in radial velocities (middle), and as a dipole pattern in $v_{\rm T}/R$ (right). 
Their analysis searches for a minimum in the rotational velocity $v_{\rm T}$ of stars along the Sun--Galactic-centre line, and determines $R_0$ by associating this minimum with the Galactic centre (Figure~\ref{fig:galactic-centre}c). Their resulting estimate, $R_0=8.23\pm0.12$~kpc, is also shown as the shaded region in Figure~\ref{fig:galactic-centre}a.
Their value would rule out the lower end of the historical measurements, as well as being inconsistent with that from orbit modelling of the Galactic centre star S2 by 
\citet{2019Sci...365..664D}.	
But it would be in good agreement with the consensus value $R_0=8.2\pm0.1$~kpc, which preceded the latest S2 measurements, given by 
\citet{2016ARA&A..54..529B}.

This improved value for $R_0$, enabled by Gaia, can be used in combination with another important result from Gaia, viz.\ the solar system's acceleration within the Galaxy using the apparent proper motions of quasars, derived by 
\citet{2021A&A...649A...9G},
and detailed in Section~\ref{sec:aberration}.
The importance of this direct measure of the Sun's centripetal acceleration was described by
\citet{2020arXiv201202169B}
as {\it `a revolutionary moment for Galactic astrophysics'}. He notes that, expressed as an acceleration
$(a_0/c) = V_0^2/(c\,R_0)=5.05\pm0.35$\muasyr\ 
which, combined with the above value for $R_0$, gives
$V_0 = 243\pm8$\kms.	
The angular frequency of the circular orbit at the Sun is, similarly, 
$\Omega_0=V_0/R_0 = 29.3\pm1.0$\kmskpc.
If Sgr~A* is assumed to be at rest at the Galactic centre, its observed proper motion $\mu_{\rm Sgr~A*}=-6.411\pm0.008$\masyr\ 
directly yields the Sun's peculiar velocity
$\mu_{\rm Sgr~A*} = \Omega_0 + (V_\odot/R_0)$, and hence $V_\Sun=8.0\pm8.4$\kms.

\subsection{Sun's height above the mid-plane}
\label{sec:height-above-disk}

For over a century it has been strongly suspected that the Sun does not lie exactly in the mid-plane of the Galactic disk. In his study of globular clusters more than a century ago,  
\citet{1918PASP...30...42S},
concluded that the Sun is {\it `\ldots some 20\,pc north of the plane'}.
\citet{2017MNRAS.465..472K} 
tabulate more than 50~estimates of this value, $Z_\odot$, that have been made since.
Out of some 30 estimates in the past 20~years, mostly based on Hipparcos and using several different tracers, values still range from as small as $Z_\odot=5.2\pm4.7$\,pc based on A/F stars 
\citep{2007ASSL..350.....V}	
to as large (and precise!)\ as $Z_\odot=34.56\pm0.56$\,pc using 93\,106 solar neighbourhood stars with a range of spectral types
\citep{2003Ap&SS.288..417B}.	
In their pre-Gaia review of the Milky Way, 
\citet{2016ARA&A..54..529B}
found that their most plausible estimates were between 20--30\,pc, while
\citet{2017MNRAS.465..472K}
noted that their own determination, $17\pm5$\,pc, was consistent with {\it `\ldots the median of 55 previous estimates published over the past century, $17\pm2$\,pc'}. Positive values of $Z$ are, incidentally, towards the North Galactic Pole, i.e.\ {\it above\/} the plane.

The distance of the Sun from the disk mid-plane is important because it affects the interpretation of numerous observations of our Galaxy's disk, such as 
asymmetries in star counts or dust emission
studies of the Galaxy warp from COBE--DIRBE
\citep{1994ApJ...429L..69F,
1997A&A...322..103P}.
The Sun's location, and its motion with respect to the Local Standard of Rest, also have consequences for comprehending our solar system's habitability, including the Galactic radiation environment, and the Sun's vertical oscillations and passages through the spiral arms.
It is also relevant in the very definition of the Galactic coordinate system. As adopted by the IAU in 1958,
the equatorial plane, $b=0^\circ$, passes through the Sun
\citep{1960MNRAS.121..132G,
1960MNRAS.121..123B}.
If the Sun were exactly in the mid-plane, the Galactic coordinate system's origin should coincide with the Galactic centre. 
The Sun's location has also become more relevant as improved observations focus the need for a more operational definition of the Galactic coordinate system
\citep[e.g.][]{2019ApJ...871..145A}.	
This is itself linked to whether the black hole at the Galactic centre, Sgr~A$^*$, precisely coincides with its dynamical centre
\citep[e.g.][]{2023MNRAS.519..948L}.  

Attempts to define $Z_0$ should be based on objects expected to trace the Galactic disk
\citep[e.g.][]{2017MNRAS.465..472K}. 
In this spirit, the (still current) IAU 1958 definition was based on the distribution of H\,{\scriptsize I} gas,  which has the advantage of tracing mass at large distances, and demonstrates a high degree of flatness in the plane.
Other suitable tracers are
young stellar objects and open clusters (which typically lie within a few hundred parsec of the plane);
infrared dark clouds (a particular class of molecular cloud known to be the sites of the earliest stages of star formation);
H\,{\scriptsize II}~regions and H\,{\scriptsize I}~shells (especially prominent in the radio and infrared);
asymptotic giant branch stars (luminous evolved stars that trace the radial and vertical structure of the disk);
and supernovae remnants.

Examples of pre-Hipparcos estimates of $Z_0$ range from 
10\,pc from interstellar dust measurements, 
10--12\,pc from several thousand OB stars within 4\,kpc; 	
15\,pc from IRAS source counts and COBE data; 
20\,pc from star counts at the Galactic poles; 			
27\,pc from the Sloan Digital Sky Survey; 				
37\,pc from Cepheids;
to as much as 42\,pc from some classical star counts.
In these studies, effects of the warp, the Gould Belt, and extinction, were often noted as complications.
Amongst many new determinations with Hipparcos were 
$22.8\pm3.3$\,pc from open clusters \citep{2005MNRAS.362.1259J},
$24.2\pm1.7$\,pc from O--B5 stars \citep{2001AJ....121.2737M},
and
$34.56\pm0.56$\,pc from a range of spectral types \citep{2003Ap&SS.288..417B}.
Amongst the more recent work {\it not\/} using Gaia, two papers have focused on very young high-mass tracers, with both finding significantly smaller values, \mbox{$Z_0\simeq5-6$\,pc}, than the past century's median.
\citet{2019ApJ...885..131R}	
analysed the distances and motions of 200 molecular masers associated with very young high-mass stars. They found that the orientation of the associated plane is consistent with the IAU 1958 definition to within $\pm$0\ddeg1, and that the Sun is offset toward the North Galactic Pole, with $Z_\odot=5.5\pm5.8$\,pc. Accounting for this offset also then places the central supermassive black hole, Sgr~A*, in the mid-plane of the Galaxy. 
\citet{2019ApJ...871..145A}		
used the WISE Catalogue of Galactic H\,{\scriptsize II} Regions to define a `high-mass star formation mid-plane', finding a similar value, \mbox{$Z_\odot=5.6\pm2.6$\,pc}. They showed that this plane is not significantly tilted or rolled with respect to the IAU mid-plane, and that the Sun is therefore near to this mid-plane. They attribute the inconsistency with many previous {\it stellar\/} studies as arising from asymmetries in the stellar distribution near the Sun.

That there was no clear consensus from Hipparcos is, today, hardly a surprise, since Gaia has since confirmed numerous morphological complexities and various non-equilibrium dynamical features (Sections~\ref{sec:moving-groups} and \ref{sec:phase-space-spiral}), making their use as accurate disk tracers more questionable. 
With the new Gaia results, at least three studies have followed the principles of the classical $K_z$ problem in Galactic dynamics, which aims to quantify the force law perpendicular to the Galactic plane (Section~\ref{sec:mass-kz}). It requires a tracer population whose number density, and vertical velocities, can be determined as a function of height, and the method then exploits the Poisson and Boltzmann equations in various forms.
Using DR1,
\citet{2019MNRAS.482..262W}	
used stars out to 160\,pc and, in addition to estimating the local matter density, found $Z_\odot=15.3\pm2.2$\,pc.
Using DR2,
\citet{2019A&A...623A..30W}	
found $Z_\odot=4.76\pm2.27$\,pc.		
\citet{2019MNRAS.482.1417B}
made a similar dynamical analysis using DR2. As well as identifying a wave-like oscillation in vertical velocity, they made the {\it `\ldots most precise and accurate determination'}, $Z_\odot=20.8\pm0.3$\,pc.
Some of the most meaningful estimates, from Gaia EDR3, were based on the Gaia Catalogue of Nearby Stars.
\citet{2021A&A...649A...6G}	
found $Z_\odot$ varying from $-4$~pc to 15~pc for young to older stellar populations, with a tilt between the Galactic and $Z=0$ plane of only 0\ddeg1.

If Gaia has not provided any ruling on a single value for $Z_\Sun$, it has perhaps made clear why, because of the rich phase-space substructure now known to characterise the disk, the challenge is almost intractable. As Eric Mamajek (priv.\ comm.) put it:  {\it `It’s like trying to measure, and argue over, the mean height of an old, lumpy mattress'}.

\subsection{Galactic rotation}
\label{sec:rotation-curve}

A galaxy's rotation curve describes the circular rotation speed as a function of radial distance. Crucially, it provides the most important constraints on the galaxy's mass distribution. Since the work of \citet{1978ApJ...225L.107R}, the flat rotation curves observed in spiral galaxies, in contrast to those  expected from the {\it observed\/} mass distribution (in gas and stars), is generally attributed to (and is one of the main arguments for) the presence of dark matter. 
For our own Galaxy, the rotation curve has provided the best constraints on the mass distribution of its major structural components (bulge, thin and thick disk, and halo), including the effects of dark matter 
\citep[e.g.][]{2009PASJ...61..227S,	
2010A&A...523A..83S,		
2010A&A...509A..25W,		
2012PASJ...64...75S,		
2015MNRAS.453..377W,		
2020SCPMA..6309801W}.	

However, defining the Galaxy's rotation curve is not straightforward. In the inner regions, it has been mainly determined from radio observations of gas in H\,{\footnotesize I} and CO
\citep{1978A&A....63....7B, 
1985ApJ...295..422C,	
1989ApJ...342..272F,	
2008ApJ...679.1288L}.	
Beyond the solar circle, it can be derived from the kinematics of suitable tracer populations, including
H\,{\footnotesize II} regions 
\citep{1989ApJ...342..272F,	
1993A&A...275...67B}, 		
planetary nebulae 
\citep{1983ApJ...274L..61S, 	
1996MNRAS.281..339A,		
2005RMxAA..41..383M},		
classical Cepheids 
\citep{1997A&A...318..416P,	
2019AN....340..787G,		
2019ApJ...870L..10M},		
blue horizontal branch stars 
\citep{2008ApJ...684.1143X, 	
2012MNRAS.424L..44D,		
2012ApJ...761...98K},		
red clump giants 
\citep{2012ApJ...759..131B,	
2014A&A...563A.128L,	
2016MNRAS.463.2623H},	
and masers 
\citep{2012PASJ...64..136H,	
2014ApJ...783..130R,		
2021AstBu..76..146G}.	
This has been extended to the Galaxy's outermost regions using globular clusters and dwarf galaxies 
\citep{2009MNRAS.392L...1S,	
2010MNRAS.406..264W,		
2013ApJ...768..140B,		
2017ApJ...850..116L},		
most recently using Gaia DR2
\citep{2020ApJ...894...10L}. 	
A compilation of 2780 measurements between \mbox{3--20\,kpc}, {\tt galkin}, is given by 
\citet{2017SoftX...6...54P}.		
Although often subject to large errors, the consensus 
\citep{2023ApJ...946...73Z}
is that the rotation curve is roughly constant at $R\sim R_0$, and fairly flat or with only a slow decline at larger radii, at least to 15--20\,kpc, implying invisible or dark matter in the outer parts.  
Before Hipparcos, such studies could only sample a small region around the Sun. Hipparcos provided better estimates of $R_0$, of the Sun's local motion with respect to the local standard of rest, of the Oort constants $A$ and $B$, and of the form of the rotation curve.
From 220 Hipparcos Cepheids, and assuming $R_0=8.5$\,kpc,
\citet{1997MNRAS.291..683F}	
estimated $(u_0,v_0,w_0)=(9.3,11.2,7.6)$\kms, and found 
$A=+14.82\pm0.84$ and $B=-12.37\pm0.64$, 
from which 
$\Omega_0=A-B=+27.19\pm0.87$ and $\Omega_0^\prime=-(A+B)=2.4\pm1.2$.

\paragraph{Gaia Cepheids}
A number of early Gaia studies also focused on Cepheids.
\citet{2017AstL...43..152B}	
used 260 Cepheids from DR1 to give 
$\Omega_0=28.84\pm0.33$\kmskpc, $\Omega_0^\prime=-4.05\pm0.10$\kms\,kpc$^{-2}$, 
yielding a rotation velocity of the Local Standard of Rest $\Theta_0=231\pm6$\kms.
\citet{2019ApJ...870L..10M}		
used 773 Cepheids from Gaia DR2 to measure the rotation curve out to $R=20$\,kpc. Assuming $R_0=8.122\pm0.031$~kpc (from the GRAVITY Collaboration), they estimated the rotation speed at the Sun of $\Theta_0=233.6\pm2.8$\kms.  They found a nearly flat rotation curve over $R=4-20$\,kpc with a small decreasing gradient of $-1.34\pm0.21$\kmskpc. 
Also using Gaia DR2, similar results were found for 218 Galactic Cepheids 
\citep{2019MNRAS.482...40K},	
for 2431 OGLE Cepheids
\citep{2019Sci...365..478S},	
and 3500 Cepheids from OGLE and Gaia over Galactocentric distances 4--19~kpc
\citep{2020ApJ...895L..12A}. 

\paragraph{Other Gaia determinations}
Other Gaia-based determinations are exploiting the much larger numbers of other tracers.
\citet{2017MNRAS.468L..63B}		
used 304\,267 main-sequence stars from DR1 (TGAS) to examine the rotation curve out to 230~pc from the Sun. The pattern of proper motions clearly displays the effects of differential rotation (their Figure~2). Along with the Oort constants $A=15.3\pm0.4$ and $B=-11.9\pm0.4$, significant (non-zero) values of $C=-3.2\pm0.4$ and $K=-3.3\pm0.6$ (all in \kmskpc), demonstrated the effects of non-axisymmetry for the velocity field of local stars.

\begin{figure}[t]
\centering
\includegraphics[width=0.61\linewidth]{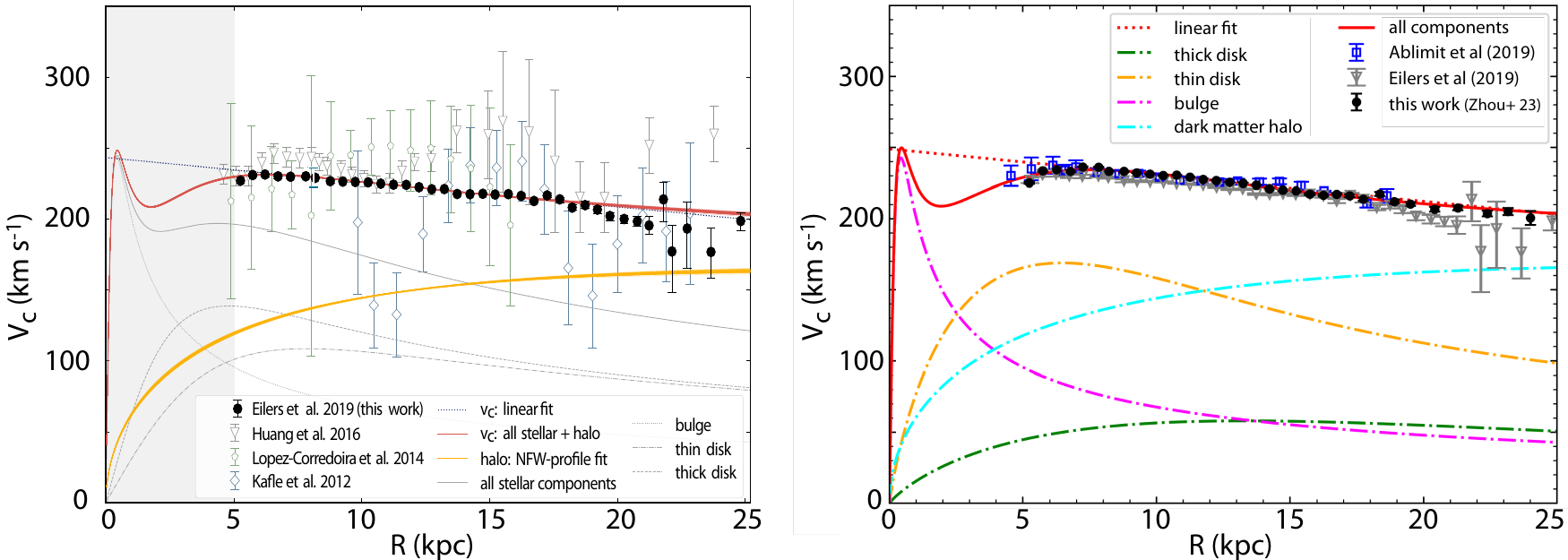}
\hspace{10pt}
\raisebox{-4pt}{\includegraphics[width=0.33\linewidth]{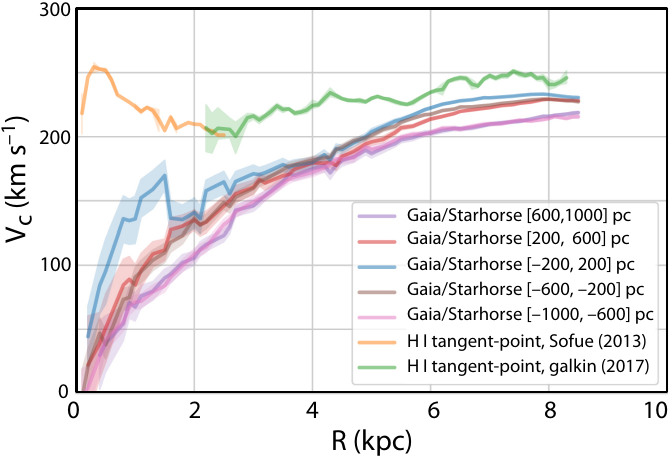}}
\caption{Three of the Gaia determinations of the Galaxy's rotation curve. 
(a, left): data from Gaia DR2 (black circles), also showing their best fit model (red) being the sum of the bulge, thin and thick disk (grey), and a dark matter NFW profile (yellow); some other determinations, and the region most influenced by the bar (grey shaded) are also shown \citep[from][Figure~3]{2019ApJ...871..120E}.
(b, middle)~data from Gaia EDR3 (black circles), with results from \citet{2019ApJ...871..120E} and \citet{2020ApJ...895L..12A} (coloured) for comparison. Their best-fit model (red) comprises the bulge (magenta), thin disk (gold), thick disk (green), and a dark matter halo (cyan) \citep[from][Figure~11]{2023ApJ...946...73Z}.
(c, right): the inner region based on Gaia stellar kinematics, labeled as Gaia/Starhorse (being the DR2 sample of \citet{2019A&A...628A..94A}), and compared with the tangent point (gas) results from \citet{2009PASJ...61..227S} and \citet{2017SoftX...6...54P} \citep[from][Figure~1]{2020RNAAS...4..165C}.
}\label{fig:rotation-curve}
\end{figure}

Using the Gaia DR2 and APOGEE DR14 data for 23\,000 thin disk red giants, 
\citet{2019ApJ...871..120E} determined the rotation curve between 5--25\,kpc, based on a Jeans model, and assuming an axisymmetric potential. Their error bars are significantly smaller than previous determinations, but still result in large uncertainties beyond 20\,kpc (Figure~\ref{fig:rotation-curve}a). The figure also shows the contributions from the thin and thick disks (assuming the model profiles of \citet{1975PASJ...27..533M}), the bulge (assume the spherical potential of \citet{1911MNRAS..71..460P}), as well as their fit to a Navarro--Frenk--White (NFW) dark matter halo profile \citep{1997ApJ...490..493N}.
Their {\it `most precise measurements of the circular velocity to date'} shows a gentle decline beyond 5\,kpc, $-1.7\pm0.1$\kmskpc, in reasonable agreement with the classical Cepheid determination by 
\citet{2019ApJ...870L..10M},	
but steeper than some earlier studies which have suggested a flat circular velocity curve
\citep[e.g.][]{2012ApJ...759..131B,2013ApJ...779..115B,2014ApJ...783..130R}.
Interpreting this rotation in the framework of the Navarro--Frenk--White (NFW) model of galaxy formation, they estimate our Galaxy's (virial) mass as $(0.822\pm0.052)\times10^{12}M_\odot$ within the corresponding (virial) radius of $191.84\pm4.12$\,kpc, and a predicted local dark matter density of $0.33\pm0.03$\,GeV\,cm$^{-3}$. They also conclude that the dark matter halo is the main contributor to the form of the Galactic rotation curve beyond distances of about 12.5--14.5~kpc.
The same rotation curve 
was used by
\citet{2019JCAP...10..037D}		
to evaluate the sensitivity of estimates of the local dark matter density to the assumed details of the dark matter halo, and the distribution of matter in the baryonic disk.

Using 54\,000 thin disk red giants from the APOGEE and LAMOST surveys, and using astrometry from Gaia EDR3,	
\citet{2023ApJ...946...73Z} 	
constructed the rotation curve from 5--25\,kpc (Figure~\ref{fig:rotation-curve}b). 
With a 2--3 times larger sample than \citet{2019ApJ...871..120E}, their rotation curve also shows a weak decline beyond 5\,kpc, with a similar gradient of $-1.83\pm0.02\pm0.07$\kmskpc, a circular velocity at $R_0$ of $234.04\pm0.08\,({\rm stat})\pm1.36\,({\rm sys})$\kms, and a significantly improved accuracy beyond 20\,kpc (see figure, top right).
They used this to construct a mass model for the Galaxy, yielding a mass of the dark matter halo $M_{200}=(8.05\pm1.15)\times10^{11}M_\Sun$, with  $R_{200}=192.37\pm9.24$\,kpc, and a local dark matter density $0.39\pm0.03$\,GeV\,cm$^{-3}$.
Further improvements came with Gaia DR3.
\citet{2023ApJ...942...12W} 	
extended the mapping to 30\,kpc, and confirmed that the rotation curve indeed shows a significant decline of $\sim$50\kms\ between 15--30\,kpc. The rotation curve (and azimuthal velocity) also presents a dependence on height above the disk for $R<15$\,kpc. 

\citet{2024MNRAS.528..693O}	
used APOGEE DR17 spectra, 2MASS and WISE photometry, and Gaia~DR3 astrometry, to construct the rotation curve to 30\,kpc. From 30\,000 red giant branch stars they confirmed the previous findings, but with an even faster decline beyond $\sim$25\,kpc, better matched to a `cored Einasto profile' \citep{1965TrAlm...5...87E} than the Navarro--Frenk--White profile. It yields a significantly lower halo virial mass, in tension with mass measurements from globular clusters, dwarf satellites, and streams. 
\citet{2023A&A...678A.208J}  
compared the methods used in these three Gaia-based determinations
\citep{2023ApJ...946...73Z,
2023ApJ...942...12W,
2024MNRAS.528..693O}. 
They also concluded that there is a sharp decrease in circular velocity of $\sim$30\kms\ between 19.5--26.5\,kpc, matching a {\it Keplerian\/} decline, and so implying the absence of significant mass beyond 20\,kpc \citep{2020ARep...64..295Z}. This would imply a total Galaxy mass $2.0\times10^{11}M_\Sun$, 4--5 times smaller than currently favoured estimates.

\paragraph{The inner rotation curve}
For the rotation curve within $R_0$, the assumption of circular orbits has underpinned use of the so-called `tangent-point method' in interpreting the gas kinematics. This assumes that the observed radial velocity extremum along any line-of-sight is at the tangent point, allowing calculation of both Galactocentric radius and circular velocity 
\citep{1954BAN....12..117V,2013PASJ...65..118S}.
Specifically, and pre-Gaia,
\citet{2015A&A...578A..14C}	
used numerical simulations to predict that, in part due to the effects of the bar, the tangent-point method overestimates the inner rotational velocity, and instead reflects local motions, concluding that {\it `the quest to determine the innermost rotation curve of the Galaxy remains open'.} 
%
Indeed, a rather radical Gaia finding along precisely these lines was reported by
\citet{2020RNAAS...4..165C}.	
They used 6~million stars from Gaia DR2 to determine the rotation curve within the solar circle, which they compared with that derived from gas kinematics using the tangent-point method (Figure~\ref{fig:rotation-curve}c).
They found significant differences between the two methods: while the tangent-point curve is mostly in the range 200--250\kms, that based on stellar kinematics increases to $\sim$150\kms\ at 2~kpc, rising to a plateau at 6--7~kpc. The peak using the tangent-point method at 1\,kpc is absent in the curve based on stellar kinematics.

\paragraph{Implications}

The slightly declining gradient at the solar circle found by 
\citet{2019ApJ...871..120E}
and others is important because, as they point out, disk galaxies in the local Universe typically show a flat or even {\it increasing\/} rotation curve 
\citep[e.g.][]{1980ApJ...238..471R,
1999ApJ...523..136S}, 
while galaxies with declining curves have only been reported at higher redshift. 
This effect seen in the early Universe has been attributed to baryons efficiently condensing at the centres of dark matter halos at a time when gas fractions were higher and dark matter less concentrated
\citep{2017Natur.543..397G, 
2017ApJ...840...92L}.		

While I say more on the Gaia-based estimates of the mass of the Galaxy in Section~\ref{sec:mass-galaxy}, the sharp decline in the rotation curve beyond 25\,kpc, found by 
\citet{2023ApJ...942...12W}
and others, is particularly noteworthy. It is probably premature to conclude that the dark matter halo actually terminates at $\sim$20\,kpc, with all its implications, and with the other inconsistencies that this would raise 
\citep[e.g.][\S6]{2023A&A...678A.208J}.
Meanwhile, work continues in investigating other effects that might be complicating the determination and interpretation of the rotation curves, including the assumptions underlying use of the Jeans equation, and alternative models of (MOND-like) gravity 
\citep{2020A&A...642A..95C,	
2020MNRAS.496.1077P,		
2023arXiv231004372N,		
2023ApJ...957...24C}.		

\subsection{Moving groups and dynamical streams}
\label{sec:moving-groups}

One of Gaia's major contributions is in the study of what are loosely referred to as `moving groups' in the solar neighbourhood. 
The earliest attempts to understand the Sun's motion through space, starting with William Herschel, assumed that the `peculiar motions' of nearby stars are random. Early in the 20th century, Kapteyn found that they actually included two preferred directions, later designated the Hyades and Ursa Major streams.
By the time the Hipparcos results appeared in 1997, and in large part due to the work of Olin Eggen over several decades
\citep[][and references]{1996AJ....112.1595E}, 
the picture had become much more complex. In addition to open clusters showing clear common spatial and kinematic properties, sparser kinematic groups of 10--100 or more members had also been identified in the solar neighbourhood.
Variously described as moving groups, kinematic groups, dynamical streams, or superclusters, these kinematically coherent groups of relatively old stars were recognisable because of their non-circular bulk motions, and their peculiar velocities well above the velocity dispersion of field stars of comparable age.

Fourteen such groups in the solar neighbourhood were listed in 1993 by 
\citet{1993AJ....105..226S},	
including
the Hyades supercluster associated with the Hyades (600\,Myr), also referred to as Eggen's Stream~I;
the Ursa Major group (or Sirius supercluster) associated with the Ursa Major cluster (300\,Myr), also referred to as Eggen's Stream~II;
the Local Association or Pleiades moving group of young stars comprising embedded clusters and associations such as the Pleiades, $\alpha$~Per, NGC~2516, IC~2602, and the Sco--Cen association (20--150\,Myr);
the IC~2391 supercluster (35--55\,Myr); 
and
the Castor moving group (200\,Myr).
Eggen hypothesised that these various groups were remnants of open clusters, and the fact that at least some have resulted from the evaporation of young clusters now seems secure.
But over the past 20--30~years, it became evident that others may be generated by a variety of dynamical mechanisms, specifically by resonances forced by the central rotating bar, or linked with the dynamics of the spiral arms.\footnote{
Certain Galaxy-scale orbital resonances are central to the physics of bars and spirals: the corotation resonance, and the inner and the outer Lindblad resonances \citep[][\S3.3.3]{2008gady.book.....B}. The former occurs at the Galactic radius where the star's circular velocity is equal to that of the rotating potential. For a bar of fixed pattern speed, the mean position of a star then remains stationary with respect to the global bar pattern. The inner Lindblad resonance occurs where the rotating star `overtakes' the potential, encountering its peak at the epicycle frequency. The outer Lindblad resonance, conversely, occurs where the potential overtakes the more slowly rotating star.}

In Eggen's day, most distances had to be inferred, indirectly, in order to construct space velocities. As a result, the reality of such old streams was often questioned. More quantitative progress became possible with the distances and space velocities provided by Hipparcos. 
\citet{1999Natur.402...53H}
found the signature of a cold stream in the velocities of halo stars from Hipparcos, which they interpreted as the tidal debris of a satellite galaxy accreted by the Milky Way. 
Similarly 
\citet{2004ApJ...601L..43N}
argued that Eggens’s Arcturus group is another debris stream, this time in the thick disk, and originating from an accretion event 5--8~Gyr ago.
More relevant in the present context, the Hipparcos data also confirmed the reality of moving groups in the velocity distribution of thin disk stars in the solar neighbourhood.
Amongst these, 
new evidence was found for the existence of the Sirius--UMa, Pleiades--Hyades, and Hercules star streams
\citep{1998AJ....115.2384D}.
The Hercules stream was associated with an outer Lindblad resonance with the Galaxy's central bar
\citep{2000AJ....119..800D,
2001A&A...373..511F},
while the Sirius--UMa and Pleiades--Hyades streams were associated to orbital resonances with spiral density waves
\citep{2005AJ....130..576Q}.

With the new Gaia data, there is a substantial story to tell about each of these groups. In the following, I will focus on Gaia's contribution to an understanding of the Hercules stream, and of the Arcturus and HR~1614 streams. 

\begin{figure}[t]
\centering
\includegraphics[width=0.27\linewidth]{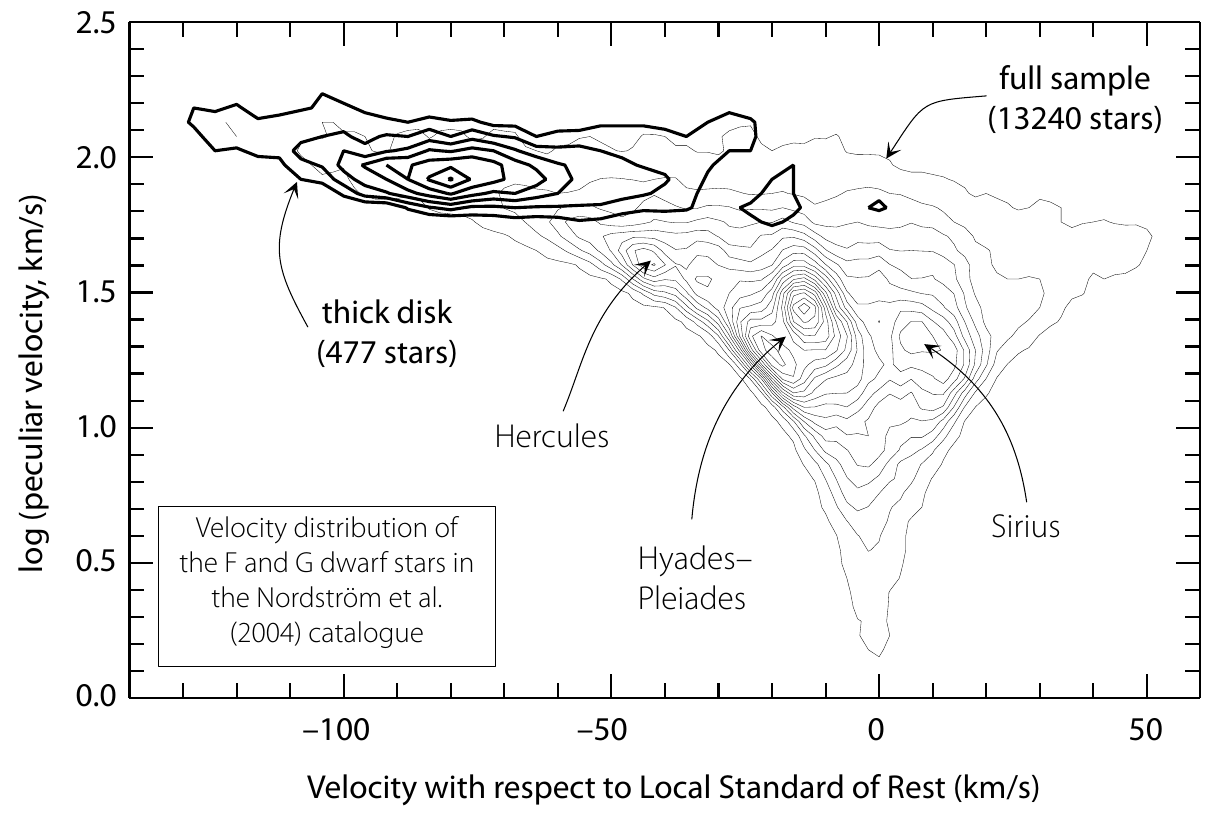}
\hspace{5pt}
\includegraphics[width=0.32\linewidth]{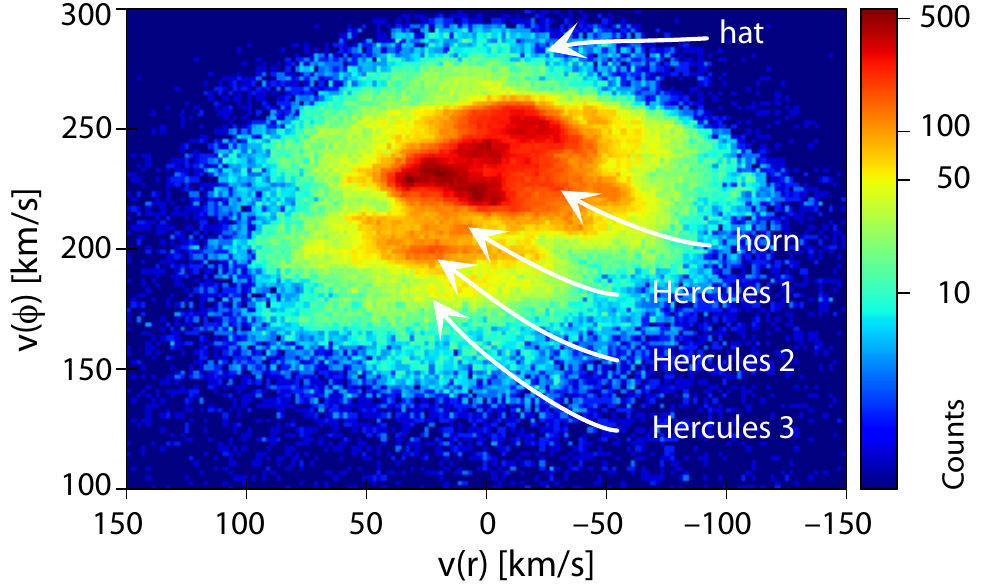}
\hspace{5pt}
\includegraphics[width=0.36\linewidth]{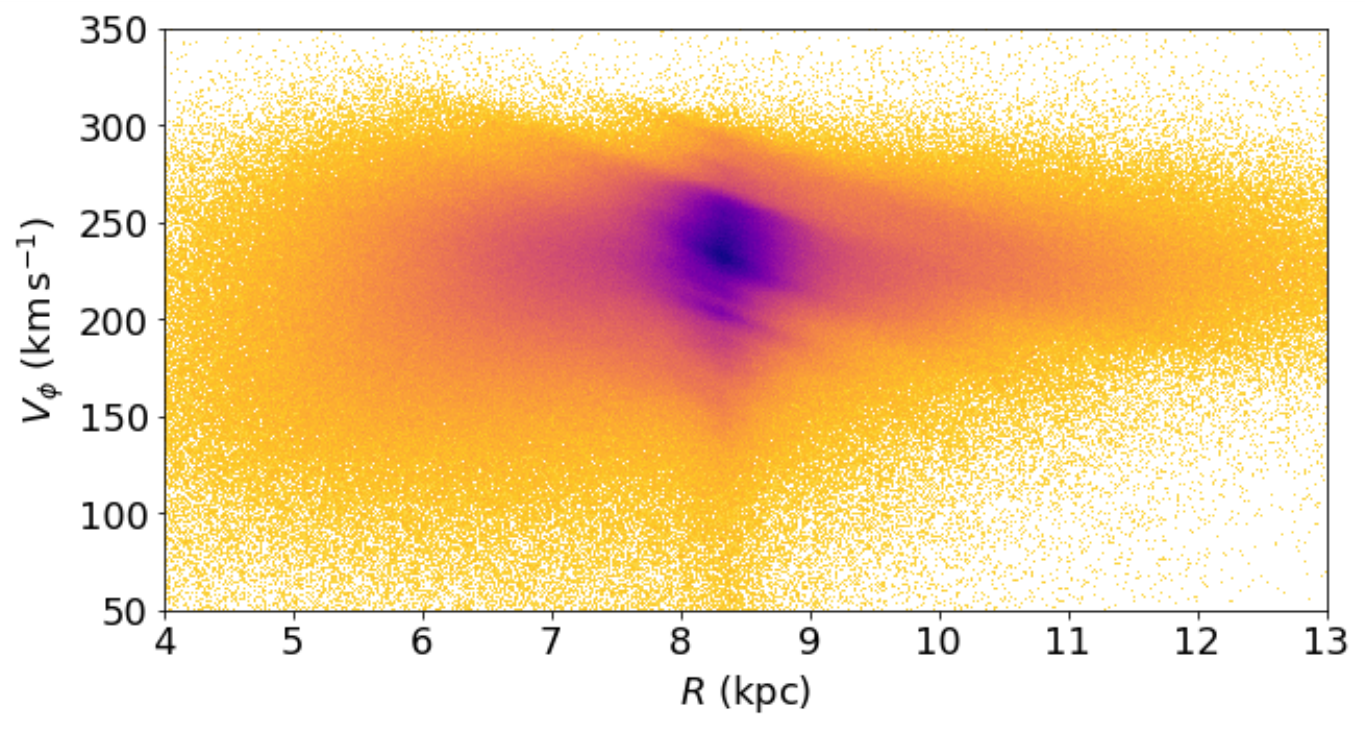}

\vspace{-5pt}
\caption{Velocity structure of the solar neighbourhood. 
(a):~pre-Gaia velocity distribution of F and G dwarfs in the Hipparcos-based catalogue of \citet{2004A&A...418..989N}, showing the Hercules stream, the Hyades--Pleiades cluster, and the Sirius moving group \citep[][Figure~1]{2007ApJ...655L..89B}.
(b):~radial and angular velocity-space distribution of Gaia DR2 stars within 200\,pc, with relative parallax errors $<$10\%, and in bins of $2\times2$\kms. The stream denoted here as `Hercules~3' is identical to Eggen's moving group HR~1614 \citep[][Figure~1]{2020MNRAS.499.2416A}.
(c):~azimuthal velocities as a function of Galactocentric radius from Gaia DR2 data, in bins of $\Delta V\phi=1$\kms, and $\Delta R=0.01$\,kpc \citep[][Figure~2]{2018Natur.561..360A}.
}\label{fig:hercules-stream}
\end{figure}

\paragraph{Hercules stream}
\label{sec:hercules-stream}

This is a prominent moving group, trailing behind the local rate of Galactic rotation, and moving outwards in the disk. Pre-Gaia, only some 100 stars were known as members
\citep[][see Figure~\ref{fig:hercules-stream}a]{2007ApJ...655L..89B}.  
With many tens of thousands of stars now in the relevant velocity slices, Gaia is transforming its characterisation, in turn allowing detailed dynamical modelling to clarify its origin. 
Gaia DR2 revealed considerable trimodal structure in velocity space
\citep[][their Figure~22]{2018A&A...616A..11G}, 	
dominated by three sub-streams at $v_{\phi} \simeq 220$, 200, and 180\kms. 
These, and other detailed structures (variously referred to as the `horn' and the `hat'), have also been identified in the Gaia DR2 
\citep[e.g.][]{2017MNRAS.465.1443M,	
2018MNRAS.477.3945H,	
2018A&A...619A..72R,	
2019MNRAS.488.3324F,	
2019MNRAS.490.1026H,		
2019MNRAS.484.3291T,	
2020MNRAS.499.2416A,	
2020ApJ...890...85L}.	
\citet{2000AJ....119..800D}
modelled the stream by calculating positions and velocities in a static $m=2$ bar potential, 
finding that stars trapped in the bar's 2:1 outer Lindblad resonance can create moving groups in the solar neighbourhood. The position of the structure in phase space\footnote{
In dynamical systems, a phase space is a space in which all possible states of a system are represented, with each possible state corresponding to one unique point in the phase space. For stellar systems, the 6-dimensional `phase space' consists of the 3d position and 3d momentum variables.}\ depends on the location of the outer Lindblad resonance, which in turn depends on the pattern speed of the bar, $\Omega_{\rm B}$. To reproduce the observed features, he required a fast-rotating bar, $\Omega_{\rm B}= 53\pm3$\kmskpc. 

Most analyses with Gaia~DR2 suggest lower bar pattern speeds in the range 35--40\kmskpc\ 
\citep{2021MNRAS.505.2412C,	
2019MNRAS.489.3519C,		
2019MNRAS.488.4552S,	
2019MNRAS.490.4740B, 
2022MNRAS.512.2171C}.		
Different values and interpretations continue to be reported from EDR3, ranging from 34--55\kmskpc\ 
\citep{2021MNRAS.508..728K,	 
2021MNRAS.507.4409M}.	
The disparity in these inferred bar pattern speeds has led to other origins of the stream being suggested. While most agree that resonant orbits due to non-axisymmetric structures, such as a bar and/or spiral arms, are responsible for the moving groups in the solar neighbourhood, the details differ. 
Amongst these are suggestions that a Hercules-like stream can be made of stars orbiting the bar's Lagrange points
\citep{2017ApJ...840L...2P,	
2024MNRAS.531L..14L},		
or as higher-order (4:1 or 6:1 outer Lindblad) resonances of a slow bar
\citep{2018MNRAS.477.3945H,		
2017MNRAS.465.1443M}, 	
or associated with the spiral's 8:1 inner Lindblad resonance
\citep{2018A&A...615A..10M},
or as a combination of resonances due to the bar plus spiral potentials 
\citep{2019MNRAS.484.4540H}.
From an N-body simulation of the entire Milky Way, using 5.1~billion particles in the stellar bulge, bar, disk, and halo, evolved over 10~Gyr,
\citet{2020MNRAS.499.2416A}	
concluded that the stream is dominated by the 4:1 and 5:1 outer Lindblad and corotation resonances, comprising some 100\,000 Gaia stars (15\% of the total) together yielding a trimodal structure, and favouring a slow pattern speed of 40--45\kmskpc. In addition, the stars in the 2:1 and 3:1 outer Lindblad resonance match the `hat' and `horn' structures respectively (Figure~\ref{fig:hercules-stream}b).
Another representation of this phase-space structure, in azimuthal velocity $V_\phi$ versus radius $R$, was given by \citet{2018Natur.561..360A}
in their study of the Gaia `phase-space spiral' (Section~\ref{sec:phase-space-spiral}), similarly revealing a number of thin diagonal ridges (Figure~\ref{fig:hercules-stream}c).

Amongst more recent models
\citep{2022MNRAS.514..460A,		
2022MNRAS.509..844T,		
2025MNRAS.539.1595L},		
those by 
\citet{2021MNRAS.500.4710C}
were used to argue that the Galaxy's central bar is {\it decelerating}, consistent with the effects of dynamical friction of a dark matter halo.  As the bar slows, its resonances sweep through phase space, dragging along a portion of previously free orbits. This leads to multiple resonances seen in the Gaia data and, in particular, reproducing the details of the Hercules stream. They derived a current slowing rate of the bar of $-4.5\pm1.4$\kmskpc\ per Gyr.

\paragraph{Arcturus and HR~1614 streams}
\label{sec:arcturus-hr1614}

Amongst other kinematic groups in the solar neighbourhood, the Arcturus stream and the HR~1614 kinematic group have both been subject to debate as to their nature. And for both, Gaia has thrown new light on their origins.

The Arcturus stream was first described by 
\citet{1971PASP...83..271E},
who discovered an over-density of some 50 stars, including Arcturus, with similar space velocity components, $V\simeq-100$\kms. Today, Eggen's idea that the stars had escaped from an open cluster is disfavoured, especially given the spread of chemical abundances. Rather, pre-Gaia, three other origins of the Arcturus stream had been proposed. 
The first attributes the stream to an accretion event, in which a small satellite galaxy merged with the Milky Way
\citep{2004ApJ...601L..43N, 	
2006MNRAS.365.1309H}.		
The arguments are that their angular momentum is too low to arise from dynamic perturbations induced by the bar, while their low metallicity and similar apocentres are consistent with them originating from an accreted satellite. 
Nonetheless, other work has not ruled out that it might be a resonance feature associated with the Galaxy's bar
\citep{2009ApJ...700L..78A,
2010MNRAS.405..545G}.
Another suggestion was that it represents an unrelaxed disk population which might have been left following a perturbing event some 1.9\,Gyr ago
\citep{2009MNRAS.396L..56M}.

\citet{2019A&A...631A..47K} used Gaia DR2
to construct space velocities, angular momenta, and actions (conserved quantities that characterise stellar orbits) for more than 5.8~million stars within 5~kpc. They used a wavelet transform to characterise kinematic over-densities in the disk, and used these to select possible group members from spectroscopic surveys (GALAH and APOGEE) to study their chemical properties.
%
They actually searched for streams in four different planes defined by combinations of velocity ($UVW$), angular momentum ($L$), and action components ($J$).\footnote{
These quantities are commonly used in kinematic searches. In Galactic $UVW$ coordinates, $U$ is in the direction of the Galactic centre, $V$ in the direction of Galactic rotation, and $W$ is towards the North Galactic Pole.
The $U-V$ plane identifies kinematic over-densities in the plane of the Galaxy, irrespective of their orbits or assumptions about the Galaxy potential.
$V$ is proportional to $L_z$, a vertical component of the angular momentum, and is an integral of motion in axisymmetric potentials, while $\sqrt{(U^2+2V^2)}$ is a measure of orbital eccentricity.
$L_z$ and $\sqrt{(L_x^2+L_y^2)}$ are integrals of motion, used in searches for stars on similar orbits. 
The most general search for kinematic structures is in action space (conserved quantities in stellar orbits), and uses the radial and azimuthal actions $J_r$ and $L_z$ that quantify orbital eccentricity and orbital angular momentum. 
Gaia is central to these studies because all of these quantities require knowledge of the stellar distances, proper motions, and radial velocities.} 
In $U-V$ velocity space, and in angular momentum space, the Arcturus stream, along with other known streams (Sirius, Pleiades, Hyades, Hercules, AF06, and KFR08) are clearly identified, all separated by about 20--30\kms\ in azimuthal velocity, $V$ (their Figure~3).
While the Hercules stream appears to be a mixture of thin and thick disk stars, the Arcturus stream (as well as the AF06 and KFR08 streams) is a high-velocity and low-angular momentum structure with chemical composition similar to the thick disk, and extending further from the Galactic plane than the Hercules stream. 
\citet{2019A&A...631A..47K}
concluded that the Arcturus stream, together with the AF06 and KFR08 streams, have properties different to those of bar-driven structures, and are instead likely to be part of a `phase-space wave', possibly caused by an ancient merger event. In any case, they convincingly exclude its origin as an evaporating cluster. 
As \citet{2004ApJ...601L..43N} reflected:
{\it `It is oddly gratifying to think of stars visible to the naked eye, such as Arcturus, as silent night-sky witnesses of the merging history of the Milky Way.'}

HR~1614 is another overdensity in velocity space, discovered and classified as a nearby moving group in 1978 by
\citet{1978ApJ...222..203E}.
He identified 40 stars (including HR~1614) with a mean rotational velocity $V\simeq-60$\kms, higher than solar metallicity, an age of $\sim$5\,Gyr, and which he attributed to an old dissolving cluster. It was re-examined using Hipparcos data by
\citet{2000A&A...357..153F}, 
and also explained as a disrupting metal-rich open cluster, but of age $\sim$2\,Gyr.
\citet{2020A&A...638A.154K}	
used Gaia DR2 astrometry, with APOGEE and GALAH spectroscopy, to refine the group membership and kinematics. They found that the group does not form an `arch' of constant energy in $U-V$ space, and is tilted in $V$. Its stars have a wide range of metallicities, ages, and abundances, similar to the thin and thick disks. They conclude that it is not a dissolving open cluster, or an accreted population, but rather has a complex origin perhaps best explained by several different mechanisms such as resonances with the Galactic bar and spiral structure, phase mixing of dissolving spiral structure, and phase mixing due to an external perturbation.

\paragraph{The future}
More recently,
\citet{2023MNRAS.519..432L}	
described an open-source 2d wavelet transformation code, {\tt MGwave}, based on similar wavelet techniques used in previous searches, but also allowing for the investigation of {\it underdensities}, presumably also relevant in probing the Milky Way's non-axisymmetric features. 
Applied to Gaia DR3, they detected 47 groups with coherent velocities, recovering the majority of the previously known moving groups, as well as identifying three additional significant candidates: one within Arcturus, and two in regions without much substructure at low $V_R$, and extending the searches to Galactocentric radii 6.5--10\,kpc.

Gaia's probing of the kinematic structures near the Sun is proving to be a powerful tool for untangling the structure of our Milky Way. With improving data, numerous complex kinematic structures are being better characterised, and new ones continue to be discovered. 
The final word on these stream has not yet been spoken, but future Gaia releases will surely further clarify these remarkable dynamical structures. Olin Eggen would have been delighted.

\subsection{The Gaia phase-space spiral}
\label{sec:phase-space-spiral}

Another addition to these phase-space structures has been the remarkable discovery of what is called the `Gaia phase-space spiral', or sometimes the `phase-plane spiral'. The description simply refers to the fact that this new, and very distinct, kinematic feature becomes evident in a graph of vertical motion in the Galaxy versus vertical position, i.e.\ in (the phase plane) $Z$ versus $V_Z$, rather than (say) the space velocities in the plane of the Galaxy, $V_R$ versus $V_\phi$.\footnote{Galactic coordinates are mostly specified as $UVW$, where $U$ is in the direction of the Galactic centre, $V$ in the direction of Galactic rotation, and $W$ is towards the North Galactic Pole. Here, $V_R$, $V_\phi$, $V_Z$ adhere to the notation used by the discovery authors.}
This particular feature was first noticed in the Gaia DR2 data by
\citet{2018Natur.561..360A},	
who found that certain phase space projections show completely new and unexpected substructures. 
They used Gaia DR2 sources with computable 6d phase space coordinates, i.e.\ all sources with available 5-parameter astrometric solution (positions, parallax and proper motions) {\it and\/} radial velocities. They selected stars with parallax errors smaller than 20\%, resulting in more than 6.3~million stars. 
They adopted a vertical distance of the Sun above the plane of 27\,pc, a distance of the Sun to the Galactic centre of 8.34\,kpc, and the Sun's circular velocity of 240\kms. 
They then selected the more than 930\,000 stars located in a local Galactic cylindrical ring, $R=8.24-8.44$~kpc, for which their median errors in $V_R$, $V_\phi$, $V_Z$ are 0.5, 0.8, and 0.6\kms\ respectively.

\begin{figure}[t]
\centering
\raisebox{2pt}{\includegraphics[width=0.26\linewidth]{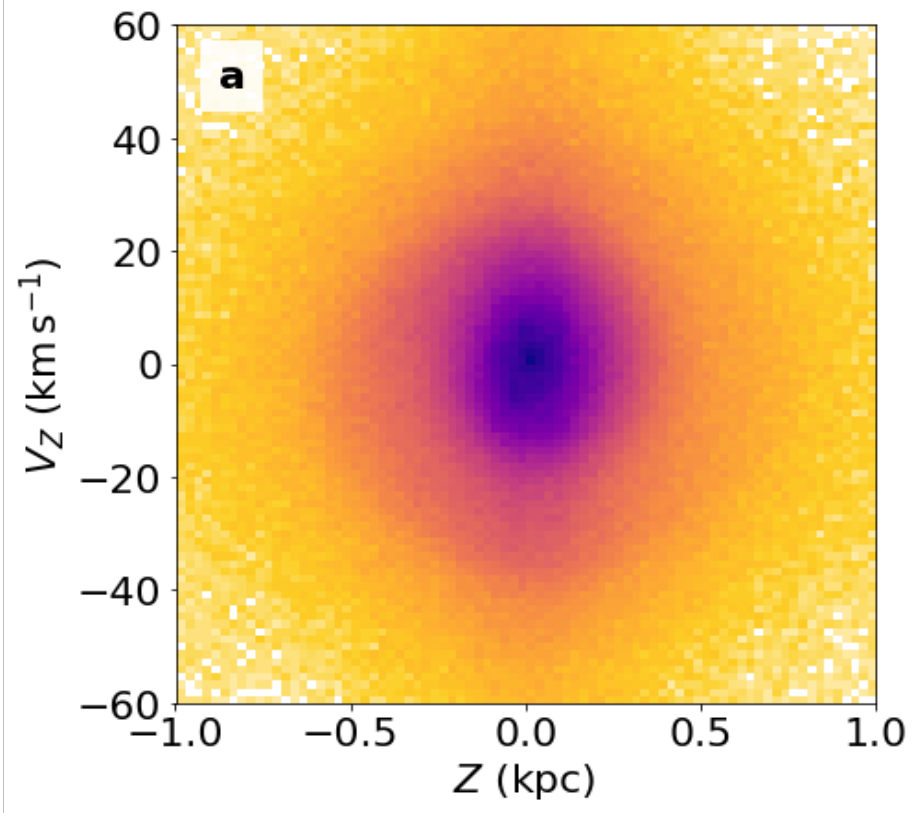}}
\hspace{10pt}
\includegraphics[width=0.66\linewidth]{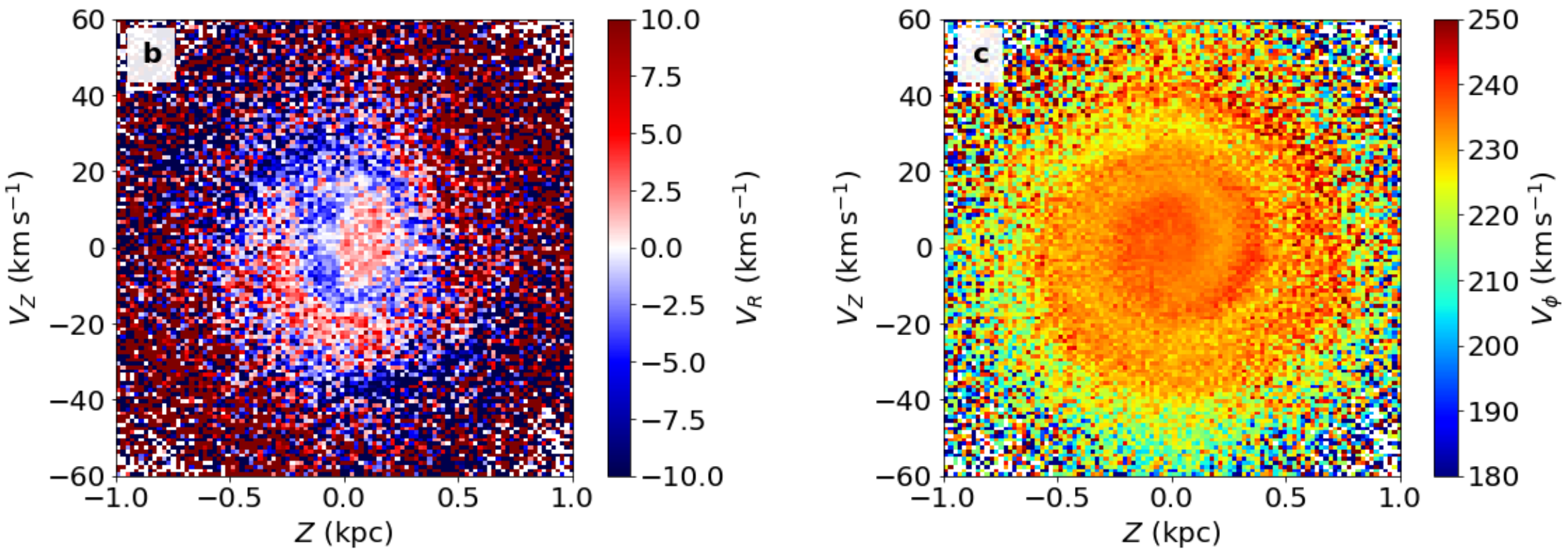}
\vspace{-5pt}
\caption{Distribution in the vertical position--velocity plane for 930\,000 stars from Gaia DR2 with parallaxes $<20$\% and full space velocities, and in the local Galactic cylindrical ring defined by $8.24<R<8.44$\,kpc. (a)~Two-dimensional histogram in bins of $\Delta Z=0.01$\,kpc and $\Delta V_Z=0.1$\kms, with intensity proportional number counts.
(b)~the $Z-V_Z$ plane coloured as a function of median $V_R$ in bins of $\Delta Z=0.02$\,kpc and $\Delta V_Z=1$\kms.
(c)~the $Z-V_Z$ plane coloured as a function of median $V_\phi$ in bins of $\Delta Z=0.02$\,kpc and $\Delta V_\phi=1$\kms\ \citep{2018Natur.561..360A}.
}\label{fig:phase-space-spiral}
\end{figure}

Figure~\ref{fig:phase-space-spiral}a shows their results in the vertical position--vertical velocity plane, $Z-V_Z$.\ The data is in bins of $\Delta Z=0.01$~kpc and $\Delta V_Z=0.1$\kms, and the intensity is proportional to the number counts. 
While spiral structure is evident in this image, albeit rather faintly, it becomes more prominent when coloured as a function of median $V_R$ (Figure~\ref{fig:phase-space-spiral}b, here with bins of $\Delta Z=0.02$~kpc and $\Delta V_Z=1$\kms), and even more so as a function of $V_\phi$ (Figure~\ref{fig:phase-space-spiral}c).
Although this kind of `phase-wrapping' had been predicted to occur in the disk after a passage of a satellite galaxy
\citep{2009MNRAS.396L..56M,		
2012MNRAS.419.2163G,	
2015MNRAS.454..933D},	
it had not been observed previously.
From numerical modelling, 
\citet{2018Natur.561..360A}
inferred that the Milky Way disk was perturbed 300--900~Myr ago, consistent with estimates of the last pericentric passage of the Sagittarius dwarf galaxy
\citep{2018MNRAS.481..286L}.		

But a consensus has not yet been reached.
Subsequent studies have continued to probe its origin, studying phase mixing after a tidal perturbation by a dwarf galaxy
\citep{2018MNRAS.481.1501B,
2019MNRAS.482.1417B,		
2019MNRAS.485.3134L},
or the result of bar buckling 
\citep{2019A&A...622L...6K,
2020A&A...638A.144K},
or as a persistent dark matter wake 
\citep{2023MNRAS.524..801G}.
\citet{2018MNRAS.481.1501B}
emphasised that the frequency at which stars oscillate vertically depends on their angular momentum about the $Z$-axis in addition to the amplitude of the star's vertical oscillations, such that spirals should form in both $V_\phi$ and $V_R$ whenever a massive substructure, such as the Sagittarius dwarf galaxy, passes through the Galactic plane. 
\citet{2019MNRAS.485.3134L}	
reached similar conclusions, further arguing that $Z-V_Z$ looks the same in all age bins, down to the youngest ages, thus ruling out a bar-buckling origin.
Meanwhile 
\citet{2022MNRAS.516L...7H}
extended the analysis, discovering a transition to two armed `breathing spirals' in the inner Milky Way, concluding that the local data actually contain signatures of {\it multiple\/} perturbations, with the prospects of using their distinct properties to infer the properties of the interactions that caused them.

\subsection{Spiral structure}	
\label{sec:spiral-arms}

The Milky Way is a barred spiral galaxy with a diameter of some 30~kpc, and a thin disk of thickness 220--450~pc. The solar system lies a little more than 8\,kpc from the Galactic centre, close to the mid-plane, on the inner edge of the Orion spiral arm. Outwards from the central bar, the disk is organised into a series of spiral arms, delineated by a higher density of gas and dust, and more pronounced ongoing star formation. 
There are, arguably, four main arms: Perseus, Norma, Scutum--Centaurus, and Carina--Sagittarius. There are at least two smaller arms or spurs, including the Orion--Cygnus arm which includes the Sun.
But its detailed structure is difficult to establish from our location deep within the disk.
Some recent work suggests that there may be two spiral arms traced out by old stars, and four arms more defined by gas and young stars.
Others have proposed that there are two different spiral patterns: the inner Sagittarius arm, and the outer more tightly wound Carina and Perseus arms with a slower rotation.
Given that other galaxies are often seen to have arms that branch, merge, and twist, with varying degrees of irregularity, it is no surprise that our Galaxy's spiral structure is complex, and that the details remain uncertain. 

Observations probing the spiral-arm structure are many and varied. They include H\,{\scriptsize I} regions at 21\,cm, H$_2$O masers with VLBA, molecular clouds in CO, H\,{\scriptsize II} regions probed by pulsars, red clump stars from 2MASS and Spitzer, cluster mapping by WISE, and many others.
Nor is there a consensus on the origin and evolution of the spiral structure: it may be that density waves (and central bars) arise as an inevitable consequence of stellar orbits, perhaps initiated by starburst shocks
\citep[e.g.][]{2013seg..book....1K}.
A bigger picture is given in recent reviews
of the overall properties of our Galaxy 
\citep{2016ARA&A..54..529B},
of its bar and spiral arms
\citep{2020RAA....20..159S},	
and of spiral arms more generally
\citep{2022ARA&A..60...73S}.
The latter argue that the spontaneous development of spiral patterns in simulations of isolated disks results from a recurrent cycle of `groove modes', which they describe as a deficiency over a narrow range of angular momentum that seeds a linear instability, itself creating new grooves (at the Lindblad resonances of the original mode), setting up a recurrent cycle. 
The alternative quasi-steady density-wave theory seeks to explain long-lived spiral patterns without the need for constant regeneration.

Choosing amongst these theories is most robustly done by matching the observed (spiral-driven) large-scale streaming motions to these predictive models.
But a major challenge of further elucidating their structure clearly rests on knowledge of the distances and dynamics of their constituent stars. Hipparcos, restricted to the nearby Perseus and Orion arms, and using young bright OB stars and Cepheids within 2--3\,kpc, allowed for some studies of their pattern speed, location of the co-rotation circle, and implications for moving groups. 

One of the early Gaia DR1 studies used the distances and motions of 77 Cepheids within 1.5\,kpc of the Perseus arm
\citep{2018ApJ...853L..23B}.	
They found that both radial and rotation velocities are correlated with distances from the arm, and found a dependency of their vertex deviation attributed to the spiral arm. Using \mbox{N-body} and hydrodynamic simulations based on two models (transient dynamic versus a static density-wave), their results favoured a model in which the Perseus arm is in the process of disruption by a transient arm.
A major observational advance came with Gaia DR2, which provided some 6.4~million FGK stars with full 6d phase space coordinates (viz.\ position and velocity), parallaxes better than 20\%, and precise Galactic velocities with uncertainties 0.9--1.4\kms\ 
\citep{2018A&A...616A..11G}.	
They used a sub-sample of 3.2~million giants to map the velocity field of the disk between 5--13~kpc from the Galactic centre (the innermost region being restricted by heavy dust extinction), and up to 2~kpc above and below the plane. The velocities of 300\,000 stars within 200~pc revealed streaming motions in all three velocity components, small-amplitude fluctuations in the velocity dispersions, and striking ridge-like features in the stellar azimuthal velocity distribution as a function of Galactocentric radius. 
This rich substructure in the phase space distribution of solar neighbourhood stars indicates that the local disk of the Milky Way is far from a settled well-mixed state.
Nevertheless, a fit to a steady-state spiral model gave an arm relative amplitude of 10\%, with a pitch angle of $12^\circ$
\citep{2020ApJ...900..186E}. 
From 8292 clusters, comoving groups, and other stellar structures, also from DR2,
\citet{2020AJ....160..279K} 	
found that the Sagittarius arm has moved by $>500$\,pc in the last 100~Myr, and that the Perseus arm has been relatively low in star formation activity over the last 25\,Myr, together confirming the transient nature of the spiral arms.

With EDR3,
\citet{2021A&A...652A.162C}
used open clusters to map the spiral arms, and to determine their pattern speed. They found a declining pattern speed with radius, along with an absence of age gradients downstream from the arms.
\citet{2023A&A...673A..99U}	
constructed counts of red clump stars extracted from 2MASS, using EDR3 astrometry and photometry to remove foreground stars. The found an overdensity of red clump stars which traces the continuous morphology of the Outer arm from the second to the third Galactic quadrant. They also found a wave-like asymmetry above and below the Galactic plane with respect to longitude, indicating a warp-like structure. 

\begin{figure}[t]
\centering
\includegraphics[width=0.90\linewidth]{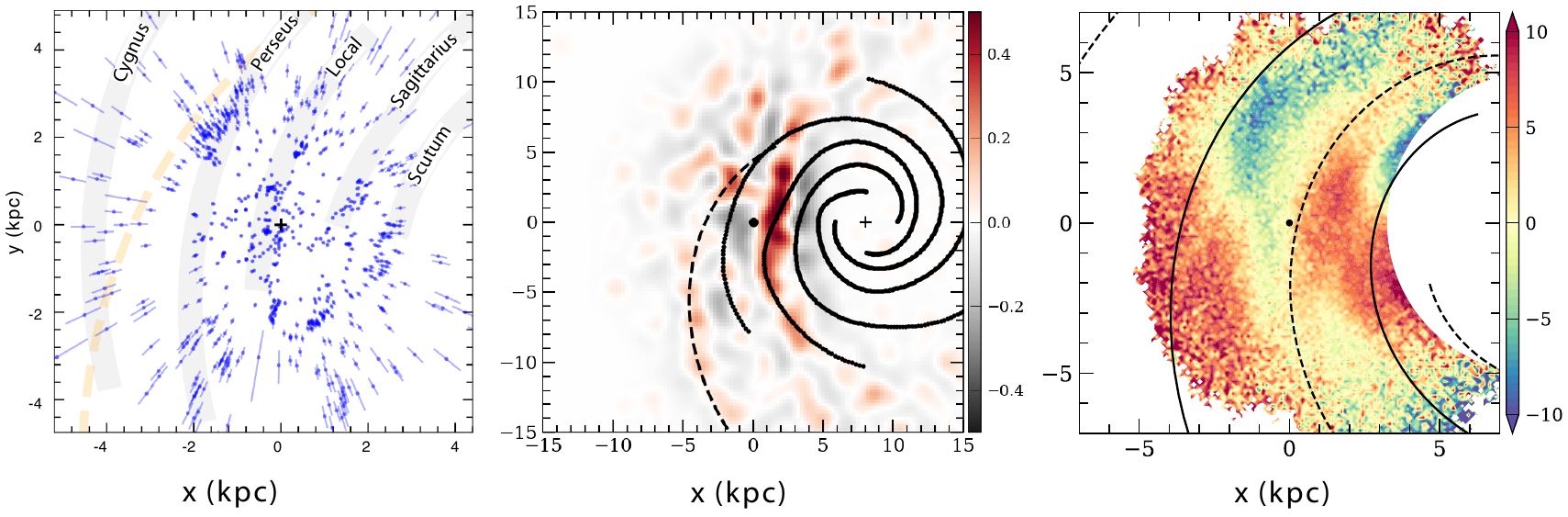}
\vspace{-5pt}
\caption{The Galaxy's spiral arm structure from Gaia DR3. 
Left: the distribution of young open clusters ($\tau<63$\,Myr) showing, amongst other features, a discontinuity in the Perseus arm.
Middle: the distribution of young Cepheids ($\tau<200$\,Myr), using a wavelet transform, compared with some literature models.
Right: inferred radial velocity for the red giant branch stars. Overlaid is an earlier 2-arm spiral model (solid lines), with the minimum inter-arm density (dashed lines). The Sun's location is marked by a black dot.
From \citet{2023A&A...674A..37G}, Figures~14, 15b, 22a.
}\label{fig:drimmel-spiral-structure}
\end{figure}

A further leap in data quantity and quality came with Gaia DR3 and, with it, major mapping of the Galaxy's asymmetric disk by   
\citet{2023A&A...674A..37G}.	
The number of sources with complete 6d phase space information increased to over 33~million, with astrophysical parameters for 470~million, and variability classification for 11~million (Figure~\ref{fig:drimmel-spiral-structure}). 
Using 580\,000 OB stars, together with 988 young open clusters ($\tau<100$\,Myr), they mapped the spiral structure associated with star formation out to 4--5\,kpc from the Sun. Beyond that, 2800 classical Cepheids ($\tau<200$\,Myr) reveal spiral features in the outer disk extending to 10\,kpc from the Sun. 
These young populations show the Local (Orion) arm to be at least 8\,kpc in length, with an outer arm consistent with that seen in H\,{\scriptsize I} surveys, which appears to be a continuation of the Perseus arm. The subset of red giant branch stars reveals the large-scale kinematic signature of the inner bar, as well as evidence of streaming motions in the outer disk that might be associated with spiral arms or bar resonances.
Evidence of streaming motions associated with spiral arms is, they conclude, less compelling than that due to the bar, but consistent with the two-armed structure seen in the near-infrared. 
Further mapping of the spiral arms using Gaia DR3 combined with LAMOST and other large-scale surveys, is continuing
\citep{2023MNRAS.525.3318H,	 
2020AJ....160..279K,		
2023ApJ...947...54X}.			

A completely different model for the generation of spiral arms attributes them to an interaction with the Sagittarius dwarf galaxy. 
\citet{2022A&A...668A..61A}	
modelled such a tidal encounter, and found a strong similarity between the resulting kinematic signatures, in the form of ridges and waves in angular momentum, to those observed in the Gaia DR2 data. Specifically, they found that an impulsive distant tidal approach generates a perturbation in velocities that leads to a two-armed spiral structure, and a `winding time' in the range 0.8--2.1~Gyr. 

In their review,
\citet{2022ARA&A..60...73S}
argue that the ubiquity of spiral patterns in the stellar disks of other galaxies appears to demand an explanation in which they largely result from self-excited disk instabilities. Other driving mechanisms, such as bars and tidal encounters, may result in spiral responses in specific cases, but such external drivers could not, they argue, account for all, or perhaps even most, spiral patterns. But they also emphasised that we still lack compelling evidence that the recurrent cycle of `groove modes', identified as the mechanism for spiral generation in numerical simulations, actually operates in real galaxies. 
Future releases of Gaia data may, they surmise, yield stronger evidence in the case of the Milky Way.

\paragraph{Breathing motion}
\label{sec:spiral-arms-breathing}

Already pre-Gaia, the existence of large-scale non-zero mean {\it vertical\/} motions with respect to the disk had been identified in various surveys, notably from
SEGUE \citep{2012ApJ...750L..41W,2014MNRAS.440.1971W}, 
LAMOST \citep{2013ApJ...777L...5C},
and RAVE \citep{2013MNRAS.436..101W}.
These fall into two distinct classes:
a `bending motion' in which stars on either side of the mid-plane move together in the same direction,
and a symmetric `breathing motion' in which stars on either side move together in opposing directions, both sides either moving towards or away from the mid-plane.
\citet{2014MNRAS.440.1971W} suggested that both could be excited by a passing satellite or dark matter subhalo, depending on the impactor's velocity. 
But certain puzzling features led
\citet{2014MNRAS.440.2564F} 	
and 
\citet{2014MNRAS.443L...1D}	
to investigate whether they could arise in the absence of such an external perturber. 
The former showed that in an equilibrium axi\-symmetric disk, in which the mean radial and vertical velocities might be expected to be zero everywhere, their 3d simulations showed instead that the global stellar response to a spiral perturbation induces both a radial velocity flow, and non-zero vertical motions, with a resulting mean velocity field similar to that observed from SEGUE, LAMOST, and RAVE.
In more detailed models, \citet{2014MNRAS.443L...1D} confirmed that spiral structures indeed induce bulk vertical velocities, as large as 10--20\kms, being compressive (pointing towards the mid-plane) as stars enter the spiral, and expanding (away from the mid-plane) as they leave it. 
Subsequent analytical models 
\citep{2016MNRAS.457.2569M}, 
further \mbox{N-body} simulations 
\citep{2016MNRAS.461.3835M,
2022MNRAS.511..784G},	 
and even large-scale cosmological simulations 
\citep{2016MNRAS.460L..94G} 
have confirmed these bending and breathing modes.

Such bending and breathing modes in the solar neighbourhood were soon identified:
in the RAVE--TGAS data \citep{2018MNRAS.475.2679C}, 
in a 3.2~million DR2 giant star sample 
\citep{2018A&A...616A..11G},	
and in the number counts (2~million stars) and velocity distributions (865\,000 stars) in DR2 \citep{2019MNRAS.482.1417B}.
For 3.1~million Gaia DR2 stars within 10\,kpc, with the distances and ages given by 
\citet{2018MNRAS.481.4093S},
\citet{2022MNRAS.511..784G} 
demonstrated that, over the region with the largest breathing motion
($x=7.6-8.1$\,kpc, $y=0.9-1.4$\,kpc),		
the amplitude increases monotonically with distance from the mid-plane, and decreases with age, i.e.\ the breathing motion is strongest for the youngest stellar populations. They concluded that the observed breathing motions are indeed driven by spiral density waves, while the bending motions are not.

These findings became still more convincing with the improved accuracies of Gaia DR3, along with more robust estimates of the various selection effects entering number density estimates.
\citet{2022A&A...668A..95W} used several million DR3 stars to distinguish three distinct structures: the Gaia phase--space spiral (Section~\ref{sec:phase-space-spiral}), a large-scale bending mode in both number density and vertical velocity, and an elongated over-density in star counts with a corresponding breathing-mode compression in vertical velocity at the location of the Local Arm.
\citet{2024MNRAS.529L...7A}		
used 26~million DR3 stars to demonstrate a clear alignment of the compressing breathing motion with the Local Arm, similar to that seen in the growth phase of spiral arms in numerical simulations. They concluded that the Local Arm's compressing breathing motion can be explained by it being in the growth phase of a transient and dynamic spiral arm. They also found tentative signatures of the expanding breathing motion associated with the Perseus arm, and a compressing breathing motion coinciding with the Outer arm, implying that the Perseus and Outer arms are in the disruption and growth phases respectively. In their face-on map of number counts (their Figure~1), the spiral arms are not evident due to the various selection biases, but the alignment between the Local Spiral Arm and the compression mode in the Gaia data is particularly striking.

\subsection{The warp}	
\label{sec:warp}

Many galaxies, including our own, display a warped structure in both gas and stars, with an outer disk twisted up at one side, and down at the other (e.g.\ as illustrated graphically 
\href{https://www.esa.int/Science_Exploration/Space_Science/Gaia/Milky_Way_s_warp_caused_by_galactic_collision_Gaia_suggests}{here}). Various explanations have been proposed, including infall of intergalactic material, or tidal forcing through a close encounter with a companion galaxy
\citep[e.g][]{2024ApJ...975...28D}.
But the existence of many {\it isolated\/} warped galaxies may signify some general misalignment of the symmetry axes of the disk and its dark matter halo. In such cases, gas in the outer disk would settle into the symmetry plane of the halo, rather than of the disk. And, importantly, such a tilted disk, embedded in a massive halo, would precess.   
Pre-Hipparcos studies mainly focused on its {\it spatial\/} structure (using bright young objects such as Cepheids and OB~stars), because its kinematic signature is most evident in the velocities tangent to the line-of-sight when viewed from in the plane.  
A warping motion, from the ACRS-based proper motions of 2000 young stars out to 0.5--3\,kpc, was detected by
\citet{1993AJ....105.2138M}.
They concluded that young stars are streaming around the Galactic centre in a tilted sheet with a velocity of 225\kms, itself rotating around the nodal line with an angular velocity of 4\kms\ per kpc.   
But
\citet{1992ARA&A..30...51B}
concluded that their detailed dynamical nature, and driving mechanism, remained uncertain. In the review by
\citet{2017A&A...601A.115P}
25~years later, this conclusion remained largely unchanged.

The Hipparcos catalogue provided proper motions with systematic and zonal errors significantly smaller than the expected warp signature.  Based on 2422 distant OB stars towards the Galactic anti-centre,
\citet{1998Natur.392..471S}
showed that the disk is flat out to approximately the solar radius, then turns up to the north in the direction of Cygnus at $l\simeq90^\circ$, and south in the direction of Vela at $l\simeq270^\circ$. But while the spatial distribution of stars was consistent with studies based on neutral hydrogen, the velocity distribution was of opposite sign.  Thus the stellar kinematics did not follow the signature of a long-lived warp, whether precessing or not. These results were largely confirmed by
\citet{2000A&A...354...67D}.
A partial explanation was put forward by
\citet{2000MNRAS.311..733I}:
they found that kinematic warps in oblate halos `wind up' and disappear within a few dynamical times, while they persist in prolate halos. 
Meanwhile, cosmological simulations were suggesting not only that dark matter halos surrounding individual galaxies are highly triaxial, but also that the fraction of prolate and oblate halos is roughly equal 
\citep{1991ApJ...378..496D}. 

Gaia has brought the possibility of probing the kinematic signature of the warp well beyond the solar circle.
With DR1, 
\citet{2017A&A...601A.115P}
selected 758 OB~stars with distances between 0.5--3~kpc. They found that the proper motions of nearby stars are consistent with a kinematic warp, while stars beyond 1~kpc are not, and that the systematic vertical disk motions cannot be explained by a stable long-lived warp. The concluded that it is either a transient feature, or that additional forces are causing systematic vertical motions that are masking the warp signal. 
\citet{2018MNRAS.478.3809S}
also used the early Gaia--TGAS distances and proper motions to estimate the vertical and azimuthal velocities in the Galactic centre and anti-centre directions. The mean vertical motions showed a linear increase with distance, consistent with the known Galactic warp. But they also revealed a previously unknown wave-like pattern, with an amplitude of 1\kms\ and a wavelength of 2.5~kpc, in both directions. 

Numerous studies followed the release of Gaia DR2 in 2018.
\citet{2018MNRAS.481L..21P}	
extended these kinematic studies (using 600\,00 upper main sequence stars and more than 12~million giants) out to a distance of 7~kpc from the Sun. Both populations showed a clear warp signature, seen as a gradient of 5--6\kms\ in the vertical velocities from 8--14~kpc in Galactic radius. 
\citet{2019A&A...627A.150R}	
used DR2 to examine the vertical motions of two populations up to Galacto\-centric distances of 16~kpc: a young bright sample of OB stars, and an older one comprising red giants. They confirmed the age dependency of the warp, both in position and kinematics, its height being around 0.2~kpc for the OB sample and 1~kpc for the older sample at a Galacto\-centric distance of 14~kpc, with an onset radius of the warp at 12--13~kpc for the former, and 10--11~kpc for the latter. But they also confirmed a high degree of complexity in both position and velocity, including a prominent wave-like pattern of a bending mode, different in the two samples. 
A study based on 2431 Cepheids from DR2 revealed asymmetric warp-like large-scale vertical motions with amplitudes 10--20\kms\
\citep{2019AcA....69..305S}. 
Similar features were found for 250\,000 OB~stars, where the flat inner disk begins to warp at $\sim$9~kpc from the Galactic centre 
\citep{2019ApJ...871..208L}, 
and in a study of 140\,000 red clump stars from LAMOST DR4 and Gaia DR2
\citep{2020ApJ...897..119W,	
2020ApJ...901...56L}. 
In later models,
\citet{2020ApJ...905...49C}	
derived a starting radius for the flare at $8.9\pm0.1$\,kpc, and a precession rate of $13.6\pm0.2$\kmskpc\ in the direction of Galactic rotation.
From their 12~million giants in DR2,
\citet{2020NatAs...4..590P}		
found a similar precession of $10.86\pm0.03$\kmskpc, about one-third the angular rotation velocity at the solar circle. 

Studies continued with EDR3/DR3, based on
Cepheids
\citep{2022A&A...668A..40L,	
2023MNRAS.523.1556D,
2024MNRAS.528.4409C,
2025A&A...699A.199P},
red clump stars
\citep{2023ApJ...943...88L,
2024MNRAS.527.4863U},		
and open clusters
\citep{2024arXiv241220344P}.	
All find complex signals in the vertical heights and vertical motions
\citep{2021ApJ...912..130C,	
2023ApJ...943...88L,			
2024MNRAS.528.4409C,		
2025A&A...699A.199P}.		
For example, with DR3,
\citet{2023MNRAS.523.1556D}	
found no warping inside $R\approx11\,$kpc, while for larger $R$ the disk is increasingly inclined, reaching an inclination $\sim\!\!3^\circ$ at $R\gtrsim14\,$kpc. With larger $R$ the azimuth of the warp's ascending node shifts by $14^\circ$~kpc$^{-1}$ in the direction of Galactic rotation, implying a leading spiral of nodes (the general behaviour of warped galaxies noted by 
\citet{1990ApJ...352...15B}). 	
And they found prograde precession of $\sim\!\!12$\kmskpc\ at 12\,kpc, decreasing to $\sim\!\!6$\kmskpc\ at 14\,kpc and beyond. This would unwind the leading spiral of nodes in $\sim\!\!100$\,Myr, suggesting that the present measurements reflect transient behaviour, and that the warp is dynamically evolving (perhaps, they suggested, as a result of the interaction with the Sgr dwarf spheroidal galaxy).  Similar behaviour was found in one of the IllustrisTNG cosmological simulations (their Figure~6). Similar precession rates were found by 
\citet{2024A&A...688A..38J}.	

There is a growing consensus that the direction and magnitude of the warp's precession, and its consistency across age groups, favour the Milky Way warp being the result of a recent or ongoing encounter with a satellite galaxy. And in this spirit, specific simulations of the Gaia Sausage--Enceladus merger (Section~\ref{sec:halo-streams}) were carried out by
\citet{2024ApJ...975...28D}.	
In their model the merger, estimated to have occurred 8--11\,Gyr ago, proceeded rapidly, taking $\sim$1\,Gyr between the first and final pericentre passages, and completed within 3\,Gyr. Their model appears to reproduce the Galactic warp amplitude and precession rate, at the same time demonstrating that the warp is indeed long-lived, non-steady, asymmetric, lopsided, and exhibiting both prograde and retrograde precession after merger completion. An oblate, tilted, and retrograde dark matter halo, also results from the merger, and appears to sustain the warp. 
Whether this finally resolves the origin and nature of the Milky Way warp remains to be seen.

\subsection{The bar}
\label{sec:bar}

Bars are rectangular or `boxy' morphological stellar structures, present in the centres of some 50--70\% of spiral galaxies, and 25--30\% of all galaxies. They vary in prominence, from those dominating the disk's appearance, to marginal distortions. The fraction of galaxies with bars depends on environment, mass, and gas content, with a lower overall rate at higher redshifts (although some have already formed at $z>2$, as seen in recent JWST images 
\citep{2023ApJ...945L..10G}).
Bars are important drivers of galaxy evolution, affecting the distribution of angular momentum, interacting with the stellar disk, the dark matter halo, and influencing gas transport and star formation.
While it was long suspected that our own Milky Way is `barred'
\citep{1957AJ.....62...19J,		
1964IAUS...20..195D},		
its structure, size and orientation, and `pattern speed' (the rotation of the bar feature, rather than the constituent stars), are still subject to debate. This is mainly because, from our edge-on location within the disk, its defining non-axisymmetric structure cannot easily be seen, with dust opacity further complicating its interpretation. 
Observational efforts to determine its morphology, ranging from the kinematics of H\,{\scriptsize I} and H$_2$ gas in the inner regions, near-infrared COBE data, red clump stars from OGLE, and infrared star counts from 2MASS, DENIS, VVV and others, is summarised by 
\citet{2012A&A...538A.106R}.	
Some of the most recent pre-Gaia bar and bulge projections, derived from near-infrared star counts, are given by 
\citet{2015MNRAS.450.4050W}.	

As to their origin, bars (and spiral arms) appear to arise as inevitable consequences of stellar orbits (and their relation to the Galaxy's inner Lindblad resonance), within a rotating asymmetric gravitational potential
\citep[e.g.][]{1977ARA&A..15..437T,	
2013seg..book....1K}.
They appear in numerical models
\citep[e.g.][]{1990A&A...233...82C,	
2011MSAIS..18...53C},			
as well as in $\Lambda$CMD cosmological simulations (Section~\ref{sec:cosmology-simulations}), such as
IllustrisTNG 
\citep{2018MNRAS.473.4077P},
NewHorizon 
\citep{2021A&A...651A.109D}
and
EAGLE 
\citep{2022MNRAS.510.5164C}.
These simulations nevertheless have a large variation in predicted bar fractions, from 5--55\%, reflecting the challenge of modelling the processes governing galaxy formation and evolution. 
While emphasising the difficulties in separating the bar from the bulge, the inner disk and inner halo, and the further complication of a `super-thin' contribution,
\citet{2016ARA&A..54..529B}	
gave its parameters as:
{\it stellar\/} mass ($7\pm1\times10^9M_\Sun$),	
Galactic orientation ($28-33^\circ$), 
half-length ($5.0\pm0.2$\,kpc),
scale height (180\,pc),
and an age of order 6--7\,Gyr. 
But estimates of these parameters still vary considerably, and their suggested pattern speed ($43\pm9$\kmskpc) and corresponding corotation radius (4.5--7\,kpc), derived from numerical and hydrodynamical simulations, remain poorly determined.
Two or more overlapping populations might also exist, complicating estimates of its shape and orientation
\citep[e.g.][]{2011ApJ...734L..20M,	
2012A&A...538A.106R,	
2020RAA....20..159S}.	
Its stellar population content, and its formation and evolutionary history, also remain unclear
\citep{2018ARA&A..56..223B}.	

Gaia distances and proper motions, combined with complementary radial velocities and spectroscopy, are revealing the barred structure in unprecedented detail. 
\citet{2019MNRAS.490.4740B} 	
used Gaia DR2 with APOGEE DR16, while 
\citet{2023MNRAS.519..948L}
updated the analysis using Gaia EDR3 with APOGEE DR17.
Combined with their determination of the distance to the Galactic centre (Section~\ref{sec:distance-galactic-centre}), their maps display the minimum in rotational velocity, and the quadrupole signature in radial velocity, expected for stars orbiting in a bar. 
The same feature was subsequently reported in the Gaia DR3 data by 
\citet{2023A&A...674A..37G}.
From N-body simulations, \citet{2023MNRAS.519..948L} estimated a pattern speed of $40.1\pm1.8$\kmskpc, and they inferred that the bar first formed $\sim$8~Gyr ago. At that epoch, the decreasing turbulence of the gas disk allowed a thin disk to form, which quickly became subject to the bar instability.

The dynamics of the inner Galaxy contains important evidence for untangling the evolutionary history of the Milky Way. But the inner Galaxy's gravitational potential is poorly constrained, in part because the bar's uncertain size, with half-length estimates ranging from 3.5--5\,kpc. 
\citet{2023MNRAS.520.4779L}	
used 210\,000 stars from Gaia EDR3 and APOGEE DR17 to model the stellar orbits as a function of assumed potential. Their results indicate a bar length of \mbox{3.4--3.9\,kpc}, and a pattern speed of 39\kmskpc.	

Although bars are physically dominant within the corotation resonance, their dynamical influence extends to the outer parts through other resonance effects. In the solar neighbourhood, the Gaia data demonstrates that the Hercules stream is one such manifestation (Section~\ref{sec:hercules-stream}).
\citet{2023A&A...670A..10D}
used Gaia DR3 data, including its compilation of 33~million stars with spectroscopic line-of-sight velocities, to identify another resonant-like feature in the angular momentum of young classical Cepheids (also seen in a much larger sample of red giant stars) which may also be related to the influence of the central bar. 
Two studies making use of the astrometry and photometry from Gaia DR2 have led to a particularly remarkable result which illustrates both the accuracy of the Gaia data, while providing further circumstantial evidence for the dark matter halo.
\citet{2021MNRAS.500.4710C}		
showed that the local stellar kinematics imply that the bar is {\it decelerating}, consistent with the effects of dynamical friction of a dark matter halo.  As the bar slows, its resonances sweep through phase space, dragging along a portion of previously free orbits. This leads to certain multiple resonances clearly seen in the Gaia data and, in particular, reproducing the details of the Hercules stream in the local 
radial--azimuthal velocity plane. They derived a current slowing rate of the bar of $-4.5\pm1.4$\kmskpc\ per Gyr.
\citet{2021MNRAS.505.2412C}		
then used Gaia photometry to show that this deceleration also results in a variation of metallicity with distance from the resonance centre. Applied to solar neighbourhood stars, this provides a new and precise measure of its {\it current\/} pattern speed, $35.5\pm0.8$\kmskpc, and placing the corotation radius at $6.6\pm0.2$~kpc. And with this pattern speed, they argue, the corotation resonance precisely explains the kinematics of the Hercules stream. The bar's deceleration, $\sim$24\% since formation, transfers angular momentum to the halo by dynamical friction, in the process adding support to the existence of a standard dark-matter halo.

A Gaia DR2-based study has had some success in imaging the bulge, at least more directly than reconstructions based on number counts.
\citet{2019A&A...628A..94A} used the {\tt StarHorse} code\footnote{
{\tt StarHorse} is widely referenced in the Gaia literature. Developed in the context of RAVE, SEGUE and APOGEE
\citep{2016A&A...585A..42S,	
2018MNRAS.476.2556Q},		
it finds the (Bayesian) posterior probability over a grid of stellar evolutionary models, distances, and extinctions, given a set of observations plus a number of priors: here, the initial mass function, density laws for the thin disk, thick disk, bulge, and halo, as well as broad metallicity and age priors for each.}
on a combination of Pan--STARRS1, 2MASS, and AllWISE photometry with Gaia~DR2 parallaxes to derive improved stellar parameters, distances, and extinctions for 137~million stars with $G\le18$~mag. Their face-on projection reveals a stellar overdensity coinciding with the expected position of the bar (their Figure~7a), inclined by about $40^\circ$ with respect to the solar azimuth, and with a semi-major axis $\sim$4~kpc. The presence of the bar is even more prominent when focusing only on the red clump stars (their Figure~8). Interestingly, its inferred shape and inclination are significantly different from their adopted prior, which was given by the Galactic bulge--bar model derived by \citet{2012A&A...538A.106R}. They also identified a kinematic imprint of the coherent motion of the Galactic bar in the proper motions.
That Gaia is probing stellar populations in the Galactic bulge and beyond is, at first sight, surprising. They reason that their distance estimates are not more {\it precise\/} than those prescribed by, e.g., \citet{2018AJ....156...58B}, using which the bar is barely evident. However, at large distances, their values (they argue) are more {\it accurate}, due to the use of more informative Galactic priors.

Two dozen or more studies have followed based on these and subsequent Gaia data releases, including interpretation in the context of numerical simulations of barred galaxies
\citep[e.g.][]{
2017MNRAS.469.1587D,		
2019MNRAS.490.4740B,		
2019MNRAS.490..797C, 		
2020MNRAS.494.5936F}.		
The present, still uncertain and incomplete picture, is of a potentially more complex morphology
\citep{2025MNRAS.538..998H};	
a current pattern speed towards the lower end of historical estimates, of perhaps 35--40\kmskpc\ 
\citep{
2021MNRAS.508..728K,	
2021MNRAS.507.4409M, 
2022MNRAS.512.2171C,	
2024MNRAS.531L..14L};	
of the bar driving complex resonances not only in the solar neighbourhood 
\citep{2024A&A...686A..92B,	
2022MNRAS.509..844T}	
but also into the halo
\citep{2023MNRAS.524.3596D,	
2024MNRAS.532.4389D};		
and of a decelerating bar driving the observed radial and azimuthal gradients in the local stellar kinematics and metallicities
\citep{
2020A&A...638A.144K,	
2021A&A...656A.156Q,	
2022ApJ...935...28W,	
2024A&A...690A..26L}.	
Some studies infer its formation around 9--10\,Gyr ago, perhaps with a much higher initial pattern speed of 60--100\kmskpc\
\citep{2024A&A...690A.147H,	
2024MNRAS.530.2972S,	
2025ApJ...983L..10Z}.	
And several draw attention to the selection effects in the current Gaia data releases that complicate the interpretation, which will improve (along with the improved accuracies) in future data releases.

\begin{figure}[t]
\centering
\raisebox{10pt}{\includegraphics[width=0.27\linewidth]{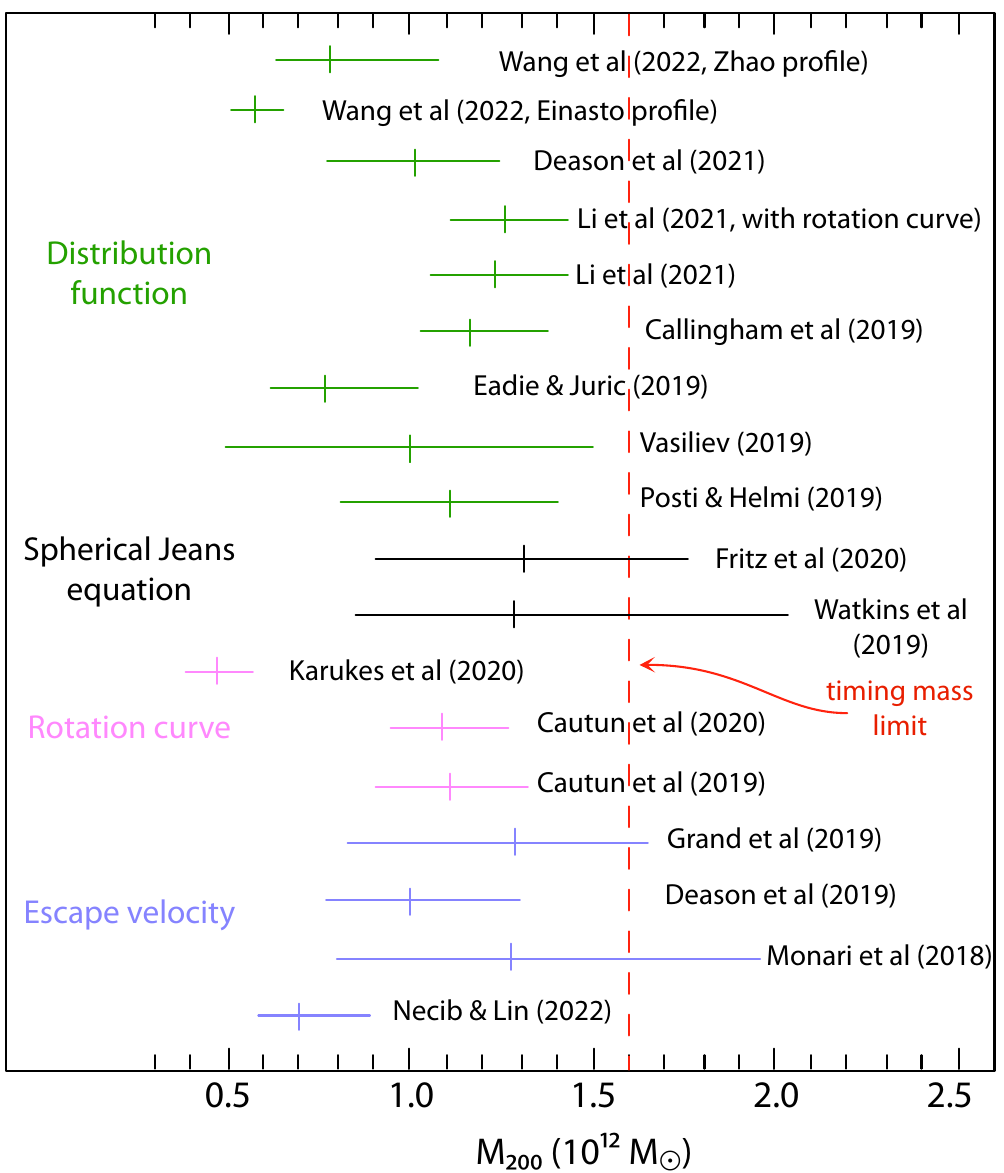}}
\hspace{30pt}
\includegraphics[width=0.57\linewidth]{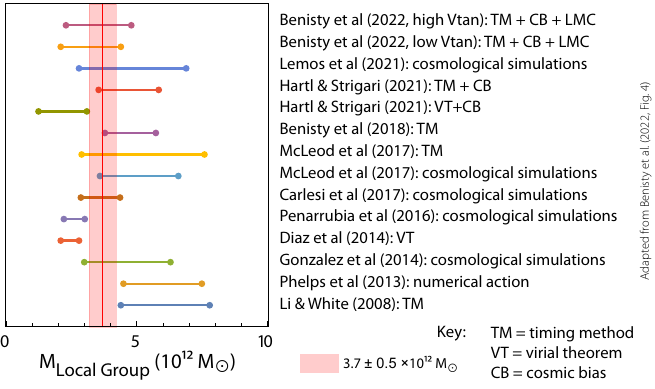}
\caption{Left: estimates of the mass of the Milky Way from Gaia DR2 and EDR3/DR3 (from \citet{2022MNRAS.510.2242W}, Figure~14).
Right: estimates of the mass of the Local Group (best fit and 68\% confidence intervals). The red vertical band shows the range $3.7^{+0.5}_{-0.5}\times10^{12}M_\Sun$, computed using recent estimates of the mass of the Milky Way and M31 (from \citet{2022ApJ...928L...5B}, Figure~4).
}\label{fig:mass-milky-way}
\end{figure}

\subsection{Mass of the Galaxy}
\label{sec:mass-galaxy}

In their review of the structural and kinematic properties of the Galaxy, and the state of knowledge pre-Gaia,
\citet{2016ARA&A..54..529B}
placed its mass at between $0.5-3\times10^{12}M_\Sun$. 
Accurate determination of its mass and mass profile has implications for understanding Galactic orbits and Galactic rotation, the dynamical history and future of the Local Group, and the role of mass in the formation and growth of structure in the Universe. Today's best estimates still span a rather wide range. The main reason is that, while the masses of the principal baryonic components (the supermassive black hole at its centre, the central bulge, and the flattened disk) are reasonably well determined, its other major component, the halo, is dominated by dark matter which contributes the greatest uncertainty. 
The distribution of dark matter controls the motion of distant `tracer' objects. These include halo stars and globular clusters as probes of the inner halo, while dwarf spheroidal galaxies give better coverage further out.  In their compilation of mass estimates between 1999--2014,
\citet[][their Table~8]{2016ARA&A..54..529B}
also quote estimates from the kinematics of halo streams, 
from the kinematics of hypervelocity stars, 
and from the escape speed (Section~\ref{sec:escape-speed}).

All of these estimates, in turn, remain sensitive to assumptions such as which of its satellites are bound, the shape and extent of the dark matter halo, and the velocity anisotropy of the halo and of its satellite system. And a key problem faced by all kinematic mass estimates has been the lack of knowledge of their complete space motion. Pre-Gaia, measurements were mostly restricted to line-of-sight velocities. And since the Sun is (relatively) close to the Galactic centre, most line-of-sight velocities probe mainly the component of motion in the Galactocentric radial direction, providing little information about their tangential motions. As a result, estimated masses pre-Gaia depended strongly on the assumptions used for the tangential motions: the so-called mass--anisotropy degeneracy.
A further problem is that these kinematic tracers are sensitive to the mass of the dark matter which has had time to virialise in the Galaxy, the so-called `virial mass' defined within some adopted `virial radius'. How far out to measure is usually expressed as the mass within a region in which the average density exceeds some multiple of the mean density of the Universe. Many recent measures refer to $M_{200}$, the mass within 200~kpc (see, e.g., \citet{2019ApJ...873..118W}).
 
In their pre-Gaia summary, 
\citet{2016ARA&A..54..529B}
found that the analysis of halo star kinematics typically resulted in relatively low values, $M_{200}\lesssim10^{12}M_\Sun$, where the main uncertainties were the lack of stellar tangential velocities from proper motions.
Estimates based on satellite and globular cluster kinematics (with the main uncertainties coming from small numbers and, again, a lack of proper motions) typically resulted in higher values, $M_{200}=1-2\times 10^{12}M_\Sun$ if the Leo~I dwarf galaxy (with its large line-of-sight velocity) is assumed to be bound to the Milky Way; or $M_{200}\lesssim10^{12}M_\Sun$ if it is considered as unbound.

One estimate of our Galaxy's total mass that does not depend on the choice or details of a tracer population is the so-called ‘timing mass’. In the original formulation by
\citet{1959ApJ...130..705K},
the Milky Way and M31 proto-galaxies are assumed to have had a small separation at the time of the Big Bang, subsequently moving apart as they participated in the Hubble flow. If there is sufficient mass, their expansion is reversed. Given an estimate of the age of the Universe, together with their present separation and approach velocity, the equations of motion can be solved to give the mass of the Galaxy, along with that of the Local Group. Timing mass estimates in subsequent work provide an upper limit to the Galaxy's virial mass of $\lesssim1.6\times 10^{12}M_\Sun$ 
\citep[e.g.][]{2008MNRAS.384.1459L,
2012ApJ...753....8V,
2022MNRAS.511.6193H,
2023ApJ...942...18C}.

Again, according to
\citet{2016ARA&A..54..529B},
the halo star kinematic studies provided (at least pre-Gaia) the largest and most reliable data sets. They gave an average of $M_{200}=(1.1\pm0.3)\times10^{12}M_\Sun$, or equivalently $M_{\rm vir}=(1.3\pm0.3)\times10^{12}M_\Sun$, consistent with the upper limit from the timing mass. This compares with a {\it baryonic\/} mass of $(6.3\pm0.5)\times10^{10}M_\Sun$ (stars and cold gas), or $(8.8\pm1.2)\times10^{10}M_\Sun$  including the hot corona. This, in turn, leads to a baryonic mass fraction (out to some chosen virial radius) of $6\pm1$\%, well short of the `universal' value of $\sim$16\%.

The Gaia data has led to several dozen papers to date, and in particular using the more complete space motions now available for halo stars, globular clusters (Section~\ref{sec:globular-clusters}), and dwarf spheroidal galaxies (Section~\ref{sec:dwarf-spheroidals}). 
Employing data from Gaia DR2,
\citet{2019MNRAS.484.5453C}
inferred the Galaxy's total mass by comparing model satellites in the EAGLE cosmological hydrodynamics simulations with the dynamics of 10 of the `classical' Milky Way satellites with six-dimensional phase-space measurements (i.e.\ positions and velocities), including updated proper motions from Gaia (including the LMC, the SMC, Draco, and Ursa Minor). In this 2-d space, the orbital properties of satellite galaxies vary according to the host halo mass, which can then be inferred from the likelihood that the observed satellite population is drawn from this distribution function.  They inferred a mass of $1.17^{+0.21}_{-0.15} \times10^{12} M_\Sun$.
\citet{2020MNRAS.494.5178F}
used a somewhat similar approach based on Gaia DR2 proper motions of 45 satellite galaxies to find a mass of  
$1.43^{+0.35}_{-0.32}\times10^{12}M_\Sun$ within 273~kpc,
extrapolating to a virial mass of 
$1.51^{+0.45}_{-0.40}\times10^{12}M_\Sun$. 
\citet{2019ApJ...873..118W}	
used the Gaia DR2 proper motions of 34 halo globular clusters within 21.1~kpc. They determined the mass inside the outermost globular cluster as ${0.21}_{-0.03}^{+0.04}\times10^{12}M_\Sun$, leading to a virial mass of ${1.28}_{-0.48}^{+0.97}\times10^{12}M_\Sun$.

Various mass determinations have followed with Gaia EDR3 in 2020, using various related methods and assumptions, including modelling halo tracers using spherical, power-law distribution functions
\citep{
2019ApJ...875..159E,	
2019A&A...621A..56P,	
2020JCAP...05..033K,	
2021MNRAS.501.5964D,	
2021ApJ...916....8L,		
2022A&A...657A..54B,	
2022MNRAS.511.2610C,	
2022ApJ...926..189N,	
2022ApJ...940..136P,	
2022MNRAS.513.4968R,	
2025MNRAS.538.1442L}.	
As one example,
\citet{2022MNRAS.510.2242W}		
used Gaia EDR3 proper motions for about 150 Milky Way globular clusters, combined with constraints from the rotation curve over Galactocentric distances 5--25~kpc. They derived a total mass, at the lower end of current estimates, in the range 
${0.536}^{+0.081}_{-0.068}\times10^{12}M_\Sun$
to 
${0.784}^{+0.308}_{-0.197}\times10^{12}M_\Sun$, depending on the assumed dark matter profile (Zhao or Einasto). 
These, and some of the other results based on Gaia DR2 and EDR3, are shown in Figure~\ref{fig:mass-milky-way}a, colour-coded by methodology (viz.\ based on an underlying distribution function, spherical Jeans equation, rotation curve, or escape velocity).

The Gaia data are providing much new insight into the complexity of estimating the Galaxy mass, with formal errors typically much smaller than they were pre-Gaia, and with a value of $\sim$$10^{12}M_\Sun$ appearing favoured. Nevertheless current estimates are still affected by the choice of tracers and model assumptions, such that the Galaxy mass, dominated by its dark matter halo, remains relatively poorly characterised.

\subsection{Local mass density} 
\label{sec:mass-kz}

In addition to the largely circular motions around the Galactic centre, stars `bounce up and down' about the Galaxy mid-plane, reaching distances of several tens of parsec above or below it, before being `hauled' back down by the gravitational force exerted by the matter concentrated in the disk itself. The stars then fly on through the mid-plane, before continuing on their indefinite oscillations, with a period of about 60--90\,Myr.
The {\it total\/} amount of matter in the solar neighbourhood determining these motions must include both visible material (mainly stars and gas, with the contribution of `cold' gas presently being the most uncertain), along with any dark matter -- mass models of the Galaxy based only on visible star counts can only be used to infer {\it lower\/} limits.
The result can be expressed in mass per unit volume, $M_\odot$\,pc$^{-3}$ (or in GeV\,cm$^{-3}$, where $1M_\odot$~pc$^{-3}$ = 37.5\,GeV\,cm$^{-3}$), or as the column density out to a given $z$~distance from the plane, expressed in $M_\odot$\,pc$^{-2}$. Present understanding is that the projected mass of disk matter corresponds to roughly 70~g\,m$^{-2}$ (about the density of typical A4 paper, spread out across our Galaxy's mid-plane).

Measurement of the {\it dynamical\/} effects of the local mass density, from the vertical motions of stars, is potentially more robust. Analysis of the density and velocity distribution of a tracer sample of stars can provide estimates of the local density of {\it all\/} disk-like matter.  This is often referred to as the $K_z$ problem in Galactic dynamics, where $K_z$~is defined as the component of the Galaxy's gravitational acceleration towards the Galactic plane in the solar neighbourhood. 
In a plane-stratified approximation out to a few kpc, $K_z$ is assumed to increase monotonically with~$z$. Studies then use some suitable `tracer' population to estimate these forces -- stars whose number density and velocities can be determined as a function of distance from the plane. Matter produces both the restoring potential (described by Poisson's equation), and is at the same time influenced by it (described by the Boltzmann or vertical Jeans' equation).

Hipparcos brought a number of observational advances to this problem: primarily an improved accuracy on distances and space velocities, but also an increased size of various tracer populations, such as A~stars and K~giants. It also provided an improved estimate of the local stellar luminosity function, such that the contribution of {\it visible\/} disk matter is better known.
Amongst these studies, 
\citet{1998A&A...329..920C}
used luminous A~stars within 125\,pc of the Sun; stars bright enough to be seen at meaningful distances, but not so young as to be influenced by velocity inhomogeneities due to recent star formation. They derived a local `dynamical' density of $0.076\pm0.015\,M_\odot\,{\rm pc}^{-3}$, and a revised assessment of the visible matter density of $0.085\,M_\odot\,{\rm pc}^{-3}$. These dynamical and visible estimates being broadly compatible, they argued that (with a dark massive halo still required to explain the Galaxy's rotation curve) the halo must be, to a first approximation, largely spherical.
Along with other Hipparcos-based studies, using different star samples and methodologies 
\citep{1999A&A...341...86B,
2000MNRAS.313..209H,
2003AJ....126.2896K,
2003A&A...398..141S,
2004MNRAS.352..440H,
2006A&A...446..933B},
another important consensus emerged: that dark matter is distributed largely in the form of the halo, with little if any concentrated in the disk \citep{2014JPhG...41f3101R}.

With its huge stellar samples and accurate space motions, Gaia is opening a new chapter in this sort of dynamical analysis. An early study using Gaia DR1 by
\citet{2019A&A...623A..30W}	
was followed by a more detailed assessment using DR2 by 
\citet{2019MNRAS.482..262W}.
Their model for the total matter density assumes that it is symmetrical, smooth, and monotonically decreasing with distance from the mid-plane. Their (Bayesian) model accounts for the position and velocity of each star, in a joint fit of the vertical velocity distribution and stellar number density. They did this for eight separate data samples, with different limits in absolute magnitude, each containing $\sim$25\,000 stars. In all cases, this resulted in a density distribution strongly peaked within 60~pc of the plane. Assuming a baryonic model and a dark matter halo of constant density, this corresponds to an excess surface density of $5-9M_\odot$\,pc$^{-2}$. They concluded that there is a surplus of matter close to the Galactic plane, perhaps due to an underestimated presence of cold gas. 

A more extensive study using DR2 was made by 
\citet{2021A&A...646A..67W},	
who used 120 stellar samples in 40~separate sub-regions of the solar neighbourhood. And by excluding areas of known open clusters, they aimed to quantify the sorts of spatially dependent systematic effects that have complicated this type of measurement in the past. Their results revealed an unexpected but clear trend for all 40 spatially separated sub-regions, implying a total matter density distribution that is highly concentrated towards the mid-plane ($<60$~pc), but decaying rapidly with height, and with a dependence on Galactic radius consistent with a disk scale length of a few~kpc.  They suggested, in particular, that the very low inferred matter density above 300~pc is inconsistent with the observed scale height of the stellar disk, and postulated a time-varying phase-space structure that is large enough to affect all stellar samples in the same way.

Gaia is also providing new possibilities for determining the total matter density of the disk
\citep{2020MNRAS.496.3112W,	
2025JCAP...01..021L}.			
\citet{2020MNRAS.496.3112W}		
proposed using `stellar streams' passing through or close to the Galactic plane. They argue that the vertical component of energy for (dynamically cold) stream stars is approximately constant, such that the vertical positions and vertical velocities of their stars are related via the disk density. This does not require the disk to be in dynamical equilibrium, and furthermore makes it possible to measure the surface density at large distances from the Sun. Such methods that avoid assumptions of dynamical equilibrium are gaining importance as other Gaia data clearly indicate such dis-equilibrium
\citep[e.g.][]{2020A&A...643A..75S},	
amongst which are the Gaia `phase-space spiral' (Section~\ref{sec:phase-space-spiral}), 
and the Galaxy's central bar (Section~\ref{sec:bar}).
First efforts in using such time-varying structures to measure the vertical gravitational potential of the Galactic disk yield a local halo dark matter density of $0.0085\pm0.0039M_\odot$~pc$^{-3}$, and an upper limit on the surface density of any thin dark disk with a scale height $<50$~pc of about $5M_\odot$~pc$^{-2}$
\citep{2021A&A...650A.124W,	
2021A&A...653A..86W,		
2022A&A...663A..15W}.		

In their review of the local estimates of the dark matter density,
\citet{2021RPPh...84j4901D}	
conclude that most local analyses coincide within a range
$\rho_{{\rm DM},\odot}=0.4-0.6$\,GeV\,cm$^{-3}=0.011-0.016M_\odot$~pc$^{-3}$,
while more global studies give a slightly lower range
$\rho_{{\rm DM},\odot}=0.3-0.5$\,GeV\,cm$^{-3}=0.008-0.013M_\odot$~pc$^{-3}$.

\begin{figure}[t]
\centering
\includegraphics[width=0.58\linewidth]{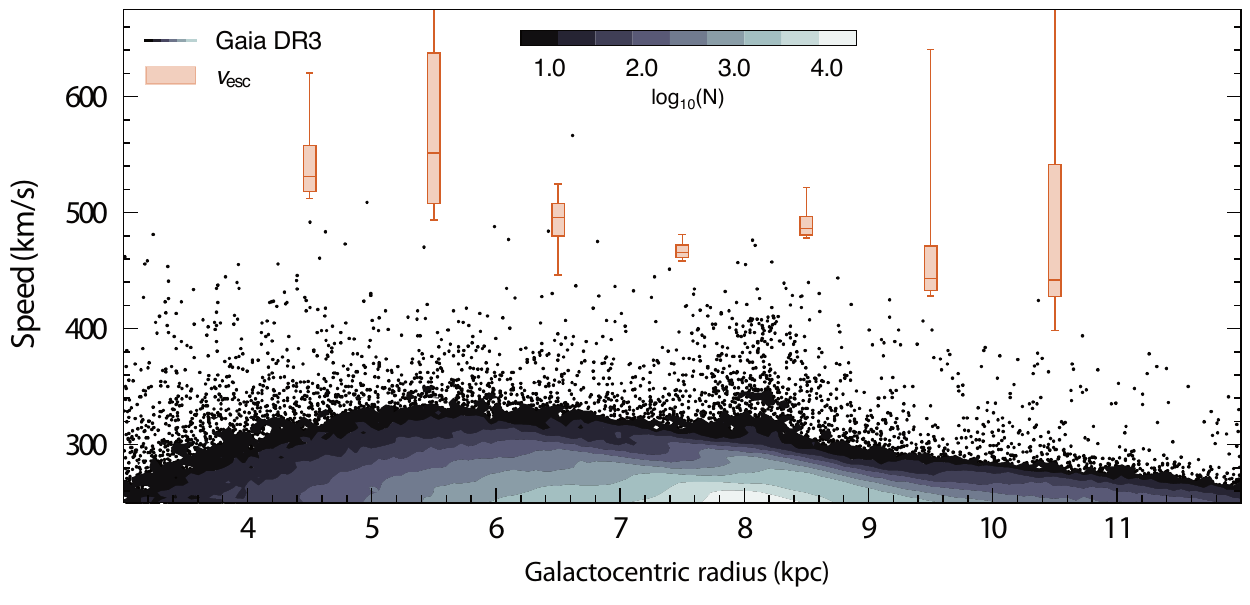}
\hspace{10pt}
\includegraphics[width=0.35\linewidth]{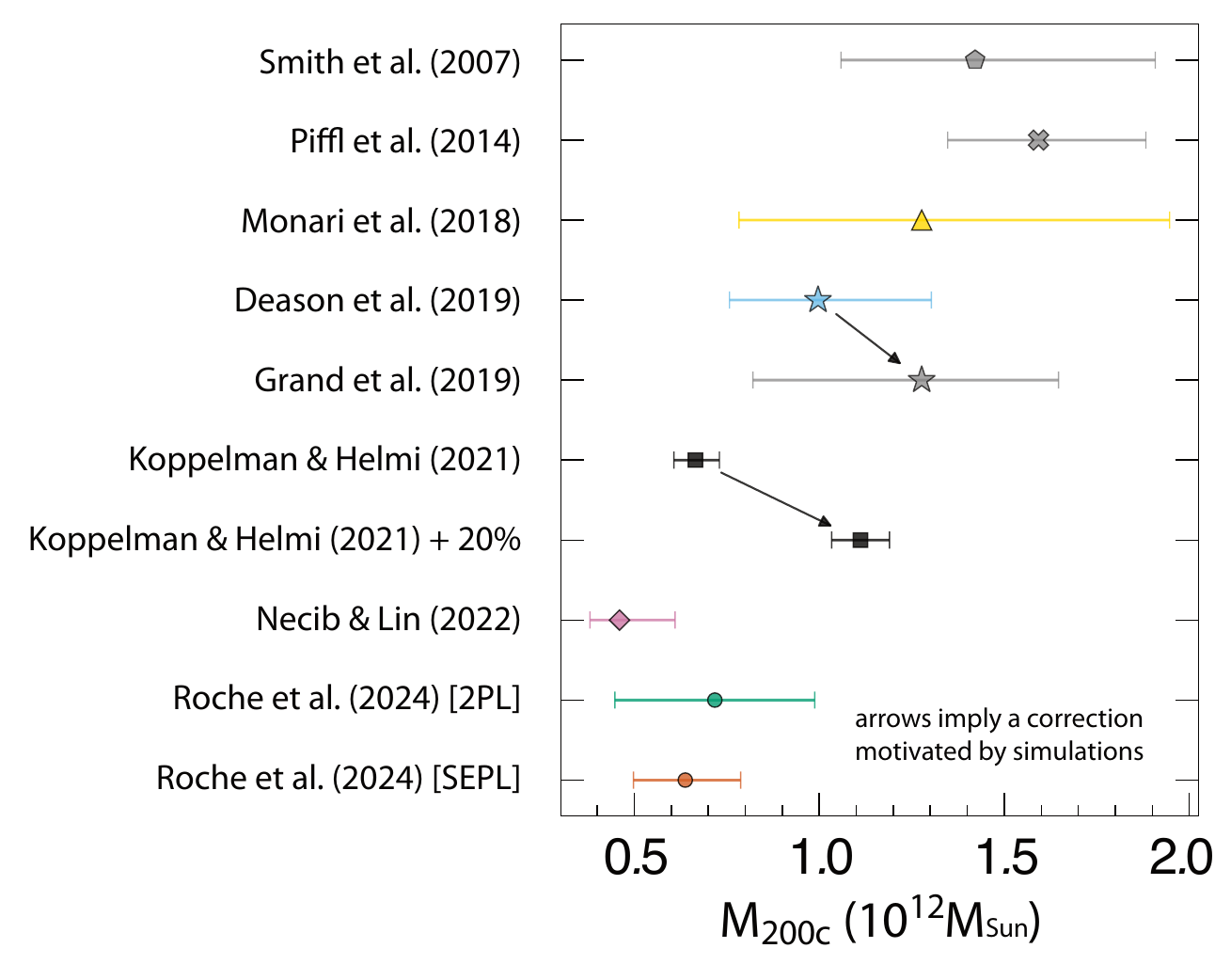}
\vspace{-5pt}
\caption{The Galactic escape speed. 
Left: High-speed and high-quality sample of stars in Gaia DR3 with 6d kinematics, and the derived escape velocity profile at 68\% (boxes) and 95\% (whiskers) confidence intervals. Escape speeds are obtained by binning the data into 1\,kpc bins, and fitting the distributions above 310\kms\ with the stretched
exponential power law model. Contours indicate logarithm of the star number at that
speed and Galactocentric radius.
Right: a selection of total Galaxy mass estimates obtained by escape speed modeling along with circular velocity constraints (2PL: 2-component power law; SEPL: stretched
exponential power law). From \citet{2024ApJ...972...70R}, Figures~1and 7.
}\label{fig:escape-speed}
\end{figure}

\subsection{Escape speed}
\label{sec:escape-speed}

Models of Galactic rotation as a function of radial distance can be used to estimate the escape speed from the solar neighbourhood via the potential \citep[e.g.][\S2.2.1]{2008gady.book.....B}.
Early estimates, $v_{\rm e}\sim50-100$\kms, were based on Kapteyn's Galaxy model in which the total mass is distributed in the same way as stars
\citep{1928BAN.....4..269O}.	
Later models of the mass distribution gave values
$v_{\rm e}>280$\kms\ \citep{1956BAN....13...15S}	
and
$v_{\rm e}=380$\kms\ \citep{1965gast.book..513S}.	
The realisation that the Galaxy has a massive dark halo led to the understanding that the local escape speed offers a direct constraint on its unknown extent. Improved measurements of high-velocity stars in the solar neighbourhood (and accounting for the dynamically ejected `runaway stars') led to better characterisation of the velocity distribution, and estimates of 400--640\kms\
\citep{1981ApJ...251...61C,	
1982MNRAS.201..579A,	
1987AJ.....93...74S,		
1988AJ.....96..560C}. 	
A statistical technique to infer the escape speed from the high-velocity tail was given by
\citet{1990ApJ...353..486L}.	
But their estimates, 450--650\kms, were strongly correlated with its {\it shape\/} and, in turn, sensitive to the errors on distances and proper motions. But they also argued that it is not possible to estimate the mass of the Galaxy using the local escape speed without a knowledge of how mass is distributed beyond the solar circle. 

Using transverse motions from Hipparcos, and radial velocities from Coravel, the technique was applied by 
\citet{1997ESASP.402..591M}
to 5307 \mbox{F--M} stars, including all known subdwarfs and high-velocity stars. Estimates from the total space velocities were in the slightly smaller range $v_{\rm esc}=440-490$\kms, probably due to the absence from the Hipparcos catalogue of the (faintest) highest velocity stars in the compilation of 
\citet{1994AJ....107.2240C}.
\citet{1999Ap&SS.265..179M}
found only 10~Hipparcos stars with $V_{\rm tot}>350$\kms. Combined with the sample of \citet{1994AJ....107.2240C}, they found four with $V_{\rm tot}\ge400$\kms, and the fastest had $V_{\rm tot}=458$\kms. They concluded that the escape velocity is unlikely to exceed 530\kms.
Estimates from the RAVE radial velocity survey 
included $v_{\rm e}\simeq544^{+64}_{-46}$\kms\ by \citet{2007MNRAS.379..755S},
and $533^{+54}_{-41}$\kms\ by \citet{2014A&A...562A..91P}.
Importantly, working the other way around, recent models of the Galaxy's structure and mass distribution gave consistent escape speeds, of around 550\kms\ 
\citep{2014ApJ...794...59K}.

The Gaia results provide improved estimates of the escape speed, reveal observable changes with Galactocentric radius, and provide new constraints on the Galaxy's mass and mass distribution.\footnote{I refer to \citet[][\S3.2]{2019MNRAS.485.3514D} for a discussion of the escape speed, and their definition as the velocity needed to bring an ejected star to the virial radius, in turn defined as 200 times the critical density, $r_{200}$, a volume containing the mass $M_{200}$. Another point to emphasise is that the properties of galaxy halos are today interpreted in the context of extensive numerical simulations, such as the Auriga suite of high-resolution Milky Way-mass halos, spanning a mass range $M_{200}=1-2\times10^{12}M_\odot$.}
Using the method of \citet{1990ApJ...353..486L},
\citet{2018A&A...616L...9M}
used the Gaia DR2 data to estimate $v_{\rm e}$ over Galactocentric radii from 5--10.5~kpc, based on an assumed  $R_0=8.34$\,kpc, and a circular velocity at the Sun $\Theta_{\rm c}=240$\kms. They used the velocity distribution of 2850 counter-rotating halo stars within 5~kpc (to ensure that the kinematic tail is representative of the stellar halo), with distance errors below 10\%, and with known line-of-sight velocities. They showed that the escape velocity decreases with Galactocentric radius (as expected in any reasonable Galactic potential), with a value at the Sun of $580\pm63$\kms. Interpreted in the context of the Navarro--Frenk--White dark matter halo profile gave $M_{200}=1.55^{+0.64}_{-0.51}\times10^{12}M_\odot$ and $c_{200}=7.93^{+0.33}_{-0.27}$.

\citet{2019MNRAS.485.3514D}		
used the inferred assembly history and phase-space distribution of halo stars from DR2 to show that the high-velocity tail of the stellar halo was strongly dependent on the velocity anisotropy and number density profile of the halo stars (with highly eccentric orbits, attributable to the early accretion of a massive dwarf satellite galaxy, having more extended high-velocity tails). From their sample of 2300 (also counter-rotating) stars within 3~kpc, of which 240 have $v_{\rm tot}>300$\kms, they determined a local ($R_0=8.3$~kpc) escape speed of $528^{+24}_{-25}$\kms. They used this to estimate the total Galaxy mass, $M_{200}=1.00^{+0.31}_{-0.24}\times10^{12}M_\odot$, and dark halo concentration parameter, $c_{200}=10.9^{+4.4}_{-3.3}$. 
A larger sample of $10^7$ halo stars was later used by 
\citet{2021A&A...649A.136K}	
to estimate $v_{\rm e}=497\pm8$\kms, which they estimate is biased low by 10\%, yielding a true escape speed closer to 550\kms. Assuming $\Theta_{\rm c}=232.8$\kms, they derived $M_{200}=1.11^{+0.08}_{-0.07}\times10^{12}M_\odot$, and $c_{200}=11.8\pm0.3$.

A more robust treatment of velocity anisotropy, accounting for kinematic substructures including the GSE halo stream, was developed by \citet{2022ApJ...926..188N,2022ApJ...926..189N}. This gave a much smaller $v_{\rm e}=484.6^{+17.8}_{-7.4}$\kms\ and, using $v_c=230\pm10$\kms, a Milky Way mass $M_{200}=7.0^{+1.9}_{-1.2}\times10^{11} M_{\odot}$ (and $c_{200}=13.8^{+6.0}_{-4.3}$), considerably lighter than previous measurements.
A further estimate from DR3 by 
\citet{2024ApJ...972...70R}, 
over the Galactocentric distance range 4--11\,kpc, gave a similar $M_{200}=6.4^{+1.5}_{‑1.4}\times10^{11}M_\Sun$ (Figure~\ref{fig:escape-speed}).
These Gaia results also have some implications for dark matter detection experiments, where exclusion limits \citep[e.g.][]{2017PhRvL.119r1301A} are affected by assumptions on the local dark matter escape velocity, the local dark matter density, and the Sun's circular velocity
\citep{2019JCAP...10..034W}.	

\subsection{Globular clusters}
\label{sec:globular-clusters}

Globular clusters are tightly bound spherical groups of a million or more stars. In contrast to the younger open clusters, which are found mainly in the disk, globular clusters are amongst the oldest populations, and are found in the disk, in the bulge, and most prominently in the halo. 
There are more than 150 known in the Milky Way, some 20\% within a few~kpc, and others extending to 30--40~kpc. The nearest, M4, is at $\sim$2.2\,kpc. Most galaxies of sufficient mass in the Local Group and beyond have their own systems of globular clusters. 
Ages of the oldest are comparable to that of the Universe.  Others appear to be remnants of the Galaxy's early accretion phase, captured from smaller galaxies during mergers or collisions. And although they represent a negligible fraction of the mass and light of the stellar halo, the total mass of the Galaxy can be determined, through Newton's laws, from their orbits. When Gaia was under consideration in 2000, the best estimates were based on the motions of 27 systems beyond 20~kpc, but the total mass of $2.3^{+3.9}_{-1.6} \times 10^{12} M_\odot$ (and within 50~kpc of $5.5^{+0.1}_{-1.1} \times 10^{11}M_\odot$), was only poorly known. Further advances have been limited by incomplete knowledge of their space velocities. 
Although no globular cluster stars were bright enough to be observed by Hipparcos, its improved reference frame allowed some progress to be made in understanding their orbits 
\citep{1999AJ....117.1792D,	
2003AJ....125.1373D}.	

In early Gaia studies,
\citet{2018A&A...616A..12G}	
used the Gaia DR2 parallaxes and proper motions to determine the space motions of 75 of the nearest, within 12--13~kpc. Radial velocities, measured by Gaia, were available for~57. Selecting according to distance and proper motions, stars out to each cluster's tidal radius, and down to $\sim$20~mag, could be identified, resulting in 20--30\,000 being identified in a number, and more than 60\,000 in 47~Tuc/NGC~104 (at 4\,kpc distance, and around 40~pc in diameter).
Mean globular cluster distances can then be derived. For example, for 47~Tucanae, they gave a mean parallax of $0.1959\pm0.0002$~milli-arcsec, i.e.\ a distance $5105\pm5$~pc. 
As they concluded: {\it `The measurements for these clusters are of outstanding quality, with the formal and systematic uncertainties being effectively negligible'.}
The bulk proper motions derived for each cluster are some one or two orders of magnitude larger than their parallaxes. The space motions could then be derived for all 75~clusters, with rotation detected in five (it was already known in three). 
Other interesting velocity structures include a pronounced `perspective contraction' in NGC~3201 (the geometrical effect due to its large radial motion and its relatively large parallax, Section~\ref{sec:perspective-acceleration}), and evidence for an expanding outer halo in the core-collapsed NGC~6397.
They then used these positions and space motions, along with various models of the Galaxy's mass distribution, to follow their Galactic orbits over the past 250~Myr, also allowing for improved searches for any tidal tails. 

With more than 250 Gaia-based papers on globular cluster studies since, I will restrict the following to just a list of topics being addressed. These cover
their orbits and anisotropy
\citep{
2019MNRAS.482.5138B,	
2019MNRAS.487.3693J,	
2019MNRAS.488..253S,	
2020MNRAS.494..983R,	
2021RAA....21..173B,	
2021MNRAS.505.5957B,	
2022ApJ...926..107M};	
halo mass and shape
\citep{
2019A&A...621A..56P,	
2019ApJ...873..118W,	
2019ApJ...875..159E,	
2019MNRAS.484.2832V,	
2022MNRAS.510.2242W};	
internal structure, radial profiles and rotation
\citep{
2019MNRAS.485.1460S,	
2019MNRAS.485.4906D,	
2020ApJ...889...18C,	
2024A&A...689A.232C};	
origin and accretion history
\citep{
2019A&A...630L...4M,	
2024OJAp....7E..23C};	
tidal tails
\citep{
2019ApJ...884..174G,	
2019MNRAS.483.1737K,	
2019MNRAS.486.1667C,	
2019MNRAS.489.4565K,	
2020MNRAS.495.2222S,	
2021MNRAS.504.2727P,	
2022MNRAS.513.3136Z};	
new discoveries: in the Galaxy
\citep{
2021MNRAS.505.5978V,	
2022A&A...659A.155G},	
in the bulge
\citep{
2019MNRAS.484L..90C,	
2021A&A...654A..39O,	
2022MNRAS.509.4962G,	
2024A&A...687A.201B},	
associated with halo streams
\citep{
2020A&A...636A.107B,	
2021A&A...650L..11M},	
with the LMC
\citep{
2019MNRAS.484L..19P,	
2024ApJ...963...84W},	
with Sagittarius
\citep{
2021A&A...647L...4M,	
2021A&A...650L..12M},	
and with other Local Group galaxies
\citep{
2021ApJ...914...16H,		
2021MNRAS.500..986H};	
and as tools for calibration of the Gaia astrometry
\citep{2021A&A...649A..13M}	
and photometry
\citep{2019MNRAS.485.3042S}.

A future synthesis of these many new observations is certain to provide a much new insight into their origin. 
Models by \citet{2024OJAp....7E..23C},	
for example, found that about 60\% of the Milky Way's clusters are likely to be {\it in situ}, 20\% associated with the Gaia Sausage--Enceladus event, 10\% with the Sagittarius dwarf galaxy, and the remaining 10\% still unassociated with any single accretion event.

\paragraph{Omega Centauri}
\label{sec:omega-centauri}

Amongst Gaia's globular clusters is Omega Centauri ($\omega$~Cen, NGC 5139), first identified as non-stellar by Edmond Halley in 1677, and visible to the naked eye. At a distance of about 5.2\,kpc, it is the largest, most massive globular cluster in the Milky Way, containing 10 million stars, a total mass of $4\times10^6M_\Sun$, a diameter of 50~pc (appearing almost as large as the full Moon), and believed to have originated as the core remnant of a disrupted dwarf galaxy. It is too distant, too crowded, and indeed too faint, for any of its stars to have been observed by Hipparcos. 

\begin{figure}[t]
\centering
\raisebox{3pt}{\includegraphics[width=0.36\linewidth]{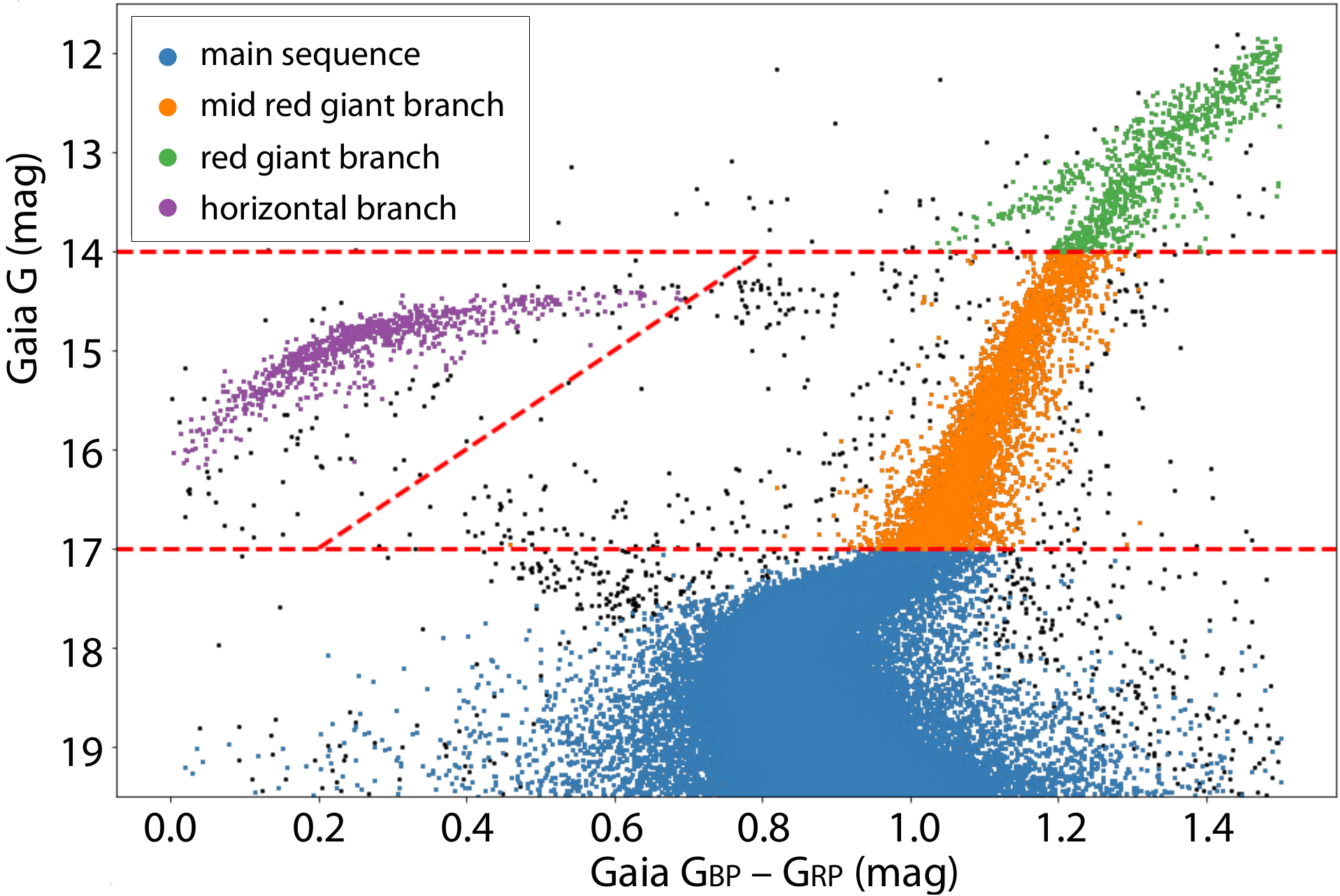}}
\hspace{2pt}
\includegraphics[width=0.26\linewidth]{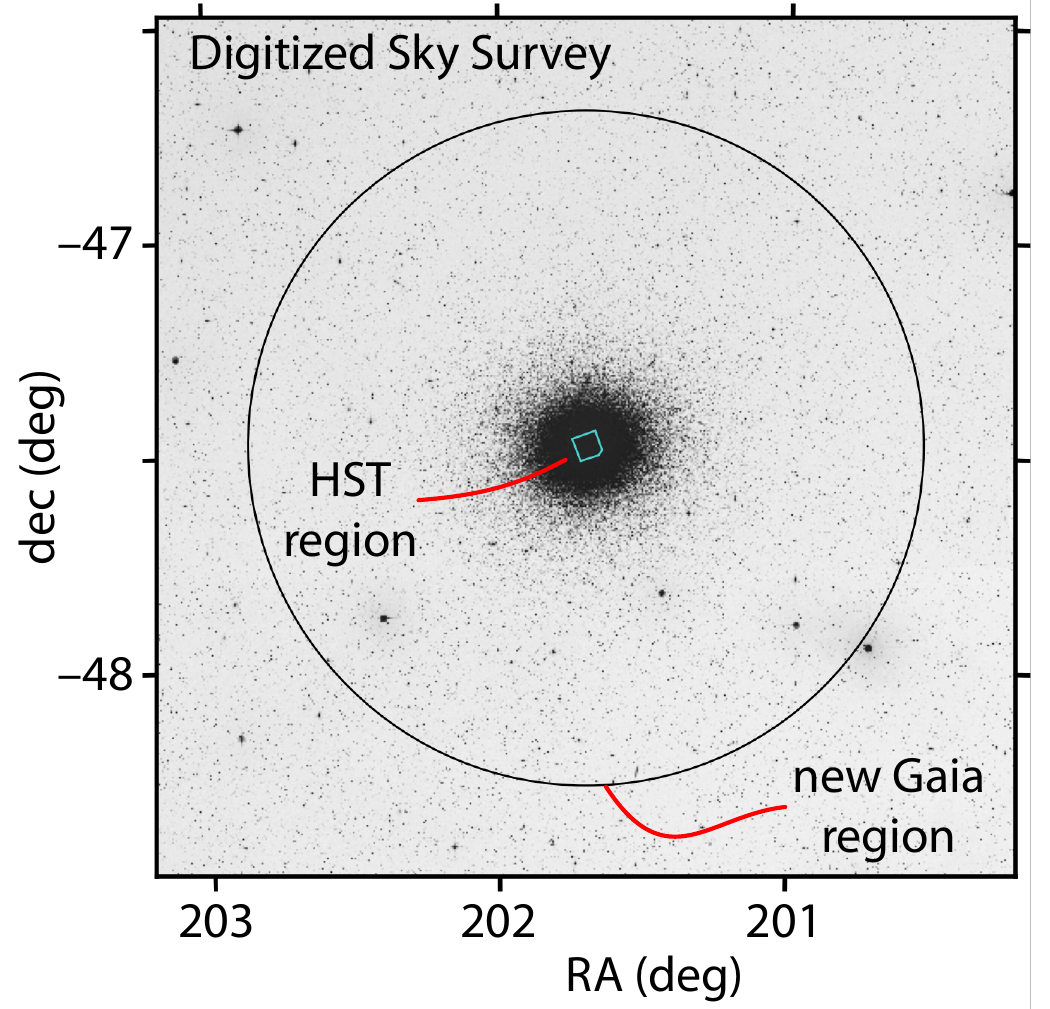}
\hspace{2pt}
\includegraphics[width=0.35\linewidth]{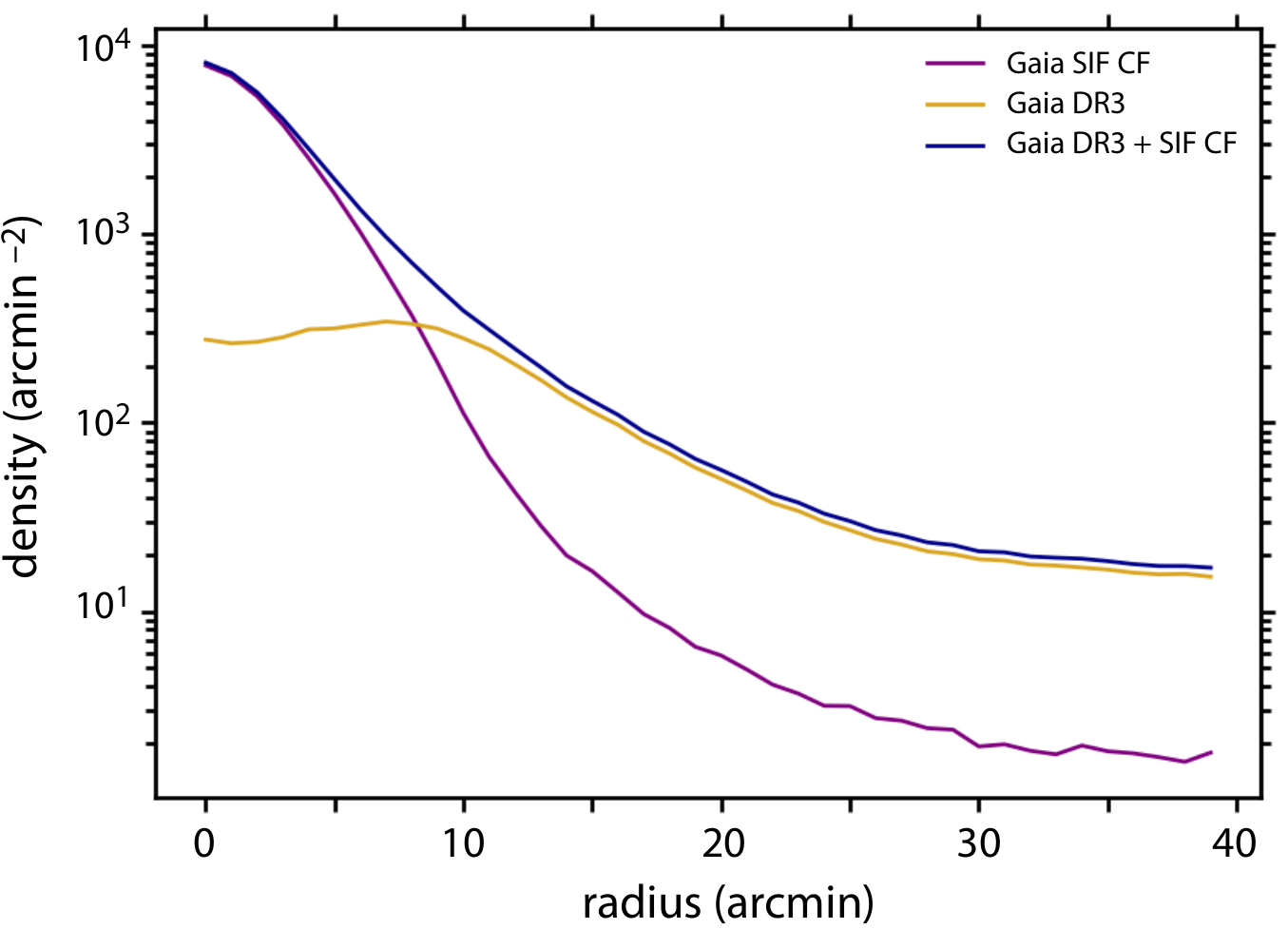}
\caption{The globular cluster $\omega$~Cen.
(a)~Colour--magnitude diagram of 66\,467 members, with the dashed lines dividing it into four main regions (from \citet{2021ApJ...908L...5S}, Figure~5).
(b) Digitized Sky Survey view, with the `Service Interface Function' crowded field region (black circle) and HST reference data (central square) indicated.
(c)~Source density versus distance the from cluster centre, showing Gaia DR3 sources (yellow), `Service Interface Function' sources (violet), and the combination (blue);
(b--c from \citet{2023A&A...680A..35G}, Figures~13 and 15).
}\label{fig:omega-cen}
\end{figure}

The number of stars measured by Gaia is in itself noteworthy (Figure~\ref{fig:omega-cen}a). 
\citet{2021ApJ...908L...5S}	%
selected 178\,548 from EDR3 within 45~arcmin of the cluster centre. They estimated a mean cluster proper motion of $\mu_\alpha=-3.25$\muas, $\mu_\delta=-6.76$\muas, in good agreement with the DR2 value found by
\citet{2019MNRAS.482.5138B}.		
Selecting stars with this space motion then resulted in 108\,054 candidate members. Further restriction according to location in the colour--magnitude diagram resulted in 66\,467~members with good Gaia astrometry and 2-colour photometry.
Their resulting mean parallax of $\omega$~Cen is $0.191\pm0.001$\,mas, corresponding to a distance of $5236\pm28$\,pc, and in good agreement, for example, with the classical photometric distance of 5.2~kpc
\citep{1996AJ....112.1487H}.	
As
\citet{2021ApJ...908L...5S} stated:
{\it `The recent Gaia EDR3 opens a new chapter in the measurement of parallaxes, placing the precise and accurate determination of the distances to nearby Galactic globular clusters within reach. None may be more prized than that of $\omega$~Cen\ldots\ A precise determination of its distance will characterise the luminosity of a broad sample of stellar types.'}

Many studies of $\omega$~Cen have since been published exploiting the Gaia data. These include:
calibration of the tip of the red giant branch (Section~\ref{sec:trgb}), and the cluster's 3-d structure and Galactic orbit  
\citep{2021ApJ...908L...5S, 2023ApJ...950...83L};
kinematics and orbital anisotropy of the various stellar sub-populations as a function of metallicity 
\citep{2019MNRAS.487.3693J, 2020ApJ...898..147C, 2020A&A...637A..46S}, 
	including retrograde substructures \citep{2018MNRAS.478.5449M},
	and internal rotation as a function of stellar mass \citep{2023MNRAS.522L..61S};
identification of stellar streams (tidal debris) torn off as it orbits the Galaxy 
\citep{2019NatAs...3..667I,2019ApJ...872..152I,2021MNRAS.507.1127K,2022A&C....4100658C};
confirmation of its disrupted dwarf galaxy origin \citep{2023MNRAS.524.2630Y};
chemical tagging (including s-process elements) of the Fimbulthul tidal stream \citep{2020MNRAS.491.3374S};
confirmation of the cluster's relation to the Gaia Sausage--Enceladus stream \citep{2022ApJ...935..109L};
the distribution of dark and luminous mass, with evidence that the dark-mass component is more centrally concentrated \citep{2022MNRAS.511.4251E};
studies of its metal-poor stars \citep{2020AJ....159..254J},
	the distribution of hot subdwarfs \citep{2021ApJS..256...28L},
	and of far ultraviolet sources \citep{2022cosp...44.2217P,2022ApJ...939L..20P};
distortion mapping of the HST instruments ACS, UVIS and WFPC2 
\citep{2018acs..rept....1K,2021PASP..133f4505C}, and  proper motions from HST 
	\citep{2018ApJ...854...45L,2021MNRAS.505.3549S};
axion and neutrino bounds from the TRGB \citep{2020PhRvD.102h3007C};
and associated constraints on the monopole--dipole potential \citep{2023PhRvD.108j3024P}.

In October 2023, between DR3 (in June 2022) and the next major release (DR4 in late 2026), the Gaia Data Processing and Analysis Consortium (DPAC) published five papers as part of a `Focused Product Release' (Section~\ref{sec:data-releases}). The first described the results of a special observing mode adopted for $\omega$~Cen, implemented to overcome some of the limitations of the extreme spatial crowding in the cluster's core. As described in detail by 
\citet{2023A&A...680A..35G}	
these new observations were derived from special scans between 1~Jan 2015 and 8~Jan 2020, and a dedicated processing pipeline.\footnote{
In Gaia's normal operational mode, the crowding limit is around 1.05 million objects deg$^{-2}$, or about one object per 12~arcsec$^{-2}$
\citep{2016A&A...595A...1G}. 
For a small number of very densely populated regions on the sky, the standard `windowing’ and readout strategy for the astrometric field becomes saturated: multiple sources begin to overlap, leading to conflicting readout windows, windows containing multiple sources, or no readout at all. 
In a program of dedicated `Service Interface Function’ scans, an alternative approach to the observations of a number of high stellar density regions was implemented: for the globular clusters $\omega$~Cen, 47~Tuc, M4, M22, and NGC~4372, as well as Baade’s Window in the bulge, the LMC and SMC, and the Sagittarius~I galaxy.
In this special mode of operation, the satellite's `scanning law' and spin rate are not modified. But sources, and their associated astrometric data, are instead derived from the binned ($2\times2$) pixels of the sky mapper. This has the advantage of preserving much more of the {\it across-scan\/} spatial information, and thereby significantly reducing source confusion.  A further improvement in the effect of crowding results from the fact that the sky mappers are only illuminated by a {\it single\/} field-of-view, while the astrometry field observations result from the superposition of the two fields of view.  
Disadvantages of these dedicated sky mapper observations are (a)~the lower astrometric accuracy compared with the astrometric field data (the latter resulting from nine along-scan CCDs, each exploiting the full along-scan pixel resolution), and (b)~the absence of colour and spectral information which, for the nominal mode of operation, is obtained from the BP/RP photometers, and the RVS fields.}
As a result, the previous `hole’ in the core of the $\omega$~Cen cluster, evident in Gaia~DR3, is now `filled' with 526\,587 additional sources, compared with the 321\,698 already present in Gaia DR3. Their fidelity, and their associated astrometry, was validated by comparing the results to a dedicated Hubble Space Telescope dataset covering the cluster core (Figure~\ref{fig:omega-cen}b--c).
Magnitude uncertainties are in the range 1--10~mmag over $G=15-20$~mag.
Astrometric standard errors are significantly higher than for Gaia DR3, with median uncertainties of $\sim\!3$\,mas in RA/Dec, $\sim\!2$\,mas in both components of proper motion, and $\sim\!4$\,mas in parallax.
This larger star sample has been used in studies of its rotation and metallicity-dependent kinematics
\citep{2024ApJ...970..192H,		
2025A&A...696A.168V,		
2025MNRAS.537.2752K}.		
It has also been used in the confirmation of the recent HST-based evidence for an intermediate-mass black hole in the cluster's centre
\citep{2024Natur.631..285H,		
2024ApJ...963...60P}.		

\begin{figure}[t]
\centering
\includegraphics[width=0.36\linewidth]{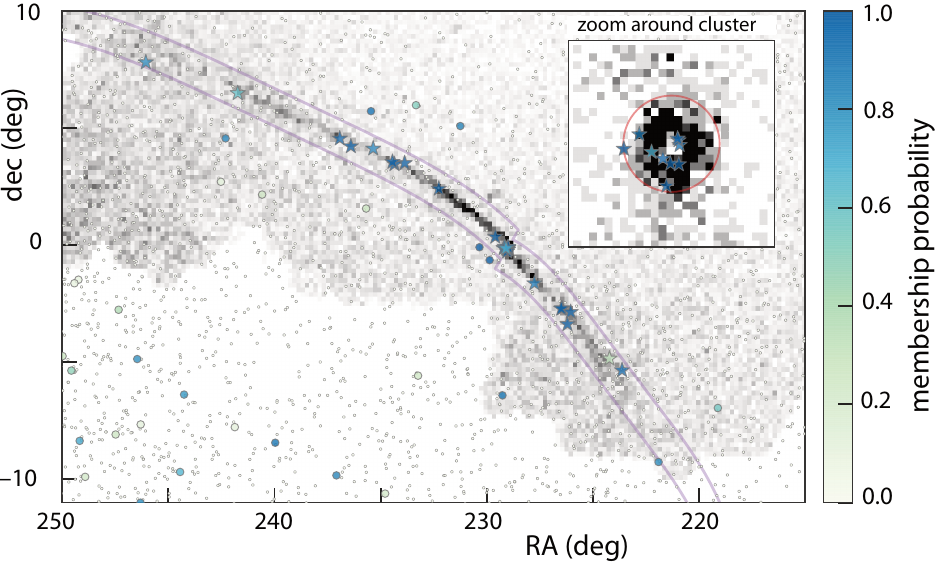}
\hspace{10pt}
\raisebox{15pt}{\includegraphics[width=0.60\linewidth]{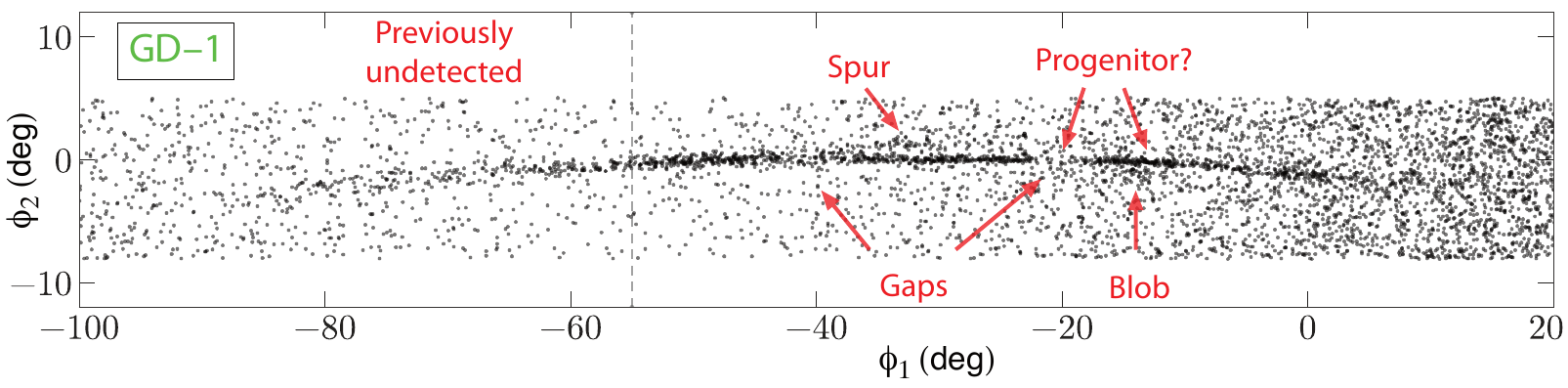}}
\caption{Left: Palomar~5, with RR~Lyrae stars tracing the stream on the sky, coloured by membership probability. The background greyscale shows the density of main sequence stars from \citet{2020ApJ...889...70B}. The inset is a $1\ddeg5\times1\ddeg5$ zoom around the cluster centre, with the Jacobi radius shown as a red circle (\citet{2019AJ....158..223P}, Figure~3).
Right: the GD--1 stream, with members selected according to proper motion and their location in the colour--magnitude diagram, showing high-contrast gaps and possible spurs (\citet{2018ApJ...863L..20P}, Figure~1a).
}\label{fig:palomar5-gd1}
\end{figure}

\paragraph{Palomar~5}	
\label{sec:palomar5}

Of the many individual globular clusters being examined with the Gaia data, Palomar~5 is one of 15 faint, distant, or obscured globular clusters in the Milky Way, discovered on the Palomar Sky Survey plates in the early 1950s, in this case by Walter Baade. Some 8~arcmin in extent, with $M\sim10^4M_\Sun$, it lies $\sim$ 18\,kpc from the Galactic centre ($\sim$20\,kpc from the Sun). It gained renewed interest twenty years ago, with the discovery, from the Sloan Digital Sky Survey, reaching well below the main sequence turnoff, of two tidal tails, one on each side of the cluster, extending over an arc of 2\ddeg6 on the sky
\citep{2001ApJ...548L.165O}.
They inferred that the cluster is experiencing heavy ongoing mass loss, mainly along a line tracing out the cluster's Galactic orbit. The strongest tidal shocks on globular clusters in the Galaxy are generated as the cluster crosses the Galactic disk, or passes through pericentre on a highly eccentric orbit. In the case of Palomar~5, the next passage through the disk will occur in $\sim$100~Myr
\citep{2001ApJ...548L.165O}.

Tidal tails (Section~\ref{sec:tidal-tails}) have now been confirmed or discovered by Gaia in many open clusters, and a number of globular clusters, and Palomar~5 is viewed as a particularly useful probe for constraining Galactic structure
\citep[e.g.][]{2017MNRAS.470...60E,
2018ApJ...858...73D,
2020ApJ...889...70B,
2021NatAs...5..957G,
2021AJ....161...12X}.
Although the cluster's main sequence is only just accessible to Gaia, its horizontal branch is bright enough, at $G\sim17.3$, to provide precise proper motions along the full extent of the tidal stream. Accordingly,
\citet{2019AJ....158..223P}
used RR~Lyrae stars to determine improved distances and kinematics, 10 as cluster members and a further~17 as members of the stream (Figure~\ref{fig:palomar5-gd1}a). They gave a revised distance of $20.6\pm0.2$~kpc, and detected gradients in distance (of 0.2~kpc per degree), and proper motion, along the stream, consistent with the simulations of 
\citet{2017NatAs...1..633P}
which, by including the rotating Galactic bar, also reproduced the observed truncation of the leading arm.
Other searches for tail members were made with DR2
\citep{2020MNRAS.493.4978S},
and EDR3 
\citep{2022MNRAS.512..315K}.
In subsequent simulations,
\citet{2023MNRAS.524.2124P}	
modelled the oblateness of the Milky Way dark matter halo from fits to the Palomar~5 stellar stream, along with two other globular cluster streams: NGC~3201 and NGC~4590 (M68), together being well fitted with a halo axis ratio $q=1.06\pm0.06$, and core radius $\sim$20~kpc. 

Gaia-based studies of halo streams and tidal tails are still in their infancy. With the tails being sensitive to the shape of the halo, and other substructures such as the bar
\citep[e.g.][]{
2012MNRAS.420.2700K,	
2018ApJ...858...73D,	
2015ApJ...799...28P},	
and with longer tails being more constraining,
\citet{2018ApJ...858...73D}
have argued that Gaia should provide {\it `\ldots\ decisive tests of the consistency with $\Lambda$CDM'}.

\paragraph{Multiple populations}

Until the 1990s, globular clusters were broadly believed to have formed from a single burst of star formation, with their early formation epochs ($z\gtrsim3$) having preceded most of the Galaxy's assembly. Observations since then have revealed growing evidence for broad or distinct main sequences, red giant branches, and sub-giant branches 
\citep[e.g.][and various others since]{
1997PhDT.........8A,
1998ApJ...500L.179G, 
1999Natur.402...55L,
2004ApJ...605L.125B, 
2005ApJ...631..868D,
2007ApJ...661L..53P,
2008ApJ...673..241M}.

Starting with studies of M4
\citep{2008A&A...490..625M,
2011ApJ...730L..16M},
these different populations are now known to be linked to their different abundances, in particular for the elements resulting from H~fusion (C, N, O, Na, Al, and in some cases Mg, Si, and K). 
In this context, stars with lower N and Na, and higher C and O, resemble Galactic field stars with similar metallicity, and are referred to as `first population' or `first generation' (and denoted~1P or 1G).  Conversely, stars enhanced in N and Na (also He and Al), and depleted in~C and~O, define one or more subsequent stellar populations, and are referred to collectively as second population (2P or 2G).
%
As a further comment on the field's nomenclature, `chromosome maps' are 2-colour diagrams derived from colours which maximise the separation of these populations. 
And clusters are often described as Type~I and Type~II, based on the distribution of stars in the chromosome maps, or on star-to-star heavy-element variations.

Understanding the origin of these multiple stellar populations has proven a challenge for models of stellar evolution, nucleosynthesis, and star-formation processes at high redshift. A simple scenario might attempt to attribute them to different generations of stars: the first burst of star formation would form stars out of pristine material, with subsequent events building stellar populations from their nucleosynthesis products, which then evolve faster. A challenge here is that 2P stars represent the majority of stars in most clusters 
\citep{2017MNRAS.464.3636M,
2022ApJ...927..207D}. 
In such a model, this would imply that the progenitors of today’s globular clusters were substantially more massive, having preferentially lost most of their initial 1P stars into the field. 
Others have suggested that they form in a single burst of star formation, with a fraction then polluted by the ejecta of more massive stars of the same generation
\citep{2013MNRAS.436.2398B,
2018MNRAS.478.2461G}.
Other explanations for the chemical anomalies have been put forward 
\citep[][\S4]{2022Univ....8..359M},
including
`hot bottom burning' in asymptotic giant branch stars;
massive stars rotating near break-up;
massive interacting binaries;	
the loss of massive stars due to their sinking into black holes;
the role of super-massive stars;
or the results of stellar mergers. 
But their conclusion was that {\it `we do not properly understand the origin of multiple stellar populations in globular clusters'}. 

Clues as to how Gaia might assist come from scenarios which predict that these populations also differ in their initial structural and kinematic properties, specifically that 2P stars may form in a sub-system more spatially concentrated than the 1P system 
\citep[e.g.][]{
2008MNRAS.391..825D,	
2011MNRAS.412.2241B,	
2017MNRAS.471.2242B,	
2019MNRAS.489.3269C,	
2022MNRAS.517.1171L,	
2024MNRAS.534.2397L}.	
Accordingly, two main aspects of the Gaia data are being used to throw further light on these multiple populations. The first is the use of the distances and proper motions to probe the morphology and kinematics of the cluster members, exploiting (and in contrast to HST) the quasi-inertial nature of Gaia's proper motion system. Another feature is that Gaia's kinematics extend to the outer cluster regions and beyond, compared with HST-based studies which tend to focus on the central regions where the star density is highest.  
Studies are also leveraging Gaia's low-resolution XP (BP/RP) spectra, which were made available in Data Release~3 (June 2022) for the 220~million brightest stars. These allow construction of any desired `synthetic photometry', subject to it lying within Gaia's optical spectral range (Section~\ref{sec:synthetic-photometry}). This is allowing the definition of optimised colour indices (and chromosome maps) to separate the different populations.

An important target in this context is NGC~1851, the well-studied prototype of the less-common Type~II systems. It shows large differences in both light- and heavy-element abundances, and has been hypothesised to be the remnant of accreted dwarf galaxies. 
\citet{2023A&A...678A.155C}	
used the low-resolution XP spectra to construct synthetic photometry in the BVI bands, and thence the pseudo-colour index $C_{\rm BVI}=(B-V)-(V-I)$, particularly effective in separating the metal-rich `anomalous' stars from the `canonical' population. 
%
They also constructed an optimised `chromosome map', $m_{\rm F415}$ versus $m_{\rm F415}-I$, where their Gaia-based F415$^{25}$ photometric filter (the notation indicating the filter's central wavelength and width, in~nm) effectively separates bright red giant branch stars into canonical and anomalous groups.
They demonstrated that both groups are found both inside and outside the cluster's tidal radius, out to $3.5r_{\rm tidal}$, or about 45~arcmin. Nonetheless, the canonical population dominates outside the tidal radius, and it exhibits a more circular on-sky morphology, in contrast to the more elliptical shape of the anomalous population. 

The Gaia proper motions also give access to the stellar kinematics. And while dynamical differences in the inner regions may be smoothed by relaxation processes, the outskirts of the cluster, characterised by much longer relaxation times, are more likely to retain evidence for any distinct dynamical evolution. And indeed they found hints of a tangentially anisotropic motion in the outer regions, indicating a preference for stars to escape on radial orbits. 
They conclude that the Gaia DR3 low-resolution XP spectra, together with Gaia DR3 astrometry and proper motions, are indeed powerful tools for investigating the multiple populations in globular clusters. 

A similar approach was followed by  
\citet{2025MNRAS.536.1077M}.	
They found that the most-efficient filters to distinguish the distinct stellar populations were F380$^{200}$ and F430$^{100}$, which are sensitive to N and C variations respectively. They were able to identify 1P and 2P stars in the cluster's outermost regions and beyond its tidal radius, the 2P stars being more centrally concentrated. Similar chromosome maps effectively distinguished multiple populations in the outer regions of four other clusters with different metallicities: NGC~3201, NGC~6121, NGC~6752, and NGC~6397.  Notwithstanding some detailed differences in these various clusters, the radial dependencies are consistent with cluster formation scenarios where the 2P stars originate in the central regions. 

\citet{2025ApJ...981...57J}		
and 
\citet{2025MNRAS.537.2342C}	
extended this approach to examine the radial distributions for 29 globular clusters, using both Gaia and HST proper motions. They found that the 1P stars transition from isotropy to slight tangential anisotropy towards the outer regions, while the 2P stars become increasingly radially anisotropic beyond the half-light radius. Statistically significant differences in the anisotropy profiles were found for dynamically young and non-relaxed clusters. And clusters with orbits closer to the Galactic centre exhibit larger dynamical differences between 1P and 2P stars than those with larger peri-Galactic radii. 
Together, their findings are again consistent with a scenario where 2P stars form in a more centrally concentrated environment, and where the interaction with the Milky Way's tidal field plays a crucial role in the dynamical evolution of both populations.

The combination of Gaia's astrometry, and its far-reaching capabilities in the area of synthetic photometry, has presumably only just started its impact on this complex problem in the understanding of globular clusters.

\subsection{Halo streams}
\label{sec:halo-streams}
%

\paragraph{Introduction}
Present theories, supported by extensive numerical simulations, argue that large galaxies, such as the Milky Way, have been built up from a series of mergers with smaller galaxies over the past several billion years, and that our Galaxy's inner stellar halo should be dominated by the debris of just a few massive progenitor galaxies merging with our own early on in its formation. One of the most spectacular and far-reaching of Gaia's contributions is the convincing observational support for these ideas. 
In this scenario, galaxies gravitationally captured by the Milky Way slowly spiral inwards over billions of years. The more distant (Sagittarius and the Magellanic Clouds) show only early signs of capture and disruption, while the most ancient may be on their second or subsequent orbital approach. While they steadily dissolve in 6d phase-space, i.e.\ progressively scattering through the halo in their positions and velocities, some preserve clear signatures of their common origin through a clustering of their abundance patterns {\it and\/} their orbital properties.  More precisely, while they may be widely scattered on the sky, their common origin is encoded in their similar `integrals of motion', e.g.\ in their total energy, $E_{\rm tot}$, and their projected orbital angular momentum
\citep[][\S3.1.1]{2000MNRAS.319..657H,
2008gady.book.....B}. 
The locus of any one tidal stream provides a record of an ancient Galaxy orbit. 
These ancient signatures of tidal infall remain accessible for study because the orbital time-scales in the outer parts of the Galaxy are several~Gyr. As a result, the halo retains kinematic evidence of the surviving remnants of accretion, as well as a chemical `memory' of early low-mass stars as a result of their very long evolutionary lifetimes.

\paragraph{Pre-Gaia discoveries}

Our Galaxy's disk has long been recognised as having been formed by the rapid collapse of a rotating galaxy-sized gas cloud, billions of years ago. But the idea that the halo has been built up over billions of years from infalling low-mass objects, such as dwarf galaxies of mass $10^7-10^8\,M_\odot$, only dates back to the seminal paper by 
\citet{1978ApJ...225..357S}.
Their study of 177 red giants in 19~globular clusters at distances $\gtrsim8$\,kpc, suggested that halo clusters originated within transient protogalactic fragments that gradually lost gas while undergoing chemical evolution, and continued to fall into the Galaxy after collapse of their central regions. This model, evidenced by
the wide range of globular cluster metallicities independent of Galactocentric radius, 
a wide age spread of halo field stars and globular clusters, 
and a subset of intermediate abundance globular clusters with retrograde mean motions, 
has been the paradigm for the halo formation for at least the past two decades
\citep[e.g.][]{2000Sci...287...79B}.

Studies in the 1990s, which considered both the chemistry, and the dynamical motions of the nearby halo stars together, hinted at the presence of such infalling stellar streams and clumps, and of correlations between the stars’ chemical abundances and their orbital parameters. 
Over the past 25~years, various of these dynamically evolved `stellar streams' have been identified, all originating from this sort of tidal stripping. Some have their origin in disrupted satellite galaxies, others in specific globular clusters.
And while globular clusters and satellite galaxies represent two very different types of stellar systems, streams can result from either. And, in the context of the following discoveries, it is useful to bear in mind that some streams can be attributed to captured dwarf galaxies, others to disrupted globular clusters, while others are still of uncertain origin due to the absence of any well-defined progenitor. 

The first of these to be identified is generally considered to be the Arcturus stream
\citep{1971PASP...83..271E,	
2004ApJ...601L..43N, 		
2006MNRAS.365.1309H,		
2019A&A...631A..47K},		
comprising some 50 ancient stars deficient in heavy elements, located in the thick disk, and attributed to some now-dissolved dwarf galaxy (although other origins are still under consideration, cf.\ Section~\ref{sec:arcturus-hr1614}).
The Sagittarius stream, tidal debris from the disrupted Sagittarius dwarf satellite, itself discovered only in 1994 
\citep{1994Natur.370..194I}, is now known to dominate the halo from 20--50\,kpc, and is described further below.
The gaseous Magellanic Stream, associated with the Large and Small Magellanic Clouds (Section~\ref{sec:magellanic-clouds}), is a more distant Local Group analogue.

The Hipparcos results in 1997 allowed the discovery of the first `disrupted event' from this type of correlated space motion and orbital angular momentum
\citep{1999Natur.402...53H}.		
Corroborated by their prominent metal deficiency, the `Helmi stream' is inferred to have started life as a dwarf galaxy of $10^7-10^8M_\Sun$, captured by the Milky Way some 6--9\,Gyr ago. It may be responsible for some 10\% of the metal-poor stars in our Galaxy's halo beyond the Sun's orbit.
Others streams have been discovered since, variously attributed to globular cluster or dwarf galaxy origins.
Several have been discovered from the Sloan Digital Sky Survey, amongst them the Acheron, Cocytos, Lethe, and Styx streams 
\citep{2009ApJ...693.1118G},	
and the polar-orbiting Cetus stream
\citep{2009ApJ...700L..61N}.	
The LAMOST~1 stream of some 20\,000 stars, was found from the LAMOST spectroscopic survey
\citep{2016ApJ...816L...2V},	
and another, the Phoenix stream, was the first found from the Dark Energy Survey
\citep{2016ApJ...820...58B}.	
By 2016, 20 streams were tabulated by 
\citet{2016ASSL..420...87G},
while there are more than 120~known today.

\paragraph{Search methods}
Various approaches have been used to search for these stellar streams. The challenge is to identify perhaps only a few hundred stars with similar orbits and chemistry, now possibly scattered widely across the sky, amongst Gaia's two billion stars. More powerful algorithms have been developed progressively to exploit improvements in the quality of distances, kinematics, and chemistry. Roughly chronologically, these include:

\noindent{\footnotesize $\bullet$}
Pole counts: this identifies high-contrast structures on great circle paths
\citep{1996ApJ...465..278J}, 
and was used in the detection of the Sagittarius stream \citep{2002MNRAS.332..921I}.

\noindent{\footnotesize $\bullet$}
Co-moving groups: this aims to identify similar star types (e.g.\ RR~Lyrae, BHB) contained in a small phase-space volume.  It was used in the discovery of the
Arcturus \citep{2006A&A...449..533A},
Virgo \citep{2006ApJ...636L..97D},
and Aquarius \citep{2011ApJ...728..102W}
streams.

\noindent{\footnotesize $\bullet$}
Matched filtering 
\citep{2002AJ....124..349R,
2011MNRAS.416..393B}:
this uses colour--magnitude filtering to find structures that belong to a given stellar population, but does not incorporate the wealth of kinematic information contained in the Gaia data. It was used in the discovery of
Palomar~5 \citep{2001ApJ...548L.165O}, 
GD--1 \citep{2006ApJ...643L..17G}, 
Orphan \citep{2006ApJ...642L.137B}, 
Lethe, Cocytos, and Styx \citep{2016ASSL..420...87G},
Eridanus and Palomar~15 \citep{2017ApJ...840L..25M},
and 11 new streams in the Dark Energy Survey data \citep{2018ApJ...862..114S}.

\noindent{\footnotesize $\bullet$}
Given full 6d phase-space coordinates, streams can be found from cuts in energy and/or angular momentum 
\citep{2019ApJ...874..138L, 
2020ApJ...900..103J}, 
or by searching for stars on similar orbits 
\citep{
2000MNRAS.319..657H,	
2019ApJ...886..154Y},	
e.g.\ using friends-of-friends type algorithms
\citep{2013ApJ...762..109B}.		

\noindent{\footnotesize $\bullet$}
{\tt STREAMFINDER} \citep{2018MNRAS.477.4063M} was developed for Gaia, and has been amongst the most successful stream-finding approach to date. It is based on the fact that stars in a (thin and dynamically cold) stream are connected through the progenitor's orbit. Starting with one or more members of a hypothesised stream, others will then be contained in a `hypertube' whose phase-space dimensions are set by the progenitor's orbit, size and velocity dispersion, moving in the Galaxy's gravitational potential. The trial orbits (and assumed potential) are adjusted to maximise the star counts in the hypertube. 
It was used to discover GD--1 from Pan--STARRS1 data
\citep{2018MNRAS.478.3862M}. 	
Gaia DR2 discoveries include
Phlegethon
\citep{2018ApJ...865...85I},	
and Slidr, Sylgr, Ylgr, Fimbulthul, Svöl, Fjörm, Gjöll, and Leiptr
\citep{2019ApJ...872..152I}, 
along with 9~new streams with EDR3
\citep{2021ApJ...914..123I}.	

\noindent{\footnotesize $\bullet$}
{\tt STARGO}
\citep{2018ApJ...863...26Y,	
2019ApJ...881..164Y} 		 
works in the 4d space of orbital energy and angular momentum, and hence also requires full 6d phase-space information. It uses unsupervised learning (a `self-organising map'), which trains a 2d neural network to learn the data set's topological structures from the energy--momentum 4-space. It has been applied to the Cetus stream
\citep{2019ApJ...881..164Y},	 
to the LMS--1 stream 
\citep{2020ApJ...898L..37Y},	
as well as several open clusters from Gaia DR3 
\citep[e.g.][]{
2022ApJ...937L...7P,		
2023A&A...675A..67Q}.	

\noindent{\footnotesize $\bullet$}
{\tt VIA MACHINAE} 
\citep{
2022MNRAS.509.5992S,		
2024MNRAS.529.4745S}:		
was also developed specifically for Gaia, but makes no assumptions about the form of the Galactic potential, orbits, or isochrones. It uses an unsupervised machine-learning method for anomaly detection ({\tt ANODE}), originally developed for the Large Hadron Collider. It first identifies stars that are, in terms of astrometric and photometric properties, `overdense' with respect to the background, thereafter restricting selection to overdensities that are broadly consistent with stellar streams. It was used to identify 102 streams in Gaia DR2, of which only 10 had been previously identified
\citep{2024MNRAS.529.4745S}. 

\noindent{\footnotesize $\bullet$}
{\tt SkyCURTAINs} is another recent model-agnostic algorithm, based on a weakly supervised machine-learning algorithm, but so far only applied to the known GD--1 stream
\citep{2025MNRAS.536.1104S}. 

\begin{figure}[t]
\centering
\includegraphics[width=0.32\linewidth]{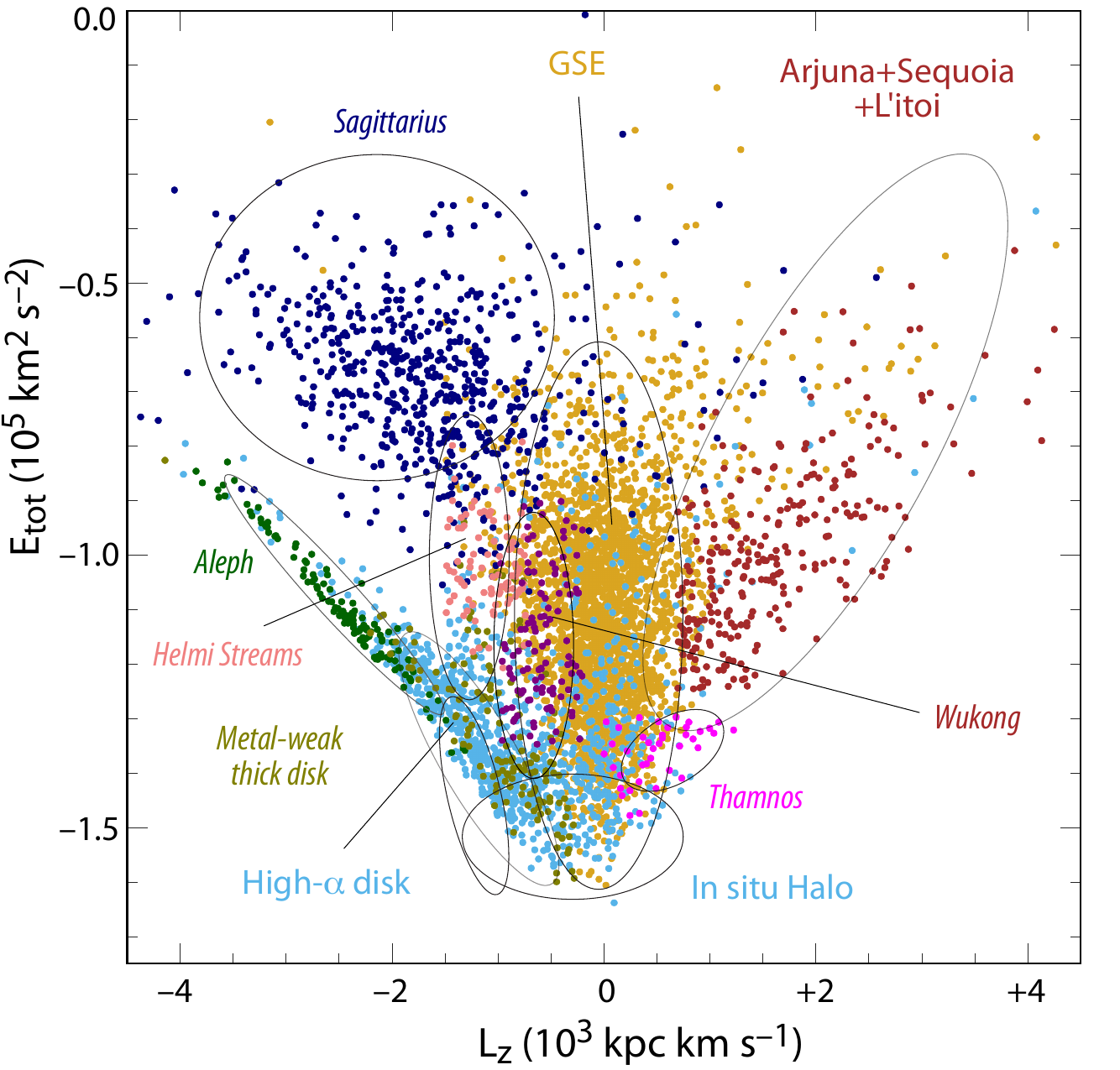}
\hspace{40pt}
\includegraphics[width=0.32\linewidth]{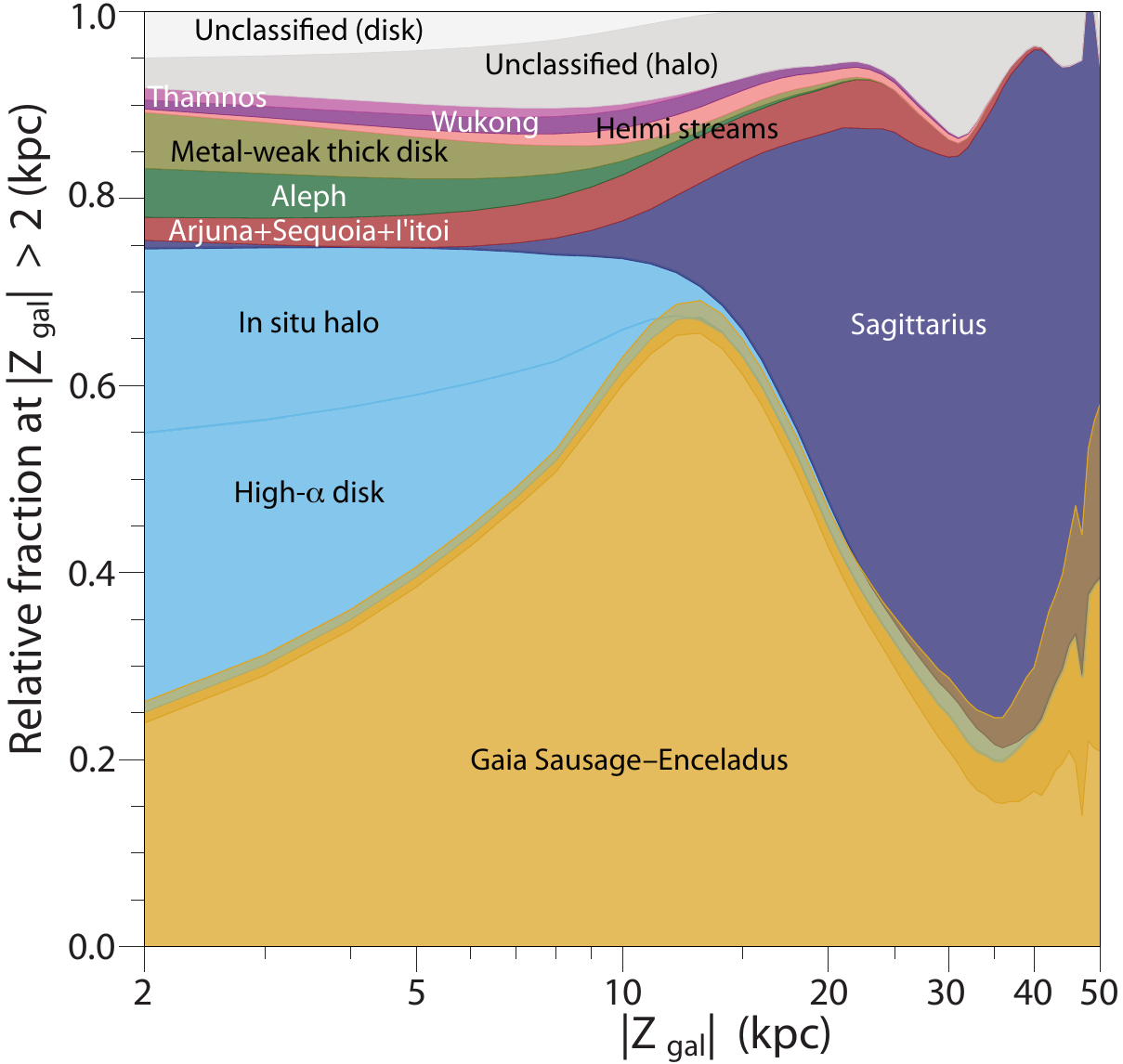}
\caption{Left: stellar streams in the $E-L_z$ plane identified by \citet{2020ApJ...901...48N}.
Right: contributions versus distance from the Galactic plane. The high-$\alpha$ disk is taken as $e<0.5$, and the {\it in~situ\/} halo as $e>0.5$, although they form a continuous distribution. GSE is the dominant component within $Z\!\sim\!10-20$\,kpc ($R_{\rm gal}\!\sim\!15-25$\,kpc), while most stars at larger distances belong to Sgr (\citet{2020ApJ...901...48N}, Figures~18 and~19b).
}\label{fig:streams-gse}
\end{figure}

\paragraph{Gaia Sausage--Enceladus, GSE}
\label{sec:gse}

By providing the distances and space motions of more than two billon stars, from which orbital distances and motions can be calculated, Gaia has opened a new chapter in the study of stellar streams. First amongst these, from Gaia DR2 in 2018, was the Gaia--Enceladus stream, subsequently more widely known as the Gaia Sausage--Enceladus stream, or GSE. In their discovery paper,
\citet{2018Natur.563...85H}
demonstrated that the inner halo, between 5--25\,kpc, is dominated by debris from an object which at infall was slightly more massive than the Small Magellanic Cloud (a mass ratio of 4:1), and which they referred to as Gaia--Enceladus.  The stars originating from the accretion event cover much of the entire sky, and are on slightly retrograde and elongated trajectories. Hundreds of RR~Lyrae stars and thirteen globular clusters following a consistent age--metallicity relation can be associated to the merger on the basis of their orbits. 
The merger would have led to the `dynamical heating' of the precursor of the Galactic thick disk, increasing their space velocities, and therefore increasing their scale height with respect to the Galaxy mid-plane. They suggested that the merger contributed to the formation of the Galaxy's thick disk some 10~Gyr ago and that, most probably, this was the last significant merger that our Galaxy experienced.

Refereed papers mentioning Gaia Sausage--Enceladus in their {\it abstract\/} is still rising: 10 in 2019, 25 in 2020, 35 in 2021, 40 in 2022, 40 in 2023, and 50 in 2024. This underscores the stream's importance in revealing the assembly history of the Milky Way -- structurally, dynamically, and chemically --  and its value in understanding our Galaxy's formation in a cosmological context. It also means that I can only give some highlights of our present knowledge. The pre-Gaia understanding of the Galaxy's halo and its formation, and more on what is known about GSE, can be found in the recent review by 
\citet{2024NewAR..9901706D}.	

It is now clear that the GSE merger, some $\sim$9.5\,Gyr ago (at $z\lesssim3$), had a major effect on the halo. 
\citet{2020ApJ...901...48N} found that, at $|Z|\gtrsim15$\,kpc, the debris of just two massive accreted dwarfs ($10^8-10^9M_\Sun$) comprise 80\% of the halo, together explaining the relatively high overall halo metallicity, [Fe/H]\,$\approx-1.2$. They attributed 95\% of their sample to one of their dozen identified structures (Figure~\ref{fig:streams-gse}a), pointing to a halo built {\it entirely\/} from accreted dwarf galaxies.
Within 25~kpc, GSE, with [Fe/H]\,=\,--1.2, was a radial merger that now dominates the local metal-poor halo (Figure~\ref{fig:streams-gse}b). The `head-on' collision resulted in rapid phase mixing, evident today only in its clustered integrals of motion. 
Beyond 25\,kpc, the halo is dominated by the Sagittarius (Sgr) dwarf galaxy and stream, [Fe/H]\,=\,--1.0, one of the first identified halo streams
\citep{1994Natur.370..194I}. 
Here, both the residual galaxy core, as well as two vast preceding and trailing streams, are clearly discernible.
Simulations by
\citet{2021ApJ...923...92N}	
suggest a GSE mass at infall of $5\times10^8M_\Sun$ in stars and $2\times10^{11}M_\Sun$ in dark matter, and that it arrived on a retrograde orbit, starting at $z\sim2$, with Arjuna being its retrograde debris, and Sequoia and I’itoi perhaps also being stripped from its outer regions. 

Beyond 10--20\,kpc lie several `overdense' regions, amongst which the Hercules--Aquila Cloud, and the Inner Virgo and the Eridanus--Phoenix overdensities, have been suggested to be GSE's apocentric pile-ups. 
From Gaia DR3 astrometry of 200\,000 red giant branch stars out to 100~kpc,
\citet{2023ApJ...951...26C}	
identified a large population of retrograde debris representing most distant `echoes' of the GSE merger. And they associated 
the southern Pisces Overdensity at 70\,kpc
\citep{2010ApJ...717..133S,
2012ApJ...753..145C},
and the more distant northern Outer Virgo Overdensity at 80\,kpc
\citep{2017ApJ...844L...4S}, 
with its successive apocentres. The majority beyond 40\,kpc follow a great-circle track consistent with the GSE orbit, and distinct from the Sagittarius stream which occupies the same plane, but orbits in the opposite sense. 
Their findings also match the N-body simulations of the GSE merger by 
\citet{2021ApJ...923...92N}.	

Mergers also have a significant effect on the disk, a prominent example being the Gaia phase-space spiral, attributed to phase-mixing through past pericentric passage of the Sagittarius dwarf (Section~\ref{sec:phase-space-spiral}).  A major local disk component is the `\mbox{high-$\alpha$} disk' and its high-eccentricity tail (or `{\it in situ\/} halo'), comprising some 15\% of the local Galaxy, and extending out to $|Z|\simeq15$\,kpc. 
\citet{2021ApJ...923...92N}
argued that the continuous eccentricity distribution supports a picture in which a primordial high-$\alpha$ disk was disturbed and `heated', possibly by the GSE merger. A similar origin was attributed to the `splash' population, stars born in the proto-disk, and subsequently `splashed' into low-angular-momentum, high-eccentricity orbits by a merger event
\citep{2020MNRAS.494.3880B,
2024NewAR..9901706D}.
But studies are still ongoing, and involve the relics of past mergers in the form of `energy wrinkles' and `phase-space folds' (also termed chevrons or caustics). 
While \citet{2023MNRAS.518.6200B} 
again attributed these to the GSE merger, other much younger events have also left their own mixed contributions
\citep[e.g.][]{
2022ApJ...932L..16D,	
2023ApJ...944..169D,	
2024MNRAS.531.1422D}.	

The demonstrated correlation between the combined mass of a galaxy’s globular clusters and its total mass 
\citep[e.g.][]{2023ApJ...950..179D}	
implies that while many tidal debris remnants are expected to be found in the stellar halo, only the most massive would be accompanied by globular clusters on similar orbits. This opens up another intriguing aspect of the properties and origin of the many halo streams now known.  
In the case of GSE, 
\citet{2018ApJ...863L..28M} 	
showed that amongst the Galaxy's {\it accreted\/} clusters, a tight grouping have very similar orbital actions (with small pericentres $\lesssim$2\,kpc, and large apocentres $\gtrsim$10\,kpc), consistent with clusters associated with the GSE merger.

These general findings are in line with simulations of galaxy formation, which predict that the inner stellar halo should be dominated by debris from just a few massive progenitors. But the agreement goes further. 
Amongst these very large-scale cosmological simulations, the EAGLE project (for example) produces a population of galaxies reproducing a broad range of observed properties. The largest of the EAGLE simulations, L100N15043, has a cubic volume of 100~Mpc in size, and includes the effects of both baryonic and dark matter. Its huge volume of simulated space--time provides numerous Milky Way-type galaxies, and with a wide range of merger histories.
Amongst these mergers, 
\citet{2019ApJ...883L...5B}	
identified one with remarkably similar properties to the GSE event, also occurring around 9~Gyr ago. These specific simulations result in merger debris on a slightly retrograde orbit (as found for GSE), bursts of star formation in the early disk, the formation of a dynamically heated thick disk (as seen in the Milky Way), and with a large fraction of the debris deposited at large heights above the Galactic disk, corresponding to the Milky Way's stellar halo.
Such $\Lambda$CDM-based simulations of structure formation are also guiding other aspects of the interpretation of Milky Way-type galaxies which have undergone major GSE-like mergers
\citep[e.g.][]{
2021MNRAS.503.5846R,
2021MNRAS.500.3750S,
2022MNRAS.513.1867D,
2023A&A...677A..89K,	
2023MNRAS.525..683O,	
2024ApJ...962...84S}.	

To conclude this outline, I refer again to 
\citet{2024NewAR..9901706D}
for a more detailed discussion of the progenitor mass, the event timing, the effects on our Galaxy disk, and the implications for understanding the ongoing Sagittarius and Magellanic Cloud mergers. While there is much more still unknown, as summarised in their conclusions, future Gaia releases should provide much greater clarity, and hopefully many more answers.

\paragraph{The Sagittarius stream}
\label{sec:sgr-stream}

The Gaia Sausage--Enceladus stream dominates the stellar halo within 25\,kpc. Beyond that, and out to some 50\,kpc, the halo is significantly populated by the Sagittarius stream. Debris of these two massive ($10^8-10^9M_\Sun$) accreted dwarfs together appears to comprise 80\% of the halo (Figure~\ref{fig:streams-gse}b). 

The Sagittarius dwarf galaxy (Sgr dSph) is a nearby satellite of our own, discovered by 
\citet{1994Natur.370..194I}.		
It was soon recognised as being tidally distorted, and inferred to have completed several close orbits around the Milky Way
\citep{1995MNRAS.277..781I,	
1996ApJ...458L..13M}.		
With a mass of $4\times10^8M_\Sun$, and distance $\sim$24\,kpc, it is a striking example of galaxy disruption and ongoing accretion.
Subsequent studies, initially with 2MASS and SDSS, progressively revealed an elongated and weakly rotating central core, with tidally stripped stars forming two long tidal tails, both bifurcated, tracing an orbit almost perpendicular to the Milky Way disk, with much other complexity
\citep{
2000ApJ...540..825Y,	
2002ApJ...569..245N,	
2003ApJ...599.1082M,	
2003ApJ...596L.191N,	
2004ApJ...601..242M,	
2006ApJ...642L.137B,	
2007ApJ...668..221N,	
2010ApJ...725.1516L,	
2012ApJ...750...80K,		
2013ApJ...762....6S}.	
The tidal debris traces out an orbit with an apogee of 50\,kpc for the leading arm, and 100\,kpc for the trailing
\citep{
2014MNRAS.437..116B,		
2017ApJ...850...96H}.		
Other work has further characterised the system's orbit, mass, chemistry, and velocity dispersion
\citep{
2012ApJ...756...74F,			
2013ApJ...777L..13M,		
2017MNRAS.464..794G,		
2017MNRAS.467.1329N},		
as well as its use in determining the Galaxy potential
\citep{
2010ApJ...714..229L,		
2016MNRAS.455.1079P,		
2019MNRAS.483.4724F}.		
The two most distant known halo stars, at 200~kpc 
\citep{2014ApJ...790L...5B}, 		
appear consistent with its maximal extent 
\citep{2017ApJ...836...92D},		
and perhaps with the `edge' of our Galaxy
\citep{2020MNRAS.496.3929D}.

With N-body simulations having to satisfy these many detailed observational constraints, one question is whether the present elongated, prolate, bar-like shape of the core is a result of tidal forces on an initially spherical galaxy  
\citep{1995ApJ...451..598J,
1997AJ....113..634I,
2001MNRAS.323..529H,
2020MNRAS.497.4162V},
or attributable to a more disk-like progenitor
\citep{2010MNRAS.408L..26P,		
2010ApJ...725.1516L,
2021ApJ...908..244D}.
\citet{2024A&A...687A..82L}
selected an Sgr analogue from the IllustrisTNG simulations to demonstrate how such a dwarf galaxy, with initial mass $>\!\!10^{11}M_\odot$, evolves around a Milky Way-like host on a tight orbit over seven pericentre passages with a period of 1\,Gyr. At the second pericentre, the disk transforms into a bar, and the bar-like shape is preserved thereafter. Strong mass-loss leaves a dwarf with final mass $\lesssim\!\!10^9M_\odot$. The gas is lost, and star formation ceases, at the third pericentre. The dwarf then retains a bar-like shape, small rotation, and a metallicity gradient, mirroring the observations.

Gaia is providing many new insights.
With DR2, investigations included identifying specific stellar types within the tidal arms, e.g.\
400 O-rich AGB stars 
\citep{2019A&A...626A.112M},	
164 M~giants
\citep{2019ApJ...874..138L},	
6000 RR~Lyrae stars with 5d phase-space coordinates
\citep{2020A&A...638A.104R},	
and
3500 RR~Lyrae with full 6d phase-space coordinates
\citep{2020ApJ...891L..19I}.	
The latter also concluded that the stream properties were still reasonably well reproduced by the earlier model of 
\citet{2010ApJ...714..229L}.		
DR2 was also used: 
to determine the mean proper motion over $2\pi$ rad along the stream, using 1500 intensity peaks comprising 100\,000 stars
\citep{2020A&A...635L...3A};	
to identify a low-metallicity population inferred to have originated in the stellar halo of the Sagittarius progenitor
\citep{2020ApJ...900..103J};	
to identify a faint globular cluster torn from the Ophiuchus stream by a close (5\,kpc) passage of the Sgr core 100~Myr ago
\citep{2020MNRAS.492.4164L};	
and to define 260\,000 members from which the mean proper motions yield a model in which the Sgr galaxy will be fully disrupted over the coming 1\,Gyr  
\citep{2020MNRAS.497.4162V}.	

Amongst these,
\citet{2021ApJ...908..244D}		
used 120\,000 stars from DR2 to reveal a bar structure 2.5\,kpc long, with the main body of the galaxy, strongly sheared by tidal forces, being a triaxial (almost prolate) ellipsoid. The inner core, of dimension $500\times330\times300$\,pc$^3$, shows no net expansion, but it is rotating, mainly about its intermediate principal axis, with a maximum velocity of $4.13\pm0.16$\kms.
\citet{2021MNRAS.501.2279V}	
found a misalignment between the stream track and the proper motion directions in the leading arm, which they interpreted as a time-dependent variation of the gravitational potential, in turn attributed to the recent passage of the LMC (of mass $1.3\pm0.3\times10^{11}M_\odot$). They argue that the stream cannot be modelled accurately in a static Galaxy potential, but rather the Milky Way is `lurching' toward the massive infalling LMC, giving the Sgr stream its peculiar shape and kinematics. 
And in a very different application, 
\citet{2018ApJ...867L..20H} 	
used the stream geometry to estimate the solar reflex velocity, and hence the velocity of the Local Standard of Rest, independently of an assumed value of $R_0$.

With Gaia EDR3, studies addressed the origin of the bifurcated tails (or parallel streams) in both northern and southern Galactic hemispheres
\citep{2006ApJ...642L.137B,	
2012ApJ...750...80K,			
2022A&A...666A..64R}.		
And while all models imply that Sgr has completed several orbits around the Milky Way, only tidal debris from the last pericentric passage had ever been detected, attributed to further phase-mixing. 
\citet{2021MNRAS.508L..26P}	
introduced a new method to identify clustering in angular momentum space, and identified 925 stars spanning $800^\circ$ on the sky, thus wrapping the Galaxy twice.
In other work, 60 high-velocity stars (including two hypervelocity stars) originating from Sgr were identified by
\citet{2022ApJ...933L..13L}.		
The effect of Sgr on other streams was investigated by 
\citet{2022MNRAS.516.1685D},	
and on the Jhelum stream in particular by
\citet{2023A&A...669A.102W}.	
Stream properties as a function of metallicity were investigated by 
\citet{2023ApJ...946...66L},		
and using Gaia DR3 by
\citet{2024ApJ...963...95C}.		
Searches for other candidates are also continuing with Gaia DR3
\citep{2024MNRAS.527.9767L}.	

That this sort of accretion process should also perturb the disk of the Milky Way had been predicted using test particle simulations 
\citep[e.g.][]{
1998A&A...336..130I,		
2009MNRAS.397.1599Q,		
2012MNRAS.419.2163G}. 	
And some evidence of such disk perturbations preceded the Gaia data
\citep[e.g.][]{
2009MNRAS.396L..56M,		
2012MNRAS.423.3727G,		
2012ApJ...750L..41W,		
2013ApJ...777L...5C,		
2015MNRAS.454..933D, 		
2018MNRAS.478.3809S, 		
2018MNRAS.481..286L}.		
Gaia DR2 provided unambiguous evidence of such gravitational disturbances of the disk with the discovery of the `Gaia phase-space spiral' (Section~\ref{sec:phase-space-spiral}), where ongoing phase mixing from an out-of-equilibrium state is evident in the phase-space projection in the $Z-V_Z$ plane 
\citep{2018Natur.561..360A}. 	
The Gaia phase-space spiral (and some low latitude overdensities such as the Monoceros Ring) has been widely attributed to pericentric passages of the Sagittarius dwarf 
\citep[e.g.][]{
2018MNRAS.481.1501B,		
2019MNRAS.489.4962K,		
2019MNRAS.483.1427L,		
2019MNRAS.485.3134L,		
2020MNRAS.492L..61L,		
2020MNRAS.497.4162V,		
2021MNRAS.504.3168B,		
2022ApJ...928...80G,		
2024MNRAS.527.4505D}.		
Others suggest that it may have arisen from several smaller disturbances, rather than a single dominant one
\citep[e.g.][]{
2022MNRAS.516L...7H,	
2023MNRAS.521..114T}. 	
A remarkable finding along similar lines, using Gaia DR2, was reported by
\citet{2020NatAs...4..965R}.		
They modelled the colour--magnitude diagram within 2\,kpc of the Sun to identify three conspicuous and narrow episodes of enhanced star formation, estimated as having occurred 5.7, 1.9 and 1.0\,Gyr ago. They found that these episodes coincide with modelled Sgr pericentre passages, with the perturbations from Sgr inferred to have repeatedly triggered major episodes of localised star formation.

A subject with its own substantial literature is the presence of globular clusters within the Sagittarius system.
\citet{1994Natur.370..194I} had already noted that M54 (NGC~6715) lies in its densest region. Later work suggested that others probably formed in its gravitational potential well, and have been stripped from it during its extended interactions with the Milky Way.
Pre-Gaia, nine globular clusters associated with Sgr were known
\citep{2003ApJ...599.1082M,	
2010ApJ...718.1128L}, 	
although several of these were only subsequently confirmed with the Gaia DR2 data
\citep{2020A&A...636A.107B}:		
M54 in the nucleus; 
Arp~2, Terzan~7 and~8 in its core;
and NGC~2419, NGC~4147, NGC~5634, Pal~12, and Whiting~1 in the tidal streams.
Using VISTA near-infrared data to identify potential RR~Lyrae members, and Gaia EDR3 proper motions to confirm membership, 
12 new clusters were identified by 
\citet{2021A&A...647L...4M},		
and a further 8 by 
\citet{2021A&A...650L..12M}.	
This rich system, 29 in total, quite likely completes the census of globular clusters associated with the Sgr system
\citep{2020AstBu..75..394A,	
2021MNRAS.508L..26P,		
2022A&A...665A...8K}.		

\begin{figure}[t]
\centering
\includegraphics[width=0.50\linewidth]{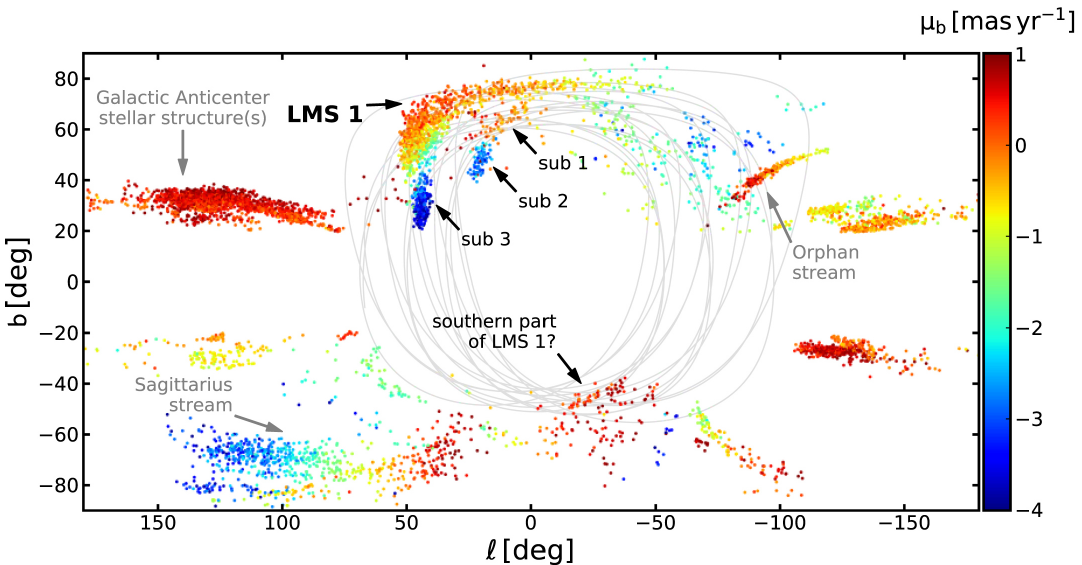}
\hspace{-5pt}
\includegraphics[width=0.50\linewidth]{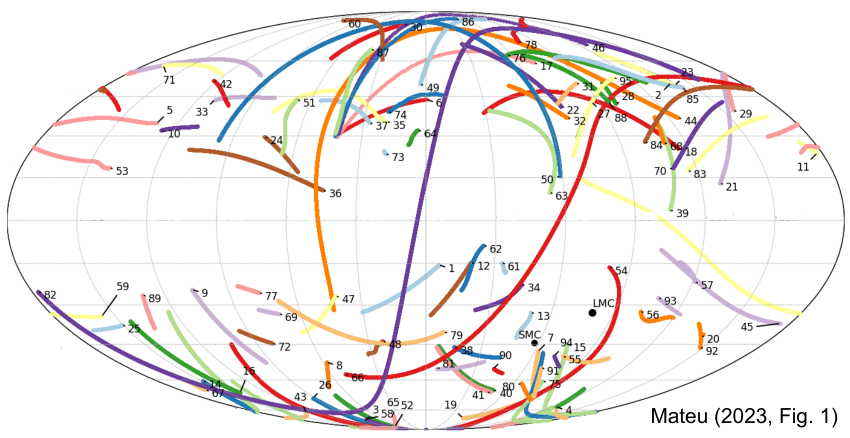}
\caption{Left: LMS--1 stream members detected in Gaia EDR3 using {\tt STREAMFINDER}, showing the proper motions in $\mu_b$ as a function of Galactic coordinates. Some other broad stellar streams are also detected (Orphan, Galactic Anticenter structure(s), and parts of Sagittarius). Some of the off-stream sub-structures are also likely associated with LMS--1 (sub-1, sub-2 and sub-3), as is the marked continuation of LMS--1 in the southern Galactic sky. The grey curve highlights the orbit of LMS--1 \citep[from][Figure~1b]{2021ApJ...920...51M}.
Right: celestial tracks for the 95 stellar streams in the {\tt galstreams} library in Galactic coordinates. The positions of the LMC and SMC are also shown \citep[from][Figure~1]{2023MNRAS.520.5225M}.
}\label{fig:streams-malhan-mateu}
\end{figure}

\paragraph{Other streams}

The GSE and Sagittarius streams dominate much of the stellar halo, but Gaia's contribution extends to the discovery of many other streams, providing many more insights to the formation and evolution of the halo.
\citet{2019ApJ...872..152I}
used their {\tt STREAMFINDER} algorithm, applied to the Gaia DR2 astrometry and photometry alone, to identify eight new structures at heliocentric distances between 1--10~kpc, which they named (from Norse mythology) Slidr, Sylgr, Ylgr, Fimbulthul, Svöl, Fj\"orm, Gj\"oll, and Leiptr. Spectroscopic measurements of seven of the streams have confirmed their reality, and have shown that they are predominantly metal-poor. One, Fimbulthul, is the trailing arm of, and the same age as, the tidal stream of the massive globular cluster $\omega$~Centauri
\citep{2019NatAs...3..667I}.
LMS--1/Wukong is another Gaia DR2 discovery, independently reported as `Low-Mass Stream~1' (based on RR~Lyrae and blue horizontal-branch stars) by 
\citet{2020ApJ...898L..37Y},
and as Wukong, using the combination of Gaia and the H3 spectroscopic survey, by 
\citet{2020ApJ...901...48N}.
Figure~\ref{fig:streams-malhan-mateu}a, from 
\citet{2021ApJ...920...51M},
shows the distribution of proper motions in Galactic latitude ($\mu_{\rm b}$, the scale ranging from 0 to just 4\masyr), illustrating the rich sub-structure in the various streams, along with their derived orbit for LMS--1.
Meanwhile, the retrograde Arjuna/Sequoia/I'itoi group appears to result from three independent mergers
\citep{2019MNRAS.488.1235M,
2020ApJ...901...48N}.
Indeed, from their study of 5684 giants within 50~kpc from the Galactic centre,
\citet{2020ApJ...901...48N}
attributed at least 95\% of their sample stars to one of their listed structures, suggesting a halo built entirely from accreted dwarfs and the associated `heating' of the disk.
Another, Nyx, comprises some 200 stars in the solar vicinity
\citep{2020ApJ...903...25N,
2020NatAs...4.1078N},
which these authors attribute to a massive dwarf galaxy dragged into the disk plane before being completely disrupted, although its extragalactic origin has been questioned
\citep{2021ApJ...907L..16R}. 

With the release of Gaia EDR3,
\citet{2022ApJ...926..107M}
assigned 170 globular clusters, 41 streams, and 46 satellite galaxies to six distinct groups
\href{https://www.youtube.com/watch?v=Pa_zS5-WuBg&ab_channel=StefanPWinc}{(depicted here)},
including the previously known Gaia Sausage--Enceladus, Sagittarius, Cetus, \mbox{LMS--1}/Wukong, Arjuna/Sequoia/I'itoi, and a new merger, Pontus. The three most-metal-poor systems (C-19 with [Fe/H]\,=\,--3.4; Sylgr with [Fe/H]\,=\,--2.9, and Phoenix with [Fe/H]\,=\,--2.7) are associated with LMS--1/Wukong, making it the most-metal-poor merger known to at the time. 
Amongst the wider studies that these results of stellar streams are starting to enable,
\citet{2021MNRAS.501..179M}
showed that the morphology and dynamics of accreted globular cluster streams are sensitive to the central dark matter density profile and mass of their parent satellites. Specifically, globular cluster that accrete within `cuspy' cold dark matter sub-halos produce streams that are physically wider and dynamically hotter than streams that accrete inside cored sub-halos. 
These very low-metallicity streams can also probe the contribution of r-process and s-process enrichment early in their formation history. 
\citet{2021ApJ...912...52G}
suggest that the progenitors of some of these streams experienced one or more r-process events (such as neutron star mergers) early on, in advance of their accretion by the Milky Way. 
\citet{2021ARep...65.1085T}	
examined the possible role of star formation and collisions of gas-rich galaxies, leading to the emergence of low-surface brightness galaxies.  Stellar streams, from the decay of satellite galaxies, may thus contain the remaining dense star clusters and include stars, exoplanets, and interstellar comets.
\citet{2021MNRAS.501.3670P}
considered their role in the formation of binary stars with extremely large separations, whose origin remains poorly understood (Section~\ref{sec:wide-binaries-origin}). He found that ultra-wide binaries can arise via chance entrapment of unrelated stars in tidal streams of disrupting clusters, suggesting that these streams may be the birthplace of hundreds of ultra-wide binaries.

By early 2023, 
\citet{2023MNRAS.520.5225M} 
had searched the literature, collated and homogenised the information in a consistent format in the form of the publicly available 
\href{https://github.com/cmateu/galstreams}{{\tt galstreams} library}, 	
and computed a set of uniform features: the stream length, its mean pole and end points, the stream's coordinate frame, its polygon footprint, and its pole and angular momentum tracks. As of early 2025, the compilation lists 126~tracks, in 95 distinct stellar streams (Figure~\ref{fig:streams-malhan-mateu}).

Gaia's large-scale kinematic information is not only enabling many new discoveries. 
It is also revealing links between well-separated streams, 
such as Orphan/Chenab \citep{2019MNRAS.485.4726K}, 
and ATLAS/Aliqa Uma \citep{2021ApJ...911..149L};
linking known streams to their globular cluster progenitors, as in the case of Fimbulthul to $\omega$~Cen \citep{2019NatAs...3..667I}, 
	and Gj\"oll to NGC~3201 \citep{2021MNRAS.504.2727P}; 
and revealing detailed features in \mbox{GD--1}, possibly attributed to dark matter sub-halos 
\citep{
2018ApJ...863L..20P,
2019ApJ...880...38B,
2019MNRAS.486.2995M,
2020MNRAS.494.5315D}.
Features such as the misalignment of velocities with stream tracks found first in the Orphan--Chenab stream by \citet{2019MNRAS.485.4726K}, and later in several of the Dark Energy Survey streams 
\citep{2018ApJ...862..114S}, are now thought to be perturbations caused by a recent close passage of the LMC.
These observations are providing significant advances in
reconstructing the assembly history of the Milky Way 
\citep[e.g.][]{2020ApJ...901...48N,
2021ApJ...909L..26B,
2022ApJ...930L...9M};
inferring the shape and mass of its dark matter halo 
\citep[e.g.][]{2019MNRAS.486.2995M,
2021MNRAS.502.4170R,
2021MNRAS.501.2279V,
2020MNRAS.494.4291C};
and constraining the dark matter sub-halos 
\citep[e.g.][]{2017MNRAS.470...60E,
2019ApJ...880...38B,
2020ApJ...889...70B,
2021JCAP...10..043B,
2021MNRAS.501..179M,
2021ApJ...911L..32G}. 

The {\tt galstreams} library does not include information on the individual stellar members, but details are given as to how the assembled `track' is constructed for each of the discoveries. The library, including the stream's tracks, is publicly available as a 
\href{https://github.com/cmateu/galstreams}{Python package}.
%
For example, Fimbulthul's celestial track, distance track, and proper motion track were implemented by fitting a 7th degree polynomial to the stream members reported by \citet[][their Table~1]{2021ApJ...914..123I}.
%
As another example, the Sagittarius stream's celestial and proper motion tracks are those derived by 
\citet{2020A&A...635L...3A}, 
supplemented by the distance track from 
\citet{2020A&A...640C...5R} for their RR~Lyrae `strip' sample. 

\citet{2023MNRAS.520.5225M} emphasises that the compilation excludes certain `classes' of previously reported stream-like structures. 
For example, despite their velocity coherence, some are too close to the Sun to result in well-defined celestial or proper motion tracks (amongst these are 
the Helmi stream, 
S1--S4, 
Nyx, 
and Icarus). 
Early accretion events, now at an advanced stage of phase-mixing, are similarly omitted (amongst these are
GSE, 
Thamnos, 
and Sequoia). 
Tidal tails (and related features) have been reported for many globular clusters, and only those where such features clearly extend several degrees beyond the tidal radius are included. 
Others excluded are the Virgo Stellar Stream (VSS) and Virgo Overdensity (VOD), whose nature remains under discussion. 

Streams associated with surviving globular clusters include 
M2, M5, M68--Fj\"orm, M92, 
NGC~288, NGC~2298, NGC~3201--Gj\"oll, NGC~5466, 
Palomar~5, 
and
$\omega$~Cen--Fimbulthul. 
Streams without known progenitors include 
AAU,
Cetus/Cetus--Palca, 
GD--1, 
Jet, 
Jhelum, 
Kwando,
LMS--1, 
Ophiuchus, 
and Orphan--Chenab.
Amongst features evident in the collective distance and velocity information are the observational bias against the detection of stellar streams near the Galactic plane.
There is also an expected bias in the detection of nearby streams, $\lesssim20$~kpc, seen in the clustering observed around the Sun's Galactocentric location.

Proper motion tracks are now available for more than half of the known streams (while less than 10\% have radial velocities), allowing the computation of angular momentum along the track. 
In an undisturbed stellar stream, stars would move predominantly along the stream. In contrast, the Gaia data have revealed several cases in which the proper motions are significantly misaligned with the stream's track. 
First observed in Orphan--Chenab, 
\citet{2019MNRAS.487.2685E} attributed this to the dynamical effect of the LMC during a recent ($<350$~Myr) close encounter. 
Similar features in the Indus and Jhelum streams, and others, have also been attributed to the perturbing effects of the LMC 
\citep{2020AJ....160..244S}.
A number of other streams, often much shorter, also show signs of misalignment at their ends  (including Gaia--8, NGC~1261, and M92), 
while some very long streams exist with no apparent misalignments (including GD--1, NGC3201--Gj\"oll, Phlegethon, Leiptr and, for most of its length, the Sagittarius track).

\paragraph{The Cetus stream}

Out of more than 120 streams now known, the Cetus stream is an example displaying multiple `wrappings' around our Galaxy. 
The stream, 
some 25--40\,kpc from the Sun, was discovered from SDSS--SEGUE by \citet{2009ApJ...700L..61N}.  From the orbit inferred from the stream's radial velocity, they suggested that the globular cluster NGC~5824 is also associated with it. \citet{2013ApJ...776..133Y} used \mbox{N-body} simulations to show that the stream could be reproduced by a disrupted dwarf galaxy of $10^8M_\odot$.
Subsequent studies used the various metallicities to investigate whether NGC~5824 is the disrupted core of the progenitor
\citep{2014MNRAS.438.3507D,
2016MNRAS.455.2417R,
2018ApJ...859...75M},
although deep photometry found no evidence of tidal tails around the cluster itself
\citep{2017AJ....154....8W,
2018MNRAS.473.2881K}.
With the arrival of the Gaia data, 
\citet{2019ApJ...881..164Y}		
used data from Gaia DR2, LAMOST, and SDSS, and the {\tt STARGO} stream-searching algorithm, to identify three groups in the metal-poor ([Fe/H]\,$<-1.5$) outer halo ($d\!>\!15$~kpc), corresponding to the Sagittarius, Orphan, and Cetus streams. The 150~members of the Cetus stream extended over both sides of the Galactic plane.
While \citet{2019ApJ...881..164Y} confirmed the stream's association with NGC~5824, its metallicity dispersion indicated that the progenitor could not have been a globular cluster. They suggested instead that NGC~5824 was associated with a low-mass dwarf galaxy involved in the merger.
\citet{2020ApJ...905..100C}		
used N-body simulations to argue that NGC~5824 was not the nuclear star cluster of a dwarf progenitor, but rather located off-centre from it.
\citet{2020ApJ...905..100C} also predicted that about half of the stream members would be in the southern sky. And their predicted location overlapped the diffuse Palca stream, which had been discovered from the Dark Energy Survey, and at a similar distance of $\sim$36\,kpc
\citep{2018ApJ...862..114S}.	
\citet{2020ApJ...905..100C} also suggested that another diffuse substructure, the Eridanus--Phoenix overdensity, was also likely to be related to the Cetus stream.
To summarise before proceeding: as of 2020, models of the Cetus stream implied that NGC~5824 is offset from the centre of the dwarf progenitor, and that the Cetus stream extends over the southern (equatorial) hemisphere, overlapping with the known Palca stream.

Further clarity came with two papers exploiting the Gaia EDR3 data.
With Gaia distances being of limited value beyond 10\,kpc,
\citet{2022A&A...660A..29T}	
used spectro-photometric distances (based on artificial neural networks) for 300\,000 SEGUE stars, extended to 6d phase-space using Gaia  proper motions, with stream members then identified in their integrals of motion. 
They confirmed that the Cetus stream and the Palca overdensity are parts of the same structure, with a combined Cetus--Palca stream mass $1.5\times10^6M_\Sun$, and a prominent distance gradient of 15~kpc over the $100^\circ$ arc on the sky. A second structure, almost parallel to the Cetus stream and extending over $50^\circ$, could be a stream resulting from the tidal disruption of a globular cluster that was orbiting {\it around\/} the Cetus stream progenitor.
Similar conclusions were reached by
\citet{2022ApJ...930..103Y},	
who combined {\tt STREAMFINDER} and {\tt StarGO} to characterise the Cetus stream as a complex, very metal-poor, nearly polar structure around the Milky Way. They confirmed the southern extensions of the northern Cetus stream as the Palca stream, and identified an additional southern stream, which overlaps on the sky but at a different distance, both extending over more than $100^\circ$ on the sky.
Their N-body model reproduces both as two wraps (of the progenitor around the Milky Way) in the trailing arm,
and yields a progenitor mass of $\gtrsim4\times10^5M_\Sun$, comparable to the Ursa Minor and Draco dwarfs.
In addition, they associated the modelled Cetus--Palca stream with the known Triangulum/Pisces stream
\citep{2012ApJ...760L...6B,	
2013ApJ...765L..39M},		
and with the Willka Yaku stream
\citep{2018ApJ...862..114S},	
as had been suggested by 
\citet{2021ApJ...909L..26B},		
and possibly with the C-20 stream discovered by 
\citet{2021ApJ...914..123I}.		
They also concluded that the globular cluster NGC~5824, of similar stellar mass, was not the main progenitor, but possibly accreted in the same group infall. The multi-wrap Cetus stream, they conclude, {\it `is a perfect example of a dwarf galaxy that has undergone several periods of stripping, leaving behind debris at multiple locations in the halo'}.

\paragraph{Stellar streams and sub-halos}

These stellar streams can place constraints on the existence and nature of the numerous dark matter sub-halos (halos within halos) that are predicted, in the standard $\Lambda$CDM cosmology, to exist surrounding the Milky Way 
\citep[e.g.][]{2008MNRAS.391.1685S,
2019Galax...7...81Z}.
An absence of such sub-halo driven perturbations in the Gaia data might be a challenge for standard $\Lambda$CDM
\citep{2017ARA&A..55..343B}.

It is now well established, observationally as well as in N-body simulations, that satellite galaxies interacting and merging with the Milky Way can result in tidal heating of the disk, in tilts and warps, and can trigger the growth of asymmetric structures such as the central bar, and the Galaxy phase-space spiral \citep{2018Natur.561..360A}.
Lower mass sub-halos, $\lesssim10^9M_\Sun$, being largely devoid of gas and dust, are far more challenging to detect
\citep{2015MNRAS.446.1000F, 
2020ApJ...891..161I}.
The morphology and flux-ratios of strongly-lensed quasars currently suggests consistency with $\Lambda$CDM predictions, albeit at low statistical significance 
\citep{2019MNRAS.485.2179R,
2020MNRAS.492.3047H,
2024MNRAS.528.1757O}.	
Various other ways of demonstrating the existence of dark sub-halos have been suggested: these include searches for $\gamma$-ray emission as a result of dark matter annihilation
\citep{1990Natur.346...39L,	
2000PhRvD..62l3005C, 		
2007ApJ...657..262D}, 		
or through detecting the coherent vertical velocities of disk stars attributable to the passage of such sub-structures through the Galactic disk
\citep{2015MNRAS.446.1000F},	
or from the gravitational scattering of stars in tidal streams
\citep{2002MNRAS.332..915I, 		
2002ApJ...570..656J,			
2008ApJ...681...40S}.			

In the latter, the underlying idea is that a narrow, dynamically cold stellar stream is susceptible to heating by repeated close encounters with the massive dark sub-halos, resulting in characteristic features such as gaps whose details depend on the sub-halo mass and distance from the progenitor
\citep[e.g.][]{2016MNRAS.457.3817S,	
2017MNRAS.466..628B},		
and which were considered detectable with Gaia
\citep{2002MNRAS.332..915I,
2015MNRAS.446.1000F}.
As described by \citet{2015MNRAS.454.3542E}:
{\it `Around a Milky Way-like galaxy, more than a thousand of these sub-halos will not be able to form stars but are dense enough to survive even deep down in the potential well of their host. There, within the stellar halo, these dark pellets will bombard tidal streams as they travel around the Galaxy, causing small but recognisable damage to the stream density distribution.'}

Today, some of the most discussed features are those in the GD--1 stream. This was discovered, from SDSS, to span $63^\circ$, with a retrograde orbit with pericentre 14~kpc and apocentre 26~kpc 
\citep{2006ApJ...643L..17G}. 
Based on their SDSS data, the discoverers considered that there was `no evidence of perturbations by large mass concentrations in the nearby halo'.
Follow-up ground-based observations, including radial velocities, provided the first evidence for gaps along the stream 
\citep{2010ApJ...712..260K, 	
2013ApJ...768..171C}.		
The distant retrograde orbit suggested that interactions with disk sub-structure were unlikely, confirming it as an excellent candidate for the study of gaps induced by dark sub-halos
\citep{2016MNRAS.463L..17A,		
2018MNRAS.477.1893D,	
2020arXiv201111684K}. 	

Further insights came with the availability of the Gaia DR2 proper motions, from which the GD--1 stream was detected as one of the highest contrast features in the Galaxy halo \citep{2018MNRAS.481.3442M}.
Additional stars surrounding the stream were detected
\citep{2018ApJ...863L..20P}, 	
including an off-track `spur', perhaps suggesting the presence of massive perturbers
\citep{2019ApJ...880...38B},		
or that the progenitor originated within a larger system
\citep{2019MNRAS.486.2995M}. 	
The `exquisite astrometry' allowed a clean separation of the stream from Milky Way stars 
\citep{2018ApJ...863L..20P}, 	
and showed clear evidence for high-contrast gaps along the stream (Figure~\ref{fig:palomar5-gd1}b).
Also working with the DR2 proper motions, and with improved filtering, 
\citet{2020MNRAS.494.5315D}	
confirmed these various `gaps' and `wiggles'. They argued that a particularly striking sinusoidal wiggle cannot be the characteristic S-shape signature of stellar debris torn off the stream’s progenitor, since it has the wrong orientation with respect to the stream's orbit. They concluded that this feature must instead come from a perturbation to the stream.

Again basing their analysis on DR2, but adding Pan-STARRS photometry, and new ground-based radial velocities,
\citet{2020ApJ...891..161I}	
reached a different conclusion.
They found that the density profile exhibits high contrast periodic peaks separated by $2.64\pm0.18$~kpc. Their N-body simulations suggested that this morphology could be modelled with simple epicyclic motion in a smooth Galactic potential, partly compounded by incompleteness in Gaia’s sky-scanning pattern in DR2.
Such epicyclic {\it overdensities\/} arise because tidal stripping mainly occurs near pericentric passages, leading to bursts of debris along the stream
\citep{2010MNRAS.401..105K,
2012MNRAS.420.2700K,
2016MNRAS.457.3817S,
2017MNRAS.466..628B}.
\citet{2020ApJ...891..161I} concluded that massive dark sub-halos do {\it not\/} appear to be required to explain the density clumping along the stream.

But, in turn,
\citet{2021MNRAS.502.2364B}	
argued that the power induced by episodic tidal stripping is far below that induced by dark matter sub-structures. They argued that the stellar density variations cannot be due to known baryonic structures, such as giant molecular clouds, globular clusters, or the Milky Way's bar or spiral arms 
\citep[a similar conclusion was drawn by][]{2022ApJ...941..129D},	
and instead (in a joint analysis of the \mbox{GD--1} and Palomar~5 streams) requires a population of dark sub-structures with $M\sim10^7-10^9M_\Sun$. 
They went on to infer a total abundance of dark sub-halos corresponding to a mass fraction in the sub-halos $f_{\rm sub}=0.14^{+0.11}_{-0.07}$~percent, compatible with hydrodynamical simulations of cold dark matter with baryons.

More is also being inferred about the likely progenitor of the GD--1 stream. The N-body simulations by 
\citet{2019MNRAS.485.5929W}
suggest that it probably lies between $-45^\circ<\phi_1<-30^\circ$ ($\phi_1,\phi_2$ are sky coordinates in a system based on the stream itself), and that it either completely disrupted $\sim$2.5\,Gyr ago, or it disrupted only 500~Myr ago, resulting in the underdensity at $\phi_1\sim -40^\circ$.
From 43 spectroscopically confirmed stream members,
\citet{2021ApJ...911L..32G}	
measured a radial velocity dispersion of $2.1\pm0.3$\kms, constant over the $15^\circ$ region surveyed. Compared with an unperturbed model of the GD--1 stream having a velocity dispersion of 0.5\kms, the observed dispersion implies that the stream has undergone dynamical heating. They hypothesise that \mbox{GD--1} originated from a globular cluster which, prior to its accretion by the Milky Way, orbited a dwarf galaxy with a core density profile, and that imprints of its original host galaxy, including the inner slope of its dark matter halo, remain observable in the stream today.
In further modelling using Gaia DR3,
\citet{2024ApJ...967...89I}		
insist on the periodicity of the high contrast peaks, and suggest a two-component stream model (a kinematically cold part with dispersion 7\kms, and a hot part with dispersion 29\kms), consistent with simulations where globular clusters form at random locations within dark matter sub-halos and are subsequently accreted onto large galaxies 
\citep{2020ApJ...889..107C,	
2021MNRAS.501..179M,		
2023ApJ...953...99C}.		
Similar studies for the Palomar~5 stream suggest the presence of two gaps due to sub-halos of $10^6-10^7M_\Sun$ and $10^7-10^8M_\Sun$
\citep{2017MNRAS.470...60E,	
2020ApJ...889...70B, 		
2021MNRAS.502.2364B,		
2021JCAP...10..043B}.		

\subsection{Age and formation}
\label{sec:age-formation}

The immense amount of new observational material from Gaia presents a rich but complicated picture of the Milky Way.  Complex structural and kinematic features have been identified in its spiral arms, the central bulge and bar, and regions of ongoing and past star formation connected with young associations and open clusters. Further out are the orbiting globular clusters, and other Local Group galaxies. Amongst new dynamical features are those attributed to resonances with the bar and dissolving clusters, and the Gaia phase-space spiral and the warp, both seemingly driven by past collisions or encounters. The rich phase-space structures in the stellar halo include some hundred or more accreted stellar streams, holding deep insights into the Galaxy's formation. 
With all this information, what can be said about the Galaxy's age, and its formation history? There is already an extensive, detailed, and rapidly evolving literature touching on these questions, and I will pick out a few central points.

\begin{figure}[t]
\centering
\includegraphics[width=0.50\linewidth]{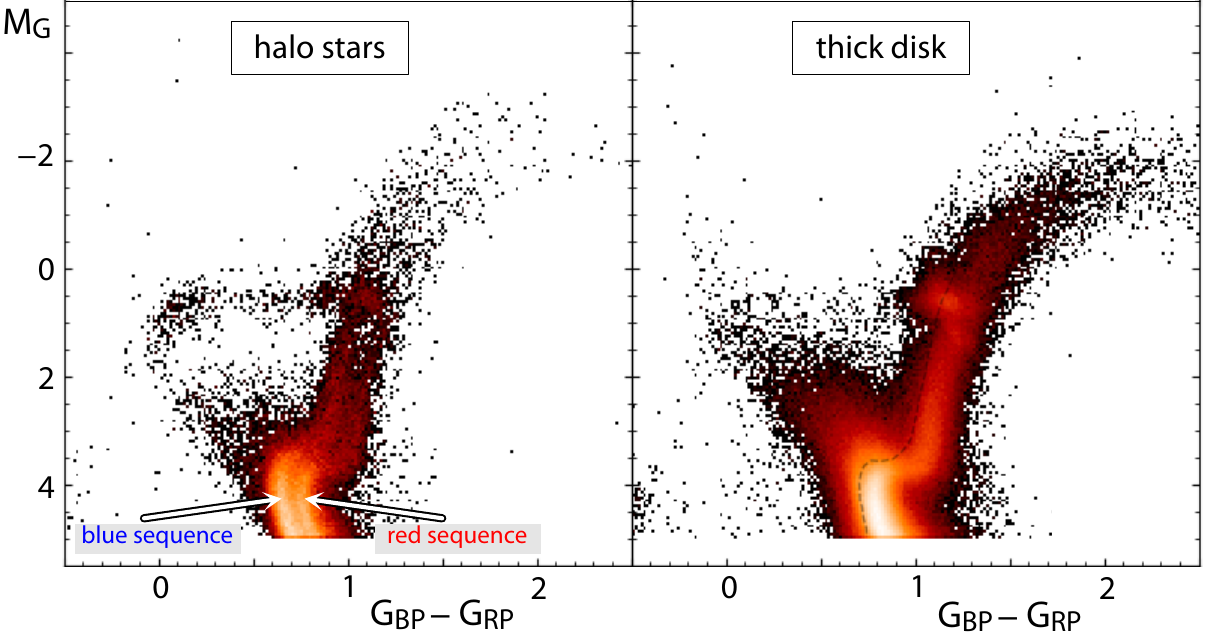}
\hspace{30pt}
\includegraphics[width=0.406\linewidth]{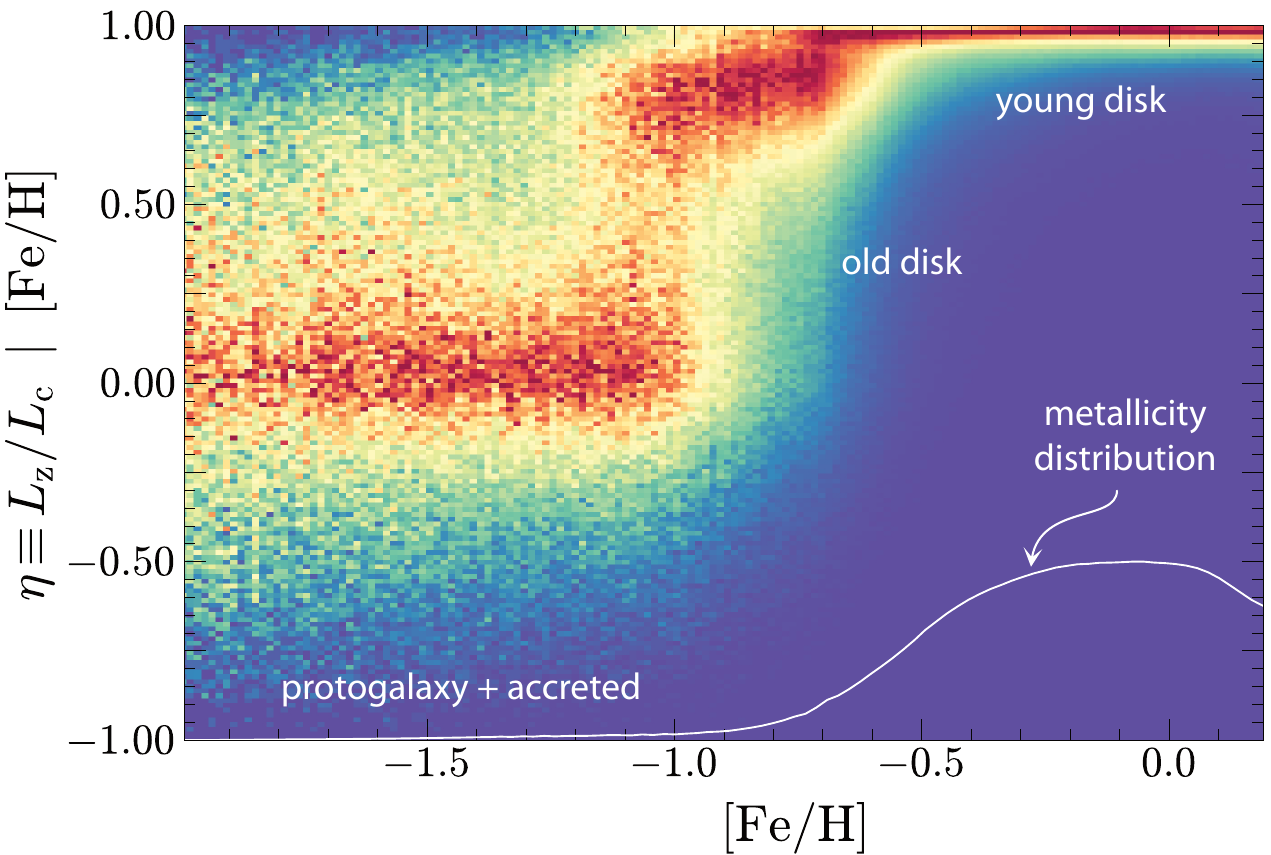}
\caption{
Left:  the observed Gaia colour--magnitude diagrams for the halo and thick disk sub-populations, using a log normalisation of the number of stars in the colour-magnitude bins: the two (red and blue) sequences are clear in the left panel \citep[][Figure~1]{2019NatAs...3..932G}.
Right: the three-phase evolution of the Milky Way revealed by 9.9~million Gaia DR3 giants. The column-normalised distribution of orbital circularity is shown as a function of Gaia metallicity, a proxy for cosmic time.  The white line shows the metallicity distribution. [Fe/H] behaves as a good cosmic clock for the high-$\alpha$ members, since their age--metallicity relation is reasonably monotonic \citep[][Figure~5a]{2024ApJ...972..112C}.
}\label{fig:age-formation}
\end{figure}

With Gaia DR2,
\citet{2019NatAs...3..932G}	
extracted two large samples of halo stars and thick disk stars, both restricted to within 2\,kpc of the Sun to ensure accurate distances and absolute magnitudes. Their kinematically defined halo population, with large tangential velocities ($>200$\kms\ relative to the Sun), contained $\sim$60\,000 stars. Their thick disk sample, of $\sim$500\,000 stars, retained only those more than 1100~pc from the Galactic plane, independent of chemical or kinematic properties, and not included in the halo sample. Their location in the Gaia colour--magnitude diagram ($M_{\rm G}$ versus $G_{\rm BP}-G_{\rm RP}$; Figure~\ref{fig:age-formation}a) shows that the halo sample consists of two distinct sub-populations: a `blue' sequence and a `red' sequence, arising from two populations with different chemical compositions, but of the same age.
Invoking the established relation between a galaxy's mass and its metallicity means that the stars in the red sequence, being more metal rich, formed in a more massive galaxy (by a ratio of about 4:1).  Also based on Gaia data, the blue sequence had already been associated with a significant merger event, Gaia Sausage--Enceladus, which occurred early during the Milky Way formation, around 9--10\,Gyr ago (Section~\ref{sec:halo-streams}).

These new stellar age distributions enabled by Gaia, aided by state-of-the-art cosmological simulations, suggest that the primitive Milky Way had been forming stars during some 3\,Gyr when a smaller galaxy, which had been forming stars on a similar timescale but was less chemically enriched owing to its lower mass, was accreted.  This merger heated a fraction of the existing stars in the main progenitor to a stellar halo-like population.  A supply of infalling gas during the merger ensured the maintenance of a disk-like configuration, with the thick disk continuing to form stars at a substantial rate. The age distributions indicate that the thick disk reached its peak star formation rate around 9\,Gyr ago, or some 4.5\,Gyr after the first stars formed in the Milky Way. Subsequently, around 8--6\,Gyr ago, the gas settled into a thin disk that has continued to form stars up to the present day.

Details of the inner regions of the Galaxy are complicated by the large distances, and by extinction. Gaia data have been used to clarify the properties of inner globular clusters as tracers of a very old Galaxy component, with
`Kraken' 	
\citep{2020MNRAS.498.2472K}		
and `Koala'
\citep{2020MNRAS.493..847F}		
identified as possible examples of early accretion. 
\citet{2022MNRAS.514..689B} 		
used Gaia EDR3 to reconstruct stellar kinematics in the inner Galaxy down to metallicities of $-1.5$ in [M/H], and identified an {\it in situ\/} proto-Galactic structure they called `Aurora'. The constituent stars show little net rotation, and may have pre-dated the Milky Way's ancient disk. 
And \citet{2022arXiv220402989C}	
used kinematics and ages from Gaia EDR3 to identify proto-Galactic stars down to $-2.5$ in [M/H] with ages $\gtrsim13$~Gyr.


In the context of the hierarchical formation of massive disk galaxies, the oldest and most metal-poor stars are expected to be a mix that either formed within one of the main overdensities which coalesced early on to form the proto-Galaxy, or formed early on in distinct satellite galaxies that eventually merged with it. The earliest phases of the Galaxy’s star-formation and enrichment history should then be reflected in the orbits and abundances of these old metal-poor stars.
To examine this early formation,
\citet{2022ApJ...941...45R}		
selected giant stars within 30$^\circ$ of the Galactic centre, derived [M/H] estimates from the BP/RP spectra, and combined them with Gaia's RVS velocities to obtain Galactic orbits for 1.5~million stars in the inner Galaxy. Their metal-poor samples were a factor 100 larger in size than previously available:  specifically, more than 4000 stars with [M/H]\,$<-2$, some 20\,000 with [M/H]\,$<-1.5$, and $70\,000$ with [M/H]\,$ <-1$.
Their samples reveal a large metal-poor population, [M/H]\,$<-1.5$, in the inner few kpc, which they show to be centrally concentrated, and most with orbital apocenters $<5$\,kpc. A minority (but nonetheless a large number) of stars with [M/H]\,$<-1.2$ currently in the inner Galaxy are on high-eccentricity orbits that can take them to $>$10\,kpc. These are presumably members of the accreted halo, currently near pericentre. They were also able to quantify their net Galactic rotation, finding that only the most metal-poor members, [M/H]\,$<-2$, show no rotation, while populations with higher [M/H] also have higher rotation.
The total stellar mass seen at [M/H]\,$ <-1.5$ within 5~kpc corresponds to $\sim 5\times10^7M_\odot$, which likely represents a lower mass limit as a result of dust obscuration.
All of this, they conclude, fits a picture in which this metal-poor heart of the Milky Way constitutes the most ancient `proto-Galactic' component of our Galaxy. This picture is supported by other details of their orbits, metallicity, and $\alpha$-element abundances, and points to its formation at a very early epoch, $z\gtrsim5$.

Using the Gaia DR3 astrometry and spectroscopy of 5.8~million stars, along with metallicities from 
\citet{2023ApJS..267....8A},
\citet{2024ApJ...964..104M}		
identified two high-contrast overdensities in the energy--angular momentum ($E-L_z$) plane of bright ($G<16$) and metal-poor ($-2.5<{\rm [M/H]}<-1.0$) stars, which they designated Shakti and Shiva. Both have $M_\star\gtrsim10^7M_\Sun$, and both follow prograde orbits inside the solar circle. But while their orbits suggest an accreted origin, their metal abundances are more typical of an {\it in situ\/} population. 
They inferred either that they may have resulted from resonant orbit trapping of the field stars by the rotating bar (along the lines described by 
\citet{2023MNRAS.524.3596D}),		
or that they were protogalactic fragments that formed stars rapidly, and coalesced early.

In another piece of the formation puzzle,
\citet{2024ApJ...972..112C} 
used the [Fe/H] and [$\alpha$/Fe] for 9.9~million red giants (derived from the Gaia data by \citet{2023ApJS..267....8A}), to characterise their angular momentum as a function of metallicity (Figure~\ref{fig:age-formation}b). Taking this as a proxy for age, they identified three distinct evolutionary phases: a disordered/chaotic protogalaxy, a (kinematically) hot old disk, and a cold young disk. 
In their interpretation, the old high-$\alpha$ disk starts at [Fe/H]\,$\simeq-1.0$, `spinning up' from the nascent protogalaxy
(as also inferred by \citet{2022MNRAS.514..689B}),	
and then exhibiting a smooth cooling down toward more ordered and circular orbits at higher metallicities.
The overlap between the protogalaxy and hot disk, $-1.2\lesssim\,{\rm [Fe/H]}\lesssim-0.9$, includes stars with intermediate orbits, extending to [Fe/H]\,$\sim-0.5$, and coinciding with a previously identified `{\it in~situ\/} halo’ or `splash’ population 
\citep{2017ApJ...845..101B,	
2020MNRAS.494.3880B}.		
They also identified an analogue from the cosmological simulations TNG50, in which the protogalaxy spins up into a thin high-$\alpha$ disk, before being heated and torqued by a major gas-rich merger (as with the Gaia Sausage--Enceladus merger, Section~\ref{sec:halo-streams}). This adds a large amount of low-metallicity gas and angular momentum, from which the kinematically cold low-$\alpha$ stellar disk is subsequently born. 

While the authors suggest that we are steadily gaining a coherent picture of our Galaxy's three-phase formation, involving spin-up, merger, and cooldown, more pieces of the puzzle are emerging from Gaia that still remain to be synthesised into an unambiguous picture of our Galaxy's early formation, both in terms of the stellar components of its central regions
\citep[e.g.][]{
2024ApJ...975..293R,	
2024arXiv241022250K,	
2024MNRAS.533..889Z,	
2025MNRAS.537.3730H},	
and its angular momentum history
\citep[e.g.][]{
2024arXiv241112165V,	
2025A&A...695A.218R,	
2025arXiv250114089C}.	

\subsection{Tumbling disk}
\label{sec:tumbling-disk}

One of the reasons that Gaia's positions and proper motions should be so carefully linked to an extragalactic reference system is that astronomy has moved on from the simple view in which our Galaxy's disk is fixed in orientation within a spherical dark matter halo. Arguments suggest that the disk--halo orientation is likely to vary with time, due to effects including 
the infall of misaligned gas 
\citep{1986MNRAS.218..743B,
2010MNRAS.408..783R},
tidal effects of the LMC
\citep{2006ApJ...641L..33W},
and those of ancient infall satellites and their associated debris 
\citep[e.g.][]{2023MNRAS.518.2870D}.
Simulations of galaxy formation also predict that most galaxy halos `tumble', in inertial space, with a rotation rate of $\sim$2~radians per Hubble time
\citep{2004ApJ...616...27B}.		
Furthermore, analytical arguments 
\citep[e.g.][]{1995MNRAS.275..897N},
as well as simulations 
\citep{1995ApJ...442..492D},
suggested that dynamical friction in the inner galaxy should couple the inner disk to the halo
\citep{2004ApJ...616...27B}.
It follows that, if the angular momentum vectors of the disk and halo remain coupled, the plane defined by the disk stars should rotate at some 30~$\mu$arcsec\,yr$^{-1}$ with respect to the quasar reference frame.
Simulations suggest that any residual rotation of the Gaia reference frame with respect to the quasar frame should be at the level of $0.2-0.5\,\mu$arcsec\,yr$^{-1}$. This should, in due course, allow for the detection of such a cosmological `tumbling' motion of our own Galaxy's disk
\citep{2014ApJ...789..166P}.

\section{Local Group galaxies and cosmology}
\label{sec:local-group-cosmology}

Over the past 20--30 years, through a combination of observations, theory, and numerical simulations, the $\Lambda$CDM paradigm has gained widespread acceptance as the best current model of the large-scale structure of the Universe.
In this context, one of Gaia's most important ongoing observational contributions has been the number and variety of new and complex phase-space features identified within our Galaxy, detailed in Section~\ref{sec:galaxy-structure}. 
These include 
the Gaia phase-space spiral (the complex phase-space structure possibly associated with a passage of the Sagittarius dwarf galaxy, Section~\ref{sec:phase-space-spiral}),
the possible imprints of a decelerating bar attributable to the Galaxy's dark matter halo (Section~\ref{sec:bar}),
and the ongoing discovery and detailed characterisation of a large numbers of ancient stellar streams making up our Galaxy's halo (Section~\ref{sec:halo-streams}).
Most of these features can only easily be interpreted within the context of our Local Group of galaxies, and indeed within the framework of $\Lambda$CDM cosmology.

More widely, Gaia is confirming many of the detailed predictions of large-scale cosmological simulations such as Millennium, Illustris, and EAGLE, and helping in their interpretation. These range from 
simulations of the merger epochs, 
the occurrence of bars, 
the orbits of globular clusters and dwarf spheroidals, 
the bulk motions of the Magellanic Clouds, 
as well as the $\Lambda$CDM `missing satellites' problem, 
	the `core--cusp' problem,
	the `too-big-to-fail' problem, 
	and the `plane of satellites' problem.
More on the determination of the Hubble constant, $H_0$, is given in the context of Cepheids in Section~\ref{sec:cepheids-h0}, and in the context of the `tip of the red giant branch' in Section~\ref{sec:trgb}.

\subsection{The Magellanic Clouds}
\label{sec:magellanic-clouds}	

The Magellanic Clouds are two `nearby' irregular dwarf galaxies, visible to the unaided eye in the dark skies of the southern hemisphere. 
The Large Magellanic Cloud (LMC), at a distance of $\sim$50~kpc ($\varpi\simeq20$\muas) has a diameter of 4.3~kpc, compared with about 30~kpc for the Milky Way.
The Small Magellanic Cloud (SMC), at a distance of $\sim$60~kpc, has a diameter of 2~kpc. The two are separated by $\sim$23~kpc ($20^\circ$ on the sky). Only the smaller Sagittarius dwarf elliptical (discovered in 1994), and the Canis Major dwarf galaxy (discovered in 2003) are closer neighbours.
The LMC is the fourth most massive of over 50 galaxies comprising the Local Group. Both have been distorted by tidal interaction with the Milky Way, as have the outer parts of the Milky Way disk.
Whether they are bound as orbital companions to our Milky Way remains uncertain. If they are, their orbital period is at least 4~Gyr. The other possibility is that they are on a first (or even second) approach, the start of a merger that may overlap with the Milky Way's expected merger with the Andromeda galaxy sometime in the future.
In neutral hydrogen (radio) images, the LMC shows clear spiral structure, characterised by an off-centre bar and one prominent spiral arm, attributed to dynamical interactions between the LMC and SMC. Streams of neutral gas connect both to the Milky Way (the Magellanic Stream) and to each other (the Magellanic Bridge). They are both gas-rich, with stars ranging from very young to very old. Both probably have large dark matter halos. 

Determining the distance to the LMC has been a long-standing challenge. It is a key step in the overall distance scale, and hence $H_0$ (Section~\ref{sec:distance-scale}). It was beyond the direct reach of Hipparcos which, because of its 11--12~mag limit, included just 36~LMC and 11~SMC stars, with accuracies 1.5--2~milli-arcsec.
Hipparcos-based LMC distances were, instead, based on various indirect methods: Population~I stars (Cepheids, red clump giants, Mira variables, and eclipsing binaries), Population~II stars (subdwarf fitting to globular clusters, horizontal branch stars, and RR~Lyrae), as well as white dwarf sequencing, and estimates based on the supernova SN~1987A. 
A critical compilation of pre-Gaia estimates has been made both for the LMC
\citep{2014AJ....147..122D},	
and for the SMC
\citep{2015AJ....149..179D}. 
Although the Magellanic Clouds are at the limits of individual Gaia measurements, the enormous improvement in accuracies, and numbers of stars, nevertheless represents a major advance in probing their structure and kinematics. 

\begin{figure}[t]
\centering
\includegraphics[width=0.543\linewidth]{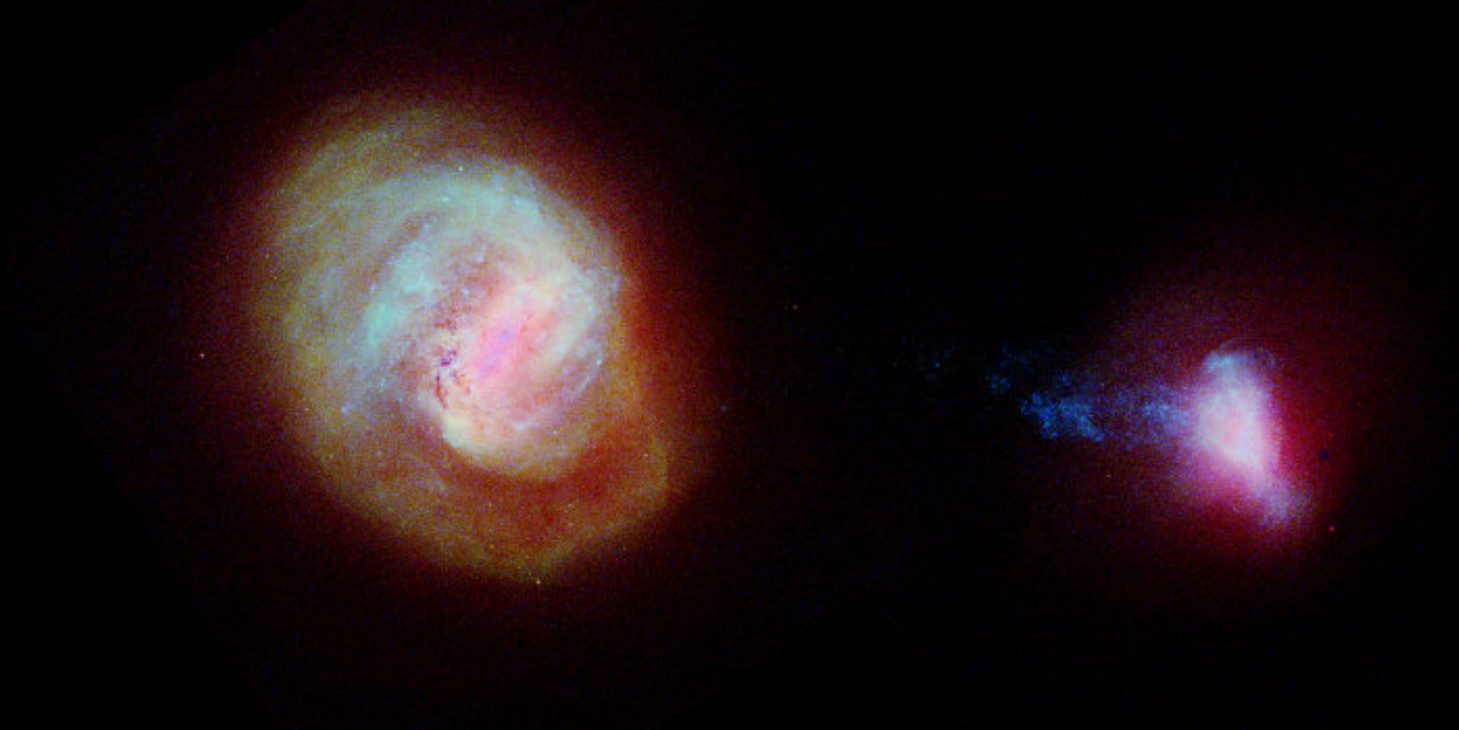}			
\hspace{30pt}
\includegraphics[width=0.27\linewidth]{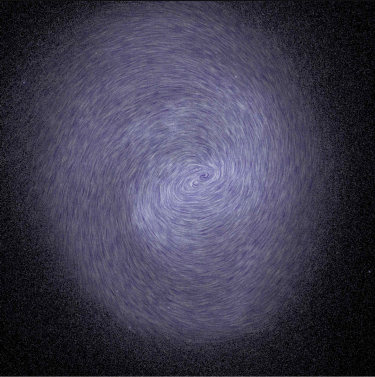}		
\caption{Left: Gaia view of the Large and Small Magellanic Clouds, with colours denoting different age populations.
Right: Rotation of the Large Magellanic Cloud visualised by Gaia. From \citet{2021A&A...649A...7G} and ESA/Gaia/DPAC.}
\vspace{-10pt}
\label{fig:lmc-smc}
\end{figure}

Gaia EDR3 in 2020 provided the first `big picture'
\citep{2021A&A...649A...7G}.	
Their membership selection used astrometric criteria to remove foreground (Milky Way) contamination. Starting from more than 27~million Gaia objects within a 20$^\circ$ radius for the LMC, and more than 4~million within a 11$^\circ$ radius for the SMC, refinement according to proper motion and parallax led to a total of 11\,156\,431 objects belonging to the LMC, and 1\,728\,303 belonging to the SMC.
Gaia's high-accuracy multi-colour photometry allows further subdivided by age, notably as
very young main sequence (ages $< 50$\,Myr),
young (50--400\,Myr),
and intermediate-age (up to 1--2\,Gyr) main-sequence populations. 
In addition, they identified red giant branch stars, the asymptotic giant branch (including long-period variables), RR~Lyrae stars, classical Cepheids, and red clump stars.
The resulting stellar density maps (Figure~\ref{fig:lmc-smc}) reveal more concentrated and clumpier distributions for younger stars in the bar and inner spiral structure than the older disk stars. Smoothed maps of the proper motion field, revealing the bulk stellar motions, shows a clear ordered rotation of the LMC, while the SMC is more chaotic. Both young stars and red clump giants trace the density and tidally-induced velocity flow of stars from the SMC towards the LMC following the Magellanic Bridge. The Gaia data shows that it appears to wrap around the LMC, connecting with the young southern arm-like structure. Additionally, the outskirts of both Magellanic Clouds reveal other well-known features, such as the north and south tidal arms of the LMC and the northern density enhancement of the SMC.

Many detailed investigations of the LMC resting on the Gaia data have followed. Amongst these are studies of
its three-dimensional structure
\citep{2021MNRAS.504....1C,		
2022MNRAS.512..563R,			
2022MNRAS.512.5423N},			
including its non-axisymmetric disk and bar pattern speed
\citep{2024A&A...683A.102J},		
the distribution of dust
\citep{2022MNRAS.511.1317C},		
its star-formation history
\citep{2021MNRAS.508..245M,		
2022MNRAS.513L..40M},			
orbits of the Milky Way--LMC--SMC system
\citep{2022A&A...657A..54B,		
2022ApJ...927..153C,		
2022ApJ...940..136P,		
2022MNRAS.517.1737C,		
2023A&A...669A..91J,		
2024MNRAS.527..437V},		
tidal stripping and formation of the Magellanic Bridge 
\citep{2021A&A...647L...9D,	
2022MNRAS.510..445C,		
2022MNRAS.512.4798C,		
2022MNRAS.514.1266P,		
2024MNRAS.527.8706M},	
its predicted orbit-induced circumgalactic medium bow shock
\citep{2023ApJ...959L..11S},	
and its hypervelocity stars
\citep{2021MNRAS.507.4997E,	
2023ApJ...952...64L}.		

\paragraph{The Small Magellanic Cloud}

The line-of-sight structure of the SMC is complex and, historically, the data have been difficult to interpret. Gaia's proper motions, radial velocities, and parallaxes (for foreground suppression), confirm this complexity. 
And amongst the most recent work, 
\citet{2024ApJ...962..120M} 
have suggested a radically new picture: that the SMC is composed of two distinct structures, with its interstellar medium arranged in two, superimposed, star-forming regions separated by $\sim$5~kpc along the line-of-sight.

From the earlier Gaia studies, the oldest stellar populations are found to be reasonably spherical within a radius of $\sim$10\,kpc, with suggestions of rotation in the central region 
\citep{2018A&A...616A..12G,	
2018A&A...613L...8N,		
2018ApJ...864...55Z,			
2021MNRAS.502.2859N}.		
But stars with estimable distances (red clump stars, Cepheids, and RR~Lyrae) extend some 20--30~kpc along the line-of-sight 
\citep{2016ApJ...816...49S,	
2017MNRAS.472..808R,		 
2021ApJ...910...36Z}. 		
In contrast, stars of the young main sequence and red giant branch display 
a radial velocity gradient indicating rotation 
\citep{2023MNRAS.523..347E}, 	
along with distinct substructures along the line-of-sight, 
both morphological 
\citep[e.g.][]{2017MNRAS.467.2980S, 
2019A&A...631A..98M, 		
2021MNRAS.504.2983T,		
2021MNRAS.500.2757O, 		
2023MNRAS.518L..25C, 		
2024MNRAS.529.3858A}, 	
as well as chemical 
\citep{2021ApJ...923..172H,	 
2022MNRAS.513L..40M, 		
2023A&A...671A.124M}. 		
Further complicating the observed morphological structure is the evidence of tidal disruption by the LMC 
\citep{2018A&A...613L...8N,	
2019ApJ...874...78Z,			
2020MNRAS.495...98D,		
2021ApJ...910...36Z,			
2021MNRAS.502.2859N,		
2023MNRAS.518L..25C}. 		
The SMC is also important for studies of the interstellar medium and star formation at low metallicity ($\sim$20\% solar), again presenting structural complexities that may have originated from gravitational interactions with the LMC. Here, Gaia data have been interpreted as an orbit on its first or second infall 
\citep{2007ApJ...668..949B,	
2018ApJ...864...55Z, 		
2020ApJ...893..121P,		
2022MNRAS.513L..40M,		
2024MNRAS.527..437V},		
or indicative of a close impact
\citep{2021ApJ...910...36Z,	 
2022ApJ...927..153C},	
with stellar debris observed in the outer regions of both 
\citep{2019A&A...631A..98M}.	 

Across these studies, the Gaia data have been used to identify the space motion of the SMC, providing evidence for a moderate rotation, some expansion, tidal stripping, and bursts of star formation.  But there has been a recurrent theme that its morphology, kinematics, and chemistry are not well replicated by a single population.  For example, from a Gaia sample of red giant stars, \citet{2021ApJ...910...36Z} suggested {\it `\ldots to treat the SMC as a series of different populations with distinct kinematics'}.
In this spirit, by comparing the average dust extinction towards nearly 2000 stars with accurate radial velocities (from Gaia and APOGEE), and with membership rigorously established on the basis of the Gaia DR3 astrometry, 
\citet{2024ApJ...962..120M}	
concluded that the inner $\pm4^{\circ}$ of the SMC is composed of two structures with distinct stellar and gaseous chemical compositions. Specifically, that the interstellar medium is organised into two, superimposed, star-forming systems with similar gas mass, separated by $\sim$5~kpc along the line-of-sight.

\subsection{Andromeda, M31}
\label{sec:andromeda}	

The Andromeda Galaxy, M31, is a barred spiral system of similar mass to the Milky Way, and the largest member of the Local Group, with a diameter 65--70\,kpc.  Various techniques have been used to estimate its distance, including the use of Cepheid variables, eclipsing binaries, and stars at the tip of the red giant branch. At around 750~kpc, or $\varpi\sim$1.3\muas\
\citep{2012ApJ...745..156R}, 
this is well beyond reach of Gaia's individual star measurements.
Its formation and subsequent evolution has been inferred from its star formation history. A close passage with M33 2--4 billion years ago, along with more recent interactions with other satellite galaxies, together may have created most of the galaxy's halo, its extended disk, and other stellar components of various ages
\citep[e.g.][]{2001Natur.412...49I,
2012ApJ...751...74D}.

A special data set made available at the same time as Gaia DR3 (and its $G$-band photometric time series for 10~million variables), the Gaia Andromeda Photometric Survey, GAPS, provides epoch photometry for all sources, variable and constant, from a region of the sky centred on M31, with a radius of 5\ddeg5. It contains epoch photometry ($G$, $B_{\rm P}$, $R_{\rm P}$) for 1\,257\,319 sources, of which the majority have 30--45 observations
\citep{2023A&A...674A...4E}.
Their Gaia source density diagram (their Figure~1) shows the prominent structure of M31, along with density enhancements due to the nearby galaxies M32 and M110. However, their classified variables in the colour--magnitude diagram (their Figure~19) belong mainly to the Milky Way.
In other work, searches for red supergiants have used the Gaia data to remove foreground stars and then using the tip of the red giant branch as distance indicator, finding some 6000 in M31, and around 3000 in M33
\citep{2021AJ....161...79M,		
2021ApJ...907...18R}.			
Searches for globular clusters associated with M31 has a particularly long history, starting with the work of Edwin Hubble
\citep{1932ApJ....76...44H}, 
and augmented to some 3000 today from 2MASS and HST surveys. From 1.85~million Gaia EDR3 sources around M31 (matched to Pan-STARRS sources), 20\,658 were flagged as extended, then inspected to result in 50 new globular cluster candidates
\citep{2023ApJ...954..206W}.		
Very different studies found that some of the hypervelocity stars discovered by Gaia (Section~\ref{sec:hypervelocity-stars}) might originate in M31 
\citep{2024MNRAS.529.3816G}.	

One of the main insights from the Gaia studies of M31 comes from defining its bulk proper motion from its member stars, then running orbit integrations and N-body simulations to determine its orbital history. 
Here,	
\citet{2019ApJ...872...24V}  
used Gaia DR2 proper motions of 1084 sources in M31 and 1518 in M33 (their Figure~2), to detect their bulk rotation, and to determine the two components of their proper motions of ($+65\pm18,-57\pm15$\muasyr) for M31, and ($+31\pm19,-29\pm16$\muasyr) for M33. They concluded that M31's orbit toward the Milky Way is less radial than previously inferred, and that M33 may be on its first infall into M31.
\citet{2020MNRAS.493.5636T}		
inferred that M31 and M33 came to with 50\,kpc of each other 6.5\,Gyr ago, although their model did not replicate a structure reminiscent of the stellar stream that is observed today.
\citet{2021MNRAS.507.2592S}		
used 18\,000 EDR3 members to derive a (zero-point corrected) bulk proper motion of $61.9\pm9.7$\muasyr, leading to a more radial orbit, and a total transverse velocity with respect to the Milky Way of $82.4\pm31.2$\kms, consistent with (but more accurate than) earlier HST measurements that predict a future merger between the two galaxies.
\citet{2023ApJ...948..104P}		
derived a new virial mass estimate for M31, $M_{\rm vir}=3.02^{+1.30}_{-0.69}\times10^{12}M_\Sun$, using the orbital angular momenta of four satellite galaxies (M33, NGC~185, NGC~147, and IC~10) derived from a combination of Gaia and HST proper motions. 
\citet{2024A&A...692A..30R}		
discussed the systematics on the proper motions across the galaxy.
\citet{2024ApJ...971...98B}		
used HST observations as first epoch, and Gaia DR3 as the second, to examine the orbital history of 12 dwarf galaxies in the Local Group with respect to the four most massive (the Milky Way, LMC, M31, and M33), finding that two thirds are on their first infall, results consistent with the absence of star-formation bursts due to previous encounters.

Another large subject of interest is the degree to which the Local Group morphology and dynamics is consistent with the large-scale numerical simulations of $\Lambda$CDM cosmology.  One particular challenge has been the `plane-of-satellites' problem
\citep[e.g.][]{2022NewAR..9501659P}, 
the fact that both major galaxies in the Local Group host planar distributions of orbiting satellite galaxies, constituting the `Vast Polar Structure' of the Milky Way and the `Great Plane of Andromeda' 
\citep{2014MNRAS.442.2362P}.
I discuss this question further in Section~\ref{sec:cosmology-challenges}, and here just finish with a reference to some of the studies that have examined the present understanding of M31 in the context of these cosmological simulations 
\citep{2023A&A...677A..89K,	
2023A&A...677A..90K,		
2024POBeo.104...59P,		
2025A&A...694A..66A}.		

\paragraph{Rotational parallaxes}
\label{sec:rotational-parallaxes}

The `rotational parallax' method of distance determination is in principle applicable to Local Group disk galaxies, such as M31 and M33. The objective is to extend distance measurements beyond the significance of individual parallaxes, potentially providing an independent check of other extragalactic distance techniques. 
For a known centre-of-mass space motion, and given rotation curve, three observables for a given star -- its two proper motion components and its radial velocity -- are, in principle, sufficient to determine the three unknowns: the orbit inclination, the rotation velocity, and the distance
\citep{2000astro.ph..5484O,	
2007MNRAS.378.1385O}. 	
The method makes use of the fact that since proper motions are distance dependent, and radial velocities are distance independent, determination of a rotation curve using both methods can provide the distance. While there are various complications,
\citet{2007MNRAS.378.1385O}	
estimated that an accuracy of a few per cent in the distances to M31, M33 and the LMC could be achieved with radial velocities at the 10\kms\ level, and proper motions around 150\muasyr.

Applied to the LMC pre-Gaia, 
\citet{2014ApJ...781..121V}		
used line-of-sight velocities and HST proper motions for 6790 stars. They derived a kinematic distance consistent but not competitive with other methods at the time. 
Using Gaia DR2,
\citet{2019ApJ...872...24V}	
studied the proper motion fields of M31 and M33. Their accuracies could not adequately constrain the shape of the rotation curve, the viewing angle, nor the centre-of-mass motion. 
A similar analysis using Gaia DR3, resulting in estimates of the bulk proper motion for both M31 and M33, was carried out by
\citet{2024A&A...692A..30R}.	
They argued that the systematics still remain the dominant source of uncertainty, being comparable to the system's proper motion. Future Gaia data releases may allow further progress.

\subsection{Dwarf spheroidal galaxies}
\label{sec:dwarf-spheroidals}	

The Local Group of galaxies, dominated by the Milky Way and Andromeda and their associated satellites, includes of order 100 members, most of which are dwarf galaxies. These come in a number of types, including dwarf ellipticals, dwarf spheroidals, dwarf irregulars, and others variously referred to as ultra-compact dwarfs, ultra-diffuse dwarfs, and ultra-faint dwarfs. 
Dwarf spheroidals (dSph) are small, low-luminosity galaxies, comprising an old stellar population, almost devoid of dust, with masses $\sim\!\!10^7M_\Sun$ and, in contrast to dwarf ellipticals, roughly spheroidal in shape. Some two dozen are known as companions to either the Milky Way or M31 (Andromeda), and named after the constellation in which they are found.
The first known, Sculptor and Fornax, were discovered by Harlow Shapley in 1938, who described them as {\it `unlike any known stellar organisation'}
\citep{1938Natur.142..715S}. 	
Further discoveries were made problematic by their low luminosities and low surface brightnesses. 
By the late 1990s, their rarity seemed in conflict with $\Lambda$CDM cosmology, which predicted that massive galaxies like the Milky Way should be surrounded by many dark-matter dominated satellite halos. 
This conflict eased with the discovery of a dozen from the Sloan Digital Sky Survey around 2000
\citep{2007ApJ...670..313S},
a similar number by the Dark Energy Survey around 2015, and others more recently
\citep{2023ApJ...953....1C}. 
%
As deduced from their stellar kinematics, and in contrast to globular clusters, they are the most dark matter-dominated systems known, often adopted as targets for indirect dark matter searches
\citep{2011ApJ...733L..46W,
2013NewAR..57...52B,
2018RPPh...81e6901S}.

\begin{figure}[t]
\centering
\raisebox{13pt}{\includegraphics[width=0.455\linewidth]{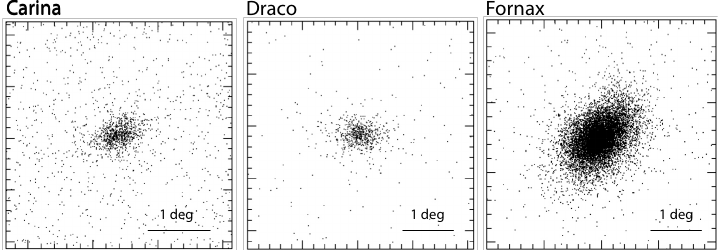}}
\hspace{15pt}
\includegraphics[width=0.50\linewidth]{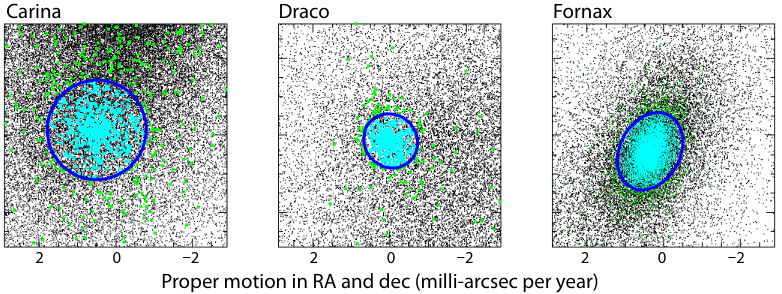}
\vspace{-15pt}
\caption{Three dwarf spheroidals with members selected by Gaia DR2 astrometry/photometry in positions (left) and proper motions (right). Blue ellipses are $3\sigma$ dispersions. Stars selected by photometric, but not astrometry, are shown in green \citep[From][Figures~2 and 12]{2018A&A...616A..12G}.
}\label{fig:dwarf-spheroidals}
\vspace{-10pt}
\end{figure}

The first Gaia DR2 study, of nine `classical' dwarf spheroidals 
\citep{2018A&A...616A..12G},
selected members based on astrometry (objects within $1-2^\circ$, and within $2\sigma$ of the system's mean proper motion) and photometry (red giants and blue horizontal branch stars). As the authors state, the possibility of selecting members according to their proper motions {\it `opens a new window for understanding the structure and extent of the dSph'}. Some showed asymmetries, with an indication of tidal streams in Carina (Figure~\ref{fig:dwarf-spheroidals}).
All are very distant, ranging from 26\,kpc (Sagittarius), to 250\,kpc (Leo~I), well beyond the Magellanic Clouds. Their bulk proper motions are consequently very small, $\sim$0.5\masyr, with that of the ultra-faint dwarf Bootes~I determined for the first time. Agreement with previous HST determinations was reasonable, but with much reduced errors and, importantly, in an absolute reference frame. 
Their positions and space motions, and models of the Galaxy's mass distribution, were used to follow their Galactic orbits over the past 250\,Myr
\citep{2018A&A...616A..12G}. 
Amongst their findings, Draco and Ursa Minor have very similar orbits, and possibly constitute a physically connected group. Most are on (slightly) prograde orbits, while Fornax is retrograde, qualitatively similar to results for globular clusters, but with smaller eccentricities.

Many Gaia-based results have followed. 
At the most basic level, Gaia data is being used to filter out foreground objects, resulting in cleaner samples of members, and more accurate distance estimates
\citep[e.g.][]{2022ApJ...929..116O}.		
Improved proper motions have been used to derive better orbits and orbital histories, with insights into their orbital alignments, infall trajectories, effects of other perturbers, and derived estimates of the masses of the Milky Way and M31
\citep{2018ApJ...863...89S,	
2020ApJ...893..121P,		
2022AJ....163....1C,			
2022ApJ...940..136P,		
2023ApJ...953....1C,			
2023MNRAS.526.1310T,		
2024ApJ...972..150M}.		
I say more on the cosmological implications of these orbits in Section~\ref{sec:cosmology-challenges}.

The three largest Milky Way satellite galaxies (LMC, SMC, and the Sagittarius dwarf spheroidal) have large populations of well-studied globular clusters. The fourth most luminous, the Fornax dSph, is now known to have six; the most recent found in DECam imaging (and inferred to be experiencing tidal destruction) is also seen as an overdensity in the Gaia data
\citep{2019ApJ...875L..13W}.		
The structure, associated globular clusters, and evidence for tidal disruption, have been investigated for Sagittarius
\citep{2020MNRAS.495.4124F,		
2021A&A...654A..23G,			
2022A&A...665A...8K},			
and in several of the other dwarf spheroidals finding, for example, extended stellar halos and evidence for escaping globular clusters
\citep{2022MNRAS.512.4171Y,		
2022MNRAS.512.5601Q,			
2022MNRAS.513.3526R,			
2023A&A...675A..49T,			
2024MNRAS.532.3713A}.			
And from these orbits, the satellite's infall time (when the galaxy first comes within the growing virial radius of the Milky Way halo), agrees with the epoch of maximum star-formation
\citep{2020ApJ...905..109M,		
2024A&A...691A.363Y}.			

With the origin of globular clusters still unclear, 
\citet{2022MNRAS.516.4560B}	
propagated the Gaia-based orbits of the 170 or so known Milky Way globular clusters, and its known satellite galaxies, backwards in time for 11\,Gyr in an attempt to identify their possible progenitors. None show clear association with any of the eight classical dwarf spheroidal galaxies, even though a large fraction of them are believed to have been accreted. They concluded that the globular clusters were either accreted as part of now-disrupted satellite galaxies, or that they may have had dark matter halos in the past, affecting their orbits.
The Gaia EDR3/DR3 proper motions are also probing the internal kinematics, and in particular the bulk rotation, for a number of dwarf spheroidals
\citep{2020A&A...633A..36M,		
2021ApJ...908..244D,			
2021MNRAS.505.5884M}.		
Effects of tidal and ram-pressure stripping have been observed and/or modelled in a number
\citep{2022MNRAS.509.5330B,		
2023ApJ...956L..37P,			
2023MNRAS.518.3083M,			
2024A&A...690A..61D,			
2024MNRAS.528.3009G}.			
Constraints on their density profiles, the inferred presence of dark matter, and the additional effects of tidal shocks at pericentre, are expected to be advanced by future data releases
\citep{2020ApJ...892....3H,			
2022MNRAS.515.2624P,			
2023ApJ...943..121G,			
2024MNRAS.532.4157A}.			
Gaia EDR3 proper motions also point to 60 Galactic high-velocity stars (including two hypervelocity stars) that were probably tidally stripped from the Sagittarius dwarf spheroidal at pericentre, 38.2\,Myr ago
\citep{2022ApJ...933L..13L}.		

\subsection{Ultra-faint dwarf galaxies}
\label{sec:ultra-faint-dwarfs} 

\paragraph{Discovery}
The distinct class of {\it ultra-faint\/} dwarf galaxies was first mentioned (I believe) in 2005, with the discovery, from SDSS, of an unusually extended object, SDSS~J1049+5103 (aka Willman~I), at 45\,kpc distance, with properties intermediate between those of globular clusters and dwarf galaxies
\citep{2005AJ....129.2692W}.	
The authors commented that: {\it `If it is a dwarf spheroidal, then it is the faintest yet known by two orders of magnitude, and is the first example of the ultra-faint dwarfs predicted by some theories.'}
A second, in Ursa Major at a distance of 100\,kpc, also from SDSS, was reported by 
\citet{2005ApJ...626L..85W}. 	
This was followed by several others, including other companions to the Milky Way,
in Canes Venatici at 220\,kpc
\citep{2006ApJ...643L.103Z},	
and in Boötes at 60\,kpc
\citep{2006ApJ...647L.111B}.
Others have since been found in the vicinity of the Magellanic Clouds
\citep{2015ApJ...805..130K}. 	
%
Crater~II, found in the VST--ATLAS survey
\citep{2016MNRAS.459.2370T}	
lies at a distance of 120~kpc, and with a surface brightness of 30.6~mag arcsec$^{-2}$, became the fourth {\it largest\/} Milky Way satellite (after the LMC, SMC and the Sgr dwarf). 

Today, more than 60 ultra-faint dwarf galaxies have been discovered, most recently with the Dark Energy Camera, DECam. Early Gaia contributions gave confirmation of several of the DECam candidates, including
Centaurus~I
\citep{2020ApJ...890..136M}
and Pegasus~IV
\citep{2023ApJ...942..111C}.
Meanwhile, cosmological $\Lambda$CDM simulations, and the fact that recent Local Group dwarf satellite discoveries pile up around the limit of surface brightness detectability, suggest that (many) others remain to be discovered
\citep[e.g.][]{2018MNRAS.481.4133G,	
2025MNRAS.540.1107S}.	

\paragraph{Cosmological context}
These ultra-faint dwarfs are now known to have the lowest surface brightness of all Local Group members. They contain from a few hundred to $\sim\!\!10^5$ stars, and have luminosities $\lesssim10^5L_\Sun$. While often resembling globular clusters in appearance, they are more extended, and with more dark matter (and more than in dwarf spheroidals), with typical $M/L\gtrsim100$. 
The current picture is that these ultra-faint dwarfs represent the faintest ($M_{\rm v}\gtrsim -6$), most metal-poor ([Fe/H] $\le -2$), and most dark matter-dominated end of the dwarf galaxy spectrum
\citep{2012AJ....144...76W}. 
As the oldest and least chemically evolved systems known, they provide {\it `unique windows into the formation of the first galaxies, and the behaviour of dark matter on small scales'}
\citep{2019ARA&A..57..375S}.	
Formed only a few million years after the Big Bang, before reionization, they provide important constraints on $\Lambda$CDM predictions, and a record of the assembly and chemical enrichment history of their host galaxies
\citep[e.g.][]{2009ApJ...693.1859B,	
2009ApJ...696.2179K,
2010MNRAS.401.2036L,
2013ApJ...765...22B,
2016MNRAS.457.1931S,
2016ApJ...827L..23W}.

Their high dark matter content is not believed to be due to baryon deficiency at the time of their formation, but rather to `dark matter heating', feedback from stellar winds and supernovae that can expel a significant fraction of the original baryonic component, causing the dark matter halo to expand
\citep{1986ApJ...303...39D,		
2016ApJ...820..131E,			
2018MNRAS.478..906C}.	
Tidal-stripping may also affect the very lowest surface brightness galaxies
\citep{2023MNRAS.521.3527C}. 	

\paragraph{Antlia~II}
The combination of Gaia's photometric, astrometric and variability information has allowed other discoveries to be made at surface brightnesses well below those from photometry alone. The first such Gaia discovery was in the `zone of avoidance', at low Galactic latitude, a region difficult to survey because of its high extinction and high density of foreground disk stars. 
\citet{2019MNRAS.488.2743T}
used Gaia DR2 parallaxes to filter out the foreground stars, using the proper motions to identify overdensities sharing common kinematics. They also used the Gaia DR2 RR~Lyrae catalogue 
\citep{2019A&A...622A..60C},	
exploiting the fact that all of the currently known Milky Way dwarfs contain at least one RR~Lyrae star 
\citep{2015AJ....150..160B},		
and yielding precise distances out to $\sim$100\,kpc.
Their discovery, an ultra-faint dwarf satellite in the constellation of Antlia, Antlia~II, is located behind the Galactic disk at latitude $b\,\sim11^\circ$, and extends over 1\ddeg26 (2.9~kpc at 130~kpc). While similar in extent to the LMC, Antlia~II  (confirmed with archival \mbox{DECam} imaging) became the lowest surface brightness system known, at 31.9\,mag arcsec$^{-2}$.
\citet{2019MNRAS.488.2743T}	
also concluded that Antlia~II has one of the least dense dark matter halos. Dynamical modelling suggests that a combination of a core dark matter profile 
\citep[e.g.][]{1996MNRAS.283L..72N},
along with strong tidal stripping, may explain its observed properties, with further support for this picture coming from Gaia
\citep{2021ApJ...921...32J,	
2022ApJ...926...78V}.		
\citet{2022ApJ...940..136P}		
used Gaia EDR3 to derive the systemic proper motion, and orbits, of 52 dwarf spheroidal satellite galaxies of the Milky Way. Again, they  predict that Antlia~II (along with Boo~III, Cra~II, Gru~II, and Tuc~III) should be undergoing tidal disruption by the Milky Way.

Antlia~II has been invoked to explain the planar warp-like disturbances seen in the Galaxy's outer H\,{\scriptsize I} disk 
\citep{2006ApJ...643..881L},		
whose origin has long been a puzzle.
\citet{2009MNRAS.399L.118C}	
suggested that it could be explained by a 1:100 mass ratio perturber on a near co-planar orbit with a pericentre of $\sim$5\,kpc, and at current distance of $\sim$90\,kpc.
\citet{2019ApJ...886...67C}		
showed that the current location of Antlia~II closely matches that prediction. If it is indeed the perturber, it would have a specific range of predicted proper motions that can be tested with future Gaia data.

\paragraph{Membership and orbits}
Over the various data releases, Gaia-based studies have been reported for numerous ultra-faint dwarf galaxies, amongst them
Boötes~I, 
Boötes~III, 
Coma~Ber, 
Eridanus~II, 
Eridanus IV,
Grus~II, 
Hercules,
Horologium~I,
Hydrus~I, 
Leo~V,
Pegasus~III,
Pegasus~IV,
Pisces~II,
Reticulum~II,
Segue~I,
Segue~II,
Tucana~II,
Tucana~III, 
Ursa Major~I, 
and 
Ursa Major~II.
These studies have targeted improved membership, as a first step in goal of determining their Galactic orbits.
Their orbits, in turn, provide fundamental constraints on 
the mass and shape of the Milky Way halo, 
on their evolution, tidal disruption and mass loss,	
on the connection between their orbits and their star formation histories, 
and ultimately how the Milky Way has assembled its population of dwarf galaxy satellites 
\citep{2018A&A...620A.155M}.

On the determination of bulk space motions of dwarf galaxies in general, Gaia DR2 {\it `has started a revolution'}
\citep{2018A&A...619A.103F}.	
Based on careful membership selection, starting with initial guesses but refined using Gaia DR2 astrometry and photometry, 
\citet{2018A&A...620A.155M}	
derived absolute proper motions for seven ultra-faint dwarf galaxies within 70\,kpc (Boötes~III, Carina~II, Grus~II, Reticulum~II, Sagittarius~II, Segue~II, and Tucana~IV).
Subsequent studies using DR2 and DR3 proper motions have now been used to establish membership, distances, and space velocities for many other systems, including 
Tucana~II
\citep{2020ApJ...891....8C},		
Hercules
\citep{2020ApJ...902..106M},	
and Ursa Major~I, Coma Berenices, and Boötes~I
\citep{2018A&A...616A..12G,	
2023MNRAS.519.1349W}.	

Rigorous membership selection is essential for characterising the morphologies, and identifying possible extensions attributable to tidal disruption. Such Gaia-derived extended morphologies have now been reported, amongst others, for
Leo~V
\citep{2019ApJ...885...53M},	
Hercules
\citep{2020ApJ...902..106M},	
Eridanus~IV 
\citep{2021ApJ...920L..44C},	
and Boötes~I 
\citep{2021ApJ...923..218F}. 	
For Crater~II, 
\citet{2022MNRAS.515.2624P}	
showed that its small velocity dispersion could result from a dwarf spheroidal that lost most of its halo mass to tidal stripping by the Milky Way, an argument supported by some Gaia DR2/EDR3 analyses 
\citep{2019ApJ...883...11F,	
2020MNRAS.492.1061V,		
2021ApJ...921...32J},		
but questioned or modelled differently by others
\citep{2022MNRAS.512.5247B,	
2024ApJ...968L..13Z}.		

Accurate bulk proper motions, and hence accurate orbits, provide the ingredients for computing their orbital histories.  Amongst these, Gaia~EDR3 proper motions for Eridanus~II suggest that it is on its first infall into the Milky Way
\citep{2022ApJ...925....6F}.		
In the case of the distant but closely separated (40\,kpc) systems Pegasus~III and Pisces~II, the study by 
\citet{2022ApJ...933..217R}	
found no morphological features indicating that a significant interaction between the two has occurred. Propagating their orbits in a combined Milky Way--LMC potential, they found a possible orbital history in which they experienced a close (10--20\,kpc) mutual passage a little more than 1\,Gyr ago, followed by a combined passage around the Large Magellanic Cloud (30--60\,kpc) just under 1\,Gyr ago. If confirmed, they would add to the rare occurrence of coherent satellite pairs within the Local Group.
For the newly discovered Pegasus~IV,
\citet{2023ApJ...942..111C}	
determined an elliptical retrograde orbit, currently near its orbital apocentre.
\citet{2025ApJ...979..171F}	
showed that the ultra-faint dwarf Segue~II had a recent ($77\pm5$\,Myr ago) close flyby with the Cetus--Palca halo stream, within the stream's $2\sigma$ width. They show that this interaction potentially enables constraints on its mass and density profile at much larger, kpc-scale, radii than are probed by its stars.

\paragraph{Use of RR~Lyrae stars}
Searching for RR~Lyrae variables in the vicinity of these satellite galaxies provides a particularly powerful method for suppressing foreground contamination, and hence identifying potential extra-tidal members
\citep[e.g.][\S2.3]{2019MNRAS.488.2743T}.
Such searches in 27~nearby ($<100$\,kpc) ultra-faint dwarfs by 
\citet{2020ApJS..247...35V}	
associated 47~Gaia RR~Lyrae stars with 14 different satellite galaxies, identifying extra-tidal RR~Lyrae in six (three with extra-tidal stars in addition to members within their tidal radius: Boötes~I, Boötes~III and Sagittarius~II; and three in which the galaxy itself does not contain any RR~Lyrae: Tucana~III, Eridanus~III, and Reticulum~III). They inferred that these galaxies may also be undergoing tidal disruption.
Similarly, 
\citet{2024AJ....167...57T}	
found seven new RR~Lyrae stars in six ultra-faint dwarf galaxies (Hydrus~I, Ursa Major~I, Ursa Major~II, Grus~II, Eridanus~II, and Tucana~II).
  
\paragraph{Constraints on star formation}
There is a demonstrated connection between the orbits of many satellite galaxies and their star-formation histories
\citep[e.g.][]{2001ApSSS.277..231G,
2009ARA&A..47..371T}.
This has been nicely illustrated in the Gaia DR2 study of the orbits of the classical dwarf spheroidals by 
\citet{2020ApJ...905..109M}. 
These authors showed that their infall time, defined as when the satellite galaxy first crosses within the growing virial radius of the Milky Way's halo, coincides well with the peak star-formation rate, as a consequence of ram pressure on their gas content. 
In contrast, the general consensus is that star-formation activity in ultra-faint dwarfs had already peaked prior to their infall times, possibly suppressed at the epoch of reionization
\citep{2008A&A...487..103O,
2012ApJ...753L..21B,
2014ApJ...796...91B,
2015ApJ...804..136W,
2019ARA&A..57..375S}.	

\paragraph{Challenges to MOND}
Ultra-faint galaxies also provide specific challenges to the gravitational theory of Modified Newtonian Dynamics (MOND) as a (non-dark matter) explanation for the rotation curve of galaxies
\citep[e.g.][]{
2021ApJ...914L..37S,	
2022ApJ...940...46S}.	
In this context, arguments against MOND relate to the observed velocity dispersion, infall history, and observed tidal features, established by Gaia, for Ursa Major~I, Willman~I, and \mbox{Tucana~III}
\citep{2021ApJ...914L..37S}.	

\subsection{Mass of the Local Group}
\label{sec:mass-local-group}

The 
\href{https://en.wikipedia.org/wiki/Local_Group}{Local Group},
a term introduced by Edwin Hubble in 1936, is a concentration of around 80 known galaxies, extending across some 3~Mpc, and dominated by the Milky Way and Andromeda (M31). Both are spiral galaxies with masses of order $10^{12}M_\Sun$, and both are accompanied by their own systems of smaller satellite galaxies, including many `ultra-faint' dwarf spheroidals. In the case of our Galaxy, these include the Large and Small Magellanic Clouds, along with Sagittarius, Draco, Carina, Sextans, Fornax, and many others.  Tied to Andromeda are M32, M110, NGC~147, and many others. 
Over the past 20~years, much new insight has been gained about the evolution and merger history of our own Galaxy, specifically as it `devoured' smaller nearby galaxies over its multi-billion year history. 

Gaia is helping to clarify the picture, revealing various debris streams, including the Gaia--Enceladus and Milky Way merger some 10~Gyr ago which contributed to the formation of our Galaxy's thick disk (Section~\ref{sec:gse}), and the `phase-space spiral' attributed to a recent crossing of our Galaxy's disk by the Sagittarius dwarf galaxy 300--900~Myr ago (Section~\ref{sec:phase-space-spiral}). In the case of Andromeda, a dominant merger some 2~Gyr ago has been proposed to explain its compact and metal-rich satellite M32 as the stripped core of the disrupted satellite, as well its rotating inner stellar halo, and its associated giant stellar stream
\citep{2018NatAs...2..737D}.

In any successful cosmological model, all of these, and many other properties of the Local Group, must also be consistent with the increasingly detailed predictions of large-scale numerical models of the formation and evolution of structure in the Universe, and mass estimates of the Local Group (amongst various other observables) continue to provide a powerful test of cosmological models.
Two of the most developed methods for determining the mass of the Local Group are via cosmological simulations, and via the so-called `timing method'. The former infers the Galaxy's total mass by comparing model satellites in (for example) the EAGLE cosmological hydrodynamics simulations with the dynamics of a number of the `classical' Milky Way satellites with six-dimensional phase-space measurements (i.e.\ positions and velocities), most recently including updated proper motions from Gaia (including the Large and Small Magellanic Clouds, Draco, and Ursa Minor). 
The timing method, mentioned already in Section~\ref{sec:mass-galaxy}, is based on the hypothesis that the Milky Way and M31 proto-galaxies had a small separation at the time of the Big Bang, subsequently moving apart as they participated in the Hubble flow. Given an estimate of the age of the Universe, together with their present separation and approach velocity, the equations of motion can be solved to give the mass of the Galaxy, along with that of the Local Group. 
Of specific interest here are the latest refinements of the orbit of M31 and M33, as a result of the improving observational accuracy of their tangential motions, most recently using Gaia DR2. 
Specifically, 
\citet{2019ApJ...872...24V}	
determined the proper motions of a number of massive stars in each galaxy, relative to surrounding background quasars, to derive a centre-of-mass proper motion of 
($65\pm18$, $-57\pm15$)\muasyr\ for M31 (1084 stars),
and
($31\pm19$, $-29\pm16$)\muasyr\ for M33 (1518 stars),
highlighting {\it `the future potential of Gaia for proper motion studies beyond the Milky Way'}.

Similar values for M31 were determined from Gaia EDR3 by 
\citet{2021MNRAS.507.2592S},
who noted that their values were consistent with (but more accurate than) earlier Hubble Space Telescope measurements that predicted a future merger between Andromeda and the Milky Way
\citep{2012ApJ...753....7S}.
Subsequently,
\citet{2022ApJ...928L...5B}
adjusted the approach velocity between the Milky Way and M31 for the effects of our Galaxy's motion towards the Large Magellanic Cloud. They gave two estimates of the total mass of the Local Group:
$3.4^{+1.4}_{-1.1} \times10^{12}M_{\Sun}$ when using a higher tangential velocity for M31 of 80\kms,
and
$3.1^{+1.3}_{-1.0} \times10^{12}M_{\Sun}$ when using a lower value of 60\kms.
This can be compared with a value of $3.7^{+0.5}_{-0.5} \times10^{12}M_{\Sun}$, based on the likely mass of the four most substantial members (the Milky Way, M31, M33 and the LMC), with the minor members contributing $\sim\!0.7 \times 10^{12} M_\Sun$.

Another complication for the timing method is due to the `fly-by' of the Large Magellanic Cloud. As the most massive satellite galaxy of the Milky Way, the LMC is just past its closest approach of about 50~kpc, and passing the Milky Way at more than 300\kms. 
\citet{2021NatAs...5..251P}
used Gaia DR2 data to show that the Milky Way disk is moving with respect to stellar tracers in the outer halo, at about 32\kms, and in a direction that points to an earlier location on the LMC's trajectory. Accounting for this perturbation lowers the inferred Local Group mass by 10--20\% compared to a static halo.  Specifically,
\citet{2023ApJ...942...18C}
gave an updated total mass of between $4.0^{+0.5}_{-0.3}\times10^{12}M_\Sun$ or $4.5^{+0.8}_{-0.6}\times10^{12}M_\Sun$ depending on their adopted kinematic measurements of M31.
A summary of these various mass determinations is given in Figure~\ref{fig:mass-milky-way}b.

\subsection{Gaia's galaxy and quasar survey}		
\label{sec:galaxy-quasar-survey}

The primary objective of Gaia is to characterise the stellar content of the Milky Way. But passing across the telescope's fields of view as it scans the sky are other objects which can be measured at the same time. Amongst them are more than 150\,000 minor bodies of the solar system (Section~\ref{sec:solar-system}), and more than a million quasars. For both types of object, their (typically) star-like images makes their detection no more complex than (and indeed operationally on-board indistinguishable from) the detection of stars. 
Early in the technology study phase of Gaia (in the late 1990s), it also became clear that distant compact galaxies could also be detected, albeit with some revision of the on-board detection and CCD sampling strategies. 

Even at 20--21 mag, most of the focal-plane CCD pixels are `empty' of target stars. Indeed, it is a basic principle of the Gaia measurements that all of the CCD pixels are {\it not\/} read out and sent to the ground: the penalty for doing so, in terms of CCD read-out noise, and the data volume to be transmitted to the ground, would have been prohibitive. 
Instead, stars and star-like objects are detected on-board, as they enter the field, by the sky mappers. Above a given signal-to-noise threshold, information on the target's brightness, location, and transverse motion are transmitted to the subsequent focal plane detectors (astrometry, photometry, and radial velocity spectrometer), where only the small region related to the source is read out, and only these `windows' are transmitted to the ground
(Section~\ref{sec:windows}).

The optimisation of a detection and sampling strategy for galaxies, i.e.\ extended objects characterised by faint surface brightness variations, was not at all straightforward, and was the subject of numerous detailed technical reports and iterations
\citep{2002EAS.....2..313V}.
%
Essentially the problem was to optimise three aspects: 
(i)~the angular areas over which the average surface brightness and local sky background values are computed;
(ii)~the detection area and signal-to-noise limit such that only useful data are transmitted to the ground, without being swamped by less interesting regions of the Milky Way or the zodiacal light;
(iii)~a specific sampling scheme for galaxies, providing a trade-off between angular resolution, CCD read-out noise and telemetry volume. 
A statistical model of galaxy number density, size and surface brightness distribution, was also developed to characterise the `typical' faint galaxy to be targeted.  Results suggested that there could be about three million galaxies brighter than this limit above $\pm15^\circ$ from the Galactic plane (i.e.\ where galaxy detection should not be compromised by the high density of stars). 
Details of the final on-board detection and sampling algorithms are given in the paper describing the overall mission
\citep{2016A&A...595A...1G},
and in the DR1 documentation 
\citep{2017gdr1.reptE...1D}.

The ultimate goal could never be a survey of {\it all\/} galaxies, but rather a high-reliability multi-colour catalogue extending to low Galactic latitudes, and with high-angular resolution imaging ($\sim$0.18\,arcsec) of all sufficiently high-surface brightness galaxies. 
The result would represent a homogeneous and well-defined sample for two main purposes: (i)~for a statistical analysis of the photometric structure of the central regions of tens of thousands of well-resolved galaxies at an unprecedented angular resolution, and (ii)~for the study of the large-scale structure of the local Universe, aiming to further clarify important features such as large filaments, `walls', and the Supergalactic Plane. 

With the adopted detection algorithm, unresolved early-type elliptical galaxies and galaxy bulges are preferentially detected, even with radii of several arcsec, while late-type spiral galaxies, even those with weak bulges, remain mostly undetected.
%
%
The on-ground processing is carried out within Coordination Unit~4 of the Gaia Data Processing and Analysis Consortium. As one of the nine coordination units, this is dedicated to the treatment of three `special' celestial source types observed by Gaia: non-single stars, solar system objects, and `extended objects'. 
This latter task, of interest here, treats all sources considered to be extended, aiming to classify the sources and to quantify their morphology. It also analyses all sources separately classified as quasars to search for underlying host galaxies, or other features that could disqualify their use in the reference-frame alignment.

Their adopted surface brightness profile fitting `pipeline' works by simulating a large set of 2d image profiles, then iteratively searching for the one that best reproduces the one-dimensional observations. 
%
As they point out, although their future list of quasar candidates will be based on Gaia's own classification algorithms, these were not available for the DR3 processing, which used instead their own compilation from the existing literature. 
The sky coverage of each of the merged source catalogues being somewhat heterogeneous, their resulting input list is similarly heterogeneous across the sky.
Their starting list of extended sources similarly contained some 1.7~million galaxies with an entry in Gaia DR3. It is reasonably homogeneous across the sky, except in the Galactic plane which is mostly empty.

For known quasars, the source structure is decomposed into two light profile components: the central quasar and a possible surrounding host galaxy. The central quasar is expected to be point-like, its apparent angular extent being essentially due to the line spread function of the Gaia instrument. It is modelled by a circular exponential profile with a fixed scale length of 39.4\,mas, corresponding to the instrument's line spread function. The adopted (S\'ersic) profile model has the flexibility to represent either spiral- or bulge-like structure.
For 1\,103\,691 previously known quasars analysed in DR3, the majority (1\,031\,607) were classified as point-like. A host galaxy was detected around 64\,498, and the surface brightness profiles of the underlying host galaxy (generally consistent with disk-like morphologies) made available for 15\,867 quasars with robust solutions.
The pipeline also analysed 940\,887 galaxies, assuming two different model profiles (a `de~Vaucouleurs' profile $\propto r^{1/4}$, and a S\'ersic profile $\propto r^{1/n}$) and provided robust solutions for 914\,837. In contrast to the underlying quasar hosts, the distribution of the S\'ersic indices confirms that Gaia indeed detects mostly elliptical galaxies, and that very few disks are detected.

The results of this entire process form just part of Gaia DR3, where they are described in detail 
\citep{2023A&A...674A..11D}.	
But in addition to this analysis of the surface brightness profiles of extended objects carried out within Coordination Unit~4
\citep{2023A&A...674A..11D},	
other modules within the DPAC Consortium also contribute to the classification and analysis of extragalactic objects. Within Coordination Unit~8, responsible for the source classification and astrophysical parameter determination (Section~\ref{sec:classification-stellar-properties}), the Discrete Source Classifier (DSC) provides classification probabilities associated with five types of sources: quasars, galaxies, stars, white dwarfs, and physical binaries, while the Quasi-Stellar Object Classifier (QSOC) and the Unresolved Galaxy Classifier (UGC) determine their redshifts from the BP/RP spectra 
\citep{2023A&A...674A..31D}.	

The combination of these various processes leads to the summary of the extragalactic content of DR3 given in Table~\ref{tab:data-release-table2}.
Of the total of 6\,649\,162 quasar candidates in DR3,
	redshifts are determined for 6\,375\,063, 
	the host galaxy is detected in 64\,498,
	and the host surface brightness profile is provided for 15\,867.
Of the total of 4\,842\,342 galaxy candidates in DR3,
	redshifts are determined for 1\,367\,153, 
	and the surface brightness profiles for 914\,837.

\subsection{Strongly lensed quasars}
\label{sec:quasars}

Quasars provide fundamental positional anchors for defining the Gaia reference frame (Section~\ref{sec:reference-frame}) although, as discussed there, there are currently unexplained differences in the vector spherical harmonic terms when divided into discrete redshift bins. Quasars are also contributing insight to two topical observational questions in $\Lambda$CDM cosmology, which I cover in Section~\ref{sec:kinematic-dipole}: the kinematic dipole anomaly, and the $S\!_8$ `tension' in the amplitude of cosmological structures.

Another important contribution of Gaia to quasar science is in detecting and classifying gravitationally lensed systems. The on-board CCD sampling provides source images with an ultimate resolution of around 0.2~arcsec in the along-scan direction, much better than can be routinely measured from the ground. Accordingly, Gaia is discovering numerous new lensed quasars with image separations below $\sim$1~arcsec.

\begin{figure}[t]
\centering
\includegraphics[width=0.60\linewidth]{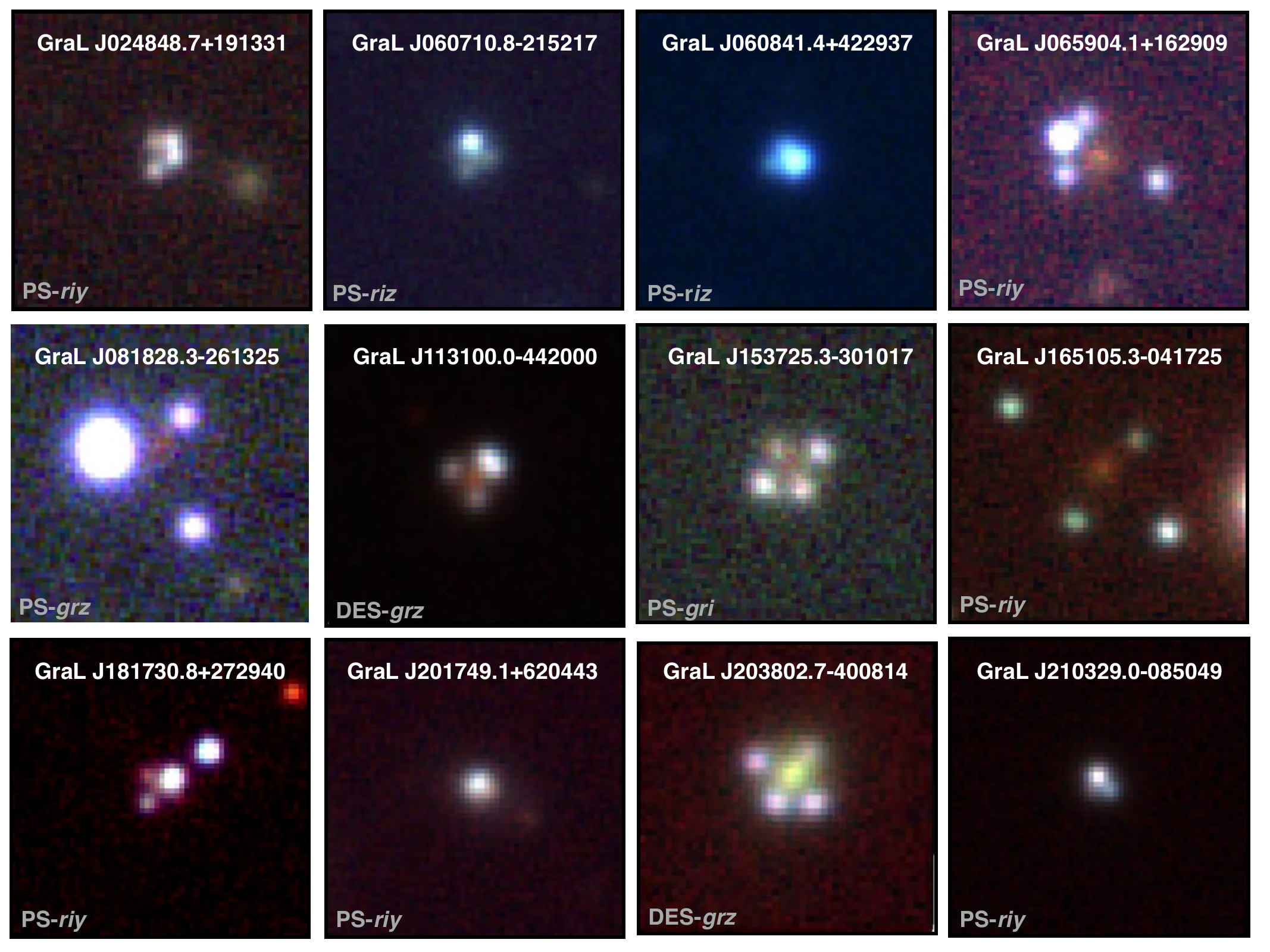}
\caption{The first 12 confirmed quadruply-imaged Gaia quasars, presented as false-colour images from PanSTARRS (PS) or the Dark Energy Survey (DES), each 15~arcsec on a side \citep[from][Figure~1]{2021ApJ...921...42S}.}
\label{fig:lensed-quasars}
\end{figure}

Strongly lensed quasars, i.e.\ those which result in multiple {\it discrete\/} images, provide an important diagnostic for many applications. These include 
measuring the Hubble constant at intermediate cosmic distances 
	\citep{2019MNRAS.490.1743C, 2020MNRAS.494.6072S},
constraining the properties of dark matter 
	\citep{2019MNRAS.487.5721G, 2020MNRAS.492.5314N},
inferring structure near the event horizon of supermassive black holes 
	\citep{2009ApJ...697.1892P, 2016AN....337..356C},
measuring the accretion disk size 
	\citep{2011ApJ...729...34B, 2021A&A...654A..70F},
constraining the geometry of the quasar broad emission-line region  
	\citep{2011A&A...528A.100S, 2017A&A...607A..32B},
measuring black hole spin 
	\citep{2015ApJ...805..161W, 2016ASSL..440...99M},
and testing general relativity in the strong-gravity regime 
	\citep{2018Sci...360.1342C, 2023MNRAS.520.1613M}.
Amongst the most keenly hunted are the quadruply-imaged `Einstein crosses', which are some of the most constraining for detailed physical modelling, and where the time delays between the image brightness variations can provide constraints on the Hubble constant (Figure~\ref{fig:lensed-quasars}).

Fortunately for the task of defining the reference system, strongly lensed quasars are rare. But their rarity and compactness makes them difficult to find.
%
An on-line compilation of known lenses 
\citep[][\S2]{2019MNRAS.483.4242L}, 
which appears to be no longer maintained, listed 220 lensed systems as of 2019, of which 50 have quadruple images. 
Dictated by the atmosphere, most lensed systems discovered from the ground have image separations above 1~arcsec 
\citep[e.g.][]{2018A&A...618A..56D}. 
Nevertheless, the expected distribution of lenses is predicted to peak {\it below\/} 1~arcsec, with estimates suggesting that some 3000 multiply imaged quasars could eventually be detected by Gaia, the majority being doubles, but with more than 250 having three or more images
\citep{2016A&A...590A..42F}.

Studies and discoveries with Gaia~DR2 (in the series titled `Gaia GraL') have dealt with
new quadruple candidates around known quasars \citep{2018A&A...616L..11K};
the statistics of previously known quadruple-image quasars \citep{2018A&A...618A..56D};
a systematic blind search for new lensed systems \citep{2019A&A...622A.165D};
ground-based confirmation and modelling of new discoveries \citep{2019A&A...628A..17W, 2019arXiv191208977K, 2021ApJ...921...42S}; 
X-ray observations of discoveries between $z=1-3$ \citep{2022ApJ...927...45C};
and a radio census of 24 lensed systems \citep{2024MNRAS.528.5880D}.
One of the special `Focused Product Release' involved the `GravLens pipeline', implemented within the Gaia DPAC Consortium (Section~\ref{sec:data-releases}). It has the goal of analysing all Gaia detections around quasars, to produce a catalogue of secondary sources around each of them, along with their mean astrometry and photometry
\citep{2024A&A...685A.130G}.	
They started with 3\,760\,480 quasars or quasar candidates with an entry in Data Release~3, analysing the 183\,368\,062 corresponding transit measurements from the first 3~years of operation, to search for nearby objects. The output is a list of 4\,760\,920 sources, including the quasars themselves, within 6\,arcsec of the quasar positions.
In most cases (87\%), the quasar is a single source, without a close neighbour. In 501\,385 cases, neighbouring sources were detected, of which 9\% have two components. There are 159\,000 systems (4\%) with more than two components. They also found 9000 multiplets with more than 10~components, generally corresponding to extended galaxies decomposed into a number of discrete sources. 

Their resulting GravLens catalogue includes 450 known or candidate lenses previously published in the literature, 76 with four images, and the rest being two-component systems. For 67 of the 76 quadruple systems, GravLens complements the existing measures from GDR3 by measuring one or more additional components, or the deflecting galaxy itself. In total it found 1293 components in the fields of known lenses, of which 1207 were present in GDR3. The 86 newly detected components in the fields of known lenses are mostly faint components surrounding lenses with a small number of existing GDR3 counterparts.
They identified 381 new lensed candidates, of which they considered 49 as the most promising. These have angular sizes from 1.03--5.97~arcsec, with a minimum image separation of 0.41~arcsec.
Several papers report follow-up campaigns, including independent searches
\citep{2025A&A...696A..51P},	
photometric monitoring enabling their use as cosmological probes, and as targets for high-resolution radio observations, X-ray observations, and near-infrared adaptive optics integral field unit spectroscopy to improve the source modelling.

\subsection{Dual active galactic nuclei}
\label{sec:dual-agn}

The consensus today is that all quasars and active galactic nuclei (AGN) host a matter-accreting massive black hole at their core. And, as a result of galaxy mergers, close pairs of supermassive black holes can result from the pre-merger progenitors. When separated by a few kpc within the merged host, they can be observed either as dual AGN (when both black holes are active), or as `offset AGN' (and displaced from the galaxy centre) either when only one is active, or when signalling a recoiling black hole after a binary merger 
\citep{2016MNRAS.458.1013S,2017MNRAS.469.4437C}.
A number of candidates have been discovered with SDSS spectroscopy
\citep{2011ApJ...737..101L,
2023MNRAS.524.5827F},
and Subaru spectroscopy
\citep{2020ApJ...899..154S},
including a few triple systems 
\citep{2011ApJ...736L...7L,
2023MNRAS.524.4482B}.
Follow-up observations have been conducted by JWST 
\citep{2025A&A...696A..59P,
2023A&A...679A..89P}, 
and at multi-wavelengths including X-ray
\citep{2023MNRAS.519.5149D}.
Only a few have been found at $z>0.5$
\citep{2022NatAs...6.1185M}.

\begin{figure}[t]
\centering
\includegraphics[width=0.80\linewidth]{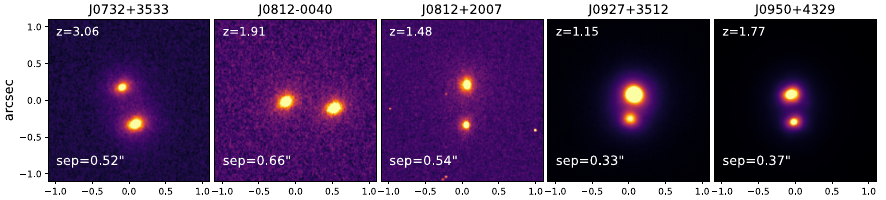}
\caption{High-resolution Ks band Large Binocular Telescope images ($2.2\times2.2$\,arcsec$^2$) of five Gaia `multi-peak' AGN discoveries, showing redshifts and separations (\citet{2022NatAs...6.1185M}, Figure~2).}
\label{fig:dual-agn}
\end{figure}

Dual AGNs remain difficult to detect and characterise because of their sub-arcsec angular separation. And since multiple AGN at small angular separations can also be due to gravitational lensing of a single source, follow-up spectroscopy is generally crucial in distinguishing the two effects. But their numbers provide important constraints for cosmological simulations 
\citep{2025A&A...696A..59P}. 
And they provide an important probe of the physical processes that drive the in-spiralling of supermassive black hole pairs in the context of gravitational wave research.

Gaia is generating its own survey of quasars and AGN (Section~\ref{sec:galaxy-quasar-survey}), and its sub-arcsec along-scan angular resolution (at  $\sim$0.18~arcsec) can in principle pinpoint in-spiralling black holes well on their way to their eventual merger. Two approaches are being followed.
The first, which the authors termed `varstrometry' for `intrinsic variability induced astrometric jitter', exploits one of the methods used for the astrometric detection of unresolved binary stars with Hipparcos, where it was termed `VIMs', for `variability induced movers' \citep{1996A&A...314..679W}: if one component of a fixed separation double varies in luminosity, it leads to the astrometric displacement of the system's {\it photocentre}. In the Hipparcos catalogue, 288 multiple star solutions were classified as VIMs \citep{1997A&A...323L..53L}, and a similar scheme is being exploited for Gaia \citep[][\S7.2.6]{2022gdr3.reptE...7P}.
As developed in the Gaia--AGN context 
\citep{2020ApJ...888...73H}, 
the idea is to detect such temporal displacements of the photocentre in unresolved (sub-kpc separation) optically selected dual active galactic nuclei. Several such studies have identified a number of candidates, often based on Gaia's `astrometric excess noise' flag (Section~\ref{sec:ruwe}), most of which are awaiting spectroscopic confirmation
\citep{2019ApJ...885L...4S, 
2022ApJ...925..162C,	
2023ApJ...943...38S}.	
In one study, 143 SDSS spectroscopically confirmed quasars were found to have multiple Gaia EDR3 detections within 1~arcsec of the quasar position
\citep{2023MNRAS.524.1909J}. 
Based on spectral-fitting, this resulted in two double quasars, and 56~other candidates.
In another study, 27~candidates were identified using DR3 astrometry, for which follow-up VLBA radio observations revealed 8~systems with either multiple radio components, or system offsets
\citep{2023ApJ...958...29C}.	

A different approach is to search for AGNs showing multiple peaks in their Gaia light profiles.
\citet{2022NatAs...6.1185M} 
identified 221 candidates at $z=0.3-4$, all of which showed multiple components with sub-arcsec separation in high-resolution archival images (26~from HST, and 5~from LBT), confirming the method's high reliability (Figure~\ref{fig:dual-agn}).
Several follow-up studies of these multiple-peak candidates have been reported.
\citet{2023A&A...671L...4C}
found two dual~AGNs out of 4~candidates observed with integral-field spectroscopy.
\citet{2023ApJ...956..117G}
established one as a lensed quasar.
\citet{2023MNRAS.524L..38W}
observed 5~candidates with \mbox{e--MERLIN}, detecting significant radio offsets in~four.
\citet{2024A&A...690A..57S}
observed 12~with VLT--MUSE, classifying 5~as dual AGNs, 2~as lensed systems, and 5~as chance alignments of foreground stars.
\citet{2024A&A...682L..15M}	
independently found 8 dual AGN candidates from the FPR release.

\subsection{Gravitational waves}
\label{sec:gravitational-waves}

Gaia is contributing to gravitational wave research in three areas: 
identifying `verification binaries' for the planned ESA--NASA Laser Interferometer Space Antenna (LISA);
searching for the electromagnetic counterparts of gravitational wave sources;
and providing constraints on the stochastic gravitational wave background.

\paragraph{Verification binaries}

The gravitational wave sources driving LISA's (and China's Tian-Qin) science goals fall into four main classes: compact neutron star--black hole binaries, massive black hole binaries, extreme mass ratio `inspirals', and compact white dwarf binaries. In this latter class, a large population of $\sim\!\!10^7$ Galactic white dwarf binaries emits gravitational waves across the whole LISA band, $10^{-1}-10^{-5}$~Hz. The majority will not be individually resolved, creating a stochastic gravitational wave `foreground' below 3--5\,mHz. 
Some 1000--10\,000 are expected to be individually resolved and characterised by LISA. Pre-Gaia, around 25 short-period compact white dwarf binaries (periods less than a few hours) were considered to be `guaranteed' LISA sources, and are often referred to as `verification binaries'. Each signal is long lived, and almost monochromatic, with just a small drift in frequency over the mission due to gravitational wave emission or/and due to mass transfer. Each signal is characterised by 7--8 parameters, with typical signal-to-noise ratios expected to be `moderate', up to about~100. 

A search in Gaia EDR3, through cross-matching with the Zwicky Transient Facility DR8, identified 429 short-period white dwarf binaries from the phase-folded light curves
\citep{2023ApJS..264...39R}: 	
220 eclipsing binaries, 
and 209 ellipsoidal variation binaries. 
From their estimated gravitational wave amplitudes, they identified 6~new LISA candidates, bringing the total number of identified verification binaries to 31 for LISA (and 18 for TianQin).
A similar search using Gaia DR3 resulted in an updated list of 48 candidate LISA binaries with properties derived from the Gaia distances
\citep{2024ApJ...963..100K}. 
This included 16 new verification binaries (considered as those detectable after 3~months), and 21~new sources detectable after 48~months. They include binary white dwarfs, AM\,CVn binaries, one ultra-compact X-ray binary, and two hot subdwarf binaries, together tripling the number of expected LISA binaries. 

\paragraph{Electromagnetic counterparts}

The ground-based LIGO--Virgo collaboration has so far made four `runs' (O1--O4), resulting in around 200 detections, mostly binary black hole mergers, but including a few neutron star--neutron star, and neutron star--black hole, mergers. While black hole mergers generate only gravitational waves, the (first) neutron star--neutron star merger, GW~170817, was also detected as a gamma-ray burst (by the Fermi Burst Monitor) and, in the optical, as a kilonova \citep{2017ApJ...848L..12A}. The 6~Feb 2025 event was the first also detected in particles (neutrinos).

Gravitational wave sources are typically located on the sky only to within a few hundred square degrees, but Gaia, as an all-sky multi-epoch photometric survey (and in common with other large-scale transient surveys), has the potential to routinely search for their optical counterparts.
An extension to the Gaia Science Alerts system (Section~\ref{sec:science-alerts}) was developed, essentially alerting on each detection of a previously unknown source, with one alert per field of view transit, rather than requiring the usual, and more secure, repeated detection in both fields. At the expense of more false alerts, their simulations suggest that some 16\% of gravitational wave events should fall in sky regions observed by Gaia within 7~days, providing about 21 candidates per day from the whole sky
\citep{2020MNRAS.493.3264K,
2021A&A...652A..76H,
2023MNRAS.525.4065B}.

\paragraph{Stochastic background detection}

Various theoretical scenarios suggest that a gravitational wave {\it background\/} is generated in the early Universe, whether as a result of inflation, cosmic strings, or `colliding bubbles' during first-order phase transitions, each predicting a specific form of the gravitational wave spectrum 
\citep{1997rggr.conf..373A,2018CQGra..35p3001C}.
The general idea underpinning the possible detection of such a stochastic background is that gravitational waves from a point source, such as a star or quasar, passing over a telescope in gravitational `free fall', will cause a time-varying shift in the apparent position of the source, i.e.\ a spurious proper motion 
\citep{1996ApJ...465..566P,1998PhRvD..58d4003D}. 
In the language of the field, the angular displacement is order of $h$, being the magnitude of the gravitational wave field at the telescope. The fact that the apparent motions are determined by the field at the telescope (rather than at the source, or over the photon trajectory) then implies that the relative motion of two nearby sources is proportional to their angular separation. 
To date, observational constraints on a stochastic background include those from 
Big Bang nucleosynthesis,	
VLBI astrometry,			
and irregularities in millisecond pulsar timing, 
with four groups announcing the detection of such a background signal in the pulsar timing data in June 2023
\citep{2023ApJ...951L...8A,
2023A&A...678A..50E,
2023ApJ...951L...6R,
2023RAA....23g5024X}.
From its early days, it was evident that Gaia could, in principle, place limits on the energy flux of primordial gravitational waves
\citep{2001A&A...369..339P}. 
Depending on the gravitational wave frequency, two regimes were later identified 
\citep{2018CQGra..35d5005K}: 
one where the effects of gravitational waves directly influence the derived proper motions, and one where the effects mostly appear in the residuals of the standard astrometric solution \citep{2014EAS....67...49K}. 
Other theoretical work on the prospects for Gaia has continued
\citep{2018PhRvD..97l4058M,		
2022PhRvD.106h4006W},	
with a number of papers reporting early results and associated upper bounds
\citep{2021PASP..133i4501L,
2021arXiv210504039A,		
2023MNRAS.524.3609J,	
2025A&A...695A.172G}.	
These early results, based on Gaia data with recognised systematics, nonetheless point to improved prospects and constraints from future data releases.

\subsection{Dark matter}
\label{sec:dark-mater}
Here, I will simply point to the sections of this review which detail Gaia results having some specific bearing (although without necessarily any specific conclusions) on the existence of `dark matter'. 
Some of the most convincing if circumstantial evidence for the existence of dark matter is the remarkable degree to which detailed numerical simulations of the $\Lambda$CDM concordance model (with its canonical prescriptions for the amount of dark matter and dark energy) successfully reproduce essentially all of the main features seen in the morphology and dynamics of both our Galaxy, and of the Local Group (Section~\ref{sec:cosmology-interpretation}). This includes, in particular, the details of the debris stellar streams seen in the halo (Section~\ref{sec:halo-streams}). 
But it also extends to the presence of the dwarf spheroidal galaxies in the Local Group, which had been predicted by $\Lambda$CDM models (Section~\ref{sec:dwarf-spheroidals}),
and by the observed occurrence of galaxy mergers evidenced by the presence of dual active galactic nuclei (Section~\ref{sec:dual-agn}). 
Further provisional support comes from the resonant motions of the local stellar population which appear to be not only consistent with the presence of the Galaxy's central bar, but in particular with a bar that is being slowed down by dynamical friction with the Galaxy's dark matter halo (Section~\ref{sec:bar}).
Together, these represent a most remarkable triumph of state-of-the-art space observations combined with leading cosmological simulations.

Other Gaia observations have some bearing on the distribution of dark matter.
The vertical force acting on stars (with respect to the Galactic plane) can be reconstructed from the vertical velocity dispersion, and the vertical density profile, both of which are benefitting from improved estimates of the distances and space velocities of a suitable tracer star population.  Observations yield a picture in which disk matter is well accounted for (by stars, and interstellar gas and dust), and in which any dark matter in the Galaxy must be distributed in the form of the spherical halo, with little if any concentrated in the form of the disk (Section~\ref{sec:mass-kz}).
Also refer to the rotation curve of the Galaxy (Section~\ref{sec:rotation-curve}), and the fact that this rotational acceleration is detected in the velocity field of quasars (Section~\ref{sec:aberration}).

Various tests may become available in the future. 
Improved Gaia astrometry may detect the rotation of the disk with respect to a tumbling triaxial dark matter halo (Section~\ref{sec:tumbling-disk}).
The more extreme hypervelocity stars (Section~\ref{sec:hypervelocity-stars}) are not gravitationally bound to the Galaxy, and will escape the confines of even its vast boundaries after some 300~Myr. Travelling in different directions, and now at a vast range of distances from their origin, they probe the Galaxy's gravitational potential, making them possible tracers of the detailed (dark) matter distribution in the Milky Way \citep[e.g.][]{2005ApJ...634..344G,2006ApJ...651..392S,2025NewAR.10001721H}. 
It has been suggested that the orbital velocities of ultra-wide binaries are inconsistent with Newtonian theory and  may instead point to some MOND-like theories of gravity, although more conventional explanations have also been put forward (Section~\ref{sec:ultra-wide-binaries}).

A highly speculative suggestion by 
\citet{2022Symm...14..721P}
would tie together some unexplained features of solar activity, and the repeated (and contested) claims of an unexplained correlation between some of these features and the orbital position of the major planets. This could, the authors propose, find a physical basis if the solar system is intercepted by a stellar stream, with dark matter particles with speeds $10^{-4}-10^{-3}c$ focused on the Sun through planetary lensing. Specific tests, which would provide direct evidence for dark matter particles, are outlined.

\paragraph{Axions}

The axion is one of a number of hypothetical elementary particles outside of the Standard Model of particle physics. It was originally proposed to solve the `strong CP' (charge--parity) problem, in order to explain the high degree of matter--antimatter symmetry observed in the Universe
\citep{1977PhRvL..38.1440P}.
 Experimental searches
(amongst them
HAYSTAC, 
ORGAN, 
CULTASK, 
and the Axion Dark Matter Experiment) 
are mostly based on the Primakoff effect, in which axions are converted to photons in strong electromagnetic fields. 
Properties of the stellar streams discovered and characterised in the Galaxy's halo by Gaia, are being used as constraints on the signal models for axions and WIMPs used, for example, in the Sudbury argon-based DEAP--3600 detector
\citep{
2019PhRvD..99b3012E, 	
2019ApJ...874....3N, 	
2020PhRvD.102h2001A,	
2020PhRvD.101f3026B, 	
2020PhRvD.101b3006O, 
2021JCAP...10..043B, 	
2023JInst..18C2046L}.	

Specific phases of stellar evolution can also provide constraints on the properties of the hypothetical axion, and several are making use of Gaia data.
I have mentioned one of these, from white dwarf cooling models and the effects on pulsation periods, in the context of limits on the variation of the gravitational constant (Section~\ref{sec:g-dot}). 
The hot end of the white dwarf luminosity function, populated by the rare short-lived DO~stars, also provides constraints on the cooling of the white dwarf population as a whole, and on the properties of any particles emitted by them
\citep{2018MNRAS.480.1211R,		
2022A&A...668A.161C,			
2022ApJ...934...34C}.			
The tip of the red-giant branch similarly provides constraints on any energy loss leading to a larger core mass at He-ignition, and thus to a brighter luminosity than predicted by standard models
\citep{2020PhRvD.102h3007C}. 

Other speculative suggestions for physical phenomena that might be responsible for some of the Gaia observations include
the ultra-faint dwarf galaxies Crater~II and Antlia~II 
\citep[][Section~\ref{sec:ultra-faint-dwarfs}]{2020PhRvD.101h3012B, 
2022MNRAS.515.2624P}; 	
triangular features on opposite sides of the Galactic centre explained by a `dark axion flow'
\citep{2021PDU....3300838C}; 
the origin of the black hole binary Gaia~BH1 
\citep[][Section~\ref{sec:black-holes}]{2023MNRAS.524.4083P}; 
as a probe of dark photon dark matter
\citep{2019JCAP...05..015G};	
the existence of `fuzzy dark matter' as an explanation for the age--velocity dispersion relation of the Milky Way disk
\citep{2023MNRAS.518.4045C}	
and for some other detailed properties of the observed halo streams
\citep{2018arXiv180800464A};
and for their effect on the low-frequency gravitational wave background generated in the early Universe 
\citep{2021arXiv210504039A, 
2023PhRvL.131q1404G}. 

\subsection{Cosmological implications}
\label{sec:cosmology}

\paragraph{Simulations}
\label{sec:cosmology-simulations}

Today's standard model of cosmology invokes the theory of cosmic inflation coupled with the `big bang' expansion. In its most successful parameterisation, $\Lambda$CDM (`Lambda cold dark matter'), some 27\% of the Universe is dark matter and 68\% is dark energy, with only a small fraction being the ordinary baryonic matter that composes stars, planets, and living organisms. 
This model provides the best available prescription for the hierarchical growth of structures as a result gravitational instabilities in the early Universe, and the progressive formation of galaxies, stars, and black holes within them. 
Over the past 20~years, very large massively parallel \mbox{N-body} simulations, resting on the $\Lambda$CDM paradigm, have been developed to investigate how dark and baryonic matter structures have evolved over time
\citep{2020NatRP...2...42V}.  
These numerical simulations are now guiding many of Gaia's advances: amongst them, interpreting stellar streams, modelling resonant motions, characterising the dynamics of local group galaxies, and comprehending the multiple manifestations of past satellite interactions.\footnote{
A key advance in these very large-scale projects was the Millennium Simulation
\citep{2005Natur.435..629S}. 
This followed the growth of dark matter structures from $z=127$ to the present. It used $2160^3$ particles, each representing $10^9M_\Sun$ of dark matter, within a cube of side 700\,Mpc. 
Millennium~II 
\citep{2009MNRAS.398.1150B}		
and Millennium~XXL 
\citep{2012MNRAS.426.2046A}		
incorporated improved parameters and input physics.
Other such simulations now include (somewhat chronologically)
%
\citep{2011ApJ...740..102K,	 
2011ApJ...742...16T},	
Eris			
\citep{2011ApJ...742...76G},	
EAGLE 		
\citep{2015MNRAS.446..521S},	
Illustris		
\citep{2014MNRAS.444.1518V,		
2014MNRAS.445..175G},	
HESTIA			
\citep{2020MNRAS.498.2968L},	
NewHorizon
\citep{2021A&A...651A.109D},	
ARTEMIS 
\citep{2020MNRAS.498.1765F},	
SIBELIUS	
\citep{2022MNRAS.512.5823M},	
and
HACC/Farpoint/New Worlds
\citep{2022ApJS..259...15F,	
2024arXiv240607276H}.		
The Illustris simulation, for example, started 12\,Myr after the Big Bang, evolved over 13~Gyr, and used 12~billion resolution elements in a cube of side 100~Mpc.
It generates elliptical and spiral galaxies, galaxy clusters, the distribution of hydrogen on large scales, and the metal and hydrogen content of galaxies on small scales
\citep{2014Natur.509..177V}.	
Galaxy formation processes include stellar evolution and feedback, gas recycling, chemical enrichment, and black hole growth and mergers
\citep{2013MNRAS.436.3031V,	
2015ARA&A..53...51S}.			
The Illustris framework has been used for other derivatives, specifically:
\href{https://www.tng-project.org}{IllustrisTNG} `The Next Generation' \citep{2018MNRAS.473.4077P};
\href{https://wwwmpa.mpa-garching.mpg.de/auriga}{Auriga}
	for high-resolution simulations of Milky Way-like dark matter halos \citep{2017MNRAS.467..179G}; 	
\href{https://www.thesan-project.com}{Thesan} for the reionisation epoch \citep{2022MNRAS.511.4005K};
\href{https://www.mtng-project.org}{MillenniumTNG} for the halo mass function \citep{2023MNRAS.524.2556H}; 
and 
\href{https://www.tng-project.org/cluster}{TNG--Cluster} for galaxy clusters \citep{2024A&A...686A.157N}.
}.  

\paragraph{Interpretation of the Gaia data}
\label{sec:cosmology-interpretation}

These simulations are guiding interpretation of the Gaia data on the largest scales, for example in the understanding of galaxy mergers, the nature of the Galaxy's central bar, and the Milky Way's cosmological evolution.

(a)~Major mergers: it is now accepted that the Gaia Sausage--Enceladus merger played a key role in the formation of our Galaxy's inner stellar halo
\citep{{2018Natur.563...85H}, 	
{2019NatAs...3..932G}},	
	as well as the disk 
\citep{{2018ApJ...863..113H}, 	
{2022Natur.603..599X}}. 	
Are these discoveries supported by cosmological simulations?
%
\citet{2022MNRAS.513.1867D} 	
showed that about one-third of galaxies from the ARTEMIS simulations contain accreted stars on highly radial orbits, similar to the Gaia Sausage--Enceladus event. The major mergers also result in disk rotation, and changes in shape and orientation of their dark matter halos. Early mergers result in retrograde stars, analogous to the `splash' or `plume' feature also discovered with Gaia 
\citep{{2019A&A...632A...4D}, 		
{2020MNRAS.494.3880B}}.		
In a study of six M31 and Milky Way analogues from the HESTIA simulations of the Local Group, all experienced between one to four mergers with stellar mass ratios between 0.2--1, with five them occurring 7--11~Gyr ago
\citep{2023A&A...677A..89K}.	
The most massive mergers result in a sharp increase in the orbit eccentricity of disk stars of the main progenitor. 

(b)~The central bar: Gaia is providing new insights into the morphology and dynamics of our Galaxy's central bar.
Arising as a consequence of stellar orbits, are they also influenced by past mergers? 
\mbox{IllustrisTNG}, NewHorizon, and EAGLE all confirm the emergence of bars, but with a large variation in bar fractions, ranging from 5--55\% 
\citep{2022MNRAS.510.5164C}.	
\citet{2020MNRAS.494.5936F} 	
showed that, in their Auriga simulations, galaxies which best reproduce the chemo-dynamical properties of the Milky Way bulge have quiescent merger histories since $z\sim3.5$. Their last major merger was more than 12~Gyr ago, with any subsequent mergers having a stellar mass ratio of 1:20 or lower. 
In turn, this suggests an upper limit of a few percent for the mass ratio of the Gaia Sausage--Enceladus merger event. They inferred that the Milky Way has had an `uncommonly quiet merger history', and hence an essentially {\it in situ\/} bulge.

\begin{figure}[t]
\centering
\vspace{-5pt}
\hbox{\hspace{0pt}
\includegraphics[width=1.0\linewidth]{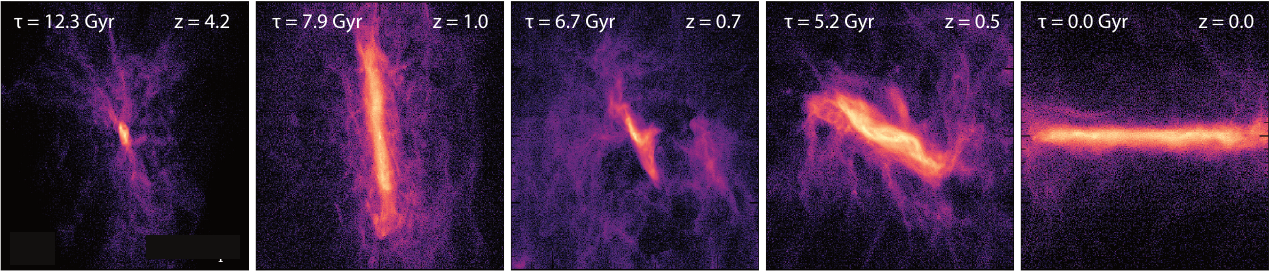}
} 
\caption{Evolution of TNG50 halo \#519311 (from the IllustrisTNG cosmological simulations) over cosmic time, covering a region $30\times20$\,kpc$^2$. The five snapshots are ordered from left to right, with the lookback time and redshift indicated at the top. This panel (one of four in the referenced paper) shows the logarithmic density of gas particles in fixed coordinates oriented edge-on to the present-day disk. This halo undergoes a major merger at $z\simeq0.7$, visible in the central panels (\citet{2024ApJ...972..112C}, Figure~10). 
An animated version of this figure is 
\href{https://www.youtube.com/watch?v=1q1TLoLUqdY&ab_channel=VedantChandra}{available here}. 
}\label{fig:chandra24}
\end{figure}

(c)~Evolution of the Galaxy:
\citet{2024ApJ...972..112C} 
used [Fe/H] and [$\alpha$/Fe] estimates for 9.9~million Gaia red giants to characterise their angular momentum as a function of metallicity. Taking this as a proxy for age, they identified three distinct evolutionary phases: a disordered/chaotic protogalaxy, a hot old high-$\alpha$ disk, and a cold young disk with more ordered and circular orbits.
Does this proposed `three-phase evolution' of the Milky Way find any support from these large-scale cosmological simulations?
%
\citet{2024ApJ...962...84S}
had already selected a representative sample of 61 Milky Way-like galaxies from the TNG50 simulations (the highest-resolution box from IllustrisTNG). Of these, 11 matched the ‘early spin-up’ previously inferred by \citet{2022MNRAS.514..689B}.
\citet{2024ApJ...972..112C} showed that halo \#519311, one of the earliest of their spin-up galaxies, exhibits a ‘three-phase’ structure in the orbit circularity versus metallicity space `remarkably similar' to the Milky Way (Figure~\ref{fig:chandra24}). This halo also experiences a major gas-rich merger 8~Gyr ago, albeit slightly later than the GSE merger. The merger adds a large amount of low-metallicity gas and angular momentum, from which the kinematically cold low-$\alpha$ stellar disk is subsequently born.

Other Gaia studies are now calling on these cosmological simulations to assist their interpretation. 
More than 20 other papers between 2018 and mid-2024 have used EAGLE or Illustris to further the understanding of the Gaia Sausage--Enceladus accretion event, with 10 in the past two years
\citep{
{2022AJ....164...41W},		
{2023MNRAS.518.6200B},	
{2023MNRAS.519L..87D},		
{2023ApJ...946...99C},		
{2023A&A...677A..89K},		
{2023A&A...677A..90K},		
{2023MNRAS.521..995R},		
{2025MNRAS.538..553L},		
{2024MNRAS.527.2165C}}.	
Various studies employ these simulations in interpreting the improved globular cluster orbits
\citep{
{2022MNRAS.514.4736C},	
{2023A&A...673A.152I},		
{2023A&A...678A..69I},		
{2024OJAp....7E..23C},		
{2024A&A...683A.146I}},		
and the improved orbits of the dwarf spheroidals 
\citep{
{2020MNRAS.492.1543P},		
{2020MNRAS.491.3042P},		
{2022MNRAS.509.5330B},		
{2023MNRAS.518.3083M}}.		
They are also being used in discussions of the mass of M31
\citep{2023ApJ...948..104P},		
warping in the orbits of Cepheids
\citep{2023MNRAS.523.1556D},		
Milky Way analogues
\citep{2018MNRAS.481.1726G},		
infall times for Local Group galaxies
\citep{2023MNRAS.520.1704B},		
hypervelocity stars and the Galactic escape speed
\citep{2019MNRAS.485.3514D},		
the Milky Way mass profile and halo mass
\citep{2020ApJ...894...10L},			
halo anisotropy
\citep{2020JCAP...07..036B},		
and the cosmological core--cusp problem
\citep{2022ApJ...941..108W}.		
Again, more accurate and complex data sets, more stringent tests, and greater accuracy and fidelity of the $\Lambda$CDM parameterisation, will come with future Gaia data releases.

\paragraph{Kinematic dipole}
\label{sec:kinematic-dipole}

Prominent amongst observational questions is the $5\sigma$ `tension' in the cosmic expansion rate $H_0$ as measured from the cosmic microwave background compared with more local distance calibrators (Section~\ref{sec:cepheids-h0}).  There is a similar discrepancy in the value of the dipole amplitude of the cosmic microwave background radiation, and the `kinematic dipole' defined by the distribution and motion of objects on scales approaching the Hubble length
\citep{2020ApJ...894...68R,		
2021EPJST.230.2051A,
2021EPJST.230.2067M,
2022AnPhy.44769159P, 					
2023CQGra..40o5005L}.					
The amplitudes of the spherical harmonic expansion of the CMB temperature of degree $\ell>1$ are interpreted as remnants of the decoupling of acoustic oscillations in the early Universe. But the dipole amplitude, $\ell=1$, is much larger than predicted from this effect, and is instead attributed to the Sun's motion relative to the CMB of 370\kms\ in the Galactic direction $l=264^\circ$, $b=48^\circ$ 
\citep{2020A&A...641A...1P}.
Adjusting for the solar motion within our Galaxy, and thereafter for the barycentric motion of the Local Group, implies that the Local Group is moving relative to the CMB at 620\kms\ in the direction $l=272^\circ$, $b=30^\circ$ 
\citep[][\S3.2.4]{2022AnPhy.44769159P}.		

Objects distant enough that their distribution should approximate to a homogeneous isotropic universe should display a similar dipole anisotropy in their number counts 
\citep{1984MNRAS.206..377E}. 
Tests at radio \citep{1998MNRAS.297..545B, 2002Natur.416..150B} and X-ray wavelengths \citep{2000ApJ...544...49S}, and based on quasars \citep{2021ApJ...908L..51S,			
2021Univ....7..107S},
have argued both for and against a discrepancy. A joint analysis of various catalogues of radio galaxies and quasars concluded that their sky distribution is {\it inconsistent\/} with the standard $\Lambda$CDM model at the $5.1\sigma$ level, similar to that of the Hubble tension 
\citep{2022ApJ...937L..31S}.

Gaia's contribution comes via the Quaia quasar catalogue, which merges the 6\,649\,162 quasar candidates from Gaia DR3 (with redshifts from its low-resolution BP/RP spectra) with the unWISE infrared catalogue, together yielding 1\,295\,502 well-defined quasars with $G<20.5$~mag
\citep{2024ApJ...964...69S}.
Taking into account selection effects, and especially contamination near the Galactic plane,
\citet{2024MNRAS.527.8497M}	
concluded that the Quaia quasar dipole {\it is\/} consistent with the CMB dipole, in terms of both amplitude and direction. That the observed quasar dipole can be (currently) attributed to a local departure from the Hubble flow lends further support to the $\Lambda$CDM model.

\paragraph{The $S\!_8$ tension}
\label{sec:s8-tension}

The `$S\!_8$~tension' refers to a distinct discrepancy, of around $3\sigma$, in the amplitude of the predicted matter density fluctuations based on the CMB data, and the many observations, at low redshifts, inferred from direct measurements of galaxy clustering data, or weak lensing (`cosmic shear') data. 
The amplitude of density fluctuations, of galaxies or clusters, is characterised by this $S\!_8$ parameter, defined as
$S\!_8=\sigma_8(\Omega_{\rm m}/0.3)^{0.5}$, 
where $\sigma_8$ is the variance of the linear matter overdensity field in spheres with a $8h^{-1}$Mpc radius, and $\Omega_{\rm m}$ is the fractional energy density of non-relativistic matter, both defined at $z=0$. 
The value of $S\!_8$ has been measured by various galaxy clustering and weak lensing surveys, resulting in various degrees of `tension' compared with the value inferred from the Planck satellite, $0.832\pm0.013$.
For example, from combined data, 
\citet{2021JCAP...10..030G} 
derived $S\!_8=0.7781\pm0.0094$, with errors comparable with the Planck CMB measurements, but deviating by $3.4\sigma$.

The Gaia constraints again make use of the Quaia catalogue, exploiting its full-sky coverage, high spatial density, and redshift coverage to $z\lesssim4$ with good redshift accuracy (e.g.\ $|\Delta z/(1+z)|<0.01$ for 62\% of the sources).
\citet{2023JCAP...11..043A} 
argue that, since it is usually not possible to fully break the degeneracy between growth and geometry with existing datasets, results are generally reported in terms of the combination $S\!_8=\sigma_8(\Omega_{\rm m}/0.3)^{0.5}$. However, the large-scale coverage of the Quaia/Gaia sample allows independent estimates of $\sigma_8$ and $\Omega_{\rm m}$. Accordingly, they derived separate constraints on the amplitude of matter fluctuations, $\sigma_8=0.766\pm0.034$, and on the fractional abundance of non-relativistic matter, $\Omega_{\rm m}=0.343^{+0.017}_{-0.019}$, together yielding $S\!_8=0.819\pm0.042$. These Quaia quasar results are compatible with Planck at the $1.4\sigma$ level, and therefore (they argue) do not provide support for the $S\!_8$ tension.

\paragraph{Some other challenges}	
\label{sec:cosmology-challenges}

While $\Lambda$CDM has passed a vast range of impressive experimental tests, there remain gaps in the theoretical understanding of its key ingredients including
the nature of dark matter and dark energy, 
the asymmetry of matter versus antimatter, 
the consistency of relativity and quantum mechanics, 
and whether the dimensionless parameters of physics evolve with time
\citep{2023CQGra..40o5005L}.
In parallel with these successes, various challenges for $\Lambda$CDM have been encountered in reconciling numerical simulations with observations on galaxy scales, amongst them
the `missing satellites' problem,
the `core--cusp' problem,
the `too-big-to-fail' problem,
and the `plane of satellites' problem 
\citep{2022NewAR..9501659P}. 
I will expand on two of these.

The `missing satellites', or `dwarf galaxy' problem, emerged from numerical simulations which demonstrated the hierarchical clustering of dark matter, in which smaller halos merge to form larger halos. In the earliest simulations, normal-sized galaxies were well reproduced, while the number of known dwarf galaxies was significantly lower than expected
\citep[e.g.][]{1999ApJ...524L..19M}. 
I say more on this subject in Section~\ref{sec:dwarf-spheroidals}.

The `plane of satellites' problem arises because both major galaxies in the Local Group host planar distributions of orbiting satellite galaxies, constituting the `Vast Polar Structure' of the Milky Way, and the `Great Plane of Andromeda'. 
%
%
Their flattened structure and coherent motions have long been considered inconsistent with $\Lambda$CDM, which instead indicate uncorrelated and nearly isotropic satellite systems, in which these sorts of spatial and kinematic coherences are very rare
\citep{
{2005MNRAS.363..146L},
{2005ApJ...629..219Z},
{2014MNRAS.442.2362P}}.
This has made {\it `\ldots the satellite planes one of the most serious small-scale problems for $\Lambda$CDM'}
\citep{2018MPLA...3330004P}, 
perhaps evidence for the modified gravity formulation MOND
\citep{{2018A&A...614A..59B},
{2018MNRAS.477.4768B}}.
Using Gaia's proper motions from DR2 to define the orbital planes and rotation of 11~satellite galaxies within the Vast Polar Structure appeared to {\it worsen\/} the tension
\citep{2020MNRAS.491.3042P}.
But with EDR3 came greatly improved proper motions of many Local Group satellites, resulting in systemic motions for some 70 galaxies out to $\sim$1.4~Mpc
\citep{2020RNAAS...4..229M,
2022A&A...657A..54B}. 
From a comparison between these Gaia observations, and the results of 60\,000 representations of Local Group analogues from the SIBELIUS simulations, 
\citet{2023NatAs...7..481S}
inferred that the high anisotropy of the Milky Way's satellite system can be attributed to its high central concentration, not previously reproduced in simulations, combined with the close but fleeting contiguity of its two most distant members.  They concluded that there was, therefore, no evidence for a `plane of satellites' incompatible with, or even particularly remarkable in, $\Lambda$CDM.

\section{The future}
\label{sec:future}

\paragraph{The end of Gaia observations}
The basic limit on the satellite's operational lifetime was the `cold gas' (nitrogen), used for the satellite's fine attitude control required for its precision scanning of the celestial sphere. On 15~January 2025, with this cold gas almost depleted, scientific operations were terminated. More than 10~years of spectacular astrometric observations of unprecedented accuracy, number, and magnitude range, came to an end. After some two months dedicated to various technology tests, the satellite was manoeuvred away from its operational Lagrange L2 orbit, minimising the chance of coming within 10~million km of Earth for at least the next century, its onboard software deliberately corrupted, and the satellite finally 
\href{https://www.esa.int/Enabling_Support/Operations/Farewell_Gaia!_Spacecraft_operations_come_to_an_end}{deactivated}
on 27~March 2025.

\paragraph{Future data releases}
Due to its substantial, iterative, and interconnected nature, the global data processing activity will continue intensively for at least the next five years. The data releases so far have made use of only a limited subset of the data acquired: 
DR1 (September 2016) was constructed from the first 14~months of satellite data;
DR2  (April 2018) covered the first 22~months;
EDR3 (December 2020) and DR3 (June 2022) covered the first 34~months.

Gaia Data Release~4 (DR4) is presently under construction based on the first 5.5~years of mission data, and is expected to be released in late 2026. With perhaps closer to 3~billion sources, this will consist of
the astrometric, photometric, and radial-velocity catalogues;
all available variable-star and non-single-star solutions;
source classifications, plus astrophysical parameters (from the astrometry, G/BP/RP photometry, and RVS spectroscopy) for stars, unresolved binaries, galaxies, and quasars;
an exoplanet list;
and the epoch and transit data for all sources.
Gaia DR5, based on all 10.5~years of data, will not be available before 2030. It will consist of the complete Gaia Legacy Archive of all data.
Recalling that the latest full catalogue, DR3, covers just the first 34~months of mission data, a substantially greater scientific harvest can be anticipated once this full mission data set becomes available.

\paragraph{Future space astrometry missions}

As indicated in Figure~\ref{fig:accuracy}, accuracies better than those achieved by Gaia should still be achievable. Fundamental noise sources (such as interstellar scintillation, gravitational microlensing, and the stochastic gravitational wave background) will perhaps limit global accuracies to 0.010--0.1\muas. But the next accuracy gains beyond Gaia will only be limited by instrumental ingenuity, and by political and financial will. 

Three mission concepts which target improvements in positional measurements are currently under study.
In Japan, JASMINE targets Gaia-level accuracies ($\sim$25\muas) in the near-infrared (1--1.6\micron), taking advantage of the significantly lower extinction at longer wavelengths to study both the obscured central regions of the Galaxy along with habitable-zone exoplanets
\citep{2024PASJ...76..386K}.	
European plans are centred on GaiaNIR, which builds on the principles of Gaia, and also focuses on the near-infrared, targeting a higher accuracy, and for some 10~billion stars
\citep{2016arXiv160907325H,	
2024eas..conf.2394H}.		
%

\section{Summary}

Building on the principles of space astrometry established by Hipparcos, Gaia is providing a revolutionary new view of the content, structure and dynamics of our Galaxy and beyond. The simultaneous provision of all-sky high-accuracy astrometry, multi-epoch multi-colour photometry, and radial velocities and medium-resolution spectroscopy, is allowing the physical characterisation of some two billion stars (nearly three million should be contained in the future data releases DR4 and DR5), a significant fraction of our Galaxy's stellar content. Model-dependent stellar properties, including temperatures, gravities, reddening, and metallicity, along with derived estimates of their masses and radii, is transforming the possibilities for understanding the content, chemistry, and evolutionary history of our Galaxy.
Not to be overlooked is the great improvement that Gaia's astrometry has brought to the pointing and operational efficiency across all ground-based and space-based observatories.

Gaia's census is greatly advancing our knowledge of the local solar neighbourhood and beyond, including the minor bodies of the solar system, involving characterisation of the space densities of different evolutionary states, and the accurate positioning of each object in the observational colour--magnitude (and theoretical Hertzsprung--Russell) diagram. More exotic stars, such as runaway and hypervelocity stars, are being measured and characterised in unprecedented numbers.
The multi-epoch photometry is yielding a vast census of the many types of variable stars, both extrinsic variables (such as eclipsing binaries), and intrinsic variables (including all classes of radial and non-radial pulsators), and, again, providing an unprecedented picture of their occurrence and properties.

The simultaneous detection and characterisation of binary (and higher multiplicity) systems, across the disparate classes of visual, astrometric, and spectroscopic binaries, is yielding vast samples of multiple systems which are being used to place constraints on models of star formation, and the occurrence and nature of the enormous range of their manifestations, amongst which are mass-transfer systems, ellipsoidal variables, equal mass `twin' binaries, eclipsing binaries, and binaries of extremely large-separation.
One of Gaia's major contributions is the accurate delineation and characterisation of huge numbers of open clusters and associations, with their important application to models of star formation and evolution, their internal dynamics including rotation and expansion, and evidence for their progressive dissolution most dramatically in the form of their tidal tails. 

On Galactic scales, Gaia has discovered or is making major contributions to a number of new phase-space features of the Milky Way, which are proving crucial to understanding its structure, dynamics, and evolutionary history. These include:	
discovering more than 100 halo streams, including the dominant Gaia Sausage--Enceladus stream, and further characterising the Sagittarius (and many other) streams;					
attributing the Hercules and Arcturus streams (and others) to dynamical resonances; 	
identifying the breathing motion of spiral arms;							
(provisionally) detecting the bar's deceleration due to the dark matter halo;	
characterising the disk warp;										
discovering the Gaia phase-space spiral, attributed to a massive galaxy--galaxy collision;	
identifying and characterising globular cluster and open cluster tidal tails;	
identifying the primordial heart of the Milky Way;						
measuring Galactic rotation, and the associated effects of Galactic aberration;	
and
discovery of the Radcliffe Wave as an instability or perturbation structure.		

On the largest scales, Gaia is providing new insights into the Local Group of galaxies (including their space motions and orbits, evidence for mergers, and mass estimates) and, beyond that, contributing a huge census of quasars (defining the quasi-inertial reference frame) and compact galaxies, amongst which are multiple-lensed systems, and binary active galactic nuclei providing evidence for galaxy mergers over cosmological time. 
All of these detailed features, especially perhaps the halo streams and the Gaia phase-space spiral, but extending to more arcane problems such as the long-standing Oosterhoff dichotomy, are providing support for the occurrence and growth of structure predicted by the current $\Lambda$CDM cosmological paradigm.

Out of just over 10~years of satellite data acquired, the latest data release (Gaia DR3) is based on data from only the first 34~months. Larger, better, and more complete data sets will be available with DR4 (expected in late 2026), and the final DR5 (expected around 2030). Along with improved parallaxes and proper motion will be more extended multi-epoch photometry for variability analyses. Availability of the multi-epoch astrometry will be transformational in characterising binary and multiple systems, including complete orbit solutions, and the identification of very large numbers of sub-stellar companions, including both brown dwarfs and exoplanets.

\paragraph{A note on source material}
This review draws on weekly essays that I have been writing on the scientific results from Gaia since January 2021, and which I post at 
\href{https://www.michaelperryman.co.uk/gaia-essays}{www.michaelperryman.co.uk}. `Legacy copies', along with interviews with some of the scientists and engineers involved in the Hipparcos and Gaia missions, are at the 
\href{https://zenodo.org/communities/gsemp}{CERN--Zenodo site}. 
\href{https://zenodo.org/records/15791236}{`Gaia Science Tree', V3} presents these essays (1--235, Jan 2021--Jun 2025) as a hyperlinked `mind map', 
with all end-nodes hyperlinked (to the Zenodo archive) to the given essay number.
%
I am also in the process of converting these essays into AI-generated `discussion-type' audio versions, available at \href{https://www.youtube.com/@GaiaEssays}{YouTube/\symbol{'100}GaiaEssays}.

\section{Acknowledgments}

Gaia was founded on the principles of the pioneering Hipparcos mission of the European Space Agency between 1980--1997, for which I was ESA's Project Scientist. I mentioned some of the leaders of this move to space in Section~\ref{sec:hipparcos}. 
Without the knowledge and experience from Hipparcos, Gaia would have been unthinkable.
The Hipparcos Science Team, and the many members of the four Hipparcos scientific consortia (INCA, NDAC, FAST, and TDAC) who worked with ESA to deliver the Hipparcos mission results, laid these foundations.

Gaia developed from its first ideas in around 1993 (due to Erik H{\o}g, Lennart Lindegren, and myself), and is therefore a project extending over more than 30~years. Both Hipparcos and Gaia are very substantial collaborations. They involve budgets of around 500\,M€, hundreds of scientists from across all ESA member countries working for many years, and some 30 industrial teams involving perhaps a thousand engineers contributing their highly specialised technologies and expertise. Overall management was by the European Space Agency, with these projects selected and supported by ESA's advisory teams, and ultimately its top-level Science Programme Committee. 

Underpinning these missions is the vision of the scientific teams in formulating the original scientific objectives and requirements, conducting the extensive and detailed studies which fed in to the satellite design, and preparing for the analysis of the huge and complex data stream sent down by the satellite. 
Here, I would like to acknowledge the Gaia Science Advisory Group which was formalised in 1997 and acted as scientific advisors to me in ESA until mission selection in 2000,\footnote{
The Gaia Science Advisory Group, 1997--2000, was Michael Perryman (chair), Klaas de Boer, Gerard Gilmore, Erik Høg, Mario Lattanzi, Lennart Lindegren, Xavier Luri, François Mignard, Siegfried Röser, and Tim de Zeeuw.}
and its post-2000 structure, the Gaia Science Team.\footnote{
The original Gaia Science Team, from 2000 until my own retirement from the project in 2006, was Michael Perryman (chair), Frédéric Arenou, Coryn Bailer-Jones, Ulrich Bastian, Erik H{\o}g, Andrew Holland, Carme Jordi, David Katz, Mario Lattanzi, Floor van Leeuwen, Xavier Luri, and François Mignard.
Subsequent members are listed at the 
\href{https://www.cosmos.esa.int/web/gaia/gaia-science-team}{ESA Gaia www pages}.
}
These teams were responsible for establishing the project's secure foundations: detailing the scientific case, guiding the overall satellite concept and its detailed payload design and optimisation, supporting the enabling technologies, coordinating the large-scale mission simulations and accuracy budget, and assisting in establishing the data reduction framework in the form of its constituent scientific working groups and subsequent coordination units. 
This included laying the foundations and developing the first prototypes of the critical and particularly challenging Astrometric Global Iterative Solution (AGIS), whose primary architects were Lennart Lindegren for its detailed mathematical formulation and Wil O'Mullane for its computational implementation.
Significant post-launch developments of AGIS, largely driven by the need to understand the real data, including the basic angle variations, was led by Uwe Lammers and José Hernández at ESAC, with fundamental contributions by Sergei Klioner and colleagues in Dresden.

The early days of Gaia involved formulating a set of clear scientific goals, community lobbying, establishing a feasible instrument concept, and obtaining funds for, and the lengthy development of, the enabling technologies. Regarding the latter, I would like to acknowledge the important role of the ESA study manager, Oscar Pace, who worked with me and the Gaia Study Team between 1995--2000 to establish a mission baseline concept, and a technology preparation roadmap, which formed the basis of the mission's acceptance by ESA's advisory committees in 2000.

I would like to emphasise the central role of the industrial prime contractor, Airbus Defence \& Space (formerly \mbox{Astrium}, Toulouse) and their subcontractors, under the leadership of Phase~B2/C/D industrial project manager Vincent Poinsignon, in the detailed design, manufacture, integration and testing of the Gaia satellite. The technologies at the heart of Gaia were at the `cutting edge', and the road to success was long and tortuous. Each system and subsystem represents the endpoint of remarkable technology development, innovation, construction, integration, testing, and management. 
Many individuals merit recognition, but I would like to acknowledge in particular the contribution of Frédéric Safa, who led the payload design within Astrium/Airbus between 1996--2007. His system engineering skills played a major part in converging on an instrument which satisfied the scientific, project, and financial constraints.

Post launch, flawless operation of the satellite and payload over more than 10~years was carried out by the operations team at ESA's operations centre, ESOC, led for most of the operational phase by the Spacecraft Operations Manager, David Milligan. 

A profound acknowledgement goes to the efforts and achievements of the Gaia Data Processing and Analysis Consortium, DPAC, which coordinates and executes the data processing leading to the successive data releases. This comprises nine Coordination Units and some 400 astronomers and computer scientists across Europe,\footnote{These nine Coordination Units grew out of a number of Working Groups set up around 2000. When first established in 2005, the managers and deputies of the six original coordination units, who laid the foundations for the subsequent work, were:
CU1 (System architecture): Wil O'Mullane/Uwe Lammers;
CU2 (Simulations): Xavier Luri/Carine Babusiaux \& François Mignard;
CU3 (Core processing): Ulrich Bastian/Jordi Torra \& Mario Lattanzi;
CU4 (Object/shell processing): Dimitri Pourbaix/Paolo Tanga \& Coryn Bailer-Jones;
CU5 (Photometric reductions): Floor van Leeuwen/Anthony Brown \& Carme Jordi;
CU6 (Spectroscopic reductions): David Katz/Mark Cropper \& Ulisse Munari.
The current managers of all nine Coordination Units,
\href{https://www.cosmos.esa.int/web/gaia/coordination-units}{listed here}, are\\
\phantom{xx}{\footnotesize$\bullet$} CU1 System architecture: A. Hutton (manager), F. Riclet (deputy manager)\\
\phantom{xx}{\footnotesize$\bullet$} CU2 Data simulations: C. Babusiaux (manager), E. Masana (deputy manager)\\
\phantom{xx}{\footnotesize$\bullet$} CU3 Core processing: M. Biermann (manager), M. Lattanzi (deputy manager), C. Fabricius (deputy manager)\\
\phantom{xx}{\footnotesize$\bullet$} CU4 Object processing: F. Arenou (manager, non-single stars), P. Tanga (manager, solar system objects),\\
	\phantom{xxxxxxxxxx}C. Ducourant (manager, extended objects), Laurent Galluccio (deputy manager, extended objects)\\
\phantom{xx}{\footnotesize$\bullet$} CU5 Photometric processing: D.W. Evans (manager), Francesca De Angeli (deputy manager and technical manager)\\
\phantom{xx}{\footnotesize$\bullet$} CU6 Spectroscopic reduction: P. Sartoretti (manager), G. Seabroke (deputy manager), D. Katz (deputy manager), P. Panuzzo (deputy manager)\\
\phantom{xx}{\footnotesize$\bullet$} CU7 Variability processing: L. Eyer (manager), D.W. Evans (deputy manager), M. Audard (deputy manager)\\
\phantom{xx}{\footnotesize$\bullet$} CU8 Astrophysical parameters: O. Creevey (manager) with M. Fouesneau, Y. Fremat, F. Pailler, R. Sordo, R. Andrae in the management team\\
\phantom{xx}{\footnotesize$\bullet$} CU9 Catalogue access: X. Luri (manager), J. Bakker (deputy manager), C. Fabricius (deputy manager)
}
with an Executive chaired by François Mignard from its start in 2006 until 2012, and thereafter by Anthony Brown. These large and talented teams are working together, and under considerable schedule and funding pressures, to deliver this remarkable 21st century view of our Galaxy.
In the end, these grand missions are made worthwhile by the remarkable contributions of the international scientific community. These many hundreds of scientists take the Gaia catalogues, and use their own expertise to reveal the remarkable complexities of Nature summarised in this review.

Finally, I would like to thank the commissioning editor, Professor Dong Lai (Cornell University), for the opportunity and encouragement to prepare this review. Two anonymous Gaia colleagues accepted the daunting task of reviewers, and I gratefully acknowledge their assessments and pointers.

This work has made extensive use of ADS and SIMBAD, and I would like to underline the significant contribution to research in astronomy and astrophysics made by these important and substantial resources.

\vfil\eject
\appendix

\section{Acronyms, symbols, and units}
\label{sec:app-a}

\noindent
{\it Acronyms}: this is an incomplete list of the more important acronyms used in this review:\\[-15pt]
\begin{tabbing}
\hspace*{0.5cm}	\=	\hspace*{3.5cm}	\= 	\kill
\>	ADS					\>	Astrophysics Data System \href{https://ui.adsabs.harvard.edu/}{(abstract service)} \\
\>	AGIS				\> 	Astrometric Global Iterative Solution (Section~\ref{sec:data-processing-agis})	\\
\>	BP/RP/XP				\>	Gaia's blue/red low-resolution spectra or derived photometry (Section~\ref{sec:photometry}) \\
\>	CDS					\> 	Centre de Données Astronomiques de Strasbourg/\href{https://cds.unistra.fr/}{Strasbourg Stellar Data Centre} \\ 
\>	CMB					\>	Cosmic microwave background \\
\>	DIB					\>	Diffuse interstellar bands (Section~\ref{sec:dib}) \\
\>	DPAC				\>	Data Processing and Analysis Consortium (Section~\ref{sec:data-processing})	\\
\>	DR1, DR2, (E)DR3	\>	Gaia Data Release~1, Data Release~2, (Early) Data Release~3 
									(Section~\ref{sec:data-releases}, and Table~\ref{tab:data-release-table1}) \\
\>	ESA					\>	European Space Agency \\
\>	ESAC				\>	European Space Astronomy Centre (Madrid) \\
\>	ESOC				\>	European Space Operations Centre (Darmstadt) \\ 
\>	FPR					\>	Focused Product Release, intermediate between DR3 and DR4 (Section~\ref{sec:data-releases}) \\
\>	GSE					\>	Gaia Sausage--Enceladus stream (Section~\ref{sec:gse}) \\
\>	GSPC				\>	Gaia Synthetic Photometry Catalogue (Section~\ref{sec:synthetic-photometry}) \\
\>	HR diagram			\>	Hertzsprung--Russell diagram (Section~\ref{sec:hr-diagram})	\\
\>	IAU					\>	International Astronomical Union		\\
\>	IFMR				\>	Initial-to-final mass relation for white dwarfs (Section~\ref{sec:wd-ifmr}) \\ 
\>	LSR					\>	Local Standard of Rest (Section~\ref{sec:solar-motion}) \\
\>	MOND				\>	Modified Newtonian Dynamics (Section~\ref{sec:ultra-wide-binaries}) \\
\>	NEA/NEO				\>	Near-Earth Asteroid/Object (Section~{sec:solar-system-asteroids})		\\
\>	NFW					\>	Navarro--Frenk--White (model profile for dark matter halos) \\
\>	PPN					\>	Parameterised Post-Newtonian metric (Section~\ref{sec:light-deflection}) \\
\>	RUWE				\>	Renormalised unit weight error (Gaia astrometry, Section~\ref{sec:ruwe}) \\
\>	RVS					\>	Radial Velocity Spectrometer (Gaia, Section~\ref{sec:rvs}) \\
\>	TGAS				\>	Tycho--Gaia Astrometric Solution (Section~\ref{sec:data-releases}) \\
\>	YSO					\>	Young stellar object (Section~\ref{sec:yso}) \\
\end{tabbing}

\vspace{-5pt}
\noindent
{\it Symbols}: some frequently used Gaia-specific, and other more widely used astronomical symbols, are as follows:\\[-15pt]
\begin{tabbing}
\hspace*{0.5cm}	\=	\hspace*{3.5cm}	\= 	\kill
\>	$\alpha, \delta$						\> 	right ascension, declination (position on the sky) \\ 
\>	$G_{\rm BP}$, $G_{\rm RP}$			\>	Gaia blue and red (optical) magnitudes \\
\>	$G$								\>	Gaia (broadband optical) magnitude \\
\>	$\log g$							\>	(log) stellar surface gravity	\\
\>	{[}M/H{]}							\>	stellar `metallicity', a measure of the composition of elements heavier than He	\\
\>	\teff								\>	stellar surface effective temperature (K)	\\
\>	$\mu$ ($\mu_\alpha$, $\mu_\delta$)		\>	proper motion, generally expressed in arcsec per year \\
\>	$\varpi$							\>	parallax, generally expressed in arcsec \\ 
\end{tabbing}

\vspace{-5pt}
\noindent
{\it Units}: while astronomy largely adheres to SI units, certain quantities are widely expressed in more convenient units:\\[-15pt]
\begin{tabbing}
\hspace*{0.5cm}	\=	\hspace*{3.5cm}	\= 	\kill
\>	arcsec					\>	second of arc (1/60 of 1~arcmin, or 1/3600 of 1~degree)	\\
\>	milli-arcsec (mas)			\>	$10^{-3}$~arcsec								\\
\>	micro-arcsec (\!\muas)		\>	$10^{-6}$~arcsec								\\
\>	au						\> 	astronomical unit, corresponding to the mean Sun--Earth distance ($1.496\times10^{11}$\,m) \\
\>	$M_\Sun$, $R_\Sun$		\>	solar mass ($1.99\times10^{30}$\,kg), solar radius ($6.96\times10^8$\,m)	\\
\>	$M_\Jupiter$, $R_\Jupiter$	\>	Jupiter mass ($1.90\times10^{27}$\,kg), Jupiter equatorial radius (71\,492\,km)	\\
\>	$M_\Earth$, $R_\Earth$		\>	Earth mass ($5.97\times10^{24}$\,kg), Earth equatorial radius (6378\,km)	\\
\>	parsec (pc)				\>	distance at which Earth's orbit subtends 1~arcsec ($3.09\times10^{16}$\,m = $2.06\times10^5$\,au = 3.26~l-y) \\
\end{tabbing}

\vfil\eject

\section{Transformations and other tools}
\label{sec:app-b}

The Gaia catalogues provide positions and proper motions in equatorial coordinates, in the ICRS reference system (Section~\ref{sec:reference-frame}), and at the reference epoch specified in each data release (Table~\ref{tab:data-release-table1}).
The transformation of equatorial coordinates to ecliptic or Galactic coordinates involves some defined quantities, which have been used in the context of both Hipparcos and Gaia, and which I summarise here 
\citep[see Chapter~4 of][by Lennart Lindegren]{2013asas.book.....V}.		
I also include considerations for propagating the Gaia positions at the specified reference epoch to any other epoch, and references to a few publicly available software tools for these and some other relevant applications.

\paragraph{Ecliptic coordinates}
The transformation between equatorial (ICRS) and ecliptic coordinates is given by:
\begin{equation}
[\,{\bf x}_{\rm K}~{\bf y}_{\rm K}~{\bf z}_{\rm K}\,] = [\,{\bf x~y~z}\,] \, {\bf A}_{\rm K} 
\end{equation}
where:
\begin{equation}
{\bf A}_{\rm K} =
\begin{pmatrix}
	1 & 0 & 0 \\
	0 & \cos\epsilon & -\sin\epsilon \\
	0 & \sin\epsilon & \phantom{-}\cos\epsilon
\end{pmatrix}
\end{equation}
and where
$\epsilon = 23^\circ\,26^\prime\,21.448^{\prime\prime}$ (exactly), or $23\ddeg439\,291\,111\,1\dots$, is the conventional value of the obliquity of the ecliptic, defined in the IAU 1976 system of astronomical constants
\citep{1980MitAG..48...59L}. 

\paragraph{Galactic coordinates}
The transformation between equatorial (ICRS) and Galactic coordinates is given by:
\begin{equation}
[\,{\bf x}_{\rm G}~{\bf y}_{\rm G}~{\bf z}_{\rm G}\,] = [\,{\bf x~y~z}\,] \, {\bf A}_{\rm G}
\end{equation}
where the matrix ${\bf A}_{\rm G}$ relates to the definition of the Galactic pole and Galactic centre in the ICRS system.  
The current system of Galactic longitudes and latitudes, ($l^{\rm II}, b^{\rm II}$), announced by the IAU in 1958, was defined by adopted directions to the north Galactic pole and the Galactic centre, based on physical features in the Galaxy, with respect to the B1950 coordinate reference frame
\citep{1960MNRAS.121..123B}.		
With no definition within the ICRS having been sanctioned by the IAU 
\citep[see, e.g.][]{2011A&A...536A.102L}, 
the Hipparcos catalogue proposed a definition consistent with previous materialisations of the optical reference frame 
\citep[][Vol.~1, \S1.5.3]{1997ESASP1200.....E}, 
viz.\ with the north Galactic pole defined by $\alpha_{\rm G}=192\ddeg859\,48$, $\delta_{\rm G}= +27\ddeg128\,25$ (exactly), and the origin of Galactic longitude $l_\Omega = 32\ddeg931\,92$ (exactly). 
The transformation matrix is then given, here to 15~decimals, adequate for nano-arcsec accuracy \citep[see Section~4.5 of][]{2013asas.book.....V},	
by:
\begin{equation}
{\bf A}_{\rm G} = 
\begin{pmatrix}
-0.054\,875\,560\,416\,215 & +0.494\,109\,427\,875\,584& -0.867\,666\,149\,019\,005 \\
-0.873\,437\,090\,234\,885 & -0.444\,829\,629\,960\,011& -0.198\,076\,373\,431\,202 \\
-0.483\,835\,015\,548\,713 & +0.746\,982\,244\,497\,219& +0.455\,983\,776\,175\,067
\end{pmatrix}
\end{equation}
\vspace{-5pt}

\paragraph{Propagation of positions}
When propagating source positions (and other astrometric parameters) provided in the Gaia catalogues from one epoch (typically the catalogue reference epoch, $T_0$) to another, $T$, the simplified linear expression
($\alpha=\alpha_0+(T-T_0)\,\mu_{\alpha_0}$, $\delta=\delta_0+(T-T_0)\,\mu_{\delta_0}$), 
is not a true physical description of how stars move on the sky, and will be significantly in error especially near the celestial poles, or when propagating over long time intervals.

Rigorous propagation, under the assumption of uniform rectilinear motion relative to the solar system barycentre, is described in detail by 
\citet{2014A&A...570A..62B}.	
The first level of complexity requires inclusion of the radial velocity, $v_{\rm R}$, to define the star's full space motion, and involves the transformation of all six astrometric parameters \mbox{($\alpha$, $\delta$, $\varpi$, $\mu_\alpha$, $\mu_\delta$, $v_{\rm R}$).} Details of the transformation, and of the corresponding $6\times6$ covariance matrix relevant for the error propagation, were given in the Hipparcos catalogue 
\citep[][Vol.~1, \S1.5.5]{1997ESASP1200.....E}, 
and are also provided by
\citet[][their Appendix~C]{2014A&A...570A..62B}.

More rigorous treatment of the propagation of source positions from one epoch to another should also take account of light travel-time effects, which were negligible at the Hipparcos level of accuracy and therefore ignored. Again, details are given by
\citet{2014A&A...570A..62B},	
along with the associated Jacobian matrix for the error propagation (their Appendix~B).
As they note, the light-time effects are generally very small, but exceed 0.1\,mas, or 0.1\ms, over 100\,yr for at least 33 stars in the Hipparcos catalogue. For high-velocity stars within a few tens of~pc from the Sun, these light-time effects are generally more important than the effects of their orbital curvature in the Galactic potential.

For the transformation of the proper motion components, I refer to Section~4.5 of 
\citet{2013asas.book.....V}.	

\paragraph{Python package {\tt PyGaia}}
The \href{https://github.com/agabrown/PyGaia}{{\tt PyGaia} package}
is a Python toolkit for `Gaia science performance simulation and astrometric catalogue data manipulation', by Anthony Brown. 
%
Amongst various modules, it includes transformations to both ecliptic and Galactic coordinates, using the above prescriptions.
The package also implements the epoch transformation, including the transformation of the astrometric covariance matrix, according to \citet{2014A&A...570A..62B}, and specifically the version neglecting light-time effects (their \S5.5). 


\vspace{5pt}
\paragraph{Python package {\tt astropy}}
At a more general (non-Gaia) level, the 
\href{https://www.astropy.org}{{\tt astropy} project}
is a community effort to develop a common core package for astronomy in Python
\citep{2022ApJ...935..167A}. 
Amongst a wide range of key functionality and common tools, the 
\href{https://docs.astropy.org/en/stable/coordinates}{{\tt coordinates} package}
includes tools for converting between common coordinate systems. It adopts a different definition for the transformation to Galactic coordinates than that described above, which results in a difference of up to 25~milli-arcsec between 
\href{https://github.com/astropy/astropy/issues/3344}{the two prescriptions}.

\vspace{5pt}
\paragraph{Python package {\tt galpy}}
The Python package {\tt galpy}, by Jo
\citet{2015ApJS..216...29B},	
provides numerous Galactic dynamics calculations, including 
orbit integration in various axisymmetric and non-axisymmetric Galactic potentials,
the determination of space velocities $UVW$ corrected for the peculiar motion of the Sun, 
and transformations between equatorial and Galactic coordinates (the latter uses the {\tt astropy} rather than the above Hipparcos/Gaia prescription).
The package also includes, for example, a simple model for the Galactic potential {\tt MWPotential2014}, based on a power-law density profile bulge, a Miyamoto--Nagai disk, and a dark-matter (NFW profile) halo.

\vspace{5pt}
\paragraph{Other Python tools}
A number of (generally Python) software tools have been made publicly available to access, visualise, and analyse the Gaia data.
Other than those already described in this review, these include:

\vspace{2pt}\hspace{5pt}$\bullet$
{\tt pygacs} for basic queries or Gaia table cross-matching
\citep{2022ascl.soft03005S};	

\vspace{2pt}\hspace{5pt}$\bullet$
{\tt catsHTM} to access and cross-match large astronomical catalogues including Gaia
\citep{2018PASP..130g5002S};	

\vspace{2pt}\hspace{5pt}$\bullet$
{\tt gaia\_tools} to read and cross-match catalogues, read RVS or XP spectra, and the Gaia archive
\citep{2023ascl.soft12032B};	

\vspace{2pt}\hspace{5pt}$\bullet$
{\tt GetGaia} for various Gaia catalogue searches 
\citep{2022zndo...6473476D};	

\vspace{2pt}\hspace{5pt}$\bullet$
{\tt pyia} providing various Gaia-related tools
\citep{2021ascl.soft09008P};	

\vspace{2pt}\hspace{5pt}$\bullet$
{\tt Firefly}, an interactive 3d visualisation tool with an example applied to Gaia DR3
\citep{2023ApJS..265...38G};	

\vspace{2pt}\hspace{5pt}$\bullet$
tools for planetarium applications
\citep{2018AAS...23222302I}.	

\vspace{10pt}\noindent
Others have been developed for specific data sets, including: 

\vspace{2pt}\hspace{5pt}$\bullet$
{\tt GaiaUnlimited}, to estimate the selection function for different Gaia samples
\citep{2023A&A...677A..37C};	

\vspace{2pt}\hspace{5pt}$\bullet$
{\tt vnohhf/XP\_Extinction\_Toolkit}
extinctions based on Gaia DR3 and LAMOST DR7
\citep{2024zndo..12621834Z};	

\vspace{2pt}\hspace{5pt}$\bullet$
{\tt lofti\_gaiaDR2} for determining binary star orbital elements
\citep{2020ApJ...894..115P};	

\vspace{2pt}\hspace{5pt}$\bullet$
{\tt GaiaHub}, to determine proper motions from a combination of Gaia and HST data
\citep{2022ApJ...933...76D}.	

\vfil\eject




\raggedright					


\end{document}